\documentclass[prx,aps,amssymb,twocolumn,nofootinbib,superscriptaddress]{revtex4-2}
\usepackage{amsmath}
\usepackage{amssymb}
\usepackage{amsthm}
\usepackage{amsfonts}
\usepackage{listings}
\usepackage[colorlinks, citecolor=blue]{hyperref}
\usepackage{dsfont}
\usepackage{enumerate}
\usepackage{latexsym}
\usepackage{longtable}
\usepackage[normalem]{ulem}
\usepackage{float}
\usepackage{multirow}

\usepackage{psfrag}

\usepackage{bm}
\usepackage{graphicx}
\usepackage{xcolor}

\newcommand{\beq}{\begin{equation}}
\newcommand{\eneq}{\end{equation}}

\newcommand{\bra}[1]{\left\langle#1\right|}
\newcommand{\ket}[1]{\left|#1\right\rangle}

\input{epsf}

\def\beq#1\eeq{\begin{equation}#1\end{equation}}
\def\beqs#1\eeqs{\begin{align}#1\end{align}}

\def\bra#1{\langle #1 |}
\def\ket#1{| #1 \rangle}

\newcommand{\bpm}{\begin{pmatrix}}
\newcommand{\epm}{\end{pmatrix}}

\newcommand{\bal}{\begin{align}}

\begin{document}
\title{Spin-Resolved Topology and Partial Axion Angles in Three-Dimensional Insulators}

\author{Kuan-Sen Lin}
\thanks{Corresponding author:~\href{mailto:kuansen2@illinois.edu}{kuansen2@illinois.edu}}
\affiliation{Department of Physics and Institute for Condensed Matter Theory, University of Illinois at Urbana-Champaign, Urbana, IL 61801, USA}
\affiliation{Kavli Institute for Theoretical Physics, University of California, Santa Barbara, CA 93106, USA}
\author{Giandomenico Palumbo}
\affiliation{School of Theoretical Physics, Dublin Institute for Advanced Studies, 10 Burlington Road, Dublin 4, Ireland}
\author{Zhaopeng Guo}
\affiliation{Beijing National Laboratory for Condensed Matter Physics and Institute of Physics, Chinese Academy of Sciences, Beijing 100190, China}
\affiliation{University of Chinese Academy of Sciences, Beijing 100049, China}
\author{Yoonseok Hwang}
\affiliation{Department of Physics and Institute for Condensed Matter Theory, University of Illinois at Urbana-Champaign, Urbana, IL 61801, USA}
\author{Jeremy Blackburn}
\affiliation{Department of Computer Science, State University of New York at Binghamton, Binghamton, NY 13902, USA}

\author{Daniel P.~Shoemaker}
\affiliation{Department of Materials Science and Engineering, University of Illinois at Urbana-Champaign, Urbana, IL 61801, USA}
\affiliation{Materials Research Laboratory, University of Illinois at Urbana-Champaign, Urbana, IL 61801, USA}

\author{Fahad Mahmood}
\affiliation{Materials Research Laboratory, University of Illinois at Urbana-Champaign, Urbana, IL 61801, USA}
\affiliation{Department of Physics, University of Illinois at Urbana-Champaign, Urbana, IL 61801, USA}

\author{Zhijun Wang}
\affiliation{Beijing National Laboratory for Condensed Matter Physics and Institute of Physics, Chinese Academy of Sciences, Beijing 100190, China}
\affiliation{University of Chinese Academy of Sciences, Beijing 100049, China}
\author{Gregory A.~Fiete}
\thanks{Corresponding author:~\href{mailto:g.fiete@northeastern.edu}{g.fiete@northeastern.edu}}
\affiliation{Department of Physics, Northeastern University, Boston, MA 02115, USA}
\affiliation{Department of Physics, Massachusetts Institute of Technology, Cambridge, MA 02139, USA}
\author{Benjamin J.~Wieder}
\thanks{Primary address for B. J. W.: Institut de Physique Th\'eorique, Universit\'{e} Paris-Saclay, CEA, CNRS, F-91191 Gif-sur-Yvette, France.  Corresponding author:~\href{mailto:benjamin.wieder@ipht.fr}{benjamin.wieder@ipht.fr}}
\affiliation{Department of Physics, Northeastern University, Boston, MA 02115, USA}
\affiliation{Department of Physics, Massachusetts Institute of Technology, Cambridge, MA 02139, USA}
\affiliation{Institut de Physique Th\'eorique, Universit\'{e} Paris-Saclay, CEA, CNRS, F-91191 Gif-sur-Yvette, France}
\author{Barry Bradlyn}
\thanks{Corresponding author:~\href{mailto:bbradlyn@illinois.edu}{bbradlyn@illinois.edu}}
\affiliation{Department of Physics and Institute for Condensed Matter Theory, University of Illinois at Urbana-Champaign, Urbana, IL 61801, USA}

\date{\today}

\begin{abstract}
Symmetry-protected topological crystalline insulators (TCIs) have primarily been characterized by their gapless boundary states.  
However, in time-reversal- ($\mathcal{T}$-) invariant (helical) 3D TCIs---termed higher-order TCIs (HOTIs)---the boundary signatures can manifest as a sample-dependent network of 1D hinge states.
We here introduce nested spin-resolved Wilson loops and layer constructions as tools to characterize the intrinsic bulk topological properties of spinful 3D insulators.
We discover that helical HOTIs realize one of three spin-resolved phases with distinct responses that are quantitatively robust to large deformations of the bulk spin-orbital texture: 3D quantum spin Hall insulators (QSHIs), ``spin-Weyl'' semimetals, and $\mathcal{T}$-doubled axion insulator (T-DAXI) states with nontrivial partial axion angles indicative of a 3D spin-magnetoelectric bulk response and half-quantized 2D TI surface states originating from a partial parity anomaly.  
Using ab-initio calculations, we demonstrate that $\beta$-MoTe$_2$ realizes a spin-Weyl state and that $\alpha$-BiBr hosts both 3D QSHI and T-DAXI regimes.
\end{abstract}

\maketitle

 {
\centerline{\bf Introduction}
\vspace{0.1in}
}

\begin{figure*}[t]
\centering
\includegraphics[width=.95\textwidth]{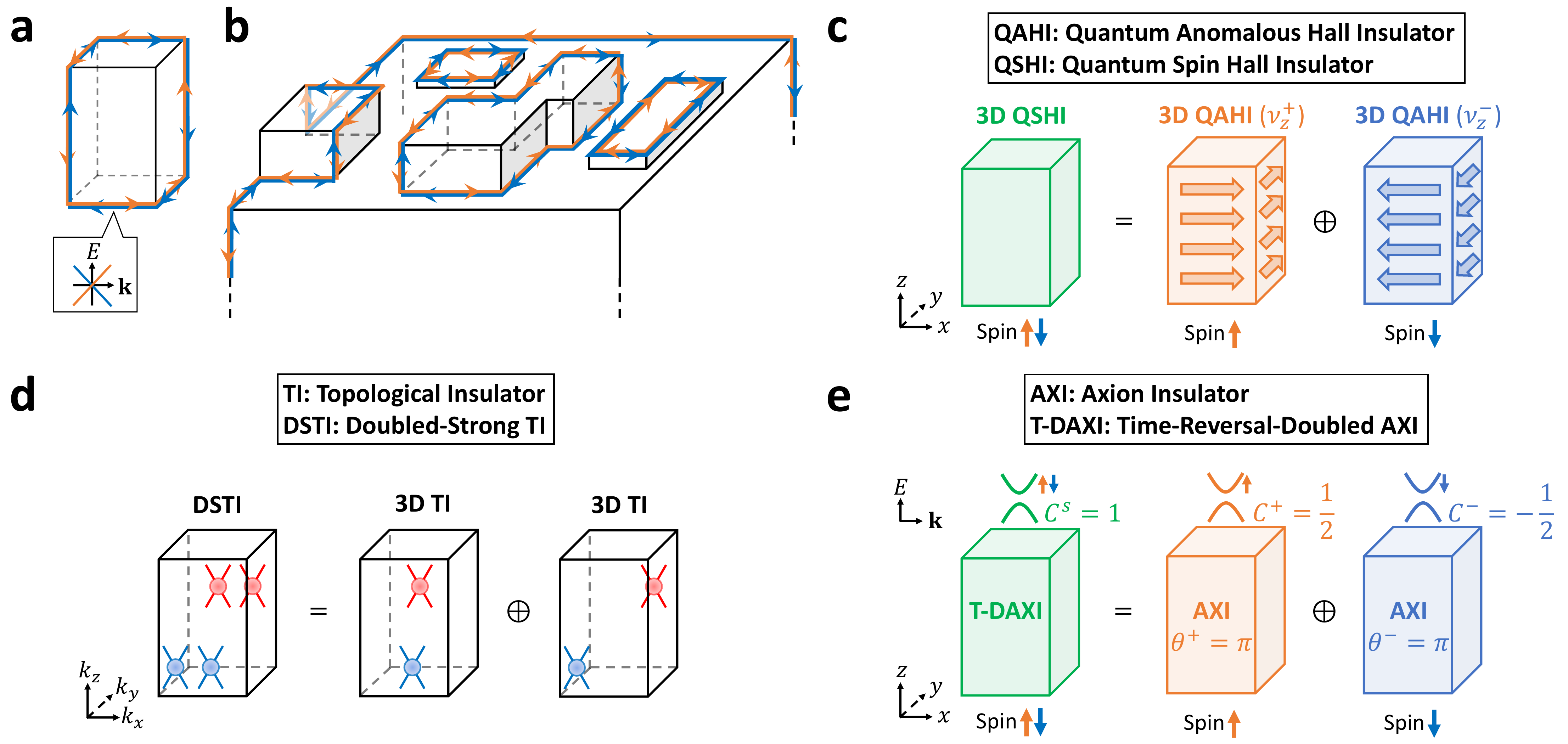}
\caption{Spin-resolving helical higher-order topological crystalline insulators.  
(a)  A helical higher-order topological crystalline insulator (HOTI) cut into a finite geometry with perfect spatial inversion ($\mathcal{I}$) symmetry. 
In (a), the configuration of intrinsic 1D helical hinge modes indicates the bulk topology~\cite{schindler2018higherorder,schindler2018higherordera,fang2019new,khalaf2018symmetry}.  
(b) A helical HOTI in a more realistic sample geometry~\cite{schindler2018higherordera} featuring irregular surfaces and broken global $\mathcal{I}$ symmetry.
The hinge modes in (b) originate from extrinsic sample details and surface physics, and are indistinguishable from the extrinsic helical modes of bulk-trivial materials~\cite{nayak2019resolving,BiBrFanHOTI}.  
By computing the gauge-invariant spin-spectrum~\cite{prodan2009robustness} (see Fig.~\ref{fig:mainPSP} and SN~2B), the bulk electronic structure of a helical HOTI with $\mathcal{I}$ and time-reversal ($\mathcal{T}$) symmetries can be further classified into one of three spin-stable phases (SN~4D).  
(c) The 3D QSHI regime of a helical HOTI, which is constructed by layering 2D TI (QSHI) states with the same spin Hall response~\cite{song2018mapping}.
The 3D QSHI state in (c) hence exhibits an extensive bulk spin Hall conductivity per unit cell (see Refs.~\cite{wang2016hourglass,SYBiBr,zhijun2021qshsquarenet} and SN~4D).
In each half of the spin spectrum, a 3D QSHI carries the topology of a 3D QAHI, as indicated by the partial weak Chern numbers $\nu_{z}^\pm$ in (c) [see SN~4C3].
(d) The spin-Weyl-semimetal [DSTI~\cite{po2017symmetry}] regime of a helical HOTI, which is constructed by superposing two 3D TIs with gapless bulk spin spectra featuring chirally-charged nodal degeneracies that we term spin-Weyl points
[red and blue circles in (d), see SN~2E].
(e) The T-DAXI regime of a helical HOTI, which is constructed by superposing time-reversed spin-polarized magnetic AXIs.
Each half of the spin spectrum in (e) is topologically equivalent to a magnetic AXI with an $\mathcal{I}$-quantized partial axion angle $\theta^{\pm}=\pi$, implying a topological bulk spin-magnetoelectric response.  
The gapped surfaces of T-DAXIs bind anomalous halves of $\mathcal{T}$-invariant 2D TIs with odd spin Chern numbers $C^{s}$---formed from summing anomalous surface states with half-integer partial Chern numbers $C^{\pm}$ and $\mathcal{T}$-related spin-orbital textures---as a consequence of a novel 2D surface partial parity anomaly (SN~5E).}
\label{fig:mainHOTI}
\end{figure*}

In recent years, the study of topological phases of matter in solid-state materials has largely focused on their anomalous gapless boundary states~\cite{wieder2021topological}.  2D and 3D topological insulators (TIs), for example, exhibit time-reversal- ($\mathcal{T}$-) symmetry-protected 1D helical modes and 2D Dirac cones on their boundaries~\cite{kane2005quantum,kane2005z,bernevig2006quantum,fu2007topologicala,fu2007topological}, respectively.  
While this focus on gapless boundary states has been validated by remarkable transport and spectroscopy experiments~\cite{konig2007quantum,BiSbScienceKane,ARPESreview}, and has revealed promising avenues for chemical applications~\cite{ClaudiaChiralCatalysis,ClaudiaCatalysisDB} and interface spintronics~\cite{SurfaceSpintronic1,SurfaceSpintronic2}, it also has drawbacks.

In particular, in 3D symmetry-protected topological crystalline insulators (TCIs), gapless boundary states only appear on 2D surfaces that preserve specific crystal symmetries, and the surface states on the remaining crystal facets are generically gapped~\cite{fu2011topological,hsieh2012topological,wang2016hourglass,wieder2018wallpaper}.  
The limitations of anomalous gapless boundary states as experimental signatures of bulk topological phases are further compounded in the class of 3D TCIs that have become known as higher-order TCIs (HOTIs), in which most---if not all---of the 2D surface states are gapped, and the 1D hinges (edges) between gapped surfaces bind gapless chiral or helical modes~\cite{benalcazar2017electric,schindler2018higherorder,schindler2018higherordera,fang2019new,khalaf2018symmetry}.  
In HOTIs, the specific configuration of hinge states can provide an indicator of the bulk topology, but only in system geometries with unrealistically high symmetry (\emph{i.e.} where the entire crystallite exhibits perfect point group symmetry)~\cite{wieder2018axion,WiederDefect} [see Fig.~\ref{fig:mainHOTI}(a,b)].  
For the subset of $\mathcal{T}$-broken HOTI phases with chiral hinge modes and relevant SOC, this issue has been overcome by recognizing that spinful chiral HOTIs are magnetic axion insulators (AXIs)~\cite{schindler2018higherorder,elcoro2021magnetic,wieder2018axion,varnava2020axion}.  
Magnetic AXIs, like 3D TIs, are characterized by a bulk topological axion angle $\theta=\pi$ (where $\theta$ is defined modulo $2\pi$)~\cite{wilczek1987two,qi2008topological,essin2009magnetoelectric,IvoAXI1}, leading to a quantized ${\bf E}\cdot {\bf B}$ bulk magnetoelectric response and anomalous quantum Hall states with half-integer Chern numbers on gapped surfaces as a consequence of the 2D surface parity anomaly.  
Importantly, the quantized bulk magnetoelectric response of AXIs can be experimentally measured without invoking gapless boundary states.  
For example, the quantized value $\theta=\pi$ was measured (in the units of the fine-structure constant) through optical experiments performed on 3D TIs with magnetically gapped surface Dirac cones~\cite{wu2016quantized,MogiOpticalAxionParity}.

Using the theoretical methods of Topological Quantum Chemistry~\cite{bradlyn2017topological,elcoro2021magnetic,gao2022magnetic} and symmetry-based indicators (SIs)~\cite{kruthoff2017topological,po2017symmetry,song2018mapping}, researchers have performed high-throughput~\cite{vergniory2019high,zhang2019catalogue,AshvinMaterials1,tang2019efficient,xu2020highthroughput} and exhaustive~\cite{AndreiMaterials2} searches for 3D topological materials, yielding thousands of candidate TIs and TCIs.  
These computational investigations have revealed candidate helical HOTI phases in readily accessible materials, including rhombohedral bismuth~\cite{schindler2018higherordera}, $\alpha$-BiBr~\cite{SYBiBr,BiBrFanHOTI}, and the transition-metal dichalcogenides MoTe$_2$ and WTe$_2$~\cite{wang2019higherorder}, in turn motivating experimental efforts to observe nontrivial topology in these materials~\cite{huang2019polar,wang2020evidence,choi2020evidence,noguchi2021evidence,FanZahidRoomTempBiBrExp}.  
However in HOTIs, the bulk topological spectral flow cannot be inferred by observing gapless surface states in photoemission experiments, because the surface states are gapped on most (if not all) 2D surfaces of 3D HOTI phases~\cite{benalcazar2017electric,schindler2018higherorder,schindler2018higherordera,fang2019new,khalaf2018symmetry}.  
Experimental investigations of helical HOTIs have therefore instead largely focused on scanning-tunneling microscopy (STM) and ballistic supercurrent signatures of 1D helical ``hinge'' channels.  
Unfortunately, the configuration of hinge states in a given material is highly dependent on sample details [Fig.~\ref{fig:mainHOTI}(b)], and hinge-state-like 1D gapless modes can also originate from crystal defects, or even manifest in materials that are topologically trivial in the bulk~\cite{nayak2019resolving,BiBrFanHOTI}.  
To better understand the existing experimental data and provide a road map for real-world applications of the topological materials identified in Refs.~\cite{vergniory2019high,zhang2019catalogue,AshvinMaterials1,tang2019efficient,xu2020highthroughput,AndreiMaterials2}, it is crucial to elucidate the geometry-independent bulk signatures of newly discovered helical TCI and HOTI phases, analogous to the characterization of chiral HOTIs as AXIs.  
Bulk topological invariants like the axion angle $\theta$ are typically robust to boundary details and disorder~\cite{ChalkerCoddington,ZhidaAXIDisorder}, and hence predictive under more realistic material conditions.

In this work we perform extensive numerical investigations to, for the first time, unravel the bulk and surface theories of helical HOTIs, and their connection to boundary-insensitive physical observables.  
Unlike 2D Chern insulators and 3D TIs, we find that the electromagnetic response of helical HOTIs does not depend solely on the electronic band topology, but also depends on additional details of the spin-orbital texture of the occupied bands.  
We start from the projected spin operator $PsP$ introduced by Prodan in Ref.~\cite{prodan2009robustness}, where $P$ projects onto a set of occupied bands, and $s$ is a choice of spin direction. 
Building on Ref.~\cite{prodan2009robustness} and the crystallographic splitting theorem of Ref.~\cite{alexandradinata2020crystallographic}, we show that topologically nontrivial $\mathcal{T}$-invariant insulators with relevant spin-orbit coupling (SOC) have topologically nontrivial spin-resolved bands.
We theoretically introduce and numerically implement a gauge-invariant (nested) Wilson loop (non-Abelian Berry phase) method~\cite{fidkowski2011model,wang2016hourglass,wieder2018wallpaper,wieder2018axion} for computing crystal-symmetry-protected, spin-resolved band topology in $\mathcal{T}$-invariant insulators [see Sections~3 and~4 of the Supplementary Notes (SN~3 and~4)].  
The extensive toy-model and real-material spin-resolved and nested Wilson loop calculations in this work (see SN~3,~4,~5,~8,~9, and~10) were performed using the freely accessible Python package~\href{https://github.com/kuansenlin/nested_and_spin_resolved_Wilson_loop}{\textsc{nested\_and\_spin\_resolved\_Wilson\_loop}}~\cite{lin2023nestedWilsonLib}, which was previously implemented and utilized for the preparation of Refs.~\cite{wieder2018axion,wieder2020strong}, and was then greatly refined and extended to spin-resolved calculations for the present work.
In SN~2C, we crucially demonstrate that gaps in the spectrum of $PsP$, termed the spin spectrum, are perturbatively robust to deformations of the spin-orbital texture of the occupied bands, leading to a controlled notion of ``spin-stable'' band topology in which spin-resolved bulk topological invariants remain quantized under symmetric deformations that neither close an energy gap nor a spin gap in the $PsP$ spectrum.  
To provide position-space physical intuition for our spin-resolved Wilson loop calculations, we introduce a spin-resolved layer construction method~\cite{song2018mapping,elcoro2021magnetic,gao2022magnetic} for enumerating and classifying 3D symmetry-protected, spin-gapped topological states.  
In SN~3H, we also introduce a formulation of a spin-resolved entanglement spectrum, which we prove to be homotopic to the spectrum of the spin-resolved Wilson loop.

\begin{figure*}[t]
\centering
\includegraphics[width=.9\textwidth]{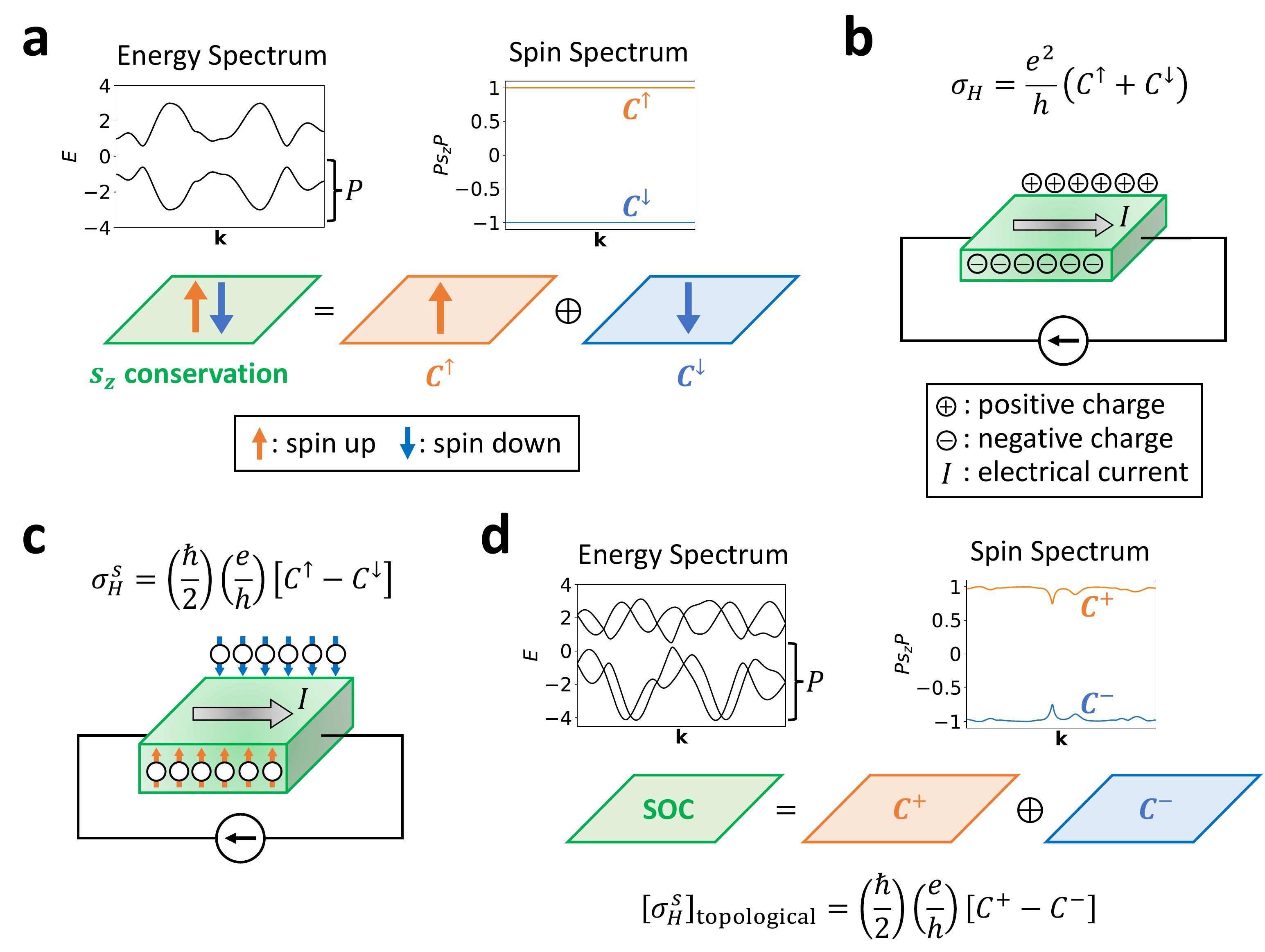}
\caption{Spin-resolved band topology.  
(a) A 2D insulator with strong $s_{z}$-preserving (\emph{e.g.} ``Ising''~\cite{wang2021isingSC} or ``Kane-Mele-like''~\cite{kane2005quantum}) spin-orbit coupling (SOC).
In (a), separate Chern numbers $C^{\uparrow,\downarrow}$ can be defined for the $s_{z}=\uparrow,\downarrow$ occupied states.  
The sum $C^{\uparrow} + C^{\downarrow}$ indicates the topological coefficient of (b) the Hall response [Eq.~(\ref{eq:HallConductivity})], whereas the difference $C^{\uparrow} - C^{\downarrow}$ indicates the topological coefficient of (c) the spin Hall response [Eq.~(\ref{eq:spinHallConductivity})]~\cite{sinova2015spin,AndreiSpinHallPRL,MurakamiSpinHall}.  
(d) A 2D insulator with $s_{z}$-breaking (\emph{e.g.} Rashba~\cite{kane2005z}) SOC.
Though the $s_{z}$ spin Hall conductivity is no longer quantized in (d), the existence of a topological contribution $[\sigma^{s}_{H}]_{\mathrm{topological}}$ to the (non-quantized) bulk $s_{z}$ spin Hall response can still be inferred from the quantized partial Chern numbers $C^{\pm}$ of spectrally isolated groupings of bands in the spin spectrum of the matrix $PsP$ with $s=s_{z}$~\cite{prodan2009robustness}  [see Eq.~(\ref{eq:PsP}) and SN~3C].  
Crucially, perturbative deformations to the system correspond to perturbative deformations of the spin spectrum (SN~2C).  
This facilitates introducing a finer notion of spin-stable topological phases in which the spin-resolved band topology of the $PsP$ spectrum indicates the existence of bulk topological contributions to (non-quantized) spin-electromagnetic response effects, which cannot be removed without closing gaps in the energy or spin spectra.  
For example, because the $s_{z}$-nonconserving $Ps_{z}P$ spin spectrum in (d) is adiabatically related to the $s_{z}$-conserving $Ps_{z}P$ spectrum in (a) without closing an energy or spin gap, then $C^{\uparrow}=C^{+}$ and $C^{\downarrow}=C^{-}$.} 
\label{fig:mainPSP}
\end{figure*}

Through our numerical calculations, we find that helical HOTIs necessarily fall into one of three regimes of spin-stable topology, each of which is characterized by a distinct spin-electromagnetic response.  
Previous studies have recognized that helical TCI phases may realize layered quantum spin Hall states in which each unit cell contributes a nonzero spin Hall conductivity~\cite{wang2016hourglass,SYBiBr,zhijun2021qshsquarenet}, and we accordingly find that some helical HOTIs for particular spin resolution directions $s$ [such as $\alpha$-BiBr for $s_{z}$ spins, see SN~10B] realize these 3D quantum spin Hall insulator (QSHI) states [Fig.~\ref{fig:mainHOTI}(c)].
However, we also discover two additional spin-stable regimes of $\mathcal{I}$- and $\mathcal{T}$-protected helical HOTIs, which are both physically distinguishable from each other, and from 3D QSHIs (SN~4D).

First, we find that helical HOTIs may exhibit (for some or all spin resolution directions) a gapless $PsP$ spectrum in which the spin-gap closing points form ``spin-Weyl fermions'' that act as monopoles of spin-resolved partial Berry curvature~\cite{wan2011topological}, a quantity that derives from the partial polarization (Berry phase) introduced by Fu and Kane in Ref.~\cite{fu2006time} [see Fig.~\ref{fig:mainHOTI}(d) and SN~2E,~3E, and~3F].  
In SN~3E and~3F, we show that all 3D TI phases realize spin-resolved spin-Weyl states, and in SN~2G, we demonstrate that spin-Weyl points in the $PsP$ spectrum can be converted into Weyl points in the energy spectrum by a strong Zeeman field, leading to the presence of topological Fermi-arc surface states.  
Through ab-initio calculations, we demonstrate in SN~9 that for all choices of $s$ in $PsP$, the candidate HOTI $\beta$-MoTe$_2$ realizes a spin-Weyl semimetal state.  
We further show that in the representative case of $s\propto s_{x} + s_{z}$, the spin gap of $\beta$-MoTe$_2$ closes at only 8 spin-Weyl points, which give rise to Fermi arcs on the experimentally accessible $(001)$-surface under a strong $(\hat{\mathbf{x}}+\hat{\mathbf{z}})$-directed Zeeman field (see Ref.~\cite{wang2019higherorder} and SN~9C).

Most intriguingly, we discover a final regime for helical HOTIs in which the bands within each sector of $PsP$ exhibit the same topology as a magnetic AXI.  
By applying nested Wilson-loop and hybrid-Wannier methods for computing $\theta$~\cite{wieder2018axion,varnava2020axion} to the spin spectrum of helical HOTIs, we specifically uncover the existence of a previously unrecognized $\mathcal{T}$-doubled AXI (T-DAXI) state characterized by bulk nontrivial partial axion angles $\theta^\pm=\pi$ [Fig.~\ref{fig:mainHOTI}(e) and SN~4D and~4E].  
We implement a spin-resolved variant of the local Chern marker~\cite{kitaev2006anyons,bianco2011mapping} to demonstrate that the gapped surfaces of T-DAXIs bind anomalous halves of 2D TI states as a consequence of a novel $\mathcal{T}$-invariant partial parity anomaly (SN~5E).   
The T-DAXI regime hence provides the first theoretical description of a helical HOTI that is free from the requirement of perfect global crystal symmetry: in T-DAXIs,  $\mathcal{I}$ symmetry pins $\theta^{\pm}=\pi$ deep in the bulk leading to a topological contribution to the bulk spin-magnetoelectric response, and 1D helical modes appear on surface (and hinge) domain walls between gapped facets hosting topologically distinct halves of 2D TI states.  
Crucially, while the spin-Weyl semimetal, QSHI, and T-DAXI spin-resolutions of a helical HOTI can be deformed into each other by closing a spin gap without closing an energy gap, we will show below that they cannot be deformed into insulators with trivial spin-stable topology without closing an energy gap.

Lastly, we remarkably discover that the T-DAXI state---as well as the aforementioned 3D QSHI state---are both realized in the quasi-1D candidate HOTI $\alpha$-BiBr.  
Through ab-initio calculations detailed in SN~10, we specifically find that $\alpha$-BiBr hosts a spin gap for nearly all spin resolution directions, which interpolates between a wide 3D QSHI regime centered around $s_{z}$ and a narrower (but still significant and numerically stable, see SN~10B) T-DAXI regime centered around $s_{x}$.  
To provide physical signatures of the spin-gapped states in $\alpha$-BiBr, we then in SN~10C use a Wannier-based tight-binding model to compute the bulk intrinsic contribution to the (non-quantized) spin Hall conductivity of $\alpha$-BiBr (per unit cell) within its 3D QSHI and T-DAXI regimes.  
Our calculations reveal a highly anisotropic bulk spin Hall response in $\alpha$-BiBr that is nearly quantized within the 3D QSHI regime ($s_{z}$ spins) and nearly vanishing within the T-DAXI regime ($s_{x}$ spins), in close agreement with the bulk spin-resolved topology (partial Chern numbers).

{
\vspace{0.14in}
\centerline{\bf Results}
\vspace{0.08in}
}

{\bf The Spin Spectrum and Spin-Stable Topology}: Spin-resolved band topology and its relationship to spin-electromagnetic response effects can most straightforwardly be understood in 2D insulators.  
To begin, we consider a 2D spinful (fermionic), noninteracting insulator lying in the $xy$-plane with $s=s_{z}$ spin-rotation symmetry (\emph{i.e.} U$(1)$ spin symmetry, in addition to the U$(1)$ charge conservation symmetry~\cite{XGWenZooReview}).  
We emphasize that the simultaneous requirements of charge and $s_{z}$ spin conservation symmetries do not require SOC to vanish, or to even be small.
Instead, $s_{z}$ symmetry only enforces that $s_{z}$-nonconserving (\emph{e.g.} Rashba~\cite{kane2005z}) contributions to the SOC vanish, whereas $s_{z}$-conserving  (\emph{e.g.} ``Ising''~\cite{wang2021isingSC} or ``Kane-Mele-like''~\cite{kane2005quantum}) SOC may be arbitrarily large.
In the band structure of the 2D insulator, each occupied Bloch eigenstate is an eigenstate of $s_{z}$, allowing separate Berry connections, curvatures, and Chern numbers $C^{\uparrow,\downarrow}$ to be defined for the occupied states [Fig.~\ref{fig:mainPSP}(a)].  
The total (charge) Hall conductivity, which characterizes the transverse voltage generated under an applied current [Fig.~\ref{fig:mainPSP}(b)], is given by:
\begin{equation}
\sigma_{H} = \frac{e^{2}}{h}C,
\label{eq:HallConductivity}
\end{equation}
where $C$ is the total Chern number:
\begin{equation}
C = C^{\uparrow} + C^{\downarrow}.
\label{eq:ChernNumber}
\end{equation}

Similarly, for each spin direction $s$, a separate spin Hall conductivity $\sigma^{s}_{H}$~\cite{sinova2015spin} can be defined to characterize the transverse $s$-polarized spin separation generated under an applied current [Fig.~\ref{fig:mainPSP}(c)]. 
For the above example of an insulator with $s_{z}$-conservation symmetry, the $s=s_{z}$ spin Hall conductivity is given by:
\begin{equation}
\sigma^{s}_{H} = \left(\frac{\hbar}{2}\right)\left(\frac{e}{h}\right)\left[ C^{\uparrow} - C^{\downarrow}\right],
\label{eq:spinHallConductivity}
\end{equation}
motivating the definition of a spin Chern number:
\begin{equation}
C^{s} = C^{\uparrow} - C^{\downarrow}.
\label{eq:sZspinChern}
\end{equation}
2D insulators with $C\neq 0$ are termed quantum Hall states~\cite{haldane1988model}, and insulators with $C=0$, $C^{s}\neq 0$ are termed quantum spin Hall states~\cite{AndreiSpinHallPRL,MurakamiSpinHall}.

As crucially emphasized by Kane and Mele~\cite{kane2005quantum,kane2005z}, $s_{z}$ symmetry is typically broken in real materials by the presence of multiple microscopic (\emph{e.g.} simultaneous Ising and Rashba) contributions to SOC, because crystal and $\mathcal{T}$ symmetries alone cannot enforce a U$(1)$ spin conservation symmetry, such as $s_{z}$.
Without $s_{z}$ symmetry, the spin Hall conductivity is no longer quantized and cannot be computed through Eq.~(\ref{eq:spinHallConductivity}), because states can no longer by labeled by the spin eigenvalues $s_{z}=\uparrow,\downarrow$.  
However, it is known that as perturbatively weak $s_{z}$-nonconserving SOC is introduced, the intrinsic bulk contribution to the spin Hall response does not instantaneously vanish, but instead remains perturbatively close to the value given by Eq.~(\ref{eq:spinHallConductivity})~\cite{shi2006proper}.

To deduce the existence of a bulk topological contribution to $\sigma^{s}_{H}$ for each spin direction $s = {\bf s}\cdot \hat{\bf n}$, Prodan introduced the projected spin operator $PsP$~\cite{prodan2009robustness}, which in this work represents a shorthand expression for the matrix:
\begin{equation}
PsP \equiv P({\bf k})\left({\bf s}\cdot \hat{\bf n}\right)P({\bf k}),
\label{eq:PsP}
\end{equation}
where $P({\bf k})$ is the projector onto an energetically isolated (typically occupied) set of electronic states at the crystal momentum ${\bf k}$ (see SN~2B).  
The eigenvalues of $PsP$ are gauge invariant and as functions of ${\bf k}$ form the spin spectrum, a physically meaningful characterization of the occupied states that is complementary to the electronic structure [Fig.~\ref{fig:mainPSP}(d)].  
When a given $s$ is a conserved symmetry (whether or not $s$-preserving SOC is present), the eigenvalues of $PsP$ in the occupied subspace are pinned to $\pm 1$, and in insulators with compensated numbers of $s=\uparrow,\downarrow$ electrons and negligible $s$-nonconserving SOC, the eigenvalues of $PsP$ are separated at all ${\bf k}$ points by a spin gap $\Delta_{s}\approx 2$.  
Importantly, as $s$-nonconserving SOC is introduced and $s$-rotation [\emph{e.g.} $s_{z}$] symmetry relaxed, the eigenvalues of $PsP$ do not fluctuate wildly, but instead, as shown in SN~2C, perturbatively deviate from $\pm 1$.  
This can be contrasted with a similar quantity, the non-Abelian Wilson loop (holonomy) matrix computed in the direction of a reciprocal lattice vector ${\bf G}$~\cite{fidkowski2011model,wang2016hourglass,wieder2018wallpaper,wieder2018axion}:
\begin{equation}
\mathcal{W}_{1,{\bf k},{\bf G}} = \prod^{{\bf k} + {\bf G}\leftarrow{\bf k}}_{\bf q} P({\bf q}),
\label{eq:WilsonLoop}
\end{equation}
for which the (hybrid Wannier) spectrum need not adiabatically change under small perturbations to the system Hamiltonian, because $\mathcal{W}_{1,{\bf k},{\bf G}}$ is non-local in ${\bf k}$ (see SN~2C).

\begin{figure*}[t]
\centering
\includegraphics[width=.9\textwidth]{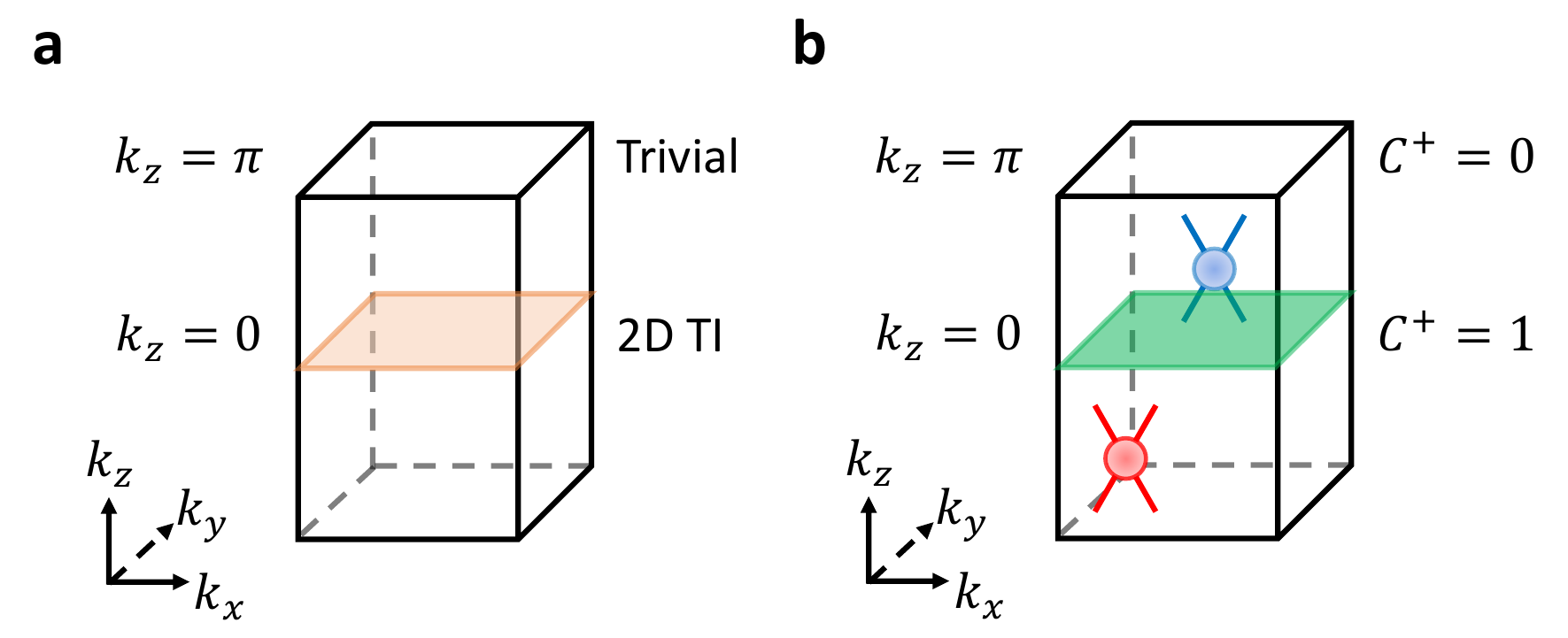}
\caption{Spin-Weyl fermions in spin-resolved 3D topological insulators.  
(a) A $\mathcal{T}$-invariant 3D strong topological insulator (TI) in crystal momentum (${\bf k}$) space.
The 3D TI in (a) can be re-expressed as a helical Thouless pump between a 2D TI (orange plane) and a trivial insulator~\cite{fu2007topologicala,fu2007topological,qi2008topological,wieder2018axion}.  
Because a 3D TI is a strong, isotropic topological phase, then we may choose the pumping parameter in (a) to be $k_{z}$ without loss of generality.  
(b) The $PsP$ spin spectrum [Eq.~(\ref{eq:PsP})] of the 3D TI in (a) is gapless for all choices of $s = {\bf s}\cdot \hat{\bf n}$ (\emph{e.g.} $s_{z}$).
In each half of the 3D Brillouin zone (BZ) in (b), the spin spectrum specifically exhibits Weyl-like~\cite{wan2011topological} nodal degeneracies with a net-odd partial chiral charge, which we term ``spin-Weyl fermions'' (see SN~2E,~3E, and~3F).
In (b), we show the simplest schematic example of a spin-resolved 3D TI [spin-Weyl state] with one positively (red) and one negatively (blue) charged spin-Weyl point in each half of the 3D BZ.
The green plane in (b) indicates that the positive $PsP$ bands at $k_{z}=0$ carry a nontrivial odd partial Chern number $C^{+}=1$ [originating from spin-resolving the 2D TI bands at $k_{z}=0$ in (a)], which stands in contrast to the trivial partial Chern number $C^{+}=0$ at $k_{z}=\pi$.
The $\Delta C^{+}\text{ mod }2=1$ difference in partial Chern numbers between $k_{z}=0,\pi$ in (b), combined with the continued validity of the $PsP$ calculation in BZ planes without $\mathcal{T}$ symmetry away from $k_{z}=0,\pi$~\cite{prodan2009robustness}, indicates the presence of an odd number of integer-charge spin-Weyl fermions per half BZ.
Spin-Weyl states like those in 3D TIs exhibit topological surface Fermi arcs under strong Zeeman fields (SN~2G and~9C), and display arc-like states along the entanglement cut in the spin-resolved entanglement spectrum (SN~3H).  
Further theoretical and numerical calculations demonstrating unremovable spin-Weyl points in 3D TIs are provided in SN~2E,~3E, and~3F.}
\label{fig:mainTISpinWeyl}
\end{figure*}

It was previously recognized in Ref.~\cite{prodan2009robustness} that even in a system without $s_{z}$ symmetry, the Chern numbers of spectrally isolated bands in the $Ps_{z}P$ spin spectrum remain gauge-invariant quantities, and can be numerically computed.  
In particular, if there exists a spin gap between the spin bands with $Ps_{z}P$ eigenvalues closer to $\pm 1$ [Fig.~\ref{fig:mainPSP}(d)], then one may compute the gauge-invariant, spin-resolved partial Chern numbers $C^{\pm}$ of the bands within each half of the spin spectrum.  
The partial Chern numbers $C^{\pm}$ importantly allow one to define the $s_{z}$ spin Chern number~\cite{sheng2006spinChern}---even when $s_{z}$ is no longer conserved---by generalizing Eq.~(\ref{eq:sZspinChern}) to the spin spectrum band topology: 
\begin{equation}
C^{s} = C^{+} - C^{-}.
\label{eq:generalSpinChern}
\end{equation}
Though Eq.~(\ref{eq:generalSpinChern}) no longer indicates the coefficient of a quantized spin Hall response away from the limit of $s_{z}$ symmetry, $C^{s}\neq 0$ still indicates the existence of a bulk topological contribution to $\sigma^{s}_{H}$ (for $s_{z}$ spins). 
Furthermore, the intrinsic bulk spin Hall conductivity $\sigma^{s}_{H}$ may even lie close to the quantized topological value given by Eq.~(\ref{eq:spinHallConductivity}) if $s_{z}$-breaking SOC is relatively weak (see SN~3C and 7B).

In an isolated 2D insulator with spinful $\mathcal{T}$ symmetry, $C^{+}=-C^{-}$, such that $C=0$, and $C^{s}\text{ mod }2=0$ for all choices of spin direction $s$.  
While $C^{s}$ can be changed by $2$ through spin band inversions between the upper and lower spin bands at a single ${\bf k}$ point in the spin spectrum, $\mathcal{T}$ symmetry enforces that spin band inversions occur in pairs at $\pm {\bf k}$ or in crossings with quadratic dispersion at time-reversal-invariant ${\bf k}$ (TRIM) points, such that $C^{s}\text{ mod }4$ cannot change without closing a gap in the energy spectrum (see SN~3C and Ref.~\cite{prodan2009robustness}).  
The spin spectrum hence also facilitates an alternative definition of the 2D $\mathbb{Z}_{2}$ invariant in $\mathcal{T}$-symmetric insulators:
\begin{equation}
z_{2} = \left(\frac{C^{s}}{2}\right)\text{ mod }2,
\label{eq:Z2for2D}
\end{equation}
for all $s$ for which $PsP$ exhibits a spin gap such that $C^{\pm}$ (and hence $C^{s}$) are well defined.
Eq.~(\ref{eq:Z2for2D}) is consistent with the crystallographic splitting theorem of Ref.~\cite{alexandradinata2020crystallographic}, and further implies that $C^{s}$ can still be nonzero in a $\mathcal{T}$-invariant insulator with $z_{2}=0$.  
More generally, in SN~3G we show that a ``fragile'' TCI, which has a less robust form of topology than stable topological phases like 2D $\mathbb{Z}_{2}$ TIs~\cite{po2018fragile,wieder2018axion}, can still carry a nonzero $C^{s}$, and hence have a nonvanishing bulk topological contribution to its (non-quantized) spin Hall response.  
Our calculations in SN~4E also imply that spin bands in the $PsP$ spectrum can exhibit a novel form of spin-resolved fragile topology.  
Lastly, unlike the 2D $\mathbb{Z}_{2}$ invariant~\cite{kane2005z}, $C^{s}$ remains well-defined when $\mathcal{T}$ symmetry is broken (SN~3C).  
We will shortly exploit the robustness of $C^{s}$ under $\mathcal{T}$-breaking potentials to analyze the topology of 3D insulators, in which ${\bf k}$-space surfaces away from TRIM points can be treated as 2D systems with broken $\mathcal{T}$ symmetry~\cite{wieder2020strong}.

In this work, we more generally recognize the partial Chern numbers $C^{\pm}$ to be members of a larger class of spin-resolved topological invariants that are stable to deformations that close neither an energy gap nor a spin gap.  
Given a 3D insulator that respects the symmetries of a nonmagnetic space group $G$, the spin bands specifically respect the symmetries of, and can carry topological invariants protected by, the magnetic space subgroup $M\subset G$ for which each element $m\in M$ commutes with the spin operator $s$ in $PsP$.  
Building off of tremendous recent progress enumerating SIs and Wilson-loop indicators for spinful magnetic topological phases~\cite{elcoro2021magnetic,gao2022magnetic,wieder2018axion}, we resolve the spin-stable topology of 3D TIs and helical HOTIs by applying the existing magnetic topological classification to the spin bands of $PsP$.  
To compute spin-resolved topological invariants, we theoretically introduce and numerically implement spin-resolved generalizations of the (nested) Wilson loop matrix [Eq.~(\ref{eq:WilsonLoop}), see SN~3B and~4B, as well as Ref.~\cite{lin2023nestedWilsonLib}].   
We further introduce in SN~3H a spin-resolved generalization of the entanglement spectrum~\cite{fidkowski2011model}, which we show to be homotopic to the spin-resolved Wilson spectrum.  
Using $PsP$ and spin-resolved Wilson loops, we discover several previously unrecognized, experimentally detectable features of well-studied 3D insulators, including spin-Weyl points in 3D TIs and nontrivial partial axion angles in helical HOTIs, which we will explore in detail below.

{\bf Spin-Weyl Fermions in 3D TIs}: 3D TIs have previously been linked to spin-orbital textures through their anomalous Dirac-cone surface states.  
Previous experimental investigations have specifically shown that the surface states of 3D TIs exhibit helical spin textures~\cite{BiSbScienceKane,ARPESreview} and can efficiently convert charge current to magnetic spin torque~\cite{SurfaceSpintronic1,SurfaceSpintronic2}.  
In this work, we find that 3D TIs additionally exhibit unremovable bulk spin textures, which are revealed by analyzing the connectivity and topology of the spin bands in $PsP$.  
We specifically find that 3D TIs must carry gapless spin spectra for all choices of $s$ in $PsP$.  
As we will show below, absent additional symmetries, the $PsP$ spectrum of a 3D TI generically exhibits an odd number of Weyl-fermion-like touching points between the $\pm$-sector spin bands in each half of the 3D BZ, where each 3D nodal point acts as a source or sink of partial Berry curvature (SN~2E,~3E, and~3F).

To see that 3D TIs for all $s$ must exhibit gapless $PsP$ spectra featuring nodal degeneracies with nontrivial chiral charge---which we term ``spin-Weyl'' points---we first note that the momentum-space band structure of a 3D TI can be re-expressed as a helical Thouless pump of a 2D TI~\cite{fu2007topologicala,fu2007topological,qi2008topological,wieder2018axion}.  
Taking $k_{z}$ to be equivalent to the Thouless pumping parameter, the occupied bands in one $\mathcal{T}$-invariant BZ plane must be equivalent to a 2D TI [$k_{z}=0$ in Fig.~\ref{fig:mainTISpinWeyl}(a)], and must be equivalent to a 2D trivial insulator in the other $\mathcal{T}$-invariant, $k_{z}$-indexed BZ plane [$k_{z}=\pi$ in Fig.~\ref{fig:mainTISpinWeyl}(a)].  
Through Eqs.~(\ref{eq:generalSpinChern}) and~(\ref{eq:Z2for2D}) and the constraint from $\mathcal{T}$ symmetry that $C^{+}=-C^{-}$ (SN~3C), this implies that for the occupied bands of the 3D TI in Fig.~\ref{fig:mainTISpinWeyl}:
\begin{equation}
C^{\pm}\text{ mod }2=1\text{ at }k_{z}=0,
\label{eq:kz0Weyl}
\end{equation}
and:
\begin{equation}
C^{\pm}\text{ mod }2=0\text{ at }k_{z}=\pi,
\label{eq:kzPiWeyl}
\end{equation}
for all choices of $s$ in $PsP$.

Crucially, unlike the $\mathbb{Z}_{2}$ invariant for 2D TIs, the partial Chern numbers $C^{\pm}$ remain well-defined when $\mathcal{T}$ is broken in 2D BZ planes away from $k_{z}=0,\pi$~\cite{prodan2009robustness}.  
Eqs.~(\ref{eq:kz0Weyl}) and~(\ref{eq:kzPiWeyl}) hence imply that $C^{\pm}$ must each change by odd numbers across each half of the 3D BZ, which can only occur if the $\pm$-sector spin bands in the spin spectrum meet in nodal degeneracies with nontrivial (partial) chiral charges [Fig.~\ref{fig:mainTISpinWeyl}(b)].  
Absent additional symmetries, nodal degeneracies with nontrivial chiral charge manifest as 3D conventional Weyl fermions with charge $\pm 1$~\cite{wan2011topological,wieder2021topological}.  
We therefore, in this work, refer to nodal points in the $PsP$ spectrum with nontrivial partial chiral charges as spin-Weyl fermions, such that a spin-resolved 3D TI realizes a spin-Weyl semimetal phase.

Because a maximally spin-gapped $PsP$ spectrum indicates the absence of $s$-nonconserving spin texture in the occupied bands (SN~2B), then the existence of unavoidable spin-Weyl points in 3D TIs implies that the occupied bands exhibit an unremovable spin texture.  
Like a Weyl semimetal state, a spin-Weyl semimetal state also exhibits forms of topological surface Fermi arcs.  
In the spin-Weyl state, the spin Fermi arcs either manifest as arc-like states along the entanglement cut in the spin-resolved entanglement spectrum (SN~3H), or as topological surface Fermi arcs in the energy spectrum under a large external Zeeman field, which we will explore in greater detail in the~Experimental Signatures and Discussion section.  
Lastly, because the occupied energy bands in a portion of the BZ in a spin-Weyl state must necessarily exhibit $C^{s}\neq 0$, then a finite sample of a spin-Weyl state, such as a 3D TI, may exhibit an extensive (though non-quantized) spin Hall conductivity.

{\bf Partial Axion Angles in Helical HOTIs}: Having deduced the spin-resolved topology of 2D and 3D TIs, we will next analyze the spin-resolved topology and response of helical HOTIs.  
In studies to date, there exist three competing theoretical constructions of an $\mathcal{I}$- and $\mathcal{T}$-protected helical HOTI state: 
\begin{enumerate}
\item{Orbital-double (superpose two identical copies of) an $\mathcal{I}$- and $\mathcal{T}$-symmetric 3D TI to form a so-called ``doubled-strong TI'' (DSTI)~\cite{po2017symmetry}.}
\item{Stack $\mathcal{I}$- and $\mathcal{T}$-symmetric 2D TIs with the same spin-orbital textures to form a layer construction with two identical 2D TIs per cell separated by a half-lattice translation~\cite{song2018mapping,elcoro2021magnetic,gao2022magnetic}.}
\item{$\mathcal{T}$-double (superpose two time-reversed copies of) an $\mathcal{I}$-symmetric magnetic AXI~\cite{wieder2018axion,wang2019higherorder}.}
\end{enumerate}
As we will show below, the three constructions of a helical HOTI in fact represent families of spin-resolved states with distinct spin-stable topology and distinct physical signatures.  
In order, the three constructions above correspond to a spin-Weyl semimetal with an even number of spin-Weyl points per half BZ [Fig.~\ref{fig:mainHOTI}(d)], a 3D QSHI state [Fig.~\ref{fig:mainHOTI}(c)], and a T-DAXI state [Fig.~\ref{fig:mainHOTI}(e)].

This result can most succinctly be understood through the language of SIs.  
An $\mathcal{I}$- and $\mathcal{T}$-symmetric HOTI is characterized by vanishing weak SIs and a nonvanishing $\mathbb{Z}_{4}$-valued strong SI $z_{4}=2$, where $z_{4}$ is defined by promoting the $\mathbb{Z}_{2}$-valued strong Fu-Kane parity ($\mathcal{I}$) criterion for 3D TIs to a $\mathbb{Z}_{4}$ invariant that further distinguishes between uninverted and doubly inverted bands~\cite{fu2007topological,po2017symmetry,khalaf2018symmetry,song2018mapping,elcoro2021magnetic,gao2022magnetic,wang2019higherorder}:
\begin{equation}
	z_4 = \frac{1}{4} \sum_{\mathbf{k}_a \in \mathrm{TRIMs}} \left( n^{a}_{+} - n^{a}_{-} \right) \text{ mod } 4,  
\label{eq:TZ4}
\end{equation}
where $n^{a}_{+}$ ($n^{a}_{-}$) is the number of occupied Bloch states at the TRIM point ${\bf k}_{a}$ with positive (negative) parity eigenvalues.  
Eq.~(\ref{eq:TZ4}) was originally obtained by performing combinatorics on the elementary (trivial) bands allowed in the nonmagnetic Shubnikov space group (SSG) $P\bar{1}1'$ (\# 2.5), which is generated by $\mathcal{I}$, $\mathcal{T}$, and 3D lattice translation symmetries~\cite{po2017symmetry,khalaf2018symmetry,song2018mapping,elcoro2021magnetic,gao2022magnetic,wang2019higherorder}.

\begin{figure*}[t]
\centering
\includegraphics[width=\textwidth]{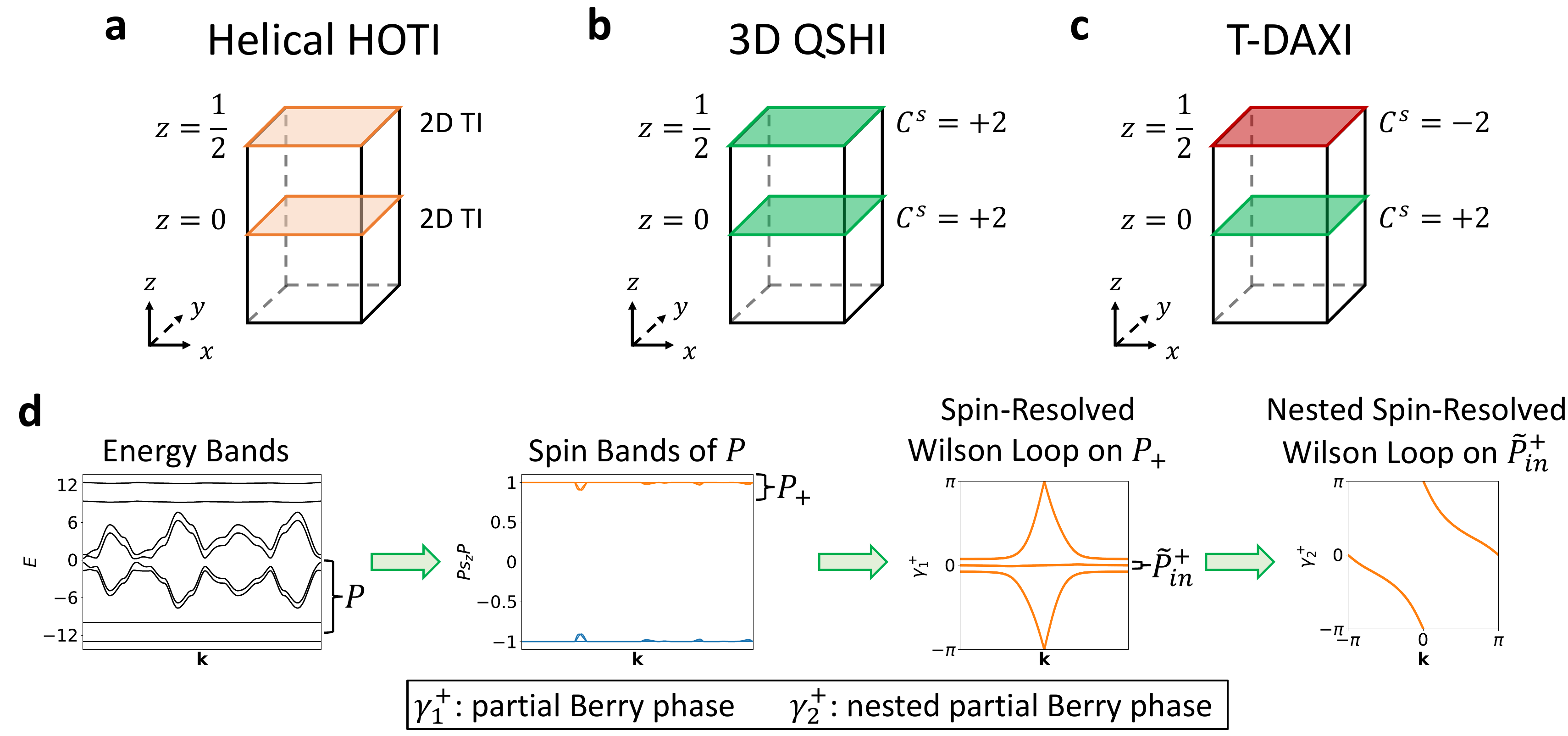}
\caption{Spin-resolved layer constructions and bulk partial axion angles in $\mathcal{T}$-doubled AXIs.  
(a) The layer construction of an $\mathcal{I}$- and $\mathcal{T}$-symmetric helical HOTI~\cite{song2018mapping,elcoro2021magnetic,gao2022magnetic}.
The HOTI in (a) is theoretically constructed by placing $\mathcal{I}$- and $\mathcal{T}$-symmetric 2D TIs (orange rectangles) in the $\mathcal{I}$-invariant $z=0$ and $z=1/2$ real-space planes in each unit cell.  
(b,c) Spin-resolved layer constructions with electronic bands that are topologically equivalent to the HOTI in (a).
Specifically if each 2D TI in (a) carries a bulk $s_{z}$ spin gap [where we have chosen $s=s_z$ for concreteness, see Fig.~\ref{fig:mainPSP}(d)], there are two ways to spin-resolve the helical HOTI layer construction in (a) while keeping a spin gap open.  
If both 2D TI layers have the same partial Chern numbers, then (b) through Eq.~(\ref{eq:generalSpinChern}), each 2D TI layer carries the same even-integer $s_{z}$ spin Chern number $C^{s}\text{ mod }4=2$ [Eq.~(\ref{eq:Z2for2D})], resulting in a 3D QSHI state with a non-quantized (but generically nonvanishing) $s_{z}$ spin Hall conductivity per bulk unit cell.  
However if the 2D TI layers in (a) have oppositely signed partial Chern numbers that are identical in magnitude, (c) the system instead realizes a T-DAXI state with a vanishing bulk $s_{z}$ spin Chern number and $\mathcal{I}$-quantized nontrivial partial axion angles $\theta^{\pm}=\pi$ (SN~4E).  
By closing and reopening the $Ps_{z}P$ spin gap, the QSHI insulator in (b) can be deformed into the T-DAXI in (c) via an intermediate spin-Weyl semimetal regime. 
Crucially, this deformation---which changes the bulk topological contribution to the spin-electromagnetic response for $s_{z}$ spins---must close an $s_{z}$ spin gap, but need not close an energy gap.
(d) Numerical workflow employed in this study to compute $\theta^{\pm}$.
We specifically extract $\theta^{\pm}$ by theoretically elucidating and numerically implementing a spin-resolved generalization of the nested Wilson loop method for computing $\theta$ that was previously introduced in Ref.~\cite{wieder2018axion}.
Documentation and details for accessing our freely available (spin-resolved) nested Wilson loop code are provided in SN~4E and~10B and Ref.~\cite{lin2023nestedWilsonLib}.
}
\label{fig:mainPartialAxion}
\end{figure*}

Importantly, when spin-resolving an insulator with the symmetries of SSG $P\bar{1}1'$ (\# 2.5), for any choice of spin direction $s$ in $PsP$ [Eq.~(\ref{eq:PsP})], the spin bands will respect the symmetries of magnetic SSG $P\bar{1}$ (\# 2.4), which is the subgroup of SSG $P\bar{1}1'$ (\# 2.5) generated by breaking $\mathcal{T}$ while preserving $\mathcal{I}$ and 3D lattice translations~\cite{elcoro2021magnetic,gao2022magnetic}.  
This can be seen by recognizing that $\{s,\mathcal{T}\}=0$ and $[s,\mathcal{I}]=0$ for all possible spin resolution directions $s$.  
Because $PsP$ splits the occupied bands in a $\mathcal{T}$-invariant insulator into halves, then the spin bands in each of the $\pm$ spin sectors will therefore each inherit half of the parity eigenvalues of the occupied bands of the original $\mathcal{T}$-invariant insulator (see SN~4D).

In magnetic SSG $P\bar{1}$ (\# 2.4) there is also a $\mathbb{Z}_{4}$-valued strong SI:
\begin{equation}
	\tilde{z}_4 = \frac{1}{2} \sum_{\mathbf{k}_a \in \mathrm{TRIMs}} \left( n^{a}_{+} - n^{a}_{-} \right) \text{ mod } 4,  
\label{eq:magneticZ4}
\end{equation}
in which the prefactor of $1/2$ differs from the prefactor of $1/4$ in Eq.~(\ref{eq:TZ4}) because spinful $\mathcal{T}$ symmetry forces states to form Kramers pairs at the TRIM points in nonmagnetic SSG $P\bar{1}1'$ (\# 2.5).  
For a helical HOTI with $z_{4}=2$, the spin bands in each $\pm$ sector will therefore carry the partial SI $\tilde{z}_{4}=2$.  
In magnetic SSG $P\bar{1}$ (\# 2.4) (see SN~4D and Refs.~\cite{elcoro2021magnetic,gao2022magnetic}), $\tilde{z}_{4}=2$ can indicate Weyl-semimetal states with even numbers of Weyl points in each half BZ, 3D quantum anomalous Hall states, and AXI states---exactly in correspondence with the possible spin-resolved topological states of a helical HOTI [Fig.~\ref{fig:mainHOTI}(c-e)].  
Importantly, it is possible for $s$-nonconserving SOC to drive spin band inversions and change the spin-stable topology without closing an energy gap.  
However, because the $\pm$-sector spin bands are related by $\mathcal{T}$ (SN~2B), and because $[\mathcal{T},\mathcal{I}]=0$, then a spin band inversion unaccompanied by an energy band inversion cannot change the value of $\tilde{z}_{4}$, and therefore cannot trivialize the spin-resolved topology of a helical HOTI.  
This represents the 3D generalization of the statement that the spin Chern number $C^{s}$ of a 2D TI can be changed without closing an energy gap, but cannot go to zero without closing an energy gap or breaking $\mathcal{T}$ symmetry [\emph{i.e.} $(C^{s}/2)\text{ mod }2=1$ for all $s$ in a 2D TI state, see Ref.~\cite{prodan2009robustness} and the text preceding Eq.~(\ref{eq:Z2for2D})].

Having established the spin-resolved partial SIs of a helical HOTI, we will now more closely analyze each family of spin-stable topological states in its spin resolution.  
Earlier, we showed that a spin-resolved 3D TI  for all $s$ necessarily has an odd number of spin-Weyl points in each half of the BZ, absent symmetries beyond $\mathcal{I}$ and $\mathcal{T}$ [Fig.~\ref{fig:mainTISpinWeyl}(b) and SN~3E and~3F].  
Building on this result, because the DSTI construction of a helical HOTI consists of superposing (orbital-doubling) two identical 3D TIs~\cite{po2017symmetry}, it follows that a spin-resolved DSTI realizes for all $s$ a spin-Weyl semimetal state with an even number of spin-Weyl points per half BZ [Fig.~\ref{fig:mainHOTI}(d)].  
Like an energy-band Weyl semimetal state, a spin-Weyl state also exhibits topological Fermi arcs, which can be detected in the spin-resolved entanglement spectrum (SN~3H), or in the surface energy spectrum in the presence of a large Zeeman field (SN~2G).  
We will shortly demonstrate in the Experimental Signatures and Discussion section that the candidate helical HOTI $\beta$-MoTe$_2$~\cite{wang2019higherorder,tang2019efficient} realizes a spin-Weyl semimetal state with an even number of spin-Weyl points per half BZ for all choices of spin direction $s$ in $PsP$, and hence lies in the DSTI regime of a helical HOTI (see SN~9B for further calculation details).

We will next consider two cases of spin-stable resolutions of helical HOTIs that can be formally expressed using a spin-resolved variant of the layer construction method for enumerating and analyzing symmetry-protected topological states~\cite{song2018mapping,elcoro2021magnetic,gao2022magnetic}.  
Given an SSG, a symmetry-protected topological state is considered to be layer-constructable if its momentum-space band topology can be completely captured in a system composed of flat, parallel layers of lower-dimensional topological states that are placed a manner in which their boundary states are pairwise gapped while preserving all system symmetries.  
For $\mathcal{T}$-invariant 3D TCI phases, the building blocks of layer constructions are 2D TIs and mirror TCIs~\cite{song2018mapping,elcoro2021magnetic,gao2022magnetic}.
In nonmagnetic SSG $P\bar{1}1'$ (\# 2.5) the layer construction of a helical HOTI consists of one $\mathcal{I}$-symmetric 2D TI at the origin of the unit cell [the $\mathcal{I}$-invariant $z=0$ plane in Fig.~\ref{fig:mainPartialAxion}(a)] and one $\mathcal{I}$-symmetric 2D TI in a real-space plane separated by a half-lattice translation from the origin [the $\mathcal{I}$-invariant $z=1/2$ plane in Fig.~\ref{fig:mainPartialAxion}(a)].

In this work, we introduce a finer distinction for layer constructions in which the spin-orbital textures of the layers, and hence their spin-resolved topology, become additional knobs in the layer construction method.  
To formulate these spin-resolved layer constructions, we begin by considering a 3D SSG that additionally carries at least one conserved spin direction $s$ at all points in space (\emph{e.g.} $s=s_{z}$ symmetry).  
Formally, the full symmetry group of SSG symmetries and at least U$(1)$ spin symmetry is isomorphic to a (nonmagnetic) ``spin space group''~\cite{BrinkmanSpinSpace}.  
The spin space groups are generally suitable for classifying the symmetry and topology of spin-wave excitations (magnons), for which minimal models represent a realistic approximation. 
However, the spin space groups are largely unsuitable for characterizing the electronic structure of solid-state materials, in which perfect spin-rotation symmetries are broken by phenomenologically distinct, symmetry-allowed contributions to the SOC, such as Ising and Rashba potentials~\cite{kane2005z,wang2021isingSC,elcoro2021magnetic,gao2022magnetic}.  
With this in mind, we next introduce $s$-nonconserving SOC to break the conserved spin symmetry, but not in a manner strong enough to close a spin gap within any of the system layers.  
Hence, we may still classify the layer construction using the partial Chern numbers $C^{\pm}$ of the occupied bands within each layer.

The simplest spin-resolved layer construction of a helical HOTI is one in which each 2D TI layer is spin-gapped for a spin direction $s$ and carries the same spin-orbital texture, such that the $\pm$-sector spin bands within each layer carry the same partial Chern numbers [Fig.~\ref{fig:mainPartialAxion}(b)].  
Through the definition of the 2D $\mathbb{Z}_{2}$ invariant in Eq.~(\ref{eq:Z2for2D}), this implies that each layer carries the same spin Chern number satisfying $C^{s}\text{ mod }4=2$.  
Because each unit cell carries a non-vanishing spin Chern number, then a helical HOTI constructed from identical 2D TI layers realizes a 3D QSHI state with an intrinsic bulk spin Hall response per unit cell that is nonvanishing (though also generically non-quantized due to the presence of $s$-nonconserving SOC).  3D QSHI states were previously predicted in the hourglass TCI KHgSb~\cite{wang2016hourglass}, distorted square-net compounds~\cite{zhijun2021qshsquarenet}, and in the helical HOTI $\alpha$-BiBr~\cite{SYBiBr}.  Indeed, our spin-resolved topological analysis of $\alpha$-BiBr, detailed below and in SN~10, reveals that $\alpha$-BiBr realizes a 3D QSHI state with a large bulk spin gap over a wide range of spin resolution directions.

\begin{figure*}[t]
\centering
\includegraphics[width=.9\textwidth]{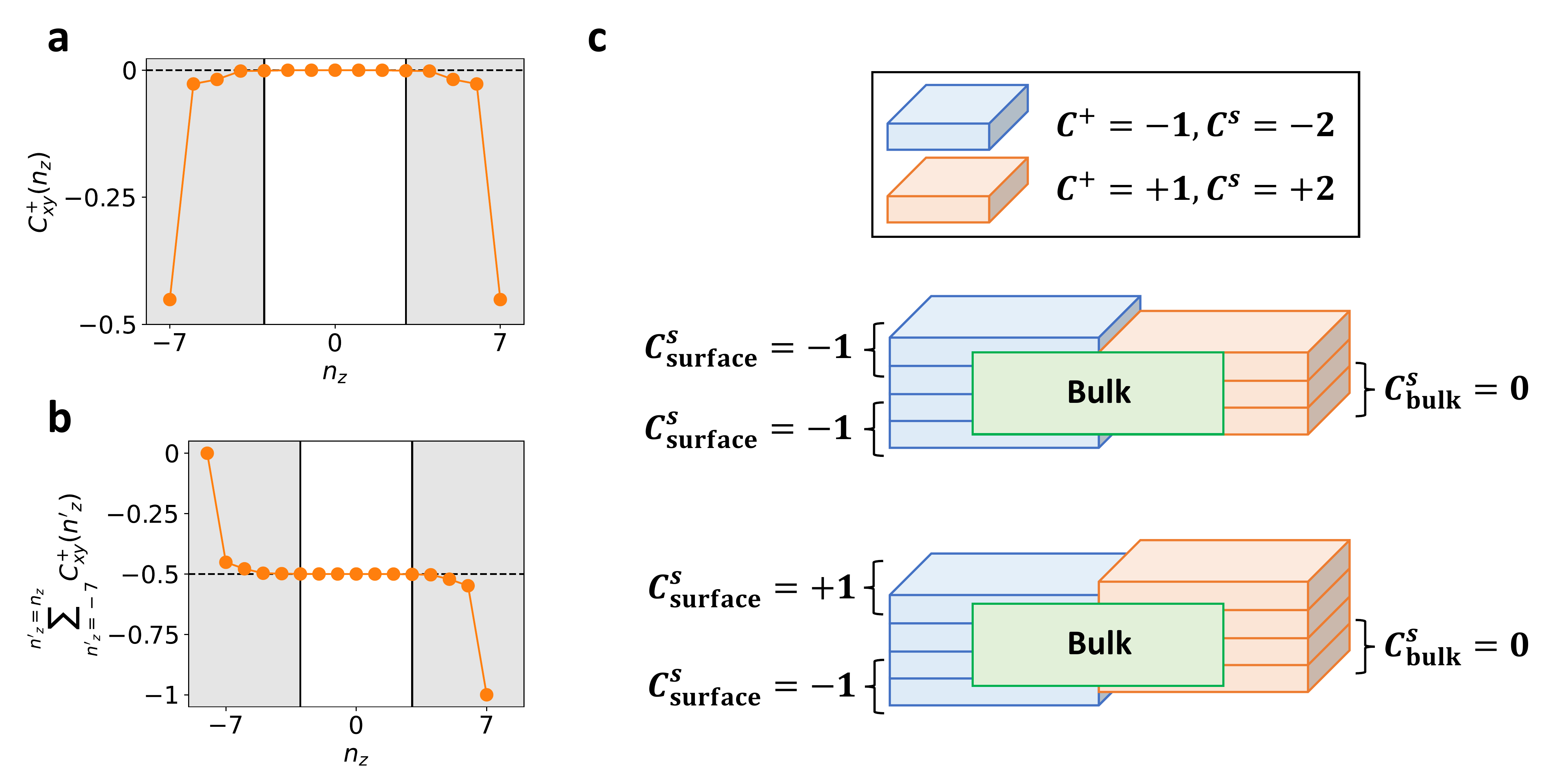}
\caption{Surface partial parity anomaly in $\mathcal{T}$-doubled axion insulators.  
(a) The layer-resolved position-space partial Chern number $C^{+}_{xy}(n_{z})$ for $s=s_{z}$ spins of an $\mathcal{I}$-symmetric finite slab of the $s_{z}$-nonconserving T-DAXI model from Fig.~\ref{fig:mainPartialAxion}(d) [adapted from Ref.~\cite{wang2019higherorder}, see SN~5E], plotted as a function of the $z$-direction slab layer index $n_{z}$.
(b) The cumulative (summed) values of $C^{+}_{xy}(n_{z})$ in (a).
In a T-DAXI, $C^{+}_{xy}(n_{z})$ is zero in the bulk of the system [white region in (a,b)] and nonvanishing on gapped surfaces [shaded regions in (a,b)].
However on each T-DAXI surface, we observe a cumulative half-integer partial Chern number [specifically $C^{+}=-0.5$ in (a,b)].
Because isolated $\mathcal{T}$-invariant noninteracting 2D insulators can only carry even spin Chern numbers (and hence integer partial Chern numbers via $C^{s}=C^{+}-C^{-}=2C^{+}$)~\cite{levin2009fractional} and because $C^{+}\text{ mod }2=1$ in 2D TIs~\cite{prodan2009robustness}, the data in (a,b) indicate that the $\mathcal{T}$-invariant gapped surfaces of T-DAXIs are not trivial, but rather carry anomalous halves of 2D TI states in a realization of a novel partial parity anomaly (SN~4D3).  
Importantly, perfect global $\mathcal{I}$ symmetry is not required to quantize $\theta^{\pm}=\pi$ in the bulk and realize anomalous surface halves of 2D TI states.
To illustrate this, in (c) we show schematic layer constructions of a finite T-DAXI slab. 
[(c), upper schematic] An $\mathcal{I}$-symmetric slab corresponding to the partial Chern number distribution in (a,b). 
[(c), lower schematic] The T-DAXI slab from the upper panel in (c).
Adding an extra (non-anomalous) layer with $C^s=2$ $(C^+=1)$ to the top surface of the system breaks global $\mathcal{I}$ symmetry, yielding a slab with a vanishing total spin Chern number.
However because each surface still carries an anomalous half of a 2D TI, each surface under an applied magnetic field still exhibits an intrinsic (non-quantized) spin Hall response unaccompanied by a bulk response, resulting overall in a 3D spin-magnetoelectric effect (see SN~7C and Refs.~\cite{JuvenSpinChargeAxion,WiederDefect}).}  
\label{fig:mainPartialParity}
\end{figure*}

However in this work, we recognize the existence of a second possible spin-resolved layer construction of an $\mathcal{I}$- and $\mathcal{T}$-symmetric helical HOTI.  
Instead of placing spin-gapped (for a spin direction $s$) 2D TI layers with the same partial Chern numbers in each $\mathcal{I}$-invariant plane, we alternatingly place layers with oppositely signed odd partial Chern numbers that are identical in magnitude [Fig.~\ref{fig:mainPartialAxion}(c)].  
In this case, the total spin Chern number within each unit cell vanishes.  
However, this does not imply a trivial spin-electromagnetic response.  
Instead we recognize that per $\pm$ spin sector, the layer construction in Fig.~\ref{fig:mainPartialAxion}(c) is identical to that of an $\mathcal{I}$-protected magnetic AXI (see Supplementary Figure~19 and Refs.~\cite{elcoro2021magnetic,gao2022magnetic,varnava2020axion}). 
This implies that taken per $\pm$ sector (which reduce to the $\uparrow,\downarrow$ spin sectors in the limit of perfect $s$ spin-rotation symmetry), the system carries an $\mathcal{I}$-quantized partial axion angle $\theta^{\pm}=\pi$, even though the total (charge) axion angle is trivial $\theta\text{ mod }2\pi=0$.  
Unlike the standard axion angle $\theta$, which can be quantized by either $\mathcal{I}$ or $\mathcal{T}$, the partial axion angles $\theta^{\pm}$ are quantized by $\mathcal{I}$ and exchanged (with a relative sign) by the action of $\mathcal{T}$ ($\theta^{\pm}\rightarrow - \theta^{\mp}$ under $\mathcal{T}$, see SN~4D).  
In the same sense that the standard axion angle $\theta$ represents the 3D generalization of the 1D Berry phase (charge polarization)~\cite{qi2008topological,essin2009magnetoelectric,IvoAXI1}, the partial axion angles $\theta^{\pm}$ therefore represent the 3D generalizations of the 1D partial Berry phases (polarization) introduced by Fu and Kane in Ref.~\cite{fu2006time}.  
We term the new spin-stable state characterized by $\theta^{\pm}=\pi$ the T-DAXI regime of a helical HOTI.  
To numerically verify the existence of $\mathcal{I}$-quantized partial axion angles in the T-DAXI state, we applied the nested Wilson loop method for computing $\theta$ previously introduced in Ref.~\cite{wieder2018axion} to the spin spectrum of a modified ($s_{z}$-nonconserving) implementation of the helical HOTI model formulated in Ref.~\cite{wang2019higherorder} (see also Ref.~\cite{lin2023nestedWilsonLib}).  
As shown in Fig.~\ref{fig:mainPartialAxion}(d) and documented in SN~4E, the $\mathcal{I}$-symmetric nested spin-resolved Wilson spectrum exhibits odd chiral winding, indicating that $\theta^{\pm}=\pi$.  
Through extensive nested spin-resolved Wilson loop calculations (see SN~10B), we find that the candidate helical HOTI $\alpha$-BiBr realizes not only the aforementioned 3D QSHI state, but also the T-DAXI state introduced in this work.

In 3D AXIs, the bulk axion angle $\theta=\pi$ also has a deep relation to the physics and response of 2D surfaces.  
Specifically in isolated 2D systems, the parity anomaly dictates that there cannot exist an odd number of symmetry-stabilized twofold Dirac cones~\cite{haldane1988model,qi2008topological,wieder2018wallpaper}.  
However the parity anomaly is circumvented on 2D interfaces (domain walls) between 3D insulators with $\theta=\pi$ (\emph{e.g.} 3D TIs) and insulators with $\theta=0$ (typically the vacuum).  
Under the preservation of specific surface (interface) symmetries (such as $\mathcal{T}$), this leads to an odd number of symmetry-stabilized surface Dirac cones~\cite{fu2007topologicala,fu2007topological,wieder2018axion,varnava2020axion}.

However, if the 2D surface does not preserve enough symmetries, then it becomes gapped. Crucially, this does not imply that the 2D surface is trivial.  
As the low-energy 2D surface theory of a 3D $\theta=\pi$ phase originates from an unpaired, parity-anomaly-violating twofold Dirac cone (integer quantum Hall critical point), then the 2D surface of a 3D TI or AXI, when gapped, realizes an anomalous (noninteracting) half quantum Hall state~\cite{qi2008topological,fu2007topological}.  
In helical HOTIs, the gapped 2D surfaces are $\mathcal{T}$-invariant, and hence have vanishing Hall conductivities [see the text preceding Eq.~(\ref{eq:Z2for2D})].  
One might therefore believe that the gapped 2D surfaces of helical HOTIs are trivial, or possibly carry integer 2D TI states, because a portion of the edges (hinges) between gapped HOTI surfaces exhibit 1D helical modes [Fig.~\ref{fig:mainHOTI}(a,b)].  
However, our discovery of $\mathcal{I}$-quantized bulk partial axion angles $\theta^{\pm}=\pi$ in the T-DAXI state suggests that instead, each partial axion angle contributes a half-integer partial Chern number to each gapped 2D surface.  
This implies that each 2D surface of a T-DAXI (with $\theta^{\pm}=\pi$ obtained for a fixed spin direction $s$) hosts a $\mathcal{T}$-invariant gapped state with an odd spin Chern number ($C^{s}\text{ mod }2=1$), a value that cannot be realized in an isolated $\mathcal{T}$-invariant noninteracting insulator with a spin gap (see Ref.~\cite{levin2009fractional} and SN~3C).
Each gapped surface of a T-DAXI is hence equivalent to an anomalous half of an isolated 2D TI as a consequence of a novel partial parity anomaly.  
To numerically verify the existence of a surface partial parity anomaly in T-DAXIs, we implemented a spin-resolved partial variant of the position-space layer-resolved Chern number~\cite{essin2009magnetoelectric,varnava2018surfaces,IvoAXI1} (see SN~5 for calculation details).
As shown in Fig.~\ref{fig:mainPartialParity}(a), the layer-resolved partial Chern number vanishes on the average in the bulk of a T-DAXI, and indeed saturates at anomalous half-integer values on its gapped surfaces.

We can draw several connections between the anomalous surfaces of T-DAXIs and previous works.  
First, anomalous halves of 2D TI states were previously predicted to occur on the top and bottom surfaces of weak TIs~\cite{liu2012half}; here, we recognize anomalous half 2D TI states to be more general features of helical HOTIs in the T-DAXI regime.  
Though T-DAXIs and globally $\mathcal{I}$-symmetric models of spin-gapped weak TIs (with odd total numbers of 2D TI layers) both exhibit anomalous surface half 2D TI states, they are still distinguishable via bulk spin Hall measurements, provided that the intrinsic spin Hall response is dominated by the bulk topological contribution.
Specifically, in weak TIs with a gap in $PsP$ for a spin direction $s$, the topological contribution to the spin Hall conductance of a finite sample (for $s$-polarized spins) is extensive and carries a nonvanishing weight in each bulk unit cell.
Conversely in a T-DAXI with a $PsP$ gap for a spin direction $s$, the topological contribution to the spin Hall conductivity vanishes in the bulk and only manifests (anomalously) on 2D surfaces, and is hence independent of sample thickness [Fig.~\ref{fig:mainPartialParity}(a,b)]. 
A magnetic field applied to a T-DAXI will therefore induce a (non-quantized) spin Hall response (for $s$-polarized spins) on spatially separated (opposing) surfaces.  
If the spin Hall responses on the opposing surfaces are oppositely signed [which necessarily breaks global $\mathcal{I}$ symmetry, see Fig.~\ref{fig:mainPartialParity}(c) and Ref.~\cite{wu2016quantized}], the magnetic field will generate a spin separation with both a transverse and a parallel component with respect to the field.  
We term the novel response originating from the field-parallel spin separation the 3D spin-magnetoelectric effect.  
As we have only demonstrated the existence of the spin-magnetoelectric effect through layering and Thouless-pump arguments (SN~3G and~4D), a linear-response formulation of the spin-magnetoelectric effect in the presence of $s$-nonconserving SOC remains an exciting and urgent direction for future study.

The anomalous odd spin Chern number of the gapped surfaces of the T-DAXI state is also reminiscent of 3D bosonic TIs, for which each 2D surface carries an odd Chern number, a value that is anomalous because isolated 2D bosonic systems without topological order are required to have even Chern numbers~\cite{AshvinBosonicTCI,senthil2013integer}.  
Additionally, 3D symmetry-protected topological phases with anomalous 2D quantum spin Hall surface responses have been proposed in field-theoretic investigations, but were not previously associated to helical HOTIs~\cite{JuvenSpinChargeAxion}.  
Lastly, the gapless surface theories of other $\mathcal{T}$-invariant 3D TCI phases, like twofold-rotation-anomaly TCIs (two twofold Dirac cones)~\cite{fang2019new,elcoro2021magnetic,gao2022magnetic} and the nonsymmorphic Dirac insulator (one fourfold Dirac cone)~\cite{wieder2018wallpaper} can be deformed into the gapped surface theory of a helical HOTI by lowering the surface crystal symmetry without breaking $\mathcal{T}$.  
This suggests that the symmetry-enhanced fermion doubling theorems circumvented in these TCI phases, which were deduced from crystal-symmetry constraints on band connectivity, may be expressible in the language of quantum field theory through the partial parity anomaly identified in this work.

\begin{figure*}[t]
\centering
\includegraphics[width=.975\textwidth]{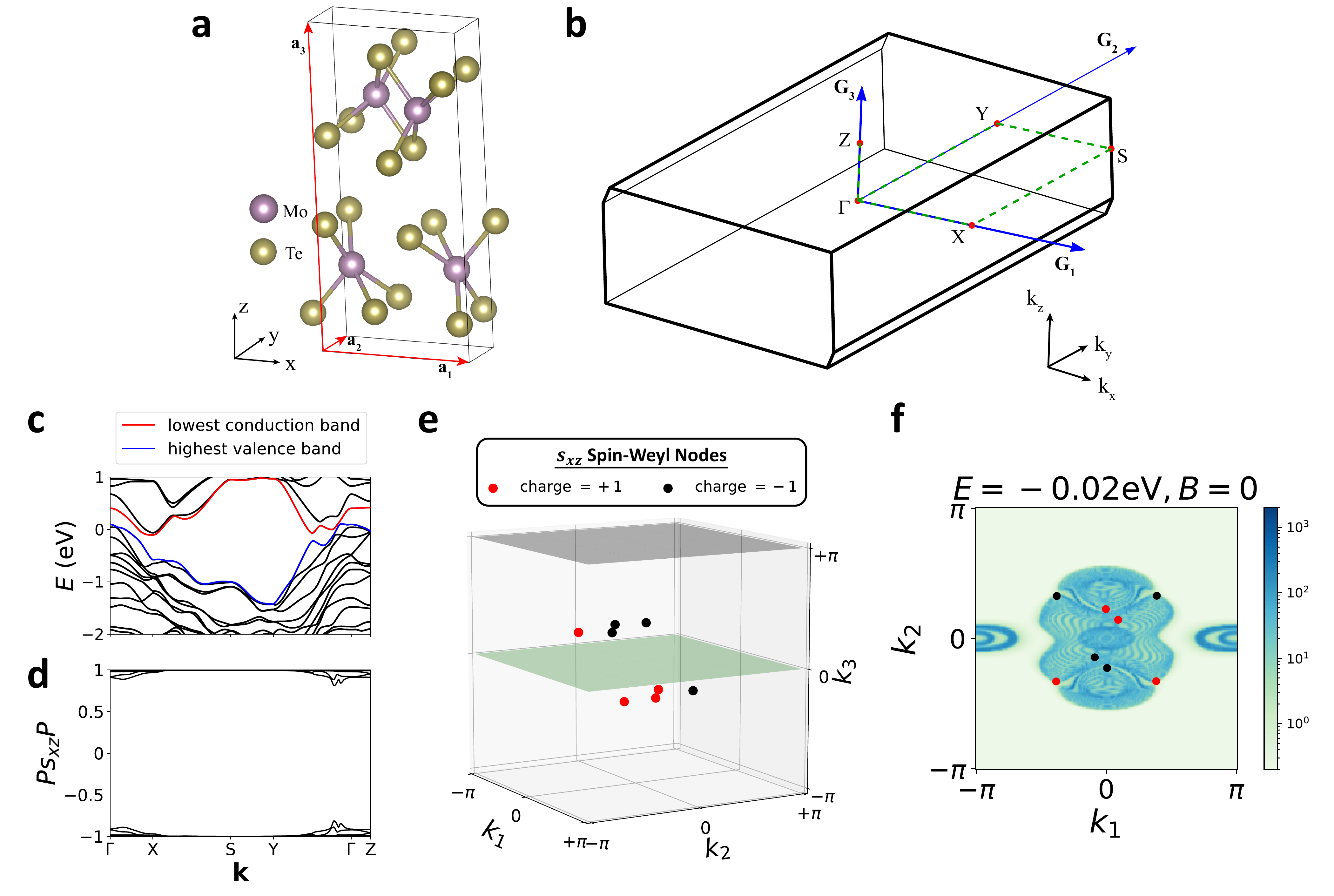}
\caption{Spin-Weyl points in $\beta$-MoTe$_2$.  
(a) Crystal structure of the candidate helical HOTI $\beta$-MoTe$_2$~\cite{wang2019higherorder,tang2019efficient}, which respects the symmetries of Shubnikov space group (SSG) $P2_1/m1'$ (\# 11.51, see SN~9A). 
The red arrows in (a) indicate the primitive lattice vectors ${\bf a}_{1,2,3}$.
(b) The bulk BZ of $\beta$-MoTe$_2$.
The blue arrows in (b) indicate the primitive reciprocal lattice vectors ${\bf G}_{1,2,3}$.
(c) Band structure of a first-principles- (DFT-) obtained, symmetric, Wannier-based tight-binding model of $\beta$-MoTe$_2$ (details provided in SN~9A), plotted along the dashed green high-symmetry ${\bf k}$-path in (b). 
In (c), we label the highest valence [lowest conduction] doubly-degenerate bands in blue [red].
(d) The $Ps_{xz}P$ spin spectrum of the occupied bands of $\beta$-MoTe$_2$ [choosing all states to be individually occupied up to and including the blue bands in (c) at each ${\bf k}$ point, see SN~9A].
Though the spin spectrum in (d) appears gapped, closer examination of the spin gap away from high-symmetry BZ lines reveals the presence of spin-Weyl points in the BZ interior.
We further find that for all choices of spin direction $s$ in $PsP$ [Eq.~(\ref{eq:PsP})], $\beta$-MoTe$_2$ realizes a spin-Weyl state with an even number of spin-Weyl nodes per half BZ (SN~9B).  
(e) The distribution of spin-Weyl nodes for the $s=s_{xz}$ [Eq.~(\ref{eq:sXZ})] spin spectrum in (d).
In (e), there are eight spin-Weyl nodes in the BZ interior with a total partial chiral charge of $|2|$ per $k_{3}$-indexed half BZ, which we have confirmed through spin-resolved Wilson loop calculations (SN~9B).
(f) The $(001)$-surface spectral function of $\beta$-MoTe$_2$ with the projected locations and partial chiral charges of the bulk $s_{xz}$ spin-Weyl points from (e) labeled with red and black circles.}
\label{fig:mainMoTe2}
\end{figure*}

\begin{figure*}[t]
\centering
\includegraphics[width=.975\textwidth]{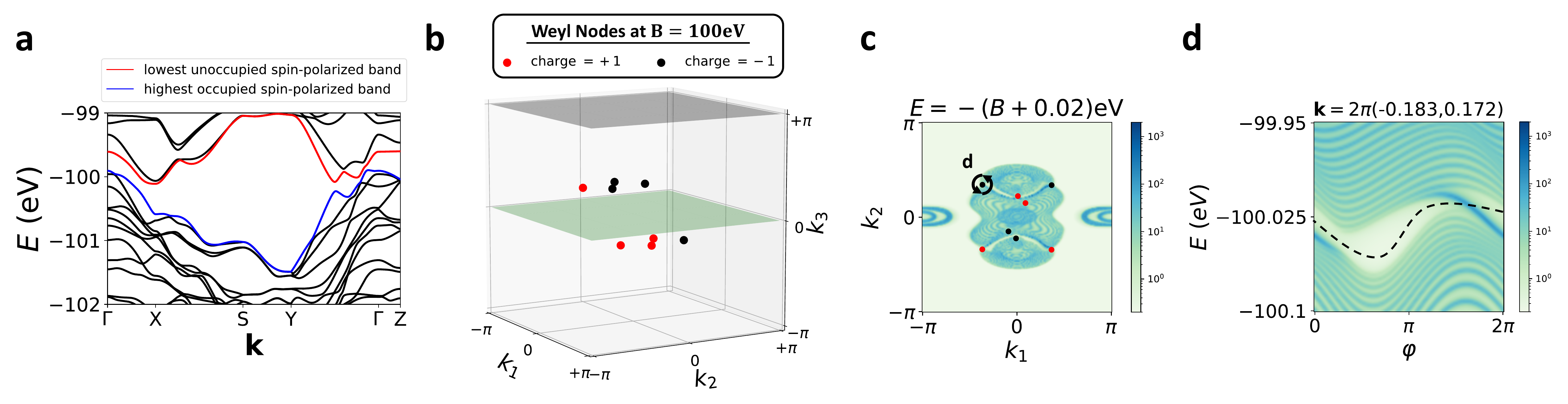}
\caption{Converting spin-Weyl fermions to Weyl fermions in $\beta$-MoTe$_2$ with an applied Zeeman field.
(a) The electronic band structure of the Wannier-based tight-binding model of $\beta$-MoTe$_2$ [Fig.~\ref{fig:mainMoTe2}(c)] in the presence of an ($\hat{\mathbf{x}}+\hat{\mathbf{z}}$)-directed $B=|{\bf B}|=100$eV (spin-) Zeeman field.
We note that the band structure in (a) within each $s_{xz}$ spin sector (here close to $E\sim -B\sim - 100$eV) exhibits nearly the same band ordering and features as the field-free band ordering at $E_{F}$ in Fig.~\ref{fig:mainMoTe2}(c).
Though the blue and red singly-degenerate [nearly spin-polarized] bands in (a) appear gapped along high-symmetry BZ lines [Fig.~\ref{fig:mainMoTe2}(b)], the blue and red bands in fact cross in the BZ interior to form Weyl fermions.
(b) The eight Weyl points connecting the blue and red bands in (a).
Remarkably, the locations and chiral charges of the Zeeman-induced Weyl nodes in (b) are nearly identical to the locations and partial chiral charges of the spin-Weyl nodes in Fig.~\ref{fig:mainMoTe2}(e).
(c) The $(001)$-surface spectral function of $\beta$-MoTe$_2$ at $E=-100$eV with the projected locations and partial chiral charges of the Zeeman-induced bulk Weyl points from (b) labeled with red and black circles.
(d) The $(001)$-surface spectral function at $E=-100$eV computed on a small, counterclockwise path encircling the $(001)$-surface projection of the bulk Weyl point circled in (c).
The bulk Weyl points in (b,c) give rise to topological Fermi-arc surface states crossing the bulk (indirect) gap, as shown in (d) and SN~9C.}
\label{fig:mainMoTe2withZeeman}
\end{figure*}

{
\vspace{0.1in}
\centerline{\bf Discussion}
\vspace{0.1in}
}

We conclude by discussing experimental signatures of spin-resolved band topology and avenues for future study.  
First, in SN~2G, we show that the spin bands of a spin-Weyl semimetal state computed for a spin direction $s = {\bf s}\cdot \hat{\bf n}$ [Figs.~\ref{fig:mainHOTI}(d) and~\ref{fig:mainTISpinWeyl}] exhibit connectivity and topology related to that of the energy bands (in each spin sector) when a large Zeeman field ${\bf B}$ is applied parallel to $s$ (${\bf B}\parallel\hat{\bf n}$).  
To explore the relationship between the energy and spin spectrum in a realistic spin-Weyl state, we performed ab-initio calculations on the layered transition-metal dichalcogenide $\beta$-MoTe$_2$ [Fig.~\ref{fig:mainMoTe2}(a,c), see SN~9A for calculation details].  
Previous theoretical works have predicted that $\beta$-MoTe$_2$ realizes an $\mathcal{I}$- and $\mathcal{T}$-protected helical HOTI phase~\cite{wang2019higherorder,tang2019efficient}, and previous experimental works have observed signatures of hinge-state-like 1D gapless channels in STM~\cite{huang2019polar} and in supercurrent oscillation~\cite{wang2020evidence} probes of MoTe$_2$. 
Through extensive spin-gap minimization calculations detailed in SN~9B, we find that for all choices of spin direction $s$, $\beta$-MoTe$_2$ realizes a spin-Weyl semimetal state with an even number of spin-Weyl nodes in each half of the BZ.  
For the particularly simple case in which $s$ is chosen to be:
\begin{equation}
s_{xz} = \frac{1}{\sqrt{2}}\left(s_{x} + s_{z}\right),
\label{eq:sXZ}
\end{equation}
we specifically find that the spin spectrum is gapped along all high-symmetry lines [Fig.~\ref{fig:mainMoTe2}(b,d)], and that in the $k_{3}>0$ half of the 3D BZ, there are three spin-Weyl points with positive charge and one spin-Weyl point with negative charge [Fig.~\ref{fig:mainMoTe2}(e,f)].  
Because pairs of oppositely charged spin-Weyl points lie close together, and because the total spin-Weyl partial chiral charge in each $k_{3}$-indexed half of the BZ is $|2|$, then we conclude that overall, $\beta$-MoTe$_2$ lies in the DSTI regime of a helical HOTI state (see SN~3E,~4D, and~9B).

When we theoretically apply a (very) large Zeeman field ${\bf B}\parallel s_{xz}$ ($|{\bf B}|=100$eV) to our Wannier-based tight-binding model of $\beta$-MoTe$_2$, we observe that the spin-Weyl nodes continuously evolve into bulk Weyl nodes at energies $E\approx \pm|{\bf B}|$, as shown in Fig.~\ref{fig:mainMoTe2withZeeman}(a,b) for energies close to $-|{\bf B}|$.  
The presence of Weyl nodes in the energy spectrum implies that the surface spectrum computed at the energy of the Weyl nodes should exhibit topological surface Fermi arcs.  
Focusing on the experimentally accessible $(001)$-surface of $\beta$-MoTe$_2$ (see Ref.~\cite{wang2019higherorder} and SN~9C), we compute the surface Green's function in the presence of a strong ($\hat{\mathbf{x}}+\hat{\mathbf{z}}$)-directed Zeeman field [Fig.~\ref{fig:mainMoTe2withZeeman}(c)]. 
Consistent with our predictions, we observe topological surface Fermi arcs crossing the bulk (indirect) gap [Fig.~\ref{fig:mainMoTe2withZeeman}(d)].

\begin{figure*}[t]
\centering
\includegraphics[width=.975\textwidth]{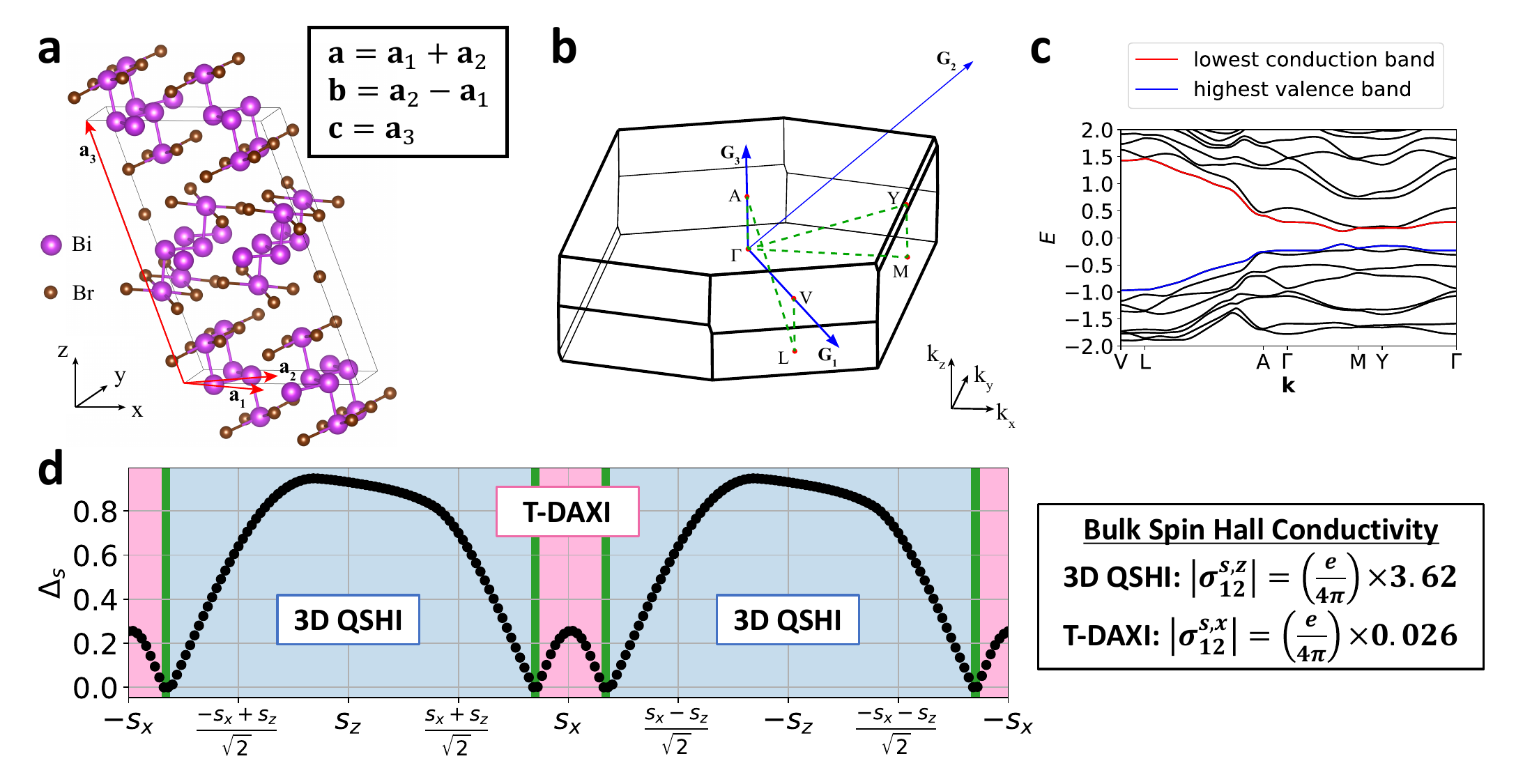}
\caption{3D quantum spin Hall and $\mathcal{T}$-doubled axion insulator states in $\alpha$-BiBr.
(a) Crystal structure of the candidate helical HOTI $\alpha$-BiBr~\cite{SYBiBr,BiBrFanHOTI}, which respects the symmetries of SSG $C 2 / m 1'$ (\# 12.59, see SN~10A).
The red arrows in (a) indicate the primitive lattice vectors ${\bf a}_{1,2,3}$, which are related to the conventional lattice vectors ${\bf a}$, ${\bf b}$, ${\bf c}$ through the equations in the inset box.
(b) The bulk BZ of $\alpha$-BiBr.
The blue arrows in (b) indicate the primitive reciprocal lattice vectors ${\bf G}_{1,2,3}$.
(c) Band structure of a DFT-obtained, symmetric, Wannier-based tight-binding model of $\alpha$-BiBr (details provided in SN~10A), plotted along the dashed green high-symmetry ${\bf k}$-path in (b). 
In (c), we label the highest valence [lowest conduction] doubly-degenerate bands in blue [red].
(d) The spin gap $\Delta_{s}$ and spin-resolved topology of $\alpha$-BiBr plotted as a function of $s$ rotated in the $xz$-plane (see SN~10B).
For nearly every spin resolution direction in (d), $\alpha$-BiBr is spin-gapped, with the largest spin gap [$\Delta_{s}\approx 0.95$, $\approx 47\%$ of its maximal value $\Delta_{s}=2$] surprisingly lying within 3 degrees of the ${\bf a}_{3}\parallel {\bf c}$ axis in (a) [see SN~10B].
The large ${\bf c}$-axis-directed spin gap indicates that the bulk spin-orbital texture in $\alpha$-BiBr is dominated by contributions almost entirely polarized along the ${\bf c}$ axis. 
Through (nested) spin-resolved Wilson loop calculations [see Fig.~\ref{fig:mainPartialAxion}(d) and~SN~10B], we obtain the spin-resolved topological phase diagram of $\alpha$-BiBr shown in (d), in which the $\pm s_{z}$-centered blue regions host 3D QSHI states, and the $\pm s_{x}$-centered pink regions host the spin-stable T-DAXI state introduced in this work.
The inset box in (d) shows the non-quantized bulk spin Hall conductivity per unit cell of $\alpha$-BiBr for the $s_{z}$ and $s_{x}$ spin directions (see SN~7 and~10C for calculation details).  For both the 3D QSHI ($s_{z}$) and T-DAXI ($s_{x}$) regimes of $\alpha$-BiBr, the bulk intrinsic spin Hall conductivity of $\alpha$-BiBr lies close to the quantized topological contribution from its nontrivial spin-resolved bulk topology.}
\label{fig:mainbibr}
\end{figure*}

To observe and characterize spin-gapped phases under real-material conditions, we next perform a detailed analysis of the (spin-resolved) topology and spin-electromagnetic response of the quasi-1D candidate HOTI $\alpha$-BiBr [Fig.~\ref{fig:mainbibr}(a,b,c), see SN~10 calculation details]~\cite{SYBiBr,BiBrFanHOTI}, for which angle-resolved photoemission spectroscopy (ARPES) and STM experiments have revealed signatures of 1D helical hinge states that persist to room temperature~\cite{noguchi2021evidence,FanZahidRoomTempBiBrExp}.  
Prior to computing the spin-resolved topology of $\alpha$-BiBr, we first probe its (hybrid) Wannier spectrum by computing the $\mathcal{I}$- and $\mathcal{T}$-symmetric nested Wilson loop of the occupied bands (SN~10B).  
Our nested Wilson loop calculations on $\alpha$-BiBr reveal the characteristic higher-order (nested Wilson) spectral flow of a helical HOTI.  
This finding itself represents a significant result, as nested Wilson loop calculations on ab-initio-derived electronic structures remain exceedingly rare, with a noteworthy previous example being the identification of a non-symmetry-indicated helical HOTI state in noncentrosymmetric ($\mathcal{I}$-broken) $\gamma$-MoTe$_2$ via a pattern of helical nested Wilson loop flow similar to that in $\alpha$-BiBr (but protected by distinct symmetries, as $\alpha$-BiBr is centrosymmetric)~\cite{wang2019higherorder}.

We next compute the spin gap for $\alpha$-BiBr over the complete range of spin resolution directions $s$.  Unlike previously for $\beta$-MoTe$_2$ (SN~9B), we find that $\alpha$-BiBr is spin-gapped for nearly all spin resolution directions (SN~10B).
In particular, when restricting $s$ to lie in the $xz$-plane [perpendicular to the $y$-directed chains in its crystal structure, see Fig.~\ref{fig:mainbibr}(a)], we observe that the spin gap in $\alpha$-BiBr only closes in four extremely narrow spin-gapless (spin-Weyl) regions, which are indicated in green in Fig.~\ref{fig:mainbibr}(d).
We observe that the $s_{z}$ spin gap in $\alpha$-BiBr is large ($\Delta_{s_{z}}\approx 0.93$, $\approx 46\%$ of the maximal value $\Delta_{s}=2$), and is much larger than the $s_{x}$ spin gap ($\Delta_{s_{x}}\approx 0.26$).
This is consistent with earlier first-principles investigations of $\alpha$-BiBr, which found the spin-electromagnetic (Rashba-Edelstein) response of its $(010)$-surface states to be strongly polarized in the $z$-direction relative to the $x$-direction~\cite{SYBiBr}.
We further find that overall, the global spin gap in $\alpha$-BiBr peaks at a similarly large value ($\Delta_{s}\approx 0.95$) and lies within $\approx 3$ degrees of the ${\bf a}_{3}\parallel {\bf c}$ lattice vector [Fig.~\ref{fig:mainbibr}(a)], indicating that the bulk spin-orbital texture in $\alpha$-BiBr is dominated by contributions that are almost entirely polarized along the $c$-axis. 
As discussed earlier, similar SOC textures that are polarized along a high-symmetry (out-of-plane) crystallographic axis in 2D materials have been termed Ising SOC~\cite{wang2021isingSC}. 
The appearance of a large bulk spin gap nearly locked to a crystallographic axis in $\alpha$-BiBr [Fig.~\ref{fig:mainbibr}(d)] suggests that it would be intriguing to investigate the microscopic mechanism of the SOC in $\alpha$-BiBr in future theoretical studies, and to study the spin-resolved response of $\alpha$-BiBr in future photoemission and transport experiments, which may exhibit an unusually high degree of spin polarization relative to other strongly spin-orbit-coupled 3D materials.

Through (nested) spin-resolved Wilson loop calculations detailed in SN~10B, we find that the four spin-gapped regions in the spin-resolved topological phase diagram of $\alpha$-BiBr [Fig.~\ref{fig:mainbibr}(d)] respectively correspond to two wide $\nu_{z}^{\pm}=\mp 2$ 3D QSHI  regions [see Fig.~\ref{fig:mainHOTI}(c)] with large spin gaps centered around $s=\pm s_{z}$, and two narrower $\nu^{\pm}_{x,y,z}=0$, $\theta^{\pm}=\pi$ T-DAXI regions with relatively smaller spin gaps centered around $s = \pm s_{x}$.
For completeness, we note that because spins lying in the $xz$-plane are left invariant under the ${\bf E}\cdot{\bf B}$-odd $C_{2y}\times\mathcal{T}$ antiunitary rotation symmetry of $\alpha$-BiBr (see SN~10A), then the nontrivial partial axion angles $\theta^{\pm}=\pi$ in the T-DAXI regime of $\alpha$-BiBr in Fig.~\ref{fig:mainbibr}(d) could alternatively be interpreted as quantized by the ``rotation-anomaly'' symmetry $C_{2y}\times\mathcal{T}$, rather than $\mathcal{I}$~\cite{fang2019new,wieder2018axion}.

To demonstrate physical signatures of nontrivial spin-resolved topology in $\alpha$-BiBr, we next compute the intrinsic bulk spin Hall conductivity (per unit cell) in the 3D QSHI ($s=s_{z}$) and T-DAXI ($s=s_{x}$) regimes.  
Formally, our calculations were performed by applying a numerical implementation of the spin-nonconserving spin Hall conductivity derived from linear response through the Kubo formula (SN~7) to a DFT-obtained, Wannier-based model of $\alpha$-BiBr (SN~10C).
Even though both the $s_{z}$ and $s_{x}$ spin gaps in $\alpha$-BiBr lie at less than half the maximal value of $\Delta_{s}=2$ [Fig.~\ref{fig:mainbibr}(d)], we find that both the 3D QSHI and T-DAXI regimes of $\alpha$-BiBr exhibit bulk (nonquantized) spin Hall conductivities that lie close to the quantized topological contribution given by $\sigma_{12}^{s} = [e/(4\pi)]\times 2\nu_{z}^{+}$.
Specifically, in the $|\nu_{z}^{+}|=2$ 3D QSHI region ($s=s_{z}$), the spin Hall conductivity is nearly quantized $|\sigma_{12}^{s_{z}}| = (e/4\pi)\times 3.62$, whereas in the T-DAXI region ($s=s_{x}$), the spin Hall conductivity is nearly vanishing $|\sigma_{12}^{s_{x}}| = (e/4\pi)\times 0.026$ [see Fig.~\ref{fig:mainbibr}(d) and SN~10C].
This result suggests a highly anisotropic spin Hall response in $\alpha$-BiBr that interpolates between a large, extensive bulk contribution for $s_{z}$ spin transport to a small, surface-dominated contribution for $s_{x}$ spin transport. 
Given that $\alpha$-BiBr is readily synthesizable~\cite{noguchi2021evidence,FanZahidRoomTempBiBrExp}, the anisotropic spin-electromagnetic response predicted in this work should be accessible through straightforward (inverse) spin Hall measurements that are achievable within a short timeframe.

In addition to the spin Hall response of QSHI states [Fig.~\ref{fig:mainPSP}(c)] and Zeeman-induced surface Fermi arcs in spin-Weyl states (Fig.~\ref{fig:mainMoTe2}), spin-resolved topology may be experimentally accessed through terahertz measurements of the spin-magnetoelectric response of helical HOTIs in the T-DAXI regime, such as $\alpha$-BiBr for $s\approx s_{x}$-polarized spins [see Fig.~\ref{fig:mainPartialParity}(b,c), Fig.~\ref{fig:mainbibr}(d), and SN~10B]. 
A spin imbalance at the surface of a T-DAXI will yield a non-quantized charge (electromagnetic) Hall response due to a lack of compensation between states with opposite partial Chern numbers. 
As depicted in Fig.~\ref{fig:mainHOTI}(e), when $s$-nonconserving SOC is weak, we can describe the surface of a helical HOTI in the T-DAXI regime in terms of a massive fourfold Dirac cone (two bands per spin) with weak spin mixing~\cite{wieder2018wallpaper,wang2019higherorder}.  
In this scenario, the spin-up surface bands hence carry an anomalous Chern number $1/2+n$, and the spin-down surface bands correspondingly carry an anomalous Chern number $-1/2-n$, where $n\in\mathbb{Z}$. 
Selectively depopulating one spin species will hence yield a surface Fermi surface with a nonvanishing anomalous Hall conductivity~\cite{IvoAXI1}. 
By analogy with previous optical experiments performed on 3D TIs with magnetically gapped surface states~\cite{wu2016quantized,MogiOpticalAxionParity}, a surface anomalous Hall conductivity can be measured through the Kerr and Faraday rotation of a terahertz probe.

A spin imbalance on the gapped surface of a T-DAXI can either be realized through selective excitation across the surface band gap using circularly polarized light (similar to optical experiments performed on monolayer transition-metal dichalcogenides and associated heterostructures~\cite{xu2014spin}), or by spin injection using an adjoining magnetic transducer layer~\cite{shiomi2014spin}. 
We expect the measured Faraday and Kerr rotation to vary linearly with the induced spin imbalance. 
Given that both the surface and the bulk are gapped, a spin-magnetoelectric response controlled by a spin imbalance could enable electric-field-tunable terahertz and infrared polarization modulators with nearly perfect transmission.

Furthermore, we note that the spin spectrum itself could be directly probed through generalizations of spin-ARPES (S-ARPES). 
Typical S-ARPES resolves the spin polarizations of individual bands, which are related to diagonal matrix elements of the $PsP$ operator. 
S-ARPES experiments have previously been performed on the candidate spin-Weyl semimetal $\beta$-MoTe$_2$ identified in this work~\cite{weber2018spin}, and should be revisited in the context of spin-resolved topology. 
To fully resolve the spin spectrum experimentally, the off-diagonal matrix elements of $PsP$ between occupied states must also be measured.  
However, this would would require measuring the spin-dependent transition probability between pairs of occupied states. 
While such a measurement is currently beyond existing photoemission methods, recent proposals on double and pair photoemission spectroscopy~\cite{mahmood2022distinguishing,su2020coincidence} may provide a promising and exciting path forward, provided that they can be developed with Mott- or very low energy electron-diffraction-based spin detection.

Further material candidates in the T-DAXI, spin-Weyl semimetal, or 3D QSHI regimes may be identifiable among centrosymmetric, exfoliable materials with narrow band gaps and the SIs of a helical HOTI [see SN~4D and the text surrounding Eq.~(\ref{eq:TZ4})], such as ZrTe$_2$~\cite{AndreiMaterials2}.
Dimensional reduction through exfoliation, as well as substitutional doping (\emph{i.e.} Br/I, Zr/Hf, and Se/Te) to tune SOC can be explored to open a gap at the Fermi level in metallic material candidates. 
Surface passivation can also be explored to drive insulating behavior in metallic candidate materials, similar to the case of Al on Bi$_2$Te$_3$~\cite{lang2012revelation}.
Though we have largely focused on solid-state realizations of spin-resolved topology, cold atoms have recently been employed to mimic 2D QSHI phases~\cite{SpinChernColdAtoms}, and hence may also serve as promising platforms for engineering the 3D T-DAXI states identified in this work.

Finally, our findings suggest several intriguing future directions.  
We have introduced a predictive framework for linking novel low-energy response theories to gauge-invariant quantities obtained from real-material calculations.  
Despite our progress unraveling the bulk and surface theories of $\mathcal{I}$- and $\mathcal{T}$-symmetric helical HOTIs, and despite other promising early efforts~\cite{TaylorTCIResponse,TaylorCompetingThouless}, there remain numerous other noninteracting TCI phases---such as SU$(2)$-doubled magnetic AXIs~\cite{wang2019higherorder,WiederDefect} and fourfold-rotation-anomaly TCIs like SnTe~\cite{schindler2018higherorder,fang2019new}---for which the bulk response theories are largely unknown.  
Additionally, while we focused on resolving band topology through the spin degree of freedom, the methods introduced in this work can straightforwardly be extended to sublattice (pseudospin), orbital, and layer degrees of freedom to predict new valleytronic and layertronic effects, such as valley- and spin-resolved generalizations of the layer Hall response recently observed in the antiferromagnetic AXI MnBi$_2$Te$_4$~\cite{SuyangLayerHall}.  
The (inverse) spin Hall effect is also measurable in magnetic systems~\cite{sinova2015spin}, and is arguably richer in magnets because it can be coupled to switchable magnetic order~\cite{Kimata19}.  
Therefore, it stands as an exciting future direction to determine whether there exist 3D magnetic materials that exhibit the axionic (inverse) spin-magnetoelectric responses introduced in this work, as well as to determine how the spin-magnetoelectric responses of such magnetic materials relate to recently introduced theories of magnetoelectric multipoles~\cite{EdererToroidal07,SpaldinMonopole13}.
Furthermore, using the position-space formulations of the partial Chern numbers and partial axion angles (via layered partial Chern numbers), one can straightforwardly extend the spin-resolved topological quantities introduced in this work to the interacting setting using twisted spin boundary conditions~\cite{sheng2006spinChern,prodan2009robustness}.  
Lastly, the spin-resolved generalizations of the axion angle and parity anomaly numerically identified in this work may also admit analytic descriptions in the languages of Berry connections and quantum field theory, which we hope to explore in future studies.

{
\vspace{0.08in}
\centerline{\bf Methods}
\vspace{0.08in}
}
We here summarize the properties of the projected spin operator and the construction of spin-resolved and nested spin-resolved Wilson loops.  
We further provide a brief summary of the computation of layer-resolved partial Chern numbers and summarize our implementation of the Kubo formula for computing the spin-Hall conductivity. 
Lastly, we review the methods used for our ab-initio calculations of the electronic and spin spectrum of $\beta$-MoTe$_2$ and $\alpha$-BiBr.  
Complete details of our research methodology can be found in the extensive Supplementary Notes.

\vspace{0.1in}
{\bf Summary of Properties of the Projected Spin Operator}: Consider a $2N\times 2N$ matrix Bloch Hamiltonian $H(\mathbf{k})$. 
$H(\mathbf{k})$ acts on a Hilbert space consisting of $N$ spin-degenerate orbitals per unit cell (see SN~2A). 
Letting the Pauli matrices $\sigma_i$ act on the spin degrees of freedom, we can define the spin operators
\begin{equation}
s_i \equiv \sigma_i\otimes \mathbb{I}_N, \label{eq:appendix-def-s-i}
\end{equation}
where $\mathbb{I}_N$ is the $N\times N$ identity matrix acting in the orbital subspace of the entire Hilbert space (including both occupied and unoccupied states). Letting $P(\mathbf{k})$ represent the projector onto a set of ``occupied'' energy eigenstates at $\mathbf{k}$, we can then form the projected spin operator 
\begin{equation}
PsP\equiv P(\mathbf{k})\hat{\mathbf{n}}\cdot\mathbf{s}P(\mathbf{k}),\label{eq:appendix-def-PsP}
\end{equation} 
for any choice of unit vector $\hat{\mathbf{n}}$ (see SN~2B).
For notational convenience, we have frequently throughout this work suppressed the $\mathbf{k}$-dependence of $PsP$ when our discussion applies to both finite and infinite systems. 
When we are considering translationally-invariant systems, the projection operator $P$ is taken to be a $2N \times 2N$ $\mathbf{k}$-dependent matrix where $2N$ is the number of spinful orbitals within each unit cell. 
The spin operator $s_i$ is hence a $\mathbf{k}$-independent $2N\times 2N$ matrix.
When we are considering finite systems with open boundary conditions, the projection operator $P$ and the spin operator $s_i$ are both taken to be $2N \times 2N$ $\mathbf{k}$-independent matrices, where $2N$ is the number of spinful orbitals in the entire finite system.

In SN~2 we prove that the spectrum of the projected spin operator $PsP$ is gauge-invariant and changes continuously under perturbations of the Hamiltonian. 
This implies that the spectrum of $PsP$ is a well-defined and perturbatively robust physical object in an insulator or for energetically isolated bands. 
For either the occupied bands of an insulator, or more generally a set of energetically isolated bands, we can write the projector $P(\mathbf{k})$ as 
\begin{equation}
P(\mathbf{k})=P_+(\mathbf{k})+P_-(\mathbf{k}),\label{eq:pdecom}
\end{equation}
where $P_+(\mathbf{k})$ is the projection operator onto a subset of $PsP$ eigenstates with largest eigenvalue, and $P_-(\mathbf{k})$ is the projector onto the remaining $PsP$ eigenstates. 
For the spin-compensated systems considered in this work, we have typically taken the rank of $P_+(\mathbf{k})$ to be equal to the rank of $P_-(\mathbf{k})$, such that the decomposition in  Eq.~\eqref{eq:pdecom} partitions the occupied states into two equal sets. 
We then define a spin gap to exist when, for every $\mathbf{k}$, the smallest $PsP$ eigenvalue for states in the image of $P_+(\mathbf{k})$ is distinct from the largest $PsP$ eigenvalue for states in the image of $P_-(\mathbf{k})$.

{\bf Summary of the Spin-Resolved and Nested Spin-Resolved Wilson Loop Methods}: For systems with a spin gap, we can use the projection operators $P_+(\mathbf{k})$ and $P_{-}(\mathbf{k})$ to define Wilson loops. 
Specifically, we can write the matrix of $P_{\pm}(\mathbf{k})$ in the tight-binding Hilbert space in a basis of eigenstates $\ket{u^\pm_{n,\mathbf{k}}}$ of $PsP$ as
\begin{align}
    [P_{\pm}(\mathbf{k})] =  \sum_{n=1}^{N_{\text{occ}}^{\pm}} \ket{u_{n,\mathbf{k}}^{\pm}} \bra{u_{n,\mathbf{k}}^{\pm}},\label{eq:P_pm_k_projector}
\end{align}
where the square brackets indicate that $[P_{\pm}(\mathbf{k})]$ is a $2N\times 2N$ matrix. 
The occupied-space matrix projector $[P(\mathbf{k})]$ is then equal to $[P_{+}(\mathbf{k})]+[P_{-}(\mathbf{k})]$ where $[P_{+}(\mathbf{k})][P_{-}(\mathbf{k})]=0$.
The corresponding holonomy matrix for $[P_{\pm}(\mathbf{k})]$ starting at a base point $\mathbf{k}$ and continuing along a straight-line path to $\mathbf{k+G}$ (where $\mathbf{G}$ is a primitive reciprocal lattice vector)---which we term the $P_{\pm}$-Wilson loop matrix (or the spin-resolved Wilson loop matrix, see SN~3B)---is then given by the path-ordered product
\begin{align}
    [\mathcal{W}_{1,\mathbf{k},\mathbf{G}}^{\pm}]_{m,n} & = \bra{u_{m,\mathbf{k}+\mathbf{G}}^{\pm}} \left(\prod_{\mathbf{q}}^{\mathbf{k}+\mathbf{G} \leftarrow \mathbf{k}} [P_{\pm}(\mathbf{q})]\right)\ket{u_{n,\mathbf{k}}^{\pm}}.
    \label{eq:partialChernMethod}
\end{align}
In SN~3 we show that $[\mathcal{W}_{1,\mathbf{k},\mathbf{G}}^{\pm}]$ is a unitary matrix with eigenvalues $e^{i(\gamma^\pm_1)_{j,\mathbf{k,G}}}$. 
From this, we define the partial Chern numbers $C^\pm$ to respectively be equal to the winding numbers of $\sum_{j}(\gamma^\pm_1)_{j,\mathbf{k,G}}$ as functions of momenta perpendicular to $\mathbf{G}$ (see SN~3C for further details).

Going further, we can write the eigenvectors of $[\mathcal{W}_{1,\mathbf{k},\mathbf{G}}^{\pm}]$ as $[{\nu}^{\pm}_{j,\mathbf{k},\mathbf{G}}]_{m}$, which satisfy
\begin{equation}
[\mathcal{W}_{1,\mathbf{k},\mathbf{G}}^{\pm}][{\nu}^{\pm}_{j,\mathbf{k},\mathbf{G}}] = e^{i(\gamma^\pm_1)_{j,\mathbf{k,G}}}[{\nu}^{\pm}_{j,\mathbf{k},\mathbf{G}}].
\end{equation}
In the Bloch basis, we can then express the $P_\pm$-Wannier band eigenstates as
\begin{equation}
    \ket{w^{\pm}_{j,\mathbf{k},\mathbf{G}}} = \sum_{m=1}^{N_{\text{occ}}^{\pm}} [{\nu}^{\pm}_{j,\mathbf{k},\mathbf{G}}]_{m} \ket{u^{\pm}_{m,\mathbf{k}}}.
\end{equation}
If there is a gap between the eigenvalues $e^{i(\gamma^\pm_1)_{j,\mathbf{k,G}}}$, we can choose a subset $j=1,\dots N_W^\pm$ of Wilson loop eigenstates on which to form the $P_\pm$-Wannier band projector
\begin{align}
    [\widetilde{P}^{\pm}_{\mathbf{G}}(\mathbf{k})] = \sum_{j=1}^{N_{W}^{\pm}} \ket{{w}^{\pm}_{j,\mathbf{k},\mathbf{G}}} \bra{{w}^{\pm}_{j,\mathbf{k},\mathbf{G}}}. \label{eq:P_pm_nested_Wilson_loop_2nd_projector}
\end{align}

Lastly from Eq.~(\ref{eq:P_pm_nested_Wilson_loop_2nd_projector}), we then define the nested spin-resolved Wilson loops as the holonomy matrices that correspond to the $P_\pm$-Wannier band projectors $[\widetilde{P}^{\pm}_{\mathbf{G}}(\mathbf{k})]$. Concretely, the nested spin-resolved Wilson loop matrix $ [{\mathcal{W}}_{2,\mathbf{k},\mathbf{G},\mathbf{G}'}^{\pm}]$ (see SN~4B) is given by
\begin{equation}
    [{\mathcal{W}}_{2,\mathbf{k},\mathbf{G},\mathbf{G}'}^{\pm}]_{i,j}  = \bra{{w}^{\pm}_{i,\mathbf{k}+\mathbf{G}',\mathbf{G}}}  \left(\prod_{\mathbf{q}}^{\mathbf{k}+\mathbf{G}' \leftarrow \mathbf{k}} [\widetilde{P}^{\pm}_{\mathbf{G}}(\mathbf{q})]\right) \ket{{w}^{\pm}_{j,\mathbf{k},\mathbf{G}}}.
\end{equation}
In SN~4C, we further show that the spectrum of the nested spin-resolved Wilson loop defines the nested partial Chern numbers, by analogy to the partial Chern numbers defined above in the text following Eq.~(\ref{eq:partialChernMethod}).

{\bf Summary of the Layer-Resolved Partial Chern Number Calculation Method}:
We begin by considering a 3D system with the primitive Bravais lattice vectors $\{\mathbf{a}_j,\mathbf{a}_l,\mathbf{a}_i \}$.
We next cut our system into a slab geometry with $N_{i}$ unit cells (slab layers) along the (now-finite) $\mathbf{a}_{i}$ direction, while keeping the system infinite along $\mathbf{a}_j$ and $\mathbf{a}_l$.
We take there to be $N_{\mathrm{sta}}=2N_{\mathrm{orb}}$ tight-binding basis states per unit cell, where the factor of $2$ accounts for the on-site (internal) spin-1/2 degree of freedom.
In this basis, the spin operator oriented in a spin direction $\hat{\mathbf{n}}$ is defined as $s \equiv \hat{\mathbf{n}} \cdot \bm{\sigma} \otimes \mathbb{I}_{N_{\mathrm{orb}}} \otimes \mathbb{I}_{N_{i}}$, where the Pauli matrices $\bm{\sigma}$ act on the spin-$1/2$ degree of freedom, and where $\mathbb{I}_{N_{\mathrm{orb}}}$ and $\mathbb{I}_{N_{i}}$ are identity matrices that respectively act on the orbital and unit cell (layer) degrees of freedom.
We next denote the projection operator onto the occupied energy bands of the finite slab at the 2D crystal momentum $\mathbf{k}=(k_j , k_l)$ as $P(\mathbf{k})$.
The projector $P(\mathbf{k})$ can then be decomposed using the projected spin operator $P s P$ via Eq.~\eqref{eq:pdecom}.

Using each projector $P_{\pm}(\mathbf{k})$, we next obtain the partial Chern number of the finite 2D slab through (see SN~3C)
\begin{equation}
	C_{jl}^{\pm} = \frac{-i}{2\pi}\int d\mathbf{k}\ \mathrm{Tr}\left( P_\pm(\mathbf{k}) \left[ \frac{\partial P_\pm(\mathbf{k})}{\partial k_j} , \frac{\partial P_\pm(\mathbf{k})}{\partial k_l} \right] \right), \label{eq:main-text-Cpmjl-def}
\end{equation}
where the integral in $\mathbf{k}$ is performed over the 2D BZ of the slab, and where the matrix trace ($\mathrm{Tr}$) is performed over both the $N_i$ unit cells (layers) and the $2N_{\mathrm{orb}}$ tight-binding basis states per unit cell.
Using Eq.~\eqref{eq:main-text-Cpmjl-def}, we may further define a layer-resolved partial Chern number $C_{jl}^{\pm} (n_i)$ by expanding the matrix trace in the tight-binding basis and then re-summing (tracing) only over the degrees of freedom within each layer (see SN~5D for further details).
The layer-resolved partial Chern number $C_{jl}^{\pm}(n_i)$ specifically quantifies how the partial Chern number of a 2D slab is distributed over the $N_{i}$ unit cells (layers) in the finite slab, and can be viewed as the spin-resolved generalization of the well-established position-space (layer-resolved) Chern number~\cite{essin2009magnetoelectric,varnava2018surfaces,IvoAXI1}.

{\bf Spin-Hall Conductivity}: 
As detailed in SN~7, the spin conductivity tensor $\sigma^{s,i}_{\mu\nu}$ parametrizes the linear response of the spin current $\mathbf{J}^{s,i}$ to an applied DC electric field $\mathbf{E}$ via
\begin{equation}
\langle J^{s,i}_\mu\rangle = \sum_\nu \sigma^{s,i}_{\mu\nu}E_\nu.
\end{equation}
Here $\mu$ and $\nu$ index spatial coordinates, and $i=x,y,z$ indexes the spin direction. 
The spin conductivity can then be evaluated using the standard Kubo formula
\begin{equation}
\sigma^{s,i}_{\mu\nu} = \lim_{\epsilon\rightarrow 0} \int_0^\infty dt \langle \left[J^{s,i}_\mu(t), X_\nu(0)\right]\rangle e^{-\epsilon t},
\label{eq:spinhallkubo}
\end{equation}
where $X_\nu$ is the $\nu$ component of  the position operator (which couples to the external electric field in the Hamiltonian), the time-dependence of operators is evaluated in the Heisenberg picture using the unperturbed ($\mathbf{E}={\bf 0}$) Hamiltonian $H_0$, and the average is computed with respect to the unperturbed ground state. 
We next define the spin current operator to be
\begin{equation}
J^{s,i}_{\mu} = \frac{\partial}{\partial t}\left(X_\mu s^i\right) = i\left[H_0, X_\mu s^i\right].
\label{eq:spincurrentdef}
\end{equation}
From Eqs.~(\ref{eq:spinhallkubo}) and~(\ref{eq:spincurrentdef}), we then define the spin Hall conductivity to be the antisymmetric part of the spin conductivity tensor. 
To numerically evaluate Eq.~\eqref{eq:spinhallkubo} for a tight-binding model, we work in a hybrid Wannier basis following the approach of Ref.~\cite{Monaco2022shc_nonconserved_spin}. 
For a semi-infinite 3D system consisting of a finite number of 2D layers, this also allows us to define a layer-resolved spin Hall conductivity as the integrand of Eq.~\eqref{eq:spinhallkubo} before taking the sum over layers (see SN~7C).

{\bf Ab-Initio Calculation Details}: 
We here detail our first-principles (DFT) calculations for $\beta$-MoTe$_2$ and $\alpha$-BiBr.
First, as detailed in SN~9, our first-principles calculations for $\beta$-MoTe$_2$ were performed within the DFT framework using the projector-augmented wave (PAW) method~\cite{blochl1994improved,kresse1994norm} as implemented in the Vienna ab-initio simulation package (VASP)~\cite{kresse1996efficiency,kresse1996efficient}. 
In our DFT calculations for $\beta$-MoTe$_2$, we adopted the Perdew-Burke-Ernzerhof (PBE) generalized gradient approximation exchange-correlation functional~\cite{perdew1996generalized}, and 
SOC was incorporated self-consistently. 
The cutoff energy for the plane-wave expansion was 400 eV, and $0.03 \times 2\pi \text{ \AA}^{-1}$ $\mathbf{k}$-point sampling grids were used in the self-consistent process.

To analyze the spin-resolved band topology, we constructed a symmetric, Wannier-based tight-binding model fit to the electronic structure of $\beta$-MoTe$_2$ obtained from our DFT calculations.
We constructed symmetric Wannier functions for the bands near the Fermi energy $E_{F}$ in $\beta$-MoTe$_2$ by using the Wannier90 package~\cite{pizzi2020wannier90} for the Mo $4d$ and the Te $5p$ orbitals,  and then performing a subsequent SG symmetrization using WannierTools~\cite{WU2018405}. 
We denote the Hamiltonian of the Wannier-based tight-binding model as $[H_{\mathrm{MoTe}_2}]$. 
The single-particle Hilbert space of $[H_{\mathrm{MoTe}_2}]$ consists of 44 spinful Wannier functions per unit cell; 
the Bloch Hamiltonian $[H_{\mathrm{MoTe}_2}(\mathbf{k})]$ is therefore an $88\times 88$ matrix. 
To reduce the computational resources required for our spin-resolved and Wilson loop calculations, we next truncated $[H_\mathrm{MoTe_2}]$ to only contain hopping terms with an absolute magnitude greater than or equal to $0.001$eV. 
We have confirmed that this truncation affects neither the band ordering nor the qualitative features of the band structure near the Fermi energy in $\beta$-MoTe$_2$ (see SN~9A for complete calculation details).

Next, as detailed in SN~10, our first-principles calculations for $\alpha$-BiBr were also performed within the DFT framework using the PAW method~\cite{blochl1994improved,kresse1994norm} as implemented in VASP~\cite{kresse1996efficiency,kresse1996efficient}.  
In our DFT calculations for $\alpha$-BiBr, we similarly adopted the PBE generalized gradient approximation exchange-correlations functional~\cite{perdew1996generalized}, and SOC was also incorporated self-consistently. 
The cutoff energy for the plane-wave expansion was 400eV, and $0.03 \times 2\pi \text{ \AA}^{-1}$ $\mathbf{k}$-point sampling grids were used in the self-consistent process.

To analyze the spin-resolved band topology of $\alpha$-BiBr, we next constructed a symmetric, Wannier-based tight-binding model fit to the electronic structure of $\alpha$-BiBr obtained from our DFT calculations.
We specifically constructed symmetric Wannier functions for the bands near $E_{F}$ in $\alpha$-BiBr by using the Wannier90 package~\cite{pizzi2020wannier90} for the Bi $6p$ and the Br $4p$ orbitals, and then performing a subsequent SG symmetrization using WannierTools~\cite{WU2018405}. 
We denote the tight-binding Hamiltonian of the Wannier-based tight-binding model as $H_{\mathrm{Bi}\mathrm{Br}}$.
The single-particle Hilbert space of $H_{\mathrm{Bi}\mathrm{Br}}$ consists of 48 spinful Wannier functions per primitive (unit) cell; the Bloch Hamiltonian $[H_{\mathrm{Bi}\mathrm{Br}}(\mathbf{k})]$ is therefore a $96\times 96$ matrix, 
To reduce the computational resources required for our spin-resolved and Wilson loop tight-binding calculations, we then truncated $[H_{\mathrm{Bi}\mathrm{Br}}(\mathbf{k})]$ to only contain hopping terms with an absolute magnitude greater than or equal to $0.001$eV. We have confirmed that the truncated Wannier-based tight-binding model exhibits the same band ordering and qualitative features as the first-principles electronic structure of $\alpha$-BiBr (see SN~10A for complete calculation details).

{
\vspace{0.08in}
\centerline{\bf Data Availability}
\vspace{0.08in}
}

The data supporting the theoretical findings of this study are available within the paper and as code examples in the \href{https://github.com/kuansenlin/nested_and_spin_resolved_Wilson_loop}{\textsc{nested\_and\_spin\_resolved\_Wilson\_loop}}~\cite{lin2023nestedWilsonLib} repository. All first-principles calculations were performed using CIF structure files with the experimental lattice parameters, which can be obtained from the Inorganic Crystal Structure Database (ICSD)~\cite{bergerhoff1983inorganic} using the accession numbers provided in SN~8.

{
\vspace{0.08in}
\centerline{\bf Code Availability}
\vspace{0.08in}
}

The spin-resolved tight-binding and (nested) Wilson loop calculations in this work were performed using the freely available Python package~\href{https://github.com/kuansenlin/nested_and_spin_resolved_Wilson_loop}{\textsc{nested\_and\_spin\_resolved\_Wilson\_loop}}~\cite{lin2023nestedWilsonLib}, which represents an extension of the~\href{https://www.physics.rutgers.edu/pythtb/}{PythTB} open-source Python tight-binding package~\cite{coh2013python} that was implemented and utilized for the preparation of Refs.~\cite{wieder2018axion,wieder2020strong}, and was then greatly expanded for the present work.

{
\vspace{0.08in}
\centerline{\bf Acknowledgments}
\vspace{0.08in}
}
We thank Andrey Gromov and Inti Sodemann for insightful discussions during the early stages of this study.  
We further acknowledge helpful discussions with Frank Schindler, Senthil Todadri, and Fan Zhang.  
Concurrent with the preparation of this work, a bulk spin-magnetoelectric response and anomalous surface halves of 2D TI states were detected in helical HOTIs in Ref.~\cite{WiederDefect} through numerical studies of the charge and spin bound to threaded magnetic flux.  
During the preparation of this work, spin-resolved topology was also explored in 2D antimonene and bismuthene~\cite{SpinChernBismuthene1,SpinChernBismuthene2}.  
Additionally, during the preparation of this work, a semiclassical treatment of a spinor-axion response was explored in relation to HOTIs in Ref.~\cite{ZilberbergSpinAxion}.  
After the initial submission of this work, nontrivial (pseudo)spin-resolved partial axion angles were also identified in magnetic helical TCIs in Ref.~\cite{YoonseokMagneticDirac} by implementing the method proposed in the present work. 
Lastly, after the submission of this work, the spin texture of $\alpha$-BiBr was measured through spin-ARPES experiments~\cite{BiBrSpinARPES}, and showed close agreement with the DFT-based spin gap calculations performed in this work.

The numerical calculations performed for and experimental proposals introduced in this work were supported by the Center for Quantum Sensing and Quantum Materials, an Energy Frontier Research Center funded by the U.~S. 
Department of Energy, Office of Science, Basic Energy Sciences under Award DE-SC0021238. 
The analytical calculations performed by K.-S.~L. and B.~B. were additionally supported by the Alfred P. Sloan foundation and the National Science Foundation under Grant DMR-1945058. 
K.-S.~L. also acknowledges the Graduate Fellowship Program at the Kavli Institute for Theoretical Physics, University of California, Santa Barbara, supported in part by the National Science Foundation under Grant No. NSF PHY-1748958 and NSF PHY-2309135, the Heising-Simons Foundation, and the Simons Foundation (216179, LB), during which this work was finalized.
The work of Y. H. on the electromagnetic response of HOTI phases was supported by the Air Force Office of Scientific Research under award number FA9550-21-1-0131.  
Y. H. received additional support from the US Office of Naval Research (ONR) Multidisciplinary University Research Initiative (MURI) Grant N00014-20-1-2325 on Robust Photonic Materials with High-Order Topological Protection.
This work made use of the Illinois Campus Cluster, a computing resource that is operated by the Illinois Campus Cluster Program (ICCP) in conjunction with the National Center for Supercomputing Applications (NCSA) and which is supported by funds from the University of Illinois at Urbana-Champaign. 
J.~B. was supported through the National Science Foundation under Grant IIS-2046590 and provided additional computing resources via iDRAMA.cloud, funded by the National Science Foundation under Grant CNS-2247867. 
Z.~G. and Z.~W. were supported by the National Natural Science Foundation of China (Grants No. 11974395, No. 12188101), the Strategic Priority Research Program of Chinese Academy of Sciences (Grant No. 
XDB33000000), the China Postdoctoral Science Foundation funded project (Grant No. 2021M703461), and the Center for Materials Genome. 
B.~J.~W. acknowledges support from the European Union’s Horizon Europe research and innovation program (ERC-StG-101117835-TopoRosetta).
G.~A.~F. and B.~J.~W. were additionally supported by the Department of Energy (BES) Award No. DE-SC0022168 and the National Science Foundation Grant No. DMR-2114825.
B.~J.~W. further acknowledges the Laboratoire de Physique des Solides, Orsay for hosting during some stages of this work.

{
\vspace{0.08in}
\centerline{\bf Author Contributions}
\vspace{0.08in}
}

The spin-resolved and nested Wilson loop formalism was developed by K.-S.~L. with guidance from B.~B., B.~J.~W., G.~P., and G.~A.~F.
The theories of partial bulk axion angles and surface partial parity anomalies were developed by G.~P. and B.~J.~W. under the guidance of G.~A.~F. and B.~B.
Material candidates were identified by B.~J.~W., B.~B., and Z.~W. with input from D.~P.~S.
The ab-initio calculations and construction of Wannier-based tight-binding models were performed by Z.~G. and Z.~W. 
The DFT-based models were then analyzed by K.-S.~L., B.~B., Y.~H., and B.~J.~W. with input from Z.~G., Z.~W., G.~P., and G.~A.~F. and computational support from J.~B.
The experimental proposals were devised by F.~M., B.~B., and B.~J.~W. with input from K.-S.~L., G.~P., and G.~A.~F.
The manuscript was written by K.-S.~L., B.~J.~W, and B.~B. with input from all the authors.
The project was conceived and directed by B.~B. and B.~J.~W.

{
\vspace{0.08in}
\centerline{\bf Competing Interests Statement}
\vspace{0.08in}
}

The authors declare no competing interests.


\bibliographystyle{naturemag}
\bibliography{refs}

\end{document}


\title{Supplementary Information for ``Spin-Resolved Topology and Partial Axion Angles in Three-Dimensional Insulators''}

\author{Kuan-Sen Lin}
\thanks{Corresponding author:~\href{mailto:kuansen2@illinois.edu}{kuansen2@illinois.edu}}
\affiliation{Department of Physics and Institute for Condensed Matter Theory, University of Illinois at Urbana-Champaign, Urbana, IL 61801, USA}
\affiliation{Kavli Institute for Theoretical Physics, University of California, Santa Barbara, CA 93106, USA}
\author{Giandomenico Palumbo}
\affiliation{School of Theoretical Physics, Dublin Institute for Advanced Studies, 10 Burlington Road, Dublin 4, Ireland}
\author{Zhaopeng Guo}
\affiliation{Beijing National Laboratory for Condensed Matter Physics and Institute of Physics, Chinese Academy of Sciences, Beijing 100190, China}
\affiliation{University of Chinese Academy of Sciences, Beijing 100049, China}
\author{Yoonseok Hwang}
\affiliation{Department of Physics and Institute for Condensed Matter Theory, University of Illinois at Urbana-Champaign, Urbana, IL 61801, USA}
\author{Jeremy Blackburn}
\affiliation{Department of Computer Science, State University of New York at Binghamton, Binghamton, NY 13902, USA}

\author{Daniel P.~Shoemaker}
\affiliation{Department of Materials Science and Engineering, University of Illinois at Urbana-Champaign, Urbana, IL 61801, USA}
\affiliation{Materials Research Laboratory, University of Illinois at Urbana-Champaign, Urbana, IL 61801, USA}

\author{Fahad Mahmood}
\affiliation{Materials Research Laboratory, University of Illinois at Urbana-Champaign, Urbana, IL 61801, USA}
\affiliation{Department of Physics, University of Illinois at Urbana-Champaign, Urbana, IL 61801, USA}

\author{Zhijun Wang}
\affiliation{Beijing National Laboratory for Condensed Matter Physics and Institute of Physics, Chinese Academy of Sciences, Beijing 100190, China}
\affiliation{University of Chinese Academy of Sciences, Beijing 100049, China}
\author{Gregory A.~Fiete}
\thanks{Corresponding author:~\href{mailto:g.fiete@northeastern.edu}{g.fiete@northeastern.edu}}
\affiliation{Department of Physics, Northeastern University, Boston, MA 02115, USA}
\affiliation{Department of Physics, Massachusetts Institute of Technology, Cambridge, MA 02139, USA}
\author{Benjamin J.~Wieder}
\thanks{Primary address for B. J. W.: Institut de Physique Th\'eorique, Universit\'{e} Paris-Saclay, CEA, CNRS, F-91191 Gif-sur-Yvette, France.  Corresponding author:~\href{mailto:benjamin.wieder@ipht.fr}{benjamin.wieder@ipht.fr}}
\affiliation{Department of Physics, Northeastern University, Boston, MA 02115, USA}
\affiliation{Department of Physics, Massachusetts Institute of Technology, Cambridge, MA 02139, USA}
\affiliation{Institut de Physique Th\'eorique, Universit\'{e} Paris-Saclay, CEA, CNRS, F-91191 Gif-sur-Yvette, France}
\author{Barry Bradlyn}
\thanks{Corresponding author:~\href{mailto:bbradlyn@illinois.edu}{bbradlyn@illinois.edu}}
\affiliation{Department of Physics and Institute for Condensed Matter Theory, University of Illinois at Urbana-Champaign, Urbana, IL 61801, USA}

\date{\today}

\maketitle

\onecolumngrid

\clearpage
\tableofcontents
\clearpage

\section*{\bf Supplementary Notes}

\section{Introduction to the Supplementary Notes}
In this section, we provide a guide to the Supplementary Notes, and summarize the main results of this work.
We begin in Supplementary Note (SN)~\ref{sec:spin_operators} with a discussion of the general properties of projected spin operators, building on the formalism introduced by Prodan in Supplementary Reference (SRef.)~\cite{prodan2009robustness}. 
In SN~\ref{appendix:TB-notation}, we first review the notation that we will use for states and operators in tight-binding models. 
The tight-binding models that we consider in this work include both physically motivated toy models as well as systematically constructed, Wannier-based tight-binding truncations that reproduce a set of bands in a real material. 
Next, in SN~\ref{appendix:properties-of-the-projected-spin-operator}, we introduce the projected spin operator $PsP$. 
We show how the projected spin operator can be defined in terms of a projector onto a set of ``occupied'' (Bloch) states and a choice of spin direction $s = \mathbf{s}\cdot\hat{\mathbf{n}}$. 
For periodic systems, we introduce the spin band structure (spin spectrum), which we define through the gauge-invariant eigenvalues of the projected spin operator. 
We examine constraints imposed on the spin operator and the spin spectrum by time-reversal and crystal symmetries, and provide a precise definition of a \emph{spin gap} in the spin spectrum. 
In SN~\ref{sec:pspperturbation}, we show how to compute changes to the spin band structure order by order in perturbations to the Hamiltonian. 
This expansion establishes the spin band structure as a well-defined, perturbatively-robust physical object in an insulator given a choice of spin direction. 
In particular in SN~\ref{sec:pspperturbation}, we show that perturbations to the Hamiltonian induce smooth and bounded changes to the spin band structure. 
Building from this result, we then show in SN~\ref{sec:physical} that the smallest nontrivial eigenvalue of the projected spin operator---which determines the spin gap in the presence of bulk inversion and time-reversal symmetries---places a bound on the relaxation time for spin-flip excitations in an insulating material.

Exploiting the fact that the energy and spin gaps are robust to perturbations, we continue in SN~\ref{appendix:explicit-calculation-of-BHZ-model} by defining the concept of \emph{spin-resolved band topology}, beginning with an explicit calculation of the spin band structure for a simple model of a three-dimensional (3D) inversion- and time-reversal-symmetric topological insulator (TI). 
We review the notion of spin-resolved partial Berry curvature as a 2D extension of the partial polarization introduced in SRefs.~\cite{fu2006time,prodan2009robustness}. 
This allows us to introduce in SN~\ref{appendix:explicit-calculation-of-BHZ-model} a formulation of partial Chern numbers, defined as the Chern numbers computed separately for the spin bands in each half of a gapped spin spectrum.
We next argue that in a 3D insulator, there can exist linearly-dispersing, twofold degeneracies in the spin spectrum---which we term spin-Weyl nodes---that act as monopole sources of partial Chern number. 
Spin-Weyl nodes are hence analogous to (energy) Weyl nodes, which act as monopole sources of ordinary (total, \emph{i.e.} charge) Chern number. 
We then prove that 3D TIs must generically have an odd number of spin-Weyl nodes in their spin band structures in each half of the Brillouin zone (BZ).

In SN~\ref{appendix:orbital-texture-and-spin-gap}, we next explore how entanglement between spin and orbital degrees of freedom can affect the spin band structure, focusing on the effect of the choice of spin direction $s$ in $PsP$. 
Then, in SN~\ref{sec:zeeman}, we derive a relationship between the spin band structure and the (energy) band structure in the presence of a strong Zeeman field. 
We rigorously show that the band structure for a spectrally flattened Hamiltonian in the presence of a strong Zeeman field is adiabatically deformable to the spin band structure. 
Going further, we argue that in many cases, we can extend the correspondence and link the spin band structure to the energy spectrum \emph{without} spectral flattening, allowing us to relate the topology and connectivity of bands in the spin spectrum to the topology and connectivity of bands in the energy system for an insulator in the presence of a strong Zeeman field.
Using this result, we argue in SN~\ref{sec:zeeman} that a 3D TI (spin-Weyl semimetal) in a strong Zeeman field will generically have low-energy Weyl nodes in the \emph{energy} spectrum, which can be adiabatically connected to spin-Weyl nodes in the spin band structure.

In SN~\ref{app:Wilson} we continue our study of the topology of spin bands by introducing a general formalism for computing spin-resolved topological invariants. 
We begin in SN~\ref{sec:P_Wilson_loop} with a short review of the ordinary Wilson loop (non-Abelian Berry phase) for a set of occupied states~\cite{yu2011equivalent,alexandradinata2014wilsonloop,taherinejad2014wannier}. 
Then, in SN~\ref{sec:P_pm_Wilson_loop} we use the spin band structure to define a spin-resolved Wilson loop for systems with a spin gap, where the ``occupied'' states are taken to be the spin bands in one half of the spin spectrum of an insulator with a spin gap.
Crucially, our formalism for spin-resolved topology \emph{does not} require the system to carry a conserved spin direction (\emph{i.e.} $s_{z}$ symmetry); the notion of spin-resolved band topology introduced in this work remains valid when spin-conservation symmetry is broken by spin-orbit coupling (SOC), which typically cannot be neglected in real materials~\cite{kane2005quantum}.
In SN~\ref{sec:general_properties_of_winding_num_of_P_pm_Wilson} we show how the partial Chern numbers first introduced in SN~\ref{appendix:explicit-calculation-of-BHZ-model} can be computed from the eigenphases of the spin-resolved Wilson loop. 
Making contact with prior work, we then show how the partial Chern numbers computed in this work can be used to compute the spin Chern numbers introduced in SRefs.~\cite{prodan2009robustness,sheng2006spinChern,fu2006time}. 
By applying our definition of the partial Chern number to the Kubo formula for spin Hall conductivity, we further show that nontrivial partial Chern numbers indicate the presence of an intrinsic, topological contribution to the spin Hall conductivity, even in systems where spin conservation is (weakly) broken by SOC.

Having established a formal definition for the spin-resolved Wilson loop, we next consider several illustrative examples of band topology identifiable from spectral flow in the (first) spin-resolved Wilson loop. 
In SN~\ref{sec:2D-spinful-TRI-system}, we begin by applying our spin-resolved Wilson loop formalism to the case of 2D time-reversal-invariant TIs. 
We specifically review the results of SRef.~\cite{prodan2009robustness}, in which it was shown that the partial Chern numbers of the occupied bands of an insulator can only change by even integers when a spin gap closes, noting that spin gap closures are typically unaccompanied by energy gap closures. 
Using this result, we demonstrate that in the presence of time-reversal symmetry, the parity of the partial Chern number cannot change without closing an energy gap, and hence provides an alternative definition of the Kane-Mele $\mathbb{Z}_2$ index for 2D insulators with spinful time-reversal symmetry. 
In SN~\ref{sec:main-text-3D-TI-P-pm} and \ref{appendix:3D-TI-with-and-without-inversion} we next compute the spin-resolved Wilson loops for models of a time-reversal-invariant 3D TI both with and without spatially inversion symmetry. 
We numerically confirm the presence of an odd number of spin-Weyl nodes in each half of the BZ, consistent with the theoretical result established in SN~\ref{appendix:explicit-calculation-of-BHZ-model}.
We additionally numerically demonstrate that the spin-Weyl nodes act as sources and sinks of partial Berry curvature. 
In SN~\ref{sec:spin_stable_topology_2d_fragile_TI}, we next examine the spin-resolved Wilson loops of the model of a 2D fourfold-rotation- and time-reversal-symmetric fragile topological insulator first introduced in SRef.~\cite{wieder2020strong}. 
We demonstrate that the topologically fragile model has a nonzero even partial Chern number, which we find to be stable to the addition of trivial bands to $PsP$ (bands with vanishing partial Chern numbers), provided that the spin gap remains open as the trivial bands are coupled to the system.

This motivates introducing a refined notion of band topology for bands within the spin spectrum, which we term ``spin-stable'' topology. 
Systems with inequivalent spin-stable topology have spin band structures that cannot be adiabatically deformed into each other without breaking a symmetry or closing either an energy gap or a spin gap. 
As we show in SN~\ref{sec:spin_stable_topology_2d_fragile_TI}, this definition is physically motivated, as systems with distinct spin-stable topology exhibit different spin-electromagnetic responses, \emph{even if they share the same electronic band topology} without spin resolution. 
In SN~\ref{sec:spin_entanglement_spectrum}, we then formulate a notion of a spin-resolved entanglement spectrum by restricting the projector onto the lower (or upper) spin bands to one half of a system in position space. 
We show that this particular formulation of the spin entanglement spectrum is homotopic to the spectrum of the spin-resolved Wilson loop taken in the direction perpendicular to the position-space bipartition. 
We demonstrate this concretely by computing the spin entanglement spectrum for a 2D TI (which features chiral modes in the entanglement spectrum), a 3D TI (which features Fermi arcs in the entanglement spectrum emanating from spin-Weyl nodes), and the spin-stable 2D fragile model from SRef.~\cite{wieder2020strong} (which features a pair of chiral modes signifying an even partial Chern number).
Our results are also relevant to ongoing experimental efforts exploring the interplay between spin and topology in quantum materials~\cite{schaoeler2020local,unzelmann2021momentum,cho2021studying,di2023flat}.

In SN~\ref{app:w2}, we use the projected spin operator to study crystal-symmetry-protected, spin-stable topology in 3D insulators, which we accomplish by computing \emph{nested} Wilson loops.
We begin in SN~\ref{sec:nested_P_Wilson_loop} by reviewing the nested Wilson loop formalism of SRefs.~\cite{benalcazar2017quantized,benalcazar2017electric,wieder2018axion}. 
We review how the Wilson loop matrix from SN~\ref{sec:P_Wilson_loop} defines a set of Wannier bands and show that if there is a gap in the Wilson loop spectrum, we can compute a second (nested) Wilson loop by projecting onto a subset of the Wannier bands. 
We pay particular attention to the numerical subtleties involved in calculating the projector onto subsets of Wannier bands, and introduce a robust method for computing the nested Wilson loop in tight-binding models. 
In SN~\ref{sec:nested_P_pm_Wilson_loop} we define a nested spin-resolved Wilson loop by first identifying the eigenstates of the spin-resolved Wilson loop as spin-resolved Wannier bands (\emph{i.e.} the Wannier bands of the Wilson loop computed using the spin bands of $PsP$). 
We then define the nested spin-resolved Wilson loop as the Wilson loop computed for a projector onto a subset of the spin-resolved Wannier bands.

We continue in SN~\ref{sec:general_properties_nested_ppm} by discussing general properties of nested (spin-resolved) Wilson loops and the relationship between the nested (spin-resolved) Wilson spectrum and (spin-stable) bulk topology.
In SN~\ref{app:W2_same_1st_2nd_G} and \ref{app:nested_Berry_phase_only_winds_in_one_direction}, we establish that in systems with an energy gap and a (spin-resolved) Wannier gap, the nested (partial) Wilson loop eigenvalues can have nonzero winding numbers as functions of at most one crystal momentum. 
This result, which was implicit in SRefs.~\cite{varnava2020axion,wieder2018axion}, allows us to define a nested (partial) Chern number independent of the base point used to define the nested Wilson loop. 
Using this result, in SN~\ref{app-relation-nested-P-pm-and-partial-weak-Chern} we show that the sum of the nested (partial) Chern numbers over all sets of (spin-resolved) Wannier bands gives the weak (partial) Chern number of the occupied bands. 
In SN~\ref{app:z2_nested_P_pm_berry_phase} we specialize to systems with both inversion and time-reversal symmetry. 
We show that for insulators with these symmetries, the nested partial Chern number can only change by an even number when a gap  in the spin-resolved Wannier spectrum closes and reopens (provided that the spin gap and energy gap remain open). 
We then relate this result to established topological phases, most notably the topological crystalline insulating states that have become known as higher-order TIs (HOTIs)~\cite{benalcazar2017electric,po2017symmetry,schindler2018higherorder,schindler2018higherordera,fang2019new,khalaf2018symmetry,song2017ensuremath2,khalaf2018higherorder,langbehn2017reflectionsymmetric,HigherOrderTIPiet2,TynerGoswamiSU2Winding,FanHOTI,EzawaMagneticHOTI,ZeroBerry,FulgaAnon,wang2019higherorder,HarukiLayers,song2018mapping,S4Guys,TitusInteractHOTI,varnava2018surfaces,wieder2018axion,ahn2019symmetry,varnava2020axion,IvoAXI1,olsen2020gapless,TaylorHingeSC,YongXuHOTIReview}.

We next consider the position-space implications of our nested (spin-resolved) Wilson loop calculations.
In SN~\ref{app:comparison-spin-stable-and-symmetry-indicated-topology} we formulate a precise relationship between the nested spin-resolved Wilson loop and a novel notion of \emph{spin-resolved layer constructions} of 3D spin-stable topological phases, generalizing the approach of SRefs.~\cite{song2018mapping,song2019topological,elcoro2021magnetic,gao2022magnetic}. 
We specifically show that the layer constructions for spin-stable topological phases with inversion and time-reversal symmetry can be obtained from the layer constructions for magnetic topological insulators with inversion symmetry, which were previously discussed in SRefs.~\cite{watanabe2018structure,elcoro2021magnetic,gao2022magnetic,peng2021topological}. 
We show how spin-stable topology refines the standard classification of topological phases by focusing on the representative example of a helical HOTI, a phase of matter whose boundary-independent bulk characterization and response have eluded earlier investigations~\cite{teo2008surface,slager2013space,khalaf2021boundaryobstructed,benalcazar2017electric,schindler2018higherorder,schindler2018higherordera,fang2019new,khalaf2018symmetry,song2017ensuremath2,khalaf2018higherorder,langbehn2017reflectionsymmetric,HigherOrderTIPiet2,TynerGoswamiSU2Winding,FanHOTI,EzawaMagneticHOTI,ZeroBerry,FulgaAnon,wang2019higherorder,HarukiLayers,song2018mapping,po2017symmetry,S4Guys,TitusInteractHOTI,varnava2018surfaces,wieder2018axion,tang2019efficient,ahn2019symmetry,varnava2020axion,IvoAXI1,olsen2020gapless,TaylorHingeSC,YongXuHOTIReview}.
A boundary-independent characterization of helical HOTIs is urgently needed to better understand ongoing experimental and theoretical studies~\cite{nayak2019resolving,SYBiBr,BiBrFanHOTI,schindler2018higherordera,AliEarlyBismuth,BismuthSawtooth,BismuthSawtooth2,ZeroHallExp,huang2019polar,wang2020evidence,choi2020evidence,WTe2HingeStep,OtherWTe2Hinge,MazNatPhysHOTIReview,noguchi2021evidence,BiBrHingeExp1,BiBrHingeExp2,FanRoomTempBiIPRX,FanZahidRoomTempBiBrExp,Bi4Bi2I2WTINatComm,BismuthSurfaceBallisticExp,FanGateTunable,BiBrFacetDependent,spinfulHingeWTe2,herveHingeSTM}, whose results prior to this study could only be interpreted through the language of 1D helical hinge modes, which manifest in configurations dependent on sample details and surface physics.
In SN~\ref{app:comparison-spin-stable-and-symmetry-indicated-topology}, we show that there exist two \emph{physically distinguishable} spin-stable, spin-resolved layer-constructable topological phases that can both be adiabatically connected to the layer construction of a helical HOTI.  
The first spin-stable, spin-gapped layer-constructable state is a 3D quantum spin Hall insulator (QSHI)~\cite{zhijun2021qshsquarenet}, which has an extensively large topological contribution to the spin Hall conductivity in a finite sample. 
Previous investigations have recognized the possibility of  3D QSHI states in time-reversal-invariant 3D topological crystalline phases~\cite{wang2016hourglass,zhijun2021qshsquarenet}.

However, there also exists a second, previously unrecognized spin-stable and spin-gapped helical HOTI formed from a time-reversal-doubled magnetic axion insulator. 
The time-reversal-doubled axion insulator (T-DAXI) state can specifically be viewed as a superposition of two magnetic axion insulators, where each magnetic axion insulator originates from bands within each half of the spin spectrum (\emph{i.e.} two superposed magnetic axion insulators with time-reversed spin-orbital textures).
We relate this result to the case of a 3D TI or magnetic axion insulator, which are both characterized by quantized magnetoelectric theta angles $\theta=\pi$ (where $\theta$ is defined modulo $2\pi$)~\cite{essin2009magnetoelectric,vanderbilt2018berry,qi2008topological,wilczek1987two,hughes2011inversionsymmetric,turner2012quantized,wieder2018axion,xu2019higherorder,varnava2018surfaces,varnava2020axion,wang2019higherorder,ahn2019symmetry,CohVDBAXI,ArisHopf,TitusRonnyKondoAXI,ahn2018band,kim2019glidesymmetric,takahashi2020bulkedge,HingeStateSphereAXI,BJYangVortex,wieder2020axionic,IvoAXI1,olsen2020gapless,S4Guys,MulliganAnomaly,yu2021dynamical,ahn2022theory,SuyangAxionOpticalExp}.  
In particular, $\theta$ transforms like $\mathbf{E}\cdot\mathbf{B}$, where $\mathbf{E}$ is the electric field and $\mathbf{B}$ is the magnetic field, and is hence quantized by any symmetry, such as inversion or time-reversal, that takes $\mathbf{E}\cdot\mathbf{B}\rightarrow-\mathbf{E}\cdot\mathbf{B}$.
Using this well-established classification of axionic insulators, we demonstrate in SN~\ref{app:comparison-spin-stable-and-symmetry-indicated-topology} that the spin bands in each half of the spin spectrum in a T-DAXI are characterized by a novel topological quantity:  a nontrivial \emph{partial} axion angle $\theta^\pm = \pi$.  
In the case of the T-DAXI regime of an inversion- and time-reversal-protected helical HOTI, we find that $\theta^{\pm}$ are individually quantized by inversion symmetry, and related by time-reversal.  
We conclude SN~\ref{app:comparison-spin-stable-and-symmetry-indicated-topology} by showing that the adiabatic deformation between a T-DAXI and a 3D QSHI involves an intermediate spin-stable (but spin-gapless) phase with an even number of spin-Weyl points in each half of the BZ; we show that this ``spin-Weyl semimetal'' regime is equivalent to the ``doubled strong TI'' (DSTI) construction of a helical HOTI introduced in SRef.~\cite{po2017symmetry}. 
In SN~\ref{sec:numerical-section-of-nested-P-pm}, we confirm our theoretical analysis by numerically computing the nested spin-resolved Wilson loops for the model of a 3D helical HOTI introduced in SRef.~\cite{wieder2018axion}. 
We first show that the simple eight-band model exhibits a novel form of ``spin-fragile'' Wilson loop winding, which implies the existence of 2D insulators with fragile spin-resolved band topology. 
We then add trivial bands to the model and explicitly compute the partial axion angles $\theta^{\pm}$ using a spin-resolved generalization of the nested Wilson loop indicator for $\theta$ introduced in SRef.~\cite{wieder2018axion}.
We find that the 3D helical HOTI model introduced in SRef.~\cite{wang2019higherorder} indeed resides in the T-DAXI regime, and demonstrate that its bulk-quantized partial axion angles remain nontrivial in the presence of spin-nonconserving SOC.  The calculations detailed in SN~\ref{sec:numerical-section-of-nested-P-pm} -- as well as the other extensive toy-model and real-material spin-resolved and nested Wilson loop calculations in this work -- were performed using the freely accessible Python package~\href{https://github.com/kuansenlin/nested_and_spin_resolved_Wilson_loop}{\textsc{nested\_and\_spin\_resolved\_Wilson\_loop}}~\cite{lin2023nestedWilsonLib}, which was previously implemented and utilized for the preparation of SRefs.~\cite{wieder2018axion,wieder2020strong}, and was then greatly refined and extended to spin-resolved calculations for the present work.

In SN~\ref{app:layer-resolved}, we next return to position space to further explore the surfaces of the T-DAXI state uncovered in this work.
We begin in SN~\ref{app:Chern-marker} by reviewing the local marker formulation of the Chern number first introduced in SRefs.~\cite{bianco2011mapping,kitaev2006anyons}. 
The Chern marker is specifically a position-space density for the Chern number in two dimensions, and gives a local contribution to the Hall conductivity. 
In SN~\ref{app:partial-Chern-marker} we use the projected spin operator to define a \emph{partial} Chern marker by evaluating the Chern marker for the bands in half of the spin spectrum in an insulator with a spin gap~\cite{chen2022optical,prodan2009robustness,monaco2021stvreda,hannukainen2022local,prodan2016bulk,gilardoni2022real}. 
The partial Chern marker gives the local contribution to the topological part of the spin Hall conductivity, building on the discussion in SN~\ref{sec:2D-spinful-TRI-system}. 
For 3D systems, we then review in SN~\ref{sec:layer-resolved-P-Chern-number} the construction of the layer-resolved Chern number introduced in SRefs.~\cite{varnava2018surfaces,pozo2019quantization}.
For a quasi-2D slab of an insulating material, it was specifically shown in SRefs.~\cite{pozo2019quantization,varnava2018surfaces} that the layer-resolved Chern number gives the contribution of each layer of the slab to the total Chern number of the slab. 
For 3D quantum anomalous Hall insulators the layer-resolved Chern number is a nonzero (integer) constant in each layer, while for magnetic axion insulators it is zero (on average) in the bulk of the slab and quantized to half integers on the top and bottom surfaces; this reflects the half-quantized Hall conductivity at a boundary where the magnetoelectric $\theta$ angle changes by $\pi$, and represents a manifestation of the parity anomaly~\cite{essin2009magnetoelectric,qi2008topological,witten2016fermion,pozo2019quantization,varnava2018surfaces}.

Building on established results for magnetic axion insulators, in SN~\ref{app:layer-resolved-partial-chern-number}, we then formulate a layer-resolved \emph{partial} Chern number for insulators with a spin gap. 
For a quasi-2D slab of an insulating material with a spin gap, the layer-resolved partial Chern number gives the contribution of each layer of the slab to the partial Chern number of the full slab. 
In SN~\ref{sec:layer-resolved-Cs-of-a-helical-HOTI}, we next numerically demonstrate that for a helical HOTI in the T-DAXI regime, the layer-resolved partial Chern number is zero (on average) in the bulk of the slab and quantized to half integers on the top and bottom surfaces. 
This is consistent with the fact that the partial axion angles $\theta^\pm$ introduced in SN~\ref{app:comparison-spin-stable-and-symmetry-indicated-topology} each change by $\pi$ at the boundary between a T-DAXI and the vacuum. 
This implies that each 2D surface of a 3D T-DAXI carries an anomalous half of the partial Chern number (and half the topological contribution to the spin Hall conductivity) of an isolated (lattice-regularized) 2D TI, representing a manifestation of a novel spin-resolved ``partial'' parity anomaly.  
The partial parity anomaly identified in this work is closely related to, and provides a deeper theoretical understanding of, the symmetry-enhanced fermion doubling theorems for topological crystalline insulator surface states formulated in SRefs.~\cite{wieder2018wallpaper,fang2019new,elcoro2021magnetic,gao2022magnetic} using the constraints imposed by crystal symmetry on noninteracting 2D lattice (tight-binding) models. 
Specifically, the gapless surface theories for the nonsymmorphic Dirac insulator introduced in SRef.~\cite{wieder2018wallpaper} and the ``rotation-anomaly'' topological crystalline insulators introduced in SRefs.~\cite{fang2019new,elcoro2021magnetic,gao2022magnetic} can be deformed to the gapped surface theories of inversion- and time-reversal-symmetric helical HOTIs by relaxing surface crystal symmetries~\cite{wang2019higherorder}, indicating a close relationship between the symmetry-enhanced fermion doubling theorems introduced in those works and the partial parity anomaly introduced in this work.

In SN~\ref{appendix:symmetry-constraints-on-Wilson-loop}, we next provide detailed proofs of the action of symmetries on the spin-resolved Wilson spectrum.
We specifically provide a comprehensive discussion of the action of inversion and time-reversal symmetry on (spin-resolved) Wilson loops and nested (spin-resolved) Wilson loops. 

Next, in SN~\ref{sec:bulk_spin_hall_conductivity} we review the theory of intrinsic spin Hall conductivity, and draw a connection between spin-resolved topology and the intrinsic bulk and surface spin Hall conductivity in spin-stable topological phases. First in SN~\ref{sec:shc_comp} we review the computational implementation of the Kubo formula for the intrinsic spin Hall conductivity, building on the formalism of SRef.~\cite{Monaco2022shc_nonconserved_spin}. With the general formalism developed, in SN~\ref{sec:2dshc} we apply our tools to compute the intrinsic spin Hall conductivity of the spin-stable 2D fragile model from SRef.~\cite{wieder2020strong} (analyzed in SN~\ref{sec:spin_stable_topology_2d_fragile_TI}). We show numerically that the intrinsic spin-$s_z$ Hall conductivity is given to leading order by the spin Chern number, with corrections that grow perturbatively in the strength of spin-$s_z$ nonconserving SOC. Finally, in SN~\ref{sec:3dshc} we define a layer-resolved spin Hall conductivity for three-dimensional systems. We compute the layer-resolved spin Hall conductivity for the model of a T-DAXI analyzed in SN~\ref{app:layer-resolved}, where we show that the anomalous half-integer partial Chern numbers at the surface of a T-DAXI imply an anomalously (approximately) half-quantized surface spin Hall conductivity.  
We thus connect spin-resolved topology to experimentally-relevant transport coefficients in spin-gapped states.

Crucially, the spin-resolved topological machinery introduced in this work can be applied to ab-initio calculations of real materials, where we provide a brief summary in SN~\ref{app:materials}. 
To demonstrate this, in SN~\ref{app:mote2}, we first analyze the spin-resolved topology of $\beta$-MoTe$_2$, which was identified in SRefs.~\cite{wang2019higherorder,tang2019efficient} as a candidate helical HOTI. 
Starting with a symmetric, Wannier-based tight-binding model obtained from ab-initio calculations (SN~\ref{sec:mote2_dft_details}), we compute the spin band structure and spin-resolved Wilson loops for $\beta$-MoTe$_2$. 
We specifically in SN~\ref{sec:mote2_spin_resolved_topology} compute the spin spectrum for all possible spin orientations $s=\mathbf{s}\cdot\hat{\mathbf{n}}$ in the projected spin operator $PsP$.
We find that for all choices of $s$, $\beta$-MoTe$_2$ is spin-gapless with an even number of spin-Weyl nodes in each half of the BZ. 
To elucidate physical signatures of the gapless spin spectrum in $\beta$-MoTe$_2$, we next compute the bulk \emph{energy} spectrum for $\beta$-MoTe$_2$ in the presence of a strong Zeeman field directed parallel to the spin direction $s\propto s_{x}+s_{z}$, for which $\beta$-MoTe$_2$ exhibits a particularly simple spin spectrum.  
We observe that in the presence of a large $(\hat{\mathbf{x}}+\hat{\mathbf{z}})$-directed Zeeman field, $\beta$-MoTe$_2$ exhibits Weyl nodes in the energy spectrum that lie at almost the same locations, and carry the same chiral charges, as the spin-Weyl nodes in the spin spectrum in the absence of a magnetic field, providing further support for the analysis in SN~\ref{sec:zeeman}. 
Lastly, we compute the surface spectral function for a finite slab of $\beta$-MoTe$_2$ in the presence of a strong Zeeman field to verify the presence of topological Fermi-arc surface states originating from the field-induced Weyl nodes, which hence serve as physical signatures of the bulk spin-Weyl semimetal state.

Finally, in addition to $\beta$-MoTe$_2$, we also in SN~\ref{sec:bibr} analyze the spin-resolved topology and physical observables of $\alpha$-BiBr.
Previous theoretical studies~\cite{SYBiBr,tang2019efficient,BiBrFanHOTI} have identified $\alpha$-BiBr as a rare bulk-insulating candidate helical HOTI (\emph{i.e.} one without bulk electron and hole pockets, in contrast to $\beta$-MoTe$_2$).
$\alpha$-BiBr is also readily experimentally synthesized, and has exhibited signatures of 1D helical hinge modes in numerous spectroscopic and transport experiments~\cite{noguchi2021evidence, FanZahidRoomTempBiBrExp,BiBrHingeExp1,BiBrHingeExp2,BiBrFacetDependent,BiBrTempLifshitz,BiBrNanowireSubstrate,BiBrNanobelt,BiBrQuantumTransport,BiBrOpticalDichotomy}.
We use first-principles calculations to construct a symmetric, Wannier-based tight-binding model of $\alpha$-BiBr as detailed in SN~\ref{sec:bibr_dft_details}.
We then in SN~\ref{sec:spin_resolved_topology_of_alpha_bibr} use our Wannier-based tight-binding model of $\alpha$-BiBr to compute the spin gap over the full range of spin resolution directions $\hat{\mathbf{n}}$.
Unlike $\beta$-MoTe$_2$, we discover that $\alpha$-BiBr is \emph{spin-gapped} over a large range of spin-resolution directions $\hat{\mathbf{n}}$.
Continuing in SN~\ref{sec:spin_resolved_topology_of_alpha_bibr}, we then compute the spin-resolved Wilson loops, nested Wilson loops, and nested spin-resolved Wilson loops for $\alpha$-BiBr within the spin-gapped regions of its $PsP$ spectra.
We remarkably find that $\alpha$-BiBr realizes \emph{all three} spin-resolved regimes of a helical HOTI, specifically interpolating as a function of $\hat{\mathbf{n}}$ between 3D QSHI states and the spin-Weyl and T-DAXI spin-stable states introduced in this work.
$\alpha$-BiBr hence represents the first known realization of a material with nontrivial, gauge-invariant partial axion angles $\theta^{\pm}=\pi$.
To provide physical signatures of the spin-resolved topology in $\alpha$-BiBr, we lastly in SN~\ref{sec:bibrshc} compute the bulk spin Hall conductivities of $\alpha$-BiBr in its 3D QSHI and T-DAXI regimes. 
We find that the intrinsic bulk contribution to the spin Hall conductivity exhibits remarkably good agreement with the spin-resolved topology in $\alpha$-BiBr, and specifically takes on nearly quantized (nearly vanishing) values in its spin-stable 3D QSHI (T-DAXI) states.

\section{Projected Spin Operators, Spin Spectra, and Spin Gaps}\label{sec:spin_operators}

In this section, we will review the properties of the projected spin operator that are necessary to analyze spin resolved (higher-order) topology. 
First, in SN~\ref{appendix:TB-notation} we will introduce the tight-binding notation used throughout this work. 
In SN~\ref{appendix:properties-of-the-projected-spin-operator} we will review general properties of the projected spin operator. 
In SN~\ref{sec:pspperturbation} we will prove that the spectrum of the spin operator changes continuously under perturbations to the Hamiltonian. 
In SN~\ref{sec:physical}
we will show how the eigenvalues of the projected spin operator impact physical observables. 
Moving to topology, in SN~\ref{appendix:explicit-calculation-of-BHZ-model} we will analytically compute the spin spectrum for a simple model of a 3D topological insulator. 
In SN~\ref{appendix:orbital-texture-and-spin-gap} we will show how entanglement between spin and orbital degrees of freedom can impact the spin spectrum. 
Finally, in SN~\ref{sec:zeeman} we will show how the spin spectrum is related to the band structure (electronic energy spectrum) for a system in a strong Zeeman field.

\subsection{Tight-Binding Notation}
\label{appendix:TB-notation}

In this section we provide the notation for our tight-binding models. 
The second-quantized Fourier-transformed Hamiltonian for an (infinite or periodic) system with discrete translation symmetry is given by
\begin{align}
    H = \sum_{\mathbf{k},\alpha,\beta} c^{\dagger}_{\mathbf{k},\alpha} [H(\mathbf{k})]_{\alpha,\beta} c_{\mathbf{k},\beta}, \label{sec:TB_convention_H}
\end{align}
where $[H(\mathbf{k})]$ is the first-quantized Bloch Hamiltonian matrix, $\mathbf{k}$ is the crystal momentum, the summation over $\mathbf{k}$ is within the first BZ of the crystal, and
\begin{align}
    & c^{\dagger}_{\mathbf{k},\alpha} = \frac{1}{\sqrt{N}} \sum_{\mathbf{R}} e^{i\mathbf{k}\cdot\left( \mathbf{R} + \mathbf{r}_{\alpha} \right)} c^{\dagger}_{\mathbf{R},\alpha},  \label{eq:TB_convention_c_dagger}\\
    & c_{\mathbf{k},\alpha} = \frac{1}{\sqrt{N}} \sum_{\mathbf{R}} e^{-i\mathbf{k}\cdot\left( \mathbf{R} + \mathbf{r}_{\alpha} \right)} c_{\mathbf{R},\alpha}, \label{eq:TB_convention_c}
\end{align}
where $N$ is the number of unit cells, $c^{\dagger}_{\mathbf{R},\alpha}$ and $c_{\mathbf{R},\alpha}$ are the creation and annihilation operators of the (spinful) orbital labeled by $\alpha$ in the unit cell $\mathbf{R}$, and $\mathbf{r}_{\alpha}$ is the position of the (spinful) orbital labeled by $\alpha$ within unit cell $\mathbf{R}$. 
The actual position of the (spinful) orbital created by $c^{\dagger}_{\mathbf{R},\alpha}$ is thus $\mathbf{R} + \mathbf{r}_{\alpha}$. 
Discrete translation symmetry implies that the Hamiltonian in Supplementary Equation (SEq.)~(\ref{sec:TB_convention_H}) is invariant if we shift the summation over $\mathbf{k}$ by a reciprocal lattice vector $\mathbf{G}$. 
In particular, we have
\begin{align}
    H & = \sum_{\mathbf{k},\alpha,\beta} c^{\dagger}_{\mathbf{k}+\mathbf{G},\alpha} [H(\mathbf{k}+\mathbf{G})]_{\alpha,\beta} c_{\mathbf{k}+\mathbf{G},\beta} \\
    & = \sum_{\mathbf{k},\alpha,\beta} c^{\dagger}_{\mathbf{k},\alpha} e^{i\mathbf{G}\cdot \mathbf{r}_{\alpha}} [H(\mathbf{k}+\mathbf{G})]_{\alpha,\beta} e^{-i\mathbf{G}\cdot \mathbf{r}_{\beta}} c_{\mathbf{k},\beta} \\
    & = \sum_{\mathbf{k},\alpha,\beta} c^{\dagger}_{\mathbf{k},\alpha} [H(\mathbf{k})]_{\alpha,\beta} c_{\mathbf{k},\beta},
\end{align}
which implies that
\begin{align}
    e^{i\mathbf{G}\cdot \mathbf{r}_{\alpha}} [H(\mathbf{k}+\mathbf{G})]_{\alpha,\beta} e^{-i\mathbf{G}\cdot \mathbf{r}_{\beta}} = [H(\mathbf{k})]_{\alpha,\beta}.
\end{align}
Defining the unitary matrix $[V(\mathbf{G})]$ with matrix elements
\begin{equation}
    [V(\mathbf{G})]_{\alpha,\beta} \equiv \delta_{\alpha \beta} e^{i \mathbf{G} \cdot \mathbf{r}_{\alpha}}, \label{eq:def-VG-1}
\end{equation}
we see that the electron operators $c^{\dagger}_{\mathbf{k},\alpha}$ satisfy the boundary condition $c^{\dagger}_{\alpha,\mathbf{k}+\mathbf{G}} = c^{\dagger}_{\beta,\mathbf{k}} [V(\mathbf{G})]_{\beta \alpha}$. 
Similarly, the Bloch Hamiltonian matrix $[H(\mathbf{k})]$ satisfies
\begin{align}
    [V(\mathbf{G})] [H(\mathbf{k}+\mathbf{G})] [V(\mathbf{G})]^{-1} = [H(\mathbf{k})].
\end{align}
Therefore, upon a shift of $\mathbf{k} \to \mathbf{k} + \mathbf{G}$, $[H(\mathbf{k})]$ transforms according to the following boundary condition 
\begin{align}
    [H(\mathbf{k}+\mathbf{G})] = [V(\mathbf{G})]^{-1} [H(\mathbf{k})] [V(\mathbf{G})], \label{sec:TB_convention_H_k_BC}
\end{align}
which is due to the gauge choices of the electron operators in SEqs.~\eqref{eq:TB_convention_c_dagger}--\eqref{eq:TB_convention_c}.
Suppose the eigenvectors of $[H(\mathbf{k})]$ are denoted as $\ket{u_{n,\mathbf{k}}}$ where $n$ is the band index, we have the eigenvalue equation
\begin{align}
    [H(\mathbf{k})]\ket{u_{n,\mathbf{k}}} = E_{n,\mathbf{k}} \ket{u_{n,\mathbf{k}}}, \label{eq:TB_convention_u_nk_eig_eqn}
\end{align}
where $E_{n,\mathbf{k}}$ is the dispersion of the $n^{\text{th}}$ energy band. 
The boundary condition in SEq.~(\ref{sec:TB_convention_H_k_BC}) then implies that we can choose the boundary condition for $\ket{u_{n,\mathbf{k}}}$ as
\begin{align}
    \ket{u_{n,\mathbf{k}+\mathbf{G}}} = [V(\mathbf{G})]^{-1}\ket{u_{n,\mathbf{k}}}.\label{eq:bcs}
\end{align}
As shown in SRefs.~\cite{benalcazar2017electric,alexandradinata2014wilsonloop}, such boundary conditions lead to the identification of the phases of the Wilson loop eigenvalues as the actual localized positions of (hybrid) Wannier functions within a unit cell.

\subsection{Properties of the Projected Spin Operator}
\label{appendix:properties-of-the-projected-spin-operator}

In this section, we derive several useful properties of the projected spin operator. 
Consider a $2N\times 2N$ matrix Bloch Hamiltonian $H(\mathbf{k})$. 
$H(\mathbf{k})$ acts on a Hilbert space consisting of $N$ spin-degenerate orbitals per unit cell. 
Letting the Pauli matrices $\sigma_i$ act on the spin degrees of freedom, we can define the spin operators
\begin{equation}
s_i \equiv \sigma_i\otimes \mathbb{I}_N, \label{eq:appendix-def-s-i}
\end{equation}
where $\mathbb{I}_N$ is the $N\times N$ identity matrix acting in the orbital subspace of the entire Hilbert space (including both occupied and unoccupied states). 
In this work we will be considering the band topology of the occupied states in spinful insulators; following SRefs.~\cite{prodan2009robustness,sheng2006spinChern,3D_phase_diagram_spin_Chern_Prodan} we will show that the eigenstates of the spin operator projected to the occupied bands provide a refinement of the usual notion of band topology. 
To formulate this precisely, let $P(\mathbf{k})$ represent the projector onto a set of ``occupied'' energy eigenstates at $\mathbf{k}$. 
We can then form the projected spin operator 
\begin{equation}
PsP\equiv P(\mathbf{k})\hat{\mathbf{n}}\cdot\mathbf{s}P(\mathbf{k}),\label{eq:appendix-def-PsP}
\end{equation} 
for any choice of unit vector $\hat{\mathbf{n}}$. 
SEq.~\eqref{eq:appendix-def-PsP} defines the projected spin operator for both infinite periodic and finite systems, where for finite systems $\mathbf{k}$ indexes only the periodic directions in the BZ (\emph{i.e.} for a system finite in three spatial directions with open boundary conditions, there is no $\mathbf{k}$ dependence). 
For notational convenience, we will suppress the $\mathbf{k}$-dependence of $PsP$ when our discussion applies to both finite and infinite systems. 
When we are considering translationally-invariant systems, the projection operator $P$ will be a $2N \times 2N$ $\mathbf{k}$-dependent matrix where $2N$ is the number of spinful orbitals within the unit cell. 
The spin operator $s_i$ will be a $\mathbf{k}$-independent $2N\times 2N$ matrix.
When we are considering finite systems with open boundary conditions, the projection operator $P$ and the spin operator $s_i$ will both be $2N \times 2N$ $\mathbf{k}$-independent matrices, where $2N$ is the number of spinful orbitals in the entire finite system.

For periodic systems, the spectrum of $PsP$ forms a set of well-defined bands; from SEq.~(\ref{eq:bcs}) we have that
\begin{equation}
P(\mathbf{k+G})sP(\mathbf{k+G}) = V^{-1}(\mathbf{G})P(\mathbf{k})[V(\mathbf{G})sV^{-1}(\mathbf{G})]P(\mathbf{k})V(\mathbf{G}).
\end{equation}
For the spectrum of $PsP$ to be periodic, we must have
\begin{equation}
[V(\mathbf{G}),s]=0. \label{eq:V-s-commute}
\end{equation}
Physically, this means that basis orbitals come in time-reversed pairs with opposite spins. 
This is consistent with the fact that the Hilbert space of a solid derives from pairs of orbitals at the same position with opposite spins (\emph{i.e.} atomic orbitals), and so the periodicity constraint in SEq.~\eqref{eq:V-s-commute} does not place any unreasonable constraints on the classes of systems we consider in this work. 
In particular, SEq.~\eqref{eq:V_G_and_s_asssumption_1} must hold for time-reversal invariant systems. 
We can thus compute the $PsP$ spectrum for any approximate model of a material [\emph{i.e.} from a density functional theory (DFT) calculation] provided the Hilbert space of the model is consistent with SEq.~\eqref{eq:V-s-commute}.

We will now derive several properties of the spectrum of $PsP$ that will be useful for the remainder of this work. 
Because $(\hat{\mathbf{n}}\cdot\mathbf{s})^2=\mathbb{I}_{2N}$, we know that the eigenvalues of $\hat{\mathbf{n}}\cdot\mathbf{s}$ are all $\pm 1$. 
This places several constraints on the eigenvalues of $PsP$. 
First, consider an eigenstate $|\psi\rangle$ of $PsP$ with eigenvalue $\lambda$, such that\begin{align}
P|\psi\rangle &= |\psi\rangle\label{eq:p_psi_action} ,\\
\langle\psi|\psi\rangle &= 1, \\
PsP|\psi\rangle &= \lambda|\psi\rangle. \label{eq:psp_psi_lambda_psi}
\end{align}
SEqs.~\eqref{eq:p_psi_action}--\eqref{eq:psp_psi_lambda_psi} imply that
\begin{align}
s|\psi\rangle = \lambda|\psi\rangle + |\tilde{\phi}\rangle, \label{eq:sunprojaction1}
\end{align}
where
\begin{align}
Q|\tilde{\phi}\rangle &\equiv (\mathbb{I}_{2N}-P)|\tilde{\phi}\rangle = |\tilde{\phi}\rangle, \\
\langle \tilde{\phi} | \tilde{\phi}\rangle &= \bra{\psi}sQs\ket{\psi}\equiv|\alpha|^2.
\end{align}
We then have
\begin{align}
1 &=\langle \psi | \psi\rangle \nonumber\\
&=\langle\psi|s^2|\psi\rangle \nonumber\\
&= (\lambda\langle\psi| + \langle\tilde{\phi}|)(\lambda|\psi\rangle + |\tilde{\phi}\rangle) \nonumber\\
&=|\lambda|^2+|\alpha|^2. \label{eq:lambda-square-plus-alpha-square}
\end{align}
From SEq.~\eqref{eq:lambda-square-plus-alpha-square} we deduce immediately that $|\lambda|\le 1$. 
Furthermore, $\lambda=1$ if and only if $|\psi\rangle$ is an eigenstate of $s$. 
Going further, if $|\lambda| < 1$, then we can define a normalized state
\begin{equation}
|\phi\rangle \equiv \frac{1}{|\alpha|}|\tilde{\phi}\rangle \label{eq:relation_tilde_phi_and_phi}
\end{equation}
such that 
\begin{equation}
    s|\psi\rangle = \lambda|\psi\rangle + |\alpha||\phi\rangle. \label{eq:action_of_s_on_psi}
\end{equation}
Acting with $s$ again we see that
\begin{align}
s^2|\psi\rangle &= |\psi\rangle = \lambda s|\psi\rangle + |\alpha| s|\phi\rangle \nonumber\\
&=|\lambda|^2|\psi\rangle + \lambda|\alpha||\phi\rangle +|\alpha| s|\phi\rangle.\label{eq:s_squared-on_psi}
\end{align}
In order to satisfy the equality in SEq.~\eqref{eq:s_squared-on_psi}, we must have that
\begin{equation}
s|\phi\rangle = |\alpha||\psi\rangle - \lambda|\phi\rangle.\label{eq:sunprojaction2}
\end{equation}
From which we deduce that 
\begin{equation}
QsQ|\phi\rangle\equiv Q(\mathbf{k})\hat{\mathbf{n}}\cdot\mathbf{s}Q(\mathbf{k}) |\phi\rangle = -\lambda|\phi\rangle,\label{eq:qsqreln}
\end{equation}\label{eq:qsqeig}
\emph{i.e.} for each eigenstate of $PsP$ within the image of $P$ with eigenvalue $\lambda$ with $|\lambda|<1$, there exists an eigenstate of $QsQ$ within the image of $Q$ with eigenvalue $-\lambda$, and vice versa.

SEq.~\eqref{eq:qsqreln} shows that given $| \psi \rangle$ in the image of $P$ with $P s P | \psi \rangle = \lambda | \psi \rangle$ where $|\lambda| < 1$, there exists a state $| \phi \rangle $ in the image of $Q$ satisfying $Q s Q | \phi \rangle = -\lambda | \phi \rangle$ where $Q = 1-P$. 
In particular, $| \phi \rangle$ satisfies $s | \phi \rangle = |\alpha| | \psi \rangle - \lambda | \phi \rangle$, where $|\alpha| = \sqrt{ 1 - \lambda^2}$.
Next, we will show that if we have $| \phi_1 \rangle$ and $| \phi_2 \rangle$---both in the image of $Q$---that are constructed from two orthogonal states $| \psi_1 \rangle$ and $| \psi_2 \rangle$ in the image of $P$ with non-unit $PsP$ eigenvalues, then $\ket{\phi_1}$ and $\ket{\phi_2}$ are orthogonal. 
To see this, we can rewrite the overlap $\bra{\phi_1}\ket{\phi_2}$ as
\begin{align}
	&\langle \phi_1 | \phi_2 \rangle  = \langle \phi_1 | s^2 | \phi_2 \rangle = \left( \langle \phi_1 | s \right) \left( s | \phi_2 \rangle \right) \\
	& = \left( |\alpha_1| \langle \psi_1 | - \lambda_1 \langle \phi_1 | \right) \left( |\alpha_2| | \psi_2\rangle - \lambda_2 | \phi_2 \rangle \right) \\
	& = |\alpha_1| |\alpha_2| \langle \psi_1 | \psi_2 \rangle - |\alpha_1| \lambda_2 \langle \psi_1 | \phi_2 \rangle - \lambda_1 |\alpha_2| \langle \phi_1 | \psi_2 \rangle + \lambda_1 \lambda_2 \langle \phi_1 | \phi_2 \rangle \\
	& = \lambda_1 \lambda_2 \langle \phi_1 | \phi_2 \rangle,
\end{align}
where we have used $\langle \psi_1 | \psi_2 \rangle=0$, $\langle \psi_1 | \phi_2 \rangle = 0$, and $\langle \phi_1 | \psi_2 \rangle=0$.
This then implies that
\begin{equation}
	\left( 1 - \lambda_1 \lambda_2 \right) \langle \phi_1 | \phi_2 \rangle = 0.
\end{equation}
Since both $\lambda_1$ and $\lambda_2$ are non-unit $PsP$ eigenvalues, we have $| \lambda_1 \lambda_2| = |\lambda_1| |\lambda_2| < 1$ such that $1 - \lambda_1 \lambda_2 \neq 0$.
This then implies that
\begin{equation}
	\langle \phi_1 | \phi_2 \rangle = 0,
\end{equation}
which is the desired result.

It will often be helpful to consider the matrix elements of $PsP$ between the $N_\mathrm{occ}$ states in the image of $P$. 
Choosing a basis $\{\ket{n} | n=1\dots N_\mathrm{occ}\}$ for the image of $P$, we can introduce the $N_\mathrm{occ}\times N_\mathrm{occ}$ {\it reduced spin matrix}
\begin{equation}
	[s_{\mathrm{reduced}}]_{m,n} = \langle m | s | n \rangle, \label{eq:s_reduced_no_k}
\end{equation}
where $N_{occ}$ is the number of states in the image of $P$, which in many realistic calculations corresponds to the number of occupied valence bands (ignoring core states).  
The nonzero eigenvalues of $[s_{\mathrm{reduced}}]$ are in one-to-one correspondence with the nonzero eigenvalues of $PsP$. 
However, $PsP$ will have more zero eigenvalues than $[s_{\mathrm{reduced}}]$, because the zero eigenspace of $PsP$ includes all states in the image of $Q$. 
Since these additional zero eigenvalues provide no new information, we will see in SN~\ref{app:Wilson} and \ref{app:w2} that it is often advantageous to diagonalize $[s_{\mathrm{reduced}}]$ to determine the spin spectrum.
Throughout this work we will use interchangeably the terms ``reduced $s$ eigenvalues'', namely the eigenvalues of the matrix $[s_{\mathrm{reduced}}]$, ``$PsP$ eigenvalues'' where we will ignore the zero $PsP$ eigenvalues arising from states in the image of $Q$, and ``(projected) spin eigenvalues''. 
In addition, for a translationally invariant system, we will call the $PsP$ eigenvalues as a function of crystal momentum $\mathbf{k}$ the ``spin ($s$) band structure''.

Going further, for each eigenvalue $\lambda$ of $PsP$ with $|\lambda|<1$, we can 
explicitly reconstruct two eigenstates of $s$. 
This will be important when we consider systems with SOC. 
In particular, combining SEqs.~(\ref{eq:sunprojaction1}) and (\ref{eq:sunprojaction2}), we can write
\begin{equation}
s(a|\psi\rangle + b|\phi\rangle) = (|\psi\rangle, |\phi\rangle)\left(\begin{array}{cc}
\lambda & |\alpha| \\
|\alpha| & -\lambda
\end{array}
\right)\left(\begin{array}{c}
a \\
b\end{array}\right) \equiv  (|\psi\rangle, |\phi\rangle) \mathbf{M} \left(\begin{array}{c}
a \\
b\end{array}\right).\label{eq:spineig}
\end{equation}
We can see that SEq.~\eqref{eq:spineig} is a $2\times2$ matrix equation, where the matrix $\mathbf{M}$ has trace $\mathrm{tr}(\mathbf{M})=0$ and determinant $\det\mathbf{M}=-1$. 
By diagonalizing $\mathbf{M}$, we obtain two eigenstates of $s$ with eigenvalues $\pm 1$ given by $\ket{\xi} = (2-2\xi\lambda)^{-1/2}( |\alpha| \ket{\psi} + (-\lambda + \xi) \ket{\phi})$ where $\xi = \pm 1$. 
This shows concretely how in the presence of SOC, spin eigenstates can be reconstructed from linear combinations of occupied and unoccupied states. 
A particularly interesting case is when $|\psi\rangle$ is a null-eigenstate of $P(\mathbf{k})sP(\mathbf{k})$, so that $\lambda=0,|\alpha|=1$. 
In this case, $|\pm\rangle =\frac{1}{\sqrt{2}}(|\psi\rangle \pm |\phi\rangle)$ is an eigenstate of $s$ with eigenvalue $\pm 1$. 
Similarly, if we have eigenstates $\ket{\pm}$ of $PsQ+QsP$ with eigenvalue $\pm1$, we can see from the block structure of SEq.~\eqref{eq:spineig} that $\frac{1}{\sqrt{2}}(\ket{+}\pm\ket{-})$ are  null eigenstates of $PsP$ and $QsQ$. 
We see then that $[s_\mathrm{reduced}]$ has a null eigenvalue if and only if $PsQ+QsP$ has a pair of eigenvalues of modulus 1.

For the bulk of this work, we will be primarily interested in time-reversal invariant systems, with time-reversal operator $\mathcal{T}$ satisfying $\mathcal{T}^2=-\mathbb{I}_{2N}$. 
Since $\mathcal{T}s_i\mathcal{T}^{-1}=-s_i$, time-reversal symmetry requires that $\mathcal{T}P(\mathbf{k})(\hat{\mathbf{n}}\cdot\mathbf{s})P(\mathbf{k})\mathcal{T}^{-1} = -P(\mathbf{-k})sP(\mathbf{-k})$. 
In particular, if $P(\mathbf{k})sP(\mathbf{k})|\psi\rangle = \lambda|\psi\rangle$, then $P(-\mathbf{k})sP(-\mathbf{k})\mathcal{T}|\psi\rangle = -\lambda \mathcal{T}|\psi\rangle$. 
An example of spin bands with $\mathcal{T}$ symmetry is shown in Supplementary Figure  (SFig.)~\ref{fig:schematic_spin_bands_with_I_T_and_IT}(b). 
If we additionally have inversion symmetry $\mathcal{I}$ satisfying $\mathcal{I}^2=\mathbb{I}_{2N}$, then $\mathcal{I}P(\mathbf{k})sP(\mathbf{k})\mathcal{I}^{-1} = P(-\mathbf{k})sP(-\mathbf{k})$. 
Combining the constraints on $PsP$ from $\mathcal{I}$ and $\mathcal{T}$ then implies that for a system with inversion and (spinful) time-reversal symmetry, the eigenvalues of $P(\mathbf{k})sP(\mathbf{k})$ come in $\pm\lambda$ pairs. 
{In other words, the spectrum of $PsP$ has an effective ``chiral'' symmetry if the bulk has $\mathcal{IT}$ symmetry, which is demonstrated in SFig.~\ref{fig:schematic_spin_bands_with_I_T_and_IT}(c).} 

\begin{figure}[ht]
\includegraphics[width=\columnwidth]{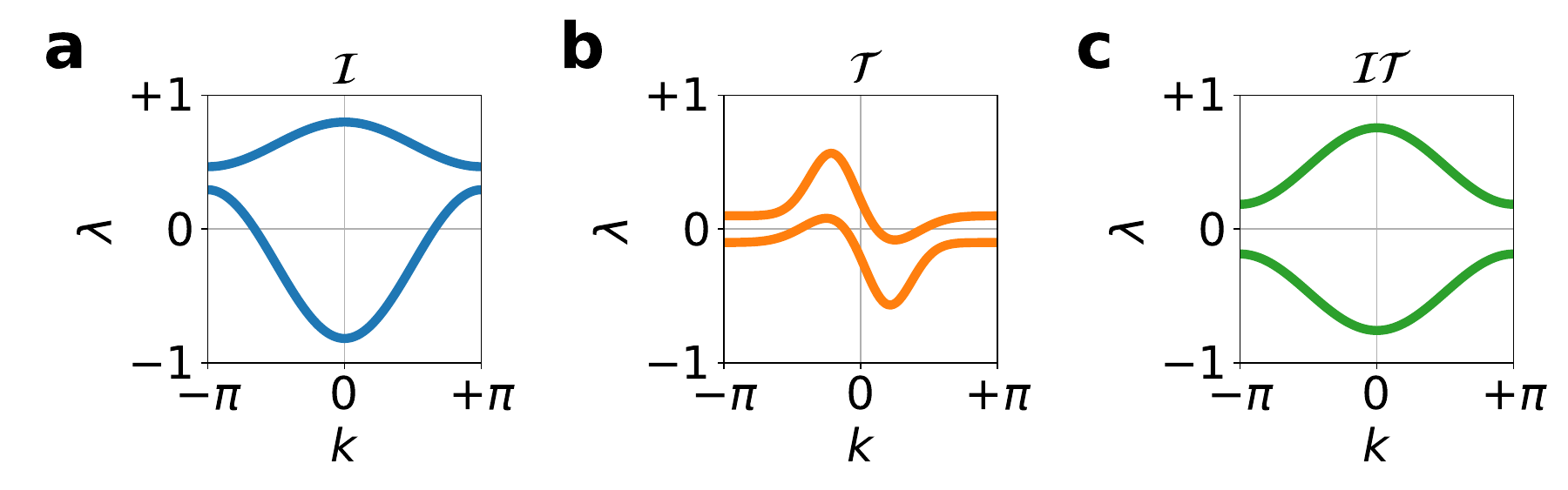}
\caption{Example of spin band structures for 1D systems with (a) inversion ($\mathcal{I}$), (b) time-reversal ($\mathcal{T}$), and (c) $\mathcal{I}\mathcal{T}$ symmetry. 
The number of occupied energy bands is assumed to be 2, hence all of (a), (b), and (c) contain two spin bands. 
We denote the set of $PsP$ eigenvalues at crystal momentum $k$ as $\{ \lambda_{n}(k) \}$ where $n=1,2$. 
In (a) the $PsP$ eigenvalues satisfy $\{ \lambda_{n}(k) \}=\{ \lambda_{n}(-k) \}$ due to $\mathcal{I}$ symmetry. 
In (b) the $PsP$ eigenvalues satisfy $\{ \lambda_{n}(k) \}=\{ -\lambda_{n}(-k) \}$ due to $\mathcal{T}$ symmetry. 
Finally, in (c) the $PsP$ eigenvalues satisfy $\{ \lambda_{n}(k) \}=\{ -\lambda_{n}(k) \}$ due to the combination of $\mathcal{I}$ and $\mathcal{T}$ symmetry. 
In all cases, there is a gap at every $\mathbf{k}$ between the two spin bands, and so the spin gap is open in (a), (b), and (c).
In particular, (b) shows that for a system with time-reversal symmetry only, the spin gap can be open throughout the BZ even though the spin bands cross zero, as long as there is no degeneracy between the $\frac{1}{2}\mathrm{rank}[P(k)] $ spin bands  with largest spin eigenvalue, and the $\frac{1}{2} \mathrm{rank}[P(k)]$ spin bands with smallest spin eigenvalue.
In contrast, for a system with $\mathcal{I}\mathcal{T}$ symmetry, the spin gap is open if and only if there is no degeneracy of spin bands at zero $PsP$ eigenvalues, as shown in (c), such that $\mathrm{rank}[P_{+}(k)] = \mathrm{rank}[P_{-}(k)] = \frac{1}{2} \mathrm{rank}[P(k)] $.
$ P_{+}(k)$ and $ P_{-}(k)$ in (c) are defined to be the projectors onto the positive and negative eigenspace of $P(k)sP(k)$, respectively.
In addition, since we have $\{ \lambda_{n}(k) \}=\{ -\lambda_{n}(k) \}$ in (c), the corresponding spin band structure has an effective ``chiral'' symmetry.
}\label{fig:schematic_spin_bands_with_I_T_and_IT}
\end{figure}

To use $PsP$ to define partial topological invariants, \emph{i.e.} topological invariants for spectrally isolated $PsP$ eigenstates~\cite{fu2006time}, we would like to divide the occupied states into two groups based on their $PsP$ eigenvalues. 
Inversion and time-reversal symmetry give us two natural ways to do this. 
For systems with time-reversal and inversion symmetry, we can define the spin gap at each crystal momentum $\mathbf{k}$ as $\Delta_s(\mathbf{k}) \equiv \min_{|\psi\rangle\in \mathrm{Image}[P(\mathbf{k})]}|\langle\psi|P(\mathbf{k})sP(\mathbf{k})|\psi\rangle|$. 
Using the Rayleigh characterization of eigenvalues~\cite{stone2009mathematics}, we see that $\Delta_s(\mathbf{k})\equiv |\lambda_{\min}(\mathbf{k})|$, where $\lambda_{\min}(\mathbf{k})$ is the eigenvalues of $P(\mathbf{k})sP(\mathbf{k})$ with the smallest absolute value. 
When the spin gap is open, $\Delta_s(\mathbf{k}) >0$ and we can partition $P(\mathbf{k}) = P_+(\mathbf{k}) + P_-(\mathbf{k})$ for all momenta $\mathbf{k}$, where $P_\pm(\mathbf{k})$ projects respectively onto the positive (negative) eigenspace of $P(\mathbf{k})sP(\mathbf{k})$. 
We remark that this process of decomposing the occupied space into spin-resolved sectors through $P(\mathbf{k})=P_{+}(\mathbf{k}) + P_{-}(\mathbf{k})$ is closely related to the sublattice or spin sector resolution employed in SRefs.~\cite{zhang2014hidden,yuan2022uncovering,wang2021isingSC} to analyze ``hidden'' spin-orbital textures. 
For systems with time-reversal and inversion symmetry, we have immediately that $\rank[P_+(\mathbf{k})]=\rank[P_-(\mathbf{k})]=\frac{1}{2}\rank[P(\mathbf{k})]$ when the spin gap is open. 
This definition of the spin gap corresponds to placing the ``spin Fermi energy'' globally at zero, and demanding that no spin bands cross it, see for instance SFig.~\ref{fig:schematic_spin_bands_with_I_T_and_IT}(c). 
Even without inversion symmetry, we have that for an insulating $\mathcal{T}$-invariant system there are an even number of spin bands. 
We can for such systems define $P_+$ to be the projector onto the $\mathrm{rank}(P)/2$ spin bands with the largest spin eigenvalues, and $P_-$ to be the projector onto the $\mathrm{rank}(P)/2$ spin bands with the smallest spin eigenvalues. 
We then say that a spin gap is open when there is no degeneracy between states in $\mathrm{Image}(P_+)$ and states in $\mathrm{Image}(P_-)$ at any $\mathbf{k}$. 
Note that for systems with both inversion and time-reversal symmetry, this construction of $P_+$ and $P_-$ coincides with that of the previous paragraph. 
Analogous to direct gaps in band structures, it is possible that with $\mathcal{T}$ symmetry alone a spin gap can be opened even if a spin band crosses zero, see for instance SFig.~\ref{fig:schematic_spin_bands_with_I_T_and_IT}(b). 
From this construction, it follows that
\begin{equation}
\mathcal{T}P_\pm(\mathbf{k}){\mathcal{T}^{-1}} = P_\mp(-\mathbf{k}).\label{eq:t_on_ppm}
\end{equation}

For simple models with additional emergent symmetries at low energies (for example the approximate chiral symmetry in graphene~\cite{semenoff1984condensed} and polyacetylene~\cite{niemi1986Fermion}, or valley conservation in twisted bilayer systems~\cite{bistritzer2011moire}) there are several additional properties of $PsP$ that we can exploit. 
First, consider insulating models with a unitary chiral symmetry $\Pi H(\mathbf{k})\Pi^{-1} = -H(\mathbf{k})$. 
In this case $P(\mathbf{k})$ projects onto the set of eigenstates with negative energies. 
In such a system, we have that $P(\mathbf{k})sP(\mathbf{k}) = \Pi Q(\mathbf{k}) \Pi^{-1} s \Pi Q(\mathbf{k}) \Pi^{-1}$. 
If, furthermore, $[\Pi,s] = 0$, then this implies that $P(\mathbf{k})sP(\mathbf{k})$ and $Q(\mathbf{k})sQ(\mathbf{k})$ are isospectral. 
To be more specific, if $| \psi \rangle \in \mathrm{Image}[P(\mathbf{k})]$ is an eigenstate of $P(\mathbf{k}) s P(\mathbf{k})$ with eigenvalue $\lambda$, then $\Pi | \psi \rangle \in \mathrm{Image}[Q(\mathbf{k})]$ is an eigenstate of $Q(\mathbf{k}) s Q(\mathbf{k})$ with the same eigenvalue $\lambda$.
Combining this with SEq.~(\ref{eq:qsqreln}), we see that chiral symmetry forces every non-unit eigenvalue $\lambda$ of $P(\mathbf{k})sP(\mathbf{k})$ to have a partner $-\lambda$. 
The spin gap in a chiral-symmetric system with $[\Pi,s]=0$ can then only close when a pair of $P(\mathbf{k})sP(\mathbf{k})$ eigenvalues crosses $0$. 
This implies that $\rank[P_+(\mathbf{k})]=\rank[P_-(\mathbf{k})]$ at $\mathbf{k}$ points that do not have $P(\mathbf{k}) s P(\mathbf{k})$ eigenvalues with modulus $1$ for a model of an insulator with chiral symmetry. 

On the other hand, since
\begin{align}
	0&=\Tr[s] = \Tr[(P+Q)s(P+Q)] \\
	& = \Tr[PsP] + \Tr[QsQ] ,
\end{align}
where we have used $PQ=0$ and the cyclic property of trace, we can obtain
\begin{equation}
	\Tr[PsP] = -\Tr[QsQ].
\end{equation}
In other words, the summations of all the eigenvalues of $PsP$ and $QsQ$ are opposite to each other. 
Combining this with the fact that if $| \psi \rangle \in \mathrm{Image}(P)$ is an eigenstate of $PsP$ with an non-unit eigenvalue $\lambda$, namely $|\lambda|< 1$, there is a corresponding eigenstate $| \phi \rangle \in \mathrm{Image}(Q)$  of $QsQ$ with eigenvalue $-\lambda$, we deduce that the summations of eigenvalues with modulus $1$, namely $|\lambda|=1$, of $PsP$ and $QsQ$ are also opposite to each other.
Denoting $n^{P/Q}_{\pm}$ as the number of $+1$/$-1$ eigenvalues in the $PsP$/$QsQ$ spectrum, this means that
\begin{equation}
	n^{Q}_{+}-n^{Q}_{-}=-\left( n^{P}_{+}-n^{P}_{-} \right). \label{eq:rank_issue_1}
\end{equation}
Furthermore, if we denote $\mathrm{rank}\left( P \right)$ and $\mathrm{rank}\left( Q \right)$ as $r_{P}$ and $r_{Q}$ respectively, using the fact that there is a one-to-one correspondence between the eigenstates of $PsP$ and $QsP$ with non-unit eigenvalues, it can be shown that
\begin{equation}
	\left( n^{Q}_{+} + n^{Q}_{-} \right) - \left( n^{P}_{+} + n^{P}_{-} \right) = r_{Q} - r_{P}. \label{eq:rank_issue_2}
\end{equation}
In other words, the total number of eigenvalues of $PsP$ and $QsQ$ with modulus $1$ will be differ by $\mathrm{rank}(P)-\mathrm{rank}(Q)$.
Combining SEqs.~\eqref{eq:rank_issue_1}--\eqref{eq:rank_issue_2}, we can show that
\begin{align}
	& n^{Q}_{+} = \frac{1}{2}\left( r_Q - r_P \right) + n^{P}_{-}, \label{eq:n_Q_plus_n_P_minus} \\
	& n^{Q}_{-} = \frac{1}{2}\left( r_Q - r_P \right) + n^{P}_{+}. \label{eq:n_Q_minus_n_P_plus}
\end{align}
Crucially, SEqs.~\eqref{eq:n_Q_plus_n_P_minus}--\eqref{eq:n_Q_minus_n_P_plus} are generic statements \emph{independent of the symmetries of the system}.
Notice that, unlike the case for non-unit eigenvalues, there is no one-to-one correspondence between the eigenstates of $PsP$ and $QsQ$ with eigenvalues of modulus $1$.
This can be understood from the fact that $n^{Q}_{+}+n^{Q}_{-}$ is not necessarily equal to $n^{P}_{+}+n^{P}_{-}$, as is demonstrated in SEq.~(\ref{eq:rank_issue_2}).
For an insulator with a unitary chiral symmetry $\Pi$ satisfying $[\Pi,s]= 0$, if we define $P$ and $Q$ as the projectors to all the negative and positive energy eigenstates respectively, we necessarily have $\mathrm{rank}\left( P \right)=\mathrm{rank}\left( Q \right)$ such that according to SEqs.~\eqref{eq:n_Q_plus_n_P_minus}--\eqref{eq:n_Q_minus_n_P_plus} we obtain
\begin{align}
	& n^{Q}_{+} =  n^{P}_{-}, \\
	& n^{Q}_{-} =  n^{P}_{+}.
\end{align}

From the perspective of non-unit $P(\mathbf{k})sP(\mathbf{k})$ eigenvalues, chiral ($\Pi$) and $\mathcal{IT}$ symmetry act in the same way provided $[\Pi,s]=0$: they both map one state $| \psi \rangle \in \mathrm{Image}[P(\mathbf{k})]$ with eigenvalue $\lambda$ to another state $| \tilde{\psi} \rangle \in \mathrm{Image}[P(\mathbf{k})]$  with eigenvalue $-\lambda$.
However, $\Pi$ and $\mathcal{IT}$ are fundamentally different symmetries: $\Pi$ is a unitary symmetry while $\mathcal{IT}$ is an antiunitary symmetry.
For example, a nearest-neighbor model of graphene in the presence of a Zeeman field without SOC~\cite{Wallace_PR_graphite}
has a $\Pi$ symmetry with $[\Pi,s]=0$, while the Kane-Mele model~\cite{kane2005quantum} has $\mathcal{I}$, $\mathcal{T}$, and thus $\mathcal{IT}$ symmetries. 
Our discussion about how unitary chiral symmetry constrains the $PsP$ spectrum is not just applicable to model systems. 
Recently, the method of topological quantum chemistry~\cite{bradlyn2017topological} has been used to identify stoichiometric crystalline materials with (nearly) flat bands~\cite{bernevig_tqc_flat_band_NatPhys2022,regnault2022catalogue,chiu2020fragile,ogata2021methods,rhim2019classification,leykam2018artificial,ZachFlatband}. 
These materials usually have a bipartite lattice, and have exact or approximate chiral symmetry.
Although the application of our method to extract out spin-resolved topology in such materials is beyond the scope of this paper, we expect that the study of spin-resolved topology in spin-orbit coupled stoichiometric solid state materials with (nearly) flat bands, as well as generalizations to sublattice-resolved topology, will be a fruitful direction to explore. 
Specifically, our approach can be generalized to any operator which, like $s$, has eigenvalues $\pm1$ only. 
As such, our methods can be applied to analyze orbital- or sublattice-resolved topology as well.  

If the Bloch Hamiltonian $H(\mathbf{k})$ can be written in ``Dirac form''
\begin{equation}
H(\mathbf{k}) = \epsilon(\mathbf{k})\sum_i \hat{d}_i(\mathbf{k})\Gamma_i \label{eq:appendix-H-k-Dirac-form}
\end{equation}
with $\{\Gamma_i,\Gamma_j\}=2\delta_{ij}$ and $\sum_{i} [\hat{d}_i(\mathbf{k})]^2=1$, then we can oftentimes analytically compute the spectrum of $PsP$. 
For a Dirac model [SEq.~(\ref{eq:appendix-H-k-Dirac-form})], the projector $P(\mathbf{k})$ onto the valence bands has the explicit form
\begin{equation}
P(\mathbf{k}) = \frac{1}{2}(\mathbb{I}_{2N} - \sum_i\hat{d}_i(\mathbf{k})\Gamma_i)
\end{equation}
such that
\begin{equation}
P(\mathbf{k})sP(\mathbf{k}) = \frac{1}{4}\left(s -\sum_i\left\{s,d_i(\mathbf{k})\Gamma_i\right\} + \sum_{i,j} d_i(\mathbf{k})d_j(\mathbf{k})\Gamma_is\Gamma_j\right).\label{eq:diracpsp}
\end{equation}
While SEq.~\eqref{eq:diracpsp} can be efficiently computed for models with small Hilbert spaces, if we are only interested in the non-unit ($|\lambda| \neq 1$) eigenspace of $P(\mathbf{k})sP(\mathbf{k})$ we can simplify the computation even further by considering 
\begin{equation}
P(\mathbf{k})sP(\mathbf{k})-Q(\mathbf{k})sQ(\mathbf{k}) = -\frac{1}{2}\sum_id_i(\mathbf{k})\left\{s,\Gamma_i\right\}.\label{eq:diracpspminusqsq}
\end{equation}
From SEq.~(\ref{eq:qsqreln}), we see that eigenvalues of $P(\mathbf{k})sP(\mathbf{k})-Q(\mathbf{k})sQ(\mathbf{k})$ with $|\lambda|<1$ are doubly degenerate, and correspond to the non-unit eigenvalues of $P(\mathbf{k})sP(\mathbf{k})$. 

Crucially, the utility of considering $PsP-QsQ$ as a proxy for $PsP$ extends beyond Dirac models and to the analysis of real materials with larger single-particle Hilbert spaces. 
This is because $PsP-QsQ$ eigenstates are all $s$ eigenstates. 
To see this, let us evaluate
\begin{align}
[s,PsP-QsQ] &= sPsP - PsPs -sQsQ + QsQs,\nonumber \\
&=(P+Q)sPsP - PsPs(P+Q), \nonumber \\
& -(P+Q)sQsQ + QsQs(P+Q) ,\nonumber\\
&=QsPsP - PsPsQ - PsQsQ + QsQsP,\nonumber \\
&=Qs(P+Q)sP - Ps(P+Q)sQ, \nonumber\\
&=Qs^2P-Ps^2Q = 0,
\end{align}
and so we can simultaneously diagonalize $PsP-QsQ$ and $s$. 
Since the non-unit spectrum of $PsP$ coincides with that of $PsP-QsQ$, we can identify points at which the spin gap closes by diagonalizing $PsP-QsQ$ within the basis of spin eigenstates. 
This is consistent with what we showed in SEqs.~(\ref{eq:qsqreln}) and (\ref{eq:spineig}), since we can take eigenstates of $PsP$ and $-QsQ$ both with eigenvalue $\lambda$ and reconstruct $s$ eigenstates.
{Moreover, to check whether the spin gap is closed in $PsP$, we {\it only} need to diagonalize $PsP-QsQ$ within the basis of spin $s$ eigenstates with either $+1$ or $-1$ eigenvalues.}

Interestingly, $PsP-QsQ$ can also be related to the (spectrally flattened) Hamiltonian projected into a single spin sector. 
To see this,  first note that
\begin{align}
    PsP-QsQ &= PsP-(1-P)s(1-P),\nonumber \\
    &= s + \{s,P\},\nonumber\\
    &=\{s,\frac{1}{2}+P\}.\label{eq:pspsimple}
\end{align}
Let us now introduce
\begin{align}
P_\uparrow &= \frac{1}{2}(1+s),\label{eq:pupdef} \\
P_\downarrow &= \frac{1}{2}(1-s),\label{eq:pdowndef}
\end{align}
which project onto spin-up and spin-down eigenstates, respectively. 
Inserting $s=P_\uparrow - P_\downarrow$ into SEq.~\eqref{eq:pspsimple}, we find
\begin{align}
PsP-QsQ &= \{P_\uparrow - P_\downarrow,1/2+P\} \\
&= P_\downarrow(1-2P)P_\downarrow-P_\uparrow(1-2P)P_\uparrow,
\end{align}
where we have used the fact that $1=P_\uparrow+P_\downarrow$. 
Let us introduce the spin-projected correlation matrices 
\begin{align}
    C_{\uparrow} &= P_\uparrow P P_\uparrow,\label{eq:cupdef} \\
    C_{\downarrow} &= P_\downarrow P P_\downarrow.
\end{align}
Then we see that 
\begin{equation}
    PsP-QsQ = P_\downarrow(1-2C_\downarrow)P_\downarrow-P_\uparrow(1-2C_\uparrow)P_\uparrow.
\end{equation}
The quantity $C_\uparrow$ was previously introduced in SRefs.~\cite{fukui2014entanglement,araki2016entanglement,araki2017entanglement} to study ``spin entanglement cuts'' in topological insulators. 
Here we see that $C_\uparrow$ is directly related to the projected spin operator; in SN~\ref{sec:zeeman} we will show that $PsP$ and $P_\uparrow-2C_\uparrow$ are in fact isospectral.

\subsection{Effects of Perturbations on the Spin Gap}\label{sec:pspperturbation}

In this section, we examine the effects of perturbations $\delta H(\mathbf{k})$ on the spectrum of $P(\mathbf{k})sP(\mathbf{k})$. 
It is crucial to establish the perturbative stability of the $P(\mathbf{k})sP(\mathbf{k})$ spectrum in order to define a robust notion of (partial) topology of the $P(\mathbf{k})sP(\mathbf{k})$. 
In particular, we will establish in this section that the spin spectrum changes smoothly and continuously in response to perturbations of the Hamiltonian (\emph{e.g.} in response to an external magnetic field or perturbations to the form of SOC).
To do so, we first need to understand the effect of a perturbation $\delta H(\mathbf{k})$ on the projector $P(\mathbf{k})$. 
Let us denote $H(\mathbf{k}) = H_0 + \delta H$, and $P(\mathbf{k}) = P_0+\delta P$. 
Assume also that the states in the image of $P_0$ are separated by an energy gap $\Delta$ from states in the image of $Q_0$, and that the perturbation $\delta H$ is sufficiently weak that no new degeneracies are created between $P s P$ bands. 
In this case, we can express the projector $P$ in terms of a contour integral
\begin{equation}
P=-\frac{1}{2\pi i}\oint_\Gamma \frac{dE}{H_0+\delta H -E} \equiv -\frac{1}{2\pi i}\oint_\Gamma G(E)dE,\label{eq:projdef}
\end{equation}
where the integral is taken around a contour $\Gamma$ in the complex plane that encloses all of energies $E^-_i$ of the states in the image of $P_0$, and none of the energies $E^+_i$ of the states in the image of $Q_0$. 
Notice that $E^-_i$ and $E^+_i$ are all of the eigenvalues of the unperturbed Hamiltonian $H_{0}$.
The resolvent (single-particle Green's function) $G(E)$ satisfies the Dyson equation
\begin{equation}
\delta H = G^{-1}(E) - G^{-1}_0(E),
\end{equation}
where the unperturbed resolvent is
\begin{equation}
G_0(E) = \frac{1}{H_0-E}.
\end{equation}
Solving the Dyson equation, we find
\begin{equation}
G(E) = G_0(1-\delta H G) = G_0(1+\delta H G_0)^{-1} = G_0\sum_{n=0}^{\infty} (-1)^n (\delta H G_0)^n.
\end{equation}
This series expansion gives us a perturbation series for the projection operator
\begin{align}
P&=\sum_{n=0}^{\infty}P_n, \\
P_n&=(-1)^{n+1}\frac{1}{2\pi i} \oint dE G_0(E)[\delta H G_0(E)]^n.\label{eq:projseries}
\end{align}
By inserting a complete set of states, each term $P_n$ can be computed by elementary contour integration. 
For instance, the first-order correction $P_1$ takes the form
\begin{align}
P_1 &= \frac{1}{2\pi i} \oint dE \frac{1}{H_0-E}\delta H \frac{1}{H_0-E} \nonumber \\
&=\frac{1}{2\pi i}\oint dE \sum_{nm}\frac{|\psi_n\rangle\langle\psi_n|\delta H|\psi_m\rangle\langle \psi_m|}{(E_n-E)(E_m-E)}\label{eq:p1contour},
\end{align}
where the sum runs over all eigenstates of $H_0$ whose eigenvalues are denoted as $E_{n}$. 
We can evaluate the contour integral in SEq.~\eqref{eq:p1contour} by first noting that if $|\psi_n\rangle$ and $|\psi_m\rangle$ are both unoccupied states or both occupied states, then the contour integral vanishes. 
In the former case, $P_0\ket{\psi_n}=P_0\ket{\psi_m}=0$ and all poles lie outside the contour of integration. 
In the latter case both poles are inside the integration contour, but have residue equal in magnitude and opposite in sign. 
Also, in the latter case, even though we will encounter $E_{n} = E_{m}$ when $n=m$ and when there are degenerate energy eigenvalues, the function $1/(E_{n}-E)^{2}$ has a second order pole and so its residue still vanishes. 
With this restriction, we can use the residue theorem to evaluate the contour integral to find
\begin{equation}
P_1=-\sum_{nm}\frac{|\psi_n\rangle\langle\psi_n|\left( Q_0\delta H P_0 - P_0\delta H Q_0 \right)|\psi_m\rangle\langle\psi_m|}{E_n-E_m}.\label{eq:proj1}
\end{equation}

Given an $M$-th order approximation $\overline{P}=\sum_{n=0}^M P_n$ to $P$, we can construct the $M$-th order perturbed projected spin operator $\overline{PsP} = \sum_{n=0}^{M}\sum_{m=n}^{M} P_n s P_{M-m}$. 
We can then use standard perturbation theory to look at corrections to the eigenvalues $\lambda = \lambda_0+\lambda_1+\dots+\lambda_M$ of $\overline{PsP}$. 

As a concrete example, let us consider the first-order correction to a non-degenerate eigenvalue $\lambda$ of $PsP$. 
Let $|\psi\rangle$ denote the corresponding unperturbed normalized eigenstate of $P_{0}sP_{0}$ satisfying $P_{0}sP_{0} |\psi\rangle = \lambda_{0} |\psi\rangle$ and $P_{0}|\psi\rangle = |\psi\rangle$. 
Then, from the non-degenerate perturbation theory, the first-order correction $\lambda_{1}$ to $\lambda_{0}$ is $\langle \psi| \left( P_{1}sP_{0} + P_{0}sP_{1} \right) |\psi\rangle$, such that
\begin{align}
\lambda_1 &= \langle \psi|P_1s|\psi\rangle +\langle \psi|sP_1|\psi\rangle \\
&=|\alpha|(\langle \psi|P_1|\phi\rangle + \langle \phi|P_1|\psi\rangle) ,
\end{align}
where $|\alpha| = \sqrt{1-|\lambda_{0}|^{2}}$, $|\phi\rangle$ is constructed using SEqs.~(\ref{eq:sunprojaction1}) and (\ref{eq:relation_tilde_phi_and_phi}), and we have used the fact that $P_1$ is block-off-diagonal [see SEq.~(\ref{eq:proj1})]. 
Simplifying further, we find
\begin{equation}
\lambda_1 = -|\alpha|\sum_{n \in \text{unocc} \atop m \in \text{occ}} \frac{\langle \phi|\psi_n\rangle\langle\psi_n|\delta H|\psi_m\rangle\langle \psi_m|\psi\rangle + \langle \psi|\psi_m\rangle\langle\psi_m|\delta H|\psi_n\rangle\langle \psi_n|\phi\rangle}{E_n-E_m}. \label{eq:lambda_1_full_expression}
\end{equation}
In particular, it can be shown that if $|\lambda_{0}|=1$, the first-order correction vanishes, namely $\lambda_{1}=0$. 
From this viewpoint, a completely spin-polarized state $| \psi \rangle$ satisfying $P_{0}| \psi \rangle  = | \psi \rangle $, $P_{0} s P_{0} | \psi \rangle = \pm | \psi \rangle $, and in fact $s|\psi\rangle = \pm 1 | \psi \rangle$, is more robust against perturbation compared with those states $| \psi \rangle$ with $|\lambda_{0}|<1$. 
Let us specialize to a mathematically interesting case where $\delta H = g \hat{\mathbf{n}}\cdot\mathbf{s} = g s$ for some small coupling constant $g$. 
In this case, we can apply the Cauchy-Schwarz inequality to derive the bound
\begin{equation}
|\lambda_1| \le 2\frac{|g|}{\Delta}(1-\Delta_s^2), \label{eq:lambda_1_upper_bound}
\end{equation}
where $\Delta$ is the smallest energy gap, $\Delta_s$ is the eigenvalue of $P_0sP_0$ with smallest absolute value, and we have made use of the fact that $|\alpha|\le \sqrt{1-\Delta_s^2}$. 
In particular, SEq.~\eqref{eq:lambda_1_upper_bound} shows that a Zeeman field produces perturbatively controllable changes to projected spin eigenvalues. 
Notice that we have assumed in the above derivation that we are considering states at a specific momentum $\mathbf{k}$. 
Therefore, the above $\delta H = gs$ can come from a $\mathcal{T}$-breaking $\mathbf{k}$-independent Zeeman field interaction such as $\delta H(\mathbf{k}) = gs$, or from a $\mathcal{T}$-invariant $\mathbf{k}$-dependent SOC such as $\delta H(\mathbf{k}) = g\sin{(k_x)} s$. 
This shows that small changes to the SOC strength in a system produce parametrically small changes to the spin band structure. 
Physically, this means that any (topological) quantity computed from the spin band structure will be a continuous function of the SOC strength. 
We note that the $\mathbf{k}$-local perturbative stability of the $PsP$ spectrum stands in contrast to the effect of perturbations on the Wilson loop (non Abelian Berry phase) spectrum. 
Recall (as we will review in SN~\ref{sec:P_Wilson_loop}) that the Wilson loop can be expressed as a product of projectors along a closed $\mathbf{k}$-path. 
From our perturbative series [SEq.~\eqref{eq:projseries}] for the changes to the projector, we then have that the eigenvalues of the Wilson loop depend on the perturbations to the Hamiltonian at \emph{every} $\mathbf{k}$ along the path. 
This significantly complicates the study of stability of gaps in the Wilson loop spectrum.

For completeness, here we prove SEq.~\eqref{eq:lambda_1_upper_bound} which states that for a perturbation $\delta H = gs$ at a given $\mathbf{k}$ point, the first-order correction to the spectrum of $P_0 s P_0$ has a controlled upper bound. 
In other words, the $P_0 s P_0$ spectrum will be stable against spin-dependent perturbations to the Hamiltonian. 
Assuming that there is an energy gap $\Delta > 0$ between the unoccupied and occupied states such that $|E_n - E_m| \geq \Delta$ for $n \in \mathrm{unocc}$ and $m\in \mathrm{occ}$, we have, from SEq.~\eqref{eq:lambda_1_full_expression}, that
\begin{equation}
	|\lambda_1| \leq \frac{|\alpha||g|}{\Delta} \left|\sum_{n \in \text{unocc} \atop m \in \text{occ}} \left(\langle \phi|\psi_n\rangle\langle\psi_n|s|\psi_m\rangle\langle \psi_m|\psi\rangle + \langle \psi|\psi_m\rangle\langle\psi_m|s|\psi_n\rangle\langle \psi_n|\phi\rangle\right) \right|,
\end{equation}
where we have substituted $\delta H = gs$.
Using the triangle inequality and the definitions of $\ket{\psi}$ and $\ket{\phi}$ which are related to each other through SEq.~\eqref{eq:action_of_s_on_psi} and are eigenstates of $P_0 s P_0$ and $Q_0 s Q_0$ respectively, we have
\begin{equation}
	|\lambda_1| \leq \frac{2|\alpha||g|}{\Delta}\left| \left\langle \phi \right|  s  \left| \psi \right\rangle \right|.
\end{equation}
Using our expression in SEq.~\eqref{eq:action_of_s_on_psi} for the action of $s$ on $\ket{\psi}$, we yield
\begin{equation}
	|\lambda_1| \leq 2 \frac{|\alpha|^2 |g|}{\Delta} = 2 \frac{ |g|}{\Delta} \cdot (1-|\lambda_0|^2),
\end{equation}
where we have also used $|\alpha| = \sqrt{1 - |\lambda_0|^2}$.
Finally, letting $\Delta_s \geq 0$ be the absolute value of the smallest-magnitude eigenvalue of the unperturbed $P_0 s P_0$ spectrum (such that $|\lambda_0| \geq \Delta_s$), we have
\begin{equation}
	|\lambda_1| \leq 2 \frac{|g|}{\Delta}(1-\Delta_s^2).
\end{equation}
We then see that all of the perturbation strength $|g|$, energy gap $\Delta$, and the absolute value of the smallest-magnitude unperturbed projected spin eigenvalue $\Delta_s$ contribute to the upper bound of $|\lambda_1|$. 
{We note that $\Delta_s$ only coincides with the spin gap at a given $\mathbf{k}$ point when the system has both inversion and time-reversal symmetry.}
Importantly, the upper bound of $|\lambda_1|$ is linearly dependent on $|g|$, which means that one can make $|\lambda_1|$ as small as possible by decreasing $|g|$.
In other words, the correction to the $P_0 s P_0$ spectrum is controllable and there is no instability where a small value of $|g|$ will induce a dramatic change of $|\lambda_1|$. 
In practice, this means that if a spin gap is open---\emph{i.e.} if it is possible to divide the spin bands into two disconnected sets---then the spin gap cannot close under infinitesimally small perturbations to the Hamiltonian.
In conclusion, the spin gap is stable against perturbations.

Before moving on, let us note that we can reformulate our perturbation theory for $PsP$ in an illuminating way. 
In particular, we see that corrections to the projected spin eigenvalues emerge from perturbative corrections to the projection operator. 
An alternative way of developing the perturbative expansion of the projection operator is in terms of the Schrieffer-Wolff transformation~\cite{Schrieffer_Wolff_1966,winkler2003spin,Bravyi_SW_2011_AoP}. 
Typically, the Schrieffer-Wolff transformation is used to find a canonical transformation $S$ that block diagonalizes the Hamiltonian, \emph{i.e.} such that
\begin{equation}
e^S (H_0+\delta H)e^{-S} = P_0H_{eff}P_0 + Q_0 H_{eff}Q_0. \label{eq:canonical_transform_on_H}
\end{equation}
The transformation $S$ can be found perturbatively by expanding the left hand side of SEq.~\eqref{eq:canonical_transform_on_H}, and recursively eliminating off-diagonal terms. 
As a consequence of our definition in SEq.~(\ref{eq:projdef}) for the projection operator, we also have that
\begin{equation}
e^{-S}P_0e^{S} = P. \label{eq:canonical_transform_on_P0}
\end{equation}
Expanding both sides of SEq.~\eqref{eq:canonical_transform_on_P0}, we recover our perturbative expansion of the projection operator. 
In particular, we can rewrite SEq.~\eqref{eq:proj1} as
\begin{equation}
P_1=[P_0,S_1],
\end{equation}
where
\begin{equation}
S_1=-\sum_{nm}\frac{|\psi_n\rangle\langle\psi_n|Q_0\delta H P_0 + P_0\delta H Q_0|\psi_m\rangle\langle\psi_m|}{E_m-E_n}
\end{equation}
is the standard form of the leading order term in the Schrieffer-Wolff transformation~\cite{winkler2003spin}. 

While the Schrieffer-Wolff computation of corrections to $P$ is equivalent to our perturbative expansion in SEq.~(\ref{eq:projseries}), it allows us to recast the computation of the perturbed spin spectrum in terms of computing modifications to $PsP$ itself. 
In particular, so long as we are concerned only with finding the eigenvalues of $PsP$, we can note that
\begin{equation}
PsP = e^{-S}P_0e^{S}se^{-S}P_0e^{S},
\end{equation}
which implies that $PsP$ is isospectral to the operator $P_0 e^{S}se^{-S}P_0$. 
This allows us to compute corrections to the $PsP$ spectrum without computing corrections to the energy eigenstates directly. 
The Schrieffer-Wolff approach is particularly well suited to analyzing perturbations to spin-conserving models that do not modify the energy spectrum, such as the changes to the spin quantization axis considered in SRef.~\cite{prodan2009robustness} to demonstrate that the spin Chern number is a $\mathbb{Z}_2$ invariant (we will review this argument in detail in SN~\ref{sec:general_properties_of_winding_num_of_P_pm_Wilson}). 
Here we describe two examples upon which the above reformulation based on Schrieffer-Wolff transformation can be useful.
Recall that the spin quantization axis of a system can be fixed by several effects.
For example, a strong external Zeeman field can (nearly) fix the spin quantization axis to be aligned with the field. 
If the Zeeman field undergoes a slow precession, the spin quantization axis will also rotate.
In addition, the spin quantization axis can also be fixed due to crystalline anisotropy. 
For example, recent experiments on $\text{WTe}_{2}$ have identified a fixed spin quantization axis parallel to a high-symmetry crystallographic direction lying in a mirror plane~\cite{zhao2021determination,garcia2020canted,kurebayashi2022magnetism}.
Therefore, for such systems, a structural distortion can change the spin quantization axis.
Finally, we note that there exists a diagrammatic method~\cite{Bravyi_SW_2011_AoP} for computing $S$ to arbitrary orders in perturbation theory, both for single-particle and many-body Hamiltonians. 
These diagrammatic methods can be leveraged for the analysis of $PsP$, though we leave this as a task for future work.

\subsection{A Physical Interpretation of the Spin Gap}\label{sec:physical}

In this section, we provide a physical interpretation of the spin gap for systems with $\mathcal{IT}$ symmetry. 
Let us consider a situation where we have an insulating, inversion and time-reversal invariant system. 
At $t=0$ we perform a quench by turning on a weak Zeeman field, $\delta H = B\hat{\mathbf{n}}\cdot\mathbf{s}\delta(t)$. 
This will create some density of excited states, and this density can be measured either through angle-resolved photoemission spectroscopy (ARPES), or by allowing the system to relax back to the ground state $| \Psi_0 \rangle$ and collecting the photons and phonons that are emitted in the process. 
The density of excited states will be proportional to the transition rate out of the ground state, which is given to lowest order in perturbation theory by
\begin{equation}
\Gamma_{ex} = \sum_{n>0} B^2|\langle \Psi_n | s |\Psi_0\rangle|^2,\label{eq:gamma_ex_1}
\end{equation}
where $|\Psi_n\rangle$ are the many-body eigenstates of the system. 
For a non-interacting system with discrete translation symmetry whose eigenstate $|\Psi_n\rangle$ can be written as a Slater determinant of single-particle states, we can rewrite SEq.~\eqref{eq:gamma_ex_1} as
\begin{equation}
\Gamma_{ex} = B^2\sum_\mathbf{k}\mathrm{Tr}\left(P(\mathbf{k})sQ(\mathbf{k})sP(\mathbf{k})\right), \label{eq:Gamma_ex_Tr_PsQsP}
\end{equation}
where the summation of $\mathbf{k}$ is within the BZ.
Note that the trace in SEq.~(\ref{eq:Gamma_ex_Tr_PsQsP}) is bounded by the maximum and minimum values of $|\alpha|^2=1-|\lambda|^2$, defined in SEq.~\eqref{eq:lambda-square-plus-alpha-square}. 
Dividing by volume to obtain an intensive quantity, we have
\begin{equation}
1-\lambda_{max}^2 \le \frac{1}{B^2N_e} \Gamma_{ex} \le 1-\lambda_{min}^2, \label{eq:bound_excitation_rate}
\end{equation}
where $\lambda_{max}$ and $\lambda_{min}$ are respectively the largest and smallest absolute values of all the $[s_\mathrm{reduced}]$ [as defined in SEq.~\eqref{eq:s_reduced_no_k}] eigenvalues in the BZ, and $N_e$ is the number of electrons in the system. 
We note that when obtaining $\lambda_{max}$ and $\lambda_{min}$ in SEq.~\eqref{eq:bound_excitation_rate} only the eigenstates of $P(\mathbf{k})sP(\mathbf{k})$ that are in the image of $P(\mathbf{k})$ are considered.
Although there may be technical challenges in disentangling the spin Zeeman interaction used here from an orbital Zeeman interaction (which could lead to the formation of Landau levels), we still see that the spin gap $\Delta_s=\lambda_{min}$ controls the upper bound on the creation of excited states. 
This shows that the spin gap---and more generally the smallest-magnitude projected spin eigenvalue---place bounds on experimental observables, and are thus measurable in principle.

\subsection{\label{appendix:explicit-calculation-of-BHZ-model}Explicit Calculations for a 3D Topological Insulator Model}

In this section, we will compute the spectrum of $P(\mathbf{k})sP(\mathbf{k})$ for the Bernevig-Hughes-Zhang (BHZ) model of a 3D TI~\cite{bernevig2006quantum,fu2007topologicala,fu2007topological,qi2008topological}. 
We will show analytically that the spin gap generically closes at isolated degeneracies with a linear dispersion reminiscent of Weyl fermions. 
In SN~\ref{app:Wilson} we will show that these ``spin-Weyl nodes'' are a general feature in the $PsP$ spectrum of 3D topological insulators.

Let us start with the Bernevig-Hughes-Zhang (BHZ) model in three dimensions,
\begin{equation}
H_0(\mathbf{k})=\tau_x\sum_{i}\sigma_i\sin k_i + (3-m-\sum_i\cos k_i)\tau_z,
\end{equation}
where $\tau_i$ are Pauli matrices acting on the orbital degrees of freedom. 
This model has $\mathcal{I}$ symmetry, $\mathcal{T}$ symmetry, and octahedral symmetries generated by fourfold rotations $C_{4i}$ about the three Cartesian axes and a threefold rotation $C_{31}$ about the $\hat{\mathbf{x}}+\hat{\mathbf{y}}+\hat{\mathbf{z}}$ cubic diagonal. 
These symmetries are represented by 
\begin{align}
[\mathcal{I}]&=\tau_x, \\
[\mathcal{T}]&=i\sigma_y\mathcal{K}, \\
[C_{4i}] &= \exp\left(-i\frac{\pi}{4}\sigma_i\right), \\
[C_{31}] &= \exp\left[-i\frac{\pi}{3\sqrt{3}}(\sigma_x+\sigma_y+\sigma_z)\right],
\end{align}
where $\mathcal{K}$ is the complex conjugation operation. 
The Hamiltonian $H_0 (\mathbf{k})$ satisfies
\begin{align}
[\mathcal{I}]H_0(\mathbf{k})[\mathcal{I}]^{-1}&=H_0(-\mathbf{k}), \\
[\mathcal{T}]H_0(\mathbf{k})[\mathcal{T}]^{-1}&=H_0(-\mathbf{k}), \\
[C_{4i}]H_0(\mathbf{k})[C_{4i}]^{-1}&=H_0(C_{4i}\mathbf{k}),  \\
[C_{31}]H_0(\mathbf{k})[C_{31}]^{-1}&=H_0(C_{31}\mathbf{k})
\end{align}
where $C_{4i}\mathbf{k}$ denotes the vector $\mathbf{k}$ rotated by $\pi/4$ radians about $k_i$, and $C_{31}(k_x,k_y,k_z)=(k_y,k_z,k_x)$.
For $0<m<2$ this model represents a strong topological insulator, while for $m<0$ it represents a trivial insulator. 

Because $H_0$ is written in Dirac form [SEq.~\eqref{eq:appendix-H-k-Dirac-form}]~\cite{qi2008topological}, we can directly compute the non-unit spectrum of $PsP$ via SEq.~(\ref{eq:diracpspminusqsq}). 
We find
\begin{equation}
PsP-QsQ=-\frac{1}{\epsilon(\mathbf{k})}\left(\sum_i \tau_x \hat{n}_i\sin k_i +(3-m-\sum_i\cos k_i)\hat{\mathbf{n}}\cdot\bm{\sigma}\tau_z \right),\label{eq:pspbhz}
\end{equation}
where we have introduced the band energy
\begin{equation}
\epsilon(\mathbf{k})=\sqrt{\sum_i \sin^2 k_i + (3-m-\sum_i\cos k_i)^2}.
\end{equation}

We would like to identify points at which the spin gap closes. 
Because SEq.~\eqref{eq:pspbhz} determining the spectrum of $PsP$ consists of anticommuting matrices, we see that the spin gap closes when the following two conditions are simultaneously satisfied:
\begin{align}
\sum_i \hat{n}_i\sin k_i &= 0, \label{eq:constraint1}\\
(3-m-\sum_i\cos k_i) & = 0.\label{eq:constraint2}
\end{align}
Note that in the topologically trivial phase, $m<0$ and so the second equation is never satisfied. 
Thus, in the topologically trivial phase the BHZ model has a spin gap for all $\mathbf{k}$. 
In the topologically nontrivial phase, we see that the first constraint [SEq.~\eqref{eq:constraint1}] defines an open surface containing origin and extending to the boundaries of the Brillouin zone, while the second constraint [SEq.~\eqref{eq:constraint2}] defines a closed surface surrounding the origin. 
These surfaces will intersect on a closed one-dimensional curve, along which the spin gap closes. 
For sufficiently small $m$, we can expand the constraint equations to lowest order in $k$ to find that the spin gap closes on a circle of radius $\sqrt{2m}$ centered at the origin and oriented normal to the vector $\hat{\mathbf{n}}$. 
As we can see from SEq.~(\ref{eq:pspbhz}), the null eigenvectors of $PsP-QsQ$ are fourfold degenerate, implying that the null eigenvectors of $PsP$ are twofold degenerate along this curve. 

Note, however, that these twofold degenerate zeros of $PsP$ are not perturbatively stable; the spin gap closes along a closed curve due to the residual symmetries of the model. 
To break these symmetries while preserving inversion and time-reversal symmetry, we can add to $H_0$ the perturbation
\begin{equation}
H_1=\eta\sum_i\hat{m}_i\sin k_i\tau_y,
\end{equation}
where $\hat{\mathbf{m}}$ is a unit vector such that $\hat{\mathbf{m}}\times\hat{\mathbf{n}}\neq 0$. 
The combined Hamiltonian $H_0+H_1$ is still of Dirac form, but with a modified band energy
\begin{equation}
\tilde{\epsilon}(\mathbf{k}) = \sqrt{\sum_i \sin^2 k_i +\eta^2(\sum_i \hat{m}_{i} \sin k_i)^2+ (3-m-\sum_i\cos k_i)^2}.
\end{equation}
Recomputing $PsP-QsQ=-1/(2\tilde{\epsilon})\{s,H_0+H_1\}$, we find
\begin{align}
PsP-QsQ&= -\frac{1}{\tilde{\epsilon}(\mathbf{k})}\sum_i \tau_x \hat{n}_i\sin k_i \nonumber \\
&-\frac{\eta}{\tilde{\epsilon}(\mathbf{k})}\sum_i \tau_y (\hat{\mathbf{n}}\cdot\bm{\sigma})\hat{m}_i\sin k_i \nonumber \\
&-\frac{1}{\tilde{\epsilon}(\mathbf{k})}(3-m-\sum_i\cos k_i)\hat{\mathbf{n}}\cdot\bm{\sigma}\tau_z.
\end{align}
Since this is still given as the sum of three anticommuting terms, we see that the spin gap closes only when each of the three terms is individually equal to zero. 
In addition to the constraints in SEqs.~\eqref{eq:constraint1} and \eqref{eq:constraint2}, we have the additional constraint that
\begin{equation}
\sum_i \hat{m}_i\sin k_i = 0.
\end{equation}
This defines a third surface in the Brillouin zone, which is open and contains the origin. 
Furthermore, the restriction $\hat{\mathbf{n}}\cross\hat{\mathbf{m}}\neq 0$ ensures that this surface is independent from the one defined in SEq.~(\ref{eq:constraint1}). 
These three surfaces will generically intersect at a set of isolated points. 
For small $m$, we can again expand our constraint equations to find that the spin gap closes at two points, which lie at the intersection of a sphere of radius $\sqrt{2m}$, and the line through the origin parallel to $\hat{\mathbf{n}}\times\hat{\mathbf{m}}$. 
These are perturbatively stable fourfold degeneracies in the spectrum of $PsP-QsQ$, and hence correspond to twofold-degenerate Weyl nodes in the spectrum of $PsP$. 
Just like Weyl nodes in the energy spectrum of a Hamiltonian, these ``spin-Weyl nodes'' are stable to perturbations of the Hamiltonian (by virtue of our results in SN~\ref{sec:pspperturbation} establishing that the $PsP$ spectrum varies continuously when the Hamiltonian is perturbed).

The fact that $PsP$ has at least one Weyl node in each half of the BZ for the $\mathcal{I}$- and $\mathcal{T}$-symmetric 3D TI can also be established on topological grounds. 
Consider two parallel $\mathcal{T}$-invariant planes in the BZ, defined as planes with constant $k_i=0$ and $k_i=\pi$ for one choice of $i=1,2,3$. 
Furthermore, assume that $PsP$ is gapped on both planes. 
Then we can compute the partial Chern numbers $C^-(k_i=0)$ and $C^-(k_i=\pi)$ of the projector $P_-(\mathbf{k})$ onto the states with negative $PsP$ eigenvalue in each of these planes. 
As shown in SRefs.~\cite{sheng2006spinChern,prodan2009robustness} (and as we will review in SN~\ref{app:Wilson}), $C^-(k_i=0,\pi) \mod 2$ gives the value of the Kane-Mele invariant $\nu_{2d}(k_i=0,\pi)$ in each plane. 
Since our entire system is a 3D TI, we must have that $|\nu_{2d}(k_i=0)-\nu_{2d}(k_i=\pi)|=1$. 
This implies that the partial Chern numbers $C^-(k_i=0)$ and $C^-(k_i=\pi)$ must differ by an odd integer. 
Since we can define the partial Chern numbers $C^-(k_i)$ for any constant-$k_i$ plane in the 3D BZ provided that the spin gap in the 2D plane is opened, and since $C^-(k_i)$ is an integer, this means that for some non-$\mathcal{T}$-invariant plane the partial Chern number must cease to be well-defined, which can only happen when the spin gap closes at a spin-Weyl point. 
Furthermore, the integer change in partial Chern number of the $N_{\mathrm{occ}}/2$ lower spin bands corresponds to the chiral charge of the spin-Weyl point. 
We will later demonstrate this numerically in SN~\ref{sec:main-text-3D-TI-P-pm} for a 3D TI with inversion symmetry, and in SN~\ref{appendix:3D-TI-with-and-without-inversion} for a 3D TI without inversion symmetry, using our spin-resolved Wilson loop formalism.

\subsection{\label{appendix:orbital-texture-and-spin-gap} Effects of Spin-Orbital Entanglement on the Spin Spectrum}

In elucidating the properties of the projected spin operator $PsP$ developed in SN~\ref{appendix:properties-of-the-projected-spin-operator}, we did not need to make a specific choice for the spin direction $\hat{\mathbf{n}}$ in $s=\hat{\mathbf{n}}\cdot\mathbf{s}$.
We can then ask the following question: for a system with a spin gap for a given $\hat{\mathbf{n}}$, must there be a spin gap for other spin directions? 
In this section we will answer this question in the negative by giving explicit examples. 
We will see shortly that although the spin operator $s$ acts as the identity in the orbital subspace, 
the {\it orbital components} of the occupied wave functions are also important when determining the $PsP$ spectrum.
In particular, entanglement between the spin and orbital degrees of freedom in the wave function can force the spin gap to close for certain choices of $\hat{\mathbf{n}}$. 
Let us consider the following two $4 \times 4$ Hamiltonians
\begin{align}
	& H_{1} = \tau_{z}\sigma_{0}, \label{eq:orbital_effect_example_H1} \\
	& H_{2} = \tau_{y}\sigma_{z}, \label{eq:orbital_effect_example_H2}
\end{align}
where the Pauli matrices $\tau_{\mu}$ and $\sigma_{\nu}$ act on orbital and spin degrees of freedom, respectively. 
$\tau_{0}$ and $\sigma_{0}$ are both $2 \times 2$ identity matrices.
Both $H_{1}$ and $H_{2}$ have energies $(-1,-1,+1,+1)$, and both are invariant under spinful $\mathcal{T}$ symmetry represented by \begin{align}
	& [\mathcal{T}] H_{1} [\mathcal{T}]^{-1} = \sigma_{y} H_{1}^{*} \sigma_{y} = H_{1}, \\
	& [\mathcal{T}] H_{2} [\mathcal{T}]^{-1} = \sigma_{y} H_{2}^{*} \sigma_{y} = H_{2}.
\end{align}
We now compute the reduced spin matrix defined in SEq.~(\ref{eq:s_reduced_no_k}) for both $H_1$ and $H_2$. 
We take for our occupied states the two states with negative energy in both cases. 
To be explicit, the two states with energy eigenvalues $-1$ are given by
\begin{align}
	& | 1 \rangle = \begin{bmatrix} 0 \\ 1 \end{bmatrix} \otimes \begin{bmatrix} 1 \\ 0 \end{bmatrix},\ | 2 \rangle = \begin{bmatrix} 0 \\ 1 \end{bmatrix} \otimes \begin{bmatrix} 0 \\ 1 \end{bmatrix} \text{ for } H_{1}, \label{eq:1_2_state_for_H1}\\
	& | 1 \rangle = \frac{1}{\sqrt{2}} \begin{bmatrix} 1 \\ -i \end{bmatrix} \otimes \begin{bmatrix} 1 \\ 0 \end{bmatrix},\ | 2 \rangle = \frac{1}{\sqrt{2}} \begin{bmatrix} 1 \\ i \end{bmatrix} \otimes \begin{bmatrix} 0 \\ 1 \end{bmatrix} \text{ for } H_{2}. \label{eq:1_2_state_for_H2}
\end{align}
The first vector in each tensor product corresponds to the orbital ($\tau$) degree of freedom, and the second vector in each tensor product corresponds to the spin ($\sigma$) degree of freedom. 
Notice that the negative energy eigenstates of $H_{1}$ and $H_{2}$ in SEqs.~\eqref{eq:1_2_state_for_H1} and \eqref{eq:1_2_state_for_H2} are all $s_z$ eigenstates. 
In addition, the {\it orbital components} of the eigenstates of $H_1$ in SEq.~\eqref{eq:1_2_state_for_H1} are {\it identical}. 
Because of this, by taking linear combinations of $\ket{1}$ and $\ket{2}$ in SEq.~\eqref{eq:1_2_state_for_H1}, the occupied eigenvectors for $H_1$ can be chosen as eigenstates of $\hat{\mathbf{n}}\cdot\mathbf{s}$ for any choice of $\hat{\mathbf{n}}$.
On the other hand, the orbital components of the  eigenstates of $H_{2}$ in SEq.~\eqref{eq:1_2_state_for_H2} are orthogonal. 
This means that for the eigenstates of $H_2$ we cannot create $\hat{\mathbf{n}}\cdot\mathbf{s}$ eigenstates for arbitrary $\hat{\mathbf{n}}$ by taking linear combinations of $\ket{1}$ and $\ket{2}$ in SEq.~\eqref{eq:1_2_state_for_H2}. 

This difference in the orbital part of the eigenfunctions is crucial in determining the spin spectrum. 
To see this, recall that the spin operator $s$ is defined as $s \equiv \tau_{0} \sigma$, where $\sigma = \bm{\sigma} \cdot \hat{\mathbf{n}}$ and $\hat{\mathbf{n}}$ is a unit vector in 3D [see SEq.~\eqref{eq:appendix-def-s-i}]. 
The reduced spin matrix from SEq.~(\ref{eq:s_reduced_no_k}), in the two-dimensional space of states with $-1$ energy eigenvalues, is then given by
\begin{equation}
	[s_{\mathrm{reduced}}]_{2\times 2} = \begin{bmatrix} \langle 1 | s | 1 \rangle & \langle 1 | s | 2 \rangle \\ \langle 2 | s | 1 \rangle & \langle 2 | s | 2 \rangle \\ \end{bmatrix} = \begin{bmatrix} \langle 1 | \tau_{0} \sigma | 1 \rangle & \langle 1 | \tau_{0} \sigma | 2 \rangle \\ \langle 2 | \tau_{0} \sigma | 1 \rangle & \langle 2 | \tau_{0} \sigma | 2 \rangle \\ \end{bmatrix} = \begin{bmatrix} \langle 1_{\tau} | \tau_{0} | 1_{\tau} \rangle \langle 1_{\sigma} |  \sigma | 1_{\sigma} \rangle & \langle 1_{\tau} | \tau_{0} | 2_{\tau} \rangle \langle 1_{\sigma} |  \sigma | 2_{\sigma} \rangle \\ \langle 2_{\tau} | \tau_{0} | 1_{\tau} \rangle \langle 2_{\sigma} |  \sigma | 1_{\sigma} \rangle & \langle 2_{\tau} | \tau_{0} | 2_{\tau} \rangle \langle 2_{\sigma} |  \sigma | 2_{\sigma} \rangle \\ \end{bmatrix} \label{eq:reduced_s_in_occ_space_orb_texture}
\end{equation}
where $| i_{\tau} \rangle $ and $| i_{\sigma} \rangle $ are the orbital and spin components of the $i^{\text{th}}$ eigenstate with energy eigenvalue $-1$ from SEqs.~\eqref{eq:1_2_state_for_H1} or \eqref{eq:1_2_state_for_H2} such that $| i \rangle  = | i_{\tau} \rangle  \otimes | i_{\sigma} \rangle $. 
Using SEq.~(\ref{eq:reduced_s_in_occ_space_orb_texture}), we find the following reduced spin matrices
\begin{align}
	& [s]_{2\times 2} = \begin{bmatrix} \sigma_{11} & \sigma_{12} \\ \sigma_{21} & \sigma_{22} \\ \end{bmatrix} = \sigma \text{ for } H_{1}, \label{eq:reduced_s_for_H1_orb_tex} \\
	& [s]_{2\times 2} = \begin{bmatrix} \sigma_{11} & 0 \\ 0 & \sigma_{22} \\ \end{bmatrix} \text{ for } H_{2}, \label{eq:reduced_s_for_H2_orb_tex}
\end{align}
where $\sigma_{ij}$ are the matrix elements of the $2 \times 2$ matrix $\sigma = \bm{\sigma} \cdot \hat{\mathbf{n}}$. 
As we can see in SEq.~(\ref{eq:reduced_s_for_H2_orb_tex}), the reduced spin matrix for $H_{2}$ is diagonal, which follows from the fact that the {\it orbital components} of $| 1 \rangle $ and $| 2 \rangle $ in SEq.~\eqref{eq:1_2_state_for_H2} are orthogonal for the occupied eigenstates of $H_{2}$. 
We can then compute the two eigenvalues of the reduced spin matrices with $\hat{\mathbf{n}} = \hat{\mathbf{x}}$, $\hat{\mathbf{y}}$, and $\hat{\mathbf{z}}$; the results are summarized in Supplementary Table~\ref{tab:eig-reduced-spin-for-H1-and-H2-orbital-vs-spin-gap}.
\begin{table}[h]
\centering
\begin{tabular}{|c|c|c|}
\hline
 & $H_{1}$ & $H_{2}$ \\
\hline
$s = \tau_{0}\sigma_{x}$ & $(-1,+1)$ & $(0,0)$ \\
\hline
$s = \tau_{0}\sigma_{y}$ & $(-1,+1)$ & $(0,0)$ \\
\hline
$s = \tau_{0}\sigma_{z}$ & $(-1,+1)$ & $(-1,+1)$ \\
\hline
\end{tabular}
\caption{The two eigenvalues of the reduced spin matrices in SEqs.~\eqref{eq:reduced_s_for_H1_orb_tex}--\eqref{eq:reduced_s_for_H2_orb_tex} with $\hat{\mathbf{n}} = \hat{\mathbf{x}}$, $\hat{\mathbf{y}}$, and $\hat{\mathbf{z}}$ in the occupied (negative energy) subspace of eigenstates of $H_{1}$ and $H_{2}$.}
\label{tab:eig-reduced-spin-for-H1-and-H2-orbital-vs-spin-gap}
\end{table}
We see that the spin gap is open for all of $\hat{\mathbf{n}} = \hat{\mathbf{x}}$, $\hat{\mathbf{y}}$, and $\hat{\mathbf{z}}$ for the eigenvectors in SEq.~\eqref{eq:1_2_state_for_H1} of $H_{1}$. 
In fact, since the reduced spin matrix for $H_{1}$ is $\sigma = \bm{\sigma} \cdot \hat{\mathbf{n}}$, as shown in SEq.~(\ref{eq:reduced_s_for_H1_orb_tex}), no matter what direction $\hat{\mathbf{n}}$ we choose, the occupied subspace of $H_{1}$ has a nonzero spin gap. 
However, in the case of $H_{2}$, we find that although the spin gap is open for $\hat{\mathbf{n}}=\hat{\mathbf{z}}$, the spin gap is closed for $\hat{\mathbf{n}} = \hat{\mathbf{x}}$ and $\hat{\mathbf{y}}$.
This is because when $\hat{\mathbf{n}} = \hat{\mathbf{x}}$ and $\hat{\mathbf{y}}$, the reduced spin matrices in SEq.~(\ref{eq:reduced_s_for_H2_orb_tex}) are zero matrices, and thus the two eigenvalues of the reduced spin operator are degenerate and equal to $0$. 
From the above example, we have demonstrated that although the spin gap in the occupied space is open for one choice of $\hat{\mathbf{n}}$, it is not guaranteed that the spin gap will be open for other choices of $\hat{\mathbf{n}}$. 
In particular we have seen that entanglement between spin and orbital degrees of freedom plays a crucial role in determining the spin gap.

We thus see that whether or not there is a spin gap for a particular choice of direction $\hat{\mathbf{n}}$ depends on the microscopic details of the spin-orbit interaction. 
Concretely, $H_2$ from SEq.~\eqref{eq:orbital_effect_example_H2} can be viewed as the spin-orbit contribution to a Bloch Hamiltonian at a time-reversal-invariant crystal momentum (TRIM). 
We see that the entanglement between spin and orbital degrees of freedom in the eigenstates [SEq.~\eqref{eq:1_2_state_for_H2}] results in a preferred choice of direction $\hat{\mathbf{n}}=\hat{\mathbf{z}}$ along which the spin gap is maximal.  We will revisit this discussion for more realistic spin and orbital textures in SN~\ref{app:mote2} and \ref{sec:bibr}, in which we compute the spin spectrum of the candidate higher-order topological insulators (HOTIs) $\beta$-MoTe$_2$~\cite{wang2019higherorder,tang2019efficient} and $\alpha$-BiBr~\cite{SYBiBr,tang2019efficient,BiBrFanHOTI}.  In SN~\ref{app:mote2} and \ref{sec:bibr}, we will specifically respectively demonstrate that while $\beta$-MoTe$_2$ does not exhibit a spin gap for any choice of spin direction, $\alpha$-BiBr exhibits a spin gap for multiple choices of $\hat{\mathbf{n}}$, including a large ($\Delta_{s}~\sim 1$ in unit of $\hbar/2$) spin gap for $\hat{\mathbf{n}}=\hat{\mathbf{z}}$.

\subsection{(Spin) Band Structure in a Strong Zeeman Field}\label{sec:zeeman}

In this section, we will show that the partial band topology topology of the spin band structure is intimately connected to the electronic band topology of an insulator in a strong Zeeman field. 
To see this, let us consider a spinful, noninteracting electron system with the Hamiltonian
\begin{equation}
H=H_0 -g\mu_B|\mathbf{B}|s + V_0,\label{eq:zeemanham}
\end{equation}
where $H_0$ determines the band structure in the absence of external perturbations, $\mathbf{B}$ is the external magnetic field, $g$ is the spin $g$-factor (where we take $g>0$ for simplicity), $\mu_B$ is the Bohr magneton, $s=\hat{\mathbf{B}}\cdot\mathbf{s}$ is the spin component along the magnetic field direction, and $V_0$ is a scalar potential which we will use to manipulate the Fermi level. 
If we take
\begin{equation}
V_0 = g\mu_B|\mathbf{B}|,
\end{equation}
then we can rewrite SEq.~\eqref{eq:zeemanham} as
\begin{align}
H &= H_0 + g\mu_B|\mathbf{B}|P_\downarrow, \\
&\equiv H_0+\delta H
\end{align}
where $P_\downarrow$
is the projection operator onto the negative eigenspace of $s$ from SEq.~\eqref{eq:pdowndef}. 
For this choice of $V_0$, the combined effects of the Zeeman and scalar potentials are then to energetically penalize spin-down electrons, while leaving spin-up electrons unaffected. 

This has particularly stark consequences in the limit $g\mu_B|\mathbf{B}|\rightarrow\infty$. 
In this limit, we can view $\delta H$ as the unperturbed Hamiltonian, and treat $H_0$ as the perturbation. 
Since $\delta H$ pushes spin-down electrons up to negative energy, the projector onto the low energy subspace of $\delta H$ is given by SEq.~\eqref{eq:pupdef}.
We then have that to first order in perturbation theory
\begin{equation}
H\approx \delta H + P_\uparrow H_0 P_\uparrow.\label{eq:zeeman1storder}
\end{equation}
In particular, since $\delta H P_\uparrow = 0$, energies in the low-energy $P_\uparrow$ subspace are completely determined by
\begin{equation}
H_{\mathrm{low}} = P_\uparrow H_0 P_\uparrow.\label{eq:hlow}
\end{equation}
This means that the effect of $\delta H$ as $g\mu_B|\mathbf{B}|\rightarrow\infty$ is to project $H_0$ onto the spin-up subspace.

Let us now focus on the low-energy $P_\uparrow$ subspace. 
Introducing a set of eigenstates $\ket{n\mathbf{k}}$ and energies $\epsilon_{n\mathbf{k}}$ for $H_0$, we can re-express SEq.~\eqref{eq:hlow} as
\begin{equation}
H_\mathrm{low}=\sum_{n\mathbf{k}}  \epsilon_{n\mathbf{k}}P_\uparrow\ket{n\mathbf{k}}\bra{n\mathbf{k}}P_\uparrow.\label{eq:hlowhomotopic}
\end{equation}
To make further progress, let us assume that $H_0$ 
has a gapped spectrum. 
By an appropriate shift of the zero of energy at each $\mathbf{k}$, we can ensure for convenience that the gap is centered at zero, such that states with $\epsilon_{n\mathbf{k}}>0$ are above the gap (unoccupied), and states with $\epsilon_{n\mathbf{k}}<0$ are below the gap (occupied). 
Before turning on a Zeeman field, we can spectrally flatten $H_0$ to have the form
\begin{equation}\label{eqref:h0flat}
    H_0(\mathbf{k})\rightarrow Q(\mathbf{k})-P(\mathbf{k})
\end{equation}
where we have introduced the projectors $P(\mathbf{k})$ and $Q(\mathbf{k})$ onto the occupied and unoccupied space of $H_0$ at each $\mathbf{k}$ respectively.
In this case, $H_\mathrm{low}$ becomes
\begin{align}
H_\mathrm{low}&\rightarrow \sum_{n\in\mathrm{unocc}} P_\uparrow \ket{n\mathbf{k}}\bra{n\mathbf{k}}P_\uparrow - \sum_{n\in\mathrm{occ}} P_\uparrow \ket{n\mathbf{k}}\bra{n\mathbf{k}}P_\uparrow \\
&= \sum_\mathbf{k}P_\uparrow Q(\mathbf{k}) P_\uparrow - P_\uparrow P(\mathbf{k}) P_\uparrow,\label{eq:flattened-zeeman}
\end{align}
Note that $P_\uparrow P(\mathbf{k}) P_\uparrow$ is exactly the spin-projected correlation matrix of SRef.~\cite{fukui2014entanglement}, and $P_\uparrow Q(\mathbf{k}) P_\uparrow$ is the analogous operator for the unoccupied states. 

Next, we can express $P$ and $Q$ in terms of eigenstates of $PsP$ and $QsQ$. 
In particular, using the results of SN~\ref{appendix:properties-of-the-projected-spin-operator}, we can write
\begin{equation}
P(\mathbf{k}) = \sum_{|\lambda_\mathbf{k}|\neq 1}\ket{\lambda_\mathbf{k}}\bra{\lambda_\mathbf{k}} + \sum_i \ket{+i\mathbf{k}}\bra{+i\mathbf{k}} +  \ket{-i\mathbf{k}}\bra{-i\mathbf{k}},\label{eq:Pinspinbasis}
\end{equation}
where $\ket{\lambda_\mathbf{k}}$ are the eigenstates of $PsP$ with eigenvalue $\lambda_\mathbf{k}\neq \pm1$, and $\ket{\pm i\mathbf{k}}$ are the eigenstates of $PsP$ with eigenvalue $\pm1$. 
As shown in SN~\ref{appendix:properties-of-the-projected-spin-operator}, the eigenstates $\ket{\lambda_\mathbf{k}}$ are in one-to-one correspondence with eigenstates 
\begin{equation}
\ket{\phi(\lambda_\mathbf{k})} = \frac{Qs\ket{\lambda_\mathbf{k}}}{\sqrt{\bra{\lambda_\mathbf{k}}sQs\ket{\lambda_\mathbf{k}}}} = \frac{Qs\ket{\lambda_\mathbf{k}}}{\sqrt{1-\lambda_k^2}}
\end{equation}
of $QsQ$ with eigenvalue $-\lambda$. 
This means we can also write
\begin{equation}
Q(\mathbf{k}) = \sum_{|\lambda_\mathbf{k}|\neq 1}\ket{\phi(\lambda_\mathbf{k})}\bra{\phi(\lambda_\mathbf{k})} + \sum_{\tilde{i}} \ket{+\tilde{i}\mathbf{k}}\bra{+\tilde{i}\mathbf{k}} +  \ket{-\tilde{i}\mathbf{k}}\bra{-\tilde{i}\mathbf{k}},\label{eq:Qinspinbasis}
\end{equation}
where $\ket{\pm\tilde{i}\mathbf{k}}$ are the eigenstates of $QsQ$ with eigenvalue $\pm1$. 
Using the definitions of $\ket{\lambda_\mathbf{k}},\ket{\phi(\lambda_\mathbf{k})},\ket{\pm i\mathbf{k}}$ and $\ket{\pm\tilde{i}\mathbf{k}}$ along with SEqs.~\eqref{eq:sunprojaction1}--\eqref{eq:sunprojaction2} we have
\begin{align}
P_\uparrow\ket{\lambda_\mathbf{k}} &= \frac{1+\lambda_\mathbf{k}}{2}\ket{\lambda_\mathbf{k}} + \frac{\sqrt{1-|\lambda_\mathbf{k}|^2}}{2}\ket{\phi(\lambda_\mathbf{k})}\label{eq:pupprops1} \\
P_\uparrow\ket{\phi(\lambda_\mathbf{k})} &=\frac{1-\lambda_\mathbf{k}}{2}\ket{\phi(\lambda_\mathbf{k})} + \frac{\sqrt{1-|\lambda_\mathbf{k}|^2}}{2}\ket{\lambda_\mathbf{k}} \\
P_\uparrow\ket{+i\mathbf{k}}&=\ket{+i\mathbf{k}} \\
P_\uparrow\ket{+\tilde{i}\mathbf{k}}&=\ket{+\tilde{i}\mathbf{k}} \\
P_\uparrow\ket{-i\mathbf{k}}&=P_\uparrow\ket{-\tilde{i}\mathbf{k}}=0.\label{eq:pupprops2}
\end{align}
We can now combine SEqs.~\eqref{eq:Pinspinbasis} and \eqref{eq:Qinspinbasis} with the definition in SEq.~\eqref{eq:hlowhomotopic}. 
Using SEqs.~\eqref{eq:pupprops1}--\eqref{eq:pupprops2} we find

\begin{align}
P_\uparrow [Q(\mathbf{k})-P(\mathbf{k})]P_\uparrow &= \sum_{|\lambda_\mathbf{k}|\neq 1} -\lambda_\mathbf{k}\left(\sqrt{\frac{1+\lambda_\mathbf{k}}{2}}\ket{\lambda_\mathbf{k}} + \sqrt{\frac{1-\lambda_\mathbf{k}}{2}}\ket{\phi(\lambda_\mathbf{k})}\right)\left(\sqrt{\frac{1+\lambda_\mathbf{k}}{2}}\bra{\lambda_\mathbf{k}} + \sqrt{\frac{1-\lambda_\mathbf{k}}{2}}\bra{\phi(\lambda_\mathbf{k})}\right) \nonumber \\
&+\sum_{\tilde{i}}\ket{+\tilde{i}\mathbf{k}} \bra{+\tilde{i}\mathbf{k}} -\sum_{i}\ket{+i\mathbf{k}} \bra{+i\mathbf{k}}.
\end{align}
Note that we can define $\ket{\phi(+1)}\equiv0$ and $\ket{\phi(-1_\mathbf{k})}\equiv\ket{+\tilde{i}\mathbf{k}}$, and that this is consistent with SEqs.~\eqref{eq:sunprojaction1}--\eqref{eq:sunprojaction2}, which allows us to write
\begin{align}
P_\uparrow [Q(\mathbf{k})-P(\mathbf{k})]P_\uparrow &= \sum_{\lambda_\mathbf{k}} -\lambda_\mathbf{k}\left(\sqrt{\frac{1+\lambda_\mathbf{k}}{2}}\ket{\lambda_\mathbf{k}} + \sqrt{\frac{1-\lambda_\mathbf{k}}{2}}\ket{\phi(\lambda_\mathbf{k})}\right)\left(\sqrt{\frac{1+\lambda_\mathbf{k}}{2}}\bra{\lambda_\mathbf{k}} + \sqrt{\frac{1-\lambda_\mathbf{k}}{2}}\bra{\phi(\lambda_\mathbf{k})}\right)\label{eq:hlow_in_terms_of_psp} \\
&\equiv\sum_{\lambda_\mathbf{k}}-\lambda_\mathbf{k}\ket{\psi_\mathbf{k}}\bra{\psi_\mathbf{k}}.\label{eq:hlow_in_terms_of_psi}
\end{align}

Since the states $\ket{\psi_\mathbf{k}}$ are orthonormal, we see that the spectrum of $P_\uparrow [Q(\mathbf{k})-P(\mathbf{k})]P_\uparrow$ coincides with the spectrum $\{\lambda_\mathbf{k}\}$ of the reduced spin operator [SEq.~\eqref{eq:s_reduced_no_k}], modulo the number of $+1$ eigenvalues corresponding to exact spin-up eigenstates of $Q$, and also modulo the number of $-1$ eigenvalues corresponding to exact spin-down eigenstates of $P$. 
Furthermore, recall that the spectrum of $P(\mathbf{k})sP(\mathbf{k})$ restricted to the occupied states coincides with the spectrum of the reduced spin operator. 
This means that the low energy spectrum of a system with spectrally flat $H_0$ [SEq.~\eqref{eqref:h0flat}] in a strong Zeeman field coincides with the nontrivial spectrum of $P(\mathbf{k})sP(\mathbf{k})$. 
As a side-effect, also note that $P_\uparrow [Q(\mathbf{k})-P(\mathbf{k})]P_\uparrow =P_\uparrow [1-2P(\mathbf{k})]P_\uparrow = P_\uparrow - 2C_\uparrow$ is directly related to the spin-projected correlation matrix of SRefs.~\cite{fukui2014entanglement,araki2016entanglement,araki2017entanglement}; SEq.~\eqref{eq:hlow_in_terms_of_psi} hence shows then that there is a one-to-one correspondence between $PsP$ eigenvalues and $C_\uparrow$ eigenvalues (modulo the number of states at the accumulation point $\lambda_\mathbf{k}=-1$).

We can go further to relate the topology of $PsP$ bands and $C_\uparrow$ bands if there is a spin gap. 
In this case, we can spectrally flatten SEq.~\eqref{eq:hlow_in_terms_of_psp} by spectrally flattening $P(\mathbf{k})sP(\mathbf{k})$, taking $\lambda_\mathbf{k}\rightarrow -1$ continuously for the bands below the spin gap and $\lambda_\mathbf{k}\rightarrow +1$ continuously for states above the spin gap, while leaving the eigenstates $\ket{\lambda_\mathbf{k}}$ and $\ket{\phi(\lambda_\mathbf{k})}$ unchanged. 
The states $\ket{\psi_\mathbf{k}}$ change continuously under this deformation, yielding 
\begin{equation}
P_\uparrow [Q(\mathbf{k})-P(\mathbf{k})]P_\uparrow\sim \sum_{\lambda_\mathbf{k}}^{\mathrm{below}}\ket{\phi(\lambda_\mathbf{k})}\bra{\phi(\lambda_\mathbf{k})}-\sum_{\lambda_\mathbf{k}}^{\mathrm{above}}\ket{\lambda_\mathbf{k}}\bra{\lambda_\mathbf{k}},
\end{equation}
where $\sum_{\lambda_\mathbf{k}}^{\mathrm{above}}$ is a sum is over $\lambda_\mathbf{k}$ in the upper spin bands, and $\sum_{\lambda_\mathbf{k}}^{\mathrm{below}}$ is a sum over $\lambda_\mathbf{k}$ in the lower spin bands. 
Finally, we see that
\begin{align}
\sum_{\lambda_\mathbf{k}}^{\mathrm{above}}\ket{\lambda_\mathbf{k}}\bra{\lambda_\mathbf{k}} &= P_+, \\
\sum_{\lambda_\mathbf{k}}^{\mathrm{below}}\ket{\phi(\lambda_\mathbf{k})}\bra{\phi(\lambda_\mathbf{k})} &= Q_+,
\end{align}
where $P_+$ is the projector onto the upper spin bands in the occupied subspace, and $Q_+$ is the projector onto the upper spin bands in the unoccupied subspace. 
This means that for systems with a spin gap, the topological properties of $P_\uparrow[Q(\mathbf{k})-P(\mathbf{k})]P_\uparrow$ bands coincide with the topological properties of the spin band structure.

Finally, we would like to determine the extent to which we can relate the topological properties of $P_\uparrow[Q(\mathbf{k})-P(\mathbf{k})]P_\uparrow$ bands to the topological properties of $H_\mathrm{low}$ directly. 
In general, we cannot do this: the process of projecting $H$ into the $P_\uparrow$ subspace does not commute with the spectral flattening process---gaps in the spectrum can close in $H_\mathrm{low}$ as we spectrally flatten $H$. 
However, for inversion symmetric systems, the theory of symmetry indicators can provide more information. 
In particular, as long as inversion eigenvalues of the occupied bands are not exchanged with inversion eigenvalues of the unoccupied bands at TRIM points as we deform $H_\mathrm{low}\rightarrow P_\uparrow[Q(\mathbf{k})-P(\mathbf{k})]P_\uparrow$, then we know that the symmetry-indicated topology of $H_\mathrm{low}$ and  $P_\uparrow[Q(\mathbf{k})-P(\mathbf{k})]P_\uparrow$ will coincide. 
We expect that for any inversion-symmetric topological system derived from a band-inverted semiconductor with weak spin orbit coupling, there will be a spin gap for some spin direction at all TRIM points. 
In these cases, applying a Zeeman field along or near that given spin direction will not invert the lowest spin band inversion eigenvalues, and hence will allow us to deduce the $H_\mathrm{low}$ band topology from the $PsP$ band topology. 
Note, furthermore, that if $H_0$ is a $\mathcal{T}$-invariant insulator, then both $H_{\mathrm{low}}$ and $PsP$ respect the symmetries of a magnetic space group. 
Since the symmetry indicators of the magnetic space groups are fully enumerated~\cite{xu2020highthroughput,elcoro2021magnetic,gao2022magnetic,peng2021topological,watanabe2018structure}, we can extend this logic to say that the magnetic symmetry indicated topology of $H_\mathrm{low}$ and $PsP$ will coincide.

As an example, we can apply this logic to 3D strong topological insulators. 
As discussed in SN~\ref{appendix:explicit-calculation-of-BHZ-model}, the $PsP$ spectrum for a 3D TI has an odd number of spin-Weyl points per half BZ. 
This implies that under the influence of a strong Zeeman field, the low energy Hamiltonian $H_\mathrm{low}$ will have an odd number of Weyl points per half BZ in its spectrum. 
This means that for a semi-infinite slab of an inversion-symmetric 3D TI under the influence of a strong Zeeman field, we expect to find Fermi arcs connecting the surface projections of these Weyl points.  
Although we can only rigorously establish the existence of Weyl points in $H_\mathrm{low}$ for inversion-symmetric systems with a spin gap, we expect that these results will hold even without inversion symmetry, due to the topological stability of Weyl points (\emph{i.e.}, the improbability of Weyl points annihilating as we adiabatically deform the spectrum of $H_\mathrm{low}$ to that of $PsP$). 
We will show that this intuition is justified in SN~\ref{app:mote2} by considering the candidate higher-order topological insulator $\beta$-MoTe$_2$~\cite{tang2019efficient,wang2019higherorder} in the presence of a strong Zeeman field. 
We will specifically show that $\beta$-MoTe$_2$ is a spin-Weyl semimetal for all choices of spin direction, with detectable Fermi arcs on the $(001)$ surface in the presence of a strong Zeeman field. Furthermore, we also show that $\alpha$-BiBr is a spin-Weyl semimetal for a neighborhood of spin directions surrounding $s_y$ in SN~\ref{sec:bibr}.

\section{Spin-Resolved Wilson Loops}\label{app:Wilson}

In this section, we will show how the projected spin operator can be used to formulate a refined notion of  band topology. 
We will start in SN~\ref{sec:P_Wilson_loop} by reviewing the definition of the Wilson loop (non-Abelian Berry phase), and its application to computing topological invariants of bands in the image of a projector $P(\mathbf{k})$. 
Next, in SN~\ref{sec:P_pm_Wilson_loop} we will show that for systems with a spin gap, we can extend the definition of the Wilson loop to compute Berry phases for $PsP$ eigenstates, which we term spin-resolved Wilson loop. 
This will allow us to introduce spin-resolved topological invariants, which classify bands that cannot be deformed into each other without closing \emph{either} an energy gap or a spin gap. 
We then study some generic properties of the spin-resolved Wilson loops, and relate their spectral flows to the 2D Kane-Mele $\mathbb{Z}_2$ invariant and the topological contribution to the spin Hall conductivity, in SN~\ref{sec:general_properties_of_winding_num_of_P_pm_Wilson}.
We will then present examples of spin-resolved topology in two (SN~\ref{sec:2D-spinful-TRI-system}) and three (SN~\ref{sec:main-text-3D-TI-P-pm} and \ref{appendix:3D-TI-with-and-without-inversion}) dimensions. 
We will explore the connection between spin-resolved and fragile topology in SN~\ref{sec:spin_stable_topology_2d_fragile_TI}. 
Finally, in SN~\ref{sec:spin_entanglement_spectrum} we will generalize the correspondence between Wilson loops and the entanglement spectrum to spin-resolved Wilson loops.

\subsection{$P$-Wilson Loop}
\label{sec:P_Wilson_loop}

We denote $[P(\mathbf{k})]$ as the matrix projector to a set of $N_{\text{occ}}$ occupied states of the Bloch Hamiltonian matrix $[H(\mathbf{k})]$, namely
\begin{align}
    [P(\mathbf{k})] = \sum_{n =1}^{N_{\text{occ}}} \ket{u_{n,\mathbf{k}}} \bra{u_{n,\mathbf{k}}}, \label{eq:P_projector}
\end{align}
where $\ket{u_{n,\mathbf{k}}}$ is defined in SEq.~(\ref{eq:TB_convention_u_nk_eig_eqn}), and $n$ denotes the band index. 
Notice that we can choose any set of occupied bands as long as there is a finite gap between the energy $E_{n,\mathbf{k}}$ of the occupied bands from the others for all $\mathbf{k}$. 
We work in a truncated tight-binding Hilbert space with $N_{\text{sta}}=2N_{\text{orb}}$ states per unit cell, where $2$ accounts for the spin degrees of freedom and $N_{\text{orb}}$ is the number of orbitals (for example $s$ and $p$ orbitals). 
Such a truncation can always be obtained for a real material through, \emph{e.g.}
Wannierization of a topologically trivial subset of electronic states in an ab-initio calculation~\cite{mostofi2008wannier90}. 
We denote by $[P(\mathbf{k})]$ the $2N_{\text{orb}} \times 2N_{\text{orb}}$ matrix representation of $P(\mathbf{k})$. 
Similarly, our Bloch eigenstates $\ket{u_{n,\mathbf{k}}}$ are $2N_{\text{orb}}$-component vectors in the tight-binding basis states. 
The $P$-Wilson loop matrix for the holonomy starting at base point $\mathbf{k}$ and going along a straight path to $\mathbf{k+G}$ where $\mathbf{G}$ is a primitive reciprocal lattice vector
is given by~\cite{yu2011equivalent,fidkowski2011model,alexandradinata2014wilsonloop}
\begin{align}
    [\mathcal{W}_{1,\mathbf{k},\mathbf{G}}]_{m,n} & = \bra{u_{m,\mathbf{k}+\mathbf{G}}} \left( \prod_{\mathbf{q}}^{\mathbf{k}+\mathbf{G} \leftarrow \mathbf{k}} [P(\mathbf{q})] \right)\ket{u_{n,\mathbf{k}}} \nonumber \\
    & = \bra{u_{m,\mathbf{k}}} [V(\mathbf{G})] \left(\prod_{\mathbf{q}}^{\mathbf{k}+\mathbf{G} \leftarrow \mathbf{k}} [P(\mathbf{q})]\right)\ket{u_{n,\mathbf{k}}}, \label{eq:P_Wilson_loop_matrix_element}
\end{align}
where both $m$ and $n$ are the indices of occupied eigenvectors, the $[V(\mathbf{G})]$ matrix is defined in SN~\ref{appendix:TB-notation}, and
\begin{align}
    & \left(\prod_{\mathbf{q}}^{\mathbf{k}+\mathbf{G} \leftarrow \mathbf{k}} [P(\mathbf{q})]\right) \\
    & \equiv \lim_{N \to \infty}
    [P(\mathbf{k}+\mathbf{G})] [P(\mathbf{k}+\frac{N-1}{N}\mathbf{G})] \ldots [P(\mathbf{k}+\frac{1}{N}\mathbf{G})] [P(\mathbf{k})]. \label{eq:P_Wilson_loop_P_k_def}
\end{align}
The eigenvalues of the $P$-Wilson loop matrix $[\mathcal{W}_{1,\mathbf{k},\mathbf{G}}]$ [SEq.~(\ref{eq:P_Wilson_loop_matrix_element})], which is an $N_{\text{occ}} \times N_{\text{occ}}$ matrix, are of the unimodular form $\exp{i(\gamma_{1})_{j,\mathbf{k},\mathbf{G}}}$ where $j = 1 \ldots N_{\text{occ}}$ index the $N_{\text{occ}}$ eigenvalues. 
We call the set $\{ (\gamma_{1})_{j,\mathbf{k},\mathbf{G}} \}$ as $P$-{\it Wannier bands} (or simply \textit{Wannier bands}) because $\{(\gamma_{1})_{j,\mathbf{k},\mathbf{G}}\}$ tells us the positions of the hybrid Wannier functions, which are localized along the lattice vector $\mathbf{a}$ dual to $\mathbf{G}$; to be specific the hybrid Wannier functions are localized at $\mathbf{a} \cdot (\gamma_{1})_{j,\mathbf{k},\mathbf{G}}/ 2\pi$.
The $P$-Wilson loop eigenphases $\{(\gamma_{1})_{j,\mathbf{k},\mathbf{G}}\}$ are the Berry phases of the $N_{\text{occ}}$ occupied states for the closed loop in $\mathbf{k}$-space parallel to $\mathbf{G}$.
If we write the base point $\mathbf{k} = \sum_{j=1}^{d}\frac{k_{j}}{2\pi}\mathbf{G}_{j}$ where $d$ is the spatial dimension of the system, and if we choose the closed loop holonomy to be along $\mathbf{G}_{i}$ with base point $\mathbf{k}$, then $\{(\gamma_{1})_{j,\mathbf{k},\mathbf{G}_{i}}\}$ will be independent of the component $k_{i} = \mathbf{k} \cdot \mathbf{a}_{i}$ where $\{\mathbf{a}_{i}\}$ is the set of position-space lattice vectors dual to the reciprocal lattice vectors $\{\mathbf{G}_{j}\}$ such that $\mathbf{a}_{i} \cdot \mathbf{G}_{j} = 2\pi \delta_{ij}$~\cite{lecture_notes_on_berry_phases_and_topology_bb}.
In addition, to directly specify the holonomy in SEq.~\eqref{eq:P_Wilson_loop_matrix_element}, we will sometimes call $\{(\gamma_{1})_{j,\mathbf{k},\mathbf{G}}\}$ as a function of $\mathbf{k}$ the {\it $\widehat{\mathbf{G}}$-directed $P$-Wannier bands} (or simply \textit{$\widehat{\mathbf{G}}$-directed Wannier bands}), where $j$ labels the band index. 
The symmetry-protected spectral flow of the eigenphases $\{(\gamma_{1})_{j,\mathbf{k},\mathbf{G}}\}$ is a powerful tool for diagnosing the nontrivial topology of energy bands~\cite{alexandradinata2014wilsonloop,soluyanov2011wannier,yu2011equivalent,fidkowski2011model,wang2016hourglass,wieder2018wallpaper,taherinejad2014wannier,gresch2017z2pack,cano2021band,po2018fragile,bouhon2019wilson,cano2018topology,bradlyn2019disconnected}.
In SN~\ref{appendix:I-constraint-on-P-wilson} and \ref{appendix:T-constraint-on-P-wilson} we prove the constraints on the values of the $P$-Wilson loop eigenphases at different $\mathbf{k}$ points due to inversion and time-reversal symmetries, respectively.

\subsection{$P_{\pm}$-Wilson Loop}
\label{sec:P_pm_Wilson_loop}

We would now like to formulate a definition of a Wilson loop applicable to spin bands. 
Recall from SEq.~\eqref{eq:s_reduced_no_k} that we can introduce the reduced spin matrix
\begin{align}
    [s_{\text{reduced}}(\mathbf{k})]_{m,n} = \bra{u_{m,\mathbf{k}}} s \ket{u_{n,\mathbf{k}}} =\bra{u_{m,\mathbf{k}}} \hat{\mathbf{n}}\cdot\mathbf{s} \ket{u_{n,\mathbf{k}}}  \label{eq:P_pm_Wilson_loop_s_reduced_def}
\end{align}
where $\hat{\mathbf{n}}$ is a unit vector, $\mathbf{s}$ is the spin operator from SEq.~\eqref{eq:appendix-def-s-i}, $m$ and $n$ index the occupied states (\emph{i.e.} states in the image of occupied-space projector $[P(\mathbf{k})]$), and where we have made the $\mathbf{k}$ dependence explicit.
Suppose that a spin gap is open (see SN~\ref{appendix:properties-of-the-projected-spin-operator}), such that we can divide the eigenvalues of $[s_{\text{reduced}}(\mathbf{k})]$ into two disjoint groups labeled by $\pm$, 
\begin{align}
    & [s_{\text{reduced}}(\mathbf{k})] \ket{\tilde{u}_{n,\mathbf{k}}^{\pm}} = \lambda_{n,\mathbf{k}}^{\pm}\ket{\tilde{u}_{n,\mathbf{k}}^{\pm}} \label{eq:eig-eqn-s-reduced-k-tilde-u-n-k},
\end{align}
where we have for all $\mathbf{k}$, $n$ and $m$ that $\lambda^+_{n,\mathbf{k}} > \lambda^-_{m,\mathbf{k}}$.
We refer to $\{ \lambda^{\pm}_{n,\mathbf{k}} \}$ as the upper and lower $P s P$ bands, respectively.
We denote by $N^+_\mathrm{occ}$ the number of upper spin bands, and by $N^-_\mathrm{occ}$ the number of lower spin bands. 
We necessarily have $N^+_\mathrm{occ}+N^-_\mathrm{occ}=N_\mathrm{occ}$. 
Furthermore, since a spin gap is open, $N^+_\mathrm{occ}$ and $N^-_\mathrm{occ}$ are independent of $\mathbf{k}$. 
Recall from SN~\ref{appendix:properties-of-the-projected-spin-operator} that in a system with $\mathcal{T}$ symmetry, it is possible to choose the upper and lower spin bands such that $N^+_\mathrm{occ}=N^-_\mathrm{occ}=N_\mathrm{occ}/2$. 
Additionally, in a system with both $\mathcal{I}$ and $\mathcal{T}$ symmetry, we can choose the upper and lower spin bands to satisfy $\lambda^+_{n,\mathbf{k}} >0$ and $ \lambda^-_{m,\mathbf{k}} < 0$, respectively.

In terms of spinful orbital basis states, we can re-express $\ket{\tilde{u}_{n,\mathbf{k}}^{\pm}}$ as
\begin{align}
    \ket{u_{n,\mathbf{k}}^{\pm}} = \sum_{m=1}^{N_{\text{occ}}} [\tilde{u}_{n,\mathbf{k}}^{\pm}]_{m} \ket{u_{m,\mathbf{k}}}, \label{eq:reexpress_the_eigenvector_of_s_reduced}
\end{align}
where $[\tilde{u}_{n,\mathbf{k}}^{\pm}]_{m}$ is the $m^{\text{th}}$ component of the $n^{\text{th}}$ $N_{\text{occ}}$-component eigenvector $\ket{\tilde{u}_{n,\mathbf{k}}^{\pm}}$ and $\ket{u_{m,\mathbf{k}}}$ is the $2N_{\text{orb}}$-component eigenvectors for the occupied bands of $[H(\mathbf{k})]$. 
We can also derive the boundary condition for $\ket{u_{n,\mathbf{k}}^{\pm}}$ as follows. 
Upon a shift $\mathbf{k} \to \mathbf{k} + \mathbf{G}$, we have
\begin{align}
    & [s_{\text{reduced}}(\mathbf{k}+\mathbf{G})]_{m,n} \\
    & = \bra{u_{m,\mathbf{k}+\mathbf{G}}} s \ket{u_{n,\mathbf{k}+\mathbf{G}}} \\
    & = \bra{u_{m,\mathbf{k}}} [V(\mathbf{G})] s [V(\mathbf{G})]^{-1} \ket{u_{n,\mathbf{k}}},
\end{align}
where the $[V(\mathbf{G})]$ matrix is defined in SN~\ref{appendix:TB-notation}.
If we assume that our orbital truncation is such that time-reversed pairs of orbitals are located at the same position (this is always possible, as it is true of atomic orbitals), we can factorize $[V(\mathbf{G})]$ into $[V(\mathbf{G})] = \sigma_{0} \otimes {[\widetilde{V}(\mathbf{G})]}$ where ${[\widetilde{V}(\mathbf{G})]}$ is a unitary $N_{\text{orb}} \times N_{\text{orb}}$ matrix (recall that $2N_\text{orb}$ is the dimension of the Hilbert space of $[H(\mathbf{k})]$) with matrix element ${[\widetilde{V}(\mathbf{G})]}_{\widetilde{\alpha},\widetilde{\beta}} = \delta_{\widetilde{\alpha}\widetilde{\beta}} e^{i \mathbf{G}\cdot \mathbf{r}_{\widetilde{\alpha}}}$. 
Notice that the indices $\widetilde{\alpha}$ and $\widetilde{\beta}$ both range from $1 \ldots N_{\text{orb}}$ and $\mathbf{r}_{\widetilde{\alpha}}$ is the position of the orbital labeled by $\widetilde{\alpha}$. 
From this we can show that the matrix spin operator $s$ is invariant under a unitary transformation of $[V(\mathbf{G})]$ 
\begin{align}
    & [V(\mathbf{G})] s [V(\mathbf{G})]^{-1} \label{eq:V_G_and_s_asssumption_1} \\
    & = \left( \sigma_{0} \otimes {[\widetilde{V}(\mathbf{G})]} \right) \left( \hat{\mathbf{n}}\cdot \bm{\sigma} \otimes \mathbb{I}_{N_{\text{orb}}} \right) \left( \sigma_{0} \otimes {[\widetilde{V}(\mathbf{G})]}^{-1} \right) \label{eq:V_G_and_s_asssumption_2} \\
    & = \hat{\mathbf{n}}\cdot \bm{\sigma} \otimes \mathbb{I}_{N_{\text{orb}}} = s. \label{eq:V_G_and_s_asssumption_3}
\end{align}
This implies
\begin{align}
    [s_{\text{reduced}}(\mathbf{k}+\mathbf{G})]_{m,n} = [s_{\text{reduced}}(\mathbf{k})]_{m,n}, \label{eq:P_pm_Wilson_loop_PBC_s_reduced}
\end{align}
and the boundary condition for $\ket{\tilde{u}_{n,\mathbf{k}}^{\pm}}$ can be chosen as
\begin{align}
    \ket{\tilde{u}_{n,\mathbf{k}+ \mathbf{G}}^{\pm}} = \ket{\tilde{u}_{n,\mathbf{k}}^{\pm}},
\end{align}
such that the component $[\tilde{u}_{n,\mathbf{k}}^{\pm}]_{m}$ in SEq.~(\ref{eq:reexpress_the_eigenvector_of_s_reduced}) has the same boundary condition $[\tilde{u}_{n,\mathbf{k}+\mathbf{G}}^{\pm}]_{m}=[\tilde{u}_{n,\mathbf{k}}^{\pm}]_{m}$. 
We now define the projectors onto the upper and lower spin bands as
\begin{align}
    [P_{\pm}(\mathbf{k})] =  \sum_{n=1}^{N_{\text{occ}}^{\pm}} \ket{u_{n,\mathbf{k}}^{\pm}} \bra{u_{n,\mathbf{k}}^{\pm}}. \label{eq:P_pm_k_projector}
\end{align}
The occupied space matrix projector $[P(\mathbf{k})]$ is then equal to $[P_{+}(\mathbf{k})]+[P_{-}(\mathbf{k})]$ where $[P_{+}(\mathbf{k})][P_{-}(\mathbf{k})]=0$.
The corresponding holonomy matrix starting at base point $\mathbf{k}$ and going along a straight line path to $\mathbf{k+G}$ for $[P_{\pm}(\mathbf{k})]$, which we term the $P_{\pm}$-Wilson loop matrix (or the spin-resolved Wilson loop matrix), is given by
\begin{align}
    [\mathcal{W}_{1,\mathbf{k},\mathbf{G}}^{\pm}]_{m,n} & = \bra{u_{m,\mathbf{k}+\mathbf{G}}^{\pm}} \left(\prod_{\mathbf{q}}^{\mathbf{k}+\mathbf{G} \leftarrow \mathbf{k}} [P_{\pm}(\mathbf{q})]\right)\ket{u_{n,\mathbf{k}}^{\pm}} \nonumber \\
    & = \bra{u_{m,\mathbf{k}}^{\pm}} [V(\mathbf{G})] \left(\prod_{\mathbf{q}}^{\mathbf{k}+\mathbf{G} \leftarrow \mathbf{k}} [P_{\pm}(\mathbf{q})]\right)\ket{u_{n,\mathbf{k}}^{\pm}}, \label{eq:P_pm_Wilson_loop_matrix_element}
\end{align}
where 
\begin{align}
    & \left(\prod_{\mathbf{q}}^{\mathbf{k}+\mathbf{G} \leftarrow \mathbf{k}} [P_{\pm}(\mathbf{q})] \right) \\
    & \equiv \lim_{N \to \infty}
    [P_{\pm}(\mathbf{k}+\mathbf{G})] [P_{\pm}(\mathbf{k}+\frac{N-1}{N}\mathbf{G})] \ldots [P_{\pm}(\mathbf{k}+\frac{1}{N}\mathbf{G})] [P_{\pm}(\mathbf{k})], \label{eq:P_pm_Wilson_loop_P_k_def}
\end{align}
and we have used the fact that, since $\ket{u_{n,\mathbf{k}+\mathbf{G}}} = [V(\mathbf{G})]^{-1} \ket{u_{n,\mathbf{k}}}$ (see SN~\ref{appendix:TB-notation}) and $[\tilde{u}_{n,\mathbf{k}+\mathbf{G}}^{\pm}]_{m}=[\tilde{u}_{n,\mathbf{k}}^{\pm}]_{m}$, we have the boundary condition for $| u_{n,\mathbf{k}}^{\pm} \rangle$ as
\begin{align}
    \ket{u_{n,\mathbf{k}+\mathbf{G}}^{\pm}} & = \sum_{m=1}^{N_{\text{occ}}} [\tilde{u}_{n,\mathbf{k}+\mathbf{G}}^{\pm}]_{m} \ket{u_{m,\mathbf{k}+\mathbf{G}}} \\
    & = \sum_{m=1}^{N_{\text{occ}}} [\tilde{u}_{n,\mathbf{k}}^{\pm}]_{m} [V(\mathbf{G})]^{-1}\ket{u_{m,\mathbf{k}}} \\
    & = [V(\mathbf{G})]^{-1}\sum_{m=1}^{N_{\text{occ}}} [\tilde{u}_{n,\mathbf{k}}^{\pm}]_{m} \ket{u_{m,\mathbf{k}}} \\
    & =[V(\mathbf{G})]^{-1} \ket{u_{n,\mathbf{k}}^{\pm}}. \label{eq:P_pm_Wilson_loop_BC_for_u_pm}
\end{align}
The $P_{\pm}$-Wilson loop matrices $[\mathcal{W}_{1,\mathbf{k},\mathbf{G}}^{\pm}]$ [SEq.~\eqref{eq:P_pm_Wilson_loop_matrix_element}] are $N_{\text{occ}}^{+} \times N_{\text{occ}}^{+}$ and $N_{\text{occ}}^{-} \times N_{\text{occ}}^{-}$ unitary matrices respectively. 
We can write the eigenvalues of $[\mathcal{W}_{1,\mathbf{k},\mathbf{G}}^{\pm}]$ as $\exp{i(\gamma_{1}^{\pm})_{j,\mathbf{k},\mathbf{G}}}$ where $j = 1\ldots N_{\text{occ}}^{\pm}$ index the $N_{\text{occ}}^{\pm}$ eigenvalues.
We call the set $\{ (\gamma_{1}^{\pm})_{j,\mathbf{k},\mathbf{G}} \}$ as $P_{\pm}$-{\it Wannier bands} (or \textit{spin-resolved Wannier bands}) because $\{(\gamma_{1}^{\pm})_{j,\mathbf{k},\mathbf{G}}\}$ tells us the positions of spin-resolved hybrid Wannier functions formed from the subspace of upper $(+)$ or lower $(-)$ spin bands; the spin-resolved hybrid Wannier functions are localized along the lattice vector dual to $\mathbf{G}$. 
Similar to the eigenphases of the ordinary Wilson loops $[\mathcal{W}_{1,\mathbf{k},\mathbf{G}}]$, the eigenphases $\{(\gamma_{1}^{\pm})_{j,\mathbf{k},\mathbf{G}}\}$ are invariant with respect to gauge transformations that do not change the boundary conditions on the Bloch wave functions. 
This follows from SEq.~\eqref{eq:V-s-commute} as well as the definition in SEq.~\eqref{eq:reexpress_the_eigenvector_of_s_reduced} of the $PsP$ eigenstates. 
The eigenphases $\{(\gamma_{1}^{\pm})_{j,\mathbf{k},\mathbf{G}}\}$ of the $P_{\pm}$-Wilson loop matrix are the Berry phases of the $P_{\pm}$-subspace of the $N_{\text{occ}}$ occupied states for the closed loop in $\mathbf{k}$-space parallel to $\mathbf{G}$. 
Again, if we write the base point $\mathbf{k} = \sum_{j=1}^{d}\frac{k_{j}}{2\pi}\mathbf{G}_{j}$ where $d$ is the spatial dimension of the system, and if we choose the closed loop holonomy to be along $\mathbf{G}_{i}$ with base point $\mathbf{k}$, then $\{(\gamma_{1}^{\pm})_{j,\mathbf{k},\mathbf{G}_{i}}\}$ will be independent of the component $k_{i} = \mathbf{k} \cdot \mathbf{a}_{i}$ where $\{\mathbf{a}_{i}\}$ is the set of position-space lattice vectors dual to the reciprocal lattice vectors $\{\mathbf{G}_{j}\}$ such that $\mathbf{a}_{i} \cdot \mathbf{G}_{j} = 2\pi \delta_{ij}$~\cite{lecture_notes_on_berry_phases_and_topology_bb}.
In addition, to directly specify the holonomy in SEq.~\eqref{eq:P_pm_Wilson_loop_matrix_element}, we will throughout this work refer to  $\{(\gamma_{1}^{\pm})_{j,\mathbf{k},\mathbf{G}}\}$ as a function of $\mathbf{k}$ as the {\it $\widehat{\mathbf{G}}$-directed $P_{\pm}$-Wannier bands} (or \textit{$\widehat{\mathbf{G}}$-directed spin-resolved Wannier bands}), where $j$ labels the band index. 
Similar to the case of {\it $\widehat{\mathbf{G}}$-directed $P$-Wannier bands} we may deduce the topological properties of the system by computing the spectral flow of $\{(\gamma_{1}^{\pm})_{j,\mathbf{k},\mathbf{G}}\}$. 
In SN~\ref{appendix:I-constraint-on-P-pm} and \ref{appendix:T-constraint-on-P-pm} we prove the constraints on the values of the $P_{\pm}$-Wilson loop eigenphases at different $\mathbf{k}$ points due to inversion and time-reversal symmetries, respectively.
In the next SN~\ref{sec:general_properties_of_winding_num_of_P_pm_Wilson} we will discuss some general properties of spectral flow in the $P_\pm$-Wilson loop.

\subsection{Spectral Flow of the $P_{\pm}$-Wilson Loop}
\label{sec:general_properties_of_winding_num_of_P_pm_Wilson}

In this section we discuss some general properties of the spectral flow of the $P_{\pm}$-Wilson loop eigenphases $\gamma^{\pm}_{1}$.
To simplify the discussion, we will in this section specify to 2D spinful systems.
We will not assume any symmetries to begin (including time-reversal symmetry) except for discrete translations. 
We take the primitive Bravais lattice vectors to be $\{\mathbf{a}_{1} ,\mathbf{a}_{2}\}$, and the corresponding primitive reciprocal lattice vectors to be $\{\mathbf{G}_{1},\mathbf{G}_2 \}$ satisfying $\mathbf{a}_{i} \cdot \mathbf{G}_{j} = 2\pi \delta_{ij}$.
The crystal momentum $\mathbf{k}$ can then be expanded using $\mathbf{k} = \sum_{i=1}^{2} \frac{k_i}{2\pi} \mathbf{G}_i$ where $k_i = \mathbf{k} \cdot \mathbf{a}_i$.
The BZ is then defined by the region with $k_i = [-\pi,\pi)$.
Toward the end of this section we will additionally specialize to time-reversal ($\mathcal{T}$) invariant systems. 
We will also generalize to 3D systems in SN~\ref{sec:main-text-3D-TI-P-pm} and \ref{appendix:3D-TI-with-and-without-inversion}.

We will suppose that both the energy and the spin gaps are open. 
This implies that the occupied-space projector $P(\mathbf{k})$ is well-defined and is a smooth function of $\mathbf{k}$, 
and that the eigenvectors of $P(\mathbf{k}) s P(\mathbf{k})$ in $\mathrm{Image}[P(\mathbf{k})]$ can be separated into {upper} and {lower} eigenspaces with projectors $P_{\pm}(\mathbf{k})$. 
The projectors $P_{\pm}(\mathbf{k})$ are similarly well-defined and smooth over the BZ.
Let us denote the $\widehat{\mathbf{G}}_2$-directed $P_{\pm}$-Wilson loop eigenphase as $\{\gamma_{1,j}^{\pm} (k_1)\}$ where $j$ is the $P_{\pm}$-Wannier band indices. 
{Notice that, to facilitate the discussion in this subsection~\ref{sec:general_properties_of_winding_num_of_P_pm_Wilson}, unlike the most general notation $\{ (\gamma_1^{\pm})_{j,\mathbf{k},\mathbf{G}} \}$ introduced in {SN}~\ref{sec:P_pm_Wilson_loop}, we have suppressed the subscripts for both $\mathbf{k}$ denoting the base point and $\mathbf{G}$ characterizing the holonomy such that only the $k_1$-dependence remains explicit in the expressions below. 
This is because the eigenphases are independent of $k_2$ for $\widehat{\mathbf{G}}_2$-directed $P_{\pm}$-Wilson loops.}
We can define the partial Chern numbers $C_{\gamma_1}^{\pm}$ as the \textit{negative} winding number of $\mathrm{Im}\log\det [\mathcal{W}^\pm_{1,\mathbf{k},{\mathbf{G}_2}}]$ as a function of {$k_1$}, or equivalently the \textit{positive} winding number of $\mathrm{Im}\log\det [\mathcal{W}^\pm_{1,\mathbf{k},{\mathbf{G}_1}}]$ as a function of {$k_2$}. 
In terms of the eigenphases $\{\gamma_{1,j}^{\pm} (k_1)\}$, we can write the partial Chern numbers as~\cite{prodan2009robustness,sheng2006spinChern}
\begin{equation}
	C_{\gamma_1}^{\pm} =-\frac{1}{2\pi} {\sum_{j=1}^{N_{\mathrm{occ}}^{\pm}}} \int_{-\pi}^{\pi} dk_1 \frac{\partial \gamma_{1,j}^{\pm}(k_1)}{\partial k_1} = -\frac{i}{2\pi}\int d^2k \mathrm{Tr}\left(P_\pm {(\mathbf{k})} \left[\frac{\partial P_\pm(\mathbf{k})}{\partial k_1},\frac{\partial P_\pm(\mathbf{k})}{\partial k_2}\right]\right).\label{eq:partial_chern_def}
\end{equation}
We see that the partial Chern numbers $C_{\gamma_1}^{\pm}$ are respectively equal to the Chern numbers of the subspaces of states in the images of $P_{\pm}(\mathbf{k})$.
Denoting the Chern number of the occupied energy bands as $C_{\gamma_1}$, we can use $P_{+}(\mathbf{k})+P_{-}(\mathbf{k}) = P(\mathbf{k})$ and $P_{+}(\mathbf{k}) P_{-}(\mathbf{k}) = 0$ in conjunction with the second equality {of SEq.~\eqref{eq:partial_chern_def}} to show that~(see for example SRef.~\cite{avron1998adiabatic})
\begin{equation}
	C_{\gamma_1}^{+} + C_{\gamma_1}^{-} = C_{\gamma_1}. \label{eq:C_plus_C_minus_C_exact_relation}
\end{equation}
As argued in SRef.~\cite{prodan2009robustness}, the partial Chern numbers $C_{\gamma_1}^{\pm}$ are topological invariants, in the sense that two systems with different partial Chern numbers cannot be adiabatically deformed into each other without closing either an energy gap or a spin gap. 
As shown in SN~\ref{sec:pspperturbation}, this is a physically motivated constraint since both the energy gap and the spin gap are robust to external perturbations and hence physically meaningful. 
During a deformation of the Hamiltonian, although $C_{\gamma_1}$ and $C_{\gamma_1}^{\pm}$ can change due to energy and spin gap closing respectively, they need to satisfy the exact relation in SEq.~\eqref{eq:C_plus_C_minus_C_exact_relation}.

Following {SRef.~\cite{prodan2009robustness}}, we may also define a {\it spin Chern number} 
\begin{equation}
	C^{s}_{\gamma_1} \equiv {C^{+}_{\gamma_1} - C^{-}_{\gamma_1}}, \label{eq:Csgamma1def}
\end{equation}
which can be viewed as the {\it relative winding number} of the Wilson loop eigenphases of the upper and lower $P(\mathbf{k})sP(\mathbf{k})$ eigenspaces. 
We have normalized $C^{s}_{\gamma_1}$ such that, for a system with conserved spin component $\hat{\mathbf{n}}\cdot\mathbf{s}$, the spin Hall conductivity is given by~\cite{sheng2006spinChern,kane2005quantum,bernevig2006quantum,konig2008quantum,murakami2004spin,sinova2004universal,sinova2015spin}
\begin{equation}\label{eq:spinHall}
\sigma^s_{H} = \left(\frac{\hbar}{2}\right)\left(\frac{e}{h}\right)C^{s}_{\gamma_1},
\end{equation}
where $e$ is the (negative) electron charge. 

SEq.~\eqref{eq:C_plus_C_minus_C_exact_relation} places constraints on how $C^{s}_{\gamma_1}$ can change when a spin gap closes. 
Suppose we deform the Hamiltonian in a way that the energy gap remains open while the spin spectrum undergoes a spin band inversion (\emph{i.e.} the spin gap between the upper and lower spin bands closes and reopens).
$C_{\gamma_1}$ will remain invariant since the energy gap is open throughout the deformation.
According to SEq.~\eqref{eq:C_plus_C_minus_C_exact_relation}, $C^{+}_{\gamma_1}$ and $C^{-}_{\gamma_1}$ can only be changed by opposite integers in order to keep $C_{\gamma_1}$ invariant.
For example, before and after the spin band inversion, we can have
\begin{align}
	& C_{\gamma_1} \to C_{\gamma_1}, \\
	& C_{\gamma_1}^{\pm} \to C_{\gamma_1}^{\pm} \pm n,
\end{align}
where $n\in \mathbb{Z}$.
Therefore, as long as the energy gap remains open, $C_{\gamma_1}^{+}$ and $C_{\gamma_1}^{-}$ cannot change independently. 
However, absent additional symmetries, the values of $C_{\gamma_1}^{+}$ and $C_{\gamma_1}^{-}$ need not to be related.
From the definition of $C_{\gamma_1}^{s}$ in SEq.~\eqref{eq:Csgamma1def}, when $C_{\gamma_1}^{\pm} \to C_{\gamma_1}^{\pm} \pm n$, we have $C_{\gamma_1}^{s} \to C_{\gamma_1}^{s} + 2n$.
Therefore, without an energy gap closing, it is possible to deform the Hamiltonian such that $C_{\gamma_1}^{s}$ is changed by an even integer $2n$ when the spin gap closes and reopens.
Therefore, $(C_{\gamma_1}^{s}) \text{ mod } 2$ is a $\mathbb{Z}_2$-stable topological invariant that cannot change even when the spin gap closes and reopens. 
This invariant specifically characterizes whether there is an even- or odd-integer difference between $C_{\gamma_1}^{+}$ and $C_{\gamma_1}^{-}$.
Note that systems with $(C_{\gamma_1}^{s}) \text{ mod } 2 = 1$ cannot be time-reversal symmetric, since the Chern number $C_{\gamma_1}=C_{\gamma_1}^{+}+C_{\gamma_1}^{-} \neq 0$ if $C_{\gamma_1}^{+} - C_{\gamma_1}^{-}$ is odd.

As an example, let us consider a toy model of a Chern insulator with two spinful orbitals per unit cell. 
We suppose that $\hat{\mathbf{n}}\cdot\mathbf{s}$ commutes with the Hamiltonian, so that spin {$\hat{\mathbf{n}}\cdot\mathbf{s}$} is a good quantum number. 
This means that we can compute the partial Chern numbers for the spin-up {(along $+\hat{\mathbf{n}}$)} and spin-down {(along $-\hat{\mathbf{n}}$)} occupied states separately. 
Suppose that the total Chern number $C_{\gamma_1}=1$, the partial Chern number $C^+_{\gamma_1}=1$ for the spin-up electrons, and the partial Chern number $C^-_{\gamma_1}=0$ for the spin-down electrons. 
In this case the spin Chern number $C_{\gamma_1}^{s} = 1$.
If we deform the Hamiltonian in a way that the energy gap does not close, but the spin gap can close and reopen, then there can be a transfer of partial Chern number between the upper and lower spin bands. 
Absent any (crystallographic or time-reversal) symmetry constraints, we can have, for instance, that
\begin{align}
	C_{\gamma_1}^{+} = 1 \to 2,\ C_{\gamma_1}^{-} = 0 \to -1,
\end{align}
such that $C_{\gamma_1}=C_{\gamma_1}^{+} + C_{\gamma_1}^{-}$ is still $1$.
The spin Chern number $C_{\gamma_1}^{s}$ has changed from $1$ to $3$.
In this example, before and after the spin gap closing, $(C_{\gamma_1}^{s}) \text{ mod } 2$ is always equal to $1$. 
Furthermore, the energy gap has remained open throughout the deformation. 
Note that if the system has conserved spin component $\hat{\mathbf{n}}\cdot\mathbf{s}$ before and after the deformation, then the spin Hall conductivity [SEq.~\eqref{eq:spinHall}] must have changed by $e/(2\pi)$ times an integer. 

For non-interacting systems with time-reversal symmetry, we have that $C^s_{\gamma_1}$ must be even, and 
\begin{equation}
	C_{\gamma_1}^{s} = 2C^{+}_{\gamma_1} = -2C^{-}_{\gamma_1}. \label{eq:C_gamma_1_relate_to_C_plus_C_minus}
\end{equation}
SEq.~\eqref{eq:C_gamma_1_relate_to_C_plus_C_minus} follows from inserting SEq.~\eqref{eq:t_on_ppm} in the definition [SEqs.~\eqref{eq:partial_chern_def} and \eqref{eq:Csgamma1def}] of the spin Chern number. 
SEq.~\eqref{eq:C_gamma_1_relate_to_C_plus_C_minus} means that for time-reversal-symmetric systems, the partial Chern number $C^{+}_{\gamma_1}$ determines the spin Chern number $C_{\gamma_1}^s$. 
In fact, it was established in SRefs.~\cite{prodan2009robustness,fu2006time,fukui2007topological} that for time-reversal-symmetric systems $C^{s}_{\gamma_1} \mod 4$ (or alternatively $C^{\pm}_{\gamma_1}\mod 2$) can be directly related to the two-dimensional strong $\mathbb{Z}_2$-valued index $\nu_{2d}$ for 2D topological insulators introduced by Kane and Mele~\cite{kane2005quantum,kane2005z}. 
To see this, let us consider a deformation of the Hamiltonian that preserves $\mathcal{T}$ symmetry and does not close an energy gap, while allowing the spin gap to close and reopen.
Due to $\mathcal{T}$ symmetry, if the spin gap closes at $\mathbf{k}$, then the spin gap will also close at $-\mathbf{k}$.
This means that $C^{\pm}_{\gamma_1}$ can in general change only by even integers, such that
\begin{equation}
	C_{\gamma_1}^{\pm} \to C_{\gamma_1}^{\pm} \pm 2n, \label{eq:delta_cpm}
\end{equation}
where $n \in \mathbb{Z}$.
{Although the states at the spin gap closing points $\mathbf{k}$ and $-\mathbf{k}$ are related by time-reversal, the same quantity of partial Chern number is transferred from the lower spin bands to the upper spin bands at each point. 
This is because unlike for energy bands, time-reversal symmetry interchanges the upper and lower spin bands.}
Furthermore, if the spin gap closes between two spin bands at a TRIM, $\mathcal{T}$ symmetry requires that the spin band dispersion near the gap closing point is quadratic; hence $C^\pm_{\gamma_1}$ can also only change by an even integer~\cite{fang2012multi}. 
SEq.~\eqref{eq:delta_cpm} implies that $C^{s}_{\gamma_1}$ can only change according to
\begin{equation}
	C^{s}_{\gamma_1} \to C^{s}_{\gamma_1} + 4n.\label{eq:trcschange}
\end{equation}
We then see that with $\mathcal{T}$ symmetry, $1/2(C^{s}_{\gamma_1} \mod 4)$ is a stable $\mathbb{Z}_2$ invariant as first shown by Prodan~\cite{prodan2009robustness}. This is consistent the crystallographic splitting theorem of SRef.~\cite{alexandradinata2020crystallographic}.

To further establish that $1/2(C^{s}_{\gamma_1} \mod 4)$ coincides with the Kane-Mele $\mathbb{Z}_2$ invariant, let us consider a $\mathcal{T}$-invariant system with $C^{\pm}_{\gamma_1}=\pm 1$ such that $C^{s}_{\gamma_1} = 2$.
Recall that $\{ \gamma_{1,j}^{\pm}(k_1) \}$ are the localized positions along $\mathbf{a}_{2}$ of the spin-resolved hybrid Wannier functions in the upper and lower $PsP$ eigenspaces, and the partial Chern numbers $C^{\pm}_{\gamma_1}$ are the corresponding \textit{negative} net winding numbers of the spin-resolved hybrid Wannier centers when $k_1 \to k_1 + 2\pi$. 
Since $\mathcal{T}P_\pm(\mathbf{k})\mathcal{T}^{-1}=P_\mp(-\mathbf{k})$, the $P_+$- and $P_-$-hybrid Wannier functions are related to each other by time-reversal symmetry.
$C^{\pm}_{\gamma_1}=\pm 1$ then means that we cannot form exponentially-localized Wannier functions using only states in the image of $P_\pm$, since the spin-resolved hybrid Wannier functions have a discontinuity when $k_1\rightarrow k_1+2\pi$~\cite{soluyanov2012smooth}. 
Thus, we see that $1/2(C^{s}_{\gamma_1} \mod 4)=1$ quantifies the obstruction to forming exponentially-localized Wannier functions related to each other by time-reversal symmetry, and so coincides with $\nu_{2d}$~\cite{soluyanov2012smooth}. 
Hence an odd winding in the $P_\pm$-Wilson loops implies {\it helical winding} protected by $\mathcal{T}$ symmetry in the centers of the ordinary hybrid Wannier functions~\cite{soluyanov2011wannier,yu2011equivalent} computed using the $P$-Wilson loop formalism (SN~\ref{sec:P_Wilson_loop}). 
We will numerically demonstrate this in {SN}~\ref{sec:2D-spinful-TRI-system}.

Note also that a nonzero spin Chern number $C^s_{\gamma_1}$ implies an intrinsic contribution to the spin Hall conductivity. 
For systems with conserved component $\hat{\mathbf{n}}\cdot\mathbf{s}$ of the spin, we have that the spin Hall conductivity is directly proportional to the spin Chern number via SEq.~\eqref{eq:spinHall}; this follows from Thouless-Komohto-Nightingale-den Nijs expression for the Kubo formula for the Hall conductivity~\cite{thouless1982quantized} applied separately to spin-up and spin-down states~\cite{sheng2006spinChern}. 
When spin is not conserved, the Kubo formula for the spin Hall conductivity takes the more complicated form
\begin{equation}\label{eq:shckubo1}
\sigma^s_H = \frac{e}{4\pi}(\sigma^s_I + \sigma^s_{II}),
\end{equation}
where (using $\mathrm{Tr} (\mathcal{O})$ to denote the trace of operator $\mathcal{O}$)
\begin{equation}\label{shckubo2}
\sigma^s_I = -\frac{i}{2\pi}\int d^2k \mathrm{Tr}\left(sP\left[\frac{\partial P}{\partial k_1},\frac{\partial P}{\partial k_2}\right]\right)
\end{equation}
describes spin transport, and $\sigma^s_{II}$ quantifies the rate of change of the spin density due to spin nonconservation in the Hamiltonian~\cite{Monaco2022shc_nonconserved_spin,shi2006proper,tokatly2008equilibrium,gorini2012onsager}. 
Our definition of $\sigma^s_{II}$ differs slightly from that in SRef.~\cite{Monaco2022shc_nonconserved_spin} in that we have chosen to include in $\sigma^s_{II}$ \emph{all} contributions to the spin Hall conductivity that vanish when spin is conserved. 
The spin Chern number gives a topological contribution to $\sigma^s_I$ even when spin is not conserved. 
Writing $P(\mathbf{k})=P_+(\mathbf{k})+P_-(\mathbf{k})$ and inserting a complete set $\ket{u^\pm_{n\mathbf{k}}}$ of states in the image of $P(\mathbf{k})$, we can write
\begin{align}
\sigma^s_I = C^s_{\gamma_1} +\frac{i}{2\pi}\int d^2k\left[2 \mathrm{Tr}\left(P_+\left[\frac{\partial P_-}{\partial k_1},\frac{\partial P_-}{\partial k_2}\right]\right) - \sum_{\alpha=\pm}\sum_n\beta_{n\mathbf{k}}^{\alpha} \bra{u^\alpha_{n\mathbf{k}}}P\left[\frac{\partial P}{\partial k_1},\frac{\partial P}{\partial k_2}\right]\ket{u^\alpha_{n\mathbf{k}}}\right],\label{eq:intrinsicspinhall}
\end{align}
where we have introduced $\beta^\pm_{n\mathbf{k}} = \pm 1 -\lambda^\pm_{n\mathbf{k}}$. 
The first term in SEq.~\eqref{eq:intrinsicspinhall} is the spin Chern number [SEq.~\eqref{eq:Csgamma1def}]. 
The second term quantifies the fact that if spin is not conserved, the lower $PsP$ eigenstates at $\mathbf{k}$ will have nonzero overlap with upper $PsP$ eigenstates at neighboring $\mathbf{k}+\delta\mathbf{k}$. 
Finally, the third term in SEq.~\eqref{eq:intrinsicspinhall} gives the integral of the diagonal matrix elements of the ordinary Berry curvature between $PsP$ eigenstates, weighted by the deviation of their $PsP$ eigenvalue from $\pm 1$. 
Thus, $\sigma^s_I-C^s_{\gamma_1}$ will be small when $\hat{\mathbf{n}}\cdot\mathbf{s}$ non-conserving SOC is weak. 
Additionally, it was established in SRefs.~\cite{Monaco2022shc_nonconserved_spin,yang2006stvreda} that $\sigma^s_{II}$ is also small when $\hat{\mathbf{n}}\cdot\mathbf{s}$ non-conserving SOC is weak. 
Hence, we have shown that the spin Chern number indicates the leading intrinsic (topological) contribution to the spin Hall conductivity for weak $\hat{\mathbf{n}}\cdot\mathbf{s}$ non-conservation. 
In SN~\ref{sec:bibrshc}, we will validate this result under realistic conditions by computing the bulk spin Hall conductivity for several choices of $\hat{\mathbf{n}}$ for the candidate helical HOTI $\alpha$-BiBr~\cite{SYBiBr,tang2019efficient,BiBrFanHOTI}, which we find to have two independent spin-gapped regimes.
Remarkably, we find that in both independent spin-gapped regimes, the intrinsic bulk contribution to the spin Hall conductivity lies in close agreement with a quantized value proportional to the spin Chern number, even when the spin gap is as small as $\Delta_{s}\sim 0.25$.

Finally, the relation $\nu_{2d} = (1/2)C^{s}_{\gamma_1} \text{ mod } 2$ in 2D allows us to deduce some general features of the spin spectrum for 3D topological insulators. 
Recall that for a 3D strong topological insulator, if we evaluate the 2D strong $\mathbb{Z}_2$ invariant $\nu_{2d}(k_i)$ over two parallel time-reversal invariant planes with $k_{i} =0$ and $k_{i} = \pi$ ($i=1,2,3$), then $|\nu_{2d}(0)-\nu_{2d}(\pi)|=1$.  
This implies that the spin Chern numbers evaluated on these planes satisfy $|C^s_{\gamma_1}(0)-C^s_{\gamma_1}(\pi)|=2+4n$ for some integer $n$. 
Since the spin Chern number $C^s_{\gamma_1}(k_i)$ can be defined over any constant-$k_i$ plane over which the spin gap is open, the only way for the $C^s_{\gamma_1}$ to change is if the spin gap closes for some $k_i$.
As such, the spin spectrum must have an odd number of gap closing points between the $k_{i} =0$ and $k_{i} = \pi$ planes~\cite{roy2010characterization}. 
Just like for energy bands, the Wigner-von Newmann theorem tells us that, absent additional crystal symmetries, the spin gap closing points at generic $\mathbf{k}$ will occur as isolated twofold degeneracies with linearly dispersing spin bands in all directions~\cite{herring1937accidental}.  
Therefore, a 3D strong topological insulator must have a \emph{gapless} spin spectrum.
For an analytic demonstration, we refer the readers to SN~\ref{appendix:explicit-calculation-of-BHZ-model} where we perform an explicit computation using a model of 3D strong topological insulator.
We will come back to this statement later in SN~\ref{sec:main-text-3D-TI-P-pm} and \ref{appendix:3D-TI-with-and-without-inversion} when we numerically demonstrate the application of our $P_{\pm}$-Wilson loop formalism to 3D strong topological insulators.

\subsection{\label{sec:2D-spinful-TRI-system}2D Spinful Time-Reversal-Invariant Systems}

\begin{figure*}[h]
\includegraphics[width=\textwidth]{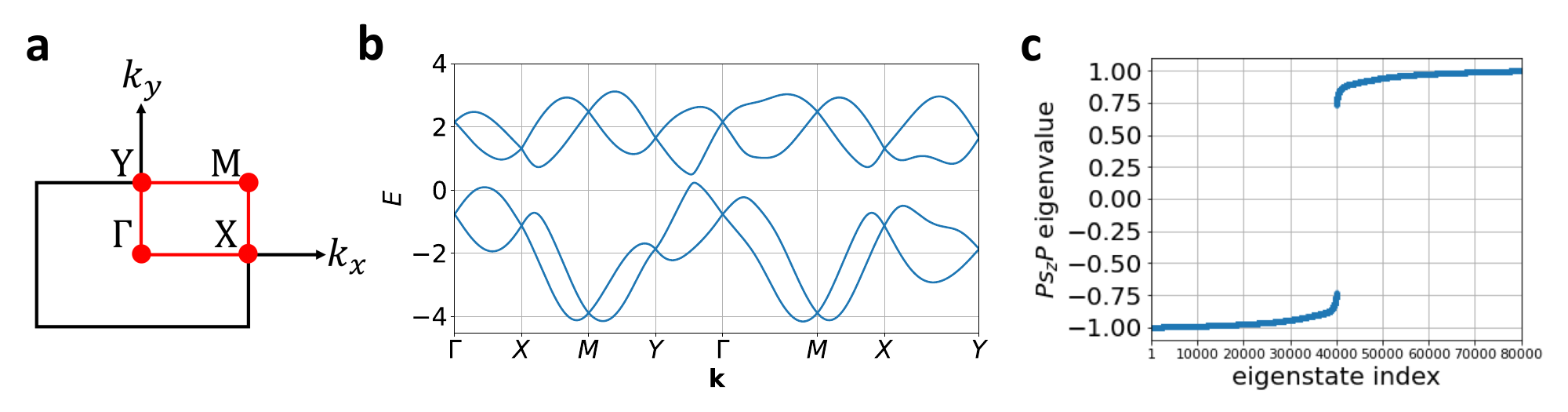}
\caption{Brillouin zone, energy spectrum, and $s_z$ spin spectrum for the 2D TI model in SEqs.~\eqref{eq:appendix-2D-TI-H} and~\eqref{eq:appendix-2D-TI-TB-parameters}. 
(a)  shows the first BZ of a 2D orthorhombic lattice. 
(b) shows the bulk energy bands of $[H(\mathbf{k})]$ [SEq.~\eqref{eq:appendix-2D-TI-H}] with tight-binding parameters in SEq.~\eqref{eq:appendix-2D-TI-TB-parameters}. 
(c) shows the $Ps_{z}P$ eigenvalues for the lowest two bands in (b) evaluated on a uniformly sampled grid of $\mathbf{k}$ points in the 2D BZ. 
The (square) grid size used in the sampling is $(\Delta k_x,\Delta k_y) = (0.01\pi,0.01\pi)$ such that the number of $\mathbf{k}$ points is $200\times 200$.
The calculations detailed in this figure were performed using the freely available Python package~\href{https://github.com/kuansenlin/nested_and_spin_resolved_Wilson_loop}{\textsc{nested\_and\_spin\_resolved\_Wilson\_loop}}~\cite{lin2023nestedWilsonLib}, which represents an extension of the~\href{https://www.physics.rutgers.edu/pythtb/}{PythTB} open-source Python tight-binding package~\cite{coh2013python} that was implemented and utilized for the preparation of SRefs.~\cite{wieder2018axion,wieder2020strong} and the present work.}
\label{fig:2D-TI-band-spin-gap-wvfn-prob}
\end{figure*}

In this section, we will numerically demonstrate the application of the $P_{\pm}$-Wilson loop formalism in SN~\ref{sec:P_pm_Wilson_loop} to resolve the nontrivial topology in the spin spectrum by considering a 2D strong TI with spinful time-reversal ($\mathcal{T}$) symmetry. 
We will also numerically confirm the symmetry constraints for the $P_{\pm}$-Wilson loop spectra we derived in SN~\ref{appendix:T-constraint-on-P-pm}.
In the following discussion, we will refer to the eigenvalues of the reduced spin matrix introduced in SEq.~\eqref{eq:P_pm_Wilson_loop_s_reduced_def} with $\hat{\mathbf{n}}$ a unit vector along the $i^{\text{th}}$ Cartesian coordinate direction as the ``$s_{i}$ spectrum'' or the ``$s_i$ eigenvalues.''
The 2D strong TI model we consider is a four-band lattice model with orbital ($s$ and $p$) and spin ($\uparrow$ and $\downarrow$) degrees of freedom represented by Pauli matrices $\tau_{\mu}$ and $\sigma_{\nu}$, respectively.
The corresponding four-band (momentum-space) Bloch Hamiltonian $[H(\mathbf{k})]$ is given by~\cite{bernevig2006quantum,fu2007topological,qi2008topological}
\begin{align}
    [H(\mathbf{k})] & = [\epsilon - t_{1,x}\cos(k_x) - t_{1,y}\cos(k_y)]\tau_{z}\sigma_{0} + t_{2,x}\sin(k_x)\tau_{y}\sigma_{0} + t_{2,y}\sin(k_y)\tau_{x}\sigma_{z}  \nonumber \\
    & + t_{R}( \sin(k_y) \tau_{0}\sigma_{x} - \sin(k_x)\tau_{0}\sigma_{y} ) + [t_{PH,x}\cos(k_x) + t_{PH,y}\cos(k_y)]\tau_{0}\sigma_{0} \nonumber \\
    & + t_{M_{x}}\cos(k_x)\tau_{y}\sigma_{x} + t_{M_{y}}\cos(k_y)\tau_{y}\sigma_{y} + t_{\mathcal{I}}\cos(k_x)\tau_{x}\sigma_{0}, \label{eq:appendix-2D-TI-H}
\end{align}
Both $\tau_{0}$ and $\sigma_{0}$ are $2\times 2$ identity matrices. 
SEq.~\eqref{eq:appendix-2D-TI-H} was formulated by adding additional terms to the Bernevig-Hughes-Zhang (BHZ) model of a 2D strong TI~\cite{bernevig2006quantum,fu2007topological,qi2008topological} to break rotation and mirror symmetries of that model. 
We will choose the following values for the parameters of the model: 
\begin{equation}
\hspace{-0.5cm}
    \epsilon = 1.0, t_{1,x} = 0.8, t_{1,y} = 1.2, t_{2,x}=1.3, t_{2,y}=0.9, t_{R} = 0.8, t_{PH,x} = 0.3, t_{PH,y} = 0.4, t_{M_{x}} = 0.3, t_{M_{y}}=0.2, t_{\mathcal{I}} = 1.0. \label{eq:appendix-2D-TI-TB-parameters}
\end{equation}
The model in SEq.~\eqref{eq:appendix-2D-TI-H} is $\mathcal{T}$-symmetric, where $\mathcal{T}$ symmetry is represented as $[\mathcal{T}]=i\sigma_y\mathcal{K}$, and $\mathcal{K}$ represents complex conjugation. 
For the parameter values given in SEq.~\eqref{eq:appendix-2D-TI-TB-parameters}, the model does not have mirror or twofold rotation symmetries. 
The spectrum and $s_z$ spin spectrum for this model are shown in SFig.~\ref{fig:2D-TI-band-spin-gap-wvfn-prob}, where we see that there is both an energy gap and a spin gap.

Instead of using $\{(\gamma_{1})_{j,\mathbf{k},\mathbf{G}}\}$ and $\{(\gamma_{1}^\pm)_{j,\mathbf{k},\mathbf{G}}\}$ to denote the set of eigenphases of the $P$- and $P_{\pm}$-Wilson loop matrices, in this section we will denote them as $\{ \gamma_{1,j}(k_{\perp})\}$ and $\{\gamma_{1,j}^{\pm}(k_{\perp})\}$ to emphasize that the eigenphases depend only on $k_\perp$. 
For example, in a 2D orthorhombic system, which is the case we will consider below, if we choose $\mathbf{G}=2\pi \hat{\mathbf{x}}$ for the holonomy, the corresponding eigenphases for the $k_x$-directed $P$- and $P_{\pm}$-Wilson loop matrices will be denoted as $\{ \gamma_{1,j}(k_y)\}$ and $\{ \gamma_{1,j}^{\pm}(k_y)\}$, respectively.

\begin{figure}[t]
\includegraphics[width=\textwidth]{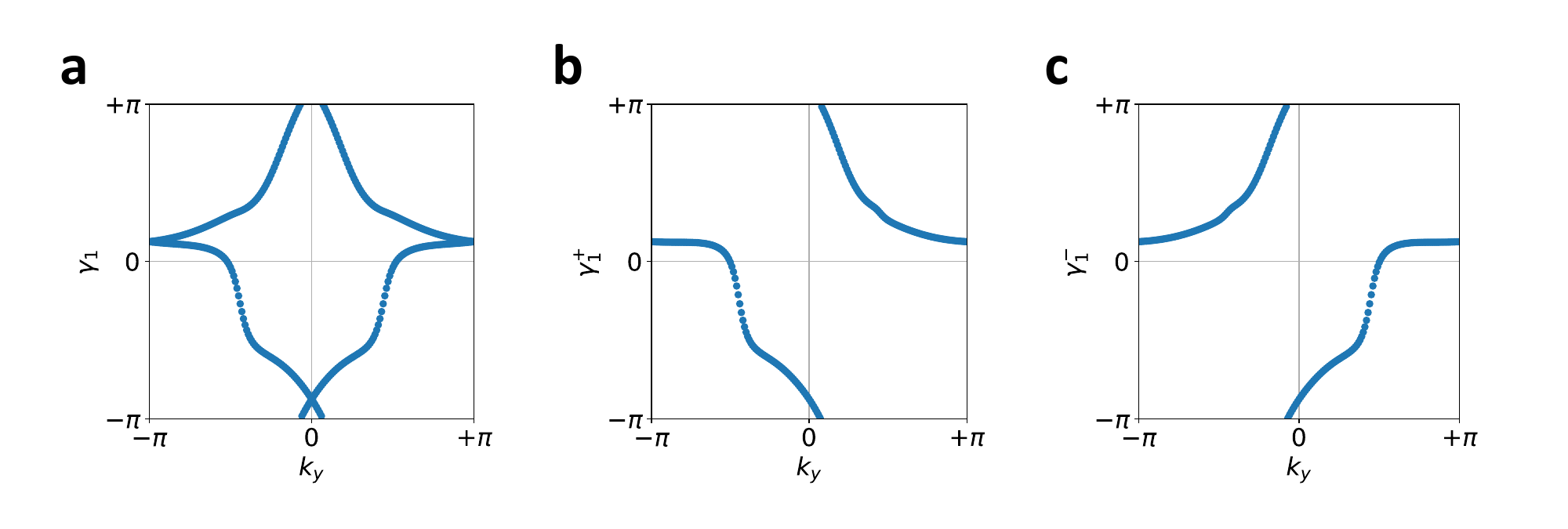}
\caption{Wilson loops and spin-resolved Wilson loops for a 2D strong TI with $\mathcal{T}$ symmetry. 
The Bloch Hamiltonian is given by SEq.~(\ref{eq:appendix-2D-TI-H}) with tight-binding parameters in SEq.~\eqref{eq:appendix-2D-TI-TB-parameters}. 
And the occupied energy bands are chosen as the lowest two energy bands.
(a) shows the eigenphases $\{ \gamma_{1,j}(k_y)\}$ of the $k_x$-directed $P$-Wilson loop matrix [SEq.~(\ref{eq:P_Wilson_loop_matrix_element})] as a function of $k_{y}$. 
There are two bands that demonstrate an odd helical winding. 
(b) and (c) show the eigenphases $\{ \gamma_{1,j}^{\pm}(k_y)\}$ of the $k_x$-directed $P_{\pm}$-Wilson loop matrix [SEq.~\eqref{eq:P_pm_Wilson_loop_matrix_element}] as a function of $k_{y}$. 
There is one band for each of (b) and (c). 
The spin-resolved Wilson loop spectra in (b) and (c) demonstrate nonzero net winding. 
For the spectrum in (b), as $k_{y} \to k_{y}+2\pi$ we have winding number $-1$ corresponding to the partial Chern number $C^{+}_{\gamma_{1}}=-1$ [using the sign convention introduced in SEq.~\eqref{eq:partial_chern_def}] while for the spectrum in (c) we have winding number $+1$ corresponding to the partial Chern number $C^{-}_{\gamma_{1}}=+1$.
The calculations detailed in this figure were performed using the freely available Python package~\href{https://github.com/kuansenlin/nested_and_spin_resolved_Wilson_loop}{\textsc{nested\_and\_spin\_resolved\_Wilson\_loop}}~\cite{lin2023nestedWilsonLib}, which represents an extension of the~\href{https://www.physics.rutgers.edu/pythtb/}{PythTB} open-source Python tight-binding package~\cite{coh2013python} that was implemented and utilized for the preparation of SRefs.~\cite{wieder2018axion,wieder2020strong} and the present work.}
\label{fig:2D-strong-TI-wilson-loop-results}
\end{figure}

In addition to helical edge states, another characteristic of a 2D strong TI is the presence of an odd helical winding in the $P$-Wilson loop spectrum~\cite{yu2011equivalent}.
In SFig.~\ref{fig:2D-strong-TI-wilson-loop-results}(a), we show the eigenphases $\gamma_1$ of the $k_x$-directed $P$-Wilson loop matrix [SEq.~(\ref{eq:P_Wilson_loop_matrix_element})] as a function of $k_y$, where we have chosen $\mathbf{G} = 2\pi \hat{\mathbf{x}}$ in SN~\ref{sec:P_Wilson_loop}.
The two-fold degeneracies at $k_{y}=0$ and $k_y=\pi$ in SFig.~\ref{fig:2D-strong-TI-wilson-loop-results}(a) are due to Kramers theorem,
and thus the helical winding, if it exists, is protected as long as the $\mathcal{T}$ symmetry is preserved~\cite{yu2011equivalent,soluyanov2011wannier}.
Since the model [SEq.~\eqref{eq:appendix-2D-TI-H}] with parameters in SEq.~\eqref{eq:appendix-2D-TI-TB-parameters} has only translation and time-reversal symmetries, the eigenphases $\{ \gamma_{1,j}(k_y) \}$ take generic values at $k_y=0,\pi$, as shown in SFig.~\ref{fig:2D-strong-TI-wilson-loop-results}(a); there are no improper rotation symmetries (such as mirror symmetries) to quantize $\{\gamma_{1,j}(k_y=0,\pi)\}$\cite{wieder2018wallpaper}.
For a detailed discussion of how crystalline symmetries can quantize the eigenphases, we refer the readers to SN~\ref{appendix:I-constraint-on-P-wilson} where we specifically consider the effect of inversion symmetry on the $P$-Wilson loop eigenphases.

To systematically resolve the topology in the spin spectrum, we first verify that there is a gap in the $s_{z}$ spectrum computed in the occupied two-band valence space throughout the 2D BZ, as shown in SFig.~\ref{fig:2D-TI-band-spin-gap-wvfn-prob}(c). 
Note that the $Ps_zP$ eigenvalues deviate from unity, indicating that $s_z$ is not conserved. 
This will enable us to decompose the occupied space into two parts and compute the $P_{\pm}$-Wilson loop eigenphases in the occupied space.
We will again choose $\mathbf{G} = 2\pi \hat{\mathbf{x}}$ in SN~\ref{sec:P_pm_Wilson_loop}.
As shown in SFig.~\ref{fig:2D-strong-TI-wilson-loop-results}(b,c), such $k_x$-directed $P_{+}$- and $P_{-}$-Wilson loop eigenphases as a function of $k_y$ are related to each other by
\begin{equation}
    \{ \gamma^{\pm}_{1,j}(k_y) \}\text{ mod } 2\pi = \{ \gamma^{\mp}_{1,j}(-k_y) \} \text{ mod } 2\pi\label{gamma-pm-t-relation}
\end{equation}
where $\{\gamma^{\pm}_{1,j}(k_y)\}$ is the set of the eigenphases of the $k_x$-directed $P_{\pm}$-Wilson loop matrix [SEq.~(\ref{eq:P_pm_Wilson_loop_matrix_element})], and $j$ is the corresponding band indices.
This is the 2D counterpart of the $\mathcal{T}$ constraints on the $P_{\pm}$-Wilson loops that we will derive in SN~\ref{appendix:T-constraint-on-P-pm}. 
Indeed, SEq.~\eqref{gamma-pm-t-relation} follows directly from the definition of the $P_\pm$-Wilson loop matrix, combined with SEq.~\eqref{eq:t_on_ppm}.
This constrains the winding numbers of $\{ \gamma_{1,j}^{+}(k_y) \}$ and $\{ \gamma_{1,j}^{-}(k_y) \}$ as $k_y \to k_y + 2\pi$ to be opposite.
In our specific examples, we see that the winding numbers of the $k_x$-directed $P_{\pm}$-Wilson loop spectra are given by $\mp 1$.
From SEq.~\eqref{eq:partial_chern_def}, the partial Chern numbers are then given by $C^\pm_{\gamma_1}=\mp 1$, where we recall that the partial Chern numbers are given by the winding number of $\{\gamma_{1,j}^{\pm}(k_y)\}$ as $k_y \to k_y + 2\pi$~\cite{gresch2017z2pack,vanderbilt2018berry,bradlyn2021lecture}.
This is consistent with the fact that the subspace of upper and lower spin bands are not invariant under $\mathcal{T}$.
However, $\mathcal{T}$ symmetry enforces the constraint that $C^{+}_{\gamma_{1}} = -C^{-}_{\gamma_{1}}$ as we have stated in SEq.~\eqref{eq:C_gamma_1_relate_to_C_plus_C_minus}.

From our definition in SEq.~\eqref{eq:Csgamma1def} we have that the spin Chern number of this model is given by
\begin{equation}
    C^{s}_{\gamma_{1}} = -2. \label{eq:C-S-theta-1-of-our-2D-strong-TI-model}
\end{equation}
As mentioned in SN~\ref{sec:general_properties_of_winding_num_of_P_pm_Wilson}, for a system with $\mathcal{T}$ symmetry,  $1/2(C^{s}_{\gamma_{1}}\mod 4)$ gives the 2D $\mathbb{Z}_{2}$ Kane-Mele invariant $\nu_{2d}$~\cite{prodan2009robustness,3D_phase_diagram_spin_Chern_Prodan}.
Therefore, SEq.~(\ref{eq:C-S-theta-1-of-our-2D-strong-TI-model}) means that our 2D model in SEq.~(\ref{eq:appendix-2D-TI-H}) with parameters in SEq.~(\ref{eq:appendix-2D-TI-TB-parameters}) is indeed a 2D strong TI with nontrivial strong $\mathbb{Z}_{2}$ invariant $\nu_{2d} = 1$.

As discussed in the text surrounding SEq.~\eqref{eq:trcschange}, the reason that $2\nu_{2d} = C^{s}_{\gamma_{1}}$ mod $4$ is because $C^{s}_{\gamma_{1}}$ can change by a multiple of $4$ under $\mathcal{T}$-symmetric adiabatic deformations of the Hamiltonian that close a spin gap but not an energy gap. 
To see this concretely, we follow the logic of SRef.~\cite{prodan2009robustness} and construct a deformation of the Hamiltonian $[H(\mathbf{k})]$ in SEq.~(\ref{eq:appendix-2D-TI-H}) through a unitary transformation
\begin{equation}
    [U(\phi)] = \exp{\frac{-i\phi \sigma_{x}}{2}} = \sigma_{0}\cos{\frac{\phi}{2}}-i\sigma_{x}\sin{\frac{\phi}{2}}, \label{eq:rotate-spin-along-x-by-theta}
\end{equation}
which corresponds to a rotation of the spin vector about the $x$-axis by an angle $\phi$.
The transformed Hamiltonian as a function of $\phi$ is denoted as $[H(\mathbf{k},\phi)]=[U(\phi)][H(\mathbf{k})][U(\phi)]^{\dagger}$.
Since $\mathcal{T}$ is represented as 
\begin{equation}
[\mathcal{T}] = i\sigma_{y}\mathcal{K}
\end{equation}
with $\mathcal{K}$ denoting the complex conjugation operator, it follows that 
\begin{equation}
[\mathcal{T}][H(\mathbf{k},\phi)][\mathcal{T}]^{-1}=[H(-\mathbf{k},\phi)]
\end{equation}
This means that $[H(\mathbf{k},\phi)]$ is still $\mathcal{T}$-invariant.
For simplicity, we will turn off all the tight-binding parameters in SEq.~(\ref{eq:appendix-2D-TI-TB-parameters}) except for $\epsilon$, $t_{1,x}$, $t_{1,y}$, $t_{2,x}$, and $t_{2,y}$. 
We will further choose $\epsilon = 1.0$, $t_{1,x}=t_{1,y}=t_{2,x}=t_{2,y} = 1.0$, which does not close the energy gap between the conduction and valence bands, and does not change the winding numbers of the $P_{\pm}$-Wilson loop eigenphases from their values in SFig.~\ref{fig:2D-strong-TI-wilson-loop-results}(b,c).
In other words, when $\phi = 0$, we have $C^{\pm}_{\gamma_{1}} = \mp 1$.
Since $[U(\pi)]^{\dagger}\sigma_{z}[U(\pi)] = -\sigma_{z}$, it can then be deduced that when $\phi = \pi$, the same eigenstates of the reduced $s_z$ matrix that we originally labeled as having positive [negative] $P s_{z} P$ eigenvalues will at $\phi=\pi$ be labeled as having negative [positive] $P s_{z} P$ eigenvalues.
Therefore, when $\phi = \pi$, we have instead $C^{\pm}_{\gamma_{1}} = \pm 1$.
This means that the spin Chern number will change from $C^{s}_{\gamma_{1}}=-2$ to $C^{s}_{\gamma_{1}}=+2$ when we deform the Hamiltonian from $\phi = 0$ to $\phi = \pi$.
As the partial Chern numbers $C^{\pm}_{\gamma_{1}}$ give the Chern numbers for the upper and lower $Ps_zP$ bands, respectively, the spin gap must close and reopen throughout the deformation from $\phi=0$ to $\phi=\pi$ in order to transfer $-2$ partial Chern number from the upper to lower spin bands such that $C^\pm_{\gamma_1}$ change from $\mp 1$ to $\pm 1$. 
In fact, with the simpler tight-binding parameter choice we mentioned above, it can be shown analytically that the gap in the $P s_{z} P$ spectrum closes when $\phi = \pi/2$.
To see this, note that since $[U(\pi/2)]^{\dagger}\sigma_{z}[U(\pi/2)] = \sigma_{y}$, the $P s_{z} P$ eigenvalues of the unitary-transformed model with $\phi = \pi / 2$ are the same as the $P s_{y} P$ eigenvalues of the model before the transformation.
Before the deformation ($\phi = 0$), with the simpler tight-binding parameter choice, we have at $(k_x,k_y) = (0,\pm \pi /2)$ the Bloch Hamiltonian
\begin{equation}
    [H(0,\pm \pi/2)] = \pm \tau_{x}\sigma_{z}.
\end{equation}
Similar to the analysis in SN~\ref{appendix:orbital-texture-and-spin-gap}, we first obtain the corresponding two valence eigenvectors
\begin{align}
    & | 1_{\pm} \rangle = \frac{1}{\sqrt{2}} \begin{bmatrix} 1 \\ \mp 1 \end{bmatrix} \otimes \begin{bmatrix} 1 \\ 0 \end{bmatrix}, \\
    & | 2_{\pm} \rangle = \frac{1}{\sqrt{2}} \begin{bmatrix} 1 \\ \pm 1 \end{bmatrix} \otimes \begin{bmatrix} 0 \\ 1 \end{bmatrix},
\end{align}
where $| 1_{\pm} \rangle$ and $| 2_{\pm} \rangle$ denote the first and second eigenvectors of $[H(0,\pm \pi/2)]$ with eigenvalue $-1$.
It can be then checked that 
the reduced $s_y$ spin matrix $[s_{y,\mathrm{reduced}}(\mathbf{k})]$ defined in SEq.~\eqref{eq:P_pm_Wilson_loop_s_reduced_def} is a $2 \times 2$ matrix of zeros at both $(k_x,k_y) = (0,\pm \pi /2)$.
This means that the $P s_{y} P$ eigenvalues for both $[H(0,\pm \pi/2)]$ 
are zeros with two-fold degeneracy.
In other words, the gap in the $P s_{y} P$ spectrum is closed at $(k_x,k_y) = (0,\pm \pi /2)$ for the Bloch Hamiltonian before the deformation ($\phi = 0$).
Therefore, the gap in the $P s_{z} P$ spectrum of the unitary-transformed model $[U(\pi/2)][H(\mathbf{k})][U(\pi/2)]^{\dagger}$ is closed.
Notice that $\mathcal{T}$ symmetry requires that if the spin gap is closed at $\mathbf{k}$, there must be another spin gap closing point at $-\mathbf{k}$.
This is because the projected spin operator $s(\mathbf{k}) = P(\mathbf{k}) s P(\mathbf{k})$ satisfies
\begin{equation}
[\mathcal{T}]s(\mathbf{k})[\mathcal{T}]^{-1}=-s(-\mathbf{k}).\end{equation}

Also note that since we are performing a unitary transformation [SEq.~(\ref{eq:rotate-spin-along-x-by-theta})] on the Hamiltonian, the energy spectrum is unchanged throughout the deformation.
However, we have changed $C^{s}_{\gamma_{1}}$ by $+4$.
The upshot is that, with $\mathcal{T}$ symmetry, if we only maintain the (Hamiltonian) energy gap, then through a closing of the spin gap we are able to change the spin Chern number [SEq.~(\ref{eq:Csgamma1def})] by a multiple of $4$~\cite{prodan2009robustness,3D_phase_diagram_spin_Chern_Prodan}.
This, correspondingly, results in a measurable change of the (topological contribution to the) spin Hall conductivity (in our example, it interchanges spin-up and spin-down electrons, which reverses the sign of the spin Hall conductivity). 

We can interpret SFig.~\ref{fig:2D-strong-TI-wilson-loop-results}(b,c) in terms of the spectral flow of hybrid Wannier centers for hybrid Wannier functions composed of upper and lower $Ps_zP$ eigenstates. 
In particular, we see that as $k_y$ is adiabatically varied from $-\pi$ to $\pi$, the centers of hybrid Wannier functions formed from $P_+$ states shift one unit cell to the left along the $x$ direction. 
Similarly, the centers of hybrid Wannier functions formed from $P_-$ states---which are the time-reversed partners of the $P_+$ hybrid Wannier functions---shift one unit cell to the right along the $x$ direction. 
Notice that such a behavior can also be seen in the ordinary $P$-Wilson loop calculation, as shown in SFig.~\ref{fig:2D-strong-TI-wilson-loop-results}(a). 
If we trace the individual bands in 
SFig.~\ref{fig:2D-strong-TI-wilson-loop-results}(a) in a smooth and continuous way, we will also obtain two distinct windings that are related to each other by time-reversal symmetry [such as the reflection with respect to the $k_y = 0$ axis in SFig.~\ref{fig:2D-strong-TI-wilson-loop-results}(b,c)], with opposite winding numbers.
We emphasize that, in this section, we have demonstrated the procedure for spin-resolving the Wilson loop spectrum by first decomposing the occupied space into two parts that are related to each other through a spinful $\mathcal{T}$ operation, and then performing the Wilson loop calculation within each of the two separate subspaces.  
Although it may seem from the simple example of a 2D TI considered in this section that the $P_{\pm}$-Wilson loop eigenvalues and general features [SFig.~\ref{fig:2D-strong-TI-wilson-loop-results}(b,c)] can be obtained via a straightforward decomposition of the $P$-Wilson loop spectrum [SFig.~\ref{fig:2D-strong-TI-wilson-loop-results}(a)], in general this is not the case.
In SN~\ref{sec:spin_stable_topology_2d_fragile_TI}, we will analyze a more complicated model of a 2D fragile topological insulator with more pronounced differences between the $P_{\pm}$-Wilson loops and naive ``halves'' of the $P$-Wilson loop spectrum.  
In subsequent sections we will further show how the spin-resolved Wilson loop can be employed to analyze 3D strong TIs (SN~\ref{sec:main-text-3D-TI-P-pm} and~\ref{appendix:3D-TI-with-and-without-inversion}), 2D fragile TIs (SN~\ref{sec:spin_stable_topology_2d_fragile_TI}), and higher-order topological phases (SN~\ref{sec:numerical-section-of-nested-P-pm}).

We conclude by summarizing the properties of 2D strong TIs shown in this section:
\begin{enumerate}
    \item The $P$-Wilson loop eigenphases have spectral flow with odd helical winding protected by $\mathcal{T}$ symmetry [SFig.~\ref{fig:2D-strong-TI-wilson-loop-results}(a)].
    \item The $P_{\pm}$-Wilson loop eigenphases have spectral flow with opposite odd winding numbers [SFig.~\ref{fig:2D-strong-TI-wilson-loop-results}(b,c)]. 
    \item The spin Chern number $C^{s}_{\gamma_{1}}$ defined in SEq.~\eqref{eq:Csgamma1def} indicates the 2D strong $\mathbb{Z}_{2}$ invariant $\nu_{2d}= 1/2 (C^s_{\gamma_1} \mod 4)$, such that for a 2D strong TI, $C^s_{\gamma_1}=2+4n$ where $n \in \mathbb{Z}$.
\end{enumerate}

\subsection{\label{sec:main-text-3D-TI-P-pm}3D Spinful Time-Reversal-Invariant Systems with Inversion Symmetry}

In this section, we will demonstrate that the $P_{\pm}$-Wilson loops can also be used to detect a 3D TI with spinful $\mathcal{T}$ symmetry. 
We will also verify that spin gap closing points in a 3D system play the role of Berry curvature monopoles for the partial Chern numbers $C^\pm_{\gamma_1}$, defined in SEq.~\eqref{eq:partial_chern_def} as the winding numbers of $P_{\pm}$-Wilson loop eigenphases. 
We term such spin gap closing points with nonzero partial Berry flux as {\it spin-Weyl nodes}.

\begin{figure}[h]
\includegraphics[width=\textwidth]{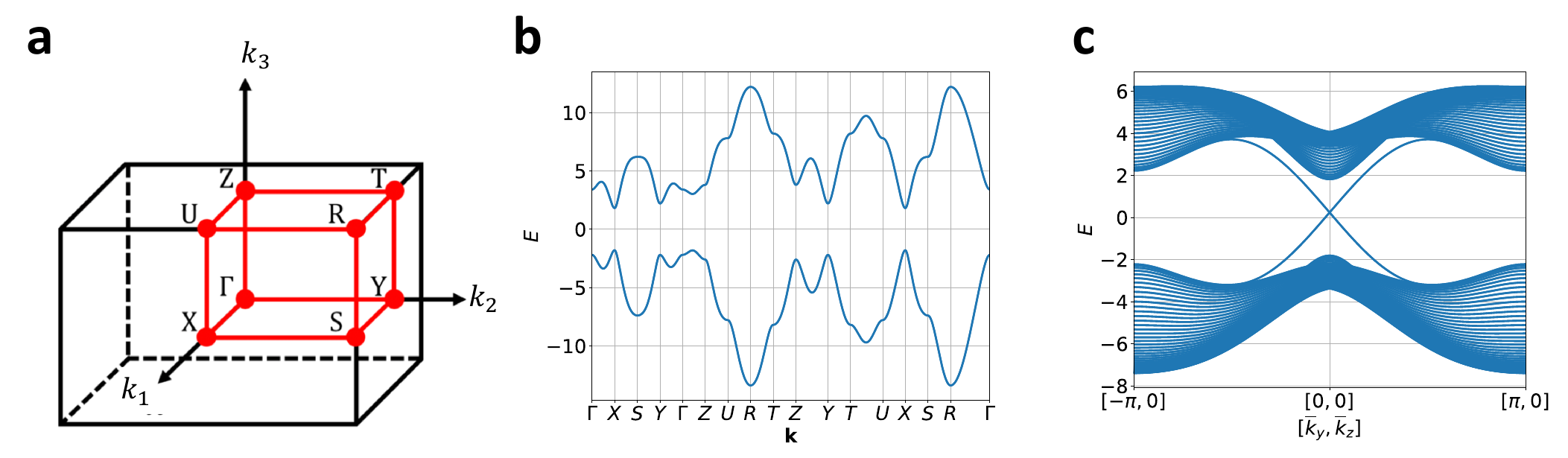}
\caption{Spectrum for the 3D $\mathcal{I}$- and $\mathcal{T}$-symmetric 3D TI given in SEqs.~\eqref{eq:3D_TI_model} and \eqref{eq:3D_TI_model_parameter_I_and_T_symmetry}. 
(a) shows the 3D Brillouin zone with high symmetry points labeled. 
(b) shows the 3D bulk band structure of $[H(\mathbf{k})]$ in SEq.~(\ref{eq:3D_TI_model}) with tight-binding parameters in SEq.~(\ref{eq:3D_TI_model_parameter_I_and_T_symmetry}). 
(c) shows the 2D band structure of a slab infinite along $\mathbf{a}_2 \parallel \hat{\mathbf{y}}$ and $\mathbf{a}_3 \parallel \hat{\mathbf{z}}$ while finite along $\mathbf{a}_1 \parallel \hat{\mathbf{x}}$ with open boundary condition and $41$ unit cells. 
We see a twofold degenerate surface Dirac cone on each surface. }
\label{fig:3D_TI_bulk_BZ_bulk_band_surface_band}
\end{figure}

We begin by introducing a $4$-band spinful $\mathcal{T}$ and $\mathcal{I}$ symmetric 3D TI with an orthorhombic lattice,
where at the origin of each unit cell we place a spinful $s$ and a spinful $p$ orbital.
The corresponding four-band (momentum-space) Bloch Hamiltonian matrix is given by~\cite{wieder2018axion}
\begin{align}
    [H(\mathbf{k})] & = m\tau_{z}\sigma_{0} + \sum_{i=x,y,z}\left( t_{1,i}\cos{(k_{i})}\tau_{z}\sigma_{0} + t_{PH,i}\cos{(k_{i})} \tau_{0}\sigma_{0} + t_{2,i}\sin{(k_{i})}\tau_{y}\sigma_{0} + v_{1,i}\sin{(k_{i})}\tau_{x}\sigma_{i} \right) \nonumber \\
    & + v_{2,xy}[\sin(k_{x})+\sin(k_{y})]\tau_{x}\sigma_{z} + v_{2,z} \sin{(k_{z})}\tau_{x}\sigma_{x}. \label{eq:3D_TI_model}
\end{align}
where again $\tau_{\mu}$ and $\sigma_{\nu}$ are Pauli matrices describing the orbital ($s$ and $p$) and spin ($\uparrow$ and $\downarrow$) degrees of freedom respectively, and both $\tau_0$ and $\sigma_0$ are $2 \times 2$ identity matrices. 
The Bloch Hamiltonian [SEq.~\eqref{eq:3D_TI_model}] is formulated by adding additional terms to the BHZ model of a 3D strong TI to break symmetries of that model.
We will choose the following parameter values:
\begin{align}
    & m = -5,\ t_{1,x}=2.3,\ t_{1,y}=2.5,\ t_{1,z} = 3,\ t_{PH,x}=t_{PH,y}=0.3,\ t_{PH,z}=0,\ t_{2,x} = 0.9,\ t_{2,y}=t_{2,z} = 0, \nonumber \\
    & v_{1,x} = v_{1,y} = 3.2,\ v_{1,z} = 2.4,\ v_{2,xy} = 1.5,\ v_{2,z} = 0.4. \label{eq:3D_TI_model_parameter_I_and_T_symmetry}
\end{align}
$[H(\mathbf{k})]$ in SEq.~(\ref{eq:3D_TI_model}) with the parameter values specified in SEq.~\eqref{eq:3D_TI_model_parameter_I_and_T_symmetry} is invariant under 3D lattice translation, inversion, and spinful time-reversal symmetries. 
In particular, inversion and time-reversal symmetries are represented as
\begin{align}
    & [\mathcal{I}][H(\mathbf{k})][\mathcal{I}]^{-1} = \tau_{z}\sigma_{0} [H(\mathbf{k})] \tau_{z}\sigma_{0} = [H(-\mathbf{k})], \\
    & [\mathcal{T}][H(\mathbf{k})][\mathcal{T}]^{-1} = \tau_{0}\sigma_{y} [H(\mathbf{k})]^{*} \tau_{0}\sigma_{y} = [H(-\mathbf{k})],
\end{align}
respectively. 
The 3D Brillouin zone, energy spectrum, and finite-sized slab spectrum for this model are shown in SFig.~\ref{fig:3D_TI_bulk_BZ_bulk_band_surface_band}.

\begin{figure}[ht]
\hspace{-0.5cm}
\includegraphics[width=0.8\textwidth]{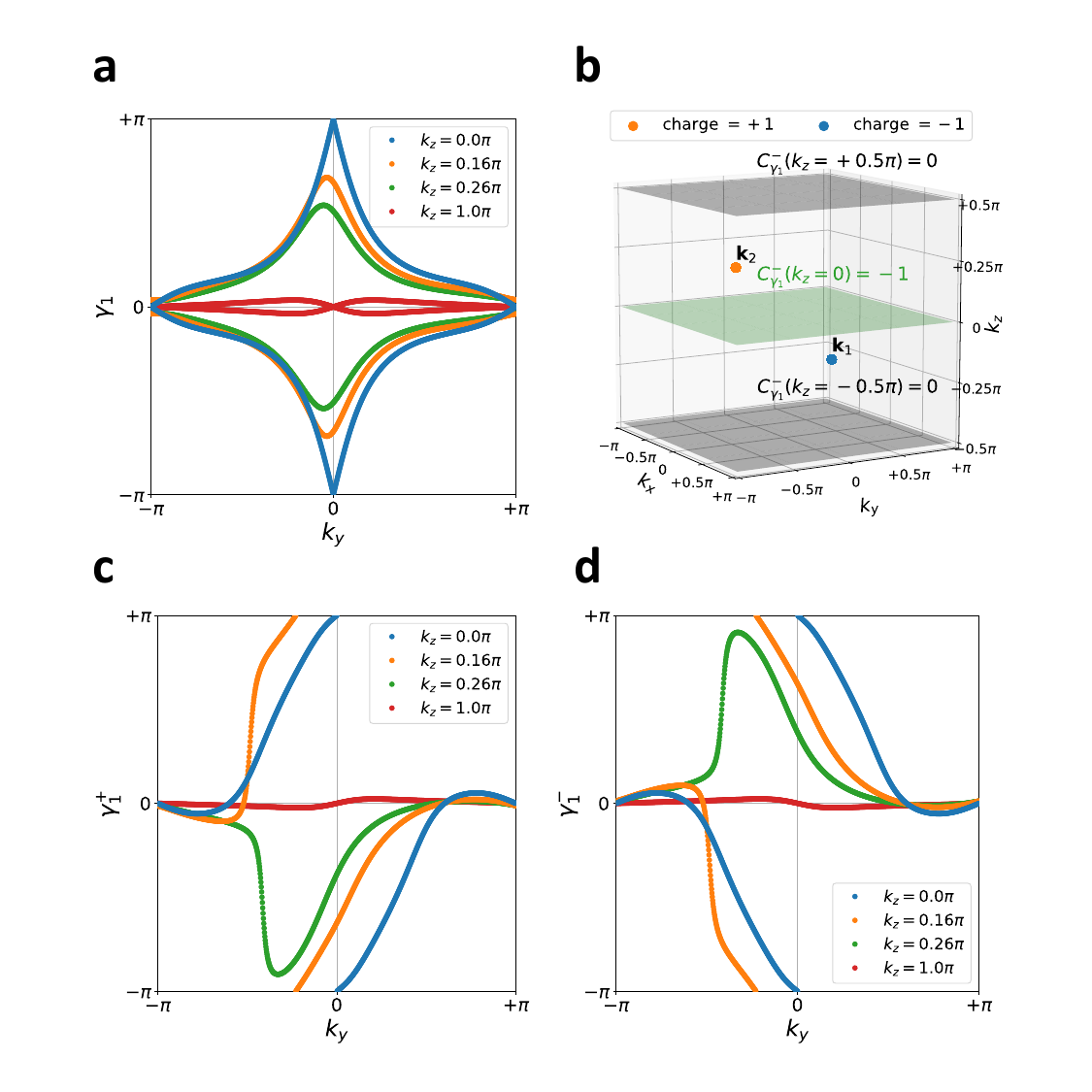}
\caption{Wilson loops for the 3D strong TI with both spinful time-reversal and inversion symmetry, with Hamiltonian given by SEqs.~\eqref{eq:3D_TI_model} and \eqref{eq:3D_TI_model_parameter_I_and_T_symmetry}.
(a) shows the $k_x$-directed $P$-Wilson loop eigenphases $\{\gamma_{1,j}(k_y,k_z)\}$ ($j=1 \sim 2$) as a function of $k_{y}$ in constant-$k_{z}$ plane. 
(b) shows the two $Ps_{z}P$ gap closing points (spin-Weyl nodes) at $\mathbf{k}_{1}=-\mathbf{k}_{2}=(0,0.456\pi,-0.212\pi)$ with their individual charges, namely the partial Berry flux, of the lower spin band indicated. 
The winding numbers $C_{\gamma_1}^{-}$ of the $P_{-}$-Wilson loop spectra at constant-$k_z$ planes separated by these spin-Weyl nodes will differ by the amount of the topological charges, as will be demonstrated in (d).
(c)$ \& $(d) show the $k_x$-directed $P_{\pm}$-Wilson loop eigenphases $\{\gamma_{1,j=1}^{\pm}(k_y,k_z)\}$ as a function of $k_{y}$ in constant-$k_{z}$ plane. 
The projectors $[P_{\pm}(\mathbf{k})]$ are defined using $s_{z}$ as the $s$ operator in SEq.~(\ref{eq:P_pm_Wilson_loop_s_reduced_def}), see SN~\ref{sec:P_pm_Wilson_loop} for further details. 
All of the $k_x$-directed $P$- and $P_{\pm}$-Wilson loop are computed by discretizing the path $\mathbf{k} + \mathbf{G} \leftarrow \mathbf{k}$ of the projector product in SEqs.~(\ref{eq:P_Wilson_loop_P_k_def}) and (\ref{eq:P_pm_Wilson_loop_P_k_def}) with 101 $\mathbf{k}$ points. 
The calculations detailed in this figure were performed using the freely available Python package~\href{https://github.com/kuansenlin/nested_and_spin_resolved_Wilson_loop}{\textsc{nested\_and\_spin\_resolved\_Wilson\_loop}}~\cite{lin2023nestedWilsonLib}, which represents an extension of the~\href{https://www.physics.rutgers.edu/pythtb/}{PythTB} open-source Python tight-binding package~\cite{coh2013python} that was implemented and utilized for the preparation of SRefs.~\cite{wieder2018axion,wieder2020strong} and the present work.
}
\label{fig:3D_TI_ord_W_pos_W_neg_W_spin_gap_closing}
\end{figure}

In SFig.~\ref{fig:3D_TI_ord_W_pos_W_neg_W_spin_gap_closing}(a), we show the $k_x$-directed $P$-Wilson loop eigenphases $\{ \gamma_{1,j}(k_y,k_z) \}$ ($j=1\sim 2$) of the two occupied bands as a function of $k_{y}$ at different constant-$k_z$ planes. 
At the $k_{z} = 0$ plane there is an odd helical winding of $\{ \gamma_{1,j}(k_y,0) \}$ as $k_y\to k_y+2\pi$ while at the $k_{z} = \pi$ plane there is no helical winding of $\{ \gamma_{1,j}(k_y,\pi) \}$ as $k_y\to k_y+2\pi$. 
In SFig.~\ref{fig:3D_TI_ord_W_pos_W_neg_W_spin_gap_closing}(a), the helical winding at the $k_{z} = 0$ plane is protected by spinful $\mathcal{T}$ symmetry, while at $k_z = \pi$ we have a trivial winding of $\{ \gamma_{1,j}(k_y,k_z) \}$ as $k_y \to k_y + 2\pi$.
This means that the Hamiltonian [SEq.~\eqref{eq:3D_TI_model}] restricted to the $k_{z} = 0$ plane is topologically equivalent to a two-dimensional Hamiltonian describing a 2D strong TI, while the Hamiltonian [SEq.~\eqref{eq:3D_TI_model}] restricted to the $k_{z} = \pi$ plane is topologically equivalent to a two-dimensional Hamiltonian describing a trivial insulator. 
SFig.~\ref{fig:3D_TI_ord_W_pos_W_neg_W_spin_gap_closing}(a) then demonstrates that our lattice model is indeed a 3D TI. 

We next numerically determine the locations of the gap closing points in the $P s_z P$ bands. 
Since our system has both $\mathcal{I}$ and $\mathcal{T}$ symmetry, we know from SN~\ref{appendix:properties-of-the-projected-spin-operator} and SFig.~\ref{fig:schematic_spin_bands_with_I_T_and_IT} that we can define the lower spin bands as those states with $[s_{z,\mathrm{reduced}}(\mathbf{k})]$ eigenvalue $\lambda_{n\mathbf{k}}^- < 0$, and similarly the upper spin bands can be defined as those states with $[s_{z,\mathrm{reduced}}(\mathbf{k})]$ eigenvalue $\lambda_{n\mathbf{k}}^+>0$. 
Spin gap closing points can only occur at points $\mathbf{k_*}$ such that $\lambda_{n\mathbf{k_*}}^+=\lambda_{n\mathbf{k_*}}^-=0$. 
We can exploit this fact to efficiently find the spin-Weyl nodes numerically by searching for points $\mathbf{k_*}$ where $\det[s_{z,\mathrm{reduced}}(\mathbf{k_*})]=0$ (which can only occur when the spin gap closes). 
We do this by applying the iterative numerical minimization algorithm to a grid of initial points in the BZ.
In SFig.~\ref{fig:3D_TI_ord_W_pos_W_neg_W_spin_gap_closing}(b) we show the two obtained spin $s_{z}$ gap closing points $\mathbf{k}_{1}=-\mathbf{k}_{2}=(0,0.456\pi,-0.212\pi)$ related to each other by $\mathcal{T}$ and $\mathcal{I}$ symmetries. 

We next compute the $k_x$-directed $P_{\pm}$-Wilson loop spectrum $\{ \gamma_{1,j}^{\pm}(k_y,k_z) \}$ as a function of $k_{y}$ for different constant-$k_z$ planes, shown in SFig.~\ref{fig:3D_TI_ord_W_pos_W_neg_W_spin_gap_closing}(c,d). 
Since our model has only two occupied bands, there is only one band in each of the positive and negative $P s_z P$ eigenspace, and hence there is only one  band $\{ \gamma_{1,j}^{\pm}(k_y,k_z) \}$ ($j=1$ only) for each of the $P_\pm$-Wilson loops.
As long as $k_{z}\neq \pm 0.212\pi$, the $s_z$ spin  gap is open, such that the $P_{\pm}$-Wilson loop is well-defined. 
We first notice that in the $\mathcal{T}$-invariant $k_{z} = 0$ plane, the $\{ \gamma_{1,j}^{+}(k_y,k_z) \}$ and $\{ \gamma_{1,j}^{-}(k_y,k_z) \}$ as a function of $k_{y}$ exhibit net winding numbers equal to $+1$ and $-1$, respectively.
This implies that the occupied bands in the $k_{z}=0$ plane have the partial Chern numbers $C^\pm_{\gamma_1}=\pm 1$ and the relative winding number ($s_z$ spin Chern number) $C^{s}_{\gamma_1} = +2$ defined in SEq.~\eqref{eq:Csgamma1def}. 
The Hamiltonian of our model restricted to the $k_z=0$ plane is thus topologically equivalent to a model of a 2D strong topological insulator. 
Since the partial Chern numbers cannot change unless the spin gap closes, we expect that $C^\pm_{\gamma_1}(k_z)=\pm 1$ for all $|k_z|<0.212\pi$. 
Indeed we see in SFig.~\ref{fig:3D_TI_ord_W_pos_W_neg_W_spin_gap_closing}(c,d) that $C^\pm_{\gamma_1}(k_z=0.16\pi)=\pm 1$, consistent with our expectation. 
Thus, away from $k_z=0$, the Hamiltonian of our model restricted to 2D planes with $|k_z|<0.212\pi$ are topologically equivalent (in the sense that they can be deformed without closing an energy gap \emph{or} a spin gap) to a model of a 2D magnetic insulator with partial Chern numbers $C^\pm_{\gamma_1}=\pm 1$. 

On the other hand, in the $\mathcal{T}$-invariant $k_{z} = \pi$ plane, both $\{ \gamma_{1,j}^{+}(k_y,k_z) \}$ and $\{ \gamma_{1,j}^{-}(k_y,k_z) \}$  exhibit zero winding as a function of $k_{y}$, which implies that the occupied bands in the $k_{z} = \pi$ plane have vanishing partial Chern numbers. 
This indicates that the Hamiltonian of our model restricted to the $k_z=\pi$ plane is topologically equivalent to a model of a 2D trivial insulator. 
Going further, since the partial Chern numbers can be defined for all BZ planes in which the spin gap is open, away from $k_z=\pi$, the occupied bands of our model restricted to 2D BZ planes with $\pi \ge |k_z|>0.212\pi$ are topologically equivalent (in the sense that they can be deformed without closing an energy gap \emph{or} a spin gap) to those of a 2D magnetic insulator with partial Chern numbers $C^\pm_{\gamma_1}= 0$. 
We numerically verify this by computing the $P_\pm$-Wilson loops at $k_z=0.26\pi$ and $k_z=\pi$ [SFig.~\ref{fig:3D_TI_ord_W_pos_W_neg_W_spin_gap_closing}(c,d)], which exhibit zero winding, indicating that the spin bands carry zero partial Chern numbers.

Since the $k_{z}$ coordinate of the spin $s_z$ gap closing points are $\pm 0.212\pi$, we expect that both $\{ \gamma_{1,j}^{+}(k_y,k_z) \}$ and $\{ \gamma_{1,j}^{-}(k_y,k_z) \}$ will have a discontinuous change of winding when the constant-$k_z$ plane passes through $k_{z} = \pm 0.212\pi$. 
As demonstrated in SFig.~\ref{fig:3D_TI_ord_W_pos_W_neg_W_spin_gap_closing}(c) [\ref{fig:3D_TI_ord_W_pos_W_neg_W_spin_gap_closing}(d)], in the $k_{z} = 0.16\pi$ plane, the $k_x$-directed $\{ \gamma_{1,j}^{+}(k_y,k_z) \}$ [$\{ \gamma_{1,j}^{-}(k_y,k_z) \}$] has a net $+1$ [$-1$] winding while in the $k_{z} = 0.26\pi$ plane it has zero winding. 
Since the partial Chern number $C^+_{\gamma_1}$ [$C^-_{\gamma_1}$] changes by $-1$ [$+1$] between planes just below and just above the spin gap closing point with $k_z=0.212\pi$, we deduce that this spin gap closing point is a monopole source of partial Berry flux with charge $-1$ [$+1$] for the upper [lower] spin bands. 
We indicate this in SFig.~\ref{fig:3D_TI_ord_W_pos_W_neg_W_spin_gap_closing}(b). 
This justifies our identification of the spin gap closing points as spin-Weyl nodes. 
Consistent with our analysis in SN~\ref{appendix:explicit-calculation-of-BHZ-model}, \ref{sec:general_properties_of_winding_num_of_P_pm_Wilson} and \ref{sec:spin_entanglement_spectrum}, we see that our 3D TI model has an odd number of spin-Weyl nodes in each half of the BZ.

\subsection{\label{appendix:3D-TI-with-and-without-inversion}3D TIs With and Without Inversion Symmetry}

In this section, we will examine the spin spectra and topology of 3D strong topological insulators with only time-reversal symmetry. 
As noted in SN~\ref{appendix:properties-of-the-projected-spin-operator}, for systems with $\mathcal{T}$ but without $\mathcal{I}$ symmetry, we can divide the spin band structure into an equal number of lower and upper spin bands. 
For $N_\mathrm{occ}$ total occupied bands, we take the projector onto the upper spin bands $P_+(\mathbf{k})$ to project onto the $N_\mathrm{occ}/2$ eigenstates of $[s_\mathrm{reduced}(\mathbf{k})]$ with largest eigenvalues; we denote the reduced spin eigenvalues for these states as $\lambda_{n\mathbf{k}}^+$, where $n=1\dots N_\mathrm{occ}/2$. 
Similarly we take the projector onto the lower spin bands $P_-(\mathbf{k})$ to project onto the $N_\mathrm{occ}/2$ eigenstates of $[s_\mathrm{reduced}(\mathbf{k})]$ with smallest eigenvalues; we denote the reduced spin eigenvalues for these states as $\lambda_{n\mathbf{k}}^-$, where $n=1\dots N_\mathrm{occ}/2$. The spin gap is open provided $\lambda_{n\mathbf{k}}^+\neq \lambda_{m\mathbf{k}}^-$ for all $m,n$ and $\mathbf{k}$.

\begin{figure*}[t]
\includegraphics[width=\textwidth]{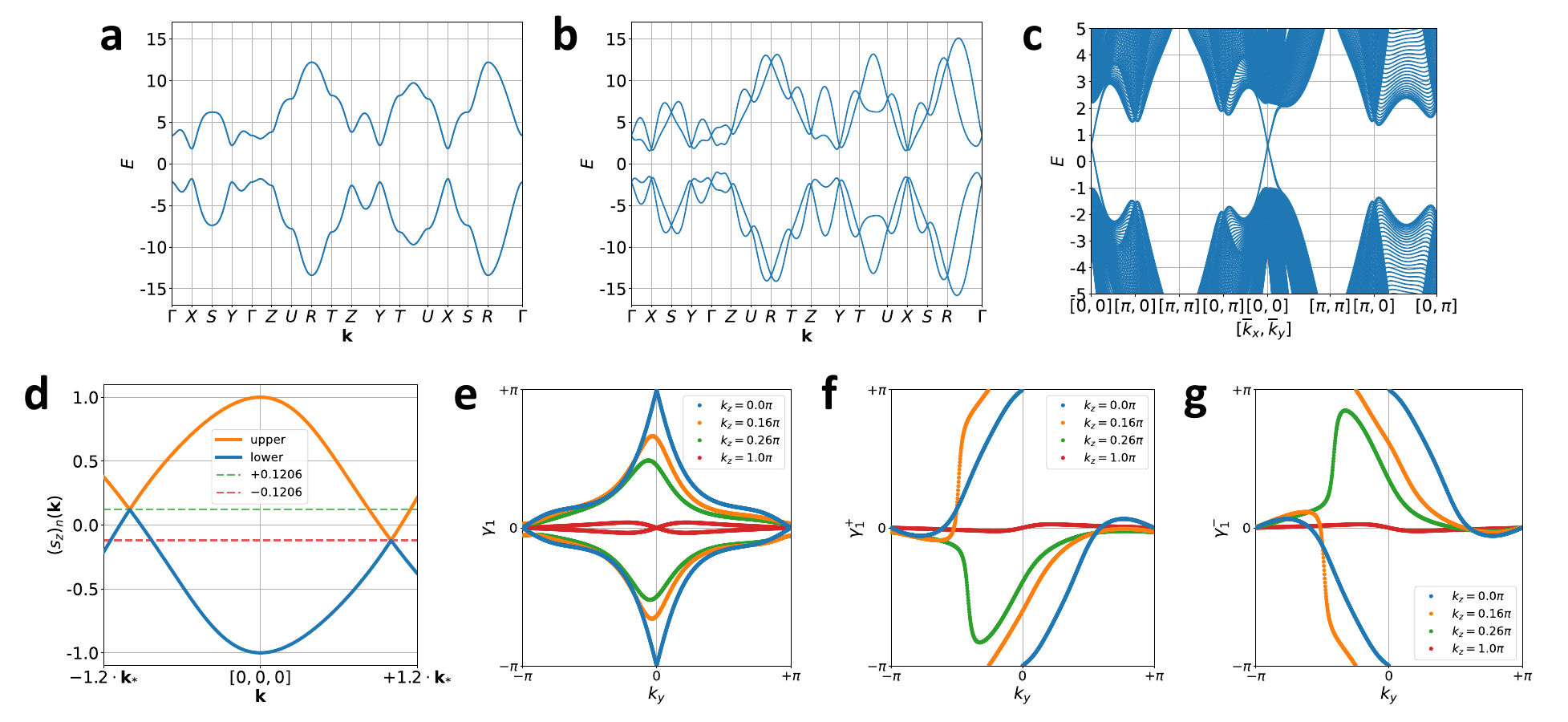}
\caption{Spectrum and Wilson loops for an inversion-breaking 3D TI. 
(a) shows the 3D bulk energy band structure of a 3D TI described by SEq.~(\ref{eq:3D_TI_model}) with parameters in SEq.~(\ref{eq:3D_TI_model_parameter_I_and_T_symmetry}), where the labels of the high-symmetry $\mathbf{k}$ points can be found in SFig.~\ref{fig:3D_TI_bulk_BZ_bulk_band_surface_band}(a). 
The energy bands are doubly degenerate at all $\mathbf{k}$ points because of the coexistence of inversion ($\mathcal{I}$) and time-reversal ($\mathcal{T}$) symmetries. 
(b) shows the 3D bulk energy band structure of a 3D TI described by SEq.~(\ref{eq:3D_TI_model}) with additional inversion-breaking terms in SEq.~(\ref{eq:inversion_breaking_terms_for_3D_TI}). 
The energy bands are no longer doubly degenerate at generic $\mathbf{k}$ points except for TRIMs. 
(c) shows the 2D energy band structure of a slab finite along $z$ with $41$ unit cells for the inversion-breaking 3D TI with the same model parameters in (b). 
We can see that there are 2D, twofold degenerate surface Dirac cones with linear dispersion at the $\bar{\Gamma}$ point ($\bar{k}_{x} = \bar{k}_{y}=0$) of the surface BZ.
(d) shows the eigenvalues $(s_z)_{n}(\mathbf{k})$ ($n=$ band index) of the reduced $s_{z}$ matrix from SEq.~(\ref{eq:P_pm_Wilson_loop_s_reduced_def}) as a function of the crystal momenta $\mathbf{k}$, which we call the $Ps_{z}P$ band structure, for the two valence bands in (b). 
The $Ps_{z}P$ band structure is plotted from $-1.2\mathbf{k}_{*}$  to $+1.2\mathbf{k}_{*}$ along a straight line. 
And $\pm\mathbf{k}_{*} \approx \pm (0,0.46\pi,-0.21\pi)$ are the positions of the spin-Weyl nodes which have {\it nonzero $Ps_{z}P$ eigenvalues} $\mp 0.1206$. 
(e) shows the $k_x$-directed $P$-Wilson loop eigenphases of the two lowest occupied bands of the inversion-breaking 3D TI plotted as a function of $k_{y}$ at different $k_{z} = $ constant planes. 
(f) \& (g) show the $k_x$-directed $P_{\pm}$-Wilson loop eigenphases of the two lowest occupied bands of the inversion-breaking 3D TI plotted as a function of $k_{y}$ at different $k_{z} = $ constant planes. 
We can see that there is a discontinuous change of the winding numbers of the spectral flow at the planes with $k_{z} < (\mathbf{k}_{*})_{z}$ and $k_{z} > (\mathbf{k}_{*})_{z}$. 
As an example, we can see that the spectral flow has a $+1$ [0] winding number at the $k_{z} = 0.16 \pi$ [$k_{z} = 0.26 \pi$] plane in (f).
And the spectral flow has a $-1$ [$0$] winding number at the $k_z = 0.16\pi$ [$k_z = 0.26\pi$] plane in (g).
The calculations detailed in this figure were performed using the freely available Python package~\href{https://github.com/kuansenlin/nested_and_spin_resolved_Wilson_loop}{\textsc{nested\_and\_spin\_resolved\_Wilson\_loop}}~\cite{lin2023nestedWilsonLib}, which represents an extension of the~\href{https://www.physics.rutgers.edu/pythtb/}{PythTB} open-source Python tight-binding package~\cite{coh2013python} that was implemented and utilized for the preparation of SRefs.~\cite{wieder2018axion,wieder2020strong} and the present work.
}
\label{fig:inversion-breaking-3D-TI}
\end{figure*}

Just as in SN~\ref{sec:main-text-3D-TI-P-pm}, the spin gap with $\mathcal{T}$ symmetry will generically close at isolated points $\mathbf{k}_*$, at which for a given choice of $n$, we have that $\lambda_{n\mathbf{k_*}}^+= \lambda_{n\mathbf{k_*}}^-$. 
However, with only $\mathcal{T}$ symmetry, $\lambda_{n\mathbf{k_*}}^+= \lambda_{n\mathbf{k_*}}^-\neq 0 $ generically at the spin gap closing points. 
Specifically, with only $\mathcal{T}$ symmetry, the spin gap for a 3D insulator will generically close at a set of spin-Weyl nodes, but those spin-Weyl nodes need not have vanishing $PsP$ eigenvalues (\emph{i.e.} the spin-Weyl nodes can appear at nonzero $PsP$ eigenvalues in the spin spectrum). 
We also have from SN~\ref{appendix:properties-of-the-projected-spin-operator} that if there exists a spin-Weyl node at $\mathbf{k}_*$ with $\lambda_{n\mathbf{k_*}}^+= \lambda_{n\mathbf{k_*}}^-=\lambda$, then by $\mathcal{T}$ symmetry, there will also be a spin-Weyl node at $-\mathbf{k}_*$ with $\lambda_{n-\mathbf{k_*}}^+= \lambda_{n-\mathbf{k_*}}^-=-\lambda$; this is illustrated in SFig.~\ref{fig:inversion-breaking-3D-TI}(d). 
As in the $\mathcal{I}$- and $\mathcal{T}$-symmetric case considered in SN~\ref{sec:main-text-3D-TI-P-pm}, spin-Weyl nodes in systems with only $\mathcal{T}$ symmetry are also monopole sources of partial Chern numbers, even though they generically carry nonzero $PsP$ eigenvalues. 
Specifically, if we consider the occupied bands in two BZ planes on either side of a spin-Weyl node in a noncentrosymmetric, $\mathcal{T}$-invariant insulator, then the partial Chern numbers $C^\pm_{\gamma_1}$ each change by $\pm 1$ (or $\mp 1$) as ${\bf k}$ crosses the spin-Weyl node.

We can now relate spin spectrum degeneracies to the strong topological index in systems with only time-reversal symmetry. 
Consider a system with a spin band structure that has an odd number of spin-Weyl points in half the BZ, none of which occur on a $\mathcal{T}$-invariant plane. 
We then have that the partial Chern numbers on the two $\mathcal{T}$-invariant planes bounding the half of the BZ must differ by $1$ (or more generally an odd integer). 
On $\mathcal{T}$-invariant planes, 2D $\mathbb{Z}_2$ Kane-Mele invariant $\nu_{2d}$ is given by the parity of the partial Chern numbers as shown in the argument following SEq.~\eqref{eq:trcschange}. 
This lets us deduce that $\nu_{2d}$ differs between these two $\mathcal{T}$-invariant planes. 
This implies immediately that the system is a strong topological insulator as defined in SRef.~\cite{fu2007topologicala}. 
It is crucial to recall in this argument that while the 2D $\mathbb{Z}_2$ invariant is only defined on $\mathcal{T}$-invariant planes, the partial Chern numbers are defined at all planes in the BZ in which the spin gap is open. 
In particular, the value of $\nu_{2d}$ for the occupied bands on a $\mathcal{T}$-invariant plane $M$ fixes parity of half the spin Chern number $C^s_{\gamma_1}/2$ [defined in SEq.~\eqref{eq:Csgamma1def}] for the occupied bands on any non-$\mathcal{T}$-invariant plane that can be reached by deforming $M$ without crossing a spin-Weyl node.  
Thus, we deduce that for a gapped Hamiltonian $H$ with spinful $\mathcal{T}$ symmetry, if the spectrum of $PsP$ has a number of Weyl nodes $n_w \mod 4 = 2$, then the system must be a 3D strong topological insulator. 
In particular, if there is a spin-Weyl node at $\mathbf{k}$ with $PsP$ eigenvalue $\lambda$, by time-reversal symmetry there will be another spin-Weyl node at $-\mathbf{k}$ with $PsP$ eigenvalue $-\lambda$. 
Crucially, the Berry flux around a spin-Weyl node reverses sign under $\mathcal{T}$ symmetry: two spin-Weyl nodes related by $\mathcal{T}$ have opposite partial Chern numbers (chiral charges). 
This might seem counter-intuitive, as we know that Weyl nodes in the energy spectrum that are related by $\mathcal{T}$ symmetry carry the same Berry flux.
However, recall that for spin bands, time-reversal not only flips the momentum but also flips the sign of the spin.
Therefore, under $\mathcal{T}$ the upper spin band of a spin-Weyl node at momentum $\mathbf{k}$ maps to the lower spin band of the image spin-Weyl node at $-\mathbf{k}$. 
Since the upper and lower bands of a Weyl node carry opposite Chern numbers {\it provided that the occupied energy bands are separated from the unoccupied energy bands by a finite energy gap}, we deduce that the spin-Weyl nodes that are related to each other by $\mathcal{T}$ symmetry will carry opposite partial Berry flux. 
On the other hand, for a system with both $\mathcal{I}$  and $\mathcal{T}$ symmetries (see SN~\ref{sec:main-text-3D-TI-P-pm}), the spin-Weyl nodes with $\lambda=0$ at $\mathbf{k}$ and $-\mathbf{k}$ are related by either $\mathcal{T}$ or $\mathcal{I}$. 
Importantly, spin-Weyl nodes related by $\mathcal{I}$ symmetry will also carry opposite Berry flux. 
This is because $\mathcal{I}$ acts on the $PsP$ eigenstates in the same way as the energy eigenstates, and $\mathcal{I}$ does not flip the sign of the spin. 
Therefore, each of spin-Weyl nodes carries nonzero partial Berry flux (even though the system has both $\mathcal{I}$  and $\mathcal{T}$ symmetry) while Weyl nodes in the energy spectrum are forbidden in systems with both $\mathcal{I}$  and $\mathcal{T}$ symmetry~\cite{armitage2018weyl}. 
As a corollary, this also constrains the perturbations to the spin band structure that can occur without closing the spectral gap of the Hamiltonian $H$. 
In particular, note that in order to have, or induce, a total number of spin-Weyl nodes $n_w \mod 4 =2$, we must invert spin bands at a TRIM point (a spin band inversion at a generic point would create two pairs of Weyl points related by $\mathcal{T}$-symmetry). 
Since this spin band inversion would change the strong $\mathbb{Z}_2$ invariant of the occupied states in a $\mathcal{T}$-invariant plane, it cannot occur without closing the gap in the Hamiltonian. 
This is consistent with the analysis of SRef.~\cite{roy2010characterization}. 

Let us now numerically verify these points.
We will consider a model of a 3D TI with only $\mathcal{T}$ symmetry and demonstrate the application of the $P_{\pm}$-Wilson loop formalism previously described in SN~\ref{sec:P_pm_Wilson_loop} to diagnose a 3D $\mathcal{T}$-invariant TI without inversion symmetry. 
To begin, we take the model of a 3D TI described in SEq.~(\ref{eq:3D_TI_model}) with parameters in SEq.~(\ref{eq:3D_TI_model_parameter_I_and_T_symmetry}) and add additional terms 
\begin{align}
	H_{\mathcal{I}\text{-breaking}}(\mathbf{k}) = \sum_{i,j=1}^{3} f_{0ij} \sin{k_{j}} \tau_{0}\sigma_{i},\ f_{0ij} \in \mathbb{R}, \label{eq:inversion_breaking_terms_for_3D_TI}
\end{align}
that break $\mathcal{I}$ while preserving $\mathcal{T}$ symmetry:
\begin{align}
	& \mathcal{I}: (\tau_{z}\sigma_{0}) f_{0ij} \sin{k_{j}} \tau_{0}\sigma_{i} (\tau_{z}\sigma_{0}) =  f_{0ij} \sin{k_{j}} \tau_{0}\sigma_{i}  \neq f_{0ij} \sin{(-k_{j})} \tau_{0}\sigma_{i},  \\
	& \mathcal{T}: (\tau_{0}\sigma_{y}) f_{0ij} \sin{k_{j}} \tau_{0}(\sigma_{i} )^{*} (\tau_{0}\sigma_{y}) = f_{0ij} \sin{(-k_{j})} \tau_{0} \sigma_{i}.
\end{align}
In SFig.~\ref{fig:inversion-breaking-3D-TI}(a,b) we show the 3D bulk band structures with $f_{0ij} = 0.0$ and $f_{0ij} = 1.0$ for all $i,j=1\sim3$, respectively. 
As we can see in SFig.~\ref{fig:inversion-breaking-3D-TI}(b), since the $\mathcal{I}$ symmetry is broken, the energy bands are no longer doubly degenerate at generic (non-time-reversal-invariant) $\mathbf{k}$ points. 
Since there is no bulk gap closing when we turn on nonzero values of $f_{0ij}$, the two valence bands of SFig.~\ref{fig:inversion-breaking-3D-TI}(b) still describe a 3D TI with a nontrivial strong $\mathbb{Z}_{2}$ invariant $\nu_{2d} = 1$.
This can be seen by computing the band structure of a 2D slab finite along $z$, which is shown in SFig.~\ref{fig:inversion-breaking-3D-TI}(c) exhibits one surface Dirac cone at the surface $\Gamma$ point of each surface. 
As mentioned above, if we denote $\{ s_n (\mathbf{k}) \}$ as the set of eigenvalues of the reduced spin matrix [SEq.~\eqref{eq:P_pm_Wilson_loop_s_reduced_def}], $\mathcal{T}$ symmetry alone only constrains $\{ s_{n}(\mathbf{k}) \} = \{ -s_{n}(-\mathbf{k}) \}$. 
Therefore, generically the spin-Weyl nodes
do not appear at zero $PsP$ eigenvalues. 
To demonstrate this, we obtain the $Ps_{z}P$ band structure of the two valence bands of SFig.~\ref{fig:inversion-breaking-3D-TI}(b), which contains two spin-Weyl nodes at momenta $\pm \mathbf{k}_{*}$ with {\it nonzero $Ps_{z}P$ eigenvalues}, as shown in SFig.~\ref{fig:inversion-breaking-3D-TI}(d). 
In SFig.~\ref{fig:inversion-breaking-3D-TI}(d), the spin-resolved projectors $P_{+}(\mathbf{k})$ and $P_{-}(\mathbf{k})$ are the projectors onto the upper (orange) and lower (blue) spin bands. 
We then compute the $k_x$-directed $P$- and $P_{\pm}$-Wilson loop spectra as a function of $k_{y}$ at various $k_{z}=$ constant planes.  
As shown in SFig.~\ref{fig:inversion-breaking-3D-TI}(e), at $k_{z} = 0$ and $k_{z} = \pi$ we have helical winding and trivial winding, respectively. 
This indicates that the two valence bands still describe a 3D TI. 
In addition, in SFig.~\ref{fig:inversion-breaking-3D-TI}(f,g) we have spectral flows with winding numbers $\pm 1$ and $0$ in the $P_{\pm}$-Wilson loop eigenphases at $k_{z} = 0$ and $k_{z} = \pi$ planes, respectively. 
This again demonstrates that the Hamiltonian restricted to $k_{z} = 0$ is topologically equivalent to a 2D $\mathcal{T}$-invariant TI. 
Similarly, the Hamiltonian restricted to the $k_{z} = \pi$ plane is topologically equivalent to a trivial 2D insulator. 
In particular, there is a discontinuous change of the spectral flow of the $P_\pm$-Wilson loops between $k_{z} < (\mathbf{k}_{*})_{z}$ and $ k_{z} > (\mathbf{k}_{*})_{z}$, as shown in SFig.~\ref{fig:inversion-breaking-3D-TI}(f,g). 
This is a clear indication that the spin-Weyl nodes at $\pm\mathbf{k}_{*}$ play the role of {\it partial Chern number monopoles}.

The above discussion reiterates the central idea first recognized in SRef.~\cite{kane2005quantum} that to diagnose a helical topological phase, one could divide the occupied space into two parts that are related to each other by the $\mathcal{T}$ operation. 
Once we have constructed $P_{\pm}(\mathbf{k})$ as a smooth function of $\mathbf{k}$ (except at the spin-Weyl nodes) without a discontinuous jump of $\text{rank}[P_{\pm}(\mathbf{k})]$, we can use one of the $P_{\pm}(\mathbf{k})$ to diagnose the helical topological phase by examining its Wilson loop spectrum. 
In SN~\ref{app:w2} we will extend this perspective to demonstrate the existence of previously unrecognized quantized spin-resolved invariants in helical HOTIs.

\subsection{Spin-Stable Topology in a 2D Fragile Topological Insulator}\label{sec:spin_stable_topology_2d_fragile_TI}

In this section, we will demonstrate that in a 2D fragile TI there can exist {\it spin-stable topology}, which is robust to perturbations provided that both the \textit{energy} and {\it spin} gaps remain open. 
We will argue that such spin-stable topological phases can exhibit robust responses to external fields and fluxes. 
Thus, we will show that even though 2D fragile topological phases have the same boundary signatures as obstructed atomic insulators~\cite{wieder2020strong,wieder2018axion,wang2019higherorder,WiederDefect,ahn2019failure,hwang2019fragile,lee2020two}, they can differ in their bulk signatures due to spin-stable topology.

Without the addition of appropriately chosen trivial bands, the occupied bands of a  fragile TI cannot form exponentially localized Wannier functions that respect all of the crystal symmetries. 
For example, let us consider the 2D fragile topological model introduced in SRef.~\cite{wieder2020strong} as a time-reversal invariant generalization of the 2D quadrupole insulator~\cite{benalcazar2017quantized,benalcazar2017electric}. 
This model has a square lattice with Bloch Hamiltonian 
\begin{align}
	[H_F(\mathbf{k})]&=  t_1 [\cos (k_x) + \cos (k_y)]\tau_z\sigma_0 \nonumber \\
	& + t_2[\cos(k_x) - \cos(k_y)]\tau_x \sigma_0  + v_m \tau_z \sigma_0 \nonumber \\
	& + t_{PH}[\cos(k_x) + \cos(k_y)]\tau_0\sigma_0 \nonumber \\
	& + v_s \sin(k_x) \sin(k_y) \tau_y\sigma_z \nonumber \\
	& + v_{M_z}[\sin(k_x) \tau_z\sigma_y - \sin(k_y)\tau_z\sigma_x] \label{eq:H_2d_fragile_original},
\end{align}
where Pauli matrices $\tau_{i}$ ($\sigma_i$) denotes the $s$ and $d$ orbital (spin) degrees of freedom that are placed at the $1a$ ($(x,y)=(0,0)$) Wyckoff position of the primitive square unit cell.
Both $\tau_0$ and $\sigma_0$ are $2 \times 2$ identity matrices.
$t_1$ and $t_2$ are the nearest-neighbor hopping energies between the same and different orbitals, $v_m$ induces an on-site orbital energy splitting, $t_{PH}$ is a spin- and orbital-independent nearest-neighbor hopping that breaks the particle-hole symmetry of the energy spectrum, and $v_s$ represents a next-nearest-neighbor spin-orbit coupling (SOC).
In addition, $v_{M_z}$ is an SOC term breaking the 3D layer group~\cite{wieder2016spinorbit} mirror reflection $M_z$ (represented as $\sigma_z$) and  inversion $\mathcal{I}$ (represented as identity) symmetries.

\begin{figure}[ht]
\includegraphics[width=\textwidth]{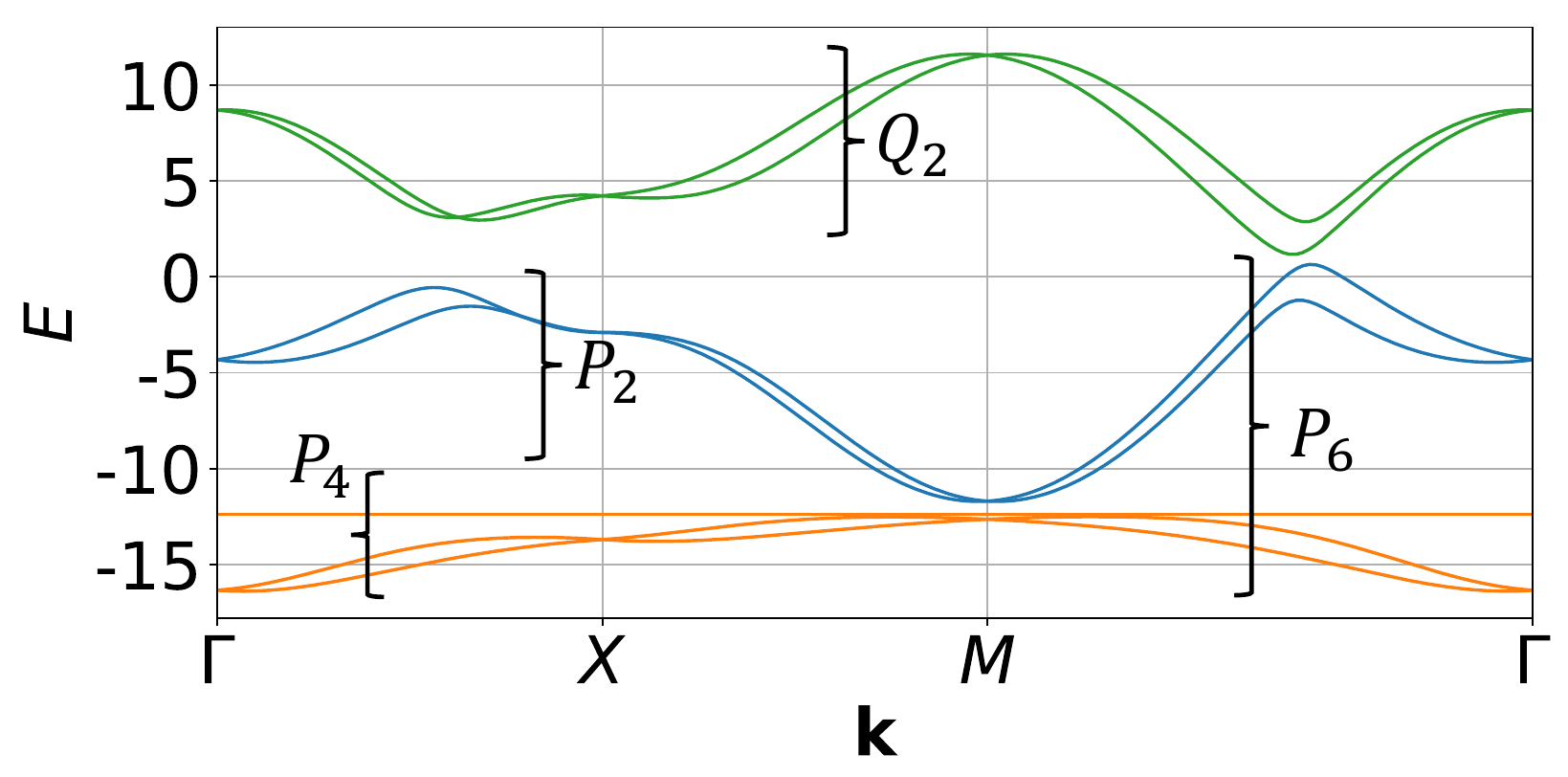}
\caption{Energy bands of a 2D fragile TI in SEq.~\eqref{eq:H_2d_fragile_original} on a square lattice where $P_2$ and $Q_2$ project respectively onto the two valence and two conduction bands. 
The high-symmetry $\mathbf{k}$ points $\Gamma$, $X$ and $M$ correspond to $(k_x,k_y)=(0,0)$, $(\pi,0)$, and $(\pi,\pi)$, respectively.
$P_4$ is the projector onto four additional energy bands induced from trivial atomic orbitals that are coupled to the 2D fragile TI [SEq.~\eqref{eq:H_2d_fragile_original}] through SEq.~\eqref{eq:V_C_coupled_to_HF} in a way that respects the symmetries of $p4m1'$ wallpaper group (see also SRef.~\cite{wieder2020strong}, where this model was originally introduced).
The energy bands are computed from Bloch Hamiltonian $[H_F (\mathbf{k})] + [V_C (\mathbf{k})]$ in SEqs.~\eqref{eq:H_2d_fragile_original} and \eqref{eq:V_C_coupled_to_HF} with tight-binding parameters $t_1 = 5.0$, $t_2 = 1.5$, $v_{m} = -1.5$, $t_{PH} = 0.1$, $v_{s} = 1.3$, $v_{M_z} = 0.4$, $v_{\mu} = 8.25\times v_{m}$, $v_{C} = 4.0$, and $v_{CS} = 0.45 \times v_{C}$.
The numbers of energy bands in the image of $Q_{2}$, $P_{2}$, and $P_{4}$ are $2$, $2$, and $4$, respectively.
}\label{fig:2d_fragile_bands_with_P2_and_P6}
\end{figure}

The Bloch Hamiltonian [SEq.~\eqref{eq:H_2d_fragile_original}] respects the symmetries of wallpaper group $p4m1'$, which is generated by mirror reflection $M_x$, the four-fold rotation $C_{4z}$, time-reversal $\mathcal{T}$, and 2D lattice translations (for further information regarding wallpaper groups and their relationship to topological phases, see SRefs.~\cite{MagneticBook,subperiodicTables,wieder2016spinorbit,SteveMagnet,wieder2018wallpaper,wieder2020strong}). 
The symmetries are represented by
\begin{align}
[M_x][H_F(\mathbf{k})][M_x]^{-1} &= \sigma_x[H_F(\mathbf{k})]\sigma_x =[H_F(M_x\mathbf{k})]  \\
[C_{4z}][H_F(\mathbf{k})][C_{4z}]^{-1} &= \tau_ze^{-i\pi/4\sigma_z}[H_F(\mathbf{k})]\tau_ze^{i\pi/4\sigma_z} =[H_F(C_{4z}\mathbf{k})]  \\
[\mathcal{T}][H_F(\mathbf{k})][T]^{-1} &= (i\sigma_y)[H_F(\mathbf{k})]^*(-i\sigma_y) =[H_F(-\mathbf{k})]  
\end{align}
The 2D bulk energy bands of SEq.~\eqref{eq:H_2d_fragile_original} correspond to the images of the projectors $P_2$ and $Q_2$ shown in SFig.~\ref{fig:2d_fragile_bands_with_P2_and_P6}. 
This model has no gapless edge states in the energy gaps between $P_2$ and $Q_2$ when placed in a ribbon geometry, but has four Kramers pairs of corner modes when placed on a finite-sized square~\cite{wieder2020strong}. 
When the bulk is half-filled, the corner mode filling is generically 3/4.

\begin{figure}[ht]
\includegraphics[width=\textwidth]{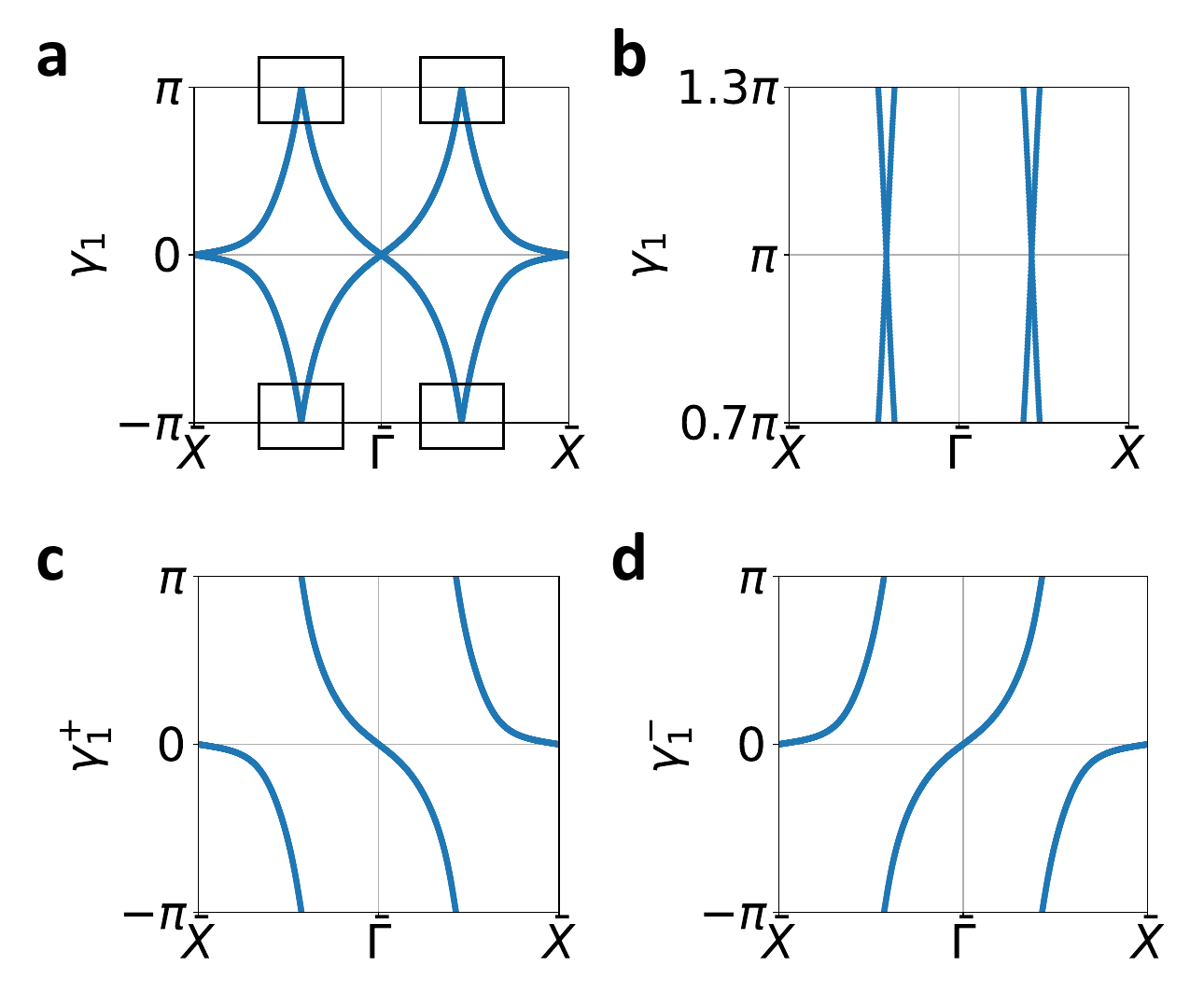}
\caption{$k_y$-directed Wilson loop eigenphases as a function of $k_{x}$ in the (a) occupied space $P_2$ of SFig.~\ref{fig:2d_fragile_bands_with_P2_and_P6}, (c) positive ($+$), and (d) negative ($-$) eigenspace of $P_2 s_z P_2$ where $s_z$ is the $z$-component of the spin vector $\mathbf{s}$, namely $s_z = \hat{\mathbf{z}} \cdot \mathbf{s}$. 
(b) is an enlarged view of the region in (a) around $\gamma_1 = \pi$, demonstrating the two-band crossings protected by $C_{2z}$ and $\mathcal{T}$.
The numbers of bands in (a), (c), and (d) are $2$, $1$, and $1$, respectively.
The calculations detailed in this figure were performed using the freely available Python package~\href{https://github.com/kuansenlin/nested_and_spin_resolved_Wilson_loop}{\textsc{nested\_and\_spin\_resolved\_Wilson\_loop}}~\cite{lin2023nestedWilsonLib}, which represents an extension of the~\href{https://www.physics.rutgers.edu/pythtb/}{PythTB} open-source Python tight-binding package~\cite{coh2013python} that was implemented and utilized for the preparation of SRefs.~\cite{wieder2018axion,wieder2020strong} and the present work.
}\label{fig:P2_wilson_loop_2d_fragile}
\end{figure}

The energy bands in the image of the projector $P_2$ have fragile topology~\cite{po2018fragile,cano2018topology,bouhon2019wilson} that can be diagnosed from symmetry indicators and the Wilson loop spectrum~\cite{wieder2020strong}.
Shown in SFig.~\ref{fig:P2_wilson_loop_2d_fragile}(a) are the $k_y$-directed $P$-Wilson loop eigenphases $\{ \gamma_{1,j}(k_x) \}$ as a function of $k_x$ computed in the $P_2$ occupied space where $j$ is the $P$-Wannier band index.
In particular, this $P$-Wilson loop spectrum has several two-band crossings at $\gamma_1 = \pi$ and generic momenta demonstrated in SFig.~\ref{fig:P2_wilson_loop_2d_fragile}(b), that are protected by $C_{2z}$ and $\mathcal{T}$ symmetries~\cite{bradlyn2019disconnected,wieder2020strong,song2019all,ahn2018band,song2021twisted}. 
To demonstrate that the two-band crossings in SFig.~\ref{fig:P2_wilson_loop_2d_fragile}(b) arise from bulk fragile topology, we introduce {\it four} additional tight-binding basis orbitals to the Hilbert space of the system consisting of spinful $s$ orbitals at the $2c$ Wyckoff positions ($(x,y)=(1/2,0)$ and $(0,1/2)$) that are coupled to $[H_F(\mathbf{k})]$ in SEq.~(\ref{eq:H_2d_fragile_original}) through
\begin{align}
	&[V_C (\mathbf{k})]  = v_{\mu} [P_{\mu^s}] \nonumber \\
	& + v_C\left[ \mu_{1,x} \cos \left( \frac{k_x}{2} \right) + \mu_{2,x} \cos \left( \frac{k_y}{2} \right) \right] \nonumber \\
	& + v_{CS} \left[ -\mu_{1,x} \sigma_y \sin\left( \frac{k_x}{2} \right) + \mu_{2,x} \sigma_x \sin\left( \frac{k_y}{2} \right) \right].\label{eq:V_C_coupled_to_HF}
\end{align}
$[P_{\mu^s}]$ is the projection matrix onto the spinful $s$ orbitals at $2c$ Wyckoff positions.
$\mu_{1,x}$ [$\mu_{2,x}$] implements the hopping matrix with nonzero coefficients being 1 that connects the spinful $d$ orbitals at $1a$ and $s$ orbitals at $(x,y)=(1/2,0)$ [$(x,y)=(0,1/2)$].
$v_{\mu}$ denotes the on-site energy (or effectively the chemical potential, up to a sign) of the spinful $s$ orbitals at $2c$.
$v_{C}$ ($v_{CS}$) represents the (spin-orbit) coupling between the spinful $d$ orbitals at $1a$ and $s$ orbitals at $2c$.
Notice that both of $v_C$ and $v_{CS}$ contain inter- and intra-cell hopping~\cite{wieder2020strong}.
Crucially, the symmetry group of the combined Hamiltonian $[H(\mathbf{k})] + [V_C(\mathbf{k})]$ remains the wallpaper group $p4m1'$, such that changes to the spectral flow of the Wilson loop cannot be attributed to symmetry breaking.

\begin{figure}[ht]
\includegraphics[width=\textwidth]{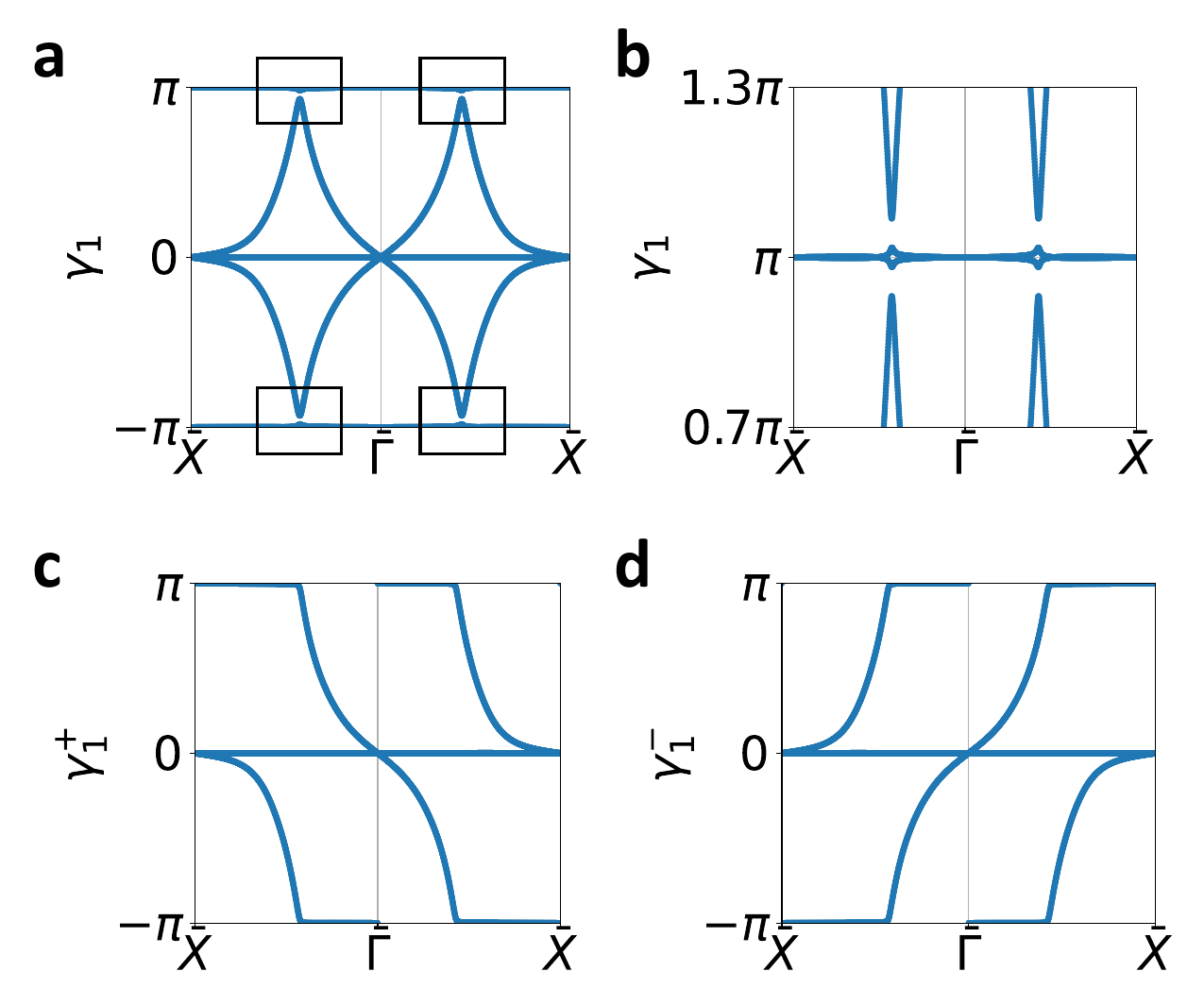}
\caption{$k_y$-directed Wilson loop eigenphases as a function of $k_{x}$ in the (a) occupied space $P_6$ of SFig.~\ref{fig:2d_fragile_bands_with_P2_and_P6}, (c) positive ($+$), and (d) negative ($-$) eigenspace of $P_6 s_z P_6$ where $s_z$ is the $z$-component of the spin vector $\mathbf{s}$, namely $s_z = \hat{\mathbf{z}} \cdot \mathbf{s}$.
(b) shows an enlarged view of (a) around $\gamma_{1} = \pi$ to demonstrate that the two-band crossings in SFig.~\ref{fig:P2_wilson_loop_2d_fragile}(b) are removed.
The numbers of bands in (a), (c), and (d) are $6$, $3$, and $3$, respectively.
The calculations detailed in this figure were performed using the freely available Python package~\href{https://github.com/kuansenlin/nested_and_spin_resolved_Wilson_loop}{\textsc{nested\_and\_spin\_resolved\_Wilson\_loop}}~\cite{lin2023nestedWilsonLib}, which represents an extension of the~\href{https://www.physics.rutgers.edu/pythtb/}{PythTB} open-source Python tight-binding package~\cite{coh2013python} that was implemented and utilized for the preparation of SRefs.~\cite{wieder2018axion,wieder2020strong} and the present work.
}\label{fig:P6_wilson_loop_2d_fragile}
\end{figure}

The additional four bands whose coupling to the original Hamiltonian [SEq.~\eqref{eq:H_2d_fragile_original}] is described by $[V_{C}(\mathbf{k})]$ [SEq.~\eqref{eq:V_C_coupled_to_HF}] lie in the image of the projector denoted as $P_4$ in SFig.~\ref{fig:2d_fragile_bands_with_P2_and_P6}.
The bands in the image of $P_4$ carry an elementary band representation of $p4m1'$ induced from spinful $s$ orbitals at the $2c$ Wyckoff position. 
As analyzed in SRef.~\cite{wieder2020strong}, although the bands in the image of $P_2$ are fragile topological, we can include the trivial bands in the image of $P_4$ and consider the combined projection operator $P_2 \oplus P_4 = P_6$. 
And the straight $P_6$-Wilson loop does not wind for any choice of direction in the 2D BZ. 
This is demonstrated in SFig.~\ref{fig:P6_wilson_loop_2d_fragile}(b) where the two-band crossings at $\gamma_1=\pi$ in SFig.~\ref{fig:P2_wilson_loop_2d_fragile}(b) are removed, and the six-band Wilson loop spectrum shown in SFig.~\ref{fig:P6_wilson_loop_2d_fragile}(a) form two groups of bands that are spectrally separated from each other--one group consists of two bands around $\gamma_1 = \pi$ mod $2\pi$ and the other group consists of four bands centered around $\gamma_1 = 0$.
We have hence demonstrated that the two-band crossings [see SFig.~\ref{fig:P2_wilson_loop_2d_fragile}(b)] of the $P$-Wilson loop spectrum in occupied space $P_{2}$ can be trivialized by adding trivial bands [see SFig.~\ref{fig:P6_wilson_loop_2d_fragile}(b)], confirming the previous calculations in SRef.~\cite{wieder2020strong}.

Although this system has a fragile topology in the occupied space $P_2$, we will now demonstrate that it also carries a {\it nontrivial spin-stable topology}. 
We can spin-resolve the occupied space according to its $P_2 s_z P_2$ eigenvalues.
As shown in SFig.~\ref{fig:P2_wilson_loop_2d_fragile}(c,d),  the $k_y$-directed $P_{\pm}$-Wilson loop eigenphases $\{ \gamma_{1,j}^{\pm}(k_x) \}$ have net $\mp 2$ winding as $k_x \to k_x + 2\pi$.
From the winding numbers in SFig.~\ref{fig:P2_wilson_loop_2d_fragile}(c,d) we can determine the partial Chern numbers of the occupied space $P_2$, using the sign convention introduced in SEq.~\eqref{eq:partial_chern_def}.
To be precise, the partial Chern numbers $C^\pm_{\gamma_1}$ in this case are determined by the negative winding numbers of the $k_y$-directed $P_\pm$-Wilson loop eigenphases as $k_x \to k_x + 2\pi$.
This means that the partial Chern numbers of the $\pm$ eigenspace of $P_2 s_z P_2 $ are $C_{\gamma_1}^{\pm}=\pm 2$, such that the {\it spin Chern number} [SEq.~\eqref{eq:Csgamma1def}] is given by $C_{\gamma_1}^{s} = +4$.
As long as both the energy and $P_2 s_z P_2$ gaps remain open, this $C_{\gamma_1}^{s} = +4$ is robust~\cite{prodan2009robustness} and can have several physical consequences when we apply external electromagnetic fields.

\begin{figure}[ht]
\includegraphics[width=\textwidth]{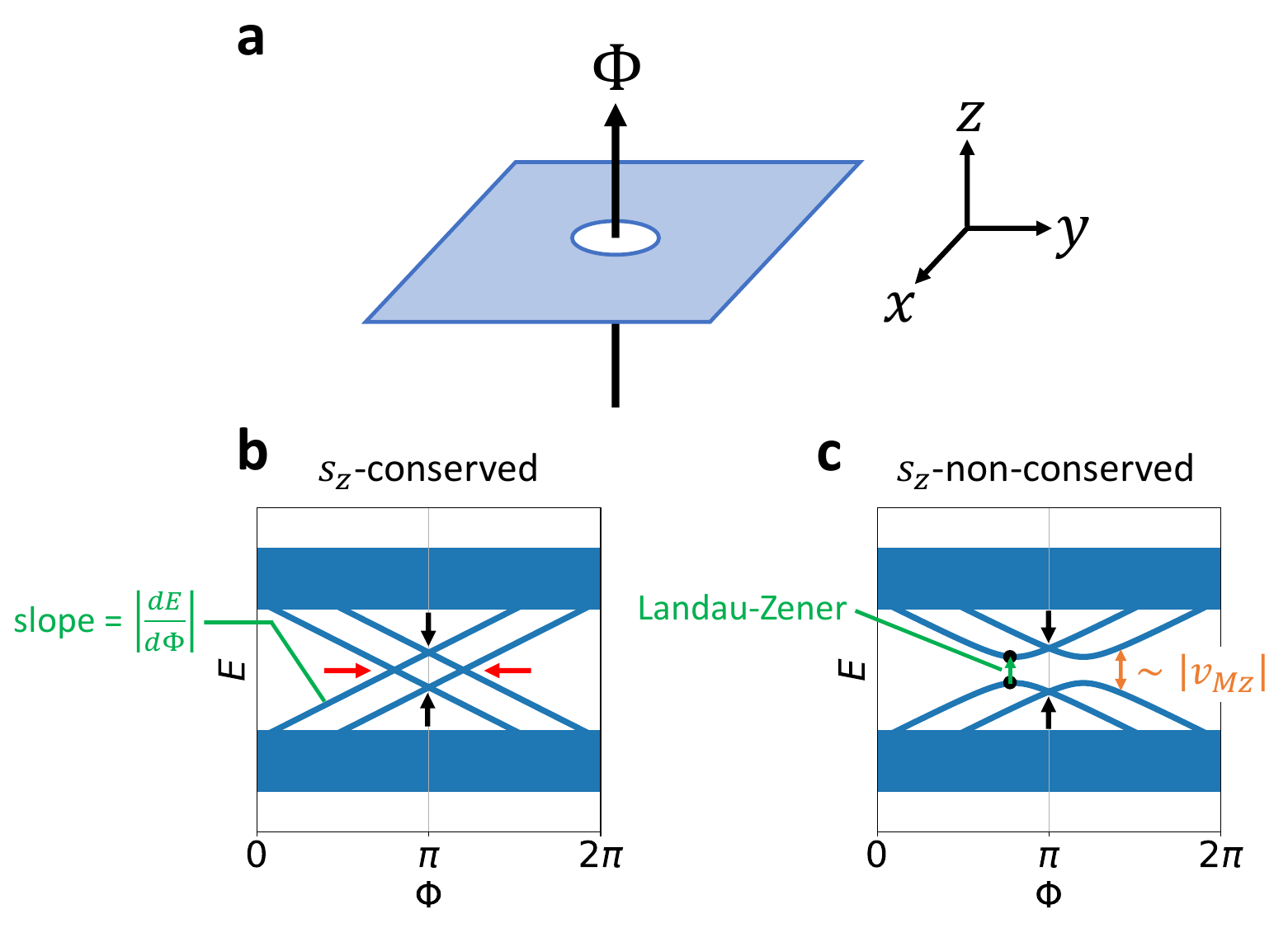}
\caption{Flux insertion for a spin-stable topological phase with spin Chern number $|C_{\gamma_1}^{s}|=4$. 
(a) shows a  schematic of a perpendicular $U(1)$ flux $\Phi$ threading through a small hole of a 2D system. 
When $\Phi = 0$ and $\Phi = \pi$ the system has time-reversal symmetry while if $\Phi \neq 0$ or $\Phi \neq \pi$ the time-reversal symmetry is broken~\cite{qi2008spincharge}.
(b) shows a schematic of the energy spectrum of the 2D sample in (a) with infinite size along $x$ and $y$ and a small hole as a function of $\Phi$ in the limit of $s_z$-conservation, which in our 2D fragile topological model [SEq.~\eqref{eq:H_2d_fragile_original}] corresponds to $v_{M_{z}} = 0$. 
There is spectral flow as a function of $\Phi$, with two bands crossing the bulk gap with positive slope, and two bands crossing the bulk gap with negative slope.
For small values of the mode velocity $\left|\frac{d E}{ d \Phi}\right|$, there will generically be four crossing points between modes with positive and negative slope.
The two crossings indicated by the red arrows at generic values of $\Phi$ are protected by $s_z$-conservation, and the two indicated by the black arrows at $\Phi = \pi$ are protected by time-reversal symmetry due to Kramers theorem.
The slope $\left|\frac{d E}{ d \Phi}\right|$ of the gap-crossing modes is also indicated. 
(c) shows a schematic of the energy spectrum of the 2D sample in (a) with infinite size along $x$ and $y$ and a small hole as a function of $\Phi$ when $s_z$ is not conserved, which in our 2D fragile topological model [SEq.~\eqref{eq:H_2d_fragile_original}] corresponds to $v_{M_{z}} \neq 0$.
There are two crossings indicated by the black arrows at $\Phi = \pi$ that are protected by $\mathcal{T}$ symmetry.
In other words, the mid-gap localized states at $\Phi = \pi$ remain doubly degenerate due to Kramers theorem.
In contrast, the crossings previously indicated by the red arrows in (b) are now gapped since $s_{z}$ is not conserved due to a nonzero $v_{M_{z}}$ in the Hamiltonian [SEq.~\eqref{eq:H_2d_fragile_original}].
In particular, the gap is proportional to $|v_{M_{z}}|$ for small  $|v_{M_{z}}|$.
A Landau-Zener transition~\cite{landau1932theorie,zener1932non} from the low energy filled state to an excited state is also indicated, whose probability is described in SEq.~\eqref{eq:Landau_Zener_P}.
}\label{fig:flux_insertion_schematics}
\end{figure}

We now discuss how this spin-stable topology with $|C^{s}_{\gamma_1}|=4$ manifests in the response of the system to external electromagnetic fields. 
We begin with the limit where $v_{M_{z}} = 0$ [see SEq.~\eqref{eq:H_2d_fragile_original}]. 
In this case, our model in SEq.~\eqref{eq:H_2d_fragile_original} retains mirror reflection $M_z$ symmetry and also $s_{z}$-conservation. 
Hence in this limit the Hamiltonian characterizes a stable mirror topological crystalline insulator (TCI) and quantum spin Hall phase with mirror Chern number $2$ and
$|C^{s}_{\gamma_1}|=4$~\cite{murakami2004spin,bernevig2006quantum}. 
This implies the presence of two pairs of fully-$s_z$-polarized helical edge states~\cite{wieder2020strong}. 
Here, ``fully-$s_z$-polarized'' means that the energy eigenstates are also eigenstates of $s_z$.
A schematic depiction of the helical states at one edge is shown in SFig.~\ref{fig:flux_insertion_schematics}(b) if we regard $\Phi$ in the $x$-axis as the crystal momentum parallel to the edge.
This will lead to a quantized spin Hall conductivity $|\sigma^s_{H}| = |C^s_{\gamma_1}| \times |e|/4\pi = |e|/\pi$ when an external electric field is applied parallel to this 2D system~\cite{sheng2006spinChern}, consistent with SEq.~\eqref{eq:spinHall}.

In addition to an in-plane electric field, we may also adiabatically thread a $U(1)$ magnetic flux $\Phi(t)$ as a function of time through the 2D system to probe the nontrivial bulk topology, see the setup as schematically depicted in SFig.~\ref{fig:flux_insertion_schematics}(a).
When $\Phi = 0$ and $\Phi = \pi$, the system has time-reversal symmetry while time-reversal is broken at generic values of $\Phi \neq 0$ and $\Phi \neq \pi$.
The corresponding energy spectrum of the 2D sample as a function of $\Phi$ when $s_z$ is conserved is shown in SFig.~\ref{fig:flux_insertion_schematics}(b).
Due to $|C_{\gamma_1}^{s}|=4$, there are four gap-crossing modes, corresponding to single-particle bound states around the magnetic flux.
Two of the modes have positive slopes and the other two have negative slopes.
Modes with opposite slopes will have opposite $s_z$ eigenvalues as a consequence of time-reversal symmetry.
Note that although time-reversal symmetry is broken when the $U(1)$ magnetic flux $\Phi \neq 0$ and $\Phi \neq \pi$, for a model with $s_z$ conservation when $\Phi =0$, the Hamiltonian continues to commute with the $s_z$ operator when $\Phi \neq 0$.
Therefore, all the energy eigenstates in SFig.~\ref{fig:flux_insertion_schematics}(b) are eigenstates of $s_z$.

Importantly, the crossings between the bands at generic $\Phi$ indicated by the red arrows in SFig.~\ref{fig:flux_insertion_schematics}(b) are protected by $s_z$ conservation, and the crossings at $\Phi = \pi$ indicated by black arrows are protected by $\mathcal{T}$.
As we gradually increase the flux $\Phi(t)$, there will be a tangential electric field induced along the azimuthal direction around the flux $\Phi(t)$.
Since we have $s_z$ conservation, $|C_{\gamma_1}^{s}|=4$ indicates that the magnetic flux induces radial currents of spin-up and spin-down electrons with same magnitudes but opposite signs.
Applying the same computation performed in SRef.~\cite{qi2008spincharge} at $\Phi = \pi$, the many-body ground state in the vicinity of the inserted flux may be described as a spinon bound state with total electric charge $Q=0$ and total spin $|S_z| = 2(\hbar/2)$~\cite{qi2008topological,AshvinFlux,AdyFlux,Wormhole_3D_TI_Franz,WormholeNumerics,MirlinFlux,CorrelatedFlux,Vlad2D,TynerFlux1,TynerFlux2,TynerFlux3}. 
Note that the accumulated spin in the vicinity of the flux (measured in the standard units of $\hbar/2$) is equal to \emph{half} the spin Chern number, consistent with the fact that $\Phi=\pi$ corresponds to half of a flux quantum. 
We also note that, correspondingly, if we were to instead insert a (fictitious or pseudo-) magnetic field that takes opposite signs in opposite spin sectors (\emph{i.e.} a spin flux), the many-body ground state in the vicinity of the flux would instead be a chargeon with zero total spin and charge $|Q|=2$~\cite{qi2008spincharge} (measured in the units of electron charge $e$).

This can also be understood as follows. 
We start from the ground state with all states in the valence continuum filled, and we fix the electron number so that the infinite system has total charge $Q=0$ (including the underlying positive ions).
If we now increase the flux $\Phi$ from $0$ to $\pi$, two states will appear at the mid-gap through the two gap-crossing modes with positive slopes.
Notice that these two states will have $s_z$ eigenvalues both equal to $+\hbar/2$ or $-\hbar/2$, depending on the sign of $C_{\gamma_1}^{s}$ (recall that we assume $|C_{\gamma_1}^{s}|=4$).
At $\Phi = \pi$, one state of the doubly degenerate Kramers pair at higher mid-gap energy will be filled, and there will be another filled state of the doubly degenerate Kramers pair at lower mid-gap energy. 
And both of the filled states have the same $s_z$ eigenvalues, which implies that the total spin of these two filled states will be $|S_z| = \hbar$.
Due to the time-reversal symmetry at $\Phi=\pi$, the valence continuum will have a total spin equal to zero.
Therefore, the spinon bound state around the magnetic flux when $\Phi = \pi$ will have total electric charge $Q = 0$ and total spin $|S_z| = \hbar$.
We then conclude that when $s_z$ is conserved, we expect to see quantized spin response.
In response to an external electric field parallel to the 2D system we have quantized spin Hall conductivity $|\sigma_{H}^s| = |e|/{\pi}$ [SEq.~\eqref{eq:spinHall}]. 
By threading a $U(1)$ magnetic flux $\Phi$ we will obtain a spinon bound state around the flux with total electric charge $Q=0$ and total spin $|S_z| = \hbar$ when $\Phi = \pi$. 
We can then ask how this response effect changes when $s_z$ is no longer conserved.

When $s_{z}$ is not conserved, which in our 2D fragile topological insulator model [SEq.~\eqref{eq:H_2d_fragile_original}] occurs when $v_{M_{z}} \neq 0$, the metallic edge states will in general be gapped due to the hybridization between fully-$s_z$-polarized edge states with opposite $s_z$ eigenvalues and opposite Fermi velocities.
For example, a schematic 1D band structure is shown in SFig.~\ref{fig:flux_insertion_schematics}(c) if we regard the $x$-axis as the momentum parallel to the edge.
In such a case, the entire system is an insulator in a ribbon geometry, including both the bulk and edge. 
In other words, when an external electric field is applied parallel to the 2D system, there will be no charge flows that could simultaneously carry spin currents.
Although there can be other mechanisms~\cite{Monaco2022shc_nonconserved_spin}, for example the local spin flipping process, that can induce spin currents without charge currents, we expect that the spin Hall conductivity $|\sigma^s_{H}|$ will no longer be quantized, as discussed in SN~\ref{sec:general_properties_of_winding_num_of_P_pm_Wilson}. 
Note that if the modes have sufficiently large velocities (slopes $|\frac{dE}{d\mathbf{k}}|$ in the energy-momentum dispersion relation), then it is possible that there are no crossings between bands with positive slope and bands with negative slope within the bulk gap. 
In this case the spectrum would be insensitive to small nonzero values of $v_{M_{z}}$. 
We do not consider this case here, as it can always be deformed to the situation in SFig.~\ref{fig:flux_insertion_schematics}(c) without closing an energy gap or a spin gap. 

On the other hand, the spin-stable topology may still be explored by threading a $U(1)$ magnetic flux $\Phi$ through the system [see SFig.~\ref{fig:flux_insertion_schematics}(a)]. 
As above, we consider the case where the velocity (slope $|\frac{dE}{d\Phi}|$ of the energy-$\Phi$ relation) of modes is low enough that there are crossings within the bulk gap as shown in SFig.~\ref{fig:flux_insertion_schematics}(b). 
Since $s_z$ is not conserved, the crossing between the counterpropagating modes indicated by the red arrows in SFig.~\ref{fig:flux_insertion_schematics}(b) will in general be gapped, with a gap size proportional to $|v_{M_{z}}|$ [see SFig.~\ref{fig:flux_insertion_schematics}(c)].
However, the doubly-degenerate localized states around the $U(1)$ flux at $\Phi = \pi$ indicated by the black arrows in SFig.~\ref{fig:flux_insertion_schematics}(b) will still be doubly-degenerate, as indicated by the black arrows in SFig.~\ref{fig:flux_insertion_schematics}(c).
This is because when $\Phi = \pi$, the system has time-reversal symmetry and the doubly-degenerate states at $\Phi = \pi$ is protected due to Kramers theorem.
If $|v_{M_{z}}|$ is small enough, we expect that the doubly-degenerate states, both at higher and lower energies, will consist of one state with a $z$-component spin angular momentum $\langle s_{z} \rangle \lesssim \hbar/2$ and another state with $\langle s_{z} \rangle \gtrsim -\hbar/2$.

Let us now consider threading a time-dependent flux $\Phi(t) = 2\pi t/T$.
The system is now described by three energy scales (with $\hbar = 1$):
\begin{enumerate}
	\item $\Delta$, the bulk energy gap,
	\item $|v_{M_{z}}|$, strength of $s_{z}$-conservation-breaking, which by our analysis in SN~\ref{sec:pspperturbation} controls the size of the spin gap,
	\item $\dot{\Phi} = d \Phi / dt = 2\pi / T$, the rate of flux insertion.
\end{enumerate}
In the following discussion, we will assume that $|v_{M_{z}}| \ll \Delta$ and $\dot{\Phi} \ll \Delta$ such that the flux insertion analysis can be carried out using SFig.~\ref{fig:flux_insertion_schematics}(c) where the bulk energy gap remains open, and the value of the bulk energy gap with $v_{M_z} \neq 0$ is close to its value with $v_{M_z}=0$.
We will start from the ground state with all states in the valence continuum filled at $\Phi = 0$, and we fix the electron number so that the infinite system has total charge $Q=0$ (including the underlying positive ions).
If $\dot{\Phi} = 2\pi /T \ll |v_{M_{z}}|$, when $t = T/2$ such that $\Phi = \pi$, 
we would have both the states in the doubly degenerate Kramers pair at the lower mid-gap energy filled. 
Since $\Phi = \pi$ preserves time-reversal symmetry, the filled valence continuum will have total spin equal to zero.
Together with the filled doubly degenerate states, which also have total spin equal to zero as the degenerate states are related to each other by a spinful time-reversal symmetry, the many-body ground state in the vicinity of the flux at $\Phi = \pi$ may be described as a bound state that carries a total charge $Q=0$ and a total spin $S_z = 0$.

Now suppose that $\dot{\Phi} = 2\pi /T \gg |v_{M_{z}}|$.
In this case, a Landau-Zener transition~\cite{landau1932theorie,zener1932non} from the filled state to the excited state [see SFig.~\ref{fig:flux_insertion_schematics}(c)] may occur when the filled state approaches the avoided crossing. 
The probability of such a transition is given by~\cite{Rubbmark1981LandauZener}
\begin{equation}
	P \sim \exp{ -2\pi \frac{1}{\hbar \left|2 \frac{d E}{d \Phi} \right|} \frac{|v_{M_{z}}|^2}{\dot{\Phi}} } = \exp{-\frac{T}{2 \hbar} \frac{|v_{M_{z}}|^2}{\left| \frac{dE}{d\Phi} \right|}}, \label{eq:Landau_Zener_P}
\end{equation}
where we have restored factors of $\hbar$, and where $\frac{d E}{d \Phi}$ is the slope of the gap-crossing mode when $v_{M_{z}} = 0$ indicated in SFig.~\ref{fig:flux_insertion_schematics}(b).
Since we have assumed $\dot{\Phi} = 2\pi /T \gg |v_{M_{z}}|$, the probability in SEq.~\eqref{eq:Landau_Zener_P} will be $P \lesssim 1$, and in such a case we in general expect that the Landau-Zener transition will occur.
Therefore, at $t = T/2$ ($\Phi = \pi$), there will be one state filled in the doubly degenerate Kramers pair at the higher mid-gap energy and another state filled in the doubly degenerate Kramers pair at the lower mid-gap energy.
Importantly, the spin expectation values $\langle s_z \rangle$ of these two filled mid-gap states will have the same signs and will both individually satisfy $|\langle s_z \rangle| \lesssim \hbar /2$.
Again, at $\Phi = \pi$ time-reversal symmetry requires the valence continuum to have total spin equal to zero.
Therefore, if we thread the $U(1)$ magnetic flux $\Phi(t)$ with $\dot{\Phi} = 2\pi /T \gg |v_{M_{z}}|$, the many-body ground state at $\Phi =\pi$ may be described by a spinon bound state with
total charge $Q= 0$ and total spin $|S_z| \lesssim \hbar$ (\emph{i.e.} less than or equal to 2 in the units of $\hbar/2$).

We hence conclude that when $s_z$ is not conserved, although the spin Hall conductivity may in principle be decreased to zero, if we thread a $U(1)$ magnetic flux $\Phi(t)$ at a rate much greater than the strength of $s_{z}$-conservation-breaking, we still expect to observe a spinon bound state with nearly spin-$\hbar$ when $\Phi = \pi$. 

{For completeness, we note that if in the $s_z$-conserving limit the slopes $\left| \frac{dE}{d\Phi} \right|$ of the midgap modes are large enough, then it can be the case that there are no crossing points between modes with opposite slopes within the bulk gap.
In this case, if we weakly break the $s_z$-conservation, there will still be two chiral midgap modes with positive slopes and two chiral midgap modes with negative slopes.
If we then insert a magnetic flux $\Phi$, there will in general be no bound states at $\Phi = \pi$.
Instead, we will observe robust counterpropagating edge states circulating around the small hole as a function of $\Phi$.
}

As such, we have established that the robust spin-stable topology indicated in SFig.~\ref{fig:P2_wilson_loop_2d_fragile} can have a physical consequence. 
In fact, such a robustness persists in the presence of a coupling to trivial atomic orbitals such as those in SEq.~(\ref{eq:V_C_coupled_to_HF}).
This is demonstrated in SFig.~\ref{fig:P6_wilson_loop_2d_fragile}(c,d), in which the $\mp 2$ winding number in the $k_y$-directed $P_{\pm}$-Wilson loop eigenphases $\{ \gamma_{1,j}^{\pm}(k_x) \}$ persists in the $P_6$ occupied space.
According to the sign convention introduced in SEq.~\eqref{eq:partial_chern_def}, this indicates that the partial Chern numbers of the occupied space still remain $C^\pm_{\gamma_1} = \pm 2$ after the addition of trivial atomic orbitals, under which the occupied space has been enlarged from the bands in $P_2$ to those in $P_6$.
We note that unlike previously for the 2D TI model analyzed in SFig.~\ref{fig:2D-strong-TI-wilson-loop-results}, the $P_{\pm}$-Wilson loop spectra for the fragile phase studied in this section [SFig.~\ref{fig:P6_wilson_loop_2d_fragile}(c,d)] is not a straightforward decomposition of the $P$-Wilson loop spectrum [SFig.~\ref{fig:P6_wilson_loop_2d_fragile}(a)].  This occurs precisely because the $P$-Wilson loop spectrum need not always wind when the $P_{\pm}$-Wilson loops wind, which is particularly well exemplified by the trivialized fragile phase analyzed in this section.

To sum up, although the occupied space of a fragile TI becomes trivial after including bands induced from trivial atomic orbitals, it is possible that each of the spin-resolved $\pm$ eigenspaces of $P s P$ have stable nontrivial \emph{spin-resolved} topology.
Importantly, such a {\it spin-resolved stable nontrivial topology} can contribute to experimentally measurable quantities such as spin Hall conductivity~\cite{sheng2006spinChern} and bound states carrying nonzero total spin around a $U(1)$ magnetic $\pi$ flux~\cite{qi2008spincharge}. 
This provides a crucial example demonstrating that, beyond the simplest 2D Chern and 3D axion insulating phases~\cite{qi2006topological,konig2008quantum,sinova2015spin}, stable electronic band topology \emph{does not} in general uniquely determine the spin-electromagnetic response of an insulating system. 
Instead, a finer treatment is required, such as the gauge-invariant spin-resolved methods introduced in this work. 

\subsection{Spin-Resolved Wilson Loops and the Spin Entanglement Spectrum}\label{sec:spin_entanglement_spectrum}

Recall that for the usual $P$-Wilson loop, there is a correspondence between the $P$-Wilson loop spectrum and the bipartite entanglement spectrum. 
In particular, SRefs.~\cite{fidkowski2011model,taherinejad2014wannier} argued that in a band insulator the spectrum of the $P$-Wilson loop in the $\widehat{\mathbf{G}}_i$ direction is the same as the spectrum of the projected position operator $P\mathbf{x}\cdot\mathbf{G}_iP$, and that this can be continuously deformed to the entanglement spectrum  defined as the spectrum of the single-particle correlation matrix $\Theta(\mathbf{x}\cdot\mathbf{G}_i)P\Theta(\mathbf{x}\cdot\mathbf{G}_i)$, where $\Theta(\mathbf{x}\cdot\mathbf{G}_i)$ is zero for states with $\mathbf{x}\cdot\mathbf{G}_i<0$, and $1$ otherwise.

The argument of SRef.~\cite{fidkowski2011model}, relying only on the geometric properties of the projection operator $P$, applies straightforwardly to our spin projection operators $P_\pm$ onto the upper/lower-half of the spin spectrum provided a spin gap is open. 
Concretely, the $P_\pm$-Wilson loops in the $\widehat{\mathbf{G}}$ direction are isospectral to the \emph{spin-projected} position operators $P_\pm \mathbf{x}\cdot \mathbf{G} P_\pm$. 
Following the homotopy arguments of SRef.~\cite{fidkowski2011model}, we can adiabatically deform $P_\pm \mathbf{x}\cdot{\mathbf{G}} P_\pm$ to the \emph{spin entanglement spectrum} given by the spectrum of $\Theta(\mathbf{x}\cdot{\mathbf{G}})P_\pm\Theta(\mathbf{x}\cdot{\mathbf{G}})$. 
As a consequence, nontrivial spectral flow in the $P_\pm$-Wilson loop implies nontrivial spectral flow in the spin entanglement spectrum, provided that the entanglement cut $\mathbf{x}\cdot\mathbf{G}>0$ does not break any symmetry that protects the spectral flow. 
This represents a generalization of the bulk-boundary correspondence to the spin spectrum and the spin entanglement spectrum.

\begin{figure}[ht]
\includegraphics[width=0.4\textwidth]{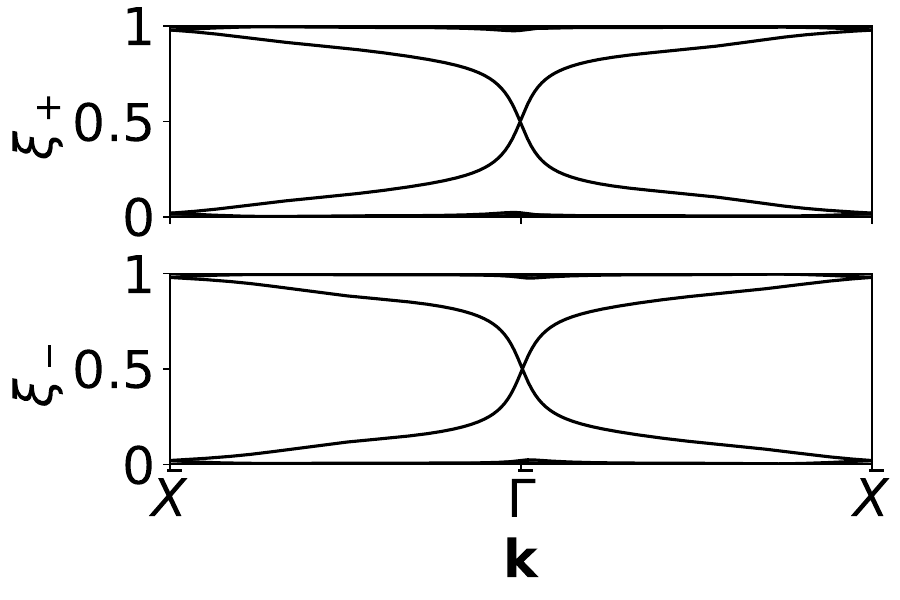}
\caption{Spin entanglement spectrum $\xi_\pm$ for the 2D strong topological insulator considered in SN~\ref{sec:2D-spinful-TRI-system}. 
The $P_+$ entanglement spectrum is shown on the top panel, and the $P_-$ entanglement spectrum is shown on the bottom panel. 
Each entanglement spectrum shows a single gap-traversing chiral mode per boundary of the entanglement region, consistent with the fact that this model has partial Chern numbers $C^\pm_{\gamma_1}$ equal to $\mp1$.}\label{fig:2d-ti-ent-spec}
\end{figure}

As a concrete example, let us consider first the model for a two-dimensional topological insulator presented in SN~\ref{sec:2D-spinful-TRI-system}. 
The spectra of the $P_+$- and $P_-$-Wilson loops for this model were discussed in SN~\ref{sec:2D-spinful-TRI-system} and shown in SFig.~\ref{fig:2D-strong-TI-wilson-loop-results}. 
This model has spin Chern number $C^s_{\gamma_1}$ equal to $-2$, which is reflected in the $P_{\pm}$-Wilson loop eigenphases (SFig.~\ref{fig:2D-strong-TI-wilson-loop-results}). 
In particular, we demonstrated that the $P_+$-Wilson loop spectrum winds one time with negative slope as a function of $k_x$, while the $P_-$-Wilson loop spectrum winds one time with positive slope as a function of $k_x$. 
We thus expect to see spectral flow in the $P_\pm$ entanglement spectrum, with one mode of negative chirality in the $P_+$ entanglement spectrum, and one mode of positive chirality in the $P_-$ entanglement spectrum at each boundary of the entanglement cut. 
To verify this, we numerically compute the spectrum of $\Theta(\mathbf{x}\cdot{\mathbf{G}})P_\pm\Theta(\mathbf{x}\cdot{\mathbf{G}})$ on a cylinder with circumference $N=60$ unit cells in the $y$-direction. 
We partition the cylinder into two regions: region $A$ has $0\le y< 30$, and region $B$ has $30\le y \le 59$. 
We then diagonalize $\Theta(y\in A)P_\pm\Theta(y\in A)$, where $\Theta(y\in A)$ is $1$ if $y\in A$, and $0$ otherwise. 
We show the results in SFig.~\ref{fig:2d-ti-ent-spec}. Since region $A$ has two boundaries, we see a pair of modes crossing the gap in each of the $P_+$ and $P_-$ entanglement spectra. 
Modes with opposite slope correspond to states localized at opposite boundaries of the entanglement region. 
Each entanglement spectrum has a spectral flow of one mode per boundary, consistent with the winding of the $P_\pm$-Wilson loops.

\begin{figure}[ht]
\includegraphics[width=0.4\textwidth]{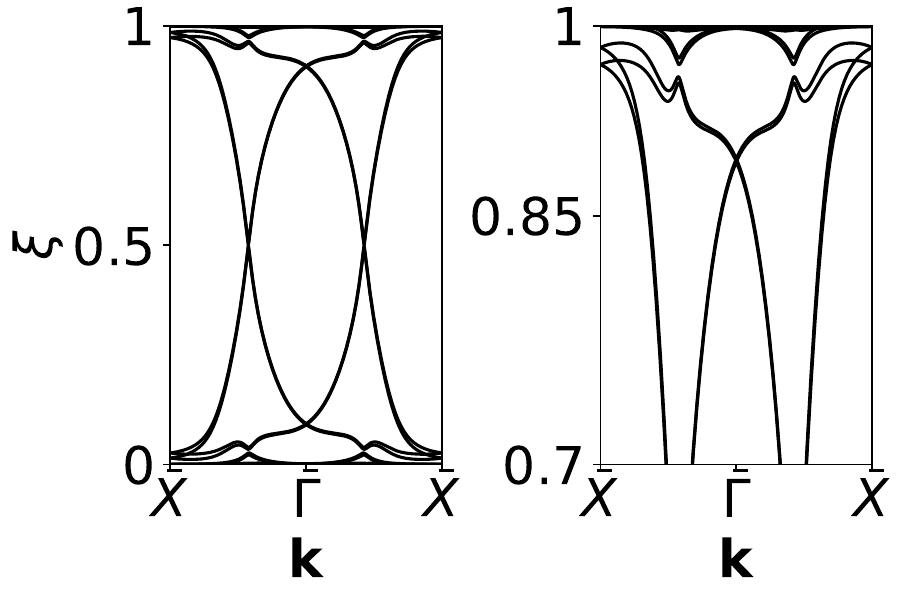}
\caption{Entanglement spectrum $\xi$ for the 2D fragile topological insulator considered in SN~\ref{sec:spin_stable_topology_2d_fragile_TI}. 
The left panel shows the full range of the entanglement spectrum. 
On the right, we zoom in on the range $0.7\le\xi\le 1$. 
We see that there is no spectral flow in the entanglement spectrum, consistent with the fact that the $P$-Wilson loop does not wind.}\label{fig:fragile_model_ent_spec}
\end{figure}

\begin{figure}[ht]
\includegraphics[width=0.4\textwidth]{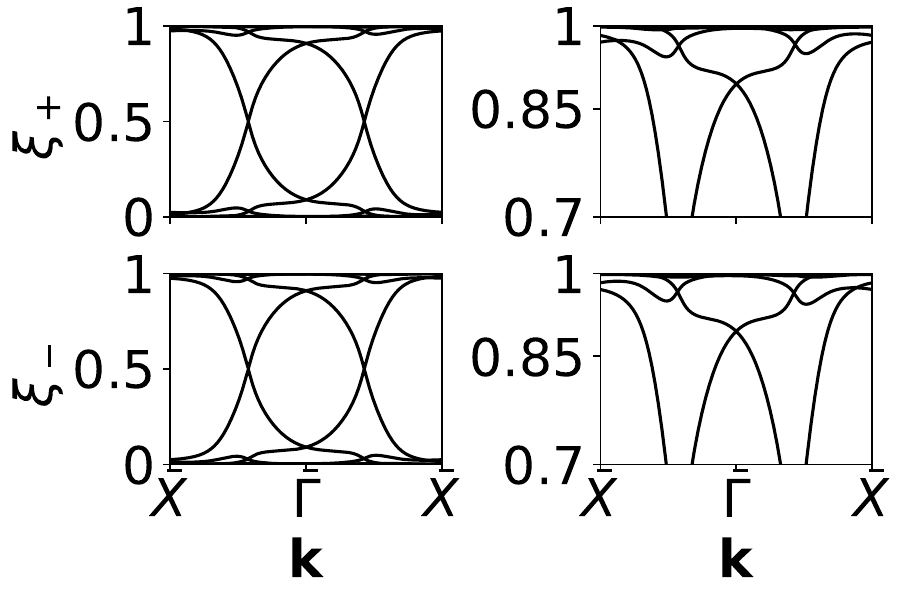}
\caption{Spin entanglement spectrum $\xi_\pm$ for the 2D fragile topological insulator considered in SN~\ref{sec:spin_stable_topology_2d_fragile_TI}. 
The $P_+$ entanglement spectrum is shown on the top, and the $P_-$ entanglement spectrum is shown on the bottom. 
On the right, we zoom in on the range $0.7\le\xi_\pm\le 1$. 
Each entanglement spectrum shows two gap-traversing chiral modes per boundary of the entanglement region, consistent with the fact that this model has partial Chern numbers $C^\pm_{\gamma_1}$ equal to $\pm2$.}\label{fig:fragile_model_spin_ent_spec}
\end{figure}

As a more interesting example, let us consider the model introduced in SRef.~\cite{wieder2020strong} and analyzed in SN~\ref{sec:spin_stable_topology_2d_fragile_TI} for a two-dimensional insulator with partial Chern numbers $C^{\pm}_{\gamma_1} = \pm 2$ and hence spin Chern number $C^{s}_{\gamma_1}=+4$. 
This model is a fragile topological insulator, and hence has no stable winding in the $P$-Wilson loop spectrum, and no gapless surface states. 
In SFig.~\ref{fig:fragile_model_ent_spec} we also verify that the ordinary entanglement spectrum computed from $\Theta(\mathbf{x}\in A)P\Theta(\mathbf{x}\in A)$ (with the region $A$ defined as above) has no spectral flow. 
Turning to the spin entanglement spectrum, however, we see in SFig.~\ref{fig:fragile_model_spin_ent_spec} that the spectra of $\Theta(\mathbf{x}\in A)P_\pm\Theta(\mathbf{x}\in A)$ each feature two chiral modes per boundary of the entanglement region, consistent with the winding of the $P_{\pm}$-Wilson loop eigenphases in SFig.~\ref{fig:P6_wilson_loop_2d_fragile}. 
We thus see that by the bulk-boundary correspondence for the spin spectrum, $C^{\pm}_{\gamma_1}=\pm 2$ implies that the $P_\pm$-Wilson loops wind twice, which implies the existence of two chiral modes per boundary in the spin entanglement spectrum.

\begin{figure}[ht]
\includegraphics[width=0.4\textwidth]{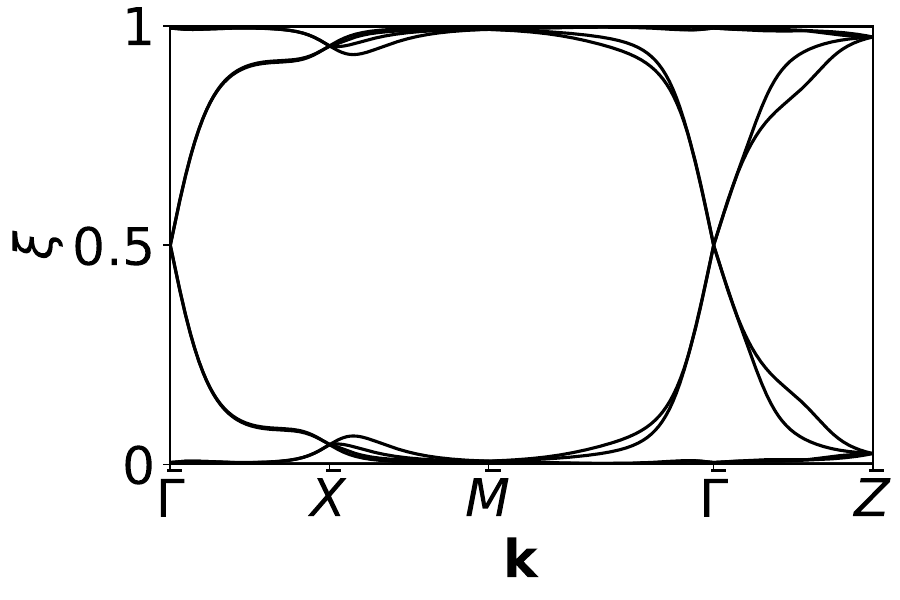}
\caption{Eigenvalues $\xi$ of the entanglement spectrum for the 3D TI model with broken inversion symmetry discussed in SN~\ref{appendix:3D-TI-with-and-without-inversion} and in SFig.~\ref{fig:inversion-breaking-3D-TI}. 
The entanglement spectrum shows a protected Dirac crossing at $\bar{\Gamma}$, with nontrivial spectral flow, in correspondence with the gapless surface states of a 3D TI.}\label{fig:3d_TI_ent_spec}
\end{figure}

As a final example, let us consider the spin entanglement spectrum for a 3D strong topological insulator. 
We consider the model for a 3D TI with broken inversion symmetry given by SEq.~\eqref{eq:3D_TI_model} with the additional inversion-symmetry breaking term [SEq.~\eqref{eq:inversion_breaking_terms_for_3D_TI}]. 
The bulk, slab, and spin band structures, as well as the $P$- and $P_\pm$-Wilson loop eigenphases for this model were shown in SFig.~\ref{fig:inversion-breaking-3D-TI}. 
The helical winding of the $P$-Wilson loop in the $k_z=0$ plane coupled with the trivial winding of the $P$-Wilson loop in every other $k_z$ plane implies that the ordinary entanglement spectrum will feature a protected Dirac crossing at $\bar{\Gamma}$ with $\xi=0.5$ on each entanglement boundary, with nontrivial helical spectral flow. 
This is indeed confirmed in SFig.~\ref{fig:3d_TI_ent_spec}, where we see that the spectrum $\{\xi\}$ of $\Theta(y) P \Theta(y)$ is homotopic to the slab spectrum in SFig.~\ref{fig:inversion-breaking-3D-TI}(c). 

\begin{figure}[ht]
\includegraphics[width=0.4\textwidth]{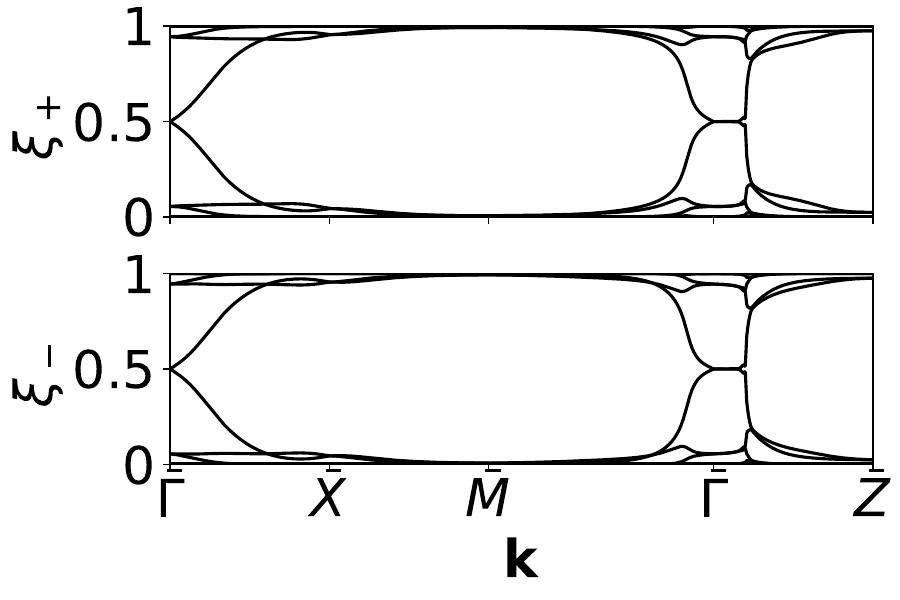}
\caption{Eigenvalues $\xi_\pm$ of the spin entanglement spectrum for the 3D TI model with broken inversion symmetry discussed in SN~\ref{appendix:3D-TI-with-and-without-inversion} and in SFig.~\ref{fig:inversion-breaking-3D-TI}. 
Each entanglement spectrum shows a ``spin Fermi arc'' originating at $\bar{\Gamma}$ and extending along the $\bar{\Gamma}-\bar{Z}$ line, in correspondence with the gapless surface states of a 3D Weyl semimetal.}\label{fig:3d_TI_spin_ent_spec}
\end{figure}

\begin{figure}[ht]
\includegraphics[width=0.4\textwidth]{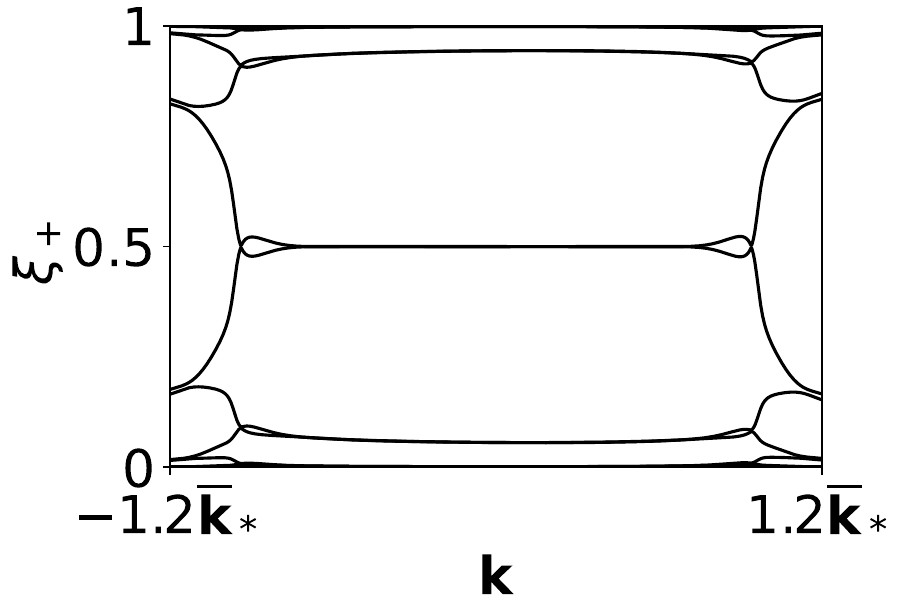}
\caption{Eigenvalues $\xi_+$ of the spin entanglement spectrum for the 3D TI model with broken inversion symmetry, plotted along a line connecting the surface projection of the spin-Weyl points. 
The flat spin Fermi arcs at $\xi_+=0.5$ are visible over the majority of the plotted range. 
Later, using the example of $\beta$-MoTe$_2$ in SN~\ref{app:mote2} we will show that spin Fermi arcs manifest as topological surface Fermi arcs in the physical energy spectrum of a spin-Weyl semimetal in a strong Zeeman field.}\label{fig:3d_TI_spin_ent_spec_between_nodes}
\end{figure}

This result can be interestingly contrasted with the spin entanglement spectrum of a 3D TI. 
Since the occupied energy bands in every constant-$k_z$ plane between the spin-Weyl nodes at $\pm\mathbf{k}_*=\pm(0,0.46\pi,-0.21\pi)$ have spin Chern number $C^s_{\gamma_1}=2$, we expect the $\Theta(y)P_\pm\Theta(y)$ entanglement spectrum to be isospectral to the boundary states (and Wilson loop) of a Weyl semimetal. 
In particular, we expect to find a set of ``spin Fermi arcs'' in the spectrum of $\Theta(y)P_\pm\Theta(y)$ between the projections $\pm\overline{\mathbf{k}}_*=\pm(0,-0.21\pi)$ of the spin-Weyl nodes onto the Brillouin zone of the entanglement cut.
In SFig.~\ref{fig:3d_TI_spin_ent_spec} we show the  spectrum $\xi_\pm$ of $\Theta(y)P_\pm\Theta(y)$ for the 3D TI model, where a Fermi arc state can be seen in the line of $\xi_\pm=0.5$ eigenvalues starting at $\bar{\Gamma}$ and proceeding along the $\bar{\Gamma}-\bar{Z}$ line. 
To make this clearer, we show in SFig.~\ref{fig:3d_TI_spin_ent_spec_between_nodes} the $P_+$ spin entanglement spectrum along a straight path between $-1.2\overline{\mathbf{k}}_*$ and $1.2\overline{\mathbf{k}}_*$, corresponding to the surface projection of the path between spin-Weyl nodes shown in SFig.~\ref{fig:inversion-breaking-3D-TI}(d), which shows the spin Fermi arc clearly. 
This thus demonstrates that, according to the generalized bulk-boundary correspondence, the spin entanglement spectrum for a 3D system with spin-Weyl nodes displays spin Fermi arcs connecting the surface projections of the spin-Weyl points.

To conclude, we have shown that the $P_\pm$-Wilson loop spectrum is homotopic to the spin entanglement spectrum, defined as the eigenvalues of the restriction of $P_\pm$ to half the system in position space. 
Our construction is similar in spirit to the symmetry-resolved entanglement spectrum studied for strongly interacting symmetry-protected topological phases in, \emph{e.g.}
SRefs.~\cite{laflorencie2014spin,goldstein2018symmetry,azses2020symmetry}. 
It should be noted that the spectrum of $\Theta(\mathbf{x}\cdot \mathbf{G})P_{\pm}\Theta(\mathbf{x}\cdot \mathbf{G})$ differs in important respects from the spin-resolved entanglement spectrum introduced in SRefs.~\cite{fukui2014entanglement,araki2016entanglement,araki2017entanglement}. 
The authors of those works considered an entanglement bipartition not in position space, but in spin space. 
That is, they considered the (nonzero) spectrum of the matrix $C_{\uparrow} = P_\uparrow P P_\uparrow$, with no position space restriction [recall we discussed $C_{\uparrow}$ matrix in SEq.~\eqref{eq:cupdef}]. 
As $C_{\uparrow}$ respects the translation symmetry of Hamiltonian, there is no bulk-boundary correspondence for the notion of spin entanglement spectrum in SRefs.~\cite{fukui2014entanglement,araki2016entanglement,araki2017entanglement}. 
By contrast, we have shown here that the spectrum of $\Theta(\mathbf{x}\cdot \mathbf{G})P_{\pm}\Theta(\mathbf{x}\cdot \mathbf{G})$ is homotopic to the spin-resolved Wilson loop spectrum.

\section{Nested Spin-Resolved Wilson Loops}\label{app:w2}

In this section, we will apply our formalism of spin-resolved topology to higher-order topological phases~\cite{benalcazar2017quantized,benalcazar2017electric,schindler2018higherorder,schindler2018higherordera,po2017symmetry,khalaf2018symmetry,song2018mapping,wieder2018axion}. 
To do so, in SN~\ref{sec:nested_P_Wilson_loop} we will start by reviewing the nested Wilson loop introduced in SRefs.~\cite{benalcazar2017quantized,wieder2018axion,varnava2018surfaces} as a diagnostic of bulk higher-order topology. 
Next, in SN~\ref{sec:nested_P_pm_Wilson_loop} we will define a nested spin-resolved Wilson loop, which we also call nested $P_{\pm}$-Wilson loop and can be understood as the nested Wilson loop computed in the subsets of the $PsP$ [SEq.~\eqref{eq:appendix-def-PsP}] spectrum. 
In SN~\ref{sec:general_properties_nested_ppm} we will discuss general properties of the nested $P_\pm$-Wilson loop, and show how its spectrum can be related to higher-order helical topological invariants described by a set of \textit{spin-stable} invariants that are protected by an energy and a spin gap. 
In particular, we derive that a 3D spin-gapped crystalline insulator with inversion and spinful time-reversal symmetry has a $\mathbb{Z}_2 \times \mathbb{Z}$ spin-resolved topology.
In SN~\ref{app:comparison-spin-stable-and-symmetry-indicated-topology} we show how the spin-resolved topology refines the classification of topological crystalline insulators and revealing the existence of a previously unrecognized bulk invariant, the \emph{partial} axion angle, which can be extracted from the nested spin-resolved Wilson loop spectrum. 
In particular, we will introduce the spin-resolved layer construction which can be deduced from the spin-stable invariants and used to enumerate spinful crystalline insulators, from which we find three different spin-resolved regimes of a helical higher-order topological insulator (HOTI).
Finally, in SN~\ref{sec:numerical-section-of-nested-P-pm} we will apply our formalism to a model of a helical HOTI, paying particular attention to practical considerations that arise in the computation.

 \subsection{\label{sec:nested_P_Wilson_loop}Nested $P$-Wilson Loop}

Before we develop the formalism of a spin-resolved version of nested Wilson loops in SN~\ref{sec:nested_P_pm_Wilson_loop}, let us review the ordinary nested $P$-Wilson loop formalism~\cite{benalcazar2017quantized,benalcazar2017electric,wieder2018axion,wang2019higherorder,wieder2020strong,NonsymmorphicNestedWilson}.
We consider a 2D or 3D lattice model with discrete translation symmetries and denote the projector to the $N_{\mathrm{occ}}$ occupied energy eigenstates at crystal momentum $\mathbf{k}$ by 
\begin{align}
    [P(\mathbf{k})] = \sum_{n=1}^{N_{\text{occ}}} \ket{u_{n,\mathbf{k}}} \bra{u_{n,\mathbf{k}}}. \label{eq:nest_P_appendix_P_energy_bands}
\end{align}
We will assume that in the following construction, the energy eigenvalues of the occupied eigenvectors of the Bloch Hamiltonian $[H(\mathbf{k})]$ are spectrally separate from the unoccupied ones such that $[P(\mathbf{k})]$ is well-defined and smooth as a function of $\mathbf{k}$. 
To form the nested $P$-Wilson loop, we must first calculate the $P$-Wilson loop along direction $\mathbf{G}$; in this section and what follows we take $\mathbf{G}$ to be a primitive reciprocal lattice vector, although the formalism applies for more general reciprocal lattice vectors as well.  
We can then find the eigenvectors of the Wilson loop, allowing us to form a basis of (Fourier-transformed) hybrid Wannier functions from the occupied states. 
We then will construct a projector $[\tilde{P}_\mathbf{G}(\mathbf{k})]$ onto a subset of these hybrid Wannier bands, and compute the eigenvalues for a Wilson loop formed as a product of the $[\tilde{P}_\mathbf{G}(\mathbf{k})]$.

To formalize this process, we begin with the $N_{\mathrm{occ}}\times N_{\mathrm{occ}}$ $P$-Wilson loop matrix $[\mathcal{W}_{1,\mathbf{k},\mathbf{G}}]$ for a holonomy starting at base point $\mathbf{k}$ and going along a straight path to $\mathbf{k}+\mathbf{G}$ where $\mathbf{G}$ is a primitive reciprocal lattice vector [see SEq.~(\ref{eq:P_Wilson_loop_matrix_element}) and surrounding text]. 
To find the eigenvectors of $[\mathcal{W}_{1,\mathbf{k},\mathbf{G}}]$, we must solve
\begin{align}
    [\mathcal{W}_{1,\mathbf{k},\mathbf{G}}] \ket{{\nu}_{j,\mathbf{k},\mathbf{G}}} = e^{i (\gamma_{1})_{j,\mathbf{k},\mathbf{G}}} \ket{{\nu}_{j,\mathbf{k},\mathbf{G}}}, \label{eq:P_nested_Wilson_loop_eig_eqn_1}
\end{align}
where $\{ | {\nu}_{j,\mathbf{k},\mathbf{G}} \rangle | j=1\ldots N_{\mathrm{occ}} \}$ is a set of $N_{\text{occ}}$-component orthonormal eigenvectors such that $\bra{{\nu}_{i,\mathbf{k},\mathbf{G}}}\ket{{\nu}_{j,\mathbf{k},\mathbf{G}}}=\delta_{ij}$. 
Recall from SN~\ref{sec:P_Wilson_loop} that the eigenvalues $(\gamma_1)_{j,\mathbf{k},\mathbf{G}}$ are invariant under shifts $\mathbf{k}\rightarrow\mathbf{k}+\Delta\mathbf{k}$ for $\Delta\mathbf{k}\parallel \mathbf{G}$. 
This means that the eigenvectors $\ket{{\nu}_{j,\mathbf{k}+\Delta\mathbf{k} ,\mathbf{G}}}$ are related to the eigenvectors $\ket{{\nu}_{j,\mathbf{k},\mathbf{G}}}$ by a unitary transformation known as parallel transport~\cite{vanderbilt2018berry,alexandradinata2014wilsonloop}.  
If we define the $P$-Wilson line matrix
\begin{equation}
	[\mathcal{W}_{1,\mathbf{k}' \leftarrow \mathbf{k}}]_{m,n} \equiv \langle u_{m,\mathbf{k}'} | \left( \prod_{\mathbf{q}}^{\mathbf{k}' \leftarrow \mathbf{k}} [P(\mathbf{q})] \right) | u_{n,\mathbf{k}} \rangle, \label{eq:wilsonline}
\end{equation}
where both $m$ and $n$ are the indices of occupied energy eigenvectors, and
\begin{equation}
    \left( \prod_{\mathbf{q}}^{\mathbf{k}' \leftarrow \mathbf{k}} [P(\mathbf{q})] \right)= \lim_{N \to \infty} [P(\mathbf{k}')] [P(\mathbf{k}+\frac{N-1}{N}(\mathbf{k}'-\mathbf{k}))] \cdots [P(\mathbf{k}+\frac{1}{N}(\mathbf{k}'-\mathbf{k}))] [P(\mathbf{k})],
\end{equation}
then the eigenvectors $\ket{{\nu}_{j,\mathbf{k},\mathbf{G}}}$ satisfy~\cite{benalcazar2017electric}
\begin{equation}
	| \nu_{j,\mathbf{k} + \Delta \mathbf{k} ,\mathbf{G}} \rangle = \exp{ -i \frac{(\gamma_{1})_{j,\mathbf{k},\mathbf{G}}}{2\pi} (\Delta \mathbf{k}) \cdot \mathbf{a}  } [\mathcal{W}_{1,\mathbf{k}+\Delta \mathbf{k} \leftarrow \mathbf{k}}] | \nu_{j,\mathbf{k},\mathbf{G}} \rangle. \label{eq:parallel-transport-nu-j-k-G}
\end{equation}
Here $\mathbf{a}$ is the real-space primitive lattice vector dual to the primitive reciprocal lattice vector $\mathbf{G}$ such that $\mathbf{a} \cdot \mathbf{G} = 2\pi$.
We also note that by definition $[\mathcal{W}_{1,\mathbf{k}+\mathbf{G} \leftarrow \mathbf{k}}]$ [SEq.~\eqref{eq:wilsonline}] is equal to $[\mathcal{W}_{1,\mathbf{k},\mathbf{G}}]$ [SEq.~\eqref{eq:P_Wilson_loop_matrix_element}].

We emphasize that in practice the computation of the eigenvectors $\ket{{\nu}_{j,\mathbf{k},\mathbf{G}}}$ requires a bit of care. 
For a given base point we can compute the $P$-Wilson loop matrix $[\mathcal{W}_{1,\mathbf{k},\mathbf{G}}]$ on a discretized $\mathbf{k}$-mesh using SEq.~\eqref{eq:P_Wilson_loop_matrix_element}. 
To account for numerical error introduced by the finite $\mathbf{k}$-mesh spacing,
we then perform a singular value decomposition (SVD) on $[\mathcal{W}_{1,\mathbf{k},\mathbf{G}}]$ to obtain $[\mathcal{W}_{1,\mathbf{k},\mathbf{G}}] = U S V^{\dagger}$ where $U$ and $V$ are unitary matrices while $S$ is a real and diagonal matrix with non-negative diagonal elements; we then 
redefine $[\mathcal{W}_{1,\mathbf{k},\mathbf{G}}]$ to be the unitary part $UV^{\dagger}$ of the decomposition~\cite{vanderbilt2018berry}. 
Being a unitary matrix, we can then diagonalize $[\mathcal{W}_{1,\mathbf{k},\mathbf{G}}]$ to obtain SEq.~(\ref{eq:P_nested_Wilson_loop_eig_eqn_1}) where the eigenvalues are of the unimodular form $e^{i (\gamma_{1})_{j,\mathbf{k},\mathbf{G}}}$ with $(\gamma_{1})_{j,\mathbf{k},\mathbf{G}} \in \mathbb{R}$. 
In practice, we use the Schur decomposition $[\mathcal{W}_{1,\mathbf{k},\mathbf{G}}] = Z T Z^{\dagger}$ to ensure that we obtain orthonormal eigenvectors for the bands of $[\mathcal{W}_{1,\mathbf{k},\mathbf{G}}]$, where $Z$ is a unitary matrix and $T$ is an upper-triangular matrix~\cite{horn2012matrix}. 
Since $[\mathcal{W}_{1,\mathbf{k},\mathbf{G}}]$ is unitary, $T$ is diagonal.  
Therefore, we can identify the diagonal elements of $T$ as $e^{i (\gamma_{1})_{j,\mathbf{k},\mathbf{G}}}$ and the corresponding columns of $Z$ as $\ket{{\nu}_{j,\mathbf{k},\mathbf{G}}}$ in SEq.~\eqref{eq:P_nested_Wilson_loop_eig_eqn_1}.

We next use the $\ket{{\nu}_{j,\mathbf{k},\mathbf{G}}}$ to form a set of ``hybrid Wannier states'' as vectors in our $N_\mathrm{sta}$-dimensional Hilbert space as~\cite{benalcazar2017quantized,benalcazar2017electric,wieder2018axion,varnava2018surfaces,varnava2020axion} (recall that $N_\mathrm{sta}$ denotes the number of basis states per unit cell in the Hilbert space of our truncated tight-binding Hamiltonian)
\begin{align}
    \ket{{w}_{j,\mathbf{k},\mathbf{G}}} = \sum_{m=1}^{N_{\text{occ}}} [{\nu}_{j,\mathbf{k},\mathbf{G}}]_{m} \ket{u_{m,\mathbf{k}}}, \label{eq:reexpressing-tilde-nu-as-nu}
\end{align}
where $[{\nu}_{j,\mathbf{k},\mathbf{G}}]_{m}$ is the $m^{\text{th}}$ ($m = 1,\ldots,N_{\text{occ}}$) component of $\ket{{\nu}_{j,\mathbf{k},\mathbf{G}}}$. 
Notice that $\ket{{w}_{j,\mathbf{k},\mathbf{G}}}$ is an $N_{\mathrm{sta}}$-component vector.

To proceed, we now derive the boundary conditions satisfied by $\ket{{w}_{j,\mathbf{k},\mathbf{G}}}$. 
Notice that, upon a shift of the crystal momentum $\mathbf{k} \to \mathbf{k} + \mathbf{G}'$ where $\mathbf{G}'$ is a reciprocal lattice vector, the matrix projector $[P(\mathbf{k})]$ [SEq.~\eqref{eq:nest_P_appendix_P_energy_bands}] to the occupied energy bands transforms as
\begin{align}
    [P(\mathbf{k}+\mathbf{G}')] & = \sum_{n \in \text{occ}} \ket{u_{n,\mathbf{k}+\mathbf{G}'}} \bra{u_{n,\mathbf{k}+\mathbf{G}'}} \\
    & = \sum_{n \in \text{occ}} [V(\mathbf{G}')]^{-1}\ket{u_{n,\mathbf{k}}} \bra{u_{n,\mathbf{k}}} [V(\mathbf{G}')] \\
    & = [V(\mathbf{G}')]^{-1} [P(\mathbf{k})] [V(\mathbf{G}')], \label{eq:P_k_plus_G_and_P_k}
\end{align}
where we have used $| u_{n,\mathbf{k}+\mathbf{G}'} \rangle = [V(\mathbf{G}')]^{-1} | u_{n,\mathbf{k}} \rangle$ [SEq.~\eqref{eq:bcs}].
Going further, we can use SEq.~\eqref{eq:P_k_plus_G_and_P_k} to show that the product of projectors along a loop starting at the base point $\mathbf{k}+\mathbf{G}'$ and going along a straight path to $\mathbf{k}+\mathbf{G}'+\mathbf{G}$ can be rewritten as
\begin{align}
    \left(\prod_{\mathbf{q}}^{\mathbf{k}+\mathbf{G}'+\mathbf{G} \leftarrow \mathbf{k}+\mathbf{G}'} [P(\mathbf{q})] \right)& = \lim_{N\to \infty} [P(\mathbf{k}+\mathbf{G}'+\mathbf{G})] [P(\mathbf{k}+\mathbf{G}'+\frac{N-1}{N}\mathbf{G})]  \ldots [P(\mathbf{k}+\mathbf{G}'+\frac{1}{N}\mathbf{G})]  [P(\mathbf{k}+\mathbf{G}')] \\
        & = \lim_{N\to \infty} [V(\mathbf{G}')]^{-1}[P(\mathbf{k}+\mathbf{G})] [P(\mathbf{k}+\frac{N-1}{N}\mathbf{G})] \ldots  [P(\mathbf{k}+\frac{1}{N}\mathbf{G})] [P(\mathbf{k})] [V(\mathbf{G}')]  \\
    & = [V(\mathbf{G}')]^{-1} \left(\prod_{\mathbf{q}}^{\mathbf{k}+\mathbf{G} \leftarrow \mathbf{k}} [P(\mathbf{q})] \right) [V(\mathbf{G}')]. \label{eq:P_product_bcs}
\end{align}
Using SEqs.~\eqref{eq:bcs} and \eqref{eq:P_product_bcs}, it follows that upon a shift of the crystal momentum $\mathbf{k} \to \mathbf{k} + \mathbf{G}'$, the matrix elements of the $P$-Wilson loop $[\mathcal{W}_{1,\mathbf{k},\mathbf{G}}]$ [SEq.~\eqref{eq:P_Wilson_loop_matrix_element}] are invariant:
\begin{align}
    [\mathcal{W}_{1,\mathbf{k}+\mathbf{G}',\mathbf{G}}]_{m,n} &= \bra{u_{m,\mathbf{k}+\mathbf{G}'}} [V(\mathbf{G})] \left(\prod_{\mathbf{q}}^{\mathbf{k}+\mathbf{G}'+\mathbf{G} \leftarrow \mathbf{k}+\mathbf{G}'} [P(\mathbf{q})]\right)\ket{u_{n,\mathbf{k}+\mathbf{G}'}} \\
    & = \bra{u_{m,\mathbf{k}}} [V(\mathbf{G}')] [V(\mathbf{G})] [V(\mathbf{G}')]^{-1} \left(\prod_{\mathbf{q}}^{\mathbf{k}+\mathbf{G} \leftarrow \mathbf{k}} [P(\mathbf{q})]\right) [V(\mathbf{G}')][V(\mathbf{G}')]^{-1}\ket{u_{n,\mathbf{k}}}\\
    & = \bra{u_{m,\mathbf{k}}}  [V(\mathbf{G})] [V(\mathbf{G}')] [V(\mathbf{G}')]^{-1} \left(\prod_{\mathbf{q}}^{\mathbf{k}+\mathbf{G} \leftarrow \mathbf{k}} [P(\mathbf{q})]\right) \ket{u_{n,\mathbf{k}}} \\
    & = \bra{u_{m,\mathbf{k}}}  [V(\mathbf{G})] \left(\prod_{\mathbf{q}}^{\mathbf{k}+\mathbf{G} \leftarrow \mathbf{k}} [P(\mathbf{q})]\right) \ket{u_{n,\mathbf{k}}} \\
    & = [\mathcal{W}_{1,\mathbf{k},\mathbf{G}}]_{m,n}. \label{eq:W_1_k_plus_Gp_G_invariant}
\end{align}
In deriving SEq.~\eqref{eq:W_1_k_plus_Gp_G_invariant} we have used the fact that both $[V(\mathbf{G})]$ and $[V(\mathbf{G}')]$ are diagonal matrices and hence they commute, \emph{i.e.}
 \begin{equation}
 [V(\mathbf{G})][V(\mathbf{G}')]=[V(\mathbf{G}')][V(\mathbf{G})].\label{eq:Vgs_commute}
 \end{equation} 
Since the $N_{\mathrm{occ}} \times N_{\mathrm{occ}}$ matrix $[\mathcal{W}_{1,\mathbf{k},\mathbf{G}}]$ [SEq.~\eqref{eq:P_Wilson_loop_matrix_element}] is {\it invariant} upon a shift of $\mathbf{k} \to \mathbf{k} + \mathbf{G}'$, this implies that we can choose 
\begin{align}
    \ket{{\nu}_{j,\mathbf{k}+\mathbf{G}',\mathbf{G}}} = \ket{{\nu}_{j,\mathbf{k},\mathbf{G}}}. \label{eq:bcs-for-tilde-nu-j-k}
\end{align}
for the eigenvectors in SEq.~\eqref{eq:P_nested_Wilson_loop_eig_eqn_1}. 
Notice that the boundary condition in SEq.~\eqref{eq:bcs-for-tilde-nu-j-k} is consistent with the eigenvalue equation in SEq.~\eqref{eq:P_nested_Wilson_loop_eig_eqn_1} and the parallel transport condition in SEq.~\eqref{eq:parallel-transport-nu-j-k-G} if we choose $\mathbf{G}' = \mathbf{G}$.
Due to the boundary condition in SEq.~\eqref{eq:bcs-for-tilde-nu-j-k}, we then have that the  $| w_{j,\mathbf{k},\mathbf{G}} \rangle $ in SEq.~\eqref{eq:reexpressing-tilde-nu-as-nu} satisfy
\begin{align}
    \ket{{w}_{j,\mathbf{k}+\mathbf{G}',\mathbf{G}}} &= \sum_{m=1}^{N_{\text{occ}}} [{\nu}_{j,\mathbf{k}+\mathbf{G}',\mathbf{G}}]_{m} \ket{u_{m,\mathbf{k}+\mathbf{G}'}} \\
    & = \sum_{m=1}^{N_{\text{occ}}} [{\nu}_{j,\mathbf{k},\mathbf{G}}]_{m} [V(\mathbf{G}')]^{-1}\ket{u_{m,\mathbf{k}}} \\
    & = [V(\mathbf{G}')]^{-1} \ket{{w}_{j,\mathbf{k},\mathbf{G}}}, \label{eq:BC_nu_j_k_G}
\end{align}
where we have again used SEq.~\eqref{eq:bcs}.
We then see that $\ket{{w}_{j,\mathbf{k},\mathbf{G}}}$ satisfies the same boundary conditions as the eigenstates of the Bloch Hamiltonian [SEq.~\eqref{eq:bcs}], and so can be regarded as physical states.
In this work, we will call the $\ket{{w}_{j,\mathbf{k},\mathbf{G}}}$ the {\it $\widehat{\mathbf{G}}$-directed $P$-Wannier band basis}~\cite{benalcazar2017quantized,benalcazar2017electric,wieder2018axion}, or simply the {\it $P$-Wannier basis}.

The $\ket{{w}_{j,\mathbf{k},\mathbf{G}}}$  are the single-particle eigenstate of the $P$-Wilson loop matrix $[\mathcal{W}_{1,\mathbf{k},\mathbf{G}}]$ with eigenphase $(\gamma_{1})_{j,\mathbf{k},\mathbf{G}}$ [SEq.~\eqref{eq:P_nested_Wilson_loop_eig_eqn_1}], expressed in terms of the tight-binding basis states. 
As long as there is a spectral gap in the $P$-Wannier bands, namely the eigenphase dispersion $\{(\gamma_{1})_{j,\mathbf{k},\mathbf{G}}|j=1\ldots N_{\mathrm{occ}}\}$ as a function of $\mathbf{k}$, we can further choose a subspace containing $N_{W}$ $P$-Wannier bands
to form the matrix projector 
\begin{align}
    [\widetilde{P}_{\mathbf{G}}(\mathbf{k})] = \sum_{j=1}^{N_{W}} \ket{{w}_{j,\mathbf{k},\mathbf{G}}} \bra{{w}_{j,\mathbf{k},\mathbf{G}}}, \label{eq:P_nested_Wilson_loop_2nd_projector}
\end{align}
where $N_{W} \leq N_{\mathrm{occ}}$, and the subscript $\mathbf{G}$ of $[\widetilde{P}_{\mathbf{G}}(\mathbf{k})]$ indicates that the $P$-Wannier basis states $\ket{{w}_{j,\mathbf{k},\mathbf{G}}}$ are obtained by diagonalizing the $P$-Wilson loop matrix [SEq.~\eqref{eq:P_Wilson_loop_matrix_element}] for a loop taken parallel to $\mathbf{G}$. 
We will refer to the states in the image of $[\tilde{P}_\mathbf{G}]$ as the ``occupied'' $P$-Wannier bands, by analogy to the occupied energy eigenstates in the image of $[P]$~[SEq.~\eqref{eq:nest_P_appendix_P_energy_bands}]. 
Being a projector onto a subspace that is spectrally isolated, we can construct the $N_{W} \times N_{W}$ $\tilde{P}_\mathbf{G}$-Wilson loop matrix along a loop $\mathbf{G}'$ with the matrix elements
\begin{align}
    [{\mathcal{W}}_{2,\mathbf{k},\mathbf{G},\mathbf{G}'}]_{i,j} & = \bra{{w}_{i,\mathbf{k}+\mathbf{G}',\mathbf{G}}}  \left(\prod_{\mathbf{q}}^{\mathbf{k}+\mathbf{G}' \leftarrow \mathbf{k}} [\widetilde{P}_{\mathbf{G}}(\mathbf{q})]\right) \ket{{w}_{j,\mathbf{k},\mathbf{G}}} = \bra{{w}_{i,\mathbf{k},\mathbf{G}}} [V(\mathbf{G}')] \left(\prod_{\mathbf{q}}^{\mathbf{k}+\mathbf{G}' \leftarrow \mathbf{k}} [\widetilde{P}_{\mathbf{G}}(\mathbf{q})]\right) \ket{{w}_{j,\mathbf{k},\mathbf{G}}}, \label{eq:nested_P_Wilson_loop_matrix_def}
\end{align}
where $i$ and $j$ range from $1\ldots N_{W}$, and where we have used the boundary condition [SEq.~\eqref{eq:BC_nu_j_k_G}] for $ \ket{{w}_{j,\mathbf{k},\mathbf{G}}}$. 
We term $[{\mathcal{W}}_{2,\mathbf{k},\mathbf{G},\mathbf{G}'}]$ the \emph{nested $P$-Wilson loop matrix}. 
From SEq.~(\ref{eq:nested_P_Wilson_loop_matrix_def}), we see that by definition, the nested $P$-Wilson loop matrix is the {\it $\tilde{P}_\mathbf{G}$-Wilson loop matrix for the ``occupied'' Wannier bands.} 
We will then denote the unimodular eigenvalues of the nested $P$-Wilson loop matrix $[{\mathcal{W}}_{2,\mathbf{k},\mathbf{G},\mathbf{G}'}]$
as $\exp{i(\gamma_{2})_{j,\mathbf{k},\mathbf{G},\mathbf{G}'}}$, where $j=1,\ldots,N_W$.
We call the set $\{ (\gamma_2)_{j,\mathbf{k},\mathbf{G},\mathbf{G}'} | j=1\ldots N_W \}$ as a function of $\mathbf{k}$ the ``{\it nested} $P$-Wannier bands,'' where $j$ is the band index. 
The eigenphases $\{ (\gamma_2)_{j,\mathbf{k},\mathbf{G},\mathbf{G}'} | j=1\ldots N_W \}$ give the positions of hybrid Wannier functions formed from the $N_W$ ``occupied'' $P$-Wannier bands and localized  along the lattice vector $\mathbf{a}'$ dual to the reciprocal lattice vector $\mathbf{G}'$~\cite{benalcazar2017quantized,benalcazar2017electric,wieder2018axion}.
The eigenphases $(\gamma_{2})_{j,\mathbf{k},\mathbf{G},\mathbf{G}'}$ of the nested $P$-Wilson loop matrix [SEq.~\eqref{eq:nested_P_Wilson_loop_matrix_def}] are also the non-Abelian Berry phases of the $N_{W}$ ``occupied ''$P$-Wannier bands for the closed loop in $\mathbf{k}$-space parallel to $\mathbf{G}'$.
In particular, the eigenphases $(\gamma_2)_{j,\mathbf{k},\mathbf{G},\mathbf{G}'}$ are independent of the momentum component $\mathbf{k} \cdot \mathbf{a}'$.
Note, however, that $(\gamma_2)_{j,\mathbf{k},\mathbf{G},\mathbf{G}'}$ in general depend on the momentum component $\mathbf{k} \cdot \mathbf{a}$ where $\mathbf{a}$ is the primitive real-space lattice vector dual to the primitive reciprocal lattice vector $\mathbf{G}$.
In SN~\ref{appendix:I-constraint-on-nested-P-wilson} and \ref{sec:T-constraint-on-nested-P} we prove the constraints on the values of the nested $P$-Wilson loop eigenphases at different $\mathbf{k}$ points due to inversion and time-reversal symmetries.

We emphasize that in defining the nested Wilson loop in SEq.~\eqref{eq:nested_P_Wilson_loop_matrix_def} it was crucial to have the eigenvectors $\ket{{\nu}_{j,\mathbf{k},\mathbf{G}}}$ be orthonormal in order for $[\widetilde{P}_{\mathbf{G}}(\mathbf{k})]$ defined in SEq.~(\ref{eq:P_nested_Wilson_loop_2nd_projector}) to be a projection matrix. 
Numerically, this was guaranteed by our use of the Schur decomposition.
Importantly, the eigenvectors $\ket{{\nu}_{j,\mathbf{k},\mathbf{G}}}$ obtained numerically via the Schur decomposition only satisfy the parallel transport equation [SEq.~\eqref{eq:parallel-transport-nu-j-k-G}] up to an arbitrary complex phase (or more generally up to an arbitrary unitary transformation between states with equal $(\gamma_1)_{j,\mathbf{k},\mathbf{G}}$.
Nevertheless, since our formalism for computing the non-Abelian (nested) Berry phases makes use of only the matrix projection operators, this numerical ambiguity will not affect our computation of the eigenphases $(\gamma_1)_{j,\mathbf{k},\mathbf{G}}$ and $(\gamma_2)_{j,\mathbf{k},\mathbf{G},\mathbf{G}'}$.
However it should be noted that in order to correctly construct the (nested) hybrid Wannier functions, one needs to numerically compute the eigenvector of the Wilson loop matrix [SEq.~\eqref{eq:P_Wilson_loop_matrix_element}] and nested Wilson loop matrix [SEq.~\eqref{eq:nested_P_Wilson_loop_matrix_def}] at a given base point $\mathbf{k}$ and obtain the eigenvectors at the other momenta via parallel transport equations such as SEq.~\eqref{eq:parallel-transport-nu-j-k-G}~\cite{benalcazar2017electric,varnava2020axion}.

Prior to this work, the nested $P$-Wilson loop formalism described in this section has been applied to identify signatures of several nontrivial insulating phases of matter, including quantized electric multipole insulators~\cite{benalcazar2017quantized,benalcazar2017electric}, magnetic axion insulators~\cite{wieder2018axion}, and higher-order topology in transition metal dichalcogenides~\cite{wang2019higherorder}. 
For example, magnetic axion insulators with 3D spatial inversion symmetry have a nontrivial $\mathbb{Z}_2$-stable spectral flow~\cite{wieder2018axion} in the eigenphases $(\gamma_2)_{j,\mathbf{k},\mathbf{G},\mathbf{G}'}$.
In the next section (SN~\ref{sec:nested_P_pm_Wilson_loop}), we will develop a spin-resolved version of nested $P$-Wilson loop formalism with an aim in the application to identify signatures of spin-stable higher-order topology.

\subsection{\label{sec:nested_P_pm_Wilson_loop}Nested $P_{\pm}$-Wilson Loop}

We will now move to one of the larger theoretical and numerical methods introduced in this work: we will apply the formalism of the previous SN~\ref{sec:nested_P_Wilson_loop} to the spin-resolved projectors $P_\pm$ in order to define a nested $P_\pm$-Wilson loop. 
As in SN~\ref{sec:P_pm_Wilson_loop}, we define the spin operator along $\hat{\mathbf{n}}$ to be $s \equiv (\hat{\mathbf{n}}\cdot \bm{\sigma}) \otimes \mathbb{I}_{N_{\text{orb}}}$, and denote $N_\mathrm{sta}=2N_\mathrm{orb}$. 
The spin operator $s$ satisfies $[V(\mathbf{G})]s[V(\mathbf{G})]^{-1} = s$, as previously proved in SEqs.~\eqref{eq:V_G_and_s_asssumption_1}--\eqref{eq:V_G_and_s_asssumption_3},
such that the reduced spin matrix $[s_{\mathrm{reduced}}(\mathbf{k})]$ [SEq.~\eqref{eq:P_pm_Wilson_loop_s_reduced_def}] satisfies the periodic boundary condition [SEq.~\eqref{eq:P_pm_Wilson_loop_PBC_s_reduced}].
We will suppose that the energy and spin gap are both open for every $\mathbf{k}$ point 
such that $[P(\mathbf{k})]$, the matrix projector to the occupied energy bands in SEq.~(\ref{eq:P_projector}), and $[P_{\pm}(\mathbf{k})]$, the matrix projector to the upper/lower spin bands in SEq.~(\ref{eq:P_pm_k_projector}), are always well-defined and smooth over the BZ. 

To begin, we consider the $N_{\text{occ}}^{\pm} \times N_{\text{occ}}^{\pm}$ $P_{\pm}$-Wilson loop matrix $[\mathcal{W}_{1,\mathbf{k},\mathbf{G}}^{\pm}]$ defined in SEq.~\eqref{eq:P_pm_Wilson_loop_matrix_element}. 
This is the holonomy for a loop starting at base point $\mathbf{k}$ and going along a straight path to $\mathbf{k}+\mathbf{G}$,
where $\mathbf{G}$ is a primitive reciprocal lattice vector, and $N_{\text{occ}}^{+} + N_{\text{occ}}^{-} = N_{\text{occ}}$ is the total number of occupied energy bands of the Bloch Hamiltonian $[H(\mathbf{k})]$. 

As in SN~\ref{sec:nested_P_Wilson_loop}, our goal will be to construct a projector onto a subset of $P_\pm$-Wannier bands. 
To do this, we must first find the eigenvectors of $[\mathcal{W}_{1,\mathbf{k},\mathbf{G}}^{\pm}]$, which satisfy
\begin{align}
    [\mathcal{W}_{1,\mathbf{k},\mathbf{G}}^{\pm}] \ket{{\nu}^{\pm}_{j,\mathbf{k},\mathbf{G}}} = e^{i(\gamma_{1}^{\pm})_{j,\mathbf{k},\mathbf{G}}}\ket{{\nu}^{\pm}_{j,\mathbf{k},\mathbf{G}}}, \label{eq:W_P_pm_eig_eqn}
\end{align}
where the $N_{\mathrm{occ}}^{\pm}$-component eigenvectors $|{\nu}^{\pm}_{j,\mathbf{k},\mathbf{G}} \rangle$ ($j=1,\ldots,N_{\mathrm{occ}}^{\pm}$) are orthonormal such that
\begin{align}
    \bra{{\nu}^{\pm}_{i,\mathbf{k},\mathbf{G}}} \ket{{\nu}^{\pm}_{j,\mathbf{k},\mathbf{G}}} = \delta_{ij}. \label{eq:W_P_pm_evec_orthonormal}
\end{align}
The eigenvectors $\ket{{\nu}^{\pm}_{j,\mathbf{k},\mathbf{G}}}$ satisfy a parallel transport condition by analogy with SEq.~\eqref{eq:parallel-transport-nu-j-k-G}: Defining the $P_{\pm}$-Wilson line matrix
\begin{equation}
	[\mathcal{W}^{\pm}_{1,\mathbf{k}' \leftarrow \mathbf{k}}]_{m,n} \equiv \langle u^{\pm}_{m,\mathbf{k}'} | \left( \prod_{\mathbf{q}}^{\mathbf{k}' \leftarrow \mathbf{k}} [P_{\pm}(\mathbf{q})] \right) | u^{\pm}_{n,\mathbf{k}} \rangle, \label{eq:Ppmwilsonline}
\end{equation}
where $m,n=1\ldots N_{\mathrm{occ}}^{\pm}$ index the upper/lower spin bands, and
\begin{equation}
    \left( \prod_{\mathbf{q}}^{\mathbf{k}' \leftarrow \mathbf{k}} [P_{\pm}(\mathbf{q})] \right)= \lim_{N \to \infty} [P_{\pm}(\mathbf{k}')] [P_{\pm}(\mathbf{k}+\frac{N-1}{N}(\mathbf{k}'-\mathbf{k}))] \cdots [P_{\pm}(\mathbf{k}+\frac{1}{N}(\mathbf{k}'-\mathbf{k}))] [P_{\pm}(\mathbf{k})],
\end{equation}
the eigenvector $\ket{{\nu}^{\pm}_{j,\mathbf{k},\mathbf{G}}}$ satisfies the parallel transport condition~\cite{benalcazar2017electric}
\begin{equation}
	| \nu^{\pm}_{j,\mathbf{k} + \Delta \mathbf{k} ,\mathbf{G}} \rangle = \exp{ -i \frac{(\gamma_{1}^{\pm})_{j,\mathbf{k},\mathbf{G}}}{2\pi} (\Delta \mathbf{k}) \cdot \mathbf{a}  } [\mathcal{W}^{\pm}_{1,\mathbf{k}+\Delta \mathbf{k} \leftarrow \mathbf{k}}] | \nu^{\pm}_{j,\mathbf{k},\mathbf{G}} \rangle \label{eq:parallel-transport-nu-pm-j-k-G}
\end{equation}
for $\Delta \mathbf{k}\parallel\mathbf{G}$, where $\mathbf{a}$ is the real-space primitive lattice vector dual to the primitive reciprocal lattice vector $\mathbf{G}$ such that $\mathbf{a} \cdot \mathbf{G} = 2\pi$.
Also notice that by definition $[\mathcal{W}^{\pm}_{1,\mathbf{k}+\mathbf{G} \leftarrow \mathbf{k}}]$ [SEq.~\eqref{eq:Ppmwilsonline}] is equal to $[\mathcal{W}^{\pm}_{1,\mathbf{k},\mathbf{G}}]$ [SEq.~\eqref{eq:P_pm_Wilson_loop_matrix_element}].
Note that we have not imposed time-reversal symmetry, and therefore $N_{\text{occ}}^{+}$ can be different from $N_{\text{occ}}^{-}$. 
Consequently, the spin-resolved Wilson loop matrices $[\mathcal{W}_{1,\mathbf{k},\mathbf{G}}^{+}]$ and $[\mathcal{W}_{1,\mathbf{k},\mathbf{G}}^{-}]$ need not generally be matrices of the same size.
As described in SN~\ref{sec:P_pm_Wilson_loop}, the eigenphases $(\gamma_{1}^{\pm})_{j,\mathbf{k},\mathbf{G}}$ are the non-Abelian Berry phases for the subset of the occupied states in the image of $P_{\pm}$. 
Denoting $\mathbf{a}$ as the real-space primitive lattice vector dual to $\mathbf{G}$, we have that the eigenphases $(\gamma_{1}^{\pm})_{j,\mathbf{k},\mathbf{G}} / (2\pi)$ correspond to the localized positions (along and measured in unit of $\mathbf{a}$) of \emph{spin-resolved} hybrid Wannier states formed from linear combinations of states in the image of $P_{\pm}$. 
The numerical methods that we employ in this work to find the eigenvectors $\ket{{\nu}^{\pm}_{j,\mathbf{k},\mathbf{G}}}$ are the same as those introduced in SN~\ref{sec:nested_P_Wilson_loop}. 
We first evaluate $[\mathcal{W}_{1,\mathbf{k},\mathbf{G}}^{\pm}]$ using the projector product in SEq.~\eqref{eq:P_pm_Wilson_loop_matrix_element}. 
We then employ a singular value decomposition to extract the unitary part of $[\mathcal{W}_{1,\mathbf{k},\mathbf{G}}^{+}]$. 
A further Schur decomposition of the result then gives us the orthonormal eigenvectors $\ket{{\nu}^{\pm}_{j,\mathbf{k},\mathbf{G}}}$ [SEq.~\eqref{eq:W_P_pm_eig_eqn}].

Analogous to SEq.~\eqref{eq:reexpressing-tilde-nu-as-nu}, we next use the $\ket{{\nu}^{\pm}_{j,\mathbf{k},\mathbf{G}}}$ to form a set of spin-resolved ``hybrid Wannier states'' as vectors in our $N_\mathrm{sta}$-dimensional Hilbert space as~\cite{benalcazar2017quantized,benalcazar2017electric,wieder2018axion}
\begin{align}
    \ket{w^{\pm}_{j,\mathbf{k},\mathbf{G}}} = \sum_{m=1}^{N_{\text{occ}}^{\pm}} [{\nu}^{\pm}_{j,\mathbf{k},\mathbf{G}}]_{m} \ket{u^{\pm}_{m,\mathbf{k}}}, \label{eq:reexpressing-tilde-nu-pm-to-nu-pm}
\end{align}
where $[{\nu}^{\pm}_{j,\mathbf{k},\mathbf{G}}]_{m}$ is the $m^{\text{th}}$ ($m=1,\ldots,N_{\text{occ}}^{\pm}$) component of $\ket{{\nu}^{\pm}_{j,\mathbf{k},\mathbf{G}}}$. 
Notice that $\ket{w^{\pm}_{j,\mathbf{k},\mathbf{G}}}$ is an $N_\mathrm{sta}=2N_{\mathrm{orb}}$-component vector.

To proceed, we now derive the boundary conditions satisfied by $\ket{{\nu}^{\pm}_{j,\mathbf{k},\mathbf{G}}}$ and $\ket{w^{\pm}_{j,\mathbf{k},\mathbf{G}}}$.  
First, we note that as previously proved in SEq.~\eqref{eq:P_pm_Wilson_loop_BC_for_u_pm}, the eigenstates $\ket{u^{\pm}_{m,\mathbf{k}}}$ of the projected spin operator $[P(\mathbf{k})]s[P(\mathbf{k})]$ satisfy the boundary condition $\ket{u^{\pm}_{m,\mathbf{k}+\mathbf{G}'}}=[V(\mathbf{G}')]^{-1} \ket{u^{\pm}_{m,\mathbf{k}}}$ under a shift of the crystal momentum from $\mathbf{k}$ to $\mathbf{k}+\mathbf{G}'$, where $\mathbf{G}'$ is any reciprocal lattice vector. 
Hence the projection operators $[P_{\pm}(\mathbf{k})]$ onto the upper/lower spin bands satisfy
\begin{align}
    [P_{\pm}(\mathbf{k}+\mathbf{G}')] & =  \sum_{n=1}^{N_{\text{occ}}^{\pm}} \ket{u_{n,\mathbf{k}+\mathbf{G}'}^{\pm}} \bra{u_{n,\mathbf{k}+\mathbf{G}'}^{\pm}} \\
    & = \sum_{n=1}^{N_{\text{occ}}^{\pm}} [V(\mathbf{G}')]^{-1}\ket{u_{n,\mathbf{k}}^{\pm}} \bra{u_{n,\mathbf{k}}^{\pm}} [V(\mathbf{G}')] \\
    & = [V(\mathbf{G}')]^{-1} [P_{\pm}(\mathbf{k})] [V(\mathbf{G}')], \label{eq:P_pm_bcs}
\end{align}
where we have used $[V(\mathbf{G}')]^{\dagger} = [V(\mathbf{G}')]^{-1}$.
Second, the product of matrix projectors $[P_{\pm}(\mathbf{k})]$ satisfies
\begin{align}
    \left(\prod_{\mathbf{q}}^{\mathbf{k}+\mathbf{G}'+\mathbf{G} \leftarrow \mathbf{k}+\mathbf{G}'} [P_{\pm}(\mathbf{q})]\right) & = \lim_{N\to \infty} [P_{\pm}(\mathbf{k}+\mathbf{G}'+\mathbf{G})] [P_{\pm}(\mathbf{k}+\mathbf{G}'+\frac{N-1}{N}\mathbf{G})]  \ldots [P_{\pm}(\mathbf{k}+\mathbf{G}'+\frac{1}{N}\mathbf{G}) ] [P_{\pm}(\mathbf{k}+\mathbf{G}')] \\
        & = \lim_{N\to \infty} [V(\mathbf{G}')]^{-1}[P_{\pm}(\mathbf{k}+\mathbf{G})] [P_{\pm}(\mathbf{k}+\frac{N-1}{N}\mathbf{G})] \ldots  [P_{\pm}(\mathbf{k}+\frac{1}{N}\mathbf{G})]  [P_{\pm}(\mathbf{k})] [V(\mathbf{G}')]  \\
    & = [V(\mathbf{G}')]^{-1} \left( \prod_{\mathbf{q}}^{\mathbf{k}+\mathbf{G} \leftarrow \mathbf{k}} [P_{\pm}(\mathbf{q})] \right) [V(\mathbf{G}')]. \label{eq:P_pm_product_bcs}
\end{align}
Putting together SEqs.~\eqref{eq:P_pm_bcs} and \eqref{eq:P_pm_product_bcs}, we can then show that,
under a shift of the crystal momentum from $\mathbf{k}$ to $\mathbf{k}+\mathbf{G}'$, the $N_{\mathrm{occ}}^{\pm} \times N_{\mathrm{occ}}^{\pm}$ $P_{\pm}$-Wilson loop matrix $[\mathcal{W}_{1,\mathbf{k},\mathbf{G}}^{\pm}]$ in SEq.~\eqref{eq:P_pm_Wilson_loop_matrix_element} is invariant:
\begin{align}
    [\mathcal{W}_{1,\mathbf{k}+\mathbf{G}',\mathbf{G}}^{\pm}]_{m,n} &= \bra{u_{m,\mathbf{k}+\mathbf{G}'}^{\pm}} [V(\mathbf{G})] \left(\prod_{\mathbf{q}}^{\mathbf{k}+\mathbf{G}'+\mathbf{G} \leftarrow \mathbf{k}+\mathbf{G}'} [P_{\pm}(\mathbf{q})]\right)\ket{u_{n,\mathbf{k}+\mathbf{G}'}^{\pm}} \\
    & = \bra{u_{m,\mathbf{k}}^{\pm}} [V(\mathbf{G}')] [V(\mathbf{G})] [V(\mathbf{G}')]^{-1} \left(\prod_{\mathbf{q}}^{\mathbf{k}+\mathbf{G} \leftarrow \mathbf{k}} [P_{\pm}(\mathbf{q})]\right) [V(\mathbf{G}')][V(\mathbf{G}')]^{-1}\ket{u_{n,\mathbf{k}}^{\pm}}\\
    & = \bra{u_{m,\mathbf{k}}^{\pm}}  [V(\mathbf{G})] [V(\mathbf{G}')] [V(\mathbf{G}')]^{-1} \left(\prod_{\mathbf{q}}^{\mathbf{k}+\mathbf{G} \leftarrow \mathbf{k}} [P_{\pm}(\mathbf{q})]\right) \ket{u_{n,\mathbf{k}}^{\pm}} \\
    & = \bra{u_{m,\mathbf{k}}^{\pm}}  [V(\mathbf{G})] \left(\prod_{\mathbf{q}}^{\mathbf{k}+\mathbf{G} \leftarrow \mathbf{k}} [P_{\pm}(\mathbf{q})]\right) \ket{u_{n,\mathbf{k}}^{\pm}} \\
    & = [\mathcal{W}_{1,\mathbf{k},\mathbf{G}}^{\pm}]_{m,n}.
\end{align}
As in SN~\ref{sec:nested_P_Wilson_loop}, we have used the fact that $[V(\mathbf{G})]$ and $[V(\mathbf{G}')]$ commute, as shown in SEq.~\eqref{eq:Vgs_commute}. 
Since the $N_{\mathrm{occ}}^{\pm} \times N_{\mathrm{occ}}^{\pm}$ matrix $[\mathcal{W}_{1,\mathbf{k},\mathbf{G}}^{\pm}]$ [SEq.~\eqref{eq:P_pm_Wilson_loop_matrix_element}] is {\it invariant} upon a shift of crystal momentum from $\mathbf{k}$ to $\mathbf{k} + \mathbf{G}'$, we can choose without loss of generality
\begin{align}
    \ket{{\nu}^{\pm}_{j,\mathbf{k}+\mathbf{G}',\mathbf{G}}} = \ket{{\nu}^{\pm}_{j,\mathbf{k},\mathbf{G}}}. \label{eq:bcs-tilde-nu-pm-j}
\end{align}
for the eigenvectors in SEq.~\eqref{eq:W_P_pm_eig_eqn}. 
Notice that the boundary condition in SEq.~\eqref{eq:bcs-tilde-nu-pm-j} is consistent with the eigenvalue equation in SEq.~\eqref{eq:W_P_pm_eig_eqn} and the parallel transport condition in SEq.~\eqref{eq:parallel-transport-nu-pm-j-k-G} if we choose $\mathbf{G}' = \mathbf{G}$.
With the boundary conditions in SEqs.~(\ref{eq:P_pm_Wilson_loop_BC_for_u_pm}) and (\ref{eq:bcs-tilde-nu-pm-j}),
we then have that the $| w^{\pm}_{j,\mathbf{k},\mathbf{G}} \rangle $ in SEq.~\eqref{eq:reexpressing-tilde-nu-pm-to-nu-pm} satisfy
\begin{align}
    \ket{w^{\pm}_{j,\mathbf{k}+\mathbf{G}',\mathbf{G}}} & = \sum_{m=1}^{N_{\text{occ}}^{\pm}} [{\nu}^{\pm}_{j,\mathbf{k}+\mathbf{G}',\mathbf{G}}]_{m} \ket{u^{\pm}_{m,\mathbf{k}+\mathbf{G}'}} \\
    & = \sum_{m=1}^{N_{\text{occ}}^{\pm}} [{\nu}^{\pm}_{j,\mathbf{k},\mathbf{G}}]_{m} [V(\mathbf{G}')]^{-1}\ket{u^{\pm}_{m,\mathbf{k}}} \\
    & = [V(\mathbf{G}')]^{-1}\ket{w^{\pm}_{j,\mathbf{k},\mathbf{G}}}. \label{eq:BC_nu_pm_j_k_G}
\end{align}
We see that $\ket{w^{\pm}_{j,\mathbf{k},\mathbf{G}}}$ satisfies the same boundary conditions as both the eigenstates of the Bloch Hamiltonian [SEq.~\eqref{eq:bcs}] and also the eigenstates $| u_{m,\mathbf{k}}^{\pm}\rangle$ of the projected spin operator $[P(\mathbf{k})]s[P(\mathbf{k})]$ [SEq.~\eqref{eq:P_pm_Wilson_loop_BC_for_u_pm}]. 
We can thus regard the $\ket{w^{\pm}_{j,\mathbf{k},\mathbf{G}}}$ as physical states. 
We will call the $\ket{{w}^{\pm}_{j,\mathbf{k},\mathbf{G}}}$ the {\it $\widehat{\mathbf{G}}$-directed $P_{\pm}$-Wannier band basis}~\cite{benalcazar2017quantized,benalcazar2017electric,wieder2018axion}, or interchangeably the {\it $P_{\pm}$-Wannier basis}. 

The $\ket{{w}^{\pm}_{j,\mathbf{k},\mathbf{G}}}$ are the single-particle eigenstate of the $P_{\pm}$-Wilson loop matrix $[\mathcal{W}^\pm_{1,\mathbf{k},\mathbf{G}}]$ with eigenphase $(\gamma_{1}^{\pm})_{j,\mathbf{k},\mathbf{G}}$, expressed in terms of the spinful tight-binding basis states. 
As long as there is a spectral gap in the $P_{\pm}$-Wannier bands, namely the eigenphase dispersion $\{(\gamma_{1}^{\pm})_{j,\mathbf{k},\mathbf{G}}|j=1\ldots N_{\mathrm{occ}}^{\pm}\}$ as a function of $\mathbf{k}$ (\emph{i.e.} as long as there is a gap in the spin-resolved Wilson loop spectrum), we can further choose a subspace containing $N_{W}^{\pm}$ $P_{\pm}$-Wannier bands to form the matrix projector
\begin{align}
    [\widetilde{P}^{\pm}_{\mathbf{G}}(\mathbf{k})] = \sum_{j=1}^{N_{W}^{\pm}} \ket{{w}^{\pm}_{j,\mathbf{k},\mathbf{G}}} \bra{{w}^{\pm}_{j,\mathbf{k},\mathbf{G}}}, \label{eq:P_pm_nested_Wilson_loop_2nd_projector}
\end{align}
where $N_{W}^{\pm} \leq N_{\mathrm{occ}}^{\pm}$, and
the subscript $\mathbf{G}$ of $[\widetilde{P}^{\pm}_{\mathbf{G}}(\mathbf{k})]$ indicates that the $P_{\pm}$-Wannier basis is obtained by diagonalizing the $P_{\pm}$-Wilson loop matrix [SEq.~\eqref{eq:P_pm_Wilson_loop_matrix_element}] for a loop taken parallel to $\mathbf{G}$. 
Again, being a projector onto a subspace that is spectrally isolated, we can construct the $N_W^\pm\times N_W^\pm$  $\tilde{P}^{\pm}_\mathbf{G}$-Wilson loop matrix along a loop parallel to $\mathbf{k}+\mathbf{G}'$ with matrix elements
\begin{align}
    [{\mathcal{W}}_{2,\mathbf{k},\mathbf{G},\mathbf{G}'}^{\pm}]_{i,j} & = \bra{{w}^{\pm}_{i,\mathbf{k}+\mathbf{G}',\mathbf{G}}}  \left(\prod_{\mathbf{q}}^{\mathbf{k}+\mathbf{G}' \leftarrow \mathbf{k}} [\widetilde{P}^{\pm}_{\mathbf{G}}(\mathbf{q})]\right) \ket{{w}^{\pm}_{j,\mathbf{k},\mathbf{G}}} = \bra{{w}^{\pm}_{i,\mathbf{k},\mathbf{G}}} [V(\mathbf{G}')] \left(\prod_{\mathbf{q}}^{\mathbf{k}+\mathbf{G}' \leftarrow \mathbf{k}} [\widetilde{P}^{\pm}_{\mathbf{G}}(\mathbf{q})]\right) \ket{{w}^{\pm}_{j,\mathbf{k},\mathbf{G}}}, \label{eq:nested_P_pm_Wilson_loop_matrix_def}
\end{align}
where $i$ and $j$ range from $1,\ldots,N_{W}^{\pm}$, and where we have used the boundary condition in SEq.~\eqref{eq:BC_nu_pm_j_k_G} for $\ket{w^{\pm}_{j,\mathbf{k},\mathbf{G}}}$. 
We term $[{\mathcal{W}}_{2,\mathbf{k},\mathbf{G},\mathbf{G}'}^{\pm}]$ the \emph{nested $P_\pm$-Wilson loop matrix}.
From SEq.~(\ref{eq:nested_P_pm_Wilson_loop_matrix_def}), we see that by definition, the nested $P_{\pm}$-Wilson loop matrix is the {\it $\tilde{P}^\pm_\mathbf{G}$-Wilson loop matrix for the ``occupied'' $P_{\pm}$-Wannier bands.} 
We then denote the unimodular eigenvalues of the $N_{W}^{\pm} \times N_{W}^{\pm}$ nested $P_{\pm}$-Wilson loop matrix $[{\mathcal{W}}_{2,\mathbf{k},\mathbf{G},\mathbf{G}'}^{\pm}]$
as $\exp{i(\gamma_{2}^{\pm})_{j,\mathbf{k},\mathbf{G},\mathbf{G}'}}$, where $j=1,\ldots,N_W^{\pm}$.
We then term the set $\{ (\gamma_2^{\pm})_{j,\mathbf{k},\mathbf{G},\mathbf{G}'}|j=1\ldots N_{W}^{\pm} \}$ as a function of $\mathbf{k}$ the ``nested $P_{\pm}$-{\it Wannier} bands,'' where $j$ is the band index. 
The eigenphases $\{ (\gamma_2^{\pm})_{j,\mathbf{k},\mathbf{G},\mathbf{G}'}|j=1\ldots N_{W}^{\pm} \}$ give the positions of spin-resolved hybrid Wannier functions formed from the $N_W^\pm$ ``occupied'' $P_\pm$-Wannier bands and localized along the lattice vector $\mathbf{a}'$ dual to the reciprocal lattice vector $\mathbf{G}'$~\cite{benalcazar2017quantized,benalcazar2017electric}.
Notice that in general $N_{W}^{+}$ can be different from $N_{W}^{-}$ if there are no constraints from symmetries such as time-reversal or $SU(2)$ spin rotation symmetry.
The eigenphases $(\gamma_{2}^{\pm})_{j,\mathbf{k},\mathbf{G},\mathbf{G}'}$ of the nested $P_{\pm}$-Wilson loop matrix [SEq.~\eqref{eq:nested_P_pm_Wilson_loop_matrix_def}] are also the non-Abelian Berry phases of the $N_{W}^{\pm}$ ``occupied'' $P_{\pm}$-Wannier bands for the closed loop in $\mathbf{k}$-space parallel to $\mathbf{G}'$. 
In particular, the eigenphases $(\gamma_2^{\pm})_{j,\mathbf{k},\mathbf{G},\mathbf{G}'}$ are independent of the momentum component $\mathbf{k} \cdot \mathbf{a}'$.
In SN~\ref{appendix:I-constraint-on-nested-P-pm} and \ref{appendix:T-constraint-on-nested-P-pm} we elucidate the constraints on the values of nested $P_{\pm}$-Wilson loop eigenphases due to inversion and time-reversal symmetries.

Similar to the case of nested $P$-Wilson loop in SN~\ref{sec:nested_P_Wilson_loop}, we may deduce the topological properties of a spinful system by computing the spectral flow of $\{(\gamma_2^{\pm})_{j,\mathbf{k},\mathbf{G},\mathbf{G}'}|j=1\ldots N_{W}^{\pm} \}$. 
We will derive general properties of the spectral flow in SN~\ref{sec:general_properties_nested_ppm} and apply these to the classification and study of spin-resolved topology in HOTIs in SN~\ref{app:comparison-spin-stable-and-symmetry-indicated-topology}. 
In SN~\ref{sec:numerical-section-of-nested-P-pm} we will numerically evaluate the nested $P_\pm$-Wilson loops for a tight-binding model of a helical HOTI, revealing the existence of previously unrecognized spin-stable partial axion angles.

\subsection{General Properties of the Nested $P_\pm$-Wilson Loop Spectra}\label{sec:general_properties_nested_ppm}

In this section, we will derive several general properties of the nested $P$- and $P_{\pm}$-Wilson loop [SEqs.~\eqref{eq:nested_P_Wilson_loop_matrix_def} and~\eqref{eq:nested_P_pm_Wilson_loop_matrix_def}] eigenphases (nested non-Abelian Berry phases) and their associated spectral flow. 
These will allow us to develop a classification of spin-resolved topology in systems with inversion and time-reversal symmetry. 
Throughout this section,  
we will use ``$(\pm)$'', to indicate that a derivation applies to both the nested $P$- and $P_{\pm}$-Wilson loops.
Unless otherwise specified, we will assume that both the energy and $PsP$ spectrum are gapped at every $\mathbf{k}$ throughout the BZ. 
We will consider 3D systems with primitive real-space lattice vectors $\{\mathbf{a}_1,\mathbf{a}_2,\mathbf{a}_3 \}$ and the dual primitive reciprocal lattice vectors $\{\mathbf{G}_1,\mathbf{G}_2,\mathbf{G}_3 \}$ such that $\mathbf{a}_i \cdot \mathbf{G}_j = 2\pi \delta_{i,j}$ ($i,j=1\ldots 3$).
The crystal momentum $\mathbf{k}$ can then be expanded using $\mathbf{k} = \sum_{i=1}^{3} \frac{k_i}{2\pi} \mathbf{G}_i$ where the momentum component $k_i = \mathbf{k} \cdot \mathbf{a}_i$.
The BZ is then defined by the region with $k_i = [-\pi,\pi)$.

We begin in SN~\ref{app:W2_same_1st_2nd_G}, by showing that the eigenphases of the nested $P_{(\pm)}$-Wilson loops [SEqs.~\eqref{eq:nested_P_Wilson_loop_matrix_def} and \eqref{eq:nested_P_pm_Wilson_loop_matrix_def}] coincide with a subset of the eigenphases of the $P_{(\pm)}$-Wilson loops [SEqs.~\eqref{eq:P_Wilson_loop_matrix_element} and \eqref{eq:P_pm_Wilson_loop_matrix_element}] when the directions of the first and second holonomy are parallel.
Using this result, we will establish in SN~\ref{app:nested_Berry_phase_only_winds_in_one_direction} that although the {\it nested} Wilson loop eigenphases depend on two momentum components of the base point $\mathbf{k}$, the eigenphases can have a nontrivial spectral flow with a nonzero winding number along at most one momentum direction.
In SN~\ref{app-relation-nested-P-pm-and-partial-weak-Chern}, we will derive an exact relation between the spectral flow of the {\it nested} (spin-resolved) Wilson loop eigenphases and the (partial) weak Chern numbers, which are the 3D generalizations of the (partial) Chern numbers from SEq.~\eqref{eq:partial_chern_def}.
While we will specialize to cases where at most one (partial) weak Chern number is nonzero, we will in SN~\ref{app-relation-nested-P-pm-and-partial-weak-Chern} show how our results extend to more general situations.
In SN~\ref{app:z2_nested_P_pm_berry_phase} we will show that for inversion- and time-reversal-symmetric systems the nested $P_{\pm}$-Wilson loops exhibit a novel variant of $\mathbb{Z}_2$-stable spectral flow, from which we classify spin-stable topology in these systems. 
Then, in the next SN~\ref{app:comparison-spin-stable-and-symmetry-indicated-topology}, based on the layer-construction of helical and magnetic (higher-order) topological insulators~\cite{song2018mapping,song2019topological,elcoro2021magnetic,gao2022magnetic}, we show how spin-stable topology refines and enriches the notion of symmetry-indicated stable band topology, and specifically reveals the existence of previously unrecognized spin-resolved (partial) axion angles in 3D insulators.

\subsubsection{Nested $P$- and $P_{\pm}$-Wilson Loops When $\mathbf{G} = \mathbf{G}'$}\label{app:W2_same_1st_2nd_G}

In the computation of the {\it nested} Wilson loop eigenphases, we were required to choose two primitive reciprocal lattice vectors $\mathbf{G}$ and $\mathbf{G}'$ in SEqs.~\eqref{eq:nested_P_Wilson_loop_matrix_def} and \eqref{eq:nested_P_pm_Wilson_loop_matrix_def}.
$\mathbf{G}$ fixes the direction of the first Wilson loop, and so determines the localization direction of the Wannier band eigenstates; $\mathbf{G}'$ gives the direction of the second (nested) Wilson loop, computed as a product of projectors onto the $\widehat{\mathbf{G}}$-directed Wannier bands.
And we refer to $\mathbf{G}$ and $\mathbf{G}'$ as the directions of the first and second closed loop holonomy. 
In this section, we consider the case where $\mathbf{G}' = \mathbf{G}$ and then prove that the resulting nested (spin-resolved or ordinary) Wilson loop eigenphases are the same as the centers of the (spin-resolved or ordinary) $\widehat{\mathbf{G}}$-directed hybrid Wannier functions.

We begin by denoting the projector onto the single-particle energy or $PsP$ eigenstates as $P_{(\pm)}$, such that
\begin{equation}
	\langle \alpha, \mathbf{k} | P_{(\pm)} | \beta , \mathbf{q} \rangle = \langle 0| c_{\alpha,\mathbf{k}} P_{(\pm)} c_{\beta , \mathbf{q}}^{\dagger} | 0 \rangle  = \delta_{\mathbf{k},\mathbf{q}} [P_{(\pm)}(\mathbf{k})]_{\alpha,\beta}, \label{eq:operator-form-P-and-P-pm}
\end{equation}
where $c_{\alpha , \mathbf{k}}^{\dagger}$ and $c_{\alpha , \mathbf{k}}$ are the Fourier-transformed creation and annihilation operators of the $\alpha^{\mathrm{th}}$ orbital defined in SEqs.~\eqref{eq:TB_convention_c_dagger} and \eqref{eq:TB_convention_c}.
The matrix projector $[P_{(\pm)}(\mathbf{k})]$ in SEq.~\eqref{eq:operator-form-P-and-P-pm} is
\begin{equation}
	[P_{(\pm)}(\mathbf{k})] = \sum_{n=1}^{N_{\mathrm{occ}}^{(\pm)}} | u_{n,\mathbf{k}}^{(\pm)} \rangle \langle u_{n,\mathbf{k}}^{(\pm)} |,
\end{equation}
where $| u_{n,\mathbf{k}}^{(\pm)} \rangle$ is the occupied energy eigenvector of the Bloch Hamiltonian matrix $[H(\mathbf{k})]$ [SEq.~\eqref{eq:TB_convention_u_nk_eig_eqn}] or the eigenvector of the reduced spin matrix $[s_{\mathrm{reduced}}(\mathbf{k})]$ [SEq.~\eqref{eq:P_pm_Wilson_loop_s_reduced_def}] in the upper/lower $PsP$ eigenspace written in terms of the tight-binding basis states [SEq.~\eqref{eq:reexpress_the_eigenvector_of_s_reduced}].
From SRefs.~\cite{benalcazar2017electric,alexandradinata2014wilsonloop}, the eigenvalues of the projected position operator $P_{(\pm)} \widehat{\mathbf{x}} \cdot \mathbf{G} P_{(\pm)}$ are the $\widehat{\mathbf{G}}$-directed $P_{(\pm)}$-Wannier bands $\{ (\gamma_1^{(\pm)})_{j,\mathbf{k},\mathbf{G}} | j = 1 \ldots N_{\mathrm{occ}}^{(\pm)} \}$ [SEqs.~\eqref{eq:P_nested_Wilson_loop_eig_eqn_1} and \eqref{eq:W_P_pm_eig_eqn}], and the corresponding eigenstates are (spin-resolved or ordinary) hybrid Wannier functions in the image of $P_{(\pm)}$ that are localized along the lattice vector $\mathbf{a}$ dual to the reciprocal lattice vector $\mathbf{G}$.
Provided that we have disjoint groups of $P_{(\pm)}$-Wannier bands (\emph{i.e.}, provided there is a gap in the $P_{(\pm)}$-Wannier bands), we can form the projector $\widetilde{P}_{\mathbf{G}}^{(\pm)}$ onto a group of $N_{W}^{(\pm)}$ $P_{(\pm)}$-Wannier bands using the corresponding single-particle hybrid Wannier functions, such that
\begin{equation}
	\langle \alpha, \mathbf{k} | \widetilde{P}_{\mathbf{G}}^{(\pm)}| \beta , \mathbf{q} \rangle = \langle 0 | c_{\alpha,\mathbf{k}}  \widetilde{P}_{\mathbf{G}}^{(\pm)} c_{\beta,\mathbf{q}}^{\dagger}   | 0 \rangle = \delta_{\mathbf{k},\mathbf{q}} [\widetilde{P}_{\mathbf{G}}^{(\pm)}(\mathbf{k})]_{\alpha,\beta}. \label{eq:operator-form-tilde-P-G-and-tilde-P-G-pm}
\end{equation}
The matrix projector $[\widetilde{P}_{\mathbf{G}}^{(\pm)}(\mathbf{k})]$ in SEq.~\eqref{eq:operator-form-tilde-P-G-and-tilde-P-G-pm} is [according to SEqs.~\eqref{eq:P_nested_Wilson_loop_2nd_projector} and \eqref{eq:P_pm_nested_Wilson_loop_2nd_projector}]
\begin{equation}
	[\widetilde{P}_{\mathbf{G}}^{(\pm)}(\mathbf{k})] = \sum_{j=1}^{N_W^{(\pm)}} | w_{j,\mathbf{k},\mathbf{G}}^{(\pm)} \rangle \langle w_{j,\mathbf{k},\mathbf{G}}^{(\pm)} |,
\end{equation}
where $| w_{j,\mathbf{k},\mathbf{G}}^{(\pm)} \rangle$ [SEqs.~\eqref{eq:reexpressing-tilde-nu-as-nu} and \eqref{eq:reexpressing-tilde-nu-pm-to-nu-pm}] is the $\widehat{\mathbf{G}}$-directed $P_{(\pm)}$-Wannier band basis corresponding to the eigenvectors of the $\widehat{\mathbf{G}}$-directed $P_{(\pm)}$-Wilson loop matrix [SEqs.~\eqref{eq:P_Wilson_loop_matrix_element} and \eqref{eq:P_pm_Wilson_loop_matrix_element}] with eigenphases $\{ (\gamma_1^{(\pm)})_{j,\mathbf{k},\mathbf{G}} | j = 1 \ldots N_{W}^{(\pm)} \}$ where $j$ indexes the ``occupied'' $N_{W}^{(\pm)}$ $P_{(\pm)}$-Wannier bands spectrally separated from the other ``unoccupied'' $N_{\mathrm{occ}}^{(\pm)} - N_{W}^{(\pm)}$ $P_{(\pm)}$-Wannier bands.

Next, we consider the projected position operator $\widetilde{P}_{\mathbf{G}}^{(\pm)} \widehat{\mathbf{x}} \cdot \mathbf{G} \widetilde{P}_{\mathbf{G}}^{(\pm)}$, whose eigenvalues are the eigenphases of the nested $P_{(\pm)}$-Wilson loop with $\mathbf{G}'=\mathbf{G}$~\cite{benalcazar2017electric,alexandradinata2014wilsonloop}
denoted by $\{ (\gamma_2^{(\pm)})_{j,\mathbf{k},\mathbf{G},\mathbf{G}} | j=1\ldots N_{W}^{(\pm)} \}$ (SN~\ref{sec:nested_P_Wilson_loop} and \ref{sec:nested_P_pm_Wilson_loop}). 
Since $P_{(\pm)}=\widetilde{P}^{(\pm)}_{\mathbf{G}}$ on states in the image of $\widetilde{P}^{(\pm)}_{\mathbf{G}}$, and since the image of $\widetilde{P}^{(\pm)}_{\mathbf{G}}$ is spanned by a set of eigenstates of $P_{(\pm)} \widehat{\mathbf{x}} \cdot \mathbf{G} P_{(\pm)}$, we have that eigenstates of $ \widetilde{P}^{(\pm)}_{\mathbf{G}} \widehat{\mathbf{x}} \cdot \mathbf{G} \widetilde{P}^{(\pm)}_{\mathbf{G}}$ are eigenstates of $P_{(\pm)} \widehat{\mathbf{x}} \cdot \mathbf{G} P_{(\pm)}$ with the same eigenvalues. 
In other words, we have established that $\{ (\gamma_2^{(\pm)})_{j,\mathbf{k},\mathbf{G},\mathbf{G}} | j=1\ldots N_{W}^{(\pm)} \}$ is the same as $\{ (\gamma_1^{(\pm)})_{j,\mathbf{k},\mathbf{G}} | j = 1 \ldots N_{W}^{(\pm)} \}$ where $j=1\ldots N_{W}^{(\pm)}$ index the $P_{(\pm)}$-Wannier bands considered in the computation of the nested $P_{(\pm)}$-Wilson loops.

In conclusion, we have proved that when the first and second closed-loop holonomy are in the same direction, the resulting nested $P_{(\pm)}$-Wilson loop eigenphases are the same as the corresponding set of $\widehat{\mathbf{G}}$-directed $P_{(\pm)}$-Wannier centers.

\subsubsection{The Nested $P$- and $P_{\pm}$-Wilson Loops Wind Along at Most One Primitive Reciprocal Lattice Direction}\label{app:nested_Berry_phase_only_winds_in_one_direction}

In this section we will further analyze the winding of the nested $P_{(\pm)}$-Wilson loop eigenphases $\{(\gamma_{2}^{(\pm)})_{n,\mathbf{k},\mathbf{G}_i,\mathbf{G}_j}| n= 1\ldots N_{W}^{(\pm)}\}$ [SN~\ref{sec:nested_P_Wilson_loop} and \ref{sec:nested_P_pm_Wilson_loop}] where the first and second closed loop holonomy are parallel to $\mathbf{G}_i$ and $\mathbf{G}_j$ respectively, and we will assume that $i \neq j$ ($i,j\in (1,2,3)$) for now, as we have discussed the case of $i=j$ in SN~\ref{app:W2_same_1st_2nd_G}.
The one remaining linearly-independent primitive reciprocal lattice vector is denoted as $\mathbf{G}_l$.
The total nested Berry phase $(\gamma_2^{(\pm)})_{\mathbf{k},\mathbf{G}_i,\mathbf{G}_j}$ is defined by summing over the nested $P_{(\pm)}$-Wilson loop eigenphases
\begin{equation}
	(\gamma_2^{(\pm)})_{\mathbf{k},\mathbf{G}_i,\mathbf{G}_j} \equiv \sum_{n=1}^{N_{W}^{(\pm)}} (\gamma_{2}^{(\pm)})_{n,\mathbf{k},\mathbf{G}_i,\mathbf{G}_j} \text{ mod } 2\pi =  \mathrm{Im} \left(\mathrm{log}\left( \mathrm{det} [\mathcal{W}^{(\pm)}_{2,\mathbf{k},\mathbf{G}_i,\mathbf{G}_j}] \right) \right). \label{eq:net_nested_Berry_pha_Gi_Gj}
\end{equation}
$(\gamma_2^{(\pm)})_{\mathbf{k},\mathbf{G}_i,\mathbf{G}_j}$ is a function of the momentum components $k_i$ and $k_l$, and $[\mathcal{W}^{(\pm)}_{2,\mathbf{k},\mathbf{G}_i,\mathbf{G}_j}] $ is the nested $P_{(\pm)}$-Wilson loop matrix [SEqs.~\eqref{eq:nested_P_Wilson_loop_matrix_def} and \eqref{eq:nested_P_pm_Wilson_loop_matrix_def}].
From $(\gamma_2^{(\pm)})_{\mathbf{k},\mathbf{G}_i,\mathbf{G}_j}$, we can define two winding numbers
\begin{align}
	& C^{(\pm)}_{\gamma_2 , i,j} (k_l;i) \equiv \int_{BZ} dk_i \frac{\partial (\gamma_2^{(\pm)})_{\mathbf{k},\mathbf{G}_i,\mathbf{G}_j}}{\partial k_i} = \int_{BZ} dk_i \frac{\partial \mathrm{Im} \left(\mathrm{log}\left( \mathrm{det} [\mathcal{W}^{(\pm)}_{2,\mathbf{k},\mathbf{G}_i,\mathbf{G}_j}] \right) \right)}{\partial k_i}, \label{eq:app-C-gamma-2-i-kl} \\
	& C^{(\pm)}_{\gamma_2 , i,j} (k_i ; l) \equiv \int_{BZ} dk_l \frac{\partial (\gamma_2^{(\pm)})_{\mathbf{k},\mathbf{G}_i,\mathbf{G}_j}}{\partial k_l} = \int_{BZ} dk_l \frac{\partial \mathrm{Im} \left(\mathrm{log}\left( \mathrm{det} [\mathcal{W}^{(\pm)}_{2,\mathbf{k},\mathbf{G}_i,\mathbf{G}_j}] \right) \right)}{\partial k_l}. \label{eq:app-C-gamma-2-l-ki}
\end{align}
$C^{(\pm)}_{\gamma_2 , i,j} (k_l;i)$ [SEq.~\eqref{eq:app-C-gamma-2-i-kl}] is the winding number of the nested Berry phase $(\gamma_2^{(\pm)})_{\mathbf{k},\mathbf{G}_i,\mathbf{G}_j}$ [SEq.~\eqref{eq:net_nested_Berry_pha_Gi_Gj}] when we fix $k_l$ and let $k_i \to k_i+2\pi$, while $C^{(\pm)}_{\gamma_2 , i,j} (k_i ; l)$ [SEq.~\eqref{eq:app-C-gamma-2-l-ki}] is the winding number of the nested Berry phase $(\gamma_2^{(\pm)})_{\mathbf{k},\mathbf{G}_i,\mathbf{G}_j}$ [SEq.~\eqref{eq:net_nested_Berry_pha_Gi_Gj}] when we fix $k_i$ and let $k_l \to k_l+2\pi$. 

We will now prove that $C^{(\pm)}_{\gamma_2 , i,j} (k_l;i)$ [SEq.~\eqref{eq:app-C-gamma-2-i-kl}] must be zero provided that the $\widehat{\mathbf{G}}_i$-directed $P_{(\pm)}$-Wannier bands can be decomposed into disjoint sets without band touching points, which is the requirement for the nested Berry phases
$(\gamma_2^{(\pm)})_{\mathbf{k},\mathbf{G}_i,\mathbf{G}_j}$ [SEq.~\eqref{eq:net_nested_Berry_pha_Gi_Gj}] to be well-defined at all $(k_i,k_l)$.

To begin, let us denote the $\widehat{\mathbf{G}}_i$-directed $P_{(\pm)}$-Wannier bands (non-Abelian Berry phases) as $\{ (\gamma_1^{(\pm)})_{n,\mathbf{k},\mathbf{G}_i}| n=1\ldots N_{\mathrm{occ}}^{(\pm)} \}$ [SEqs.~\eqref{eq:P_nested_Wilson_loop_eig_eqn_1} and \eqref{eq:W_P_pm_eig_eqn}].
Since $(\gamma_1^{(\pm)})_{n,\mathbf{k},\mathbf{G}_i}$ is independent of the momentum component $k_i$, we may view $\{ (\gamma_1^{(\pm)})_{n,\mathbf{k},\mathbf{G}_i}| j=1\ldots N_{\mathrm{occ}}^{(\pm)} \}$ as a 3D band structure that is flat as a function of $k_i$, with the caveat that the eigenvectors of the $P_{(\pm)}$-Wilson loop matrix must satisfy the parallel transport conditions [SEqs.~\eqref{eq:parallel-transport-nu-j-k-G} and \eqref{eq:parallel-transport-nu-pm-j-k-G}].
We will also assume that $\{ (\gamma_1^{(\pm)})_{n,\mathbf{k},\mathbf{G}_i}| n=1\ldots N_{\mathrm{occ}}^{(\pm)} \}$ consists of spectrally separated groups of bands so that the computation of $(\gamma_2^{(\pm)})_{\mathbf{k},\mathbf{G}_i,\mathbf{G}_j}$ is well-defined over the 2D BZ spanned by $\mathbf{G}_i$ and $\mathbf{G}_l$.
Using the sign convention in SEq.~\eqref{eq:partial_chern_def} that relates the Chern number and the Berry phase winding number, if we choose $(j,i,l)$ to index a right-handed coordinate system, $C^{(\pm)}_{\gamma_2 , i,j} (k_l;i)$ indicates the \textit{Chern number} in the $k_l$-constant BZ plane for the $N_{W}^{(\pm)}$ $\widehat{\mathbf{G}}_i$-directed, spectrally separated $P_{(\pm)}$-Wannier bands.
On the other hand, the same Chern number can also be obtained by first computing the nested Berry phases
\begin{equation}
	(\gamma_2^{(\pm)})_{\mathbf{k},\mathbf{G}_i , \mathbf{G}_i} \equiv \sum_{n=1}^{N_{W}^{(\pm)}} (\gamma_2^{(\pm)})_{n,\mathbf{k},\mathbf{G}_i , \mathbf{G}_i} \text{ mod } 2\pi =  \mathrm{Im} \left(\mathrm{log}\left( \mathrm{det} [\mathcal{W}^{(\pm)}_{2,\mathbf{k},\mathbf{G}_i,\mathbf{G}_i}] \right) \right), \label{eq:net_nested_Berry_pha_Gi_Gi}
\end{equation}
which depend on the momentum components $k_j$ and $k_l$.
Notice that on the right-hand side of SEq.~\eqref{eq:net_nested_Berry_pha_Gi_Gi}, both the first and second closed-loop holonomy are parallel to $\mathbf{G}_{i}$.
We next obtain the corresponding winding number
\begin{equation}
	C^{(\pm)}_{\gamma_2 ,i,i} (k_l;j) = \int_{BZ} dk_j \frac{\partial (\gamma_2^{(\pm)})_{\mathbf{k},\mathbf{G}_i , \mathbf{G}_i}}{\partial k_j} = \int_{BZ} dk_j \frac{\partial \mathrm{Im} \left(\mathrm{log}\left( \mathrm{det} [\mathcal{W}^{(\pm)}_{2,\mathbf{k},\mathbf{G}_i,\mathbf{G}_i}] \right) \right)}{\partial k_j}. \label{eq:app-C-gamma-2-i-i-kl-j},
\end{equation} 
in which $C^{(\pm)}_{\gamma_2 ,i,i} (k_l;j)$ is the winding number of the nested Berry phase $(\gamma_2^{(\pm)})_{\mathbf{k},\mathbf{G}_i , \mathbf{G}_i}$ [SEq.~\eqref{eq:net_nested_Berry_pha_Gi_Gi}] for a fixed value of $k_l$ and taking $k_j \to k_j + 2 \pi$.
Notice that using the sign convention in SEq.~\eqref{eq:partial_chern_def} that relates the Chern number and the Berry phase winding number, if we choose $(i,j,l)$ to index a right-handed coordinate system, $C^{(\pm)}_{\gamma_2 ,i,i} (k_l;j)$ also indicates the \textit{Chern number} of the $\widehat{\mathbf{G}}_i$-directed $P_{(\pm)}$ Wannier bands in a constant-$k_l$ BZ plane. 
In particular, $C^{(\pm)}_{\gamma_2 ,i,i} (k_l;j)$ [SEq.~\eqref{eq:app-C-gamma-2-i-i-kl-j}] and $C^{(\pm)}_{\gamma_2 ,i,j} (k_l;i)$ [SEq.~\eqref{eq:app-C-gamma-2-i-kl}]  satisfy
\begin{equation}
	C^{(\pm)}_{\gamma_2 ,i,j} (k_l;i) = - C^{(\pm)}_{\gamma_2 ,i,i} (k_l;j). \label{eq:opposite-chern-ijli-iilj}
\end{equation}

Next, from the discussion in SN~\ref{app:W2_same_1st_2nd_G}, we can deduce that the nested Berry phase $(\gamma_2^{(\pm)})_{\mathbf{k},\mathbf{G}_i , \mathbf{G}_i}$ [SEq.~\eqref{eq:net_nested_Berry_pha_Gi_Gi}] is equal to $ \sum_{n=1}^{N_{W}^{(\pm)}} (\gamma_1^{(\pm)})_{n,\mathbf{k},\mathbf{G}_i}$ mod $2\pi$, where $n = 1\ldots N_{W}^{(\pm)}$ indexes the $N_{W}^{(\pm)}$ Wannier bands.
Since we have assumed that the $N_W^{(\pm)}$ Wannier bands are spectrally separated from the other Wannier bands, the winding number of $\sum_{n=1}^{N_{W}^{(\pm)}} (\gamma_1^{(\pm)})_{n,\mathbf{k},\mathbf{G}_i}$ mod $2\pi$ must be zero.
Since $(\gamma_2^{(\pm)})_{\mathbf{k},\mathbf{G}_i , \mathbf{G}_i} = \sum_{n=1}^{N_{W}^{(\pm)}} (\gamma_1^{(\pm)})_{n,\mathbf{k},\mathbf{G}_i}$ mod $2\pi$, this then implies the winding number $C^{(\pm)}_{\gamma_2 , i , i }(k_l ; j)$ [SEq.~\eqref{eq:app-C-gamma-2-i-i-kl-j}] is also zero. 
From the relation $C^{(\pm)}_{\gamma_2 ,i,j} (k_l;i) = - C^{(\pm)}_{\gamma_2 ,i,i} (k_l;j)$ [SEq.~\eqref{eq:opposite-chern-ijli-iilj}], we then conclude that $C^{(\pm)}_{\gamma_2 ,i,j} (k_l;i)$ [SEq.~\eqref{eq:app-C-gamma-2-i-kl}] satisfies
\begin{equation}
	C^{(\pm)}_{\gamma_2 ,i,j} (k_l;i)=0.
\end{equation}
Therefore, the winding number of the nested Berry phase $(\gamma_2^{(\pm)})_{\mathbf{k},\mathbf{G}_i , \mathbf{G}_j}$ [SEq.~\eqref{eq:net_nested_Berry_pha_Gi_Gj}] must be zero if we fix $k_l$ and let $k_i \to k_i + 2\pi$.

We emphasize that, employing the sign convention in SEq.~\eqref{eq:partial_chern_def} and choosing $(i,j,l)$ to index a right-handed coordinate system, since the winding number $C_{\gamma_2 , i , j }^{(\pm)}(k_i ; l)$ [SEq.~\eqref{eq:app-C-gamma-2-l-ki}] is equal to the \textit{Chern number} of the $N_{W}^{(\pm)}$ spectrally separated $\widehat{\mathbf{G}}_i$-directed $P_{(\pm)}$-Wannier bands in a constant-$k_i$ BZ plane, we must have that $C_{\gamma_2 , i , j }^{(\pm)}(k_i ; l)$ [SEq.~\eqref{eq:app-C-gamma-2-l-ki}] is a constant independent of $k_i$. 
Therefore, the nested Berry phase $(\gamma_2^{(\pm)})_{\mathbf{k},\mathbf{G}_i , \mathbf{G}_j}$ [SEq.~\eqref{eq:net_nested_Berry_pha_Gi_Gj}] can have at most one nonzero winding number $C_{\gamma_2 , i , j }^{(\pm)}(k_i ; l)$ [SEq.~\eqref{eq:app-C-gamma-2-l-ki}], and its value is independent of $k_i$.

To conclude, if we compute the nested Berry phase $(\gamma_2^{(\pm)})_{\mathbf{k},\mathbf{G}_i , \mathbf{G}_j}$ through SEq.~\eqref{eq:net_nested_Berry_pha_Gi_Gj} with $\mathbf{G}_i \neq \mathbf{G}_j$, even though the eigenphases depend on $k_{i}$ and $k_{l}$ (noting that $\mathbf{G}_i$, $\mathbf{G}_j$, $\mathbf{G}_l$ are linearly-independent), there can be no net winding as a function of $k_i$.
As such, nonzero winding in the spectrum of $(\gamma_2^{(\pm)})_{\mathbf{k},\mathbf{G}_i , \mathbf{G}_j}$ [SEq.~\eqref{eq:net_nested_Berry_pha_Gi_Gj}] can occur only as a function of $k_l$, and in addition such a winding number must be independent of $k_i$ provided that there exists a Wannier gap.
Therefore, without loss of generality, when we refer to the winding number of the nested Berry phase $(\gamma_2^{(\pm)})_{\mathbf{k},\mathbf{G}_i,\mathbf{G}_j}$ [SEq.~\eqref{eq:net_nested_Berry_pha_Gi_Gj}], we exclusively are referring to the winding number $C^{(\pm)}_{\gamma_2 , i,j} (k_i;l)$ [SEq.~\eqref{eq:app-C-gamma-2-l-ki}]. 
As such, in later sections we will introduce a simpler notation for $C^{(\pm)}_{\gamma_2 , i,j} (k_i;l)$. 
Lastly, we note that $\mathbf{G}_l$ is the only  primitive reciprocal lattice vector that we have not yet used either for the first or the second closed-loop holonomy in the computation of the nested Berry phases $(\gamma_2^{(\pm)})_{\mathbf{k},\mathbf{G}_i,\mathbf{G}_j}$ [SEq.~\eqref{eq:net_nested_Berry_pha_Gi_Gj}].

The winding number of the nested $P$-Wilson loop has been shown in SRefs.~\cite{wieder2018axion,varnava2020axion} to be $\mathbb{Z}_2$-stable, and is nontrivial for 3D strong magnetic axion insulators (\emph{i.e.} magnetic insulators with bulk quantized Chern-Simons axion angles $\theta = \pi$ and vanishing weak Chern numbers). 
In particular, the winding number modulo 2 has been shown to be an indicator of $\theta$~\cite{qi2008topological,essin2009magnetoelectric}.  
In the next section (SN~\ref{app-relation-nested-P-pm-and-partial-weak-Chern}), we will generalize this observation and derive a relationship between the winding number of the nested $P_{(\pm)}$-Wilson loop spectrum and the (partial) weak Chern number, which we will then in SN~\ref{app:comparison-spin-stable-and-symmetry-indicated-topology} use to extract a novel spin-resolved (partial) variant of the axion angle. 

\subsubsection{Relation Between Nested $P$- and $P_{\pm}$-Wilson Loops and (Partial) Weak Chern Numbers}\label{app-relation-nested-P-pm-and-partial-weak-Chern}

In this section, we will derive exact relations between the winding number of the nested (spin-resolved) Wilson loop eigenphases and the (partial) weak Chern numbers, which are the 3D generalization of the (partial) Chern numbers defined in SEq.~\eqref{eq:partial_chern_def}.
To be specific, we here consider the nested Berry phases $(\gamma_2^{(\pm)})_{\mathbf{k},\mathbf{G}_i,\mathbf{G}_j}$ in SEq.~\eqref{eq:net_nested_Berry_pha_Gi_Gj}, where the first and second holonomy are parallel to the linearly-independent primitive reciprocal lattice vectors $\mathbf{G}_i$ and $\mathbf{G}_j$, respectively, and the one remaining linearly-independent primitive reciprocal lattice vector is denoted as $\mathbf{G}_l$.
In order for $(\gamma_2^{(\pm)})_{\mathbf{k},\mathbf{G}_i,\mathbf{G}_j}$ to be well-defined over the 2D BZ spanned by $\mathbf{G}_i$ and $\mathbf{G}_l$, we also assume that the $\widehat{\mathbf{G}}_i$-directed $P_{(\pm)}$-Wannier bands described by $\{(\gamma_1^{(\pm)})_{j,\mathbf{k},\mathbf{G}_i} | j=1\ldots N_{\mathrm{occ}}^{(\pm)} \}$ contain spectrally separated $N_{\mathrm{group}}^{(\pm)}$ groups of bands.
As previously proved in SN~\ref{app:nested_Berry_phase_only_winds_in_one_direction}, the winding number of the nested Berry phase $(\gamma_2^{(\pm)})_{\mathbf{k},\mathbf{G}_i,\mathbf{G}_j}$ for a given group of $P_{(\pm)}$-Wannier bands can only be nonzero if computed as as a function of $k_{l}$. 
{We will hence in this section specifically focus on the winding number in SEq.~\eqref{eq:app-C-gamma-2-l-ki} of the nested Berry phase $(\gamma_2^{(\pm)})_{\mathbf{k},\mathbf{G}_i,\mathbf{G}_j}$ computed as a function of $k_l$ for a given group of $P_{(\pm)}$-Wannier bands.}

To begin, we can decompose the matrix projector $[P_{(\pm)}(\mathbf{k})]$ onto the occupied energy bands or the upper/lower spin bands into
\begin{equation}
	[P_{(\pm)}(\mathbf{k})] = \sum_{n=1}^{N_{\mathrm{group}}^{(\pm)}}[\widetilde{P}^{(\pm)}_{\mathbf{G}_i,n}(\mathbf{k})], \label{eq:Ppm_multiple_separate_base_ki0}
\end{equation}
where 
\begin{equation}
	[\widetilde{P}^{(\pm)}_{\mathbf{G}_i,n}(\mathbf{k})] = \sum_{j \in n^{\mathrm{th}} \text{ group}} | w_{j,\mathbf{k},\mathbf{G}_i}^{(\pm)} \rangle \langle w_{j,\mathbf{k},\mathbf{G}_i}^{(\pm)} | \label{eq:nth-group-wannier-basis-projector}
\end{equation}
is the matrix projector [SEqs.~\eqref{eq:P_nested_Wilson_loop_2nd_projector} and \eqref{eq:P_pm_nested_Wilson_loop_2nd_projector}] formed from the $\widehat{\mathbf{G}}_i$-directed  $P_{(\pm)}$-Wannier band basis $| w_{j,\mathbf{k},\mathbf{G}_i}^{(\pm)} \rangle$ [SEqs.~\eqref{eq:reexpressing-tilde-nu-as-nu} and \eqref{eq:reexpressing-tilde-nu-pm-to-nu-pm}] in the $n^{\mathrm{th}}$ ($n=1\ldots N_{\mathrm{group}}^{(\pm)}$) group of $\widehat{\mathbf{G}}_i$-directed $P_{(\pm)}$-Wannier bands.
In addition, we also have the orthogonality conditions
\begin{equation}
	[\widetilde{P}_{\mathbf{G}_i,n_1} (\mathbf{k})][\widetilde{P}_{\mathbf{G}_i,n_2}(\mathbf{k})] = \delta_{n_1,n_2} [\widetilde{P}_{\mathbf{G}_i,n_1}(\mathbf{k})] \label{eq:P_multiple_separate_base_ki0_orthogonal}
\end{equation}
for the projectors onto the subspaces within the image of $[P(\mathbf{k})]$, and
\begin{equation}
	[\widetilde{P}^{\sigma_1}_{\mathbf{G}_i,n_1}(\mathbf{k})] [\widetilde{P}^{\sigma_2}_{\mathbf{G}_i,n_2}(\mathbf{k})] = \delta_{\sigma_1,\sigma_2} \delta_{n_1 , n_2} [\widetilde{P}^{\sigma_1}_{\mathbf{G}_i,n_1}(\mathbf{k})] \label{eq:Ppm_multiple_separate_base_ki0_orthogonal}
\end{equation}
for the projectors onto the subspaces within the image of $[P_{\pm}(\mathbf{k})]$ where $\sigma_1$, $\sigma_2 = \pm$ and $n_1$, $n_2$ denotes the group.

Since we have assumed that the energy and $PsP$ spectrum are gapped throughout the 3D BZ, we can define (partial) weak Chern numbers $\nu_{1}^{(\pm)}$, $\nu_{2}^{(\pm)}$, and $\nu_{3}^{(\pm)}$.
$\{\nu_1 , \nu_2 , \nu_3 \}$ specifically denote the weak Chern numbers of the occupied energy bands, while $\{\nu_1^{\pm} , \nu_2^{\pm} , \nu_3^{\pm} \}$ are the partial weak Chern numbers of the upper ($+$) and lower ($-$) spin bands.
To be precise, $\nu^{(\pm)}_{1}$, $\nu^{(\pm)}_{2}$, and $\nu^{(\pm)}_{3}$ are the Chern numbers of the single-particle states in the image of $[P_{(\pm)}(\mathbf{k})]$ in BZ planes of constant-$k_1$, constant-$k_2$, and constant-$k_3$, respectively.
We note that in the literature, the term ``weak indices'' is generally used to refer to topological invariants computed in the $k_i = \pi$ BZ planes~\cite{elcoro2021magnetic,gao2022magnetic,song2018mapping,fu2007topologicala,ran2009onedimensional,teo2010topological}. 
However in  systems with an energy and a spin gap, the (partial) Chern numbers cannot change as functions of $k_{i}$ (which would require energy or spin gap-closing points), and hence the (partial) Chern numbers are constant and independent of $k_i$.
Since we require the $\widehat{\mathbf{G}}_{i}$-directed $P_{(\pm)}$-Wannier bands to consist of disjoint groups that are spectrally separated, we must have that $\nu^{(\pm)}_{j} = \nu^{(\pm)}_{l} =0 $.
Conversely, $\nu^{(\pm)}_{i}$ can take any integral values. 
The partial weak Chern numbers $\nu^{+}_{i}$ and $\nu^{-}_{i}$ are generically independent, but can be related to each other by symmetries that reverse the spin direction, as we will see below.
Recall that throughout this section, we are employing a notation in which $i$, $j$, $l$ to denote the three linearly-independent primitive reciprocal lattice vectors $\mathbf{G}_i$, $\mathbf{G}_j$, and $\mathbf{G}_l$.
We can express the (partial) weak Chern number $\nu_{i}^{(\pm)}$ from $[P_{(\pm)}(\mathbf{k})]$ [SEq.~\eqref{eq:Ppm_multiple_separate_base_ki0}] computed in a BZ plane of constant-$k_i$ as
\begin{equation}
	\nu_i^{(\pm)} = {-\frac{i}{2\pi}}\int_{2D\  BZ \atop k_{i}=const} dk_j dk_l \mathrm{Tr}\left([P_{(\pm)} (\mathbf{k})]\left[\frac{\partial [P_{(\pm)}(\mathbf{k})]}{\partial k_j},\frac{\partial [P_{(\pm)}(\mathbf{k})]}{\partial k_l}\right]\right). \label{eq:app_partial_nu_pm}
\end{equation}
Similarly, the (partial) Chern number of the $n^{\mathrm{th}}$ group of the $\widehat{\mathbf{G}}_i$-directed $P_{(\pm)}$-Wannier bands in a constant-$k_i$ BZ plane can be obtained through
\begin{equation}
	C_{\gamma_{2},n}^{(\pm)} = {-\frac{i}{2\pi}}\int_{2D\ BZ \atop k_{i}=const} dk_j dk_l \mathrm{Tr}\left([\widetilde{P}^{(\pm)}_{\mathbf{G}_i,n}(\mathbf{k})]\left[\frac{\partial [\widetilde{P}^{(\pm)}_{\mathbf{G}_i,n}(\mathbf{k})]}{\partial k_j},\frac{\partial [\widetilde{P}^{(\pm)}_{\mathbf{G}_i,n}(\mathbf{k})]}{\partial k_l}\right]\right), \label{eq:C_gamma_2_n_pm_k_i}
\end{equation}
in which we have chosen $(j,l,i)$ to index a right-handed coordinate system.
$C_{\gamma_{2},n}^{(\pm)}$ in SEq.~\eqref{eq:C_gamma_2_n_pm_k_i} is identical to the winding number $C^{(\pm)}_{\gamma_2 , i,j} (k_i;l)$ defined in SEq.~\eqref{eq:app-C-gamma-2-l-ki} provided that, as specified above, $(j,l,i)$ align with the right-handed coordinate system in SEq.~\eqref{eq:app-C-gamma-2-l-ki}.
The fact that different groups of $\widehat{\mathbf{G}}_i$-directed $P_{(\pm)}$-Wannier bands are spectrally separated allows us to deduce that $C_{\gamma_{2},n}^{(\pm)}$ in SEq.~\eqref{eq:C_gamma_2_n_pm_k_i} does not depend on the choice of constant-$k_i$ BZ plane in which we perform the calculation on the right-hand side of SEq.~\eqref{eq:C_gamma_2_n_pm_k_i}, as is discussed in SN~\ref{app:nested_Berry_phase_only_winds_in_one_direction}.

Combining SEqs.~\eqref{eq:Ppm_multiple_separate_base_ki0}, \eqref{eq:P_multiple_separate_base_ki0_orthogonal},  \eqref{eq:Ppm_multiple_separate_base_ki0_orthogonal}, \eqref{eq:app_partial_nu_pm}, and \eqref{eq:C_gamma_2_n_pm_k_i}, it follows in analogy with SEq.~\eqref{eq:C_plus_C_minus_C_exact_relation} that~\cite{avron1998adiabatic}
\begin{equation}
	\nu_{i}^{(\pm)} = \sum_{n=1}^{N_{\mathrm{group}}^{(\pm)}} C_{\gamma_2 , n}^{(\pm)}. \label{eq:general_condition_diff_group_C_gamma_2_pm}
\end{equation}
This implies that the sum of the nested (partial) Chern numbers $C_{\gamma_2 , n}^{(\pm)}$ [SEq.~\eqref{eq:C_gamma_2_n_pm_k_i}] over all of the $N_{\mathrm{group}}^{(\pm)}$ groups of $P_{(\pm)}$-Wannier bands gives the (partial) weak Chern number of the occupied energy bands (or the upper/lower spin bands). 
In addition, from $[P(\mathbf{k})]=[P_{+}(\mathbf{k})] + [P_{-}(\mathbf{k})]$ where $[P_{+}(\mathbf{k})] [P_{-}(\mathbf{k})] = 0$, it also follows that
\begin{equation}
	\nu_{i} = \nu_{i}^{+} + \nu_{i}^{-}, \label{eq:relation-nu-and-nu-plus-nu-minus}
\end{equation}
such that the total weak Chern number of the occupied bands is the sum of the partial weak Chern numbers of the upper and lower spin bands. 
Before we apply SEq.~\eqref{eq:general_condition_diff_group_C_gamma_2_pm} to specific cases, we pause to briefly restate the essential assumptions made in the above derivations:
\begin{enumerate}
	\item The energy and spin gaps are open throughout the 3D BZ.
	\item The $\widehat{\mathbf{G}}_i$-directed $P_{(\pm)}$-Wannier bands can be separated into disjoint groups such that the decomposition in SEq.~\eqref{eq:Ppm_multiple_separate_base_ki0} is well-defined and smooth over the 3D BZ, which implies that two of the (partial) weak Chern numbers are zero, namely $\nu_j^{(\pm)}=\nu_l^{(\pm)} = 0$. 
    However, we note that $\nu^{(\pm)}_{j} = \nu^{(\pm)}_{l} = 0$ does not imply that the $\widehat{\mathbf{G}}_i$-directed $P_{(\pm)}$-Wannier bands can be separated into disjoint groups.
\end{enumerate}

Let us now consider some specific implications of SEq.~\eqref{eq:general_condition_diff_group_C_gamma_2_pm}.
First, consider a spinful system with time-reversal symmetry, and gaps in both the energy and $PsP$ spectrum at every $\mathbf{k}$ point throughout the BZ. 
The weak Chern number of the occupied bands at any constant-$k_{i}$ plane must be zero, which implies that $\nu_{i} = 0$.
From SEq.~\eqref{eq:relation-nu-and-nu-plus-nu-minus}, this further implies that 
{\begin{equation}
    \nu^{+}_{i} = -\nu^{-}_{i} \label{eq:partial_weal_Cherns_with_T_symmetry}
\end{equation}
in the presence of spinful time-reversal symmetry.}
Then, if we have $\nu^{\pm}_{i} = \pm m \neq 0$ where $m\in\mathbb{Z}$, we can deduce that
\begin{equation}
	\sum_{n=1}^{N_{\mathrm{group}}^{\pm}} C_{\gamma_2 , n}^{\pm} = \pm m. \label{eq:app:sum-of-C-pm-gamma-2-with-TR}
\end{equation}
This case occurs, for example, in the 3D quantum spin Hall insulator considered in SRef.~\cite{zhijun2021qshsquarenet}, in which both the energy and spin gaps are open, and the partial weak Chern numbers are $\nu_{1}^{\pm} = \pm 2$, $\nu_{2}^{\pm} = \nu_{3}^{\pm} = 0$.

{Next, consider a 3D insulator with an energy gap and $\nu_i = 0$, or a 3D insulator with energy and spin gaps and $\nu^{\pm}_{i} = 0$. 
The former can be realized, for instance, by a 3D magnetic axion insulator~\cite{wieder2018axion}, and the latter can be realized by a time-reversal-symmetric helical HOTI, as we will see later in SN~\ref{app:comparison-spin-stable-and-symmetry-indicated-topology}. 
The vanishing (partial) weak Chern number in either case implies that}
\begin{equation}
	\sum_{n=1}^{N_{\mathrm{group}}^{(\pm)}} C_{\gamma_{2},n}^{(\pm)} = 0.
\end{equation}
If we then choose to decompose the $\widehat{\mathbf{G}}_i$-directed $P_{(\pm)}$-Wannier bands into two groups (which is a convenient choice if the system has inversion symmetry, as will be discussed in detail in SN~\ref{app:z2_nested_P_pm_berry_phase}), we have that the winding numbers $C_{\gamma_{2},1^{\mathrm{st}}\ \mathrm{group}}^{(\pm)}$ and $C_{\gamma_{2},2^{\mathrm{nd}}\ \mathrm{group}}^{(\pm)}$ are opposite to each other:
\begin{equation}
	C_{\gamma_{2},1^{\mathrm{st}}\ \mathrm{group}}^{(\pm)} = -C_{\gamma_{2},2^{\mathrm{nd}}\ \mathrm{group}}^{(\pm)}. \label{eq:app-two-groups-C-gamma2-pm-relation-with-zero-nu-i-pm}
\end{equation}

Now that we have enumerated useful properties of nested $P_{(\pm)}$-Wilson loops, we can proceed to use them as tools for diagnosing (spin-resolved) band topology. 
In particular, in the next section we will show that  in inversion- and time-reversal symmetric systems, the winding number of the nested $P_\pm$-Wilson loop is a $\mathbb{Z}_2$-valued invariant that characterizes properties of a system that are robust as long as neither an energy gap nor a spin gap closes (termed ``spin stability'' in this work). 
This will allow us to deduce the existence of nonzero spin-electromagnetic response coefficients in 3D helical HOTIs from their spin-resolved electronic band topology.

As a final remark, we will now generalize our analysis to cases where all of the (partial) weak Chern numbers can be nonzero, provided that the $P_{(\pm)}$-Wannier bands can be divided into disjoint groupings. 
Using the (partial) weak Chern numbers $\{ \nu_1^{(\pm)} , \nu_2^{(\pm)} , \nu_3^{(\pm)} \}$ and the primitive reciprocal lattice vectors $\{ \mathbf{G}_1 , \mathbf{G}_2 , \mathbf{G}_3 \}$, we can construct a (partial) Chern vector~\cite{halperin1987possible,kohmoto1992diophantine,haldane2004berry,wieder2020axionic}
\begin{equation}
    \bm{\nu}^{(\pm)}= \nu_1^{(\pm)} \mathbf{G}_1 + \nu_2^{(\pm)} \mathbf{G}_2 + \nu_3^{(\pm)} \mathbf{G}_3 \label{eq:partial_Chern_vector_def}
\end{equation} 
which is related (for weak Chern numbers in the energy spectrum) to the Hall conductivity via
\begin{equation}
    \sigma_{ij}^{H} = \frac{e^2}{h} (\bm{\nu})_{k} \epsilon_{kij},\label{eq:3dhc} 
\end{equation}
or (for partial weak Chern numbers in the $PsP$ spectrum) to the topological contribution to the spin Hall conductivity [$C^s_{\gamma_1}$ in SEq.~\eqref{eq:intrinsicspinhall}] via
\begin{equation}
    \sigma_{ij,\mathrm{top}}^{s} = \frac{e}{4\pi} (\bm{\nu}^{+}-\bm{\nu}^{-})_{k} \epsilon_{kij}.\label{eq:3dshc}
\end{equation}
In SEqs.~\eqref{eq:3dhc} and \eqref{eq:3dshc}, $i,j=x,y,z$ index \emph{Cartesian} directions, and $\epsilon_{ijk}$ is the antisymmetric Levi-Civita symbol. 
From the primitive position-space lattice vectors $\{\mathbf{a}_1 , \mathbf{a}_2 , \mathbf{a}_3 \}$, we can form a (non-unique) set of supercell position-space lattice vectors $\{\mathbf{a}_1^{sc} , \mathbf{a}_2^{sc} , \mathbf{a}_3^{sc} \}$ where
\begin{align}
    & \mathbf{a}_1^{sc} = \nu_1^{(\pm)}\mathbf{a}_3 - \nu_3^{(\pm)} \mathbf{a}_1, \\ 
    & \mathbf{a}_2^{sc} = \nu_2^{(\pm)}\mathbf{a}_3 - \nu_3^{(\pm)} \mathbf{a}_2, \\
    & \mathbf{a}_3^{sc} = \mathbf{a}_3,
\end{align}
are all linear combinations of primitive lattice vectors $\{\mathbf{a}_1 , \mathbf{a}_2 , \mathbf{a}_3 \}$ with integral coefficients.
A semi-infinite slab formed from the supercell with finite thickness along $\mathbf{a}_3^{sc}$ has a surface normal vector parallel to the (partial) Chern vector $\bm{\nu}^{(\pm)}$ since~\cite{vanderbilt2018berry}
\begin{equation}
    \mathbf{a}_1^{sc} \cdot \bm{\nu}^{(\pm)} = \mathbf{a}_2^{sc} \cdot \bm{\nu}^{(\pm)} = 0.
\end{equation}
Assuming without loss of generality that the (partial) weak Chern number $\nu_3^{(\pm)} \neq 0$ [if not we can always permute our coordinates to facilitate this provided that $\bm{\nu}^{(\pm)}\neq {\bf 0}$ in SEq.~\eqref{eq:partial_Chern_vector_def}], the dual supercell primitive reciprocal lattice vectors  are $\{\mathbf{G}_1^{sc} , \mathbf{G}_2^{sc} , \mathbf{G}_3^{sc} \}$ where
\begin{align}
    & \mathbf{G}_1^{sc} = -\frac{1}{\nu_3^{(\pm)}} \mathbf{G}_1, \\
    & \mathbf{G}_2^{sc} = -\frac{1}{\nu_3^{(\pm)}} \mathbf{G}_2, \\
    & \mathbf{G}_3^{sc} = \frac{\nu_1^{(\pm)}}{\nu_3^{(\pm)}} \mathbf{G}_1 + \frac{\nu_2^{(\pm)}}{\nu_3^{(\pm)}} \mathbf{G}_2 + \mathbf{G}_3.
\end{align}
Recall that the (partial) Chern vector is a property of a set of occupied states in a 3D translationally-invariant system, which is independent of the description whether we use the primitive lattice vectors $\{\mathbf{a}_1 , \mathbf{a}_2 , \mathbf{a}_3 \}$ or the supercell lattice vectors $\{\mathbf{a}_1^{sc} , \mathbf{a}_2^{sc} , \mathbf{a}_3^{sc} \}$.
Therefore, the (partial) Chern vector of the 3D translationally-invariant system described by the supercell lattice vectors $\{\mathbf{a}_1^{sc} , \mathbf{a}_2^{sc} , \mathbf{a}_3^{sc} \}$ is the same as SEq.~\eqref{eq:partial_Chern_vector_def}.
To obtain the (partial) weak Chern numbers of the supercell system, we rewrite the (partial) Chern vector $\bm{\nu}^{(\pm)}$ [SEq.~\eqref{eq:partial_Chern_vector_def}] in terms of $\{\mathbf{G}_1^{sc} , \mathbf{G}_2^{sc} , \mathbf{G}_3^{sc} \}$ and obtain
\begin{align}
    \bm{\nu}^{(\pm)} &= \nu_1^{(\pm),sc} \mathbf{G}_1^{sc} + \nu_2^{(\pm),sc} \mathbf{G}_2^{sc} + \nu_3^{(\pm),sc} \mathbf{G}_3^{sc} \nonumber \\
    &= \nu_{3}^{(\pm)} \mathbf{G}_3^{sc}. \label{eq:partial_Chern_vectors_in_terms_of_G_sc}
\end{align}
SEq.~\eqref{eq:partial_Chern_vectors_in_terms_of_G_sc} implies that the supercell (partial) weak Chern numbers $\{\nu_1^{(\pm),sc},\nu_2^{(\pm),sc},\nu_3^{(\pm),sc}\}$ of the 3D translationally-invariant system whose Bravais lattice vectors are $\{\mathbf{a}_1^{sc},\mathbf{a}_2^{sc},\mathbf{a}_3^{sc} \}$ are given by
\begin{equation}
    \nu_1^{(\pm),sc} = 0,\quad \nu_2^{(\pm),sc} = 0,\quad \nu_3^{(\pm),sc} = \nu_{3}^{(\pm)}.
\end{equation}
Therefore, provided that the (partial) Chern vector $\bm{\nu}^{(\pm)}$ in SEq.~\eqref{eq:partial_Chern_vector_def} is a nonzero vector, we can choose a supercell spanned by $\{\mathbf{a}_1^{sc},\mathbf{a}_2^{sc},\mathbf{a}_3^{sc} \}$ such that only one of the supercell (partial) weak Chern numbers is nonzero .
As such, the results derived in this section can be directly generalized and applied to the supercell system to obtain the relation between the supercell (partial) weak Chern numbers $\nu_{3}^{(\pm),sc}$ and the winding numbers of the nested $P_{(\pm)}$-Wilson loop spectra, provided that the $\widehat{\mathbf{G}}_3^{sc}$-directed $P_{(\pm)}$-Wannier bands of the supercell system contain disjoint groupings.
We emphasize again that different choices of the non-unique supercell Bravais lattice vectors can lead to distinct supercell (partial) weak Chern numbers. 
However, physical observables like the (spin) Hall conductivity [SEqs.~\eqref{eq:3dhc} and \eqref{eq:3dshc}] only depend on the invariant (partial) Chern vector [SEq.~\eqref{eq:partial_Chern_vector_def}]. 
As long as one can obtain a set of supercell lattice vectors for which the $P_{(\pm)}$-Wannier bands of the supercell system computed along one of the supercell reciprocal lattice vectors can be separated into disjoint groupings such that two of the supercell (partial) weak Chern numbers are zero, the results derived in this section can be directly extended to the relation between the supercell (partial) weak Chern numbers and the winding numbers of the nested $P_{(\pm)}$-Wilson loop spectra of the supercell system.

\subsubsection{$\mathbb{Z}_2$-Stable Spectral Flow in the Nested $P_{\pm}$-Wilson Loop Eigenphases in the Presence of Bulk $\mathcal{I}$ and Spinful $\mathcal{T}$ Symmetries, and the Corresponding Spin-Stable Bulk Topology}\label{app:z2_nested_P_pm_berry_phase}

In this section, we will show that with inversion ($\mathcal{I}$) and time-reversal ($\mathcal{T}$) symmetry, the spectral flow of the nested Berry phases in the positive and negative $PsP$ eigenspaces is $\mathbb{Z}_2$-stable to deformations of the spin-resolved Wannier band structure, provided that both the energy and spin gap remain open throughout the deformation. 
To be specific, as in SN~\ref{app-relation-nested-P-pm-and-partial-weak-Chern}, we consider the nested Berry phases in SEq.~\eqref{eq:net_nested_Berry_pha_Gi_Gj} where the first and second closed-loop holonomy in the computation of the nested Berry phases are parallel to $\mathbf{G}_i$ and $\mathbf{G}_j$ ($i\neq j$), respectively.
We again denote the one remaining linearly-independent primitive reciprocal lattice vector as $\mathbf{G}_l$.
As was previously proved in SN~\ref{app:nested_Berry_phase_only_winds_in_one_direction}, the winding numbers of the nested Berry phases [SEq.~\eqref{eq:net_nested_Berry_pha_Gi_Gj}] for a given group of Wannier bands can only have nonzero winding as a function of $k_l$. 
{We will thus in this section exclusively consider the winding numbers of the nested Berry phases [SEq.~\eqref{eq:net_nested_Berry_pha_Gi_Gj}] for a given group of Wannier bands as a function of $k_l$.}

To begin, we consider 3D insulators with $\mathcal{I}$ and $\mathcal{T}$ symmetry whose energy and spin spectra are both gapped.
As shown in SN~\ref{appendix:properties-of-the-projected-spin-operator}, in the presence of $\mathcal{I}\mathcal{T}$ symmetry, the upper ($+$) and lower ($-$) eigenspace of $PsP$ are the positive and negative eigenspace of $PsP$.
In the presence of $\mathcal{I}$ symmetry, the $\widehat{\mathbf{G}}_i$-directed $P_{\pm}$-Wannier band eigenphases $(\gamma_1^\pm)_{j,\mathbf{k},\mathbf{G}_i}$ satisfy 
\begin{equation}
	\{ (\gamma_{1}^{\pm})_{j,\mathbf{k},\mathbf{G}_i } | j=1\ldots N_{\mathrm{occ}}^{\pm} \} = \{ -(\gamma_{1}^{\pm})_{j,-\mathbf{k},\mathbf{G}_i} | j=1\ldots N_{\mathrm{occ}}^{\pm} \}, \label{eq:app-ph-symmetry-gamma1pmkjkl}
\end{equation}
and in the presence of $\mathcal{T}$ symmetry, we have
\begin{equation}
	\{ (\gamma_{1}^{\pm})_{j,\mathbf{k},\mathbf{G}_i } | j=1\ldots N_{\mathrm{occ}}^{\pm} \} = \{ (\gamma_{1}^{\mp})_{j,-\mathbf{k},\mathbf{G}_i} | j=1\ldots N_{\mathrm{occ}}^{\pm} \}, \label{eq:app-TR-symmetry-gamma1pmkjkl}
\end{equation}
{where $(\gamma_{1}^{\pm})_{j,\mathbf{k},\mathbf{G}_i }$ is defined modulo $2\pi$.}
Note that $(\gamma_{1}^{\pm})_{j,\mathbf{k},\mathbf{G}_i }$ depends only on the momentum components $k_j$ and $k_l$.
We refer the readers to SN~\ref{appendix:I-constraint-on-P-pm} and \ref{appendix:T-constraint-on-P-pm} for the detailed proof of SEqs.~\eqref{eq:app-ph-symmetry-gamma1pmkjkl} and \eqref{eq:app-TR-symmetry-gamma1pmkjkl}, respectively.

\begin{figure}[ht]
\includegraphics[width=\columnwidth]{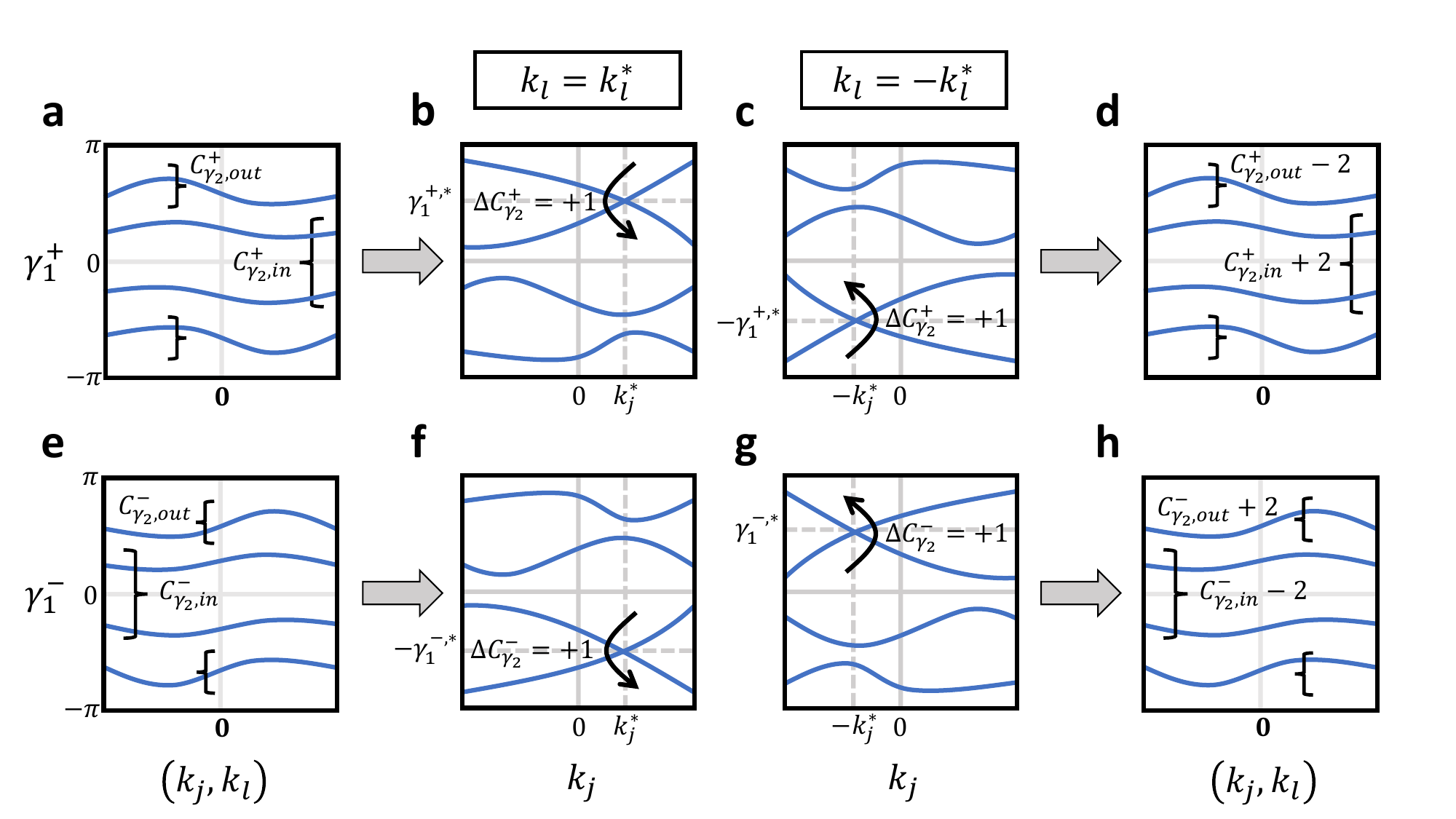}
\caption{Effect of Wannier gap closings on the nested $P_\pm$-Wilson loop spectrum in the presence of spinful time-reversal and 3D inversion symmetries. 
(a)--(d) schematically show a deformation of the Hamiltonian that induces $P_{+}$-Wannier band gap closings between the inner and outer set of $P_{+}$-Wannier bands.
The outer set of $P_{+}$-Wannier bands transfers $+2$ partial Chern number to the inner set of $P_{+}$-Wannier bands via the two $P_{+}$-Wannier band gap closings in (b) and (c) related by inversion symmetry.
After the deformation, the winding numbers of the nested Berry phases in the $P_+$-eigenphase change according to $C_{\gamma_2,in}^{+} \to C_{\gamma_2,in}^{+} + 2$ and $C_{\gamma_2,out}^{+} \to C_{\gamma_2,out}^{+} - 2$, as indicated in (a) and (d).
(e)--(h) show the corresponding deformation of the $P_{-}$-Wannier bands related to (a)--(d) by the operation of time-reversal. 
In particular, (a) and (e), (b) and (g), (c) and (f), (d) and (h) are related to each other by time-reversal.
The inner set of $P_{-}$-Wannier bands transfers $+2$ partial Chern number to the outer set of $P_{-}$-Wannier bands via the two $P_{-}$-Wannier band gap closings in (f) and (g) related by inversion symmetry.
After the deformation, the winding numbers of the nested Berry phases in the $P_{-}$-eigenspace change according to $C_{\gamma_2,in}^{-} \to C_{\gamma_2,in}^{-} - 2$ and $C_{\gamma_2,out}^{-} \to C_{\gamma_2,out}^{-} + 2$, as indicated in (e) and (h).
Both (b) and (f) are taken in the $k_l = k_l^{*}$ plane, and both (c) and (g) are in the $k_l = -k_l^{*}$ plane, in which the $P_\pm$-Wannier band gap closes during the deformation.
}\label{fig:schematics-of-Z2-classification-based-on-relative-winding}
\end{figure}

Assuming that there a gap in the $P_{\pm}$-Wannier bands, the inversion symmetry constraint in SEq.~\eqref{eq:app-ph-symmetry-gamma1pmkjkl} allows us to divide the $P_{\pm}$-Wannier bands into two disjoint groups---one centered around $\gamma_1^{\pm} = 0$, which we term the ``inner set'', and another around $\gamma_1^{\pm} = \pi$ (or equivalently $-\pi$), which we term the ``outer set''~\cite{wieder2018axion,wang2019higherorder,schindler2018higherorder,schindler2018higherordera,WiederDefect,benalcazar2017quantized,benalcazar2017electric,varnava2020axion}. 
A schematic example is shown in SFig.~\ref{fig:schematics-of-Z2-classification-based-on-relative-winding}(a,e).
The inner and outer sets of spin-resolved ($P_\pm$-) Wannier bands are effectively inversion-symmetric in the sense that they individually satisfy SEq.~\eqref{eq:app-ph-symmetry-gamma1pmkjkl}. 
In the following, we will use interchangeably the ``winding number of the nested Barry phase $(\gamma_2^{\pm})_{\mathbf{k},\mathbf{G}_i,\mathbf{G}_j}$ [SEq.~\eqref{eq:net_nested_Berry_pha_Gi_Gj}]'', and the term ``nested partial Chern number [SEq.~\eqref{eq:C_gamma_2_n_pm_k_i}]'', where the sign convention in SEq.~\eqref{eq:partial_chern_def} has been implicitly applied.
From SEqs.~\eqref{eq:general_condition_diff_group_C_gamma_2_pm}  and \eqref{eq:app:sum-of-C-pm-gamma-2-with-TR} in SN~\ref{app-relation-nested-P-pm-and-partial-weak-Chern}, we have that the nested partial Chern numbers for a system with time-reversal symmetry satisfy 
\begin{equation}
	C^{\pm}_{\gamma_2 , in} + C^{\pm}_{\gamma_2 , out} = \nu_i^{\pm} = \pm m, \label{eq:app-z2-discussion-sum-over-in-out}
\end{equation}
where $\nu_i^{\pm}$ is the partial weak Chern number (defined in SN~\ref{app-relation-nested-P-pm-and-partial-weak-Chern}) of the positive/negative spin bands at constant-$k_i$ planes, $m \in \mathbb{Z}$, and we have used ``in'' and ``out'' to denote the disjoint inner and outer groups of spin-resolved Wannier bands.
As is discussed in SN~\ref{app-relation-nested-P-pm-and-partial-weak-Chern}, by construction, we are here specializing (without loss of generality) to systems in which the other two partial weak Chern numbers $\nu_j^{\pm}$ and $\nu_{l}^{\pm}$ are zero due to the fact that the $\widehat{\mathbf{G}}_{i}$-directed $P_{\pm}$-Wannier bands contain disjoint groups.

In the presence of $\mathcal{T}$ symmetry, since the nested Berry phases $(\gamma_2^{\pm})_{\mathbf{k},\mathbf{G}_i,\mathbf{G}_j}$ for a given group of $P_{\pm}$-Wannier bands satisfy $(\gamma_2^{\pm})_{\mathbf{k},\mathbf{G}_i,\mathbf{G}_j} \text{ mod } 2\pi = (\gamma_2^{\mp})_{-\mathbf{k},\mathbf{G}_i,\mathbf{G}_j}$ mod $2\pi$ (we refer the readers to SN~\ref{appendix:T-constraint-on-nested-P-pm} for a detailed proof), we can deduce that the nested partial Chern numbers [SEq.~\eqref{eq:C_gamma_2_n_pm_k_i}] must satisfy
\begin{align}
	& C^{\pm}_{\gamma_2 , in} = -C^{\mp}_{\gamma_2 , in}, \label{eq:z2-gamma2-in-Cpm-is-minus-Cmp} \\
	& C^{\pm}_{\gamma_2 , out} = -C^{\mp}_{\gamma_2 , out}. \label{eq:z2-gamma2-out-Cpm-is-minus-Cmp}
\end{align}
Similar to SN~\ref{sec:general_properties_of_winding_num_of_P_pm_Wilson}, we may then define the {\it relative winding numbers} of the nested Berry phases between the positive and negative $PsP$ eigenspaces as
\begin{align}
	& C^{s}_{\gamma_2 , in} \equiv  C^{+}_{\gamma_2 , in} - C^{-}_{\gamma_2 , in},  \\
	& C^{s}_{\gamma_2 , out} \equiv C^{+}_{\gamma_2 , out}  - C^{-}_{\gamma_2 ,out}. 
\end{align}
We will also call $C^{s}_{\gamma_2 , in}$ and $C^{s}_{\gamma_2 , out}$ as the \textit{nested spin Chern numbers}.
From SEqs.~\eqref{eq:z2-gamma2-in-Cpm-is-minus-Cmp} and \eqref{eq:z2-gamma2-out-Cpm-is-minus-Cmp}, in the presence of $\mathcal{T}$ symmetry, we have
\begin{align}
	& C^{s}_{\gamma_2 , in} = 2 C^{+}_{\gamma_2 , in} = -2 C^{-}_{\gamma_2 , in}, \label{eq:z2-gamma2-in-Cs-is-even} \\
	& C^{s}_{\gamma_2 , out} = 2 C^{+}_{\gamma_2 , out} = -2 C^{-}_{\gamma_2 , out} . \label{eq:z2-gamma2-out-Cs-is-even}
\end{align}
In addition, since the partial weak Chern number $\nu_{i}^{+} = -\nu_{i}^{-}$ in the presence of $\mathcal{T}$ symmetry, SEqs.~\eqref{eq:app-z2-discussion-sum-over-in-out}, \eqref{eq:z2-gamma2-in-Cs-is-even}, and \eqref{eq:z2-gamma2-out-Cs-is-even} also imply that
\begin{equation}
	C^{s}_{\gamma_2 , in} + C^{s}_{\gamma_2 , out} = 2 \nu^{+}_i = -2 \nu^{-}_i. \label{eq:app-z2-sum-of-Csgamma2-in-and-out}
\end{equation}
From SEqs.~\eqref{eq:app-z2-discussion-sum-over-in-out}, \eqref{eq:z2-gamma2-in-Cpm-is-minus-Cmp}, and \eqref{eq:z2-gamma2-out-Cpm-is-minus-Cmp}, we see that the four winding numbers $C^{\pm}_{\gamma_2 , in}$ and $C^{\pm}_{\gamma_2 , out}$ of the nested Berry phases are fully specified given any of the four winding numbers and the partial weak Chern number
\begin{equation}
	\nu_{i}^{\pm} = \pm m\in \mathbb{Z}. \label{eq:nu_in_m}
\end{equation}
To be explicit, using SEqs.~\eqref{eq:app-z2-discussion-sum-over-in-out}, \eqref{eq:z2-gamma2-in-Cpm-is-minus-Cmp}, \eqref{eq:z2-gamma2-out-Cpm-is-minus-Cmp}, and \eqref{eq:nu_in_m}, it can be shown that if 
\begin{equation}
	C^{+}_{\gamma_2 , in} = C \in \mathbb{Z}, \label{eq:C_gamma_2_in_C_1}
\end{equation}
then
\begin{align}
	& C^{+}_{\gamma_2 , out} = -C + m, \label{eq:C_gamma_2_in_C_2} \\
	& C^{-}_{\gamma_2 , in} = -C, \label{eq:C_gamma_2_in_C_3} \\
	& C^{-}_{\gamma_2 , out} = C-m, \label{eq:C_gamma_2_in_C_4}
\end{align}
which from SEqs.~\eqref{eq:z2-gamma2-in-Cs-is-even} and \eqref{eq:z2-gamma2-out-Cs-is-even} also imply that the nested spin Chern numbers are 
\begin{align}
	& C^{s}_{\gamma_2 , in} = 2 C, \\
	& C^{s}_{\gamma_2 , out} = -2 C+2m.
\end{align}

As an example, we can consider an inversion-symmetric 3D quantum spin Hall insulator with $s_z$ conservation and $\nu_i^{\pm}=\pm 2$~\cite{zhijun2021qshsquarenet}, as will be detailed in SN~\ref{app:comparison-spin-stable-and-symmetry-indicated-topology}. 
If there is a Wannier gap, then such a system can have $C^{+}_{\gamma_2 , in} = +1$, from which we can deduce that $C^{+}_{\gamma_2 , out} = +1$, $C^{-}_{\gamma_2 , in} = -1$, $C^{-}_{\gamma_2 , out} = -1$, $C^{s}_{\gamma_2 , in} = +2$, and $C^{s}_{\gamma_2 ,out}=+2$. 
We will show in SN~\ref{app:comparison-spin-stable-and-symmetry-indicated-topology} that this configuration of nested partial Chern numbers can be realized through a layer construction with two 2D quantum spin Hall insulators in each unit cell. 
{For introductions to and further discussions of the layer-construction method for enumerating and analyzing symmetry-protected topological states, interested readers may consult SRefs.~\cite{elcoro2021magnetic,gao2022magnetic,song2019topological,huang2017building,song2018mapping,fang2019new}.}
Another example of a system that has an open spin gap and an open energy gap is a 3D helical higher-order topological insulator (HOTI)~\cite{wang2019higherorder} with $\nu_i^{\pm}=0$ and $C^{+}_{\gamma_2 , in} = +1$, which can also be obtained from a layer construction (detailed in SN~\ref{app:comparison-spin-stable-and-symmetry-indicated-topology} and SRefs.~\cite{song2018mapping,song2020real}) and will be the main numerical focus in SN~\ref{sec:numerical-section-of-nested-P-pm}. 
$\nu_i^{\pm}=0$ and $C^{+}_{\gamma_2 , in} = +1$ implies that for this helical HOTI $C^{+}_{\gamma_2 , out} = -1$, $C^{-}_{\gamma_2 , in} = -1$, $C^{-}_{\gamma_2 , out} = +1$, $C^{s}_{\gamma_2 , in} = +2$, and $C^{s}_{\gamma_2 , out} = -2$.
{We will show in SN~\ref{app:comparison-spin-stable-and-symmetry-indicated-topology} that $C^{\pm}_{\gamma_2 , out} = \mp 1$ and $C^{\pm}_{\gamma_2 , in} = \pm 1$ can be realized via a layer construction by placing one $\mathcal{I}$- and spinful $\mathcal{T}$-symmetric 2D TI with partial Chern numbers $C_{\gamma_1}^{\pm} = \pm 1$ in the $\mathcal{I}$-symmetric planes containing the origin of each unit cell, along with placing one $\mathcal{I}$- and spinful $\mathcal{T}$-symmetric 2D TI with partial Chern numbers $C_{\gamma_1}^{\pm} = \mp 1$ in the inversion-symmetric planes at the boundary of each unit cell (defined to be a half-lattice translation from the origin)~\cite{wang2019higherorder,schindler2018higherorder,schindler2018higherordera}.} 
We will explicitly demonstrate this in SN~\ref{sec:numerical-section-of-nested-P-pm} through a numerical example implementing our nested spin-resolved Wilson loop formalism.

Recall from SEq.~\eqref{eq:C_gamma_2_n_pm_k_i} that the winding numbers of the nested Berry phases in the $P_{\pm}$-eigenspaces are identical to the partial Chern numbers of the $P_{\pm}$-Wannier bands. 
Let us now consider a deformation of the Hamiltonian that preserves $\mathcal{I}$ and spinful $\mathcal{T}$ symmetries and keeps the energy and spin gaps open.
Focusing on the $P_{+}$-Wannier bands, from SEq.~\eqref{eq:app-ph-symmetry-gamma1pmkjkl}, if there is a Wannier gap closing between the inner and outer set of bands at $\gamma_1^{+,*} = \gamma$ and crystal momentum $(k_j^{*},k_{l}^{*})$ [see SFig.~\ref{fig:schematics-of-Z2-classification-based-on-relative-winding}(b)], there must be another gap closing at $-\gamma_1^{+,*}=-\gamma$ and crystal momentum $(-k_j^{*},-k_{l}^{*})$ [see SFig.~\ref{fig:schematics-of-Z2-classification-based-on-relative-winding}(c)] related by inversion symmetry.
These two gap closings transfer the same partial Chern numbers between the inner and outer set of $P_{+}$-Wannier bands, such that after the deformation, the winding numbers can change according to
\begin{align}
	& C^{+}_{\gamma_2 , in} \to C^{+}_{\gamma_2 , in}  + 2n, \label{eq:app-C-plus-gamma-2-in-change-2n} \\
	& C^{+}_{\gamma_2 , out} \to C^{+}_{\gamma_2 , out}  - 2n, \label{eq:app-C-plus-gamma-2-out-change-2n}
\end{align}
where $n \in \mathbb{Z}$.
As an example, in SFig.~\ref{fig:schematics-of-Z2-classification-based-on-relative-winding}(a--d) we schematically show the case of SEqs.~\eqref{eq:app-C-plus-gamma-2-in-change-2n} and \eqref{eq:app-C-plus-gamma-2-out-change-2n} with $n=1$.

On the other hand, due to $\mathcal{T}$ symmetry, SEq.~\eqref{eq:app-TR-symmetry-gamma1pmkjkl} implies that simultaneously there will be also be a pair of gap closings at $\pm \gamma_1^{-,*} = \pm \gamma$ and crystal momentum $(\mp k_j^{*},\mp k_{l}^{*})$ between the inner and outer set of $P_{-}$-Wannier bands, see for instance SFig.~\ref{fig:schematics-of-Z2-classification-based-on-relative-winding}(f,g).
Since this pair of band crossings in the $P_{-}$-Wannier bands are related to the previous pair of band crossings in the $P_{+}$-Wannier bands by time-reversal, we deduce that after the deformation, we have
\begin{align}
	& C^{-}_{\gamma_2 , in} \to C^{-}_{\gamma_2 , in}  - 2n, \label{eq:app-C-minus-gamma-2-in-change-2n} \\
	& C^{-}_{\gamma_2 , out} \to C^{-}_{\gamma_2 , out}  + 2n. \label{eq:app-C-minus-gamma-2-out-change-2n}
\end{align}
As an example, in SFig.~\ref{fig:schematics-of-Z2-classification-based-on-relative-winding}(e--h) we have schematically shown the case of SEqs.~\eqref{eq:app-C-minus-gamma-2-in-change-2n} and \eqref{eq:app-C-minus-gamma-2-out-change-2n} with $n=1$.

From SEqs.~\eqref{eq:app-C-plus-gamma-2-in-change-2n}, \eqref{eq:app-C-plus-gamma-2-out-change-2n}, \eqref{eq:app-C-minus-gamma-2-in-change-2n}, and \eqref{eq:app-C-minus-gamma-2-out-change-2n}, the relative winding numbers $C^{s}_{\gamma_2 , in/out}$, which are even numbers from SEqs.~\eqref{eq:z2-gamma2-in-Cs-is-even} and \eqref{eq:z2-gamma2-out-Cs-is-even}, will also change according to
\begin{align}
	& C^{s}_{\gamma_2 , in} \to C^{s}_{\gamma_2 , in} + 4n, \label{eq:app-C-s-gamma-2-in-change-4n} \\
	& C^{s}_{\gamma_2 , out} \to C^{s}_{\gamma_2 , out} - 4n. \label{eq:app-C-s-gamma-2-out-change-4n}
\end{align}
{In particular SEqs.~\eqref{eq:app-C-plus-gamma-2-in-change-2n}--\eqref{eq:app-C-s-gamma-2-out-change-4n} ensure that the nested partial Chern numbers satisfy SEqs.~\eqref{eq:app-z2-discussion-sum-over-in-out}--\eqref{eq:z2-gamma2-out-Cpm-is-minus-Cmp}, and \eqref{eq:app-z2-sum-of-Csgamma2-in-and-out} both before and after the deformation.}
Therefore, we can deduce that the quantities $C_{\gamma_2 , in}^{\pm}$, $C_{\gamma_2 , out}^{\pm}$, $\frac{1}{2} C_{\gamma_2 , in}^{s}$, and $\frac{1}{2} C_{\gamma_2 , out}^{s}$ are all $\mathbb{Z}_2$-stable when a spin-resolved Wannier gap closes and reopens. 
In fact all of $C_{\gamma_2 , in}^{\pm}$, $C_{\gamma_2 , out}^{\pm}$, $\frac{1}{2} C_{\gamma_2 , in}^{s}$, and $\frac{1}{2} C_{\gamma_2 , out}^{s}$ are uniquely specified given $(C_{\gamma_{2},in}^{+},\nu_{i}^{+})$.
{In all the following discussion, we refer to $(C_{\gamma_{2},in}^{+}\mod 2,\nu_{i}^{+})$ as the \emph{spin-stable invariant} for a 3D inversion- and spinful time-reversal-invariant system with an energy gap, a spin gap, and $\nu_j^+=\nu^+_l=0$.}

If both the energy gap and the gap between the positive and negative $PsP$ eigenspaces are required to be opened during a deformation of the bulk that preserves $\mathcal{I}$ and $\mathcal{T}$ symmetry, $\nu_{i}^{+} \in \mathbb{Z}$ cannot change.
Therefore, we have that
\begin{equation}
	(C_{\gamma_{2},in}^{+}\mod 2,\nu_{i}^{+}) \in \mathbb{Z}_2 \times \mathbb{Z} \label{eq:z2xzphasesdef}
\end{equation}
classifies the spin-stable topology of a 3D insulator with $\mathcal{I}$ and spinful $\mathcal{T}$ symmetry and $\nu^\pm_j=\nu^\pm_l=0$ (recalling that we have taken $\nu_j^{\pm}$ and $\nu_{l}^{\pm}$ to be zero in order to guarantee the possibility of spectrally separated $\widehat{\mathbf{G}}_i$-directed $P_{\pm}$-Wannier bands).
In SN~\ref{app:comparison-spin-stable-and-symmetry-indicated-topology} we will compare this $\mathbb{Z}_2 \times \mathbb{Z}$ spin-stable topology with symmetry-indicated (energy) band topology.
We will then show that a symmetry-indicated helical HOTI can be further resolved into (at least) two families of spin-stable (spin-gapped) phases with different spin-response effects, {which includes the experimentally observable spin-magnetoelectric response and the topological contribution to the bulk spin Hall conductivity.}
{After we introduce the notion of \emph{spin-resolved elementary layer construction} in SN~\ref{app:comparison-spin-stable-and-symmetry-indicated-topology}, we will also discuss how to systematically enumerate and analyze various families of 3D spin-stable (spin-gapped) phases with inversion and spinful time-reversal symmetries by stacking 2D insulators in 3D.
The stacking approach has the merit of making the spin-response effects physically manifest.
}

\subsection{Comparison Between Spin-Stable Topology and Symmetry-Indicated Topology in $\mathcal{I}$- and Spinful $\mathcal{T}$-Symmetric 3D Insulators}
\label{app:comparison-spin-stable-and-symmetry-indicated-topology}

In this section, we will show that the spin-stable topology (SN~\ref{app:z2_nested_P_pm_berry_phase}) characterized by the spin-resolved Wannier band configurations $(C^+_{\gamma_2,in},\nu_i^+)$ [SEq.~\eqref{eq:z2xzphasesdef}] provides a finer topological classification than the usual stable (symmetry-indicated) (energy) band topology~\cite{xu2020highthroughput,po2017symmetry,song2018mapping,khalaf2018symmetry,wang2019higherorder}.
We will introduce a formal notion of spin-resolved layer constructions and the corresponding symmetry indicators (SIs) to enumerate and analyze insulators in the nonmagnetic space group $P \bar{1} 1'$ (\# $2.5$) [generated by 3D translations, inversion ($\mathcal{I}$), and spinful time-reversal ($\mathcal{T}$)] with a spin gap, which can include the $\mathcal{I}$- and $\mathcal{T}$-protected helical higher-order topological insulators (HOTIs) studied in this work.
We will then discuss the different spin-electromagnetic responses of two families of spin-stable (spin-gapped) insulators that are both symmetry-indicated helical HOTIs in the nonmagnetic space group $P \bar{1} 1'$ (\# $2.5$).

In this section, we consider 3D insulators in the nonmagnetic space group $P\bar{1} 1'$ (\# $2.5$) with a spin gap.
Provided that the spin gap is open, we can spin-resolve the occupied electronic bands into positive ($+$) and negative $(-)$ spin bands, as the system has both $\mathcal{I}$ and spinful $\mathcal{T}$ symmetries [see also SN~\ref{appendix:properties-of-the-projected-spin-operator} and SFig.~\ref{fig:schematic_spin_bands_with_I_T_and_IT}(c)].
We denote the 3D partial weak Chern numbers as $\{\nu_1^{\pm},\nu_2^{\pm},\nu_3^{\pm} \}$ [SEq.~\eqref{eq:app_partial_nu_pm}].
Following the discussion in SN~\ref{app-relation-nested-P-pm-and-partial-weak-Chern} and \ref{app:z2_nested_P_pm_berry_phase}, we assume that in the presence of $\mathcal{I}$ symmetry, without loss of generality, the $\widehat{\mathbf{G}}_3$-directed $P_{\pm}$-Wannier bands can be separated into inner and outer groups centered around the partial Berry phase values $\gamma_1^{\pm} = 0$ and $\gamma_1^\pm = \pi$ (or equivalently $-\pi$) respectively, which by construction implies that the partial weak Chern numbers
\begin{equation}
    \nu_1^\pm = \nu_2^{\pm} = 0. \label{eq:say_again_that_nu_1_pm_nu_2_pm_are_zero}
\end{equation}
Recall from SN~\ref{app-relation-nested-P-pm-and-partial-weak-Chern} that we can always choose a supercell such that SEq.~\eqref{eq:say_again_that_nu_1_pm_nu_2_pm_are_zero} holds. 
The nested partial Chern numbers, namely the partial Chern numbers of the inner and outer groups of $\widehat{\mathbf{G}}_3$-directed $P_{\pm}$-Wannier bands, satisfy $C_{\gamma_2,in}^{\pm} + C_{\gamma_2,out}^{\pm} = \nu_3^\pm$ from SEq.~\eqref{eq:general_condition_diff_group_C_gamma_2_pm}.
Spinful $\mathcal{T}$ symmetry implies from SEqs.~\eqref{eq:partial_weal_Cherns_with_T_symmetry}, \eqref{eq:z2-gamma2-in-Cpm-is-minus-Cmp}, and \eqref{eq:z2-gamma2-out-Cpm-is-minus-Cmp} that $C_{\gamma_2,in}^{+} = -C_{\gamma_2,in}^{-}$, $C_{\gamma_2,out}^{+} = -C_{\gamma_2,out}^{-}$, and $\nu_3^{+} = -\nu_3^{-}$. 
We then have that the pair $(C_{\gamma_2,in}^{+},\nu_3^+)$ determines the remaining nested partial Chern numbers as
\begin{align}
	& C_{\gamma_2,in}^{\pm} = \pm C_{\gamma_2,in}^{+}, \label{eq:compare-spin-stable-C-gamma-2-in-pm}\\
	& C_{\gamma_2,out}^{\pm} = \pm \left( -C_{\gamma_2,in}^+ + \nu_3^+ \right). \label{eq:compare-spin-stable-C-gamma-2-out-pm}
\end{align}

We next consider deforming the Hamiltonian in a manner that respects the symmetries of the nonmagnetic space group $P\bar{1}1'$ (\# $2.5$) and does not close either the energy gap or the spin gap. 
Under such a deformation, two spin-resolved Wannier band configurations can be deformed into each other provided that
\begin{equation}
	(C_{\gamma_2 , in}^{+},\nu_3^{+})_{\mathrm{configuration}\ 1} - (C_{\gamma_2 , in}^{+},\nu_3^{+})_{\mathrm{configuration}\ 2} = (2n,0), \label{eq:spin-stable-condition-1}
\end{equation}
where $n \in \mathbb{Z}$. 
The deformation proceeds via pairs of gap closings and reopenings related by $\mathcal{I}$ between the inner and outer groups of the $\widehat{\mathbf{G}}_3$-directed $P_{\pm}$-Wannier bands, as described in SFig.~\ref{fig:schematics-of-Z2-classification-based-on-relative-winding}.
For example, the trivial spin-resolved Wannier band configuration $(C_{\gamma_2 , in}^{+},\nu_3^{+})=(0,0)$ can be deformed into $(C_{\gamma_2 , in}^{+},\nu_3^{+})=(2,0)$ as depicted in SFig.~\ref{fig:schematics-of-Z2-classification-based-on-relative-winding}, where the $P_{\pm}$-Wannier bands transfer $\pm 2$ partial Chern number from the outer to inner $P_{\pm}$-Wannier bands.
Hence $(C_{\gamma_2 , in}^{+},\nu_3^{+})=(2,0)$ is equivalent to the trivial $(C_{\gamma_2 , in}^{+},\nu_3^{+})=(0,0)$ in the $\mathbb{Z}_2 \times \mathbb{Z}$ classification of the spin-stable topological phases with vanishing partial weak Chern numbers $\nu_1^\pm=\nu_2^\pm=0$, as per SEq.~\eqref{eq:z2xzphasesdef}. 
On the other hand, the trivial spin-resolved Wannier band configuration $(C_{\gamma_2 , in}^{+},\nu_3^{+})=(0,0)$ cannot be deformed into $(C_{\gamma_2 , in}^{+},\nu_3^{+})=(1,1)$ or $(0,1)$ since the difference between their $\mathbb{Z}_2 \times \mathbb{Z}$ spin-stable invariants violates SEq.~\eqref{eq:spin-stable-condition-1}.
In fact, the spin-resolved Wannier band configuration with $(C_{\gamma_2 , in}^{+},\nu_3^{+})=(1,1)$ corresponds to a 3D weak topological insulator (WTI)~\cite{fu2007topological,fu2007topologicala}. 
This is shown schematically in SFig.~\ref{fig:elementary-layer-construction-WTI-oWTI}(a). 
Similarly, the configuration $(C_{\gamma_2 , in}^{+},\nu_3^{+})=(0,1)$ corresponds to a 3D ``obstructed'' weak topological insulator (oWTI), which is related to WTI by a half-lattice translation~\cite{wieder2020axionic}. 
This is shown schematically in SFig.~\ref{fig:elementary-layer-construction-WTI-oWTI}(b). 
Both $(C_{\gamma_2 , in}^{+},\nu_3^{+})=(1,1)$ and $(0,1)$ are of particular importance in this work, as they serve as the building blocks of spin-stable topological crystalline phases in the nonmagnetic space group $P\bar{1}1'$ (\# $2.5$) with a spin gap and partial weak Chern numbers $\nu_1^\pm = \nu_2^\pm  = 0$. 
We will shortly discuss these two cases in greater detail.

In Supplementary Table~\ref{tab:four-winding-num-given-C-m}, we summarize values of nested partial Chern numbers $C_{\gamma_2,in}^{\pm}$ [SEq.~\eqref{eq:compare-spin-stable-C-gamma-2-in-pm}] and $C_{\gamma_2,out}^{\pm}$ [SEq.~\eqref{eq:compare-spin-stable-C-gamma-2-out-pm}] consistent (up to the addition of a trivial spin-resolved Wannier band configuration) with the $\mathbb{Z}_2 \times \mathbb{Z}$ spin-stable invariants $(C_{\gamma_2,in}^+\mod 2,\nu_3^+)$ in SEq.~\eqref{eq:z2xzphasesdef}.
A spin-resolved Wannier band configuration has trivial invariant $(C_{\gamma_2,in}^+ \mod 2,\nu_3^+)=(0,0)$ if the inner and outer groups of spin-resolved hybrid Wannier functions have nested partial Chern numbers $C_{\gamma_2,in}^\pm = \pm 2n$ and  $C_{\gamma_2,out}^\pm = \mp 2n$ where $n \in \mathbb{Z}$, respectively [as shown in SEq.~\eqref{eq:spin-stable-condition-1}].
From Supplementary Table~\ref{tab:four-winding-num-given-C-m}, a spin-resolved Wannier band configuration has $(C_{\gamma_2,in}^+ \mod 2,\nu_3^+)=(1,1)$ if the inner and outer groups of the spin-resolved hybrid Wannier functions have nested partial Chern numbers $C_{\gamma_2,in}^\pm = \pm 1$ and  $C_{\gamma_2,out}^\pm = 0$, respectively, modulo the addition of a trivial spin-resolved Wannier band configuration.
On the other hand, up to the addition of a trivial configuration $(C_{\gamma_2,in}^+ \text{ mod } 2,\nu_3^+)=(0,0)$, a spin-resolved Wannier band configuration has $(C_{\gamma_2,in}^+ \mod 2,\nu_3^+)=(0,1)$ if the inner and outer groups of the spin-resolved hybrid Wannier functions have nested partial Chern numbers $C_{\gamma_2,in}^\pm = 0$ and  $C_{\gamma_2,out}^\pm = \pm 1$, respectively.
Importantly, $(C_{\gamma_2,in}^+ \mod 2,\nu_3^+)=(1,1)$ and $(0,1)$ can be taken as the generators of the $\mathbb{Z}_2 \times \mathbb{Z}$ spin-stable invariants in nonmagnetic space group $P\bar{1}1'$ (\# $2.5$) with a spin gap and partial weak Chern numbers $\nu_1^\pm = \nu_2^\pm = 0$.
Any spin-resolved Wannier band configurations can be derived from linear combinations of these two generators where a ``negative'' generator is obtained by reversing the signs of $(C_{\gamma_2,in}^+,\nu_3^+)$, which according to SEqs.~\eqref{eq:compare-spin-stable-C-gamma-2-in-pm} and \eqref{eq:compare-spin-stable-C-gamma-2-out-pm} amounts to reversing the signs of the corresponding (nested) partial Chern numbers.
Several examples of spin-stable (spin-gapped) phases in the nonmagnetic space group $P\bar{1}1'$ (\# $2.5$) with  $\nu_1^\pm = \nu_2^\pm = 0$ built using the configurations $(C_{\gamma_2,in}^+ ,\nu_3^+)=(1,1)$ and $(0,1)$ are shown and discussed in Supplementary Table~\ref{tab:four-winding-num-given-C-m}.

\begin{table}[ht]
\begin{tabular}{ |c|c|c|c||c|c|c| } 
 \hline
 ~ & \multicolumn{6}{c|}{$(C_{\gamma_2 , in}^{+}\mod 2 , \nu_3^{+})\in \mathbb{Z}_2 \times \mathbb{Z}$ Spin-Stable Invariants} \\
 \hline
   & $(0,0)$ & $(1,1)$ & $(0,1)$ & $(0,2)$ & $(1,2)$ & $(1,0)$   \\ 
 \hline
 $C_{\gamma_2,in}^{\pm}$ & $\pm 2n$ & $\pm 1$ & $0$ & $0$ & $\pm 1$  & $\pm 1$    \\ 
 \hline
 $C_{\gamma_2,out}^{\pm}$ & $\mp 2n$ & $0$ & $\pm 1$ & $\pm 2$ & $\pm 1$  & $\mp 1$   \\ 
 \hline
\end{tabular}
\caption{\label{tab:four-winding-num-given-C-m}Values of the nested partial Chern numbers (spin-resolved Wannier band configurations) $C_{\gamma_2,in}^\pm$ and $C_{\gamma_2,out}^\pm$ consistent with the given $\mathbb{Z}_2 \times \mathbb{Z}$ spin-stable invariants $(C_{\gamma_2 , in}^{+}\mod 2,\nu_3^{+})$, assuming that the partial weak Chern numbers $\nu_1^\pm=\nu_2^\pm=0$ as in SEq.~\eqref{eq:say_again_that_nu_1_pm_nu_2_pm_are_zero} (which is required to admit the possibility of a gap in the $\widehat{\mathbf{G}}_3$-directed $P_\pm$-Wilson loop spectrum). 
The first column indicates that the trivial configuration $(0,0)$ is consistent with  even-integer nested partial Chern numbers $C_{\gamma_2,in}^\pm=\pm 2n$ for the inner $P_{\pm}$-Wanner bands, and opposite even-integer nested partial Chern numbers $C_{\gamma_2,out}^\pm=\mp 2n$ for the outer $P_{\pm}$-Wannier bands. 
The configurations $(1,1)$ and $(0,1)$ can be taken as the generators of the $\mathbb{Z}_2\times\mathbb{Z}$ group of spin-stable (spin-gapped) phases in the nonmagnetic space group $P\bar{1}1'$ (\# $2.5$) with $\nu_1^\pm=\nu_2^\pm=0$, where we define the negative of the generators by reversing the signs of the nested partial Chern numbers in the table. 
In particular, $(1,1)$ corresponds to the $\nu_3^\pm = \pm 1$ weak topological insulator~\cite{fu2007topological,fu2007topologicala} depicted in SFig.~\ref{fig:elementary-layer-construction-WTI-oWTI}(a), and $(0,1)$ corresponds to the $\nu_3^\pm = \pm 1$ obstructed weak topological insulator~\cite{wieder2020axionic} depicted in SFig.~\ref{fig:elementary-layer-construction-WTI-oWTI}(b).
After the double-vertical line in the table, we give three examples of spin-stable phases with spin-resolved Wannier band configurations derived from linear combinations of the generators. 
To add two configurations we add their inner and outer nested partial Chern numbers, or equivalently add their $(C_{\gamma_2,in}^+,\nu_3^+)$.
The $(0,2)$ phase can be realized with the configuration $C_{\gamma_2 , in}^{\pm}=0$ and $C_{\gamma_2 , out}^{\pm}=\pm2$, corresponding to the linear combination $(0,2)=(0,1)\oplus(0,1)$. 
The $(1,2)$ spin-stable phase can be realized with the configuration $C_{\gamma_2 , in}^{\pm}=\pm1$ and $C_{\gamma_2 , out}^{\pm}=\pm1$, corresponding to the linear combination $(1,2)=(0,1)\oplus(1,1)$. 
This corresponds to the $\nu_3^\pm=\pm 2$ quantum spin Hall insulator (QSHI), depicted in SFig.~\ref{fig:layer-construction-QSHI-DAXI}(a). 
The $(1,0)$ spin-stable phase can be realized with the configuration $C_{\gamma_2 , in}^{+}=\pm1$ and $C_{\gamma_2 , out}^{+}=\mp1$, corresponding to the linear combination $(1,0)=(1,1)\ominus(0,1)$. 
This phase corresponds to the $\mathcal{T}$-doubled axion insulator (T-DAXI) depicted in SFig.~\ref{fig:layer-construction-QSHI-DAXI}(b). 
We note that we can add a configuration in the trivial phase $(0,0)$ for any integer $n$ to any of these representative configurations  without changing the spin-stable phase.
This corresponds to the fact that, as shown in SEq.~\eqref{eq:z2xzphasesdef} and the surrounding text, $C_{\gamma_2,in}^+$ is only stable modulo $2$. } 
\end{table}

\subsubsection{Spin-Resolved Layer Constructions}

\begin{figure}[ht]
\includegraphics[width=\textwidth]{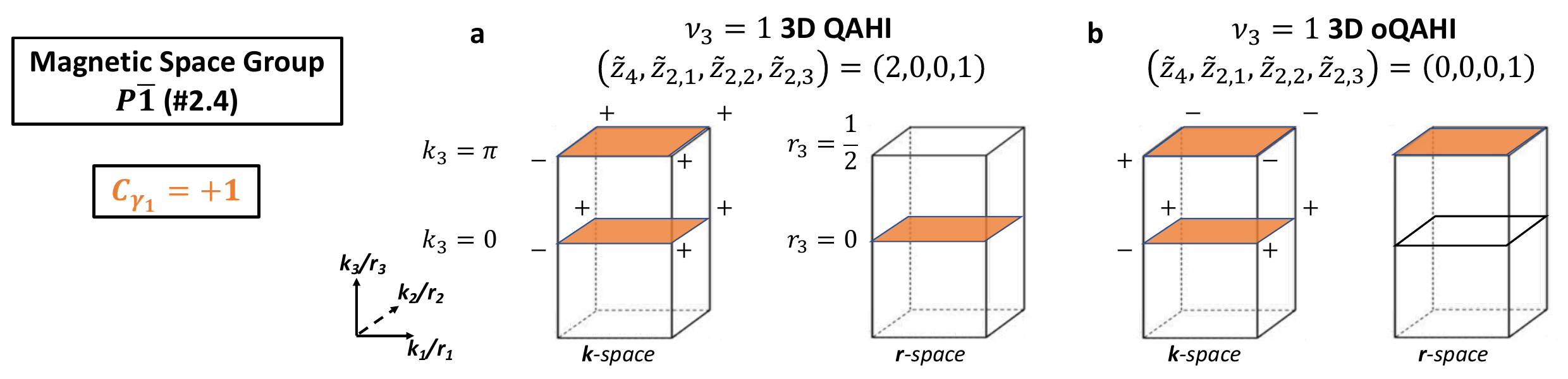}
\caption{Elementary layer constructions for topological crystalline phases in the magnetic space group $P\bar{1}$ (\# $2.4$) with weak Chern numbers $\nu_1 = \nu_2 = 0$.
The coordinates $k_{i}$ and $r_{i}$ ($i=1,2,3$) in momentum($\mathbf{k}$)-space and position($\mathbf{r}$)-space are given by $\mathbf{k} = \sum_{i=1}^{3} \frac{k_i}{2\pi} \mathbf{G}_i$ and $\mathbf{r} = \sum_{i=1}^{3} r_{i} \mathbf{a}_i$ respectively. 
$\{ \mathbf{a}_1 , \mathbf{a}_2 , \mathbf{a}_3 \}$ are the $\mathbf{r}$-space primitive lattice vectors and $\{ \mathbf{G}_1 , \mathbf{G}_2 , \mathbf{G}_3 \}$ are the dual $\mathbf{k}$-space primitive reciprocal lattice vectors such that $\mathbf{a}_i \cdot \mathbf{G}_j = 2\pi \delta_{ij}$ ($i,j=1,2,3$).
The $\mathcal{I}$-invariant constant-$r_3$ planes contain $r_3 = 0$ and $r_3 = 1/2$, which correspond to the center and the boundary of the primitive unit cell along $\mathbf{a}_3$.
Similarly, the $\mathcal{I}$-symmetric constant-$k_3$ planes contain $k_3 = 0$ and $k_3 = \pi$, which correspond to the center and the boundary of the BZ along $\mathbf{G}_3$.
(a) shows the layer construction of a $\nu_3 = 1$ 3D quantum anomalous Hall insulator (QAHI) in both $\mathbf{r}$-space (right) and $\mathbf{k}$-space (left).
In $\mathbf{r}$-space, the $\nu_3 = 1$ 3D QAHI can be constructed by tiling 3D space with 2D $\mathcal{I}$-symmetric Chern insulators with Chern number $C_{\gamma_1} = 1$ placed in the $\mathcal{I}$-symmetric $r_3 = 0$ plane within each unit cell, as shown on the right-hand side of (a).
The occupied energy bands in momentum space for this layer construction each have Chern number $C_{\gamma_1}=1$ in both the $\mathcal{I}$-symmetric $k_3 = 0$ and $k_3 = \pi$ planes.
(b) shows the layer construction of a $\nu_3 = 1$ 3D obstructed quantum anomalous Hall insulator (oQAHI) in both $\mathbf{r}$-space (right) and $\mathbf{k}$-space (left)~\cite{wieder2020axionic}.
In $\mathbf{r}$-space, the $\nu_3 = 1$ 3D oQAHI can be constructed by tiling 3D space with 2D $\mathcal{I}$-symmetric Chern insulators with Chern number $C_{\gamma_1} = 1$ placed in the $\mathcal{I}$-symmetric $r_3 = 1/2$ plane within each unit cell, as shown on the right-hand side of (b).
The occupied energy bands in momentum space for this layer construction have Chern number $C_{\gamma_1}=1$ in both the $\mathcal{I}$-symmetric $k_3 = 0$ and $k_3 = \pi$ planes.
The 3D oQAHI in (b) and 3D QAHI in (a) are related to each other by a half-lattice translation along $\mathbf{a}_3$.
The inversion eigenvalues at the eight TRIMs compatible with the $\mathbf{r}$-space layer construction for (a) and (b) with one occupied electronic energy band are also shown individually on the left-hand side of each panel, from which the $\mathbb{Z}_4 \times \left( \mathbb{Z}_2 \right)^3$ SIs $(\tilde{z}_4 , \tilde{z}_{2,1} , \tilde{z}_{2,2} , \tilde{z}_{2,3})$ [SEqs.~\eqref{eq:strong-z4-def_bb} and \eqref{eq:weak-z2i-def_bb}] are determined to be $(2,0,0,1)$ for (a) and $(0,0,0,1)$ for (b).
Any topological crystalline phases in the magnetic space group $P\bar{1}$ (\# $2.4$) with weak Chern numbers $\nu_1 = \nu_2 = 0$ can be derived from linear combinations of (a) and (b), where the negative of a layer construction is obtained by reversing the signs of the Chern numbers of the 2D layers (as opposed to subtracting bands from the occupied subspace)~\cite{elcoro2021magnetic,gao2022magnetic}. 
}\label{fig:elementary-layer-construction-QAH-oQAH}
\end{figure}

The enumeration of spin-stable phases based on their spin-resolved Wannier band configurations characterized by $(C_{\gamma_2,in}^+,\nu_3^+)$, or equivalently $C_{\gamma_2,in}^\pm$ and $C_{\gamma_2,out}^\pm$ (Supplementary Table~\ref{tab:four-winding-num-given-C-m}), allows us to formulate position-space \emph{layer constructions} of spin-stable (spin-gapped) topological crystalline phases with $\nu_1^\pm=\nu_2^\pm=0$ in the nonmagnetic space group $P\bar{1}1'$ (\# $2.5$) generated by 3D lattice translations, $\mathcal{I}$, and spinful $\mathcal{T}$. 
Layer constructions were introduced in SRefs.~\cite{elcoro2021magnetic,gao2022magnetic,song2019topological,huang2017building,song2018mapping} to build and classify magnetic and nonmagnetic topological phases protected by crystal symmetries. 
Given a 3D (magnetic) space group, the layer construction approach builds a topological crystalline phase by tessellating space with flat layers of 2D topological crystalline phases placed in high-symmetry planes within the unit cell in a manner consistent with the crystal symmetries, and coupled such that the bulk remains gapped. 
Two layer constructions correspond to the same topological phase if they can be deformed into each other by sliding layers in a manner consistent with the crystal symmetries.
As a representative example, let us consider the 3D magnetic space group $P\bar{1}$ (\# $2.4$), generated by $\mathcal{I}$ and 3D lattice translations with Bravais lattice vectors $\{ \mathbf{a}_1,\mathbf{a}_2,\mathbf{a}_3 \}$.
There are six $\mathcal{I}$-invariant planes in the unit cell perpendicular to one of the primitive reciprocal lattice vectors $\mathbf{G}_i$: they are the constant $r_i=0,1/2$ planes (where $\mathbf{r}=r_1\mathbf{a}_1+r_2\mathbf{a}_2+r_3\mathbf{a}_3$). 
It was argued in SRefs.~\cite{elcoro2021magnetic,gao2022magnetic} that all gapped, noninteracting topological crystalline phases in magnetic space group $P\bar{1}$ (\# $2.4$) can be generated from four elementary configurations of layers, termed \emph{elementary layer constructions} (eLCs).
In SFig.~\ref{fig:elementary-layer-construction-QAH-oQAH}(a,b) we show two of the  elementary layer constructions for topological crystalline phases in magnetic space group $P\bar{1}$ (\# $2.4$) with weak Chern numbers $\nu_1 = \nu_2 = 0$, where we tile 3D space by $C_{\gamma_1}=1$ inversion-symmetric Chern insulators in the $r_3 = 0$ or $r_3 = 1/2$ planes respectively.
These two elementary layer constructions correspond to $\nu_3=1$ 3D quantum anomalous Hall insulator (QAHI) and $\nu_3 = 1$ 3D obstructed quantum anomalous Hall  insulator (oQAHI), where the $\nu_3 = 1$ oQAHI in SFig.~\ref{fig:elementary-layer-construction-QAH-oQAH}(b) is related to the $\nu_3 = 1$ QAHI in SFig.~\ref{fig:elementary-layer-construction-QAH-oQAH}(a) by a half-lattice translation along $\mathbf{a}_3$~\cite{wieder2020axionic}.
For completeness, we note that the two remaining elementary layer constructions are obtained by placing an inversion-symmetric Chern insulator with Chern number $C_{\gamma_1}=1$ in either the $r_1=1/2$ or $r_2=1/2$ plane. 
They have, respectively, $(\nu_1,\nu_2,\nu_3)=(1,0,0)$ and $(\nu_1,\nu_2,\nu_3)=(0,1,0)$. 
There are four eLCs rather than 3 to account for the fact that the ``Chern number polarization'' relative to the origin is a strong $\mathbb{Z}_2$ invariant protected by inversion symmetry~\cite{song2018mapping,varnava2020axion,elcoro2021magnetic,peng2021topological,wieder2020axionic}.

\begin{figure}[ht]
\includegraphics[width=\textwidth]{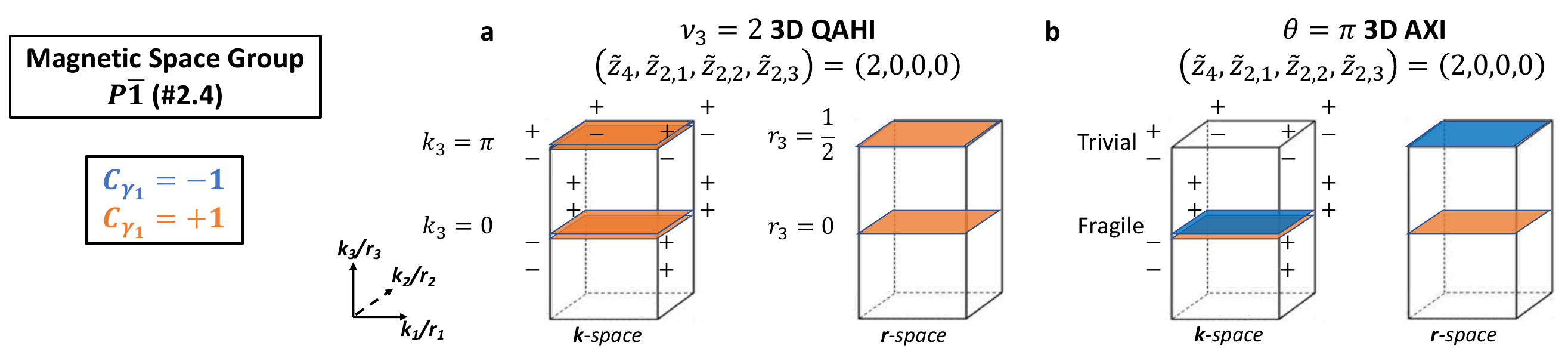}
\caption{Layer constructions for (a) a 3D quantum anomalous Hall insulator (QAHI) with weak Chern numbers $\nu_1= \nu_2 = 0$ and $\nu_3 = 2$, and (b) an $\mathcal{I}$-protected 3D axion insulator (AXI) characterized by $\nu_1 = \nu_2 = \nu_3 = 0$ and an $\mathcal{I}$-quantized bulk axion angle $\theta = \pi$~\cite{elcoro2021magnetic,gao2022magnetic,song2019topological,huang2017building,song2018mapping,wieder2018axion}.
Both (a) and (b) are topological phases in the magnetic space group $P\bar{1}$ (\# $2.4$) with $\nu_1 = \nu_2 = 0$.
The coordinates $k_{i}$ and $r_{i}$ ($i=1,2,3$) in momentum($\mathbf{k}$)-space and position($\mathbf{r}$)-space are given by $\mathbf{k} = \sum_{i=1}^{3} \frac{k_i}{2\pi} \mathbf{G}_i$ and $\mathbf{r} = \sum_{i=1}^{3} r_{i} \mathbf{a}_i$ respectively. 
$\{ \mathbf{a}_1 , \mathbf{a}_2 , \mathbf{a}_3 \}$ are the $\mathbf{r}$-space primitive lattice vectors and $\{ \mathbf{G}_1 , \mathbf{G}_2 , \mathbf{G}_3 \}$ are the dual $\mathbf{k}$-space primitive reciprocal lattice vectors such that $\mathbf{a}_i \cdot \mathbf{G}_j = 2\pi \delta_{ij}$ ($i,j=1,2,3$).
The $\mathcal{I}$-invariant constant-$r_3$ planes contain $r_3 = 0$ and $r_3 = 1/2$, which correspond to the center and the boundary of the primitive unit cell along $\mathbf{a}_3$.
Similarly, the $\mathcal{I}$-symmetric constant-$k_3$ planes contain $k_3 = 0$ and $k_3 = \pi$, which correspond to the center and the boundary of the BZ along $\mathbf{G}_3$.
The right-hand side of (a) shows that to layer-construct a 3D QAHI with $\nu_3 = 2$, we place $\mathcal{I}$-symmetric Chern insulators with Chern numbers $C_{\gamma_1} = +1$ in both the $r_3 = 0$ and $r_3 = 1/2$ planes within each unit cell in the $\mathbf{r}$-space.
The left-hand side of (a) shows that the occupied energy bands in $\mathbf{k}$-space for this layer construction have Chern number $C_{\gamma_1} = +2$ in both the $k_3 = 0$ and $k_3 = \pi$ planes. 
The right-hand side of (b) shows that to layer-construct a 3D AXI, we place an $\mathcal{I}$-symmetric Chern insulator with a Chern number $C_{\gamma_1} = +1$ in the $r_3 = 0$ plane of each unit cell, and a $\mathcal{I}$-symmetric Chern insulator with a Chern number $C_{\gamma_1} = -1$ in the $r_3 = 1/2$ plane of each unit cell in $\mathbf{r}$-space. 
The left-hand side of (b) shows that the two occupied energy bands in $\mathbf{k}$-space for this layer construction individually carry the Chern numbers $C_{\gamma_1} = \pm1$ in the $k_3=0$ plane, such that the total Chern number in the $k_{3}=0$ plane vanishes.  Along with the vanishing Chern number $C_{\gamma_1} = 0$ in the $k_3 = \pi$ plane, this is consistent with the fragile-phase (instead of strong 2D-phase) pumping picture of an AXI introduced in SRef.~\cite{wieder2018axion}.
The inversion eigenvalues at the eight TRIMs compatible with the layer construction for (a) and (b) with two occupied electronic energy bands are also shown on left-hand side of each panel. 
The $\mathbb{Z}_4 \times \left( \mathbb{Z}_2 \right)^3$ SIs $(\tilde{z}_4 , \tilde{z}_{2,1} , \tilde{z}_{2,2} , \tilde{z}_{2,3})$ [SEqs.~\eqref{eq:strong-z4-def_bb} and \eqref{eq:weak-z2i-def_bb}] are  $(2,0,0,0)$ for both (a) and (b).
In particular, on the left-hand side of (b), the two occupied energy bands in the $k_3 = 0$ plane for a 3D AXI have 2D fragile topology in the magnetic space group $P\bar{1}$ (\# $2.4$), though the 3D (strong axionic) topology is stable~\cite{wieder2018axion}.
}\label{fig:layer-construction-QAH-AXI}
\end{figure}

Let us now consider the topological crystalline phases built from the two $(\nu_1,\nu_2,\nu_3)=(0,0,1)$ elementary layer constructions in magnetic space group $P\bar{1}$ (\# $2.4$) discussed above.
Focusing on topological configurations with weak Chern numbers $\nu_1=\nu_2=0$, in SFig.~\ref{fig:layer-construction-QAH-AXI} we show the layer construction of two representative phases. 
In SFig.~\ref{fig:layer-construction-QAH-AXI}(a) we show how a 3D QAHI with weak Chern number $\nu_3 = 2$ in magnetic space group $P\bar{1}$ (\# $2.4$) can be constructed by tiling space with a $C_{\gamma_1}=1$ Chern insulator placed in the $r_3=0$ plane in each unit cell and a $C_{\gamma_1}=1$ Chern insulator placed in the $r_3=1/2$ plane in each unit cell. 
This corresponds to the sum of the two elementary layer constructions depicted in SFig.~\ref{fig:elementary-layer-construction-QAH-oQAH}.
In SFig.~\ref{fig:layer-construction-QAH-AXI}(b) we show how an $\mathcal{I}$-protected 3D magnetic axion insulator (AXI)  in magnetic space group $P\bar{1}$ (\# $2.4$), which is characterized by zero weak Chern numbers $\nu_1 = \nu_2 = \nu_3 = 0$ and a quantized bulk axion angle $\theta = \pi$~\cite{qi2008topological,essin2009magnetoelectric}, can be constructed by tiling space with a $C_{\gamma_1}=1$ inversion-symmetric Chern insulator placed in the $r_3=0$ plane in each unit cell and a $C_{\gamma_1}=-1$ inversion-symmetric Chern insulator placed in the $r_3=1/2$ plane in each unit cell. 
This corresponds to the {\it difference} of the two elementary layer constructions depicted in SFig.~\ref{fig:elementary-layer-construction-QAH-oQAH}, where the negative of an elementary layer construction is obtained by reversing the signs of the Chern numbers of the 2D layers that tile 3D space. 
We emphasize that both adding and subtracting elementary layer constructions increases the number of occupied bands in the system.

The layer construction method allows us to efficiently identify symmetry-protected topological phases that span the SI group of a given (magnetic or nonmagnetic) space group.  
Recall that the magnetic space group $P\bar{1}$ (\# $2.4$) has the SIs $(\tilde{z}_4,\tilde{z}_{2,1},\tilde{z}_{2,2},\tilde{z}_{2,3})$ which can be computed from the inversion eigenvalues of the occupied energy bands at the high-symmetry points in the BZ~\cite{elcoro2021magnetic,gao2022magnetic,xu2020highthroughput,peng2021topological,watanabe2018structure}. 
The strong $\mathbb{Z}_4$ index $\tilde{z}_4$ is specifically determined by the half of the difference between the number of occupied energy bands of positive ($n_+^a$) and negative ($n_-^a$) parity (inversion eigenvalues), modulo $4$ at all $8$ TRIMs $\mathbf{k}_a$, 
\begin{equation}
    \tilde{z}_4 \equiv \frac{1}{2} \sum_{\mathbf{k}_a \in \mathrm{TRIMs}} \left( n^{a}_{+} - n^{a}_{-} \right) \text{ mod } 4.
\label{eq:strong-z4-def_bb}
\end{equation}
The three weak $\mathbb{Z}_2$ indices are determined by the half of the difference between the number of occupied energy bands of positive and negative parity at all $4$ TRIMs in the $\mathbf{k}_a\cdot\mathbf{a}_i=\pi$ planes, modulo $2$,
\begin{equation}
    \tilde{z}_{2,i} \equiv \frac{1}{2} \sum_{\mathbf{k}_a \in \mathrm{TRIMs} \atop \mathbf{k}_a \cdot \mathbf{a}_i = \pi} \left( n^{a}_{+} - n^{a}_{-} \right) \text{ mod } 2, \label{eq:weak-z2i-def_bb}
\end{equation}
where $i=1,2,3$.
We see in SFig.~\ref{fig:elementary-layer-construction-QAH-oQAH} that the 3D QAHI with $\nu_3 = 1$ and the 3D oQAHI with $\nu_3 = 1$ have the SIs $(\tilde{z}_4,\tilde{z}_{2,1},\tilde{z}_{2,2},\tilde{z}_{2,3})=(2,0,0,1)$ and $(0,0,0,1)$, respectively.
Similarly, we see in SFig.~\ref{fig:layer-construction-QAH-AXI} that the 3D QAHI with $\nu_3 = 2$ and the 3D AXI have the same SIs $(\tilde{z}_4,\tilde{z}_{2,1},\tilde{z}_{2,2},\tilde{z}_{2,3})=(2,0,0,0)$.

The layer construction of topological crystalline insulators has a direct connection to the nested (partial) Berry phase classifications that we introduced in SN~\ref{sec:nested_P_Wilson_loop} and \ref{sec:nested_P_pm_Wilson_loop}. 
This is because the inner nested (partial) Chern number $C_{\gamma_2,in}^{(\pm)}$ gives the (partial) Chern number for the (spin-resolved) hybrid Wannier bands centered at an $\mathcal{I}$-symmetric plane at the origin of the unit cell. 
Similarly, the outer nested (partial) Chern number $C_{\gamma_2,out}^{(\pm)}$ gives the (partial) Chern number for the (spin-resolved) hybrid Wannier bands centered at an $\mathcal{I}$-symmetric plane at the boundary of the unit cell. 
In particular, as we have shown in SEq.~\eqref{eq:general_condition_diff_group_C_gamma_2_pm} of SN~\ref{app-relation-nested-P-pm-and-partial-weak-Chern}, the summation of $C_{\gamma_2,in}^{(\pm)}$ and $C_{\gamma_2,out}^{(\pm)}$ is the (partial) weak Chern number along the direction normal to the layers (in a layer construction with parallel layers).
This allows us to associate each layer construction to a configuration of $C_{\gamma_2,in}^{(\pm)}$ and $C_{\gamma_2,out}^{(\pm)}$, where $C_{\gamma_2,in}^{(\pm)}$ gives the (partial) Chern number of the layer at the origin of the unit cell, and $C_{\gamma_2,out}^{(\pm)}$ gives the (partial) Chern number at the boundary of the unit cell.

We can now adapt these observations to formulate a \emph{spin-resolved} layer construction for the spin-stable (spin-gapped) topological crystalline phases with $\nu^\pm_1=\nu^\pm_2=0$ in the nonmagnetic space group $P\bar{1}1'$ (\# $2.5$).
First, although the projector $P(\mathbf{k})$ onto the occupied electronic energy bands is invariant under the nonmagnetic space group $P\bar{1}1'$ (\# $2.5$), the projectors $P_{\pm}(\mathbf{k})$ onto the positive/negative spin bands are invariant only under the magnetic subgroup $P\bar{1}$ (\# $2.4$) of symmetries that commute with $PsP$. 
From this it follows that the spin-resolved topology of the projectors $P_{\pm}(\mathbf{k})$ correspond to the topology of bands in the magnetic subgroup $P\bar{1}$ (\# $2.4$) of $P\bar{1}1'$ (\# $2.5$).
(We note that in systems with conserved spin, spin-resolved band topology can also be analyzed using spin space groups~\cite{BrinkmanSpinSpace,elcoro2021magnetic,yang2021symmetryprotected,corticelli2022spin,liu2022spin,SpinSpaceChina1,SpinSpaceChina2,SpinSpaceChina3,SpinSpaceChina4,spinSpaceChenNoSOC}).

\begin{figure}[ht]
\includegraphics[width=\textwidth]{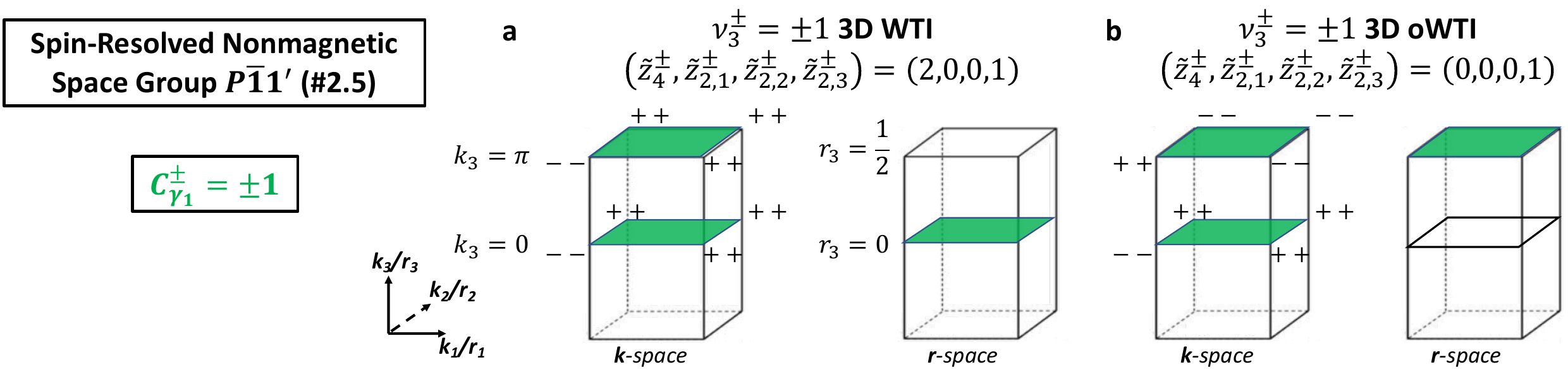}
\caption{Elementary spin-resolved layer constructions for topological crystalline phases in the nonmagnetic space group $P\bar{1}1'$ (\# $2.5$) with a spin gap and partial weak Chern numbers $\nu_1^\pm = \nu_2^\pm = 0$.
The coordinates $k_{i}$ and $r_{i}$ ($i=1,2,3$) in momentum($\mathbf{k}$)-space and position($\mathbf{r}$)-space are given by $\mathbf{k} = \sum_{i=1}^{3} \frac{k_i}{2\pi} \mathbf{G}_i$ and $\mathbf{r} = \sum_{i=1}^{3} r_{i} \mathbf{a}_i$ respectively. 
$\{ \mathbf{a}_1 , \mathbf{a}_2 , \mathbf{a}_3 \}$ are the $\mathbf{r}$-space primitive lattice vectors and $\{ \mathbf{G}_1 , \mathbf{G}_2 , \mathbf{G}_3 \}$ are the dual $\mathbf{k}$-space primitive reciprocal lattice vectors such that $\mathbf{a}_i \cdot \mathbf{G}_j = 2\pi \delta_{ij}$ ($i,j=1,2,3$).
The $\mathcal{I}$-invariant constant-$r_3$ planes contain $r_3 = 0$ and $r_3 = 1/2$, which correspond to the center and the boundary of the primitive unit cell along $\mathbf{a}_3$.
Similarly, the $\mathcal{I}$-symmetric constant-$k_3$ planes contain $k_3 = 0$ and $k_3 = \pi$, which correspond to the center and the boundary of the BZ along $\mathbf{G}_3$.
(a) shows the layer construction of a $\nu_3^\pm = \pm 1$ 3D weak topological insulator (WTI) in both $\mathbf{r}$-space (right) and $\mathbf{k}$-space (left).
In $\mathbf{r}$-space, the $\nu_3^\pm = \pm 1$ 3D WTI can be constructed by tiling 3D space by 2D translation-, $\mathcal{I}$-, and spinful $\mathcal{T}$-symmetric layers with partial Chern numbers $C_{\gamma_1}^\pm = \pm 1$ in the $\mathcal{I}$-symmetric $r_3 = 0$ plane within each unit cell, as shown on the right-hand side of (a).
As we showed in SN~\ref{sec:general_properties_of_winding_num_of_P_pm_Wilson}, a 2D spinful $\mathcal{T}$-invariant system with partial Chern numbers $C_{\gamma_1}^\pm = \pm 1$ is a 2D strong topological insulator, hence by construction (a) is a WTI~\cite{fu2007topological,fu2007topologicala}.
The occupied energy bands for the layer construction of this 3D WTI have partial Chern numbers $C_{\gamma_1}^\pm=\pm 1$ in both the $\mathcal{I}$-symmetric $k_3 = 0$ and $k_3 = \pi$ planes, as shown in the left of panel (a).
(b) shows the layer construction of a $\nu_3^\pm = \pm 1$ 3D obstructed weak topological insulator (oWTI) in both $\mathbf{r}$-space (right) and $\mathbf{k}$-space (left)~\cite{wieder2020axionic}.
In $\mathbf{r}$-space, the $\nu_3^\pm = \pm 1$ 3D oWTI can be constructed by tiling 3D space by 2D layers with partial Chern numbers $C_{\gamma_1}^\pm = \pm 1$ in the $\mathcal{I}$-symmetric $r_3 = 1/2$ plane within each unit cell, as shown on the right-hand side of (b).
The occupied energy bands for the layer construction of 3D oWTI have partial Chern numbers $C_{\gamma_1}^\pm=\pm 1$ in both the $\mathcal{I}$-symmetric $k_3 = 0$ and $k_3 = \pi$ planes, as shown in the left of panel (b).
The 3D oWTI in (b) and 3D WTI in (a) are related to each other by a half-lattice translation along $\mathbf{a}_3$~\cite{wieder2020axionic}.
The inversion eigenvalues at the eight TRIMs compatible with the $\mathbf{r}$-space layer construction for (a) and (b) with two occupied electronic energy band are also shown on the left-hand side of each panel. 
The $\mathbb{Z}_4 \times \left( \mathbb{Z}_2 \right)^3$ partial SIs (defined in SN~\ref{sec:defs_partial_sis}) $(\tilde{z}^\pm_4 , \tilde{z}^\pm_{2,1} , \tilde{z}^\pm_{2,2} , \tilde{z}^\pm_{2,3})$  are $(2,0,0,1)$ for (a) and $(0,0,0,1)$ for (b).
Any topological crystalline phases in the nonmagnetic space group $P\bar{1}1'$ (\# $2.5$) with a spin gap and partial weak Chern numbers $\nu_1^\pm = \nu_2^\pm = 0$ can be derived from linear combinations of (a) and (b), where the negative of a layer construction is obtained by reversing the signs of the partial Chern numbers of 2D layers tiling 3D space in the layer construction~\cite{song2018mapping,song2020real}. 
In particular, (a) and (b) correspond to the $L_1$ and $L_2$ elementary spin-resolved layer constructions in Supplementary Table~\ref{tab:layer-construction-given-C-m}, respectively.
}\label{fig:elementary-layer-construction-WTI-oWTI}
\end{figure}

Spinful $\mathcal{T}$ symmetry only constrains that the states in the image of $P_{+}(\mathbf{k})$ are related to the states in the image of $P_{-}(\mathbf{k})$ by spinful $\mathcal{T}$, such that states in the image of $P(\mathbf{k})=P_+(\mathbf{k})+P_-(\mathbf{k})$ are symmetric under the symmetries of $P\bar{1}1'$ (\# $2.5$).
Recall that the four elementary layer constructions for the magnetic subgroup $P \bar{1}$ (\# $2.4$) correspond to tiling 3D space with $\mathcal{I}$-symmetric Chern insulators with Chern number $C_{\gamma_1} = 1$ in $\mathcal{I}$-invariant planes within the unit cell.
We then deduce that the four elementary spin-resolved layer constructions for spin-stable phases in nonmagnetic group $P\bar{1}1'$ (\# $2.5$) correspond to tiling 3D space with 2D spin-gapped, translation-, $\mathcal{I}$-, and spinful $\mathcal{T}$-invariant insulators with partial Chern numbers $C_{\gamma_1}^\pm = \pm 1$ in $\mathcal{I}$-invariant planes within the unit cell.
Restricting to the cases with $\nu_1^{\pm} = \nu_2^\pm = 0$, this yields two elementary spin-resolved layer constructions $L_1$ and $L_2$, as shown in Supplementary Table~\ref{tab:layer-construction-given-C-m}; two other elementary spin-resolved layer constructions can be obtained by permuting the crystalline axes in $L_2$ such that $\nu_1^{\pm} = \nu_3^\pm = 0$ or $\nu_2^{\pm} = \nu_3^\pm = 0$.
$L_1$ [$L_2$] are constructed by tiling 3D space with 2D $\mathcal{I}$- and $\mathcal{T}$-symmetric spin-gapped insulators with partial Chern numbers $C_{\gamma_1}^\pm = \pm 1$ in the $\mathcal{I}$-invariant plane $r_3 = 0$ [$r_3 = 1/2$] within each unit cell, as shown in SFig.~\ref{fig:elementary-layer-construction-WTI-oWTI}(a) [SFig.~\ref{fig:elementary-layer-construction-WTI-oWTI}(b)].
Recall from SN~\ref{sec:general_properties_of_winding_num_of_P_pm_Wilson} that a $\mathcal{T}$-symmetric spin-gapped 2D system with odd partial Chern numbers is a 2D strong topological insulator.
By construction, $L_1$ is then a WTI~\cite{fu2007topological,fu2007topologicala} while $L_2$ is an oWTI~\cite{wieder2020axionic} related to $L_1$ by a half-lattice translation along $\mathbf{a}_3$.
$L_1$ and $L_2$ are \emph{elementary} in the sense that any spin-resolved layer constructions with spin-resolved Wannier band configurations described by nested partial Chern numbers in SEqs.~\eqref{eq:compare-spin-stable-C-gamma-2-in-pm} and \eqref{eq:compare-spin-stable-C-gamma-2-out-pm} can be derived from linear combinations of $L_1$ and $L_2$, where the negative of a spin-resolved layer construction can be obtained by reversing the signs of all the partial Chern numbers $C_{\gamma_1}^\pm$ of the 2D layers.
In particular, as indicated in Supplementary Table~\ref{tab:layer-construction-given-C-m}, we have that $L_1$ and $L_2$ have spin-stable invariants $(C_{\gamma_2,in}^+\mod 2,\nu_3^+) = (1,1)$ and $(0,1)$ respectively, which are the generators of the $\mathbb{Z}_2 \times \mathbb{Z}$ group of spin-stable topology of nonmagnetic space group $P\bar{1}1'$ (\# $2.5$) with $\nu^{\pm}_1=\nu_2^{\pm}=0$.

\begin{table}[ht]
\begin{tabular}{ |c|c|c||c|c|c|c|c| } 
 \hline
  & $L_1$ & $L_2$  & \multicolumn{5}{c|}{built from elementary spin-resolved layer constructions} \\ 
 \hline
 $(C_{\gamma_2 , in}^{+},\nu_3^{+})$ & $(1,1)$ & $(0,1)$ & $(0,0)$ & $(0,2)$ &$(0,2)$ & $(1,2)$ & $(1,0)$  \\ 
 \hline
 $C_{\gamma_1}^{+}$ in $r_3 = 0$ & $1$ & $0$ & $2n$ & $0$ &$2$ & $ 1$   & $ 1$    \\ 
 \hline
 $C_{\gamma_1}^{+}$ in $r_3 = 1/2$ & $0$ & $1$ & $-2n$ & $2$ &$0$  & $ 1$  &   $- 1$   \\ 
 \hline
\end{tabular}
\caption{\label{tab:layer-construction-given-C-m}Spin-resolved layer constructions for the spin-stable phases in the nonmagnetic space group $P\bar{1}1'$ (\# $2.5$) with a spin gap and partial weak Chern numbers $\nu_1^\pm = \nu_2^\pm = 0$ in Supplementary Table~\ref{tab:four-winding-num-given-C-m}. 
Since $C_{\gamma_1}^- = -C_{\gamma_1}^+$ by time-reversal symmetry, we only list the values of partial Chern number $C_{\gamma_1}^+$ for the 2D layers.
All the spin-stable phases in the nonmagnetic space group $P\bar{1}1'$ (\# $2.5$) with a spin gap and $\nu_1^\pm = \nu_2^\pm = 0$ in Supplementary Table~\ref{tab:four-winding-num-given-C-m} can be derived from linear combinations of the elementary spin-resolved layer constructions $L_1$ and $L_2$, where we define the negative of a spin-resolved layer construction by reversing the signs of the partial Chern numbers of all the 2D layers.
The trivial spin-stable phase is realized by any linear combination of the form $2n L_1 \ominus 2n L_2$ where $n \in \mathbb{Z}$.
The spin-resolved layer construction with $\mathbb{Z}_2\times\mathbb{Z}$ invariant $(C_{\gamma_2,in}^+\mod 2,\nu_3^+)=(0,2)$ can be obtained both from $2L_1$ or $2L_2= 2L_1 \oplus (2L_2\ominus 2L_1)$.
This is because $2 L_1$ corresponds to $(C_{\gamma_2,in}^+\mod 2,\nu_3^+)=2 \times (1\mod 2,1)  = (0,2)$, and $2 L_2$ corresponds to $(C_{\gamma_2,in}^+\mod 2,\nu_3^+)=2 \times (0 \mod 2,1) = (0,2)$.
$2L_1$ and $2L_2$ can be constructed by placing a 2D layer with partial Chern numbers $C_{\gamma_1}^\pm = \pm 2$ in the $r_3 = 0$ or $r_3 = 1/2$ plane, respectively.
Such a 2D layer with $C_{\gamma_1}^\pm = \pm 2$ and two occupied energy bands can be realized, for example, using the model of a fragile topological insulator introduced in SRef.~\cite{wieder2020strong} (and analyzed in detail in SN~\ref{sec:spin_stable_topology_2d_fragile_TI}) in the $s_z$-conserved limit [$v_{Mz} =0$ in SEq.~\eqref{eq:H_2d_fragile_original}], or also in the presence of $s_z$-non-conserving SOC provided that the spin gap is open [$v_{Mz} \neq 0$ in SEq.~\eqref{eq:H_2d_fragile_original}].
The 3D quantum spin Hall insulator with $\nu_3^\pm=\pm 2$ in SFig.~\ref{fig:layer-construction-QSHI-DAXI}(a), which has the spin-stable invariants $(C_{\gamma_2,in}^+\mod 2,\nu_3^+)=(1,2)$, can be constructed as $L_1 \oplus L_2$.
The 3D $\mathcal{T}$-doubled axion insulator with $\nu_3^\pm=0$ and quantized bulk partial axion angles $\theta^\pm = \pi$ [SEq.~\eqref{eq:quantized-theta-pm-DAXI}] in SFig.~\ref{fig:layer-construction-QSHI-DAXI}(b), which has the spin-stable invariants $(C_{\gamma_2,in}^+\mod 2,\nu_3^+)=(1,0)$, can be constructed as $L_1 \ominus L_2$.
}
\end{table}

\begin{figure}[ht]
\includegraphics[width=0.7\textwidth]{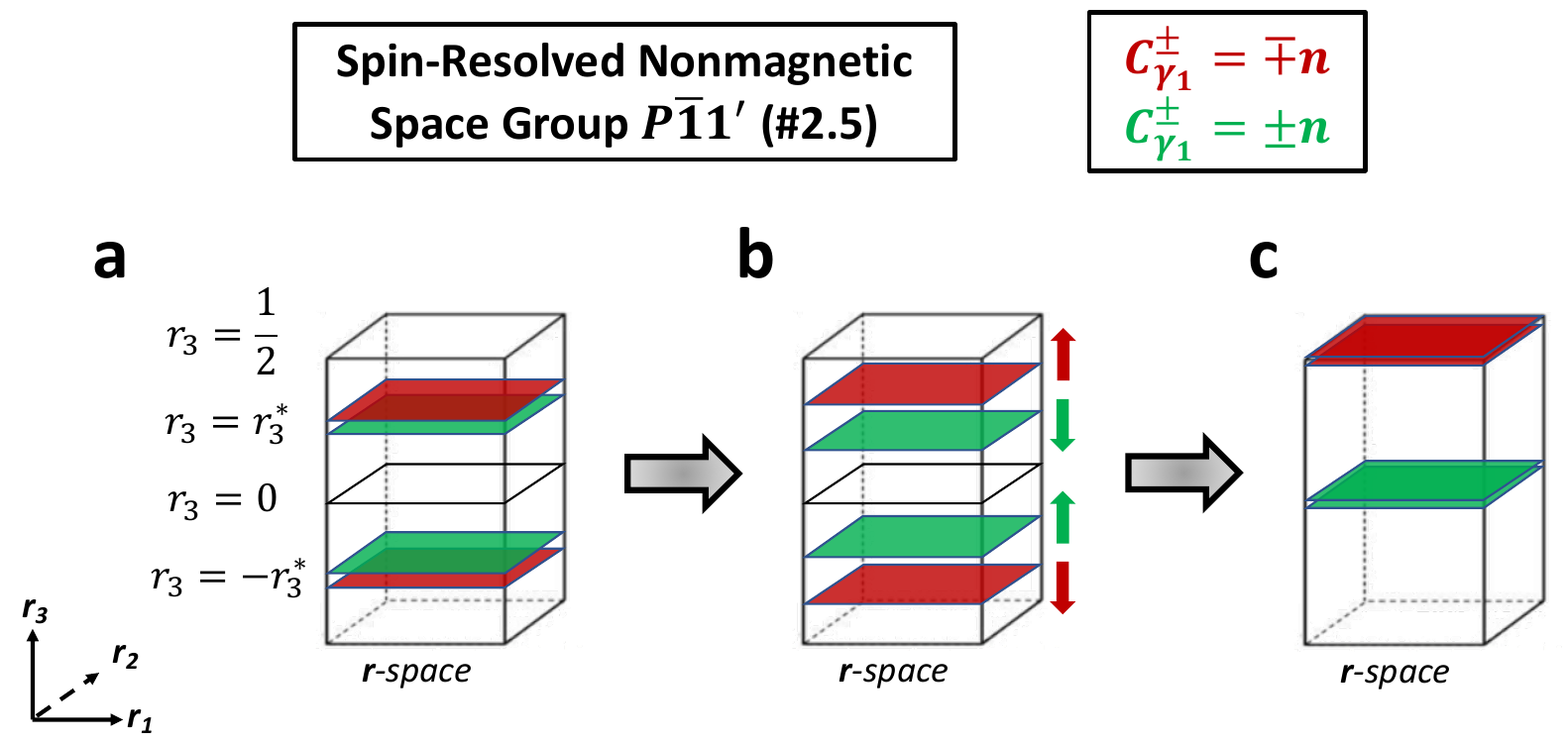}
\caption{Schematic demonstration of the \emph{spin-resolved bubbling process}~\cite{song2020real} in position- ($\mathbf{r}$-)space for spin-resolved layer constructions in the nonmagnetic space group $P\bar{1}1'$ (\# $2.5$) with a spin gap and partial weak Chern numbers $\nu_1^\pm = \nu_2^\pm = 0$.
For any spin-resolved layer construction with $\nu_1^\pm = \nu_2^\pm = 0$, we can always add two layers within each unit cell in a $\mathcal{I}$-symmetric manner, one with partial Chern number $C_{\gamma_1}^\pm = \pm n$ and the other with $C_{\gamma_1}^\pm = \mp n$ where $n \in \mathbb{Z}$, to each of the generic $\mathcal{I}$-related $r_3 = r_3^*$ and $r_3 = -r_3^*$ planes  as shown in (a).
Notice that the slight offset between the two layers with $C_{\gamma_1}^\pm = \pm n$ and $C_{\gamma_1}^\pm = \mp n$ is a guide to the eye; the two layers should be taken to be in the same $r_{3} = r_3^*$ (or $-r_3^*$) plane.
In (a), in each of the $r_3 = r_3^*$ and $r_3 = -r_3^*$ planes, we thus have 2D layers with in total zero (trivial) partial Chern numbers.
For any spin-resolved layer construction of spin-stable (spin-gapped) phases with $\nu_1^\pm = \nu_2^\pm  = 0$ in the nonmagnetic space group $P\bar{1}1'$ (\# $2.5$), we can always insert 2D layers in the way depicted in (a) without changing the spin-stable topology by closing and reopening a set of spin-resolved Wannier gaps.
Next, we can slide the 2D layers in a manner that respects the symmetries (namely $\mathcal{I}$ and $\mathcal{T}$) of $P\bar{1}1'$ (\# $2.5$), by sliding the 2D layers with $C_{\gamma_1}^\pm = \pm n$ and $C_{\gamma_1}^\pm = \mp n$ toward the $r_3 = 0$ and $r_3 = 1/2$ plane, respectively. 
This is shown schematically in (b).  
Finally, in (c), we have moved the two  2D layers with $C_{\gamma_1}^\pm = \pm n$ to  $\mathcal{I}$-invariant $r_3  = 0$ plane, while also moving the two layers  with $C_{\gamma_1}^\pm = \mp n$ to the $\mathcal{I}$-invariant $r_3  = 1/2$ plane.
In terms of the spin-resolved Wannier band configuration, (c) corresponds to $C_{\gamma_2,in}^\pm = \pm 2n$ and $C_{\gamma_2,out}^\pm = \mp 2n$, which as we have shown in Supplementary Table~\ref{tab:four-winding-num-given-C-m} is consistent with the trivial $\mathbb{Z}_2 \times \mathbb{Z}$ spin-stable invariants $(C_{\gamma_2,in}^+,\nu_3^+) = (0,0)$.
We term the process described by $(a) \to (b) \to (c)$ the \emph{spin-resolved bubbling process}~\cite{peng2021topological,song2020real}.
In particular, inserting 2D layers with (in total) trivial spin-resolved topology in generic $r_3 = r_3^*$ and $r_3 = -r_3^*$ planes in a manner respecting the symmetries of $P\bar{1}1'$ (\# $2.5$) cannot change the spin-resolved topology, and therefore corresponds to the trivial spin-resolved Wannier band configuration with $C_{\gamma_2,in}^\pm = \pm 2n$ and $C_{\gamma_2,out}^\pm = \mp 2n$.
We refer to spin-stable phases with spin-resolved layer constructions related by the spin-resolved bubbling process as \emph{spin-resolved bubble equivalent}~\cite{song2020real,peng2021topological}.
}\label{fig:spin-resolved-bubble-equivalence}
\end{figure}

From the two elementary spin-resolved layer constructions $L_1$ and $L_2$ in Supplementary Table~\ref{tab:layer-construction-given-C-m} and SFig.~\ref{fig:elementary-layer-construction-WTI-oWTI}, we can build the trivial $\mathbb{Z}_2 \times \mathbb{Z}$ spin-stable phases with $(C_{\gamma_2,in}^+ \mod 2,\nu_3^+)=(0,0)$ by forming the linear combinations $2n L_1 \ominus 2n L_2$, since
\begin{equation}
	2n (1\mod 2,1) - 2n(0\mod 2,1) = (0,0).
\end{equation}
Therefore, any spin-resolved layer construction with trivial $\mathbb{Z}_2 \times \mathbb{Z}$ invariants $(C_{\gamma_2,in}^+\mod 2,\nu_3^+)=(0,0)$ corresponds to having 2D layers with partial Chern numbers $C_{\gamma_1}^\pm = \pm 2n$ and $C_{\gamma_1}^\pm = \mp 2n$ in the $r_3 = 0$ and $r_3 = 1/2$ planes, respectively, consistent with Supplementary Table~\ref{tab:four-winding-num-given-C-m}. 
We can always add this trivial layer construction to any spin-stable phase without changing the spin-stable topology.
We can make contact with SN~\ref{app:z2_nested_P_pm_berry_phase} by noting that the trivial spin-resolved layer constructions $2nL_1\ominus 2nL_2$ can also be obtained by tiling 3D space by 2D layers with \emph{in total} zero partial Chern numbers at generic constant-$r_3$ planes related by $\mathcal{I}$. 
These layers can be moved to inversion-symmetric constant-$r_3$ planes in a manner respecting the nonmagnetic space group symmetry $P\bar{1}1'$ (\# $2.5$); such a \emph{spin-resolved bubbling process}~\cite{song2020real,peng2021topological} is demonstrated in SFig.~\ref{fig:spin-resolved-bubble-equivalence}.
Two spin-resolved layer constructions in the nonmagnetic space group $P\bar{1}1'$ (\# $2.5$) describe the same spin-stable phases if they can be related by such a spin-resolved bubbling process.
We term such an equivalence as the \emph{spin-resolved bubble equivalence}~\cite{song2020real}. 
The spin-resolved bubbling process requires a gap closing and reopening in the spin-resolved Wannier spectrum, consistent with the change in nested partial Chern numbers [SEqs.~\eqref{eq:app-C-plus-gamma-2-in-change-2n} and \eqref{eq:app-C-plus-gamma-2-out-change-2n}].

\subsubsection{The Quantum Spin Hall Insulator and the $\mathcal{T}$-Doubled Axion Insulator}
\label{sec:defs_partial_sis}

\begin{figure}[ht]
\includegraphics[width=\textwidth]{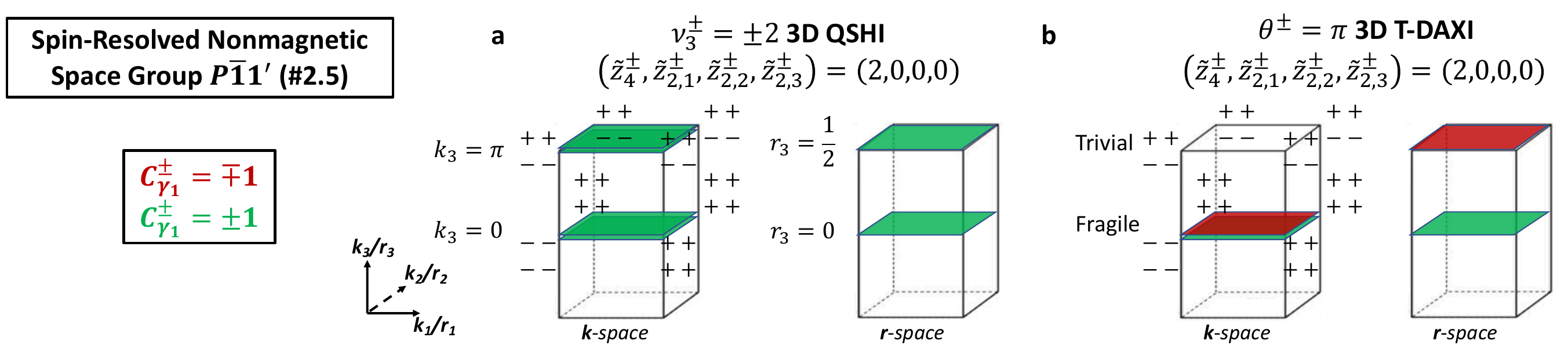}
\caption{Spin-resolved layer constructions for (a) a 3D quantum spin Hall insulator (QSHI) with partial weak Chern numbers [SEq.~\eqref{eq:app_partial_nu_pm}] $\nu_1^\pm= \nu_2^\pm = 0$ and $\nu_3^\pm = \pm 2$, and (b) a 3D $\mathcal{T}$-doubled axion insulator (T-DAXI) characterized by $\nu_1^\pm = \nu_2^\pm = \nu_3^\pm = 0$ and quantized bulk partial axion angles $\theta^\pm = \pi$ [SEq.~\eqref{eq:quantized-theta-pm-DAXI}].
Both (a) and (b) are in the nonmagnetic space group $P\bar{1}1'$ (\# $2.5$) with $\nu_1^\pm = \nu_2^\pm = 0$ and a spin gap.
The coordinates $k_{i}$ and $r_{i}$ ($i=1,2,3$) in momentum($\mathbf{k}$)-space and position($\mathbf{r}$)-space are given by $\mathbf{k} = \sum_{i=1}^{3} \frac{k_i}{2\pi} \mathbf{G}_i$ and $\mathbf{r} = \sum_{i=1}^{3} r_{i} \mathbf{a}_i$ respectively. 
$\{ \mathbf{a}_1 , \mathbf{a}_2 , \mathbf{a}_3 \}$ are the $\mathbf{r}$-space primitive lattice vectors and $\{ \mathbf{G}_1 , \mathbf{G}_2 , \mathbf{G}_3 \}$ are the dual $\mathbf{k}$-space primitive reciprocal lattice vectors such that $\mathbf{a}_i \cdot \mathbf{G}_j = 2\pi \delta_{ij}$ ($i,j=1,2,3$).
The $\mathcal{I}$-invariant constant-$r_3$ planes contain $r_3 = 0$ and $r_3 = 1/2$, which correspond to the center and the boundary of the primitive unit cell along $\mathbf{a}_3$.
Similarly, the $\mathcal{I}$-symmetric constant-$k_3$ planes contain $k_3 = 0$ and $k_3 = \pi$, which correspond to the center and the boundary of the BZ along $\mathbf{G}_3$.
The right-hand side of (a) shows that to layer-construct a 3D QSHI with $\nu_3^\pm =\pm 2$, we can place translation-, $\mathcal{I}$-, and spinful $\mathcal{T}$-symmetric 2D layers with partial Chern numbers $C_{\gamma_1}^\pm = \pm 1$ in both the $r_3 = 0$ and $r_3 = 1/2$ planes within the primitive unit cell in $\mathbf{r}$-space.
The left-hand side of (a) shows that the occupied energy bands in $\mathbf{k}$-space for such a layer construction in $\mathbf{r}$-space on the right-hand side of (a) are two spin-gapped insulators, each with partial Chern numbers $C_{\gamma_1}^\pm = \pm 1$, in both the $k_3 = 0$ and $k_3 = \pi$ planes. 
The right-hand side of (b) shows that to layer-construct a 3D T-DAXI, we place a 2D layer with partial Chern numbers $C_{\gamma_1}^\pm = \pm 1$ in the $r_3 = 0$ plane and another 2D layer with partial Chern numbers $C_{\gamma_1}^\pm = \mp 1$ in the $r_3 = 1/2$ plane within the primitive unit cell in the $\mathbf{r}$-space. 
The left-hand side of (b) shows that the occupied energy bands in $\mathbf{k}$-space for such a layer construction in $\mathbf{r}$-space on the right-hand side of (b) are two 2D spin-gapped layers in the $k_3 = 0$ plane, one with partial Chern numbers $C_{\gamma_1}^\pm = \pm 1$ and the other with $C_{\gamma_1}^\pm = \mp 1$, such that the total partial Chern numbers are zero in the $k_3 = 0$ plane.  In the $k_3 = \pi$ plane of panel (b), the occupied spin bands carry trivial stable and fragile topology.
The inversion eigenvalues at the eight TRIMs compatible with the layer constructions for (a) and (b) with four occupied electronic energy bands are also shown individually on the  left-hand side of each panel. 
The $\mathbb{Z}_4 \times \left( \mathbb{Z}_2 \right)^3$ partial SIs $(\tilde{z}^\pm_4 , \tilde{z}^\pm_{2,1} , \tilde{z}^\pm_{2,2} , \tilde{z}^\pm_{2,3})$ are determined to be $(2,0,0,0)$ for both (a) and (b).
Notably, on the left-hand side of (b), the four occupied energy bands in the $k_3 = 0$ plane for a 3D T-DAXI have 2D fragile topology in the nonmagnetic space group $P\bar{1}1'$ (\# $2.5$), as will be numerically demonstrated in SN~\ref{sec:numerical-section-of-nested-P-pm}. 
However, as emphasized in SFig.~\ref{fig:layer-construction-QAH-AXI}, the four occupied bands taken throughout the 3D BZ in \emph{both} (a) and (b) carry the \emph{stable} topology of a helical HOTI. 
We additionally note that (a) and (b) correspond to the $L_1 \oplus L_2$ and $L_1\ominus L_2$ spin-resolved layer constructions in Supplementary Table~\ref{tab:layer-construction-given-C-m}, respectively.
}\label{fig:layer-construction-QSHI-DAXI}
\end{figure}

Using the two elementary spin-resolved layer constructions $L_1$ and $L_2$ in Supplementary Table~\ref{tab:layer-construction-given-C-m}, we will now build two representative families of spin-stable (spin-gapped) topological crystalline phases in nonmagnetic space group $P\bar{1}1'$ (\# $2.5$) with $\nu_1^\pm = \nu_2^\pm = 0$.
In SFig.~\ref{fig:layer-construction-QSHI-DAXI}(a), we show that a 3D quantum spin Hall insulator (QSHI) with the partial weak Chern numbers $\nu_3^\pm = \pm 2$ can be constructed by tiling 3D space with $C_{\gamma_1}^\pm=\pm 1$ 2D layers in the $\mathcal{I}$-symmetric $r_3=0$ plane within each unit cell, and $C_{\gamma_1}^\pm=\pm 1$ 2D layers in the $\mathcal{I}$-symmetric $r_3=1/2$ plane within each unit cell. 
SFig.~\ref{fig:layer-construction-QSHI-DAXI}(a) corresponds to the sum $L_1 \oplus L_2$ of the two elementary spin-resolved layer constructions defined in Supplementary Table~\ref{tab:layer-construction-given-C-m} and SFig.~\ref{fig:elementary-layer-construction-WTI-oWTI}.
In SFig.~\ref{fig:layer-construction-QSHI-DAXI}(b) we show that a 3D phase in nonmagnetic space group $P\bar{1}1'$ (\# $2.5$) that we term in this work the $\mathcal{T}$-doubled magnetic axion insulator (T-DAXI)  can be constructed by tiling 3D space with $C_{\gamma_1}^\pm=\pm 1$ 2D layers in the $r_3=0$ plane within each unit cell, and $C_{\gamma_1}^\pm=\mp 1$ 2D layers in the $r_3=1/2$ plane within each unit cell. 
This corresponds to the difference $L_1 \ominus L_2$ of the two elementary spin-resolved layer constructions (Supplementary Table~\ref{tab:layer-construction-given-C-m} and SFig.~\ref{fig:elementary-layer-construction-WTI-oWTI}).
Recall that the elementary spin-resolved layer constructions in the nonmagnetic space group $P\bar{1}1'$ (\# $2.5$) are obtained from time-reversed pairs of the elementary layer construction in the magnetic subgroup $P\bar{1}$ (\# $2.4$).
The QSHI with $\nu_3^\pm = \pm 2$ in SFig.~\ref{fig:layer-construction-QSHI-DAXI}(a) can be spin-resolved into QAHI states [SFig.~\ref{fig:layer-construction-QAH-AXI}(a)] in both the positive and negative $PsP$ eigenspaces that are related to each other by spinful $\mathcal{T}$.
On the other hand, the T-DAXI in SFig.~\ref{fig:layer-construction-QSHI-DAXI}(b) can be spin-resolved into AXI states [SFig.~\ref{fig:layer-construction-QAH-AXI}(b)] in both the positive and negative $PsP$ eigenspaces that are related to each other by spinful $\mathcal{T}$. 
This justifies our choice to term this spin-stable phase as the {\it $\mathcal{T}$-doubled} AXI. 
The T-DAXI is characterized by zero partial weak Chern numbers $\nu_1^\pm = \nu_2^\pm = \nu_3^\pm = 0$ and, as we will show in SEq.~\eqref{eq:quantized-theta-pm-DAXI} and the surrounding text, $\mathcal{I}$-quantized bulk \emph{partial} axion angles $\theta^\pm = \pi$.

Note that the projectors $P_{\pm}(\mathbf{k})$ onto the positive/negative spin bands are symmetric under the action of the magnetic subgroup $P\bar{1}$ (\# $2.4$) of $P\bar{1}1'$ (\# $2.5$). 
This implies that we can assign parity eigenvalues to the spin bands at TRIMs, which are inherited from the parity eigenvalues of the occupied bands. 
We can then compute the corresponding SIs [SEqs.~\eqref{eq:strong-z4-def_bb} and \eqref{eq:weak-z2i-def_bb}] within each of the positive and negative $PsP$ eigenspaces. 
This allows us to define SIs within the positive/negative $PsP$ eigenspace, which we term the \emph{partial} SIs. 
To be specific, we define the partial SIs $(\tilde{z}_4^\pm,\tilde{z}_{2,1}^\pm,\tilde{z}_{2,2}^\pm,\tilde{z}_{2,3}^\pm)$ in the same way as SEqs.~\eqref{eq:strong-z4-def_bb} and \eqref{eq:weak-z2i-def_bb} except that the numbers of the $+1$ and $-1$ inversion eigenvalues at TRIMs are taken from $PsP$ eigenstates in the positive ($+$) and negative ($-$) spin bands. 
We can compute the partial SIs for each spin-stable (spin-gapped) phase in space group $P\bar{1}1'$ (\# $2.5$) with $\nu^{\pm}_1=\nu^{\pm}_2=0$ using our spin-resolved layer constructions. 

Let us first consider the partial SIs for the elementary spin-resolved layer constructions (Supplementary Table~\ref{tab:layer-construction-given-C-m} and SFig.~\ref{fig:elementary-layer-construction-WTI-oWTI}).
The elementary spin-resolved layer construction $L_1$ (WTI) in SFig.~\ref{fig:elementary-layer-construction-WTI-oWTI}(a) consists of two copies of the magnetic elementary layer construction for a 3D QAHI with $\nu_3=1$ shown in SFig.~\ref{fig:elementary-layer-construction-QAH-oQAH}(a), one in each spin subspace, related by time-reversal symmetry. 
Since the magnetic 3D QAHI with $\nu_3=1$ has the magnetic SIs $(\tilde{z}_4,\tilde{z}_{2,1},\tilde{z}_{2,2},\tilde{z}_{2,3})=(2,0,0,1)$, we deduce that the elementary spin-resolved layer construction $L_1$ has the partial SIs $(\tilde{z}_4^\pm,\tilde{z}_{2,1}^\pm,\tilde{z}_{2,2}^\pm,\tilde{z}_{2,3}^\pm)=(2,0,0,1)$. 
On the other hand, the elementary spin-resolved layer construction $L_2$ (oWTI) in SFig.~\ref{fig:elementary-layer-construction-WTI-oWTI}(b) consists of two copies of the magnetic elementary layer construction for a 3D oQAHI with $\nu_3=1$ shown in SFig.~\ref{fig:elementary-layer-construction-QAH-oQAH}(b), one in each spin subspace, related by time-reversal symmetry. 
Since the magnetic 3D oQAHI with $\nu_3=1$ has the magnetic SIs $(\tilde{z}_4,\tilde{z}_{2,1},\tilde{z}_{2,2},\tilde{z}_{2,3})=(0,0,0,1)$, we deduce that the elementary spin-resolved layer construction $L_1$ has the partial SIs $(\tilde{z}_4^\pm,\tilde{z}_{2,1}^\pm,\tilde{z}_{2,2}^\pm,\tilde{z}_{2,3}^\pm)=(0,0,0,1)$.

Next, let us consider the two families (SFig.~\ref{fig:layer-construction-QSHI-DAXI}) of spin-stable phases built from these two elementary spin-resolved layer constructions $L_1$ and $L_2$.
The 3D QSHI with $\nu_3^\pm = \pm 2$  in SFig.~\ref{fig:layer-construction-QSHI-DAXI}(a), which corresponds to $L_1 \oplus L_2$, has partial SIs $(\tilde{z}_4^\pm,\tilde{z}_{2,1}^\pm,\tilde{z}_{2,2}^\pm,\tilde{z}_{2,3}^\pm)=(2,0,0,0)$ as the bands in the positive and negative $PsP$ eigenspaces are equivalent to a time-reversed pair of QAHIs with $\nu_3 =\pm 2$ depicted in SFig.~\ref{fig:layer-construction-QAH-AXI}(a) with magnetic symmetry indicators $(\tilde{z}_4,\tilde{z}_{2,1},\tilde{z}_{2,2},\tilde{z}_{2,3})=(2,0,0,0)$.
Similarly, the 3D T-DAXI in SFig.~\ref{fig:layer-construction-QSHI-DAXI}(b), which corresponds to $L_1 \ominus L_2$, also has the partial SIs $(\tilde{z}_4^\pm,\tilde{z}_{2,1}^\pm,\tilde{z}_{2,2}^\pm,\tilde{z}_{2,3}^\pm)=(2,0,0,0)$ as the bands in the positive and negative $PsP$ eigenspaces are equivalent to the time-reversed pair of AXIs depicted in SFig.~\ref{fig:layer-construction-QAH-AXI}(b), with the magnetic symmetry indicators $(\tilde{z}_4,\tilde{z}_{2,1},\tilde{z}_{2,2},\tilde{z}_{2,3})=(2,0,0,0)$.

We now investigate the relationship between the spin-stable phases and the stable symmetry-indicated electronic band topology.
For  topological crystalline phases in the nonmagnetic space group $P\bar{1}1'$ (\# $2.5$) with a spin gap, we can obtain the $\mathbb{Z}_4 \times \left( \mathbb{Z}_2\right)^3$ SIs $(z_4,z_{2,1},z_{2,2},z_{2,3})$ for electronic band topology based on the partial SIs $(\tilde{z}^{\pm}_4 , \tilde{z}^{\pm}_{2,1} , \tilde{z}^{\pm}_{2,2} , \tilde{z}^{\pm}_{2,3})$ within the positive/negative $PsP$ eigenspaces.
The strong $\mathbb{Z}_4$ index for the nonmagnetic space group $P\bar{1}1'$ (\# $2.5$) is given by~\cite{po2017symmetry,khalaf2018symmetry,song2018mapping}
\begin{equation}
	z_4 \equiv \frac{1}{4} \sum_{\mathbf{k}_a \in \mathrm{TRIMs}} \left( n^{a}_{+} - n^{a}_{-} \right) \text{ mod } 4,  \label{eq:strong-z4-def}
\end{equation}
and the three weak $\mathbb{Z}_2$ indices for the nonmagnetic space group $P\bar{1}1'$ (\# $2.5$) are given by~\cite{fu2007topological}
\begin{equation}
	z_{2,i} \equiv \frac{1}{4} \sum_{\mathbf{k}_a \in \mathrm{TRIMs} \atop \mathbf{k}_a \cdot \mathbf{a}_i = \pi} \left( n^{a}_{+} - n^{a}_{-} \right) \text{ mod } 2, \label{eq:weak-z2i-def}
\end{equation}
where $i=1,2,3$.
Using SEqs.~\eqref{eq:strong-z4-def_bb}, \eqref{eq:weak-z2i-def_bb}, \eqref{eq:strong-z4-def}, and \eqref{eq:weak-z2i-def}, together with the facts that $[\mathcal{I},\mathcal{T}]=0$, the positive and negative $PsP$ eigenspaces in the occupied energy bands are orthogonal to each other, and that the numbers of positive and negative inversion eigenvalues are additive, we deduce that for topological crystalline phases in the nonmagnetic space group $P\bar{1}1'$ (\# $2.5$) with a spin gap we have
\begin{equation}
	(z_4,z_{2,1},z_{2,2},z_{2,3}) = (\tilde{z}^{+}_4 , \tilde{z}^{+}_{2,1} , \tilde{z}^{+}_{2,2} , \tilde{z}^{+}_{2,3}) = (\tilde{z}^{-}_4 , \tilde{z}^{-}_{2,1} , \tilde{z}^{-}_{2,2} , \tilde{z}^{-}_{2,3}).
\end{equation}
Therefore, we can tabulate the SIs and partial SIs of the elementary spin-resolved layer constructions $L_1$ and $L_2$ [SFig.~\ref{fig:elementary-layer-construction-WTI-oWTI}(a,b)] for topological crystalline phases in nonmagnetic $P\bar{1}1'$ (\# $2.5$) with a spin gap and $\nu_1^\pm = \nu_2^\pm = 0$, as shown in Supplementary Table~\ref{tab:SI_and_partial_SI}. 
The additivity of the symmetry indicators and of the $(C_{\gamma_2,in}^+\mod 2, \nu_3^+)$ $\mathbb{Z}_2\times\mathbb{Z}$ invariant allows us to deduce that the SIs $(z_4,z_{2,1},z_{2,2},z_{2,3})$ [SEqs.~\eqref{eq:strong-z4-def} and \eqref{eq:weak-z2i-def}] for the topological crystalline phases in the nonmagnetic space group $P\bar{1}1'$ (\# $2.5$) with $\nu_1^\pm = \nu_2^\pm = 0$ and a spin gap are given by
\begin{align}
    & z_{4} = 2(C_{\gamma_2 , in}^{+} \mod 2), \label{eq:z4-from-C-m-final-form} \\
    & z_{2,1} = 0, \label{eq:z2j-from-C-m-final-form} \\
    & z_{2,2} = 0, \label{eq:z2l-from-C-m-final-form} \\
    & z_{2,3} = \nu_3^{+} \mod 2. \label{eq:z2i-from-C-m-final-form}
\end{align}
Note that the existence of a spin gap requires that $z_4$ be even; this is because $z_4=1,3$ correspond to 3D strong topological insulators, which are spin-Weyl semimetals as detailed in SN~\ref{appendix:explicit-calculation-of-BHZ-model}, \ref{sec:main-text-3D-TI-P-pm}, and \ref{appendix:3D-TI-with-and-without-inversion}. 
This is consistent with the fact that the partial symmetry indicators $\tilde{z}_4^\pm$ and the magnetic symmetry indicator $\tilde{z}_4$ cannot be odd for layer-constructable (\emph{i.e.} spin-gapped or insulating) phases. 

We emphasize that spin-resolved topology is a refinement of electronic (energy) band topology for systems with a spin gap. 
Systems with distinct spin-stable topology but equivalent electronic (energy) band topology can be deformed into each other without closing an energy gap provided a spin gap closes. 
We summarize this in Supplementary Table~\ref{tab:symmetry-indicator-given-C-m}, where we show how the spin-stable invariants $(C_{\gamma_2,in}^+\mod 2, \nu_3^+)$ characterizing spin-stable topological phases with $\nu_1^{\pm}=\nu_2^{\pm}=0$ in space group $P\bar{1}1'$ (\# $2.5$) collapse onto the symmetry indicators of (energy) band topology [SEqs.~\eqref{eq:z4-from-C-m-final-form}--\eqref{eq:z2i-from-C-m-final-form}] when a spin gap is allowed to close. 
We emphasize that since $[\mathcal{I},\mathcal{T}]=0$, a spin band inversion without an energy band inversion cannot trivialize the partial SIs. 
This represents the 3D generalization of the statement that the spin Chern number $C^{s}_{\gamma_1}$ of a 2D TI can be changed without closing an energy gap, but cannot go to zero without closing an energy gap or breaking $\mathcal{T}$ symmetry (\emph{i.e.} $(C^{s}_{\gamma_1}/2)\text{ mod }2=1$ in a 2D TI state, see SRef.~\cite{prodan2009robustness} and SN~\ref{sec:general_properties_of_winding_num_of_P_pm_Wilson}).

Let us now examine the $L_1\oplus L_2$ $\nu_3^\pm = \pm 2$ QSHI and the $L_1\ominus L_2$ T-DAXI in SFig.~\ref{fig:layer-construction-QSHI-DAXI} in more detail. 
From Supplementary Table~\ref{tab:SI_and_partial_SI} we find that both the QSHI and T-DAXI have the SIs $(z_4,z_{2,1},z_{2,2},z_{2,3})=(2,0,0,0)$.
This implies that the QSHI and T-DAXI represent distinct spin resolutions of the same symmetry-indicated stable topological crystalline phase in the nonmagnetic space group $P\bar{1}1'$ (\# $2.5$).
Note that $(z_4,z_{2,1},z_{2,2},z_{2,3})=(2,0,0,0)$ corresponds to a symmetry-indicated helical HOTI.
On the other hand, in terms of the $\mathbb{Z}_2 \times \mathbb{Z}$ spin-stable invariants $(C_{\gamma_2,in}^+,\nu_3^+)$, the QSHI has $(C_{\gamma_2,in}^+ \mod 2,\nu_3^+)=(1,2)$ while the T-DAXI has $(C_{\gamma_2,in}^+ \mod 2,\nu_3^+)=(1,0)$ (Supplementary Tables~\ref{tab:four-winding-num-given-C-m} and \ref{tab:layer-construction-given-C-m}).
Hence, although the QSHI [SFig.~\ref{fig:layer-construction-QSHI-DAXI}(a)] and T-DAXI [SFig.~\ref{fig:layer-construction-QSHI-DAXI}(b)] collapse to the same symmetry-indicated stable topological crystalline phase in the nonmagnetic space group $P \bar{1} 1'$ (\# $2.5$) without spin resolution, they realize distinct spin-stable phases.
We will discuss how these differences manifest in physical observables in the remainder of this section below.

\begin{table}[ht]
\begin{tabular}{ |c|c|c|c| } 
 \hline
 ~ & \multicolumn{3}{c|}{(Partial) Symmetry Indicators and Spin-Stable Invariants} \\
 \hline
 ~ & $(\tilde{z}^{\pm}_4 , \tilde{z}^{\pm}_{2,1} , \tilde{z}^{\pm}_{2,2} , \tilde{z}^{\pm}_{2,3})\}$ & $(z_4,z_{2,1},z_{2,2},z_{2,3})$ & $(C_{\gamma_2,in}^+ \mod 2,\nu_3^+) $ \\ 
 \hline
 $L_1$ [WTI in SFig.~\ref{fig:elementary-layer-construction-WTI-oWTI}(a)] & $(2,0,0,1)$ & $(2,0,0,1)$ & $(1,1)$  \\ 
 \hline
 $L_2$ [oWTI in SFig.~\ref{fig:elementary-layer-construction-WTI-oWTI}(b)] & $(0,0,0,1)$ & $(0,0,0,1)$ & $(0,1)$   \\ 
 \hline
 \hline
 $2nL_1\ominus 2nL_2$ ($n\in \mathbb{Z}$) (trivial spin-stable phase in SFig.~\ref{fig:spin-resolved-bubble-equivalence}) & $(0,0,0,0)$ & $(0,0,0,0)$ & $(0,0)$   \\ 
 \hline
 $L_1 \oplus L_2$ [QSHI in SFig.~\ref{fig:layer-construction-QSHI-DAXI}(a)] & $(2,0,0,0)$ & $(2,0,0,0)$ & $(1,2)$  \\ 
 \hline
 $L_1 \ominus L_2$ [T-DAXI in SFig.~\ref{fig:layer-construction-QSHI-DAXI}(b)] & $(2,0,0,0)$ & $(2,0,0,0)$ & $(1,0)$   \\ 
 \hline
\end{tabular}
\caption{\label{tab:SI_and_partial_SI}Partial SIs $(\tilde{z}^{\pm}_4 , \tilde{z}^{\pm}_{2,1} , \tilde{z}^{\pm}_{2,2} , \tilde{z}^{\pm}_{2,3})$, SIs $(z_4,z_{2,1},z_{2,2},z_{2,3})$, and the $\mathbb{Z}_2 \times \mathbb{Z}$ spin-stable invariants $(C_{\gamma_2,in}^+\mod 2,\nu_3^+)$ for different spin-resolved layer constructions of topological crystalline phases in the nonmagnetic space group $P\bar{1}1'$ (\# $2.5$) with a spin gap and partial weak Chern numbers $\nu_1^\pm = \nu_2^\pm = 0$.
We see that $L_1 \oplus L_2$ and $L_1 \ominus L_2$ have distinct spin-resolved topology, as their invariants $(C_{\gamma_2,in}^+\mod 2,\nu_3^+)$ do not satisfy SEq.~\eqref{eq:spin-stable-condition-1}. 
However, we also see that $L_1\oplus L_2$ and $L_1\ominus L_2$ have the same SIs $(z_4,z_{2,1},z_{2,2},z_{2,3})=(2,0,0,0)$.
The corresponding symmetry-indicated electronic band topology for the spin-stable phases are given in Supplementary Table~\ref{tab:symmetry-indicator-given-C-m}.
}
\end{table}

\begin{table}[ht]
\begin{tabular}{ |c|c|c||c|c|c|c| } 
 \hline
 ~ & \multicolumn{6}{c|}{$(C_{\gamma_2 , in}^{+},\nu_3^{+})$} \\
 \hline
 ~ & $(1,1)$  & $(0,1)$ & $(0,0)$ & $(0,2)$ & $(1,2)$ & $(1,0)$ \\ 
 \hline
 $(z_4,z_{2,1},z_{2,2},z_{2,3})$  & $(2,0,0,1)$ & $(0,0,0,1)$ & $(0,0,0,0)$  & $(0,0,0,0)$ & $(2,0,0,0)$  & $(2,0,0,0)$   \\ 
 \hline
 Symmetry-indicated topological phase  & WTI & oWTI & trivial & trivial & helical HOTI  & helical HOTI  \\ 
 \hline
\end{tabular}
\caption{\label{tab:symmetry-indicator-given-C-m}SIs $(z_4,z_{2,1},z_{2,2},z_{2,3})$ [SEqs.~\eqref{eq:z4-from-C-m-final-form}, \eqref{eq:z2j-from-C-m-final-form}, \eqref{eq:z2l-from-C-m-final-form}, and \eqref{eq:z2i-from-C-m-final-form}] and the corresponding stable symmetry-indicated topological crystalline phases in nonmagnetic space group $P\bar{1}1'$ (\# $2.5$) with a spin gap and $\nu_1^\pm = \nu_2^\pm = 0$ for the spin-stable phases with $(C_{\gamma_2,in}^+,\nu_3^+)$ given in Supplementary Tables~\ref{tab:four-winding-num-given-C-m} and \ref{tab:layer-construction-given-C-m}.
The stable symmetry-indicated topological crystalline phases include a trivial insulator, weak topological insulator (WTI)~\cite{fu2007topological,fu2007topologicala}, obstructed weak topological insulator (oWTI)~\cite{wieder2020axionic}, and a helical higher-order topological insulator (HOTI).
As we can see, although the spin-stable phases with $(C_{\gamma_2,in}^+,\nu_3^+)=(1,2)$ and $(1,0)$ are both symmetry-indicated helical HOTIs, they have distinct spin-resolved topology, as the difference between their $(C_{\gamma_2,in}^+,\nu_3^+)$ violates SEq.~\eqref{eq:spin-stable-condition-1}.
Similarly, although both $(C_{\gamma_2,in}^+,\nu_3^+)=(0,0)$ and $(0,2)$ are symmetry-indicated trivial insulators, they have distinct spin-resolved topology, as $(0,2)$ corresponds to tiling  3D space with 2D layers carrying the partial Chern numbers $C_{\gamma_1}^\pm = \pm 2$ in the $r_3 = 1/2$ plane within the primitive unit cell. 
}
\end{table}

\subsubsection{Responses of 3D QSHIs and T-DAXIs, and the Deduction of an Intermediate Spin-Weyl Regime}
\label{app:response_QSHI_TDAXI}

Both the 3D QSHI and T-DAXI in SFig.~\ref{fig:layer-construction-QSHI-DAXI}(a,b) respectively are symmetry-indicated helical HOTIs, as they both have zero weak $\mathbb{Z}_2$ invariants ($z_{2,1}=z_{2,2}=z_{2,3}=0$) and the strong $\mathbb{Z}_4$ invariant $z_4 = 2$ (Supplementary Table~\ref{tab:symmetry-indicator-given-C-m}). 
In addition, both have odd nonzero winding numbers $C^{\pm}_{\gamma_2 ,in/out}$ in their nested partial Berry phases (Supplementary Table~\ref{tab:symmetry-indicator-given-C-m}), indicating that we cannot form time-reversed pairs of inversion-symmetric exponentially localized Wannier functions~\cite{bradlyn2017topological}.
However, the QSHI and T-DAXI in SFig.~\ref{fig:layer-construction-QSHI-DAXI} differ from each other in their spin-resolved response to external fields.
{Let us again restrict to the spin-stable (spin-gapped) phases in the nonmagnetic space group $P\bar{1}1'$ (\# $2.5$) with partial weak Chern numbers $\nu_1^\pm = \nu_2^\pm = 0$.}
To see that their responses differ, let us first consider the limit in which the spin component $\hat{\mathbf{n}} \cdot \mathbf{s}$ is conserved.
In this limit, the bulk spin Hall conductivity for the QSHI [SFig.~\ref{fig:layer-construction-QSHI-DAXI}(a)] and T-DAXI [SFig.~\ref{fig:layer-construction-QSHI-DAXI}(b)] per unit cell are given by, according to SEq.~\eqref{eq:3dshc},
\begin{align}
	& \sigma^{s}_{ij,\mathrm{top, QSHI}} = \frac{e}{4\pi}\epsilon_{ijk} \cdot \left( \nu_{3}^{+} - \nu_{3}^{-} \right) (\mathbf{G}_3)_{k}= \frac{e}{4\pi} \cdot \left( 2\nu_{3}^{+} \right)\epsilon_{ijk}(\mathbf{G}_3)_{k} = \frac{e}{4\pi} \cdot 4\epsilon_{ijk}(\mathbf{G}_3)_{k}, \label{eq:sigma_s_QSHI}  \\
	& \sigma^{s}_{ij,\mathrm{top, T-DAXI}} = 0, \label{eq:sigma_s_DAXI}
\end{align}
where we have used SEq.~\eqref{eq:partial_weal_Cherns_with_T_symmetry} in the presence of spinful $\mathcal{T}$ symmetry.
As noted in SEq.~\eqref{eq:intrinsicspinhall}, in the presence of weak $\hat{\mathbf{n}} \cdot \mathbf{s}$ non-conserving SOC, we expect SEqs.~\eqref{eq:sigma_s_QSHI} and \eqref{eq:sigma_s_DAXI} to give the leading-order bulk topological contributions to $\sigma^s_{ij,\mathrm{QSHI}}$ and $\sigma^s_{ij,\mathrm{T-DAXI}}$, respectively.
As such, the spin Hall conductivity between QSHI and T-DAXI can be greatly different, even when spin is not conserved. 
{We will later numerically demonstrate that a T-DAXI has a vanishing bulk topological contribution to the spin Hall conductivity by computing its layer-resolved partial Chern numbers in SN~\ref{sec:layer-resolved-Cs-of-a-helical-HOTI}.}

On the other hand, from the relation between the winding numbers of the nested Berry phases and the bulk axion angle introduced in SRef.~\cite{wieder2018axion}, our nested spin-resolved Wilson loop formalism (SN~\ref{sec:nested_P_pm_Wilson_loop}) allows us to introduce {\it partial axion angles} $\theta^{\pm}$, which respectively correspond to the 3D bulk axion angle~\cite{essin2009magnetoelectric} within each of the positive and negative $PsP$ eigenspaces. 
In analogy with the ordinary axion angle, we can use the layer constructions in Supplementary Table~\ref{tab:layer-construction-given-C-m} to define the partial axion angle as $2\pi/|\mathbf{a}_3|$ times the partial Chern number polarization per unit cell (taken modulo $2\pi$)~\cite{varnava2020axion,wieder2018axion}. 
For the T-DAXI in SFig.~\ref{fig:layer-construction-QSHI-DAXI}(b), this gives
\begin{align}
\theta^{\pm}_{\mathrm{T-DAXI}} =2\pi[0(\pm1) + \frac{1}{2}(\mp 1)] \mod 2\pi = \pi. \label{eq:quantized-theta-pm-DAXI}
\end{align}
The bulk contribution to the isotropic magnetoelectric polarizability~\cite{essin2009magnetoelectric} {\it separately in the positive and negative eigenspace} of $PsP$ is then given by
\begin{equation}
	\alpha^{\pm} = \theta^{\pm} \frac{ e^2}{2\pi h}, \label{eq:partial-magnetoelectric-polarizability-def}
\end{equation}
where $\alpha^{\pm}$ is the {\it partial magnetoelectric polarizability}, which represents the 3D analog of the 1D partial polarization introduced by Fu and Kane in SRef.~\cite{fu2006time}.  
In the T-DAXI state [SFig.~\ref{fig:layer-construction-QSHI-DAXI}(b)], $\theta^{\pm} \mod 2\pi = \pi$, a result that is {\it origin-independent}, because the bulk partial weak Chern numbers all vanish~\cite{varnava2020axion,varnava2018surfaces,wieder2020axionic}.  
Conversely in a 3D QSHI state [SFig.~\ref{fig:layer-construction-QSHI-DAXI}(a)], the partial Chern numbers per unit cell are nonvanishing, and the partial axion angles $\theta^\pm$ are hence origin-dependent.

We can combine this result with our arguments about flux insertion from SN~\ref{sec:spin_stable_topology_2d_fragile_TI} to understand the response of QSHIs and T-DAXIs in SFig.~\ref{fig:layer-construction-QSHI-DAXI}(a,b) respectively to the insertion of a magnetic flux. 
Viewing a QSHI via the layer construction from Supplementary Table~\ref{tab:layer-construction-given-C-m}, we see that the spin Hall conductivity in SEq.~\eqref{eq:sigma_s_QSHI} implies that the intrinsic spin Hall conductivity of a quasi-2D slab of a QSHI is proportional to the thickness of the slab. 
Inserting a $\pi$ magnetic flux tube along the $\mathbf{a}_3$ axis into a $\nu^{\pm}_{3}\neq 0$ QSHI will thus bind an extensive number of mid-gap states localized at the flux tube. 
Conversely, the vanishing partial Chern number in each unit cell of a T-DAXI implies that a $\pi$-flux tube in a T-DAXI will not bind any states in the bulk. 
Instead, we expect the flux-insertion response of a T-DAXI to manifest on surfaces. 
Specifically by analogy to magnetic AXIs, the partial Chern number polarization in a T-DAXI implies that gapped surfaces with normal vectors parallel to $\pm \mathbf{G}_3$ have the response associated with a half-quantized partial Chern number~\cite{WiederDefect,armitage2019matter,varnava2018surfaces,wieder2020axionic}.  
From the results of SRefs.~\cite{qi2008spincharge,AshvinFlux,MirlinFlux}, we hence expect a $\pi$-flux tube in a finite slab of a T-DAXI to bind one spinon between the top and bottom surface at half filling.
This is consistent with the fact that, due to the presence of a sample-encircling helical hinge mode, a finite crystallite of a T-DAXI with global $\mathcal{I}$-symmetry, when viewed as a quasi-2D system, is a 2D strong TI. 
In SN~\ref{app:layer-resolved} we will formalize the notion of a half-quantized surface partial Chern number by developing a layer-resolved marker for the partial Chern number in position space, {which reveals that the gapped surfaces of helical HOTIs in the T-DAXI regime bind anomalous quantum spin Hall states with half-integral ($n+1/2$ where $n\in \mathbb{Z}$) partial Chern numbers, which are  equivalent to anomalous halves of 2D TIs.}
{Concurrent with the preparation of this work, a numerical investigation of $\pi$-flux insertion in helical HOTIs revealed that a $\pi$-flux tube, on the average, binds half a spinon per surface in a helical HOTI~\cite{WiederDefect}, which is consistent with the above argument and the results of this work.}
{This suggests it would be interesting to explore the partial axion angles $\theta^\pm$ and the associated spin-magnetoelectric response from the perspective of topological quantum field theory, which we leave as an exciting direction for future investigation.}

To summarize, our spin-resolved formalism of (nested) Wilson loops in SN~\ref{sec:P_pm_Wilson_loop} and \ref{sec:nested_P_pm_Wilson_loop} has allowed us to identify two distinct forms of spin-stable topology in symmetry-indicated helical HOTIs {in the nonmagnetic space group $P\bar{1}1'$ (\# $2.5$) with a spin gap, $\nu_1^\pm=\nu_2^\pm = 0$, and spin-resolved Wannier band configurations described by} $(C_{\gamma_2, in}^{+} \mod 2,\nu_{3}^{+})=(1,2)$ and $(1,0)$ [SFig.~\ref{fig:layer-construction-QSHI-DAXI}(a,b)]:
\begin{enumerate}
	\item The spin-resolved Wilson loop formalism from SN~\ref{sec:P_pm_Wilson_loop} allows us to compute the partial weak Chern numbers, which characterize distinct topological contributions to the bulk and surface spin Hall conductivity that distinguish between 3D QSHIs and T-DAXIs [SEqs.~\eqref{eq:sigma_s_QSHI} and \eqref{eq:sigma_s_DAXI}]. 
	\item The nested spin-resolved Wilson loop formalism from SN~\ref{sec:nested_P_pm_Wilson_loop} allows us to deduce that both 3D QSHIs [SFig.~\ref{fig:layer-construction-QSHI-DAXI}(a)] and T-DAXIs [SFig.~\ref{fig:layer-construction-QSHI-DAXI}(b)] with odd winding numbers $C_{\gamma_2 , in/out}^{\pm}$ in their nested partial Berry phases {(Supplementary Table~\ref{tab:four-winding-num-given-C-m})} cannot form 3D inversion- and time-reversal-symmetric exponentially localized Wannier functions. 
    Furthermore, since the nested $P_{\pm}$-Wilson loop winding provides information regarding the partial Chern numbers of the spin-resolved hybrid Wannier bands in position space, it also allows us to {discover that beyond the spin Hall conductivity, there also exists a partial 3D magnetoelectric polarizability in spinful $\mathcal{T}$-invariant 3D insulators with vanishing weak $\mathbb{Z}_2$ indices, which can be considered the 3D generalization of the 1D partial polarization introduced by Fu and Kane in SRef.~\cite{fu2006time}.} We can then deduce the spin-electromagnetic response of 3D QSHIs [SFig.~\ref{fig:layer-construction-QSHI-DAXI}(a)] and T-DAXIs [SFig.~\ref{fig:layer-construction-QSHI-DAXI}(b)] to external magnetic fields via the spin-resolved layer construction method introduced in this work {(Supplementary Table~\ref{tab:layer-construction-given-C-m}, Supplementary Table~\ref{tab:SI_and_partial_SI}, SFig.~\ref{fig:elementary-layer-construction-WTI-oWTI}, and SFig.~\ref{fig:layer-construction-QSHI-DAXI})}.
\end{enumerate}
We summarize in Supplementary Table~\ref{tab:summary-for-qshi-and-daxi} the properties of 3D QSHIs [SFig.~\ref{fig:layer-construction-QSHI-DAXI}(a)] and 3D T-DAXIs [SFig.~\ref{fig:layer-construction-QSHI-DAXI}(b)] derived in this section.
In SN~\ref{sec:bibr} we will demonstrate that the candidate helical HOTI $\alpha$-BiBr~\cite{SYBiBr,tang2019efficient,BiBrFanHOTI} realizes both 3D QSHI and T-DAXI states, depending on the spin resolution direction. 
We will specifically show that if $\alpha$-BiBr is spin-resolved along $s_z$, it realizes a 3D QSHI state with a partial Chern vector $\bm{\nu}^\pm = \mp 2 \mathbf{G}_3$ and the nested spin-resolved Wilson loop winding numbers (identified as the nested partial Chern numbers [SEq.~\eqref{eq:C_gamma_2_n_pm_k_i}]) $C^{\pm}_{\gamma_2 , in} = \mp 1$, and $C^{\pm}_{\gamma_2 , out} = \mp 1$.
Furthermore, we excitingly find that if $\alpha$-BiBr is spin-resolved along $s_x$, it realizes a T-DAXI state with a partial Chern vector $\bm{\nu}^\pm = \mathbf{0}$ and the nested spin-resolved Wilson loop winding numbers (identified as the nested partial Chern numbers [SEq.~\eqref{eq:C_gamma_2_n_pm_k_i}]) $C^{\pm}_{\gamma_2 , in} = \mp 1$, $C^{\pm}_{\gamma_2 , out} = \pm 1$, which are indicative of origin-independent nontrivial partial axion angles $\theta^\pm = \pi$.

\begin{table}[ht]
\hspace*{-1cm}
\begin{tabular}{ |c|c|c| } 
 \hline
 Spin-Stable Phases  & 3D QSHI with $\nu_3^\pm = \pm 2$ [SFig.~\ref{fig:layer-construction-QSHI-DAXI}(a)] & 3D T-DAXI [SFig.~\ref{fig:layer-construction-QSHI-DAXI}(b)]  \\ 
 \hline
 \hline
 $(C_{\gamma_2 , in}^{+},\nu_{3}^{+})$  & $(1,2)$ & $(1,0)$   \\ 
 \hline
 $C_{\gamma_2 , in}^{\pm}$ & $\pm 1$ & $\pm 1$ \\
 \hline
 $C_{\gamma_2 , out}^{\pm}$ & $\pm 1$ & $\mp 1$ \\
 \hline
 Spin-Resolved Layer Construction & $C_{\gamma_1}^{\pm} = \pm 1$ in $r_3=0$, $C_{\gamma_1}^{\pm} = \pm 1$ in $r_3=\frac{1}{2}$ & $C_{\gamma_1}^{\pm} = \pm 1$ in $r_3=0$, $C_{\gamma_1}^{\pm} = \mp 1$ in $r_3=\frac{1}{2}$ \\
 \hline
 Partial Symmetry Indicators $(\tilde{z}^\pm_4,\tilde{z}^\pm_{2,1},\tilde{z}^\pm_{2,2},\tilde{z}^\pm_{2,3})$ & $(2,0,0,0)$ & $(2,0,0,0)$ \\
 \hline
 Symmetry Indicators $(z_4,z_{2,1},z_{2,2},z_{2,3})$ & $(2,0,0,0)$ & $(2,0,0,0)$ \\
 \hline
 Symmetry-Indicated Topological Phase & helical HOTI & helical HOTI \\
 \hline
 Topological Contribution to the Spin Hall Conductivity & $\frac{e}{4\pi} \cdot 4\epsilon_{ijk}(\mathbf{G}_3)_{k}$  & $0$ \\
 \hline
 Partial Axion Angles $\theta^{\pm}$ mod $2\pi$ & origin-dependent & $\pi$ \\
 \hline
\end{tabular}
\caption{\label{tab:summary-for-qshi-and-daxi}
Spin-resolved properties of helical HOTIs in nonmagnetic space group $P\bar{1}1'$ (\# $2.5$) with a spin gap, $\nu_1^\pm = \nu_2^\pm = 0$ (SFig.~\ref{fig:layer-construction-QSHI-DAXI}), and different spin-stable topology.  
}
\end{table}

Finally, let us make two further remarks. 
First, since the 3D QSHI and 3D T-DAXI spin-stable phases discussed in this section (see also the summary in Supplementary Table~\ref{tab:summary-for-qshi-and-daxi}) are both symmetry-indicated helical HOTIs {in the nonmagnetic space group $P\bar{1}1'$ (\# $2.5$) with a spin gap and $\nu_1^\pm=\nu_2^\pm = 0$}, they can be deformed into each other while keeping the energy gap open and preserving  inversion and time-reversal symmetry. 
However, since their partial weak Chern numbers $\nu_3^{+}$ differ by $2$, a gap must still close in the \emph{spin} spectrum during such a deformation. 
In particular, the 3D QSHI [3D T-DAXI] considered in this section carries the partial Chern numbers $C_{\gamma_1}^{\pm}(k_3)=\pm 2$ [$C_{\gamma_1}^{\pm}(k_3)=0$] in both the $k_{3} = 0$ and $k_{3} = \pi$ planes (see also SFig.~\ref{fig:layer-construction-QSHI-DAXI}).
The intermediate spin-gapless phase during the deformation will hence have the partial Chern numbers $C_{\gamma_1}^{\pm}(k_3=0)=\pm 2$ and $C_{\gamma_1}^{\pm}(k_3=\pi)=0$, or vice-versa, {but we expect it to have non-vanishing, nonquantized partial axion angles (by analogy to the nonquantized axion angles in Weyl semimetals~\cite{burkov2015negative,wang2013chiral,you2016response}).} 
This implies that the intermediate phase is a spin-Weyl semimetal state with an even number of spin-Weyl nodes in each half of the BZ. 
We will show in SN~\ref{app:mote2} that the candidate helical HOTI $\beta$-MoTe$_2$ exhibits a spin band structure that lies within the spin-Weyl semimetal regime for all choices of $\hat{\mathbf{n}} \cdot \mathbf{s}$.  
We will also show in SN~\ref{sec:bibr} that the candidate helical HOTI $\alpha$-BiBr interpolates between 3D QSHI and T-DAXI regimes through an intermediate spin-Weyl state as the spin resolution direction is rotated from $s_z = \hat{\mathbf{z}} \cdot \mathbf{s}$ to $s_x = \hat{\mathbf{x}} \cdot \mathbf{s}$.

To understand the intermediate spin-Weyl semimetal phase that lies between 3D QSHI and T-DAXI states, recall from SN~\ref{sec:main-text-3D-TI-P-pm} that a 3D strong topological insulator has an odd number of spin-Weyl nodes within each half of the 3D BZ. 
This implies that the intermediate spin-Weyl semimetal phase between a 3D QSHI and a 3D T-DAXI can be constructed by superposing two identical 3D strong topological insulators, in fact realizing the doubled strong topological insulator (DSTI) introduced in SRef.~\cite{po2017symmetry}, which as discussed in the main text, represents one construction of a symmetry-indicated helical HOTI. 
Therefore, through the analysis in this section of spin-stable topology and deformations between distinct spin-stable phases, we have found that the minimal model of a helical HOTI in the nonmagnetic space group $P\bar{1}1'$ (\# $2.5$) with a spin gap and $\nu_1^\pm = \nu_2^\pm =0$ realizes one of three spin-resolved phases with distinct spin-stable topology: a 3D QSHI with $\nu_3^\pm = \pm 2$ [SFig.~\ref{fig:layer-construction-QSHI-DAXI}(a)], a T-DAXI with $\theta^\pm \mod 2\pi = \pi$ [SFig.~\ref{fig:layer-construction-QSHI-DAXI}(b) and SEq.~\eqref{eq:quantized-theta-pm-DAXI}], or a DSTI with an even number of spin-Weyl points per half BZ.

Second, although we have in this section considered a 3D QSHI with $(C_{\gamma_2, in}^{+} \mod 2,\nu_3^{+}) = (1,2)$, there can also exist 3D QSHIs with $(C_{\gamma_2 , in}^{+},\nu_3^{+}) = (0,2)$. 
{Spin-resolved Wannier band configurations consistent with $(C_{\gamma_2 , in}^{+} \mod 2,\nu_3^{+}) = (1,2)$ and $(0,2)$} have the same topological contribution to the spin Hall conductivity, as their partial weak Chern numbers $\nu_3^{\pm}$ are equal. 
However, although a 3D QSHI with $(C_{\gamma_2, in}^{+} \mod 2,\nu_3^{+}) = (1,2)$ is a symmetry-indicated helical HOTI, 3D QSHIs with $(C_{\gamma_2 , in}^{+} \mod 2,\nu_3^{+}) = (0,2)$ are symmetry-indicated trivial insulators, as shown in Supplementary Table~\ref{tab:symmetry-indicator-given-C-m}.
This means that in the presence of $s$-nonconserving SOC, the $(1,2)$ QSHI remains topologically nontrivial even if a spin gap closes, while the $(0,2)$ state can be trivialized by deformations that close and reopen the spin gap (and include sufficient trivial bands to trivialize any fragile topology that may arise in few-band models). 

In conclusion, we have shown that spin-resolved band topology provides a refinement of the classification of symmetry-indicated helical HOTIs and the layer construction method. 
Using the (nested) $P_{\pm}$-Wilson loop formalism developed in SN~\ref{sec:P_pm_Wilson_loop} and \ref{sec:nested_P_pm_Wilson_loop},
we have identified distinct spin-stable helical HOTI phases that cannot be deformed into each other without closing either an energy gap or a spin gap. 
We have used the method of spin-resolved layer constructions introduced in this work to demonstrate that insulators with distinct spin-stable topology, even if they share the same symmetry-indicated electronic band topology without spin resolution, exhibit distinct spin-electromagnetic responses.
In the next section (SN~\ref{sec:numerical-section-of-nested-P-pm}), we will apply the theoretical techniques developed in this section to numerically analyze spin-resolved topology in a model of a helical HOTI.

\subsection{Numerical Calculations for Nested $P_{\pm}$-Wilson Loops of an $\mathcal{I}$- and $\mathcal{T}$-Symmetric Helical HOTI}
\label{sec:numerical-section-of-nested-P-pm}

We now apply the formalism of nested $P$- and $P_{\pm}$-Wilson loop from SN~\ref{sec:nested_P_Wilson_loop} and \ref{sec:nested_P_pm_Wilson_loop} to analyze the 3D bulk topology of a helical HOTI with inversion ($\mathcal{I}$) and spinful time-reversal ($\mathcal{T}$) symmetry.
Provided that the spin gap defined in SN~\ref{appendix:properties-of-the-projected-spin-operator} is open, we may compute the bulk spin-resolved topology of the helical HOTI by considering either the positive (upper) or negative (lower) $PsP$ eigenspace. 
Through the analysis of spin-resolved topology we will show that our model of the helical HOTI is in the T-DAXI regime.

\begin{figure*}[t]
\includegraphics[width=0.9\linewidth]{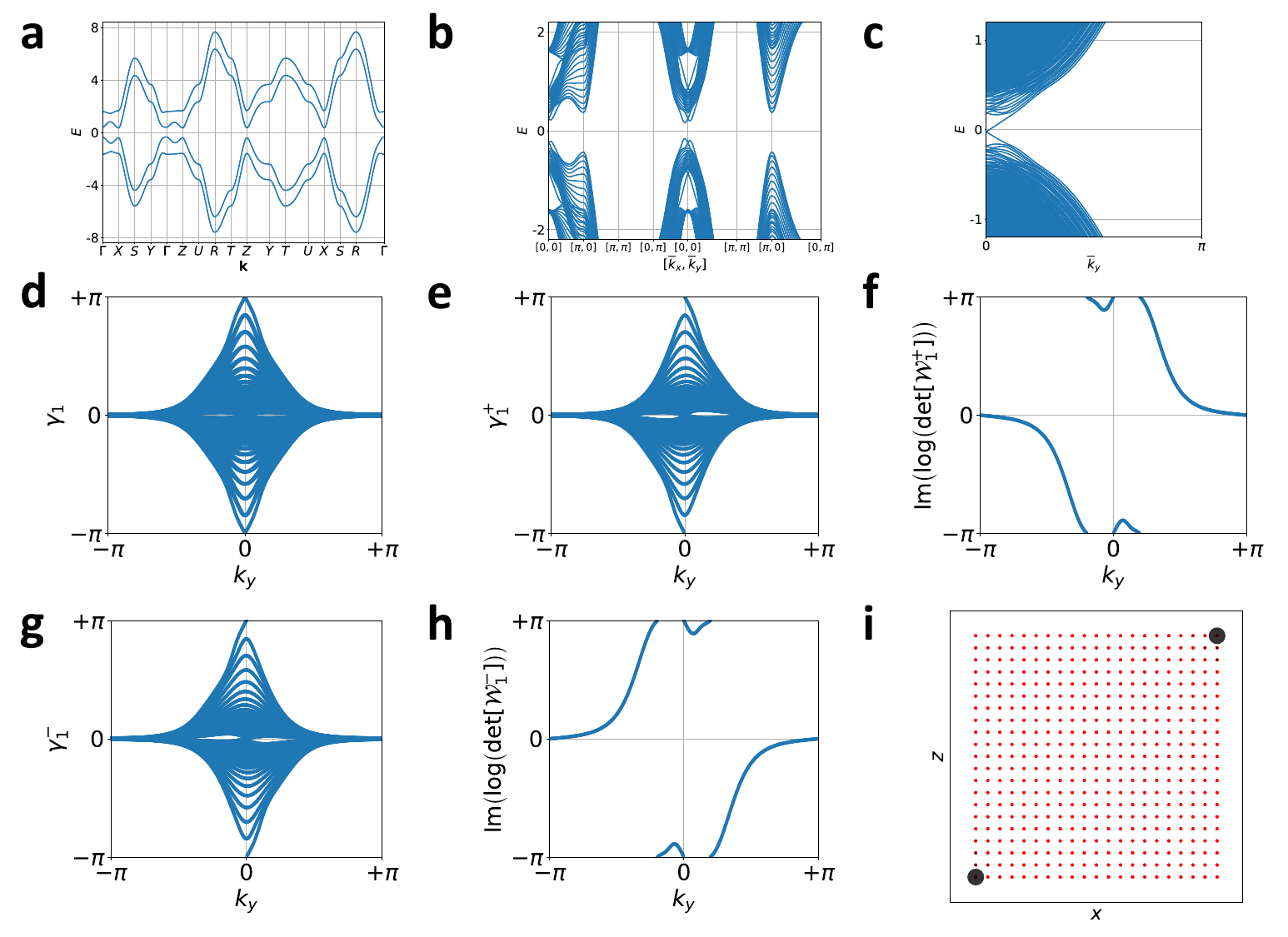}
\caption{Bulk and boundary spectrum of the eight-band helical HOTI tight-binding model introduced in SRef.~\cite{wang2019higherorder}, which has been reproduced in a slightly modified form in SEqs.~\eqref{eq:helical_HOTI_TB_model}, \eqref{eq:parameter_choice_of_helical_HOTI} and \eqref{eq:A_spin_mixing_of_helical_HOTI}.
(a) shows the 3D bulk band structure. 
The labels of the high-symmetry points are given in SFig.~\ref{fig:3D_TI_bulk_BZ_bulk_band_surface_band}(a). 
(b) shows the 2D band structure of a slab infinite along $x$ and $y$ while finite along $z$ with $21$ unit cells. 
(c) shows the 1D band structure of a rod infinite along $y$ while finite along $x$ and $z$ with size $21\times 21$. 
Here ``$k_x$-directed'' means that we have chosen $\mathbf{G} = 2\pi \hat{\mathbf{x}}$ in the $P$- and $P_{\pm}$-Wilson loop matrices [SEqs.~\eqref{eq:P_Wilson_loop_matrix_element} and \eqref{eq:P_pm_Wilson_loop_matrix_element}].
(d) shows the $k_x$-directed $P$-Wilson loop eigenphases as a function of $k_{y}$ for a 2D slab infinite along $x$ and $y$ while finite along $z$ with $21$ unit cells. 
There is an odd helical winding with band crossings at $k_y=0$ and $\pi$ protected by the spinful $\mathcal{T}$ symmetry in (d).
(e) shows the $k_x$-directed $P_{+}$-Wilson loop eigenphases as a function of $k_{y}$ for the same 2D slab model in (d). 
(f) shows the sum of the $k_x$-directed $P_{+}$-Wilson loop eigenphases in (e), which exhibit an overall $-1$ winding number.
(g) shows the $k_x$-directed $P_{-}$-Wilson loop eigenphases as a function of $k_{y}$ for the same 2D slab in (d). 
(h) shows the sum of the $k_x$-directed $P_{-}$-Wilson loop eigenphases of (g), which exhibit an overall $+1$ winding number. 
(d)--(h) collectively establish that the inversion-symmetric 2D slab of a helical HOTI is a time-reversal invariant strong topological insulator with nontrivial strong $\mathbb{Z}_2$ invariant $\nu_{2d}=1$~\cite{yu2021dynamical,wieder2020axionic,wieder2018axion}. 
(i) shows the averaged probability distribution of the four in-gap modes with $k_{y} = 0 $ in (c). 
In particular, when the system is terminated in a $\hat{\mathbf{y}}$-directed rod geometry preserving both $\mathcal{I}$ and $\mathcal{T}$ symmetries as in (c), the 1D metallic states at the same hinge are related to each other by $\mathcal{T}$ symmetry; 
states at opposite hinges are related by $\mathcal{I}$ symmetry.
These then demonstrate the existence of helical hinge modes.
The calculations detailed in this figure were performed using the freely available Python package~\href{https://github.com/kuansenlin/nested_and_spin_resolved_Wilson_loop}{\textsc{nested\_and\_spin\_resolved\_Wilson\_loop}}~\cite{lin2023nestedWilsonLib}, which represents an extension of the~\href{https://www.physics.rutgers.edu/pythtb/}{PythTB} open-source Python tight-binding package~\cite{coh2013python} that was implemented and utilized for the preparation of SRefs.~\cite{wieder2018axion,wieder2020strong} and the present work.
}
\label{fig:helical-HOTI-before-adding-trivial-bands}
\end{figure*}

As in the case of magnetic axion insulators (AXIs), whose bulk axion angles $\theta$ can be computed from the spectral flow of the {\it nested} $P$-Wilson loop eigenphases (see SN~\ref{sec:nested_P_Wilson_loop} and SRef.~\cite{wieder2018axion}), we will here show how the nested $P_\pm$-Wilson loop allows us to compute the spin-resolved (partial) bulk topological invariants of time-reversal and inversion symmetric helical HOTIs.
As demonstrated in SN~\ref{app:comparison-spin-stable-and-symmetry-indicated-topology} and SRef.~\cite{WiederDefect}, the spin-resolved (partial) bulk topological invariants in helical HOTIs manifest as response coefficients of bulk spin-electromagnetic effects. 
Just like with AXIs, we will see that care must be taken to add extra trivial bands to our models in order to remove fragile winding in the $P$- and $P_\pm$-Wilson loops~\cite{wieder2018axion}. 
Specifically, although AXIs and helical HOTIs are stable topological crystalline insulators~\cite{elcoro2021magnetic,gao2022magnetic,schindler2018higherorder,po2017symmetry,watanabe2018structure,song2018mapping,witten2016fermion,vanderbilt2018berry,varnava2020axion,wieder2018axion,schindler2018higherordera}, minimal models of AXIs and helical HOTIs exhibit gapless Wilson loops, whereas models with larger number of (occupied) bands generally do not~\cite{wieder2018axion,varnava2020axion}. 
We consider an eight-band helical HOTI with a 3D orthorhombic lattice, formed by placing two spinful $s$ and two spinful $ip$ orbitals at the $1a$ Wyckoff position [$(x,y,z)=(0,0,0)$] of the primitive unit cell. 
Symmetry-breaking hopping terms are included to break accidental symmetries while preserving $\mathcal{I}$ and  $\mathcal{T}$ symmetries~\cite{wang2019higherorder}.
Normalizing the lattice constants to one, we have that the primitive reciprocal lattice vectors are $\mathbf{G}_1 = 2\pi \hat{\mathbf{x}}$, $\mathbf{G}_2 = 2\pi \hat{\mathbf{y}}$, and $\mathbf{G}_3 = 2\pi \hat{\mathbf{z}}$. 
The corresponding eight-band (momentum space) Bloch Hamiltonian is given by~\cite{wang2019higherorder}
\begin{align}
    [H(\mathbf{k})] = & \left( m_{1} + \sum_{i=x,y,z} v_{i}\cos{(k_{i})} \right) \tau_{z}\mu_{0}\sigma_{0} + m_{2}\tau_{z}\mu_{x}\sigma_{0} + m_{3}\tau_{z}\mu_{z}\sigma_{0} + u_{x}\sin{(k_{x})} \tau_{y}\mu_{y}\sigma_{0} + u_{z}\sin{(k_{z})}\tau_{x}\mu_{0}\sigma_{0} \nonumber \\
    & + m_{v_{1}}\tau_{0}\mu_{z}\sigma_{0} + m_{v_{2}}\tau_{0}\mu_{x}\sigma_{0} + v_{H}\sin{(k_{y})}\tau_{y}\mu_{z}\sigma_{z} + A_{\text{spin-mixing}} \sin{(k_{z})} \tau_{y} \mu_{0} \sigma_{x} + f_{323}\tau_{z}\mu_{y}\sigma_{z}. \label{eq:helical_HOTI_TB_model}
\end{align}
The Bloch Hamiltonian in SEq.~(\ref{eq:helical_HOTI_TB_model}) has both $\mathcal{I}$ and $\mathcal{T}$ symmetries such that
\begin{align}
    & [\mathcal{I}]  [H(\mathbf{k})] [\mathcal{I}]^{-1} = (\tau_{z}\mu_{0}\sigma_{0}) [H(\mathbf{k})] (\tau_{z}\mu_{0}\sigma_{0}) = [H(-\mathbf{k})],\\
    & [\mathcal{T}] [H(\mathbf{k})] [\mathcal{T}]^{-1} = (\tau_{z}\mu_{0}\sigma_{y}) [H(\mathbf{k})]^{*} (\tau_{z}\mu_{0}\sigma_{y}) = [H(-\mathbf{k})],
\end{align}
respectively.
We will be using the following set of parameters:
\begin{align}
    m_{1} = -3.0, v_{x}=v_{z}=u_{x}=u_{z} = 1.0, v_{y} = 2.0, m_{2} = 0.3, m_{3} = 0.2, m_{v_{1}}=-0.4, m_{v_{2}} = 0.2, v_{H} = 1.2, f_{323} = 0.25. \label{eq:parameter_choice_of_helical_HOTI}
\end{align}
When $A_{\mathrm{spin-mixing}}\neq 0$ the Hamiltonian [SEq.~\eqref{eq:helical_HOTI_TB_model}] does not conserve $s_z$. 
In this section we take
\begin{equation}
A_{\mathrm{spin-mixing}}=0.5. \label{eq:A_spin_mixing_of_helical_HOTI}
\end{equation}
Although $s_z$ is not conserved, the gap in the $P(\mathbf{k}) s_z P(\mathbf{k})$ spectrum remains open throughout the 3D BZ.

The 3D band structure [shown in SFig.~\ref{fig:helical-HOTI-before-adding-trivial-bands}(a)] consists of four doubly degenerate bands due to the coexistence of $\mathcal{I}$ and spinful $\mathcal{T}$ symmetries. 
The 2D surfaces of this helical HOTI are fully gapped [SFig.~\ref{fig:helical-HOTI-before-adding-trivial-bands}(b)].
In particular, the higher-order topology manifests in the existence of an odd number of gapless 1D hinge states in highly symmetric {\it finite-sized} model geometries (for any choice of Miller index, since the wallpaper group symmetries $p11'$ for any smooth surface termination cannot stabilize degeneracies between surface bands at generic momenta~\cite{wieder2018wallpaper,elcoro2021magnetic,gao2022magnetic,schindler2018higherorder}) that preserves $\mathcal{I}$ and spinful $\mathcal{T}$ symmetries. 
The gapless modes in SFig.~\ref{fig:helical-HOTI-before-adding-trivial-bands}(c) are helical hinge states [SFig.~\ref{fig:helical-HOTI-before-adding-trivial-bands}(i)] that are related to each other by $\mathcal{I}$ at $\mathcal{I}$-related hinges, and spinful $\mathcal{T}$ symmetry at the same hinges.

Employing the formalism developed in SN~\ref{sec:P_Wilson_loop} and \ref{sec:P_pm_Wilson_loop}, we next compute the $k_x$-directed $P$- and $P_{\pm}$-Wilson loop eigenphases for a highly-symmetric 2D slab of this helical HOTI infinite along $x$ and $y$ while finite along $z$ at half-filling.
Since the system has both $\mathcal{I}$ and spinful $\mathcal{T}$ symmetries, the set of $PsP$ eigenvalues $\{ (s_{z})_{n}(\mathbf{k})\}$ with $n=1,\ldots,N_{\text{occ}}$ at momentum $\mathbf{k}$ will be the same as the set $\{ -(s_{z})_{n}(\mathbf{k})\}$, as is discussed in SN~\ref{appendix:properties-of-the-projected-spin-operator}.
The spin gap in our model is open throughout the 3D BZ, such that we can define the $[P_{\pm}(\mathbf{k})]$ matrix projectors onto the positive and negative $P s_z P$ eigenspaces.  
In particular, we have $\text{rank}( [P_{+}(\mathbf{k})])=\text{rank}( [P_{-}(\mathbf{k})])=N_{\mathrm{occ}}/2$ for all $\mathbf{k}$. 
As shown in SFig.~\ref{fig:helical-HOTI-before-adding-trivial-bands}(d), the $k_x$-directed $P$-Wilson loop eigenphases for the occupied energy bands of the 2D slab with half-filling exhibit a helical winding when $k_y \to k_y + 2\pi$.
In addition, its $k_x$-directed $P_{\pm}$-Wilson loop eigenphases have winding numbers equal to $\mp 1$ when $k_y \to k_y + 2\pi$ as shown in SFig.~\ref{fig:helical-HOTI-before-adding-trivial-bands}(e--h). 
According to the sign convention in SEq.~\eqref{eq:partial_chern_def} relating the winding number of the Wilson loop spectrum and the Chern number, we deduce that the partial Chern numbers of this 2D slab are $C^\pm_{\gamma_1} = \mp 1$.
This confirms that the 2D slab of a helical HOTI is a 2D $\mathcal{T}$-invariant topological insulator with nontrivial $\mathbb{Z}_2$ invariant $\nu_{2d} = 1$~\cite{prodan2009robustness,3D_phase_diagram_spin_Chern_Prodan}.

\begin{figure*}[t]
\includegraphics[width=\linewidth]{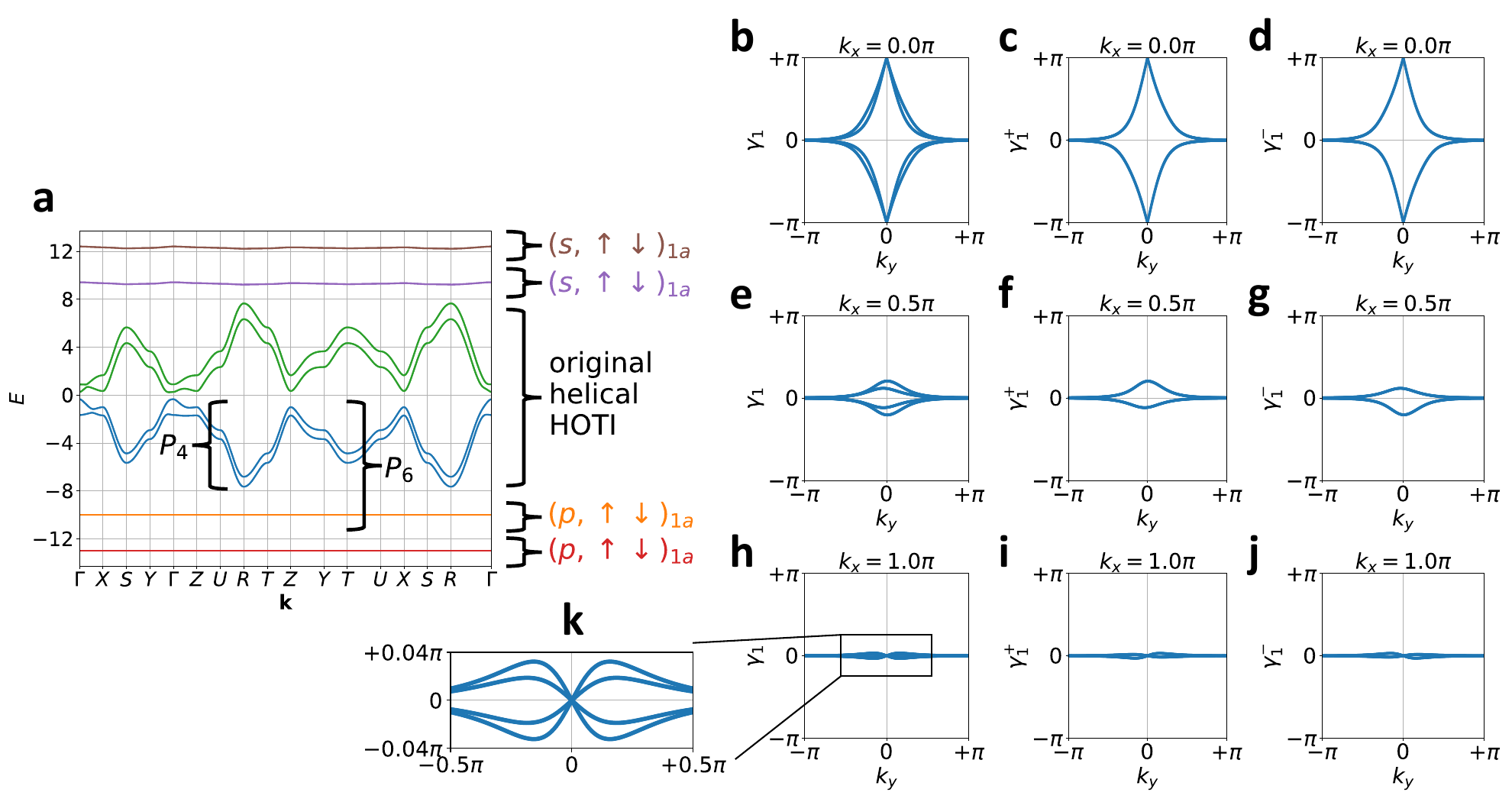}
\caption{$P$- and $P_\pm$-Wilson loops for four bands in the sixteen-band helical HOTI model. 
The model is obtained by coupling the eight-band helical HOTI model introduced in SRef.~\cite{wang2019higherorder} [and described in SEqs.~\eqref{eq:helical_HOTI_TB_model}, \eqref{eq:parameter_choice_of_helical_HOTI}, and \eqref{eq:A_spin_mixing_of_helical_HOTI}] to eight additional trivial bands, as described in SEqs.~\eqref{eq:H_sp_helical}, \eqref{eq:H_C_helical} and \eqref{eq:t_a_t_b_t_c_choice}. 
(a) shows the 3D bulk band structure where we have labeled the elementary band representation $(\rho,\uparrow\downarrow)_{\mathbf{q}}$ of the additional trivial bands induced from spinful $\rho$ orbitals at Wyckoff position $\mathbf{q}$~\cite{bradlyn2017topological,cano2018building}. 
The middle four doubly-degenerate bands (eight bands in total) are also indicated as originating from the ``original'' eight band helical HOTI in SFig.~\ref{fig:helical-HOTI-before-adding-trivial-bands}. 
$P_{4}$ denotes the projector onto the four valence bands of the original helical HOTI and $P_{6}$ is the projector constructed from the direct sum of $P_{4}$ and the projector onto the two additional trivial bands induced from spinful $p$ orbitals at the $1a$ Wyckoff position [whose bands are shown in orange in (a)]. 
(b), (c) and (d) show the $k_z$-directed $P$-, $P_{+}$- and $P_{-}$-Wannier bands as a function of $k_{y}$ for $P_{4}$ in the $k_{x} = 0$ plane. 
For our helical HOTI model with a 3D orthorhombic lattice, ``$k_z$-directed'' refers to our choice of $\mathbf{G} = 2\pi \hat{\mathbf{z}}$ in the $P$- and $P_{\pm}$-Wilson loop matrices in SEqs.~\eqref{eq:P_Wilson_loop_matrix_element} and \eqref{eq:P_pm_Wilson_loop_matrix_element}.
In (b) we see that the $P$-Wilson loop eigenphases evaluated with the projector $P_4$ have an even helical winding. 
On the other hand, in (c) and (d) we see that the $P_\pm$-Wilson loop eigenphases constructed from the projector $P_4$ have an odd helical winding.
As explained in the text, such windings can be removed and are fragile, as demonstrated in SFig.~\ref{fig:helical-HOTI-after-adding-trivial-bands-2}.
(e), (f) and (g) are the $k_z$-directed $P$-, $P_{+}$- and $P_{-}$-Wannier bands as a function of $k_{y}$ for $P_{4}$ in the $k_{x} = 0.5\pi$ plane. 
(h), (i) and (j) are the $k_z$-directed $P$-, $P_{+}$- and $P_{-}$-Wannier bands as a function of $k_{y}$ for $P_{4}$ in the $k_{x} = \pi$ plane. 
For (e)--(j), there is no spectral flow as opposed to (b)--(d). 
(k) is an enlarged plot of (h) showing that there are in total four Wannier bands in (h).
The numbers of bands for the $P$- [panels (b), (e), and (h)], $P_{+}$- [panels (c), (f), and (i)], and $P_{-}$ -Wannier bands [panels (d), (g), and (j)] are $4$, $2$ and $2$ respectively.
The calculations detailed in this figure were performed using the freely available Python package~\href{https://github.com/kuansenlin/nested_and_spin_resolved_Wilson_loop}{\textsc{nested\_and\_spin\_resolved\_Wilson\_loop}}~\cite{lin2023nestedWilsonLib}, which represents an extension of the~\href{https://www.physics.rutgers.edu/pythtb/}{PythTB} open-source Python tight-binding package~\cite{coh2013python} that was implemented and utilized for the preparation of SRefs.~\cite{wieder2018axion,wieder2020strong} and the present work.
}
\label{fig:helical-HOTI-after-adding-trivial-bands-1}
\end{figure*}

Next, anticipating the possibility of fragile Wilson loop winding at TRIM planes in the 3D BZ (see SRefs.~\cite{wieder2020strong,wieder2018axion}) with $\mathcal{I}$ and spinful $\mathcal{T}$ symmetries, we couple our helical HOTI model to eight additional trivial bands which also respect $\mathcal{I}$ and spinful $\mathcal{T}$ symmetries. 
We place two spinful $s$-like orbitals at a generic position $\mathbf{r}_{3} = ( r_{3,x}, r_{3,y} , r_{3,z} ) = (0.35,0.15,0.31)$ in the unit cell (in reduced coordinates) and another two spinful $s$-like orbitals at $-\mathbf{r}_{3}$. 
We will use $d^{\dagger}_{\mathbf{R},\mu,\sigma}$ [$f^{\dagger}_{\mathbf{R},\mu,\sigma}$] to denote the second-quantized creation operators for the $\mu^{\text{th}}$ ($\mu = 1,2$) spin $\sigma$ ($\sigma = \uparrow,\downarrow$) $s$-like orbital at $\mathbf{r}_{3}$ [$-\mathbf{r}_{3}$] in unit cell $\mathbf{R}$. 
There are in total {\it eight} additional tight-binding basis orbitals.
In order to induce spinful $p$-like orbitals at the $1a$ Wyckoff position, we couple these eight additional degrees of freedom through the following Hamiltonian:
\begin{align}
    H_{sp} = & \sum_{\mathbf{R},\mu,\sigma}\left[ t_{a} \left( d^{\dagger}_{\mathbf{R},\mu,\sigma} f_{\mathbf{R},\mu,\sigma} + \text{H.c.} \right) \right] + \sum_{\mathbf{R},\mu,\sigma} \left[ t_{b}\left( d^{\dagger}_{\mathbf{R},\mu,\sigma}d_{\mathbf{R},\mu,\sigma} + f^{\dagger}_{\mathbf{R},\mu,\sigma} f_{\mathbf{R},\mu,\sigma} \right) \right]  \nonumber \\
    & + \sum_{\mathbf{R},\sigma} \left[ t_{c}\left( d^{\dagger}_{\mathbf{R},1,\sigma} d_{\mathbf{R},1,\sigma} - d^{\dagger}_{\mathbf{R},2,\sigma} d_{\mathbf{R},2,\sigma} + f^{\dagger}_{\mathbf{R},1,\sigma} f_{\mathbf{R},1,\sigma} - f^{\dagger}_{\mathbf{R},2,\sigma} f_{\mathbf{R},2,\sigma} \right) \right], \label{eq:H_sp_helical}
\end{align}
where $t_{a}$, $t_{b}$ and $t_{c}$ are all real numbers, and H.c. means the Hermitian conjugate.
A nonzero value of $t_{a}$ will hybridize the single-particle states $| d_{\mathbf{R},\mu,\sigma}\rangle = d_{\mathbf{R},\mu,\sigma}^{\dagger}|0\rangle $ and $| f_{\mathbf{R},\mu,\sigma}\rangle = f_{\mathbf{R},\mu,\sigma}^{\dagger}|0\rangle $. 
Next, we couple these additional orbitals to the original helical HOTI model [SEq.~(\ref{eq:helical_HOTI_TB_model})]. 
The creation [annihilation] operators of the spinful $s$ and $ip$ orbitals in the original helical HOTI model will be denoted as $c^{\dagger}_{\mathbf{R},s,\mu,\sigma}$ [$c_{\mathbf{R},s,\mu,\sigma}$] and $c^{\dagger}_{\mathbf{R},ip,\mu,\sigma}$ [$c_{\mathbf{R},ip,\mu,\sigma}$]. 
$c^{\dagger}_{\mathbf{R},s,\mu,\sigma}$ [$c^{\dagger}_{\mathbf{R},ip,\mu,\sigma}$] creates an electron in the $\mu^{\text{th}}$ ($\mu = 1,2$) $s$ [$ip$] orbital with spin $\sigma$ ($\sigma = \uparrow,\downarrow$) at the origin of the unit cell $\mathbf{R}$.
The Hamiltonian that couples the additional eight tight-binding basis states created by $d^{\dagger}_{\mathbf{R},\mu,\sigma}$  and $f^{\dagger}_{\mathbf{R},\mu,\sigma}$ to the original helical HOTI model [SEq.~(\ref{eq:helical_HOTI_TB_model})] is given by
\begin{align}
    H_{C} = \sum_{\mathbf{R},\sigma}\sum_{\mu,\nu=1}^{2} \left[ t_{d}\left( c^{\dagger}_{\mathbf{R},s,\mu,\sigma} \left\{ d_{\mathbf{R},\nu,\sigma} + f_{\mathbf{R},\nu,\sigma}\right\} + \text{H.c.} \right) \right]. \label{eq:H_C_helical}
\end{align}
Both $H_{sp}$ in SEq.~(\ref{eq:H_sp_helical}) and $H_{C}$ in SEq.~(\ref{eq:H_C_helical}) are $\mathcal{I}$- and $\mathcal{T}$-symmetric. 
We take
\begin{align}
    t_{a} = 11.0,\ t_{b} = -0.5,\ t_{c} = 1.5, t_d=1.0. \label{eq:t_a_t_b_t_c_choice}
\end{align}
The 3D bulk band structure of the resulting sixteen-band model has eight pairs of doubly-degenerate bands and is shown in SFig.~\ref{fig:helical-HOTI-after-adding-trivial-bands-1}(a) where the middle four doubly degenerate bands (eight bands in total) are from the original helical HOTI model [SEq.~\eqref{eq:helical_HOTI_TB_model}], as indicated in the figure. 
Our choice $|t_a|\gg |t_b|, |t_c|,|t_d|$ ensures that the lowest four bands in SFig.~\ref{fig:helical-HOTI-after-adding-trivial-bands-1}(a) are topologically trivial.

In order to explicitly see the fragile Wilson loop winding,
we first compute the $k_z$-directed $P$- and $P_{\pm}$-Wilson loop eigenphases as a function of $k_{y}$ for different constant-$k_{x}$ planes for the four valence bands from the original helical HOTI model denoted by $P_{4}$ in SFig.~\ref{fig:helical-HOTI-after-adding-trivial-bands-1}(a). 
For our model with a 3D orthorhombic lattice, taking the $P$- and $P_\pm$-Wilson loops along the $k_z$ direction ensures that our (spin-resolved) Wannier bands correspond to states localized in the $\hat{\mathbf{z}}$ direction, allowing us to make contact with the layer constructions of SN~\ref{app:comparison-spin-stable-and-symmetry-indicated-topology}.  
Again, we construct the $[P_{\pm}(\mathbf{k})]$ matrix projectors such that they project to the positive and negative $P s_z P$ eigenspace, respectively.
At constant-$k_{x}$ planes, the $P$-Wannier bands [SFig.~\ref{fig:helical-HOTI-after-adding-trivial-bands-1}(b,e,h)] can be spin-resolved into $P_{+}$-Wannier bands [SFig.~\ref{fig:helical-HOTI-after-adding-trivial-bands-1}(c,f,i)] and $P_{-}$-Wannier bands [SFig.~\ref{fig:helical-HOTI-after-adding-trivial-bands-1}(d,g,j)].
In particular, the number of $P_{+}$-Wannier bands is the same as the $P_{-}$-Wannier bands, as $\text{rank}( [P_{+}(\mathbf{k})])=\text{rank}( [P_{-}(\mathbf{k})])$.
Notice that the $P$-Wannier bands are not a direct superposition of the corresponding $P_{\pm}$-Wannier bands, which would be the case in a system with $s_z$ conservation.
At the $k_{x} = 0$ plane, the $P$-Wannier bands [SFig.~\ref{fig:helical-HOTI-after-adding-trivial-bands-1}(b)] consist of two sets of helically winding bands.
This is the pattern of Wilson loop winding that one would obtain from superposing two identical copies of a 3D spinful $\mathcal{T}$-invariant strong topological insulator (TIs) with $\mathcal{I}$ symmetry~\cite{po2017symmetry,song2018mapping,khalaf2018symmetry}. 
This doubled helical winding in the $P$-Wilson loop spectrum in SFig.~\ref{fig:helical-HOTI-after-adding-trivial-bands-1}(b), however, is fragile; below we will remove it by considering the projector $P_6$ indicated in SFig.~\ref{fig:helical-HOTI-after-adding-trivial-bands-1}(a). 

Turning now to the spin-resolved topology, we can consider the $P_{\pm}$-Wilson loops constructed from the projector $P_4$. 
At the $k_{x} = 0$ plane, each of the $P_{\pm}$-Wannier bands [SFig.~\ref{fig:helical-HOTI-after-adding-trivial-bands-1}(c,d)] exhibits fragile winding (we will explicitly demonstrate the fragility shortly) that is protected by $\mathcal{I}$ symmetry, while at other constant-$k_{x}$ planes there is no such fragile winding [SFig.~\ref{fig:helical-HOTI-after-adding-trivial-bands-1}(f,g,i,j)]. 
Notice that at the $k_{x} = 0$ and $\pi$ planes, each of the positive and negative $P s_z P$ eigenspace has $\mathcal{I}$ symmetry while there is no spinful $\mathcal{T}$ symmetry---the spinful $\mathcal{T}$ operation maps between the positive and negative $P s_z P$ eigenspace.
The $P_\pm$-Wannier band crossing in SFig.~\ref{fig:helical-HOTI-after-adding-trivial-bands-1}(c,d) is fragile, in the sense that they can be removed by  adding additional trivial degrees of freedom to the occupied subspace.   
This implies that in 2D, there may also exist phases of matter with \emph{fragile spin-resolved topology}. 
We leave the further exploration of this intriguing possibility for future work. 
By considering the bands in $P_{4}$ of SFig.~\ref{fig:helical-HOTI-after-adding-trivial-bands-1}(a) together with trivial bands induced from spinful $p$-orbitals at the $1a$ Wyckoff position, we may remove the fragile winding in both the $P$- and $P_{\pm}$-Wannier bands at the $k_{x} = 0$ planes.

\begin{figure*}[t]
\includegraphics[width=0.9\linewidth]{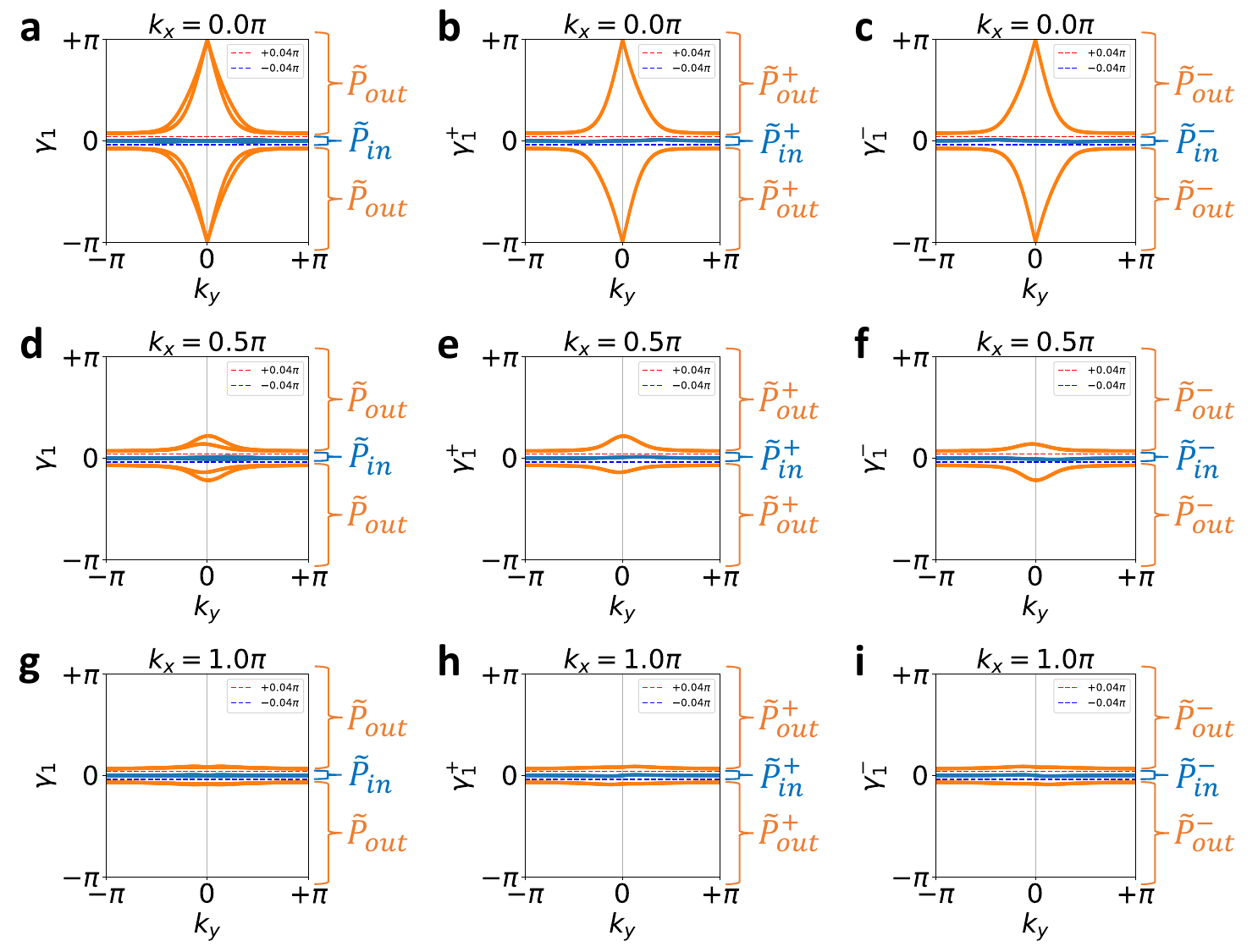}
\caption{$P$- and $P_{\pm}$-Wilson loops with $\mathbf{G} = 2\pi \hat{\mathbf{z}}$ [SEqs.~\eqref{eq:P_Wilson_loop_matrix_element} and \eqref{eq:P_pm_Wilson_loop_matrix_element}] of the sixteen-band helical HOTI model whose energy band structure is shown in SFig.~\ref{fig:helical-HOTI-after-adding-trivial-bands-1}(a). 
The model is obtained by coupling the eight-band helical HOTI model [SEqs.~\eqref{eq:helical_HOTI_TB_model}, \eqref{eq:parameter_choice_of_helical_HOTI}, and \eqref{eq:A_spin_mixing_of_helical_HOTI}] to eight additional trivial bands described by SEqs.~\eqref{eq:H_sp_helical}, \eqref{eq:H_C_helical} and \eqref{eq:t_a_t_b_t_c_choice}. 
The occupied energy band projector considered here is $P_6$ in SFig.~\ref{fig:helical-HOTI-after-adding-trivial-bands-1}(a), which projects onto the valence bands of the original eight-band model of a helical HOTI together with two additional trivial bands induced from spinful $p$ orbitals at the $1a$ Wyckoff position.
(a), (b) and (c) are the $k_z$-directed $P$-, $P_{+}$- and $P_{-}$-Wannier bands as functions of $k_{y}$ for $P_6$ in the $k_{x} = 0$ plane.
(d), (e) and (f) are the $k_z$-directed $P$-, $P_{+}$- and $P_{-}$-Wannier bands as functions of $k_{y}$ for $P_{6}$ in the $k_{x} = 0.5\pi$ plane.
(g), (h) and (i) are the $k_z$-directed $P$-, $P_{+}$- and $P_{-}$-Wannier bands as functions of $k_{y}$ for $P_{6}$ in the $k_{x} = \pi$ plane.
We can see that the spectral flow in SFig.~\ref{fig:helical-HOTI-after-adding-trivial-bands-1}(b--d) has been trivialized in panels (a)--(c).
In particular, for each panel (a)--(i) of this figure, the $P$- and $P_{\pm}$-Wannier bands can be separated into inner and outer sets separated by a Wannier band gap size of at least $\approx 0.04\pi$.
For each panel, we have also labeled the projectors onto the inner and outer set of Wannier bands by $\widetilde{P}_{in/out}$ and $\widetilde{P}_{in/out}^{\pm}$ explicitly, following the notation of SRef.~\cite{wieder2018axion}.
We can view the inner and outer sets of Wannier bands here as representing hybrid Wannier states localized around the inversion-symmetric center ($z=0$) and boundary ($z=1/2$) of the unit cell [SN~\ref{app:z2_nested_P_pm_berry_phase}].
The numbers of bands for the $P$- [panels (a), (d), and (g)], $P_{+}$- [panels (b), (e), and (h)], and $P_{-}$-Wannier bands [panels (c), (f), and (i)] in here are $6$, $3$ and $3$, respectively.
The calculations detailed in this figure were performed using the freely available Python package~\href{https://github.com/kuansenlin/nested_and_spin_resolved_Wilson_loop}{\textsc{nested\_and\_spin\_resolved\_Wilson\_loop}}~\cite{lin2023nestedWilsonLib}, which represents an extension of the~\href{https://www.physics.rutgers.edu/pythtb/}{PythTB} open-source Python tight-binding package~\cite{coh2013python} that was implemented and utilized for the preparation of SRefs.~\cite{wieder2018axion,wieder2020strong} and the present work.
}
\label{fig:helical-HOTI-after-adding-trivial-bands-2}
\end{figure*}

To remove the fragile winding in SFig.~\ref{fig:helical-HOTI-after-adding-trivial-bands-1}(b--d) while---crucially---not changing the bulk stable (spin-resolved) topology, we add the doubly degenerate bands right below the image of $P_{4}$ to construct the projector $P_{6}$ onto the set of bands denoted in SFig.~\ref{fig:helical-HOTI-after-adding-trivial-bands-1}(a), which contains in total six bands.
We then compute the $k_z$-directed $P$- and $P_{\pm}$-Wannier bands as a function of $k_{y}$ at different constant-$k_{x}$ planes for the six bands in the image of $P_{6}$. 
As shown in SFig.~\ref{fig:helical-HOTI-after-adding-trivial-bands-2}(a--c), there is no longer any winding of the $P$- and $P_{\pm}$-Wannier bands at the $k_{x} = 0$ plane. 
Moreover, for all constant-$k_{x}$ constant planes, the $P$- and $P_{\pm}$-Wannier bands can be divided into two disjoint sets of {\it inner} and {\it outer} bands [SN~\ref{app:z2_nested_P_pm_berry_phase}] that are spectrally separated from each other by a {\it Wannier gap} of size $\approx 0.04\pi$, as indicated in SFig.~\ref{fig:helical-HOTI-after-adding-trivial-bands-2}(a--i).
The inner and outer set of $k_z$-directed Wannier bands here represent hybrid Wannier states localized at the inversion-symmetric center ($z=0$) and boundary ($z=1/2$) of the unit cell respectively.
Importantly, as discussed in SN~\ref{app:z2_nested_P_pm_berry_phase}, the inner and outer sets of Wannier bands both obey SEq.~\eqref{eq:app-ph-symmetry-gamma1pmkjkl} due to bulk $\mathcal{I}$ symmetry. 
We refer the readers to SN~\ref{appendix:I-constraint-on-P-wilson} and \ref{appendix:I-constraint-on-P-pm} for the rigorous proof of the inversion symmetry constraint on the $P$- and $P_{\pm}$-Wannier bands.
Together with the fact that the sixteen-band helical HOTI is $\mathcal{I}$- and $\mathcal{T}$-symmetric with gapped energy and $P s_z P$ spectrum, We can next apply the formalism of SN~\ref{sec:nested_P_Wilson_loop} and \ref{sec:nested_P_pm_Wilson_loop} to compute the nested $P$- and nested $P_\pm$-Wilson loops for the occupied bands in the image of the projector $P_6$.

Before we move on to compute the nested $P$- and $P_{\pm}$-Wilson loop eigenphases of our helical HOTI model, we here propose that a 3D helical HOTI can be viewed as a pump of a 2D $\mathcal{I}$- and $\mathcal{T}$-symmetric fragile (or obstructed atomic) phase, following the results of SRef.~\cite{wieder2018axion}. 
We emphasize first that since we use a $k_z$-directed Wilson loop to define our Wannier bands (to make contact with the layer constructions of SN~\ref{app:comparison-spin-stable-and-symmetry-indicated-topology}), the nested Wilson loop spectra varies as a function of $k_x$. 
Note first that the fragile winding of the $P$-Wannier bands at the $k_x = 0$ plane can be removed by adding trivial degrees of freedom that respect $\mathcal{I}$ and $\mathcal{T}$ symmetries. 
In the sixteen-band helical HOTI model the trivial bands are induced from spinful $p$ orbitals at the $1a$ Wyckoff position. 
Next, since there is a spin gap, we can decompose the occupied space into two parts---the positive and negative $P s_z P$ eigenspaces related to each other by $\mathcal{T}$. 
The fragile winding of the $P_{\pm}$-Wannier bands at the $k_x = 0$ plane can be removed by adding trivial degrees of freedom that respect $\mathcal{I}$ symmetry within each of the $P_{\pm}$-eigenspace. 
In our case such trivial bands within each of the $P_{\pm}$-eigenspaces are induced from spinless $p$ orbitals at the $1a$ Wyckoff position.
If we view $k_x$ as a pumping parameter, we can say that a helical HOTI demonstrates a pump from a 2D fragile topological insulator with both $\mathcal{I}$ and $\mathcal{T}$ symmetries and (filling-) anomalous spin-charge-separated corner modes~\cite{wieder2018axion,wang2019higherorder,WiederDefect} at the $k_x = 0$ plane to a 2D trivial insulator at the $k_x = \pi$ plane. 
This is the spinful time-reversal-symmetric generalization of the concept that a magnetic AXI can be viewed as a pump from a 2D fragile topological insulator to a 2D trivial insulator with $\mathcal{I}$ symmetry~\cite{wieder2018axion}.
In other words, from this pumping perspective, a helical HOTI can be viewed as superposing (\emph{i.e.}, ``stacking'' in the language of SRef.~\cite{po2017symmetry}) two orbital (effectively spinless or spin-polarized) magnetic AXIs related to each other by spinful $\mathcal{T}$ symmetry. 
This picture is the momentum-space analogue of the position-space, spin-resolved layer construction of a helical HOTI in the T-DAXI regime introduced in SN~\ref{app:comparison-spin-stable-and-symmetry-indicated-topology}.

\begin{figure}[ht]
\includegraphics[width=0.695\linewidth]{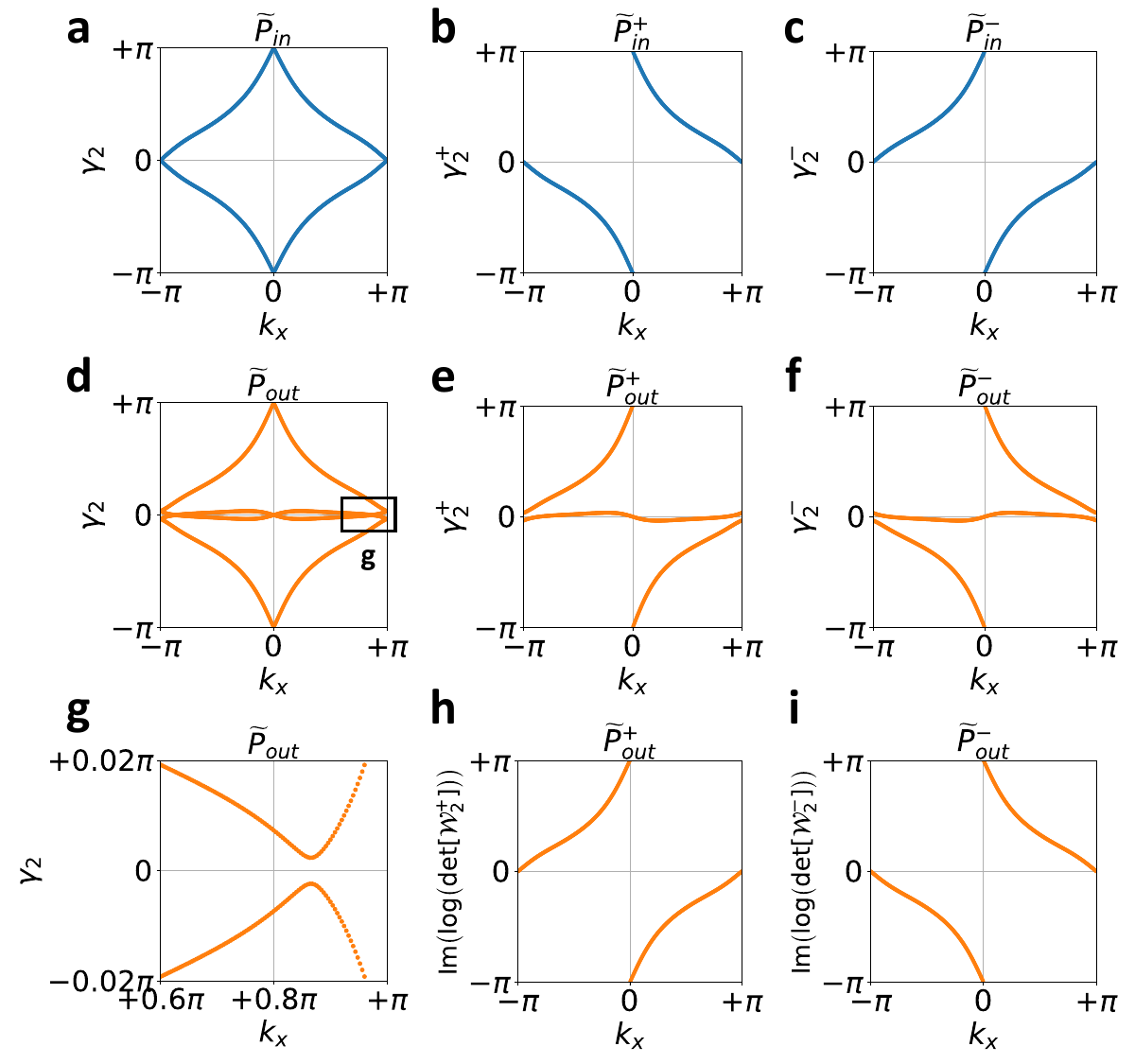}
\caption{Nested $P$- and $P_{\pm}$-Wilson loops with $\mathbf{G} = 2\pi \hat{\mathbf{z}}$ and $\mathbf{G}' = 2\pi \hat{\mathbf{y}}$ [SEqs.~\eqref{eq:nested_P_Wilson_loop_matrix_def} and \eqref{eq:nested_P_pm_Wilson_loop_matrix_def}] as the first and second closed-loop holonomy of the sixteen-band helical HOTI model whose energy band structure is shown in SFig.~\ref{fig:helical-HOTI-after-adding-trivial-bands-1}(a). 
The model is obtained by coupling the eight-band helical HOTI model [SEqs.~\eqref{eq:helical_HOTI_TB_model}, \eqref{eq:parameter_choice_of_helical_HOTI}, and \eqref{eq:A_spin_mixing_of_helical_HOTI}] to eight additional trivial bands described by SEqs.~\eqref{eq:H_sp_helical}, \eqref{eq:H_C_helical} and \eqref{eq:t_a_t_b_t_c_choice}.
We use the projector $P_6$ indicated in SFig.~\ref{fig:helical-HOTI-after-adding-trivial-bands-1}(a) to construct the nested Wilson loops.
(a), (b) and (c) are the nested $P$-, $P_{+}$- and $P_{-}$-Wilson loop eigenphases of the inner Wannier bands denoted by $\widetilde{P}_{in}$, $\widetilde{P}_{in}^{+}$ and $\widetilde{P}_{in}^{-}$ in SFig.~\ref{fig:helical-HOTI-after-adding-trivial-bands-2}, respectively.
We observe an odd helical winding in the spectrum of the nested $P$-Wilson loop shown in (a). 
In the nested spin-resolved Wilson loop spectra shown in (b) and (c), we see a corresponding odd chiral winding with winding number $-1$ in the nested $P_+$-Wilson loop spectrum and winding number $+1$ in the nested $P_-$-Wilson loop spectrum.
(d), (e) and (f) are the nested $P$-, $P_{+}$- and $P_{-}$-Wilson loop eigenphases of the outer Wannier bands denoted by $\widetilde{P}_{out}$, $\widetilde{P}_{out}^{+}$ and $\widetilde{P}_{out}^{-}$ in SFig.~\ref{fig:helical-HOTI-after-adding-trivial-bands-2}, respectively.
(g) is an enlarged view of (d) which demonstrates that the band crossings at generic momentum $k_x \neq 0$ and $k_x \neq \pi$ are not protected and are gapped.
There is also a helical winding in (d), similar to that shown in panel (a). 
Correspondingly, the nested spin-resolved Wilson loop spectra for the outer spin-resolved Wannier bands shown in (e) and (f) have odd chiral winding. 
(h) and (i) are the total nested partial Berry phases (each of (e) and (f) contains two bands) of the outer $P_{\pm}$-Wannier bands denoted by $\widetilde{P}_{out}^{+}$ and $\widetilde{P}_{out}^{-}$ in SFig.~\ref{fig:helical-HOTI-after-adding-trivial-bands-2}, respectively.
From (h) and (i), we see that there is spectral flow with winding numbers $+1$ and $-1$ in (e) and (f), respectively. 
Together with (b) and (c) which carry opposite winding numbers of (h) and (i) respectively, this indicates that the helical HOTI in SFig.~\ref{fig:helical-HOTI-after-adding-trivial-bands-1} lies in the T-DAXI regime and is hence characterized by the nontrivial \emph{partial} axion angles $\theta^\pm=\pi$ [see SEq.~\eqref{eq:daxi_theta_from_model} and the surrounding text]. 
The number of nested (spin-resolved) Wannier bands in (a), (b), and (c) is $2$, $1$, and $1$, respectively,
and the number of nested (spin-resolved) Wannier bands in (d), (e), and (f) is $4$, $2$, and $2$, respectively.
The calculations detailed in this figure were performed using the freely available Python package~\href{https://github.com/kuansenlin/nested_and_spin_resolved_Wilson_loop}{\textsc{nested\_and\_spin\_resolved\_Wilson\_loop}}~\cite{lin2023nestedWilsonLib}, which represents an extension of the~\href{https://www.physics.rutgers.edu/pythtb/}{PythTB} open-source Python tight-binding package~\cite{coh2013python} that was implemented and utilized for the preparation of SRefs.~\cite{wieder2018axion,wieder2020strong} and the present work.
}
\label{fig:helical-HOTI-after-adding-trivial-bands-3}
\end{figure}

We next compute the nested $P$- and nested $P_{\pm}$-Wilson loop eigenphases for our sixteen-band HOTI model
[SEqs.~(\ref{eq:helical_HOTI_TB_model}),~(\ref{eq:H_sp_helical}), and~(\ref{eq:H_C_helical})].
After presenting the numerical results, we will then discuss how the coexistence of both $\mathcal{I}$ and spinful $\mathcal{T}$ symmetries can constrain the various nested (partial) Berry phases.
For the $P$-Wannier bands in SFig.~\ref{fig:helical-HOTI-after-adding-trivial-bands-2}(a,d,g), the disjoint inner and outer set of bands contain two and four bands, respectively.
For the $P_{\pm}$-Wannier bands in SFig.~\ref{fig:helical-HOTI-after-adding-trivial-bands-2}(b,c,e,f,h,i), the disjoint inner and outer set of bands contain one and two bands, respectively.
We then construct the projectors onto the disjoint inner and outer set of $P$- [$P_{\pm}$-] Wannier bands, which we label as $\widetilde{P}_{in}$ [$\widetilde{P}_{in}^{\pm}$] and $\widetilde{P}_{out}$ [$\widetilde{P}_{out}^{\pm}$], respectively, as shown in SFig.~\ref{fig:helical-HOTI-after-adding-trivial-bands-2}(a--i).
As the decomposition of the occupied band and positive/negative spin band projectors into $\widetilde{P}_{in/out}$ and $\widetilde{P}_{in/out}^{\pm}$ are well-defined and smooth over the full Brillouin zone, we can compute the corresponding {\it nested} $P$- and $P_{\pm}$-Wilson loop eigenphases.
Following the formalism introduced in this work in SN~\ref{sec:nested_P_Wilson_loop} and \ref{sec:nested_P_pm_Wilson_loop}, we compute the nested $P$- and $P_\pm$-Wilson loops by choosing the primitive reciprocal lattice vectors $\mathbf{G} = 2\pi \hat{\mathbf{z}}$ and $\mathbf{G}'= 2 \pi \hat{\mathbf{y}}$ [SEqs.~\eqref{eq:nested_P_Wilson_loop_matrix_def} and \eqref{eq:nested_P_pm_Wilson_loop_matrix_def}] as the first and second closed-loop holonomy, and 
the resulting eigenphases, which we also term nested Berry phases generally in this work, are plotted as functions of $k_{x}$.
In SFig.~\ref{fig:helical-HOTI-after-adding-trivial-bands-3}(a,d), the nested $P$-Wilson loop eigenphases for $\widetilde{P}_{in}$ and $\widetilde{P}_{out}$ both have an odd {\it helical} winding. 
And as shown in SFig.~\ref{fig:helical-HOTI-after-adding-trivial-bands-3}(g), accidental crossing points in the nested $P$-Wilson loop eigenphases at generic momentum $k_x \neq 0$ and $k_x \neq \pi$ are not protected by symmetries or local topological invariants (like those protecting Weyl points~\cite{armitage2018weyl}), and are hence generically gapped.
SFig.~\ref{fig:helical-HOTI-after-adding-trivial-bands-3}(a,d,g) therefore indicate that the 2D hybrid Wannier states localized at the $\mathcal{I}$-symmetric center ($z=0$) and boundary ($z=1/2$) planes of the unit cell both carry the same stable topology, and are specifically both topologically equivalent to $\mathcal{T}$-invariant 2D strong TIs.
This can be compared with magnetic AXIs, whose spectral flow in the nested $P$-Wilson loop eigenphases has a nonzero odd chiral winding that is protected by bulk crystal symmetries, such as $\mathcal{I}$~\cite{wieder2018axion}.
By analogy, we hence establish that odd, helical winding in the nested $P$-Wilson loop spectrum is one bulk signature of a helical HOTI.
As established in Supplementary Table~\ref{tab:layer-construction-given-C-m}, this is consistent with the recognition that a helical HOTI can be layer-constructed as an $\mathcal{I}$-symmetric stack of $\mathcal{T}$-invariant 2D strong TIs, with two parallel TI layers per unit cell separated by a half-lattice translation and pinned to $\mathcal{I}$-invariant real-space planes~\cite{song2018mapping,elcoro2021magnetic,gao2022magnetic}.

Due to the presence of a spin gap and a $P_\pm$-Wannier gap (see SFig.~\ref{fig:helical-HOTI-after-adding-trivial-bands-2}), we can further examine the spin-resolved topology of the sixteen-band helical HOTI model [SEqs.~(\ref{eq:helical_HOTI_TB_model}),~(\ref{eq:H_sp_helical}), and~(\ref{eq:H_C_helical})] by computing the nested $P_{\pm}$-Wilson loop eigenphases.
Let us denote the nested partial Chern numbers computed from $\widetilde{P}_{in}^{\pm}$ and $\widetilde{P}_{out}^{\pm}$ as $C_{\gamma_2 , in}^{\pm}$ and $C_{\gamma_2 , out}^{\pm}$, respectively.
We will here employ the sign convention used in SEq.~\eqref{eq:partial_chern_def} to relate the winding number of the (nested) Wilson loop spectrum to the Chern number (which relies on fixing a right-handed coordinate system) to determine the signs of $C_{\gamma_2 , in}^{\pm}$ and $C_{\gamma_2 , out}^{\pm}$ consistently.  In this convention, $C_{\gamma_2 , in}^{\pm}$ and $C_{\gamma_2 , out}^{\pm}$ are also equal to the partial Chern numbers of the spin-resolved hybrid Wannier states localized at the center and the boundary of the unit cell.
As discussed in SN~\ref{app:z2_nested_P_pm_berry_phase}, $C_{\gamma_2 , in}^{\pm}$ and $C_{\gamma_2 , out}^{\pm}$ are only $\mathbb{Z}_2$-stable after accounting for Wannier gap closures without accompanying closures of the spin gap or the energy gap.
From SFig.~\ref{fig:helical-HOTI-after-adding-trivial-bands-3}(b,c,e,f,h,i), we observe that for the sixteen-band helical HOTI model, the nested partial Chern numbers are
\begin{align}
    & C_{\gamma_2 , in}^{\pm} = \pm 1, \label{eq:C_gamma_2_in_pm_our_helical_HOTI} \\
    & C_{\gamma_2 , out}^{\pm} = \mp 1. \label{eq:C_gamma_2_out_pm_our_helical_HOTI}
\end{align}
We emphasize that $C_{\gamma_2 , in}^{\pm}$ and $C_{\gamma_2 , out}^{\pm}$ in SEq.~(\ref{eq:C_gamma_2_out_pm_our_helical_HOTI}) are the \textit{negatives} of the winding numbers of the corresponding nested spin-resolved Wilson loop eigenphases in SFig.~\ref{fig:helical-HOTI-after-adding-trivial-bands-3}(b,c,e,f,h,i), owing to the sign convention in SEq.~(\ref{eq:partial_chern_def}), in which the Chern number is equal to the negative [positive] winding number of the $k_y$-directed [$k_x$-directed] Wilson loop eigenphases as $k_x \to k_x + 2\pi$ [$k_y \to k_y + 2\pi$].
SEqs.~(\ref{eq:C_gamma_2_in_pm_our_helical_HOTI}) and~(\ref{eq:C_gamma_2_out_pm_our_helical_HOTI}) are consistent with the spin-stable topology of a T-DAXI characterized by the $\mathbb{Z}_2 \times \mathbb{Z}$ spin-stable topological indices 
\begin{equation}
    (C_{\gamma_2 , in}^{+}\mod 2,\nu_{z}^{+})=(1 \mod 2,0) = (1,0) 
\end{equation}
using the classification previously introduced in SN~\ref{app:comparison-spin-stable-and-symmetry-indicated-topology}, in which $\nu_{z}^{+}$ is the partial weak Chern number in the $xy$ plane, and where we have specialized to cases in which the other, unspecified $\mathbb{Z}$-valued partial weak Chern numbers vanish $\nu_x^\pm=\nu_y^\pm=0$ (see also Supplementary Table~\ref{tab:summary-for-qshi-and-daxi}).

Next, we recall that as established in SRef.~\cite{wieder2018axion}, the nontrivial axion angle $\theta=\pi$ of a magnetic AXI can be inferred from the presence of an odd winding number in the nested $P$-Wilson loop eigenphases, provided that the weak Chern numbers $\nu_i$ all vanish (and that the bulk contains magnetic crystal symmetries that quantize $\theta$, namely those that change the sign of ${\bf E}\cdot{\bf B}$).
From this perspective, SFig.~\ref{fig:helical-HOTI-after-adding-trivial-bands-3}(b,c,e,f,h,i), together with SEqs.~\eqref{eq:C_gamma_2_in_pm_our_helical_HOTI} and \eqref{eq:C_gamma_2_out_pm_our_helical_HOTI}, therefore demonstrate that the positive and negative $P s_z P$ eigenspaces can each be viewed as orbital (effectively spinless or spin-polarized) AXIs. 
Hence our helical HOTI model lies in the T-DAXI regime.  
From the odd winding numbers $C_{\gamma_2 , in/out}^{\pm}$ [SEqs.~\eqref{eq:C_gamma_2_in_pm_our_helical_HOTI} and \eqref{eq:C_gamma_2_out_pm_our_helical_HOTI}], we specifically deduce that the partial axion angles [SEq.~\eqref{eq:quantized-theta-pm-DAXI}] for the sixteen-band helical HOTI model are 
\begin{equation}
\theta^\pm = \pi. \label{eq:daxi_theta_from_model}
\end{equation}
We further note that the winding numbers $C_{\gamma_2 , in}^{\pm}$ and $C_{\gamma_2 , out}^{\pm}$ are opposite to each other since $\nu_z^{\pm} = 0$ [SEq.~\eqref{eq:general_condition_diff_group_C_gamma_2_pm}] in our helical HOTI model, which most clearly differentiates it from a spin-stable 3D QSHI as discussed in SN~\ref{app:comparison-spin-stable-and-symmetry-indicated-topology} (see specifically Supplementary Table~\ref{tab:summary-for-qshi-and-daxi}). 
For completeness, it is also useful to note that $\mathcal{T}$ symmetry enforces that the winding numbers $C_{\gamma_2 , in}^{+}$ [$C_{\gamma_2 , out}^{+}$] and $C_{\gamma_2 , in}^{-}$ [$C_{\gamma_2 , out}^{-}$]
carry opposite signs [SN~\ref{app:z2_nested_P_pm_berry_phase} and \ref{appendix:T-constraint-on-nested-P-pm}].

As discussed in SN~\ref{app:comparison-spin-stable-and-symmetry-indicated-topology}, since our helical HOTI model is a spin-stable T-DAXI with an $s_z = \hat{\mathbf{z}} \cdot \mathbf{s}$ spin gap, we have that the bulk topological contribution to its $s_z$ spin Hall conductivity in the $xy$-plane per unit cell along $z$ vanishes $(\sigma_{\mathrm{T-DAXI}}^{s})_{\mathrm{topological}} = 0$ [SEq.~\eqref{eq:sigma_s_DAXI}].  However, we importantly find that it instead carries a bulk topological contribution to the partial magnetoelectric polarizability $\alpha^{\pm} = \pi \cdot \frac{e^2}{2\pi h}$ [SEq.~\eqref{eq:partial-magnetoelectric-polarizability-def}].

To summarize, we have explicitly demonstrated the following bulk signatures of a symmetry-indicated, spin-gapped helical HOTI lying in the T-DAXI regime:
\begin{enumerate}
    \item Independent of its spin resolution, a helical HOTI is characterized by an odd helical winding in the nested $P$-Wilson loop eigenphases as demonstrated in SFig.~\ref{fig:helical-HOTI-after-adding-trivial-bands-3}(a,d), whenever there is a Wannier gap as demonstrated in SFig.~\ref{fig:helical-HOTI-after-adding-trivial-bands-2}(a,d,g).
    \item Independent of its spin-resolution, a helical HOTI can be viewed as a momentum-space pump from an $\mathcal{I}$- and spinful $\mathcal{T}$-symmetric 2D fragile topological insulator (or obstructed atomic insulator) to a 2D trivial insulator, building on the description of $\mathcal{I}$-symmetric magnetic AXIs established in SRef.~\cite{wieder2018axion}.
    \item A helical HOTI in the T-DAXI regime with vanishing partial weak Chern numbers $\nu_x^\pm=\nu_y^\pm=0$ has a $\mathbb{Z}_2 \times \mathbb{Z}$ invariant $(C_{\gamma_2 , in}^{+} \mod 2,\nu_i^{+})=(1 \mod 2,0) = (1,0)$ , which is indicated by odd winding numbers in the nested Berry phases in the $P_{\pm}$-eigenspace, as demonstrated in SFig.~\ref{fig:helical-HOTI-after-adding-trivial-bands-3}(b,c,e,f,h,i).
    \item A helical HOTI in the T-DAXI regime is characterized by the quantized nontrivial partial axion angles $\theta^{\pm} = \pi$, which give rise to a bulk topological contribution to the partial magnetoelectric polarizability, and represent a 3D generalization of the partial polarization introduced by Fu and Kane in SRef.~\cite{fu2006time}.
    \end{enumerate}

\section{Position-Space Signatures of Spin-Resolved Topology}
\label{app:layer-resolved}

In this section, we extend the formalism and numerical methods for computing layer-resolved Chern numbers~\cite{essin2009magnetoelectric,pozo2019quantization,varnava2018surfaces} in position space to compute layer-resolved partial Chern numbers. 
This will allow us to quantitatively search for anomalies on the gapped 2D surfaces of 3D helical HOTIs in the T-DAXI regime, and to explore the implications of the partial axion angle introduced in SN~\ref{app:comparison-spin-stable-and-symmetry-indicated-topology}. 

To begin, in this section, we will consider $d$-dimensional translation-invariant systems with primitive Bravais lattice vectors $\{\mathbf{a}_1 \ldots \mathbf{a}_d \}$, where $d=2$ or $3$. 
Each lattice point $\mathbf{R}$ (\emph{i.e.} the center of each unit cell indexed by $\mathbf{R}$), can be written as $\mathbf{R}=\sum_{i=1}^{d} n_{i} \mathbf{a}_i$ where $n_{i} \in \mathbb{Z}$ ($i=1\ldots d$).
Given $\{\mathbf{a}_1 \ldots \mathbf{a}_d \}$, we can construct a set of dual primitive reciprocal lattice vector $\{\mathbf{G}_1 \ldots \mathbf{G}_d \}$ satisfying $\mathbf{a}_i \cdot \mathbf{G}_j = 2\pi \delta_{ij}$ ($i=1\ldots d$).
We denote the number of (spinful) tight-binding basis orbitals within a primitive unit cell as $N_{\mathrm{sta}}$.
From a 3D translationally-invariant system we can form a 2D slab infinite along $\mathbf{a}_j$ and $\mathbf{a}_l$ while finite along $\mathbf{a}_i$ with $N_i$ unit cells, where $\{ \mathbf{a}_j , \mathbf{a}_l , \mathbf{a}_i \}$ is the linearly independent set of 3D real-space primitive lattice vectors.
Throughout this section, we define a \emph{layer} to consist of a 2D plane of the finite slab that is one \emph{unit cell} thick in the finite direction of the slab~\cite{essin2009magnetoelectric,pozo2019quantization}.
For example, if we use $N_i = 15$ \emph{unit cells} of the original 3D translation-invariant system to form a slab that is finite along $\mathbf{a}_i$, we say this slab has $15$ \emph{layers} along $\mathbf{a}_i$.
This is distinct from the usage of ``layer'' in the context of layer constructions (\emph{e.g.} in SRefs.~\cite{varnava2018surfaces,elcoro2021magnetic,gao2022magnetic,song2019topological,huang2017building,song2018mapping} and SN~\ref{app:comparison-spin-stable-and-symmetry-indicated-topology}), but is consistent with the terminology of SRefs.~\cite{essin2009magnetoelectric,pozo2019quantization}.
For such a slab, we will see below that the layer-resolved partial Chern numbers tell us how the partial Chern numbers [SEq.~\eqref{eq:partial_chern_def}] of the 2D slab is distributed over the $N_i$ unit cells, or equivalently over the $N_i$ layers.
We will argue that the layer-resolved partial Chern numbers are related to the response of a 2D slab to the insertion of a magnetic flux. 
For example, in SN~\ref{app:comparison-spin-stable-and-symmetry-indicated-topology} we predicted that a $\Phi=\pi$ magnetic flux tube in a finite slab of a T-DAXI binds one spinon between the top and bottom surface~\cite{WiederDefect,armitage2019matter,varnava2018surfaces,wieder2020axionic}.
We will show that this is a consequence of the fact that in a T-DAXI the gapped surfaces have half-integer partial Chern numbers, while in the bulk the layer-resolved partial Chern numbers vanish when averaged over the degrees of freedom within each unit cell (layer).
In this section we will first develop a formulation for the layer-resolved position-space marker~\cite{essin2009magnetoelectric,bianco2011mapping,caio2019topological,prodan2009robustness,varnava2018surfaces} for the partial Chern number. 
We will then compute the layer-resolved partial Chern numbers for the model of a T-DAXI introduced in SN~\ref{sec:numerical-section-of-nested-P-pm} and numerically identify the {\it partial parity anomaly} (defined properly in SN~\ref{sec:layer-resolved-Cs-of-a-helical-HOTI}) realized on its gapped surfaces.

In SN~\ref{app:Chern-marker} and \ref{app:partial-Chern-marker}, we will review the Chern marker~\cite{kitaev2006anyons,bianco2011mapping,caio2019topological} and partial Chern markers~\cite{prodan2009robustness}, which characterize for a 2D system how the Chern number and partial Chern numbers vary as functions of real-space position, respectively.
As an extension of SN~\ref{app:Chern-marker}, in SN~\ref{sec:layer-resolved-P-Chern-number} we will review the formalism for the layer-resolved Chern number, which describes how the Chern number of a 2D slab formed from a 3D translation-invariant system is distributed along the finite direction in position space.
Based on SN~\ref{sec:layer-resolved-P-Chern-number}, we then develop the formalism of layer-resolved partial Chern numbers in SN~\ref{app:layer-resolved-partial-chern-number}, which describes how the \emph{partial} Chern numbers of a 2D slab formed from a 3D spinful translation-invariant system are distributed along the finite direction in position space.
Finally, in SN~\ref{sec:layer-resolved-Cs-of-a-helical-HOTI} we compute the layer-resolved partial Chern numbers of a T-DAXI, and show that a T-DAXI has vanishing partial Chern numbers in its bulk (when averaged over the degrees of freedom within each layer, or equivalently unit cell), and half-integer partial Chern numbers at its gapped surfaces---the hallmark of a {\it partial parity anomaly} arising from the bulk-quantized partial axion angle. 
We will also distinguish between the layer-resolved partial Chern numbers of a 3D quantum spin Hall insulator (QSHI) and a T-DAXI, which represent two physically distinct spin-stable topological regimes of spin-resolved helical HOTIs, as discussed in SN~\ref{app:comparison-spin-stable-and-symmetry-indicated-topology}. 

\subsection{Chern Marker}\label{app:Chern-marker}

In this section, we will review the construction of the position-space Chern marker $C(\mathbf{R})$, first introduced by Resta in SRef.~\cite{bianco2011mapping}. 
This will allow us to develop intuition for topological markers in position space. 
We begin by considering a 2D tight-binding lattice Hamiltonian $H$ with a finite size, and define the projector onto a set of occupied energy eigenstates as
\begin{equation}
    P \equiv \sum_{n \in \text{occ}} \ket{n} \bra{n}, \label{eq:P-for-2d-finite-size-system}
\end{equation}
where $\ket{n}$ is the eigenstate of $H$ satisfying $H \ket{n} = E_{n} \ket{n}$.
The matrix representatives of the position operators $\widehat{x}$ and $\widehat{y}$ in the (spinful) tight-binding basis states have matrix elements
\begin{align}
    & \left[ \widehat{x} \right]_{\mathbf{R}\alpha,\mathbf{R}'\beta} = \langle \mathbf{R} , \alpha | \widehat{x} | \mathbf{R}' , \beta \rangle = \langle 0 | c_{\mathbf{R},\alpha} \widehat{x} c_{\mathbf{R}',\beta}^{\dagger} | 0 \rangle = \delta_{\mathbf{R}\mathbf{R}'} \delta_{\alpha\beta} (\mathbf{R}+\mathbf{r}_{\alpha})_{x}, \label{eq:x-operator-matrix-elements}\\ 
    & \left[ \widehat{y} \right]_{\mathbf{R}\alpha,\mathbf{R}'\beta} = \langle \mathbf{R} , \alpha | \widehat{y} | \mathbf{R}' , \beta \rangle = \langle 0 | c_{\mathbf{R},\alpha} \widehat{y} c_{\mathbf{R}',\beta}^{\dagger} | 0 \rangle  = \delta_{\mathbf{R}\mathbf{R}'} \delta_{\alpha\beta} (\mathbf{R}+\mathbf{r}_{\alpha})_{y}, \label{eq:y-operator-matrix-elements}
\end{align}
where $c^{\dagger}_{\mathbf{R},\alpha}$ and $c_{\mathbf{R},\alpha}$ represent creation and annihilation operators for the (spinful) orbital labeled by $\alpha$ in the unit cell $\mathbf{R}$, and $\mathbf{r}_{\alpha}$ is the position of the (spinful) orbital labeled by $\alpha$ within the unit cell $\mathbf{R}$. 
Using SEq.~\eqref{eq:P-for-2d-finite-size-system}, we can then compute the position operators projected onto the occupied states as
\begin{equation}
    \widehat{x}_{P}=P \widehat{x}P,\ \widehat{y}_{P}=P \widehat{y} P.\label{eq:pxpdef}
\end{equation}
From SEq.~\eqref{eq:pxpdef} we can define the position-space-resolved Chern number termed the (local) Chern marker~\cite{kitaev2006anyons,bianco2011mapping,caio2019topological} in the unit cell $\mathbf{R}$ as
\begin{equation}
    C(\mathbf{R}) = \frac{2\pi i}{A_{\text{cell}}}  \sum_{\alpha=1}^{N_{\mathrm{sta}}} \bra{\mathbf{R},\alpha} [\widehat{x}_{P},\widehat{y}_{P}] \ket{\mathbf{R},\alpha}, \label{eq:P_Chern_maker_definition_Cxy}
\end{equation}
where $A_{\text{cell}}$ is the unit cell area of the 2D lattice, and $N_{\mathrm{sta}}$ is the number of tight-binding basis state within a unit cell of the 2D lattice. 
In SEq.~(\ref{eq:P_Chern_maker_definition_Cxy}), we have performed a trace of $[\widehat{x}_{P},\widehat{y}_{P}]$ over the tight-binding basis states within the unit cell $\mathbf{R}$, and $\ket{\mathbf{R},\alpha}$ is the tight-binding basis state labeled by $\alpha$ in the unit cell $\mathbf{R}$. 
$C(\mathbf{R})$ represents the position-space resolution of the Chern number~\cite{bianco2011mapping,caio2019topological} and physically corresponds to the local contribution to the Hall conductivity~\cite{pozo2019quantization}.
For a 2D gapped homogeneous system without disorder, the values of $C(\mathbf{R})$ for $\mathbf{R}$ within the bulk region of a finite-sized sample are equal to the Chern number of the system~\cite{bianco2011mapping,caio2019topological}. 
In particular, in the thermodynamic limit, far from the (gapless) edges of a bulk-gapped system, $C(\mathbf{R})$ is a constant independent of $\mathbf{R}$ and equal to the Chern number. 
The local Chern marker can be used to study single-particle topology in the absence of translation invariance, such as for disordered systems, quasiperiodic systems and at interfaces between bulk topological phases~\cite{tran2015topological,tran2017probing,loring2019bulk,marra2020topologically,privitera2016quantum}. 
In particular, formulated as a function of  position, the Chern marker $C(\mathbf{R})$ allows one to compute how the Chern number varies as a function of position, either near the interface between two insulators with different Chern numbers~\cite{bianco2011mapping}, or in disordered systems~\cite{marsal2020topological}.
By studying the scaling of the Chern marker as a function of the parameters inducing a band inversion, critical properties of (weakly-disordered) Chern insulators near topological quantum phase transition can be quantitatively computed~\cite{caio2019topological,kibblezurek_disordered_chern_insulator_2020}.
Furthermore, it is also possible to study the dynamics of the Chern marker $C(\mathbf{R},t)$ as a function of both position $\mathbf{R}$ and time $t$, from which a Chern marker current can be defined~\cite{caio2019topological}.

\subsection{The Partial Chern Marker and the Spin Chern Marker}\label{app:partial-Chern-marker}

In this work, we are primarily focusing on $\mathcal{T}$-invariant insulators, in which the Chern number, and hence local Chern marker, vanish. 
However, this does not imply that $\mathcal{T}$-invariant insulators are featureless.  
Starting from the Chern marker formalism in SN~\ref{app:Chern-marker},
we will here formulate a partial Chern marker, which resolves partial Chern number [SEq.~\eqref{eq:partial_chern_def}] in position space. 
The partial Chern marker formalism that we will describe in this section builds off
related constructions in SRefs.~\cite{prodan2009robustness,Feng-Liu-spin-Bott-PRL,Feng-Liu-spin-Bott-PRB,monaco2021stvreda,chen2022optical,gilardoni2022real}.
Similar to SN~\ref{app:Chern-marker}, we begin by considering a 2D spinful tight-binding lattice Hamiltonian $H$ with a finite size.
We take the finite-sized system to have $N$ unit cells with $N_{\mathrm{sta}}=2N_{\mathrm{orb}}$ spinful orbitals in each unit cell, where the factor of $2$ accounts for the spin-$1/2$ degree of freedom (where we recall from the discussion following SEq.~\eqref{eq:V-s-commute} that the Hilbert space of a solid derives from pairs of orbitals at the same position with opposite spin, for systems both with and without $\mathcal{T}$-symmetry.)
The spin operator along the direction $\hat{\mathbf{n}}$ in terms of the tight-binding basis states is defined in SEq.~\eqref{eq:appendix-def-s-i} as $s \equiv \hat{\mathbf{n}} \cdot \bm{\sigma} \otimes \mathbb{I}_{N_{\mathrm{orb}}} \otimes \mathbb{I}_{N}$, where $\bm{\sigma}$ acts on the spin-$1/2$ degree of freedom, while $\mathbb{I}_{N_{\mathrm{orb}}}$ and $\mathbb{I}_{N}$ are identity operators acting on the orbital and unit cell degrees of freedom, respectively. 
We can form the reduced spin operator in the occupied subspace as [see also SEq.~\eqref{eq:s_reduced_no_k}]
\begin{equation}
    [s_{\text{reduced}}]_{m,n} = \bra{m} s \ket{n},\label{eq:reducedspindef}
\end{equation}
where $| n \rangle$ is the $n^{\mathrm{th}}$ energy eigenstate of the Hamiltonian $H$, such that $H | n \rangle = E_n | n \rangle$, and $m,n=1\ldots N_{\mathrm{occ}}$ with $N_{\mathrm{occ}}$ representing the number of occupied energy eigenstates.
Therefore, $[s_{\mathrm{reduced}}]$ is an $N_{\mathrm{occ}} \times N_{\mathrm{occ}}$ matrix. 
The matrix elements of $[s_{\mathrm{reduced}}]$ are the matrix elements of the projected spin operator $P s P$ between occupied energy eigenstates, where $P$ is given in SEq.~\eqref{eq:P-for-2d-finite-size-system}.
For a finite system with spinful $\mathcal{T}$ symmetry, recall from SN~\ref{appendix:properties-of-the-projected-spin-operator} that the eigenvalues $\lambda_{n}$ of $[s_{\mathrm{reduced}}]$ satisfy $\{ \lambda_{n} | n = 1\ldots N_{\mathrm{occ}} \} = \{ -\lambda_{n} | n = 1\ldots N_{\mathrm{occ}} \}$.
We can then define the spin gap $\Delta_s$ for a spinful $\mathcal{T}$-invariant system with a finite size as twice the smallest magnitude among the $[s_{\mathrm{reduced}}]$ eigenvalues such that $\Delta_{s} = 2|\lambda_{\mathrm{min}}|$, where $\lambda_{\mathrm{min}}$ is the eigenvalue of $[s_{\mathrm{reduced}}]$ with smallest absolute value.
We remind the reader that here we are considering a finite-sized system, such that both the reduced spin matrix $[s_{\text{reduced}}]$ and the $PsP$ eigenvalues $\lambda_{n}$ are not functions of crystal momentum.
In particular, when $\Delta_{s} > 0$, we say that the spin gap of this spinful $\mathcal{T}$-invariant finite-sized system is open.

When the spin gap is open, we can divide the occupied subspace into the positive and negative $PsP$ eigenspace as follows. 
We first write the eigenvalue equation of $[s_{\mathrm{reduced}}]$ as
\begin{equation}
    [s_{\mathrm{reduced}}] \ket{\tilde{u}_{n}^{\pm}} = \lambda_{n}^{\pm} \ket{\tilde{u}_{n}^{\pm}},
\end{equation}
where $\lambda_{n}^{+} > 0$ and $\lambda_{n}^{-} < 0$, and $|\lambda_n^+|=|\lambda_n^{-}|$ due to $\mathcal{T}$ symmetry. 
We may re-express the $N_{\mathrm{occ}}$-component eigenvectors $\ket{\tilde{u}_{n}^{\pm}}$ of $[s_{\text{reduced}}]$ in the spinful tight-binding orbital basis states using
\begin{equation}
    \ket{u_{n}^{\pm}} = \sum_{m=1}^{N_{\text{occ}}} [\tilde{u}_{n}^{\pm}]_{m} \ket{m},
\end{equation}
where $\ket{m}$ are each $(2\times N_{\mathrm{orb}}\times N)$-component occupied energy eigenvectors of $H$. 
We can then construct the projectors onto the positive and negative $PsP$ eigenspaces as
\begin{equation}
    P_{\pm} = \sum_{n=1}^{N_{\mathrm{occ}}/2} \ket{u_{n}^{\pm}} \bra{u_{n}^{\pm}}.
\end{equation}
The position operators $\widehat{x}$ and $\widehat{y}$ [SEqs.~\eqref{eq:x-operator-matrix-elements} and \eqref{eq:y-operator-matrix-elements}] projected onto the subspace of $\mathrm{Image}(P_{\pm})$ are then given by
\begin{equation}
    \widehat{x}_{P_{\pm}} = P_{\pm} \widehat{x} P_{\pm},\ \widehat{y}_{P_{\pm}} = P_{\pm} \widehat{y} P_{\pm}.
\end{equation}
From this we can compute the position-space-resolved \emph{partial Chern marker}~\cite{prodan2009robustness} at the unit cell $\mathbf{R}$ using
\begin{equation}
    C^{\pm}(\mathbf{R}) = \frac{2\pi i}{A_{\text{cell}}} \sum_{\alpha=1}^{N_{\mathrm{sta}}} \bra{\mathbf{R},\alpha} [\widehat{x}_{P_{\pm}},\widehat{y}_{P_{\pm}}] \ket{\mathbf{R},\alpha}, \label{eq:P_pm_Chern_maker_definition_Cxy}
\end{equation}
in analogy with  SEq.~(\ref{eq:P_Chern_maker_definition_Cxy}). 
Note that the partial Chern marker in SEq.~(\ref{eq:P_pm_Chern_maker_definition_Cxy}) is defined in terms of position operators projected onto the subspace of $\mathrm{Image}(P_{\pm})$.
We can then define the {\it spin Chern maker} in a similar manner as SEq.~\eqref{eq:Csgamma1def}:
\begin{equation}
    C^{s}(\mathbf{R}) = C^{+}(\mathbf{R})-C^{-}(\mathbf{R}). \label{eq:spin_Chern_marker_def}
\end{equation}
$C^{s}(\mathbf{R}) $ is then the position-space resolution of the spin Chern number $C_{\gamma_1}^{s}$~\cite{sheng2006spinChern,prodan2009robustness,3D_phase_diagram_spin_Chern_Prodan} in SEq.~\eqref{eq:Csgamma1def}. 
In particular, for a clean system with an energy gap and a spin gap in the bulk, in the thermodynamic limit the spin Chern marker $C^s(\mathbf{R})$ is independent of $\mathbf{R}$ and converges to the spin Chern number [SEq.~\eqref{eq:Csgamma1def}] when evaluated away from the boundaries.
For a system with spinful $\mathcal{T}$ symmetry and a gapped spin spectrum, we have that $\frac{1}{2}\left( C^{s}_{\gamma_1} \text{ mod } 4 \right)$ is equal to the strong $\mathbb{Z}_2$ Kane-Mele invariant for 2D TIs~\cite{kane2005quantum,kane2005z,prodan2009robustness,3D_phase_diagram_spin_Chern_Prodan}.
Crucially, we note for our later discussion that in a 2D non-interacting system in the thermodynamic limit with both an energy and a spin gap, the partial Chern markers $C^\pm(\mathbf{R})$ are integers in the bulk, and hence with spinful $\mathcal{T}$ symmetry the spin Chern marker $C^s(\mathbf{R})$ [SEq.~\eqref{eq:spin_Chern_marker_def}] is restricted to be an even integer in the bulk.

\subsection{Layer-Resolved Chern Number}\label{sec:layer-resolved-P-Chern-number}

The formalism of resolving the Chern number in position space [SN~\ref{app:Chern-marker}] can be extended to semi-infinite systems to study how the Chern number of a 2D slab is distributed over the finite direction~\cite{varnava2018surfaces}. 
To do so, we consider a 2D slab finite along $\mathbf{a}_{i}$ with $N_{i}$ unit cells and infinite along $\mathbf{a}_j$ and $\mathbf{a}_l$ where $\{\mathbf{a}_j,\mathbf{a}_l,\mathbf{a}_i \}$ are the primitive lattice vectors of a 3D Bravais lattice.
Expressing the Chern number in terms of projection operators~\cite{vanderbilt2018berry}, we can write~\cite{essin2009magnetoelectric}
\begin{align}
    C_{jl} & = -\frac{i}{2\pi} \int_{-\pi}^{\pi} dk_j \int_{-\pi}^{\pi} dk_l \Tr{  [P(k_j,k_l)] \left[ \frac{\partial [P(k_j,k_l)]}{\partial k_{j}}, \frac{\partial [P(k_j,k_l)]}{\partial k_{l}} \right] } \label{eq:P_Chern_slab_1} \\
    & = -\frac{i}{2\pi} \int_{-\pi}^{\pi} dk_j \int_{-\pi}^{\pi} dk_l\sum_{n_{i}=1}^{N_{i}}\sum_{\alpha=1}^{N_{\mathrm{sta}}} \bra{n_{i},\alpha , k_j,k_l} \left( [P(k_j,k_l)] \left[ \frac{\partial [P(k_j,k_l)]}{\partial k_{j}}, \frac{\partial [P(k_j,k_l)]}{\partial k_{l}} \right] \right) \ket{n_{i},\alpha , k_j,k_l}, \label{eq:P_Chern_slab_2}
\end{align}
where the integral is over the 2D BZ of the slab. 
In SEqs.~\eqref{eq:P_Chern_slab_1} and \eqref{eq:P_Chern_slab_2}, $[P(k_j,k_l)] $ is the projector
\begin{equation}
    [P(k_{j},k_{l})] = \sum_{n \in \text{occ}} \ket{u_{n,k_{j},k_{l}}} \bra{u_{n,k_{j},k_{l}}} \label{eq:matrix-projector-of-2d-slab-occupied-energy-eigenvector}
\end{equation}
onto the subspace spanned by the $N_{\mathrm{occ}}$ occupied energy eigenvectors $|u_{n,k_j,k_l} \rangle$ of the $(N_{i}N_{\mathrm{sta}} \times N_{i}N_{\mathrm{sta}})$-dimensional Bloch Hamiltonian matrix $[H(k_{j},k_{l})]$ for the 2D slab. 
In SEq.~\eqref{eq:P_Chern_slab_2}, $\ket{n_{i},\alpha, k_j,k_l}$ are the Fourier-transformed Bloch basis functions at the crystal momentum $(k_j,k_l)$ for the Hilbert space of the system. 
The index $\alpha$ runs over the (spinful) tight-binding orbitals within each unit cell, and  $n_i$ indexes the unit cells along the finite $\mathbf{a}_i$ direction.
By interchanging the order of the summation over $n_i$ and the integration over the Brillouin Zone in SEq.~\eqref{eq:P_Chern_slab_2}, we can define the \emph{layer-resolved Chern number}
\begin{align}
    C_{jl}(n_{i})
     & = -\frac{i}{2\pi} \int_{-\pi}^{\pi} dk_j \int_{-\pi}^{\pi} dk_l \sum_{\alpha=1}^{N_{\mathrm{sta}}} \bra{n_{i},\alpha,k_j,k_l} \left( [P(k_j,k_l)] \left[ \frac{\partial [P(k_j,k_l)]}{\partial k_{j}}, \frac{\partial [P(k_j,k_l)]}{\partial k_{l}} \right] \right) \ket{n_{i},\alpha,k_j,k_l}, \label{eq:P_layer_resolved_Chern_1}
\end{align}
such that
\begin{equation}
C_{jl} =\sum_{n_i=1}^{N_i} C_{jl}(n_i).
\end{equation}

$C_{jl}(n_{i})$ quantifies how the Chern number of the 2D slab is distributed over the $N_i$ unit cells along the finite direction parallel to $\mathbf{a}_{i}$. 
Note that according to our sign convention in SEq.~\eqref{eq:partial_chern_def}, to actually identify $C_{jl}(n_{i})$ [SEq.~\eqref{eq:P_layer_resolved_Chern_1}] as the Chern number of the 2D slab distributed over the finite direction $\mathbf{a}_i$, $(jli)$ should be chosen to align with the right-handed coordinate system, namely a cyclic permutation of $(123)$.
Physically, $C_{jl}(n_{i})$ tells us how each of the $N_i$ unit cells along the finite direction parallel to $\mathbf{a}_{i}$ contribute to the Hall conductivity of the slab. 
It encodes the topology of layers in the bulk of the slab as well as at gapped surfaces. 
We remind the readers that throughout this section we define a \emph{layer} to consist of a 2D plane of the finite slab that is one \emph{unit cell} thick in the $\mathbf{a}_i$ direction~\cite{essin2009magnetoelectric,pozo2019quantization}. 
As such, for the slab we are considering here, we say that it has $N_i$ layers along $\mathbf{a}_i$.
Since the matrix projector $[P(k_{j},k_{l})]$ in SEq.~\eqref{eq:matrix-projector-of-2d-slab-occupied-energy-eigenvector} is invariant under any $U(N_{\mathrm{occ}})$ gauge transformation of the occupied energy eigenvectors, together with SEq.~\eqref{eq:P_layer_resolved_Chern_1} we deduce that the layer-resolved Chern number $C_{jl}(n_{i})$  is also invariant under any $U(N_{\mathrm{occ}})$ gauge transformation.
In SRefs.~\cite{essin2009magnetoelectric,varnava2018surfaces}, $C_{jl}(n_{i})$ was used to show that the gapped surfaces of magnetic axion insulators have half-integer Hall conductivity.

\subsection{Layer-Resolved Partial Chern Numbers and Layer-Resolved Spin Chern Number}\label{app:layer-resolved-partial-chern-number}

Building on the formalism for the partial and spin Chern markers in 2D developed in SN~\ref{app:partial-Chern-marker}, and the formalism for the layer-resolved local Chern marker in 3D insulators developed in SN~\ref{sec:layer-resolved-P-Chern-number}, we will next here formally introduce layer-resolved \emph{partial} and \emph{spin} Chern numbers.

We begin by again considering a 2D slab finite along $\mathbf{a}_{i}$ with $N_{i}$ unit cells and infinite along $\mathbf{a}_j$ and $\mathbf{a}_l$ where $\{\mathbf{a}_j,\mathbf{a}_l,\mathbf{a}_i \}$ are the primitive lattice vectors for a 3D Bravais lattice. 
We take there to be $N_{\mathrm{sta}}=2N_{\mathrm{orb}}$ tight-binding basis states per unit cell, where the factor of $2$ accounts for the spin-1/2 degree of freedom. 
As in SN~\ref{app:partial-Chern-marker}, the spin operator along $\hat{\mathbf{n}}$ is defined as $s \equiv \hat{\mathbf{n}} \cdot \bm{\sigma} \otimes \mathbb{I}_{N_{\mathrm{orb}}} \otimes \mathbb{I}_{N_{i}}$ where $\bm{\sigma}$ acts on the spin-$1/2$ degree of freedom, and $\mathbb{I}_{N_{\mathrm{orb}}}$ and $\mathbb{I}_{N_{i}}$ are identity matrices acting on the orbital and layer (unit cell) degrees of freedom, respectively. 
We suppose that the slab is insulating both in the bulk and on the surfaces with $N_{\mathrm{occ}}$ occupied energy bands.
To spin-resolve the occupied energy eigenspace, we first construct the $N_{\mathrm{occ}} \times N_{\mathrm{occ}}$ reduced spin matrix [SEq.~\eqref{eq:P_pm_Wilson_loop_s_reduced_def}] as a function of momentum in the 2D Brillouin zone,
\begin{equation}
    [s_{\text{reduced}}(k_{j},k_{l})]_{m,n} = \bra{u_{m,k_{j},k_{l}}} s \ket{u_{n,k_{j},k_{l}}}.
\end{equation}
Here $m,n \in \mathrm{occ}$ index the occupied single-particle eigenstates.
From SN~\ref{appendix:properties-of-the-projected-spin-operator}, if $N_{\mathrm{occ}}$ is an even number (as is required in insulators with spinful $\mathcal{T}$ symmetry), the spin gap is defined to be open when the $N_{\mathrm{occ}} / 2$ spin bands with the largest $[s_{\text{reduced}}(k_{j},k_{l})]$ eigenvalues (the upper spin bands) are disjoint from the $N_{\mathrm{occ}} / 2$ spin bands with the smallest $[s_{\text{reduced}}(k_{j},k_{l})]$ eigenvalues (the lower spin bands).  
We refer the readers to SFig.~\ref{fig:schematic_spin_bands_with_I_T_and_IT}(b) for a schematic demonstration of the spin band structure of a spinful $\mathcal{T}$-symmetric system with a spin gap.

Let us assume that the $N_{\mathrm{occ}}$ is even and that the spin gap is open.
The eigenvalue equation of $[s_{\text{reduced}}(k_{j},k_{l})]$ can then be written as
\begin{equation}
    [s_{\text{reduced}}(k_{j},k_{l})] \ket{\widetilde{u}_{n,k_{j},k_{l}}^{\pm}} = \lambda_{n,k_{j},k_{l}}^{\pm} \ket{\widetilde{u}_{n,k_{j},k_{l}}^{\pm}}, \label{eq:eig-eqn-s-k1-k2-2D-slab}
\end{equation}
where $n=1\ldots N_{\mathrm{occ}}/2$ index the upper and lower spin bands, and $\lambda_{n,k_{j},k_{l}}^{+} > \lambda_{m,k_{j},k_{l}}^{-}$ for all $n$, $m$, $k_j$, and $k_l$.
We may re-express the $N_{\mathrm{occ}}$-component eigenvectors $\ket{\widetilde{u}_{n,k_{j},k_{l}}^{\pm}}$ of $[s_{\text{reduced}}(k_{j},k_{l})]$ in terms of the spinful tight-binding basis functions using 
\begin{align}
    \ket{u_{n,k_{j},k_{l}}^{\pm}} = \sum_{m=1}^{N_{\mathrm{occ}}} [\tilde{u}_{n,k_{j},k_{l}}^{\pm}]_{m} \ket{u_{m,k_{j},k_{l}}}, \label{eq:reexpression-2D-slab-tilde-u-n-pm}
\end{align}
as in SEq.~\eqref{eq:reexpress_the_eigenvector_of_s_reduced}, where $\ket{u_{m,k_{j},k_{l}}}$ are the $(2\times N_{\mathrm{orb}} \times N_{i})$-component occupied energy eigenvectors of the Bloch Hamiltonian matrix of the slab.
We can then construct the $(2N_{\mathrm{orb}}N_{i} \times 2N_{\mathrm{orb}}N_{i})$-dimensional matrix projector $[P_{\pm}(k_{j},k_{l})]$ onto the upper ($+$) and lower ($-$) spin bands as
\begin{align}
    [P_{\pm}(k_{j},k_{l})] = \sum_{n=1}^{N_{\mathrm{occ}}/2} \ket{u_{n,k_{j},k_{l}}^{\pm}} \bra{u_{n,k_{j},k_{l}}^{\pm}}. \label{eq:P-pm-k1-k2-2D-slab}
\end{align}
Extending SEq.~\eqref{eq:P_Chern_slab_1}  to compute the partial Chern numbers [SEq.~\eqref{eq:partial_chern_def}] within the upper and lower spin bands, the partial Chern numbers of the 2D slab are
\begin{align}
    C_{jl}^{\pm} & = -\frac{i}{2\pi} \int_{-\pi}^{\pi} dk_j \int_{-\pi}^{\pi} dk_l \Tr{  [P_{\pm}(k_{j},k_{l})] \left[ \frac{\partial [P_{\pm}(k_{j},k_{l})]}{\partial k_{j}}, \frac{\partial [P_{\pm}(k_{j},k_{l})]}{\partial k_{l}} \right] } \label{eq:P_pm_Chern_slab_1} \\
    & = -\frac{i}{2\pi} \int_{-\pi}^{\pi} dk_j \int_{-\pi}^{\pi} dk_l \sum_{n_{i}=1}^{N_{i}}\sum_{\alpha=1}^{2N_{\mathrm{orb}}} \bra{n_{i},\alpha,k_j,k_l} \left( [P_{\pm}(k_{j},k_{l})] \left[ \frac{\partial [P_{\pm}(k_{j},k_{l})]}{\partial k_{j}}, \frac{\partial [P_{\pm}(k_{j},k_{l})]}{\partial k_{l}} \right] \right) \ket{n_{i},\alpha,k_j,k_l}. \label{eq:P_pm_Chern_slab_2}
\end{align}
In analogy to our logic in SEqs.~\eqref{eq:P_Chern_slab_2} and \eqref{eq:P_layer_resolved_Chern_1}, we can interchange the summation over the layer index $n_i$ with the Brillouin zone integration to define the  {\it layer-resolved partial Chern numbers} 
\begin{align}
    C_{jl}^{\pm}(n_i) & = -\frac{i}{2\pi} \int_{-\pi}^{\pi} dk_j \int_{-\pi}^{\pi} dk_l \sum_{\alpha=1}^{2N_{\mathrm{orb}}} \bra{n_{i},\alpha,k_j,k_l} \left( [P_\pm(k_j,k_l)] \left[ \frac{\partial [P_\pm(k_j,k_l)]}{\partial k_{j}}, \frac{\partial [P_\pm(k_j,k_l)]}{\partial k_{l}} \right] \right) \ket{n_{i},\alpha,k_j,k_l}, \label{eq:P_pm_layer_resolved_Chern_1} 
\end{align}
which satisfy
\begin{equation}
C_{jl}^{\pm}= \sum_{n_i=1}^{N_i} C_{jl}^{\pm}(n_i).
\end{equation}
$C_{jl}^{\pm}(n_i)$  quantifies how the partial Chern numbers [SEq.~\eqref{eq:partial_chern_def}] of the 2D slab are distributed over the $N_{i}$ unit cells along the finite direction parallel to $\mathbf{a}_i$.
Similar to SEq.~\eqref{eq:Csgamma1def}, we can define the \textit{layer-resolved spin Chern number} as
\begin{align}
    C^{s}_{jl}(n_{i}) = C^{+}_{jl}(n_{i}) - C^{-}_{jl}(n_{i}). \label{eq:def-Cs-n3}
\end{align}

We emphasize that it is important to re-express the $N_{\mathrm{occ}}$-component eigenvectors $\ket{\widetilde{u}_{n,k_{j},k_{l}}^{\pm}}$ of $[s_{\text{reduced}}(k_{j},k_{l})]$ in terms of the spinful tight-binding basis functions using SEq.~\eqref{eq:reexpression-2D-slab-tilde-u-n-pm} in order to perform the layer resolution in SEq.~\eqref{eq:P_pm_layer_resolved_Chern_1}.
Since the matrix projector $[P_{\pm}(k_{j},k_{l})] $ [SEq.~\eqref{eq:P-pm-k1-k2-2D-slab}] is invariant under any $U(N_{\mathrm{occ}}/2)$ gauge transformation in the space of (upper or lower) spin bands, then together with SEq.~\eqref{eq:P_pm_layer_resolved_Chern_1} we deduce that the layer-resolved partial Chern number $C_{jl}^{\pm}(n_{i})$  is also invariant under any $U(N_{\mathrm{occ}}/2)$ gauge transformation within the upper or lower spin bands.

Physically, $C_{jl}^{s}(n_i)$ tells us how the topological contribution to the spin Hall conductivity in SEq.~\eqref{eq:intrinsicspinhall} of a 2D slab is distributed over the $N_i$ unit cells along the finite direction parallel to $\mathbf{a}_i$. 
The layer-resolved partial and spin Chern numbers encode the topology of the layers in the bulk of the slab as well as at gapped surfaces, in analogy to the layer-resolved Chern number in magnetic insulators (see SN~\ref{sec:layer-resolved-P-Chern-number} and SRefs.~\cite{essin2009magnetoelectric,varnava2018surfaces}). 
In the next section, SN~\ref{sec:layer-resolved-Cs-of-a-helical-HOTI}, we will numerically apply the layer-resolved partial and spin Chern numbers developed in this section to a model of a helical HOTI in the T-DAXI regime.
We will find that the layer-resolved partial Chern numbers surprisingly do not vanish on the gapped surfaces of T-DAXIs, but instead saturate at an anomalous half-integer value, indicating that the gapped surfaces of T-DAXIs exhibit a novel \emph{partial parity anomaly}. 

\subsection{Layer-Resolved Spin Chern Number of a Helical HOTI}\label{sec:layer-resolved-Cs-of-a-helical-HOTI}

In this section, we will apply the formalism developed in SN~\ref{app:layer-resolved-partial-chern-number} to compute
the layer-resolved partial Chern numbers $C_{jl}^{\pm}(n_i)$ [defined in SEq.~\eqref{eq:P_pm_layer_resolved_Chern_1}] for the model of a symmetry-indicated helical higher-order topological insulator (HOTI) with inversion ($\mathcal{I}$) and time-reversal ($\mathcal{T}$) symmetries introduced in SRef.~\cite{wieder2018axion} and analyzed in SN~\ref{sec:numerical-section-of-nested-P-pm}.
In SN~\ref{app:comparison-spin-stable-and-symmetry-indicated-topology} and Supplementary Table~\ref{tab:summary-for-qshi-and-daxi} we demonstrated that this model has a spin gap and realizes the $\mathcal{T}$-doubled axion insulator (T-DAXI) spin-stable topological phase.
Now we will explicitly compute the layer-resolved partial Chern numbers and will demonstrate that a semi-infinite 2D slab of a T-DAXI with gapped surfaces satisfies the following properties:
\begin{enumerate}
	\item It is a 2D time-reversal-invariant strong topological insulator (TI) for highly symmetric surface terminations (\emph{i.e.} in the case in which the entire slab has local $\mathcal{T}$ and global $\mathcal{I}$ symmetry).
	\item The partial Chern numbers do not scale with the thickness of the slab, which physically distinguishes the T-DAXI from the 3D QSHI state discussed in SN~\ref{app:comparison-spin-stable-and-symmetry-indicated-topology}.
	In particular, see SFig.~\ref{fig:layer-construction-QSHI-DAXI} for their individual spin-resolved layer constructions. 
    This implies that the topological contribution to the spin Hall conductivity of a T-DAXI vanishes in the bulk of the slab averaged over each unit cell (layer) along the finite direction.
	\item The T-DAXI slab has half-quantized partial Chern numbers at the gapped surfaces, which cannot be realized for an isolated 2D system with energy and spin gaps. 
    Hence the gapped surfaces of a T-DAXI exhibit a novel \emph{partial parity anomaly}, which is the spinful time-reversal-symmetric generalization of the parity anomaly encountered on the gapped surfaces of magnetic axion insulators~\cite{armitage2019matter,qi2008topological,witten2016fermion}.
   
\end{enumerate}

We begin by constructing a semi-infinite slab of a T-DAXI. 
Recall that our model has an orthorhombic lattice.
Normalizing the lattice constants to $1$, we take the three position-space primitive lattice vectors to be $\mathbf{a}_1 = \widehat{x}$, $\mathbf{a}_2 = \widehat{y}$, and $\mathbf{a}_3 = \widehat{z}$. 
The dual primitive reciprocal lattice vectors are $\mathbf{G}_1 = 2\pi \widehat{x}$, $\mathbf{G}_2 = 2\pi \widehat{y}$, and $\mathbf{G}_3 = 2\pi \widehat{z}$, satisfying $\mathbf{a}_i \cdot \mathbf{G}_j = 2\pi \delta_{ij}$ ($i,j=1\ldots 3$).
Our model has eight bands, for which the matrix Bloch Hamiltonian is given explicitly in SEq.~\eqref{eq:helical_HOTI_TB_model}.
The tight-binding parameters, specified in SEq.~\eqref{eq:parameter_choice_of_helical_HOTI}, together with spin-non-conserving SOC term [$A_{\mathrm{spin-mixing}}$ in SEq.~\eqref{eq:helical_HOTI_TB_model}] are chosen such that the energy spectrum of the surfaces with normal vectors $\pm \widehat{x}$, $\pm \widehat{y}$, and $\pm \widehat{z}$ are all gapped.
We construct 2D inversion- and time-reversal-symmetric helical HOTI slabs that are finite along $\mathbf{a}_i$ with $15$ unit cells and infinite along $\mathbf{a}_j$ and $\mathbf{a}_l$ where $(ijl)$ are cyclic permutations of $(123)$. 
We below compute the spin spectrum and spin-resolved topology using  $Ps_zP$ [SEq.~\eqref{eq:appendix-def-s-i}], where the occupied energy bands used to form the projector in SEq.~\eqref{eq:matrix-projector-of-2d-slab-occupied-energy-eigenvector} are the lower half of all the energy bands of the slab. 
In other words, we consider the slab at half-filling.
We will compare the layer-resolved partial Chern numbers of our model with both spin-$s_z$ conservation ($A_{\mathrm{spin-mixing}}=0.0$) and with large spin-$s_z$-non-conserving SOC ($A_{\mathrm{spin-mixing}}=0.5$).

We next compute the layer-resolved partial Chern numbers $C_{jl}^{\pm}(n_i)$ for each $(ijl)$ a cyclic permutation of $(123)$ using SEq.~\eqref{eq:P_pm_layer_resolved_Chern_1}.
In order to numerically evaluate the integrand in SEq.~\eqref{eq:P_pm_layer_resolved_Chern_1}, we discretize the 2D BZ of each slab into a $200 \times 200$ grid. 
We use a symmetric finite difference approximation to numerically evaluate derivatives; specifically for a function $f(x)$, we approximate $\frac{d f(x)}{d x} \approx \frac{f(x + \frac{1}{2} \Delta x) - f(x - \frac{1}{2} \Delta x)}{\Delta x}$ where $\Delta x$ is the grid spacing.

We will examine two quantities derived from the layer-resolved partial Chern number. 
First, we compute the partial Chern number $C_{jl}^{\pm}$ of the entire slab, which is given by summing $C_{jl}^{\pm}(n_i)$ over all of the layers (unit cells):
\begin{equation}
	C_{jl}^{\pm} = \sum_{n_i \in \text{ all layers}} C_{jl}^{\pm}(n_i). \label{eq:partial_Chern_by_summing_over_layers}
\end{equation}
Second, we define the surface partial Chern number, which is given by summing $C_{jl}^{\pm}(n_i)$ over only the layers near the surface:
\begin{equation}
	C_{jl,\mathrm{surface}}^{\pm} = \sum_{n_i \in \text{ layers near a surface}} C_{jl}^{\pm}(n_i). \label{eq:surface_partial_Chern_numbers_def}
\end{equation}
We will see shortly that for our T-DAXI model, $C_{jl,\mathrm{surface}}^{\pm}$ is well-defined since $C_{jl}^\pm(n_i)\rightarrow 0$ for $n_i$ in the bulk of the slab. 
In particular, if $C_{jl}^\pm(n_i)\approx 0$ for $N_{\textrm{bottom}}<n_i<N_{\textrm{top}}$, then the sum in SEq.~\eqref{eq:surface_partial_Chern_numbers_def} is over all $n_i<N_\textrm{bottom}$ for the bottom surface, and over all $n_i>N_\textrm{top}$ for the top surface.

\begin{figure}[ht]
\includegraphics[width=0.75\textwidth]{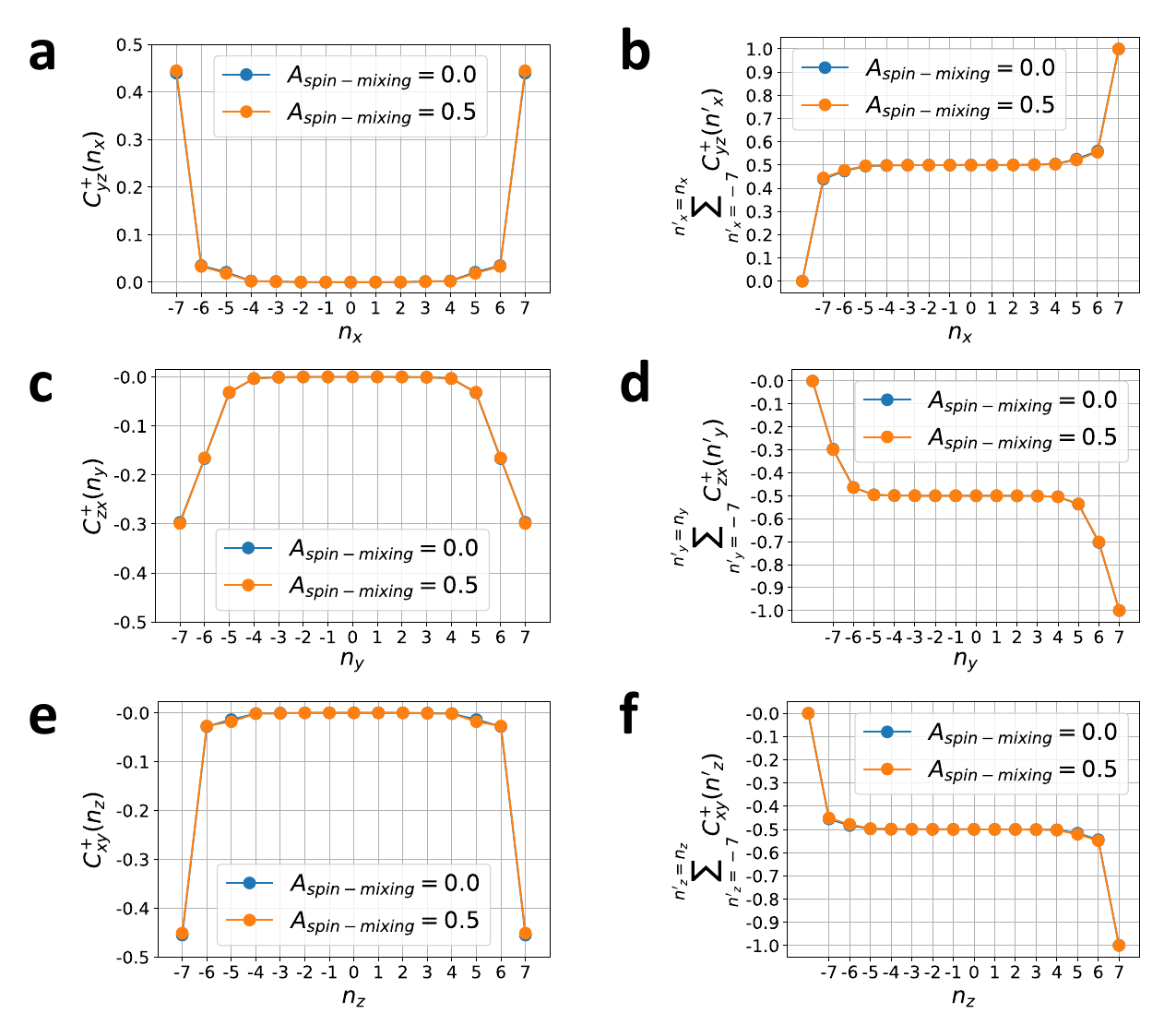}
\caption{Layer-resolved partial Chern numbers $C_{jl}^{+}(n_i)$ of 2D slabs formed from our model of an $\mathcal{I}$- and $\mathcal{T}$-symmetric helical HOTI [SEq.~\eqref{eq:helical_HOTI_TB_model}] in the T-DAXI regime (see SN~\ref{sec:numerical-section-of-nested-P-pm}).
We consider T-DAXIs both with ($A_{\mathrm{spin-mixing}}=0.0$) and without ($A_{\mathrm{spin-mixing}}=0.5$) spin-$s_z$ conservation. 
Due to time-reversal symmetry we have $C_{jl}^{+}(n_i)=-C_{jl}^{-}(n_i)$, hence we show only $C_{jl}^{+}(n_i)$ in this figure. 
(a) shows the layer-resolved partial Chern number $C_{yz}^{+}(n_x)$ for a 2D slab finite along $x$ with $15$ unit cells and infinite along $y$ and $z$.
(b) shows the cumulative partial Chern number $\sum_{n_{x}'=-7}^{n_{x}'=n_{x}} C_{yz}^{+}(n_{x}')$ as a function of $n_x$ beginning from the bottom layer in (a).
(c) shows the layer-resolved partial Chern number $C_{zx}^{+}(n_y)$ for a 2D slab finite along $y$ with $15$ unit cells and infinite along $z$ and $x$.
(d) shows the cumulative partial Chern number $\sum_{n_{y}'=-7}^{n_{y}'=n_{y}} C_{zx}^{+}(n_{y}')$ as a function of $n_y$ beginning from the bottom layer in (c).
(e) shows the layer-resolved partial Chern number $C_{xy}^{+}(n_z)$ for a 2D slab finite along $z$ with $15$ unit cells and infinite along $x$ and $y$.
(f) shows the cumulative partial Chern number $\sum_{n_{z}'=-7}^{n_{z}'=n_{z}} C_{xy}^{+}(n_{z}')$ as a function of $n_z$ beginning from the bottom layer in (e).
As we can see in (a), (c) and (e), the nonzero values of $C_{jl}^{+}(n_{i})$ are concentrated around the top and bottom layers, and this feature persists when spin-$s_{z}$ conservation is broken due to a nonzero value of the SOC term $A_{\mathrm{spin-mixing}}$ in SEq.~\eqref{eq:helical_HOTI_TB_model}. 
In addition, (b), (d), and (f) demonstrate the appearance of anomalous, half-quantized partial Chern numbers around the gapped surfaces of the T-DAXI, as the cumulative values of $C_{jl}^{+}(n_i)$ beginning from the bottom layer quickly converge to $\pm\frac{1}{2}$ and remain constant in the bulk.
The numerical values of the partial Chern numbers obtained by summing $C_{jl}^{+}(n_i)$ over all the layers for (b), (d), and (f) are given in Supplementary Table~\ref{tab:summed_layer_resolved_partial_Chern_numbers}.
The calculations detailed in this figure were performed using the freely available Python package~\href{https://github.com/kuansenlin/nested_and_spin_resolved_Wilson_loop}{\textsc{nested\_and\_spin\_resolved\_Wilson\_loop}}~\cite{lin2023nestedWilsonLib}, which represents an extension of the~\href{https://www.physics.rutgers.edu/pythtb/}{PythTB} open-source Python tight-binding package~\cite{coh2013python} that was implemented and utilized for the preparation of SRefs.~\cite{wieder2018axion,wieder2020strong} and the present work.
}
\label{fig:layer-resolved-partial-Chern-number-of-helical-HOTI}
\end{figure}

In SFig.~\ref{fig:layer-resolved-partial-Chern-number-of-helical-HOTI}(a,c,e), we show the layer-resolved partial Chern numbers $C_{jl}^{+}(n_{i})$ of our three T-DAXI slabs with normal vectors $\mathbf{a}_1$, $\mathbf{a}_2$, and $\mathbf{a}_3$, respectively.
As discussed in SN~\ref{sec:general_properties_of_winding_num_of_P_pm_Wilson}, using the fact that time-reversal acts locally in the position space, we have in the presence of time-reversal symmetry that
\begin{equation}
	C_{jl}^{+}(n_{i})=-C_{jl}^{-}(n_{i}),
\end{equation}
in analogy to SEq.~\eqref{eq:C_gamma_1_relate_to_C_plus_C_minus}.
There are three main features of $C_{jl}^{+}(n_{i})$ in SFig.~\ref{fig:layer-resolved-partial-Chern-number-of-helical-HOTI}(a,c,e).
First, from $C_{jl}^{\pm}(n_{i})$, we can compute the partial Chern numbers $C_{jl}^{\pm}$ using SEq.~\eqref{eq:partial_Chern_by_summing_over_layers}.
As shown in Supplementary Table~\ref{tab:summed_layer_resolved_partial_Chern_numbers}, all of the partial Chern numbers $C_{jl}^{\pm}$ of our helical HOTI slabs, with or without spin-$s_z$ conservation, satisfy $C_{jl}^{+}=-C_{jl}^{-}$ and $C_{jl}^{+}= +1$ or $-1$. 
This implies that each slab is a 2D TI with a spin gap~\cite{prodan2009robustness} [SN~\ref{sec:general_properties_of_winding_num_of_P_pm_Wilson}]. 
This can be viewed as a consequence of the global inversion symmetry of our semi-infinite slabs coupled with the spin-resolved layer construction description of the T-DAXI presented in SN~\ref{app:comparison-spin-stable-and-symmetry-indicated-topology}. 
This result can also be compared to the analogous case of $\mathcal{I}$-symmetric finite slabs of $\mathcal{I}$-protected magnetic AXI phases,
which have integer Chern numbers~\cite{armitage2019matter,varnava2018surfaces,wieder2018axion,wieder2020axionic}. 
However, the integer partial Chern number of the finite slab can be changed by adding a 2D topological insulator with a spin gap to the surface of the slab, and so a priori is not a robust invariant characterizing a T-DAXI phase, as real-material samples generically do not exhibit perfect global $\mathcal{I}$ symmetry.
\begin{table}[ht]
\begin{tabular}{ |c|c|c| } 
 \hline
  & $A_{\mathrm{spin-mixing}}=0.0$ & $A_{\mathrm{spin-mixing}}=0.5$  \\
 \hline
  $C^{\pm}_{yz}$ & $\pm 0.999404 \approx \pm 1$ & $\pm 0.999401 \approx \pm 1$  \\ 
 \hline
 $C^{\pm}_{zx}$ & $\mp 0.999573 \approx \mp 1$ & $\mp 0.999469 \approx \mp 1$  \\ 
 \hline
 $C^{\pm}_{xy}$ & $\mp 0.999373 \approx \mp 1$ & $\mp 0.999399 \approx \mp 1$   \\ 
 \hline
\end{tabular}
\caption{\label{tab:summed_layer_resolved_partial_Chern_numbers}
Numerically computed partial Chern numbers $C_{jl}^{\pm}$ [SEq.~\eqref{eq:partial_Chern_by_summing_over_layers}] for our T-DAXI slabs over different slab stacking directions.
$C_{yz}^{\pm}$ are the partial Chern numbers of the slab finite along $x$, $C_{zx}^{\pm}$ are the partial Chern numbers of the slab finite along $y$, and $C_{xy}^{\pm}$ are the partial Chern numbers of the slab finite along $z$.
$A_{\mathrm{spin-mixing}}=0.0$ and $0.5$ correspond to the computations performed using the T-DAXI model with and without spin-$s_z$ conservation, respectively.
In each entry we provide the numerical value as well as the closest integer. 
We see that in all cases $|C_{jl}^{\pm}|\approx 1$, and the deviation from unity can be attributed to numerical error coming from the finite discretization of the BZ.}
\end{table}

To contrast the T-DAXI with the 3D QSHI regime of a helical HOTI, we can examine how the partial Chern number is distributed over the layers of the slab. 
In SFig.~\ref{fig:layer-resolved-partial-Chern-number-of-helical-HOTI}(a,c,e), we see that the nonzero values of $C_{jl}^{+}(n_{i})$ are concentrated around the top and bottom gapped surfaces of the slab.
In other words, we have $C_{jl}^{+}(n_{i}) = 0$ for layers $n_{i}$ far enough from the gapped surfaces.
Therefore, the partial Chern numbers $C_{jl}^{\pm}$ of our helical HOTI slabs, which are in the T-DAXI regime, only have contributions from the gapped surfaces, and thus $|C_{jl}^{\pm}|=1$ independent of the thickness of the slab.
This should be contrasted with a slab obtained from a 3D QSHI state discussed in SN~\ref{app:comparison-spin-stable-and-symmetry-indicated-topology}, which has a nonzero and quantized value of $C_{jl}^{+}(n_{i})$ for layers $n_i$ in the bulk region far from the gapped surfaces.
For example, the layer-constructed 3D QSHI in Supplementary Table~\ref{tab:summary-for-qshi-and-daxi} [see also SFig.~\ref{fig:layer-construction-QSHI-DAXI}(a)] has $|C_{jl}^{+}(n_{i})|=2$ in the bulk.
This shows that the T-DAXI is physically distinguishable from a QSHI state. 
In particular, this implies that the topological contribution to the spin Hall conductivity of a slab of a T-DAXI is constant and independent of the thickness of the slab, unlike in a QSHI. 
Furthermore, the topological contribution to the spin Hall conductivity is only nonzero near the boundary of the slab; in the limit in which $A_{\mathrm{spin-mixing}}$ is zero, this implies that the spin current in response to an electric field only flows near the boundary of the system. 
Similarly, when $A_{\mathrm{spin-mixing}}\neq 0$, the topological contribution to the spin current will still flow near the boundary of the system. 

Furthermore, we find that
the gapped surface of a slab of a T-DAXI is qualitatively distinct from an isolated 2D system. 
Since $C_{jl}^{+}(n_{i})$ in SFig.~\ref{fig:layer-resolved-partial-Chern-number-of-helical-HOTI}(a,c,e) is concentrated within the vicinity of the top and bottom gapped surfaces of our slabs, then from global inversion symmetry we can obtain the surface partial Chern numbers $C^{\pm}_{jl,\mathrm{surface}}$ [SEq.~\eqref{eq:surface_partial_Chern_numbers_def}] as
\begin{equation}
	C^{\pm}_{jl,\mathrm{surface}} = \frac{1}{2}C_{jl}^{\pm}\label{eq:half_partial_chern}
\end{equation}
for both the top and bottom gapped surfaces.
Therefore, from Supplementary Table~\ref{tab:summed_layer_resolved_partial_Chern_numbers}, we deduce that our helical HOTI slabs have
\begin{equation}
	C_{yz,\mathrm{surface}}^{\pm} = \pm \frac{1}{2},\ C_{zx,\mathrm{surface}}^{\pm} = \mp \frac{1}{2},\ C_{xy,\mathrm{surface}}^{\pm} = \mp \frac{1}{2}. \label{eq:surface_partial_C_yz_C_zx_C_xy}
\end{equation}
SEq.~\eqref{eq:surface_partial_C_yz_C_zx_C_xy} is also verified in SFig.~\ref{fig:layer-resolved-partial-Chern-number-of-helical-HOTI}(b,d,f), where we plot the cumulative layer-resolved partial Chern number
\begin{equation}
\sum_{n_i'=\text{bottom}}^{n_i} C_{jl}^+(n_i') \label{eq:cumulative-partial-chern-number}
\end{equation}
summed over layers beginning at the bottom of the slab. 
We see that the cumulative layer-resolved partial Chern number [SEq.~\eqref{eq:cumulative-partial-chern-number}] quickly converges to $\pm\frac{1}{2}$ for $n_i$ in the bulk region of the slab. 
It is then constant throughout the bulk, and finally converges to $\pm1$ when $n_i$ reaches the top layer of the slab.

While the global inversion symmetry of our slabs allowed us to quickly deduce SEqs.~\eqref{eq:half_partial_chern} and \eqref{eq:surface_partial_C_yz_C_zx_C_xy}, the fact that the gapped surface of a T-DAXI slab carries a half-integer partial Chern number is independent of the global inversion symmetry of the slab. 
To see this, first note that we can always break global inversion symmetry by ``gluing'' a layer of 2D TI to just one surface of our helical HOTI slab without breaking time-reversal symmetry.
Recall from SN~\ref{sec:general_properties_of_winding_num_of_P_pm_Wilson} that a 2D TI with a spin gap has the partial Chern numbers $C_{jl}^{+}=-C_{jl}^{-}$ with $C_{jl}^{+}$ an odd integer. 
Hence, if we add any number of 2D TIs to the top surface of our T-DAXI slab, the partial Chern numbers at the top surface will satisfy
\begin{equation}
	C_{jl,\mathrm{surface}}^{+} = -C_{jl,\mathrm{surface}}^{-}, \label{eq:C_jl_surface_plus_vs_C_jl_surface_minus}
\end{equation}
and
\begin{equation}
	C_{jl,\mathrm{surface}}^{+} = \frac{1}{2} + n \label{eq:half-quantized-C-jl-surface-plus}
\end{equation}
where $n \in \mathbb{Z}$. 
Furthermore, the fact that the entire slab of a T-DAXI must have integer partial Chern number implies that local perturbations to the top surface of the sample cannot modify SEq.~\eqref{eq:half-quantized-C-jl-surface-plus} without closing either a spin gap or an energy gap. 
The half-integer surface partial Chern number hence provides a robust surface signature of the spin-stable T-DAXI phase, \emph{even} when spin-conservation symmetry is broken by SOC.

The half-integer partial Chern number at the surface of a T-DAXI distinguishes the surface of a T-DAXI from an isolated 2D insulator with an energy and spin gap (which can only have an integer partial Chern number). 
In that sense, the gapped surface of a T-DAXI is an anomalous 2D system. 
There is a direct analogy between the anomaly at the surface of a T-DAXI and the parity anomaly. 
Recall that the parity anomaly in a magnetic axion insulator (AXI) is manifested through a half-quantized Chern number on gapped surfaces~\cite{essin2009magnetoelectric,wieder2018axion,varnava2018surfaces,varnava2020axion}. 
In a magnetic AXI this is a direct consequence of the bulk-quantized electromagnetic theta (axion) angle $\theta=\pi$, which implies a half-integer Chern number on gapped surfaces, which represent ``axion domain walls''~\cite{wilczek1987two} between the $\theta=\pi$ bulk and the $\theta=0$ vacuum.  
The quantized value of $\theta$ in the bulk compensates for the parity anomaly on the surface. 
The half-quantized \emph{partial} Chern numbers in SEq.~\eqref{eq:half-quantized-C-jl-surface-plus} on the gapped surfaces of a T-DAXI similarly exhibit a {\it partial} parity anomaly, which is a spinful and time-reversal-symmetric generalization of the parity anomaly.
In other words, when we examine the spin-stable topology of a T-DAXI, each of the (positive and negative) $PsP$ eigenspaces exhibits a half-quantized partial Chern numbers on gapped surfaces.
For an isolated, lattice-regularized 2D spinful system with time-reversal symmetry and a spin gap, the partial Chern numbers must be integral.
Therefore, the half-quantized partial Chern numbers on the 2D surface of a T-DAXI are anomalous, exhibiting a \emph{partial} parity anomaly in each of the positive and negative $PsP$ eigenspaces. 
The anomaly is compensated by the bulk, which we showed in SN~\ref{app:comparison-spin-stable-and-symmetry-indicated-topology} and SEq.~\eqref{eq:quantized-theta-pm-DAXI} to have an $\mathcal{I}$-quantized \emph{partial} axion angle $\theta^{\pm} = \pi$. 
We thus see that a T-DAXI is uniquely characterized by a bulk partial axion angle quantized by inversion symmetry to $\theta^{\pm} = \pi$ and corresponding anomalous surfaces with half-integer partial Chern numbers [SEq.~\eqref{eq:half-quantized-C-jl-surface-plus}] satisfying SEq.~\eqref{eq:C_jl_surface_plus_vs_C_jl_surface_minus} due to time-reversal symmetry.
In particular, the $1/2$ in SEq.~\eqref{eq:half-quantized-C-jl-surface-plus} implies that the anomalous surfaces of a T-DAXI have partial Chern numbers that are halves of the partial Chern numbers of lattice-regularized 2D TIs, which are always odd integers~\cite{prodan2009robustness}. 
Going further, SEqs.~\eqref{eq:C_jl_surface_plus_vs_C_jl_surface_minus} and \eqref{eq:half-quantized-C-jl-surface-plus} together imply that the surface spin Chern number of a T-DAXI satisfies
\begin{equation}
C^s_{jl,\textrm{surface}} = C_{jl,\mathrm{surface}}^{+} - C_{jl,\mathrm{surface}}^{-} = 2n+1 \label{eq:daxi_surface_cs}
\end{equation}
for some integer $n$. 
As discussed in SN~\ref{sec:2D-spinful-TRI-system}, an isolated lattice-regularized 2D system with an energy gap and a spin gap cannot have an odd spin Chern number without interactions~\cite{levin2009fractional}. 
Thus SEq.~\eqref{eq:daxi_surface_cs} further demonstrates that the surface of a T-DAXI is anomalous.

We further note that SRef.~\cite{WiederDefect} recently demonstrated that upon the adiabatic insertion of a U(1) magnetic flux, the gapped surface of a T-DAXI exhibits half of the response of an isolated 2D TI. 
The layer-resolved partial Chern numbers of our T-DAXI slabs in SFig.~\ref{fig:layer-resolved-partial-Chern-number-of-helical-HOTI} are consistent with this result, since the half-quantized partial Chern numbers at the gapped surfaces imply that the topological contribution to the spin Hall conductivity [SEq.~\eqref{eq:intrinsicspinhall}] at the gapped surfaces will be 
\begin{equation}
    [\sigma^s_{H}]_{\mathrm{surface},\mathrm{topological}} = \frac{e}{4\pi}C^s_{jl,\textrm{surface}} = \frac{e}{4\pi} \left( 1+2n \right) \label{eq:sigmasHsurfacetopo}
\end{equation}
where we have used SEqs.~\eqref{eq:intrinsicspinhall} and \eqref{eq:daxi_surface_cs} with $n\in \mathbb{Z}$. 
On the contrary, for an isolated, lattice regularized, and spin-gapped 2D TI the topological contribution to the spin Hall conductivity [SEq.~\eqref{eq:intrinsicspinhall}] must be 
\begin{equation}
    [\sigma^s_{H}]_{\mathrm{2D\ TI},\mathrm{topological}} = \frac{e}{4\pi} \cdot 2 \cdot \left( 1+2m \right) \label{eq:sigmasH2dTItopo}
\end{equation}
where $m \in \mathbb{Z}$, and we have used the fact that an isolated, spinful $\mathcal{T}$-invariant, lattice-regularized 2D TI with an energy gap and a spin gap has $C_{\gamma_1}^+ =- C_{\gamma_1}^-$ and $C_{\gamma_1}^s \mod 4 = 2$.
In other words, the topological contribution $[\sigma^s_{H}]_{\mathrm{surface},\mathrm{topological}}$ to the surface spin Hall conductivity of a T-DAXI is half-integer quantized in units of the spin Hall conductivity of an isolated 2D system.
In particular, combining the results of Supplementary Table~\ref{tab:summed_layer_resolved_partial_Chern_numbers} with our analysis of flux response in SN~\ref{sec:spin_stable_topology_2d_fragile_TI}, we deduce that in the presence of a $\pi$ magnetic flux a slab of T-DAXI will bind one spinon between the top and bottom surface, with on average ``half'' a spinon per surface.
From the point of view of responses to external applied fields, we hence conclude that the gapped surfaces of a T-DAXI may be viewed in a precise sense as a half of an isolated 2D TI.
We remark that it was previously predicted in SRef.~\cite{liu2012half} that the gapped top surfaces of weak TI states exhibit similar halves of 2D QSHIs---the position-space calculations performed in this section reveal the surface half-2D TI state to be more generic, appearing on all the gapped surfaces of helical HOTIs in the T-DAXI regime.
As a final remark, note that in SFig.~\ref{fig:layer-resolved-partial-Chern-number-of-helical-HOTI}(a--f), we observe that the surface partial Chern numbers in a T-DAXI remain nearly quantized to half integers, even when $s_z$-conservation is broken by SOC, as long as the energy and the spin gap remains open both in the bulk and on the surfaces.
Therefore, we see that the \emph{partial} parity anomaly remains robust as long as the spin gap both in the bulk and on the gapped surfaces remains open.

\section{Symmetry Constraints on Wilson Loop Spectra} 
\label{appendix:symmetry-constraints-on-Wilson-loop}

In this section we derive constraints  that unitary inversion ($\mathcal{I}$) and antiunitary time-reversal ($\mathcal{T}$) symmetry place on the  Wilson loop spectra and their nested and spin-resolved generalizations.
Throughout this section, we will consider 3D translation-invariant systems with position-space primitive lattice vectors $\{\mathbf{a}_1,\mathbf{a}_2,\mathbf{a}_3\}$ and dual primitive reciprocal lattice vectors $\{\mathbf{G}_1,\mathbf{G}_2,\mathbf{G}_3\}$ such that $\mathbf{a}_i \cdot \mathbf{G}_j = 2\pi \delta_{ij}$ ($i,j=1\ldots 3$).
Any crystal momentum $\mathbf{k}$ can be expanded as $\sum_{j=1}^{3} \frac{k_j}{2\pi} \mathbf{G}_j$ with reduced $k_j = \mathbf{k} \cdot \mathbf{a}_j$.
In the text below, the first BZ is taken to be the region defined by $k_j \in [-\pi,\pi)$ for all $j=1,2,3$.  
We will denote $\mathbf{k}$ points by their coordinates as $(k_1 , k_2, k_3)$.
We will denote the number of (spinful) tight-binding basis states in the primitive unit cell as $N_{\mathrm{sta}}$.
When the tight-binding model has spin-1/2 degree of freedom, we will assume
$N_{\mathrm{sta}}=2N_\mathrm{orb}$,
where $N_{\mathrm{orb}}$ is the number of orbitals (such as $s$, $p$, and $d$ orbitals) and the factor of $2$ accounts for the spin-$1/2$ degree of freedom.
In addition, we will assume that both the energy spectrum and $PsP$ spectrum are gapped at every $\mathbf{k}$ in the BZ such that projectors onto the energy and spin bands are well-defined and are smooth functions of $\mathbf{k}$.

Throughout the derivation in this section, we will focus on the eigenstates and the eigenvalues of (nested) (spin-resolved) Wilson loop operators represented as products of $N_\mathrm{sta}\times N_\mathrm{sta}$ projection matrices acting in the tight-binding Hilbert space (see  SEqs.~\eqref{eq:P_Wilson_loop_P_k_def}, \eqref{eq:P_pm_Wilson_loop_P_k_def} and the surrounding text, and where we have included an additional $[V(\mathbf{G})]$ at the end of the products in order to close the loop in $\mathbf{k}$-space in a gauge-invariant manner~\cite{alexandradinata2014wilsonloop,wieder2018wallpaper}).
This will allow us to derive symmetry constraints on the Wilson loop spectra in terms of the matrix representatives of symmetries in the tight-binding basis states, which can be generated from the symmetry data of the crystal~\cite{cano2018building}.
The spectrum of an $N_\mathrm{sta}\times N_\mathrm{sta}$ (nested) (spin-resolved) Wilson loop operator contains a set of unimodular eigenvalues that coincide with the eigenphases of the (nested) (spin-resolved) Wilson loop matrices defined in SEqs.~\eqref{eq:P_Wilson_loop_matrix_element}, \eqref{eq:P_pm_Wilson_loop_matrix_element}, \eqref{eq:nested_P_Wilson_loop_matrix_def}, and \eqref{eq:nested_P_pm_Wilson_loop_matrix_def}.
In addition, the spectrum of an $N_\mathrm{sta}\times N_\mathrm{sta}$ (nested) (spin-resolved) Wilson loop operator also contains a set of zero eigenvalues whose eigenvectors correspond to the unoccupied (energy, $PsP$, Wannier band, or spin-resolved Wannier band) eigenstates.

We will in this section construct $P$- and $P_\pm$-Wilson loops parallel to the ${\mathbf{G}}_1$ direction in the BZ (see SN~\ref{sec:P_Wilson_loop} and \ref{sec:P_pm_Wilson_loop}).
Similarly, we will use ${\mathbf{G}}_1$ and ${\mathbf{G}}_2$ as the directions for the first and second closed loop needed to define nested $P-$ and nested $P_{\pm}$-Wilson loops (see SN~\ref{sec:nested_P_Wilson_loop} and \ref{sec:nested_P_pm_Wilson_loop}).
We will introduce a simplified notation throughout the derivation (in SN~\ref{appendix:I-constraint-on-P-wilson}--\ref{sec:T-constraint-on-nested-P}) to make the calculation explicit, while we will generalize the relevant results for (nested) (spin-resolved) Wilson loops oriented along {\it arbitrary} primitive reciprocal lattice vectors using the general notation of SN~\ref{sec:P_Wilson_loop}, \ref{sec:P_pm_Wilson_loop}, \ref{sec:nested_P_Wilson_loop}, and \ref{sec:nested_P_pm_Wilson_loop}.

The remainder of this section is organized as follows.
In SN~\ref{appendix:I-constraint-on-P-wilson} and \ref{appendix:I-constraint-on-nested-P-wilson}, we will derive the constraints from $\mathcal{I}$---a unitary symmetry---on the $P$- and nested $P$-Wilson loop spectra, respectively.
We will show that $\mathcal{I}$ places a quantization condition on the $P$- and nested $P$-Wilson loop eigenphases at the time-reversal invariant momenta.
In SN~\ref{appendix:T-constraint-on-P-wilson} and \ref{sec:T-constraint-on-nested-P}, we will derive the constraints from $\mathcal{T}$---an antiunitary symmetry---on the $P$- and nested $P$-Wilson loop spectra, respectively, and a set of general results that hold for both spinless $\mathcal{T}$ ($\mathcal{T}^2 = 1$, where $1$ is shorthand for the identity operator) and spinful $\mathcal{T}$ ($\mathcal{T}^2 = -1$, where $1$ is shorthand for the identity operator) will be presented.
We will then show that if we specialize to the case with $\mathcal{T}^2 = -1$, there will be Kramers' degeneracies in the $P$- and nested $P$-Wilson loop spectra at TRIMs.
In SN~\ref{appendix:I-constraint-on-P-pm} and \ref{appendix:I-constraint-on-nested-P-pm}, we will derive the constraints from $\mathcal{I}$ on the $P_{\pm}$- and nested $P_{\pm}$-Wilson loop spectra respectively for a spinful system, in which we also derive a quantization condition enforced by $\mathcal{I}$ at TRIM points.
In SN~\ref{appendix:T-constraint-on-P-pm} and \ref{appendix:T-constraint-on-nested-P-pm}, we will derive the constraints from $\mathcal{T}$ on the $P_{\pm}$- and nested $P_{\pm}$-Wilson loop spectra respectively for a spinful system with $\mathcal{T}^2=-1$, where $\mathcal{T}$ acts to flip the spin.
In SN~\ref{app:short_summary_for_symmetry_constraints}, we will present a short summary of the symmetry constraints on the (nested) (spin-resolved) Wilson loop spectra derived throughout this section.
In SN~\ref{appendix:eigval_eigvec_W}, we will briefly review the relation between the eigenvalues and eigenvectors of Wilson loop operators and their Hermitian conjugates, which will be used throughout this section.
Throughout this section, we will denote $[\mathcal{I}]$ and $[\mathcal{T}]=[U_{\mathcal{T}}] \mathcal{K}$ as the representatives of $\mathcal{I}$ and $\mathcal{T}$ ($\mathcal{T}^2$ can be either $+1$ or $-1$) in the tight-binding basis states. 
We will present the construction of the representatives for $\mathcal{I}$ and $\mathcal{T}$ in SN~\ref{sec:proof-IVGI-dag} and \ref{sec:proof-TVGT-inverse} respectively, in which we will also derive how the matrix $[V(\mathbf{G})]$ in SEq.~\eqref{eq:def-VG-1} transforms under $\mathcal{I}$ and $\mathcal{T}$. 
Although the constraints presented in SN~\ref{appendix:I-constraint-on-P-wilson}, \ref{appendix:I-constraint-on-nested-P-wilson} \ref{appendix:T-constraint-on-P-wilson}, \ref{appendix:T-constraint-on-nested-P-pm}, \ref{sec:proof-IVGI-dag} and \ref{sec:proof-TVGT-inverse} are not new (see for example SRefs.~\cite{wieder2018wallpaper,alexandradinata2014wilsonloop,alexandradinata2016topological,benalcazar2017quantized,benalcazar2017electric,wieder2018axion}), we will review them here for completeness, and to complement the results of SN~\ref{appendix:I-constraint-on-P-pm}, \ref{appendix:I-constraint-on-nested-P-pm}, \ref{appendix:T-constraint-on-P-pm}, and \ref{appendix:T-constraint-on-nested-P-pm}, which are derived for the first time in the present work.

\subsection{Unitary $\mathcal{I}$ Constraint on the $P$-Wilson Loop} \label{appendix:I-constraint-on-P-wilson}
We begin by constructing the $P$-Wilson loop associated to a projector $P$ onto a set of spectrally isolated occupied bands. 
Recall from SN~\ref{sec:P_Wilson_loop} and SRefs.~\cite{yu2011equivalent,alexandradinata2014wilsonloop,gresch2017z2pack} that the $\widehat{\mathbf{G}}_1$-directed $P$-Wilson loop operator with base point $\mathbf{k}=(k_1,k_2,k_3)$ can be written as
\begin{equation}
    \mathcal{W}_{1,\mathbf{k},\mathbf{G}_1} = [V(\mathbf{G}_1)] \lim_{N\to \infty} \left( [P(k_1+2\pi,k_2,k_3)][P(k_1+\frac{2\pi(N-1)}{N},k_2,k_3)] \cdots [P(k_1+\frac{2\pi}{N},k_2,k_3)] [P(k_1,k_2,k_3)]\right), \label{eq:def-W1}
\end{equation}
where the product of projectors $[P(\mathbf{k})]$ is taken along a straight path from $\mathbf{k}$ to $\mathbf{k}+\mathbf{G}_1$, or equivalently from $(k_1,k_2,k_3)$ to $(k_1+2\pi,k_2,k_3)$ in reduced coordinates.
In SEq.~\eqref{eq:def-W1} the matrix $[V(\mathbf{G})]$ defined in SEq.~\eqref{eq:def-VG-1} encodes the positions of the tight-binding basis orbitals, and $[P(\mathbf{k})]$ is the projector onto the occupied eigenvectors $|u_{n,\mathbf{k}}\rangle$ of the Bloch Hamiltonian as defined in SEq.~\eqref{eq:P_projector}. 
For completeness, we summarize the important properties of $[V(\mathbf{G})]$ and $[P(\mathbf{k})]$ below:
\begin{align}
    & [V (\mathbf{G})]_{\alpha \beta} = e^{i\mathbf{G}\cdot \mathbf{r}_{\alpha}}\delta_{\alpha \beta}, \label{eq:def-VG} \\
    & [P(\mathbf{k})] = \sum_{n=1}^{N_{\text{occ}}} |  u_{n,\mathbf{k}}\rangle \langle u_{n,\mathbf{k}}|, \label{eq:def-Pk} \\
    & |  u_{n,\mathbf{k}+\mathbf{G}}\rangle = [V(\mathbf{G})]^{-1} |  u_{n,\mathbf{k}}\rangle = [V(\mathbf{G})]^{\dagger} |  u_{n,\mathbf{k}}\rangle , \label{eq:bc-of-u-nk}
\end{align}
where $\alpha,\beta = 1\ldots N_{\mathrm{sta}}$ in SEq.~\eqref{eq:def-VG} label the tight-binding basis states within the primitive unit cell, and $N_{\text{occ}}$ in SEq.~\eqref{eq:def-Pk} is the number of occupied energy bands. 
From the definition in SEq.~\eqref{eq:def-Pk}, the matrix projector $[P(\mathbf{k})]$ is Hermitian, such that 
\begin{equation}
	[P(\mathbf{k})]^{\dagger} = [P(\mathbf{k})]. \label{eq:P-is-hermitian}
\end{equation}
For a system with inversion symmetry, $[V(\mathbf{G})]$ and $[P(\mathbf{k})]$ transform under $\mathcal{I}$ according to
\begin{align}
    & [\mathcal{I}] [V(\mathbf{G})] [\mathcal{I}]^{\dagger} = [V(-\mathbf{G})], \label{eq:inversion-act-on-VG} \\
    & [\mathcal{I}] [P(\mathbf{k})] [\mathcal{I}]^{\dagger} = [P(-\mathbf{k})]. \label{eq:inversion-act-on-P}
\end{align}
For the proof of SEq.~\eqref{eq:inversion-act-on-VG} and also the explicit construction of $[\mathcal{I}]$ as the unitary matrix representative of $\mathcal{I}$ in the tight-binding basis states, see SN~\ref{sec:proof-IVGI-dag} and SRefs.~\cite{alexandradinata2014wilsonloop,wieder2018wallpaper}.
Note that $[P(\mathbf{k})]$ must project onto a set of energy bands that is mapped into itself under $\mathcal{I}$ in order for SEq.~\eqref{eq:inversion-act-on-P} to be satisfied. 
This is guaranteed if $[P(\mathbf{k})]$ projects onto a set of gapped energy bands for an $\mathcal{I}$-symmetric Hamiltonian. 
Using SEqs.~\eqref{eq:def-VG}, \eqref{eq:def-Pk},  \eqref{eq:bc-of-u-nk}, \eqref{eq:P-is-hermitian}, \eqref{eq:inversion-act-on-VG}, and \eqref{eq:inversion-act-on-P}, it follows that under the action of $\mathcal{I}$, the $\widehat{\mathbf{G}}_1$-directed $P$-Wilson loop operator $\mathcal{W}_{1,\mathbf{k},\mathbf{G}_1}$ transforms according to
\begin{align}
    [\mathcal{I}] \mathcal{W}_{1,\mathbf{k},{\mathbf{G}}_1} [\mathcal{I}]^{\dagger} & = [\mathcal{I}] [V({\mathbf{G}}_1)] [P(k_1+2\pi,k_2,k_3)] \cdots [P(k_1,k_2,k_3)] [\mathcal{I}]^{\dagger} \nonumber \\
    & = [\mathcal{I}] [V({\mathbf{G}}_1)] [\mathcal{I}]^{\dagger} [\mathcal{I}] [P(k_1+2\pi,k_2,k_3)] [\mathcal{I}]^{\dagger} \cdots [\mathcal{I}] [P(k_1,k_2,k_3)] [\mathcal{I}]^{\dagger} \nonumber \\
    & = [V(-{\mathbf{G}}_1)]  [P(-k_1-2\pi,-k_2,-k_3)]  \cdots  [P(-k_1,-k_2,-k_3)] \nonumber \\
    & = [V(-{\mathbf{G}}_1)]  [V(-{\mathbf{G}}_1)]^{-1} [P(-k_1,-k_2,-k_3)] [V(-{\mathbf{G}}_1)]  \cdots [V(-{\mathbf{G}}_1)]^{-1} [P(-k_1+2\pi,-k_2,-k_3)] [V(-{\mathbf{G}}_1)] \nonumber \\
    & = [P(-k_1,-k_2,-k_3)]  \cdots  [P(-k_1+2\pi,-k_2,-k_3)] [V(-{\mathbf{G}}_1)] \nonumber \\
    & = \left( [V(-{\mathbf{G}}_1)]^{\dagger} [P(-k_1+2\pi,-k_2,-k_3)] \cdots [P(-k_1,-k_2,-k_3)] \right)^{\dagger} \nonumber \\
    & = \left( [V({\mathbf{G}}_1)] [P(-k_1+2\pi,-k_2,-k_3)] \cdots [P(-k_1,-k_2,-k_3)] \right)^{\dagger} \nonumber \\
    & = \mathcal{W}_{1,-\mathbf{k},{\mathbf{G}}_1}^{\dagger}, \label{eq:I_transform_W1}
\end{align}
where $[P(\mathbf{k}_f)] \cdots [P(\mathbf{k}_i)]$ is a shorthand notation for the product of projection matrices along a straight path in $\mathbf{k}$-space from $\mathbf{k}_i$ to $\mathbf{k}_f$. 
$\mathcal{W}_{1,\mathbf{k},{\mathbf{G}}_1}$ has $N_{\mathrm{sta}}-N_{\mathrm{occ}}$ zero eigenvalues corresponding to the number of unoccupied eigenstates. 
More importantly, $\mathcal{W}_{1,\mathbf{k},{\mathbf{G}}_1}$ has $N_{\mathrm{occ}}$ unimodular eigenvalues that are independent of $k_1$~\cite{alexandradinata2014wilsonloop,vanderbilt2018berry,lecture_notes_on_berry_phases_and_topology_bb}. 
For simplicity, we denote the set of unimodular eigenvalues of $\mathcal{W}_{1,\mathbf{k},{\mathbf{G}}_1}$ as $\{ e^{i \gamma_{1,j}(k_2,k_3)} | j = 1 \ldots N_{\mathrm{occ}} \}$ where $j$ is the Wannier band index. 
SEq.~\eqref{eq:I_transform_W1} implies that $\mathcal{W}_{1,\mathbf{k},{\mathbf{G}}_1}$ and $\mathcal{W}_{1,-\mathbf{k},{\mathbf{G}}_1}^{\dagger}$ are isospectral since they are related to each other by a unitary transformation.
According to SEq.~\eqref{eq:I_transform_W1}, and using the fact that the eigenvalues of $\mathcal{W}_{1,-\mathbf{k},{\mathbf{G}}_1}^{\dagger}$ coincide with the complex conjugates of the eigenvalues of $\mathcal{W}_{1,-\mathbf{k},{\mathbf{G}}_1}$ (see SN~\ref{appendix:eigval_eigvec_W} for more details), we have that 
\begin{equation}
    \{ e^{i\gamma_{1,j}(k_2,k_3)} | j = 1 \ldots N_{\mathrm{occ}}\} = \{ e^{-i\gamma_{1,j}(-k_2,-k_3)} | j = 1 \ldots N_{\mathrm{occ}}\}. \label{eq:I-constraint-on-exp-i-theta-1j}
\end{equation}
In terms of the set of eigenphases $\gamma_{1,j}(k_2,k_3)$, SEq.~\eqref{eq:I-constraint-on-exp-i-theta-1j} implies that
\begin{equation}
    \{ \gamma_{1,j}(k_2,k_3) | j = 1 \ldots N_{\mathrm{occ}} \} \text{ mod } 2\pi = \{ -\gamma_{1,j}(-k_2,-k_3) | j = 1 \ldots N_{\mathrm{occ}}\} \text{ mod } 2\pi. \label{eq:I-constraint-on-theta-1j}
\end{equation}
We refer to the $\mathcal{I}$ constraint in SEq.~\eqref{eq:I-constraint-on-theta-1j} as an effective ``particle-hole'' symmetry in the $P$-Wilson loop eigenphases.
A more operational way to interpret SEq.~\eqref{eq:I-constraint-on-theta-1j} is that the $P$-Wannier spectrum $\{\gamma_{1,j}(k_2,k_3) | j=1\ldots N_{\mathrm{occ}} \}$ is invariant under a simultaneous sign-change of the  momenta $(k_2,k_3)\rightarrow(-k_2,-k_3)$ and the phase  $\gamma_1\rightarrow-\gamma_1$. 
Recall from SN~\ref{sec:P_Wilson_loop} that the eigenphases [SEq.~\eqref{eq:I-constraint-on-theta-1j}] of the $P$-Wilson loop operator [SEq.~\eqref{eq:def-W1}] correspond to the localized positions of hybrid Wannier functions formed from the set of occupied states~\cite{alexandradinata2014wilsonloop,benalcazar2017electric,benalcazar2017quantized,lecture_notes_on_berry_phases_and_topology_bb,vanderbilt2018berry}. 
With this in mind, SEq.~\eqref{eq:I-constraint-on-theta-1j} then follows from the fact that inversion flips both position and momentum of a hybrid Wannier function.

We can also compute the sum over $j$ of the $\gamma_{1,j}(k_2,k_3)$ which defines the total Berry phase~\cite{yu2011equivalent,alexandradinata2014wilsonloop,soluyanov2012smooth,marzari1997maximally}
\begin{equation}
    \gamma_{1}(k_2,k_3) \equiv \sum_{j=1}^{N_{\text{occ}}} \gamma_{1,j}(k_2,k_3) \text{ mod } 2\pi , \label{eq:def-total-theta-1}
\end{equation}
where SEq.~\eqref{eq:I-constraint-on-theta-1j} indicates that
\begin{equation}
    \gamma_{1}(k_2,k_3) \text{ mod } 2\pi = -\gamma_{1}(-k_2,-k_3) \text{ mod } 2\pi. \label{eq:I-constraint-on-total-theta-1}
\end{equation}
Recall that SEq.~\eqref{eq:W_1_k_plus_Gp_G_invariant} implies that the eigenphases $\gamma_1(k_2,k_3)$ are invariant if we shift the base point by a reciprocal lattice vector. 
Using this fact, SEq.~\eqref{eq:I-constraint-on-total-theta-1} implies that at the four time-reversal invariant momenta (TRIMs) $(k_2^{TRIM},k_3^{TRIM})=(0,0)$, $(\pi,0)$, $(0,\pi)$, and $(\pi,\pi)$, we have 
\begin{equation}
    \left( \gamma_{1}(k_2^{TRIM},k_3^{TRIM}) \mod \pi \right)= 0. \label{eq:I-quantization-of-theta-1}
\end{equation}

Since inversion symmetry treats all momentum components $k_i$ in the same way---it flips the sign of all the components $k_i$---the results of this section can be generalized to $\widehat{\mathbf{G}}_2$- and $\widehat{\mathbf{G}}_3$-directed $P$-Wilson loops. 
The results of SEqs.~\eqref{eq:I-constraint-on-exp-i-theta-1j}, \eqref{eq:I-constraint-on-theta-1j}, \eqref{eq:I-constraint-on-total-theta-1}, and \eqref{eq:I-quantization-of-theta-1}, written in the general notation in SN~\ref{sec:P_Wilson_loop}, therefore generalize to
\begin{align}
	& \{ e^{i (\gamma_1)_{j,\mathbf{k},\mathbf{G}}} | j = 1 \ldots N_{\mathrm{occ}}\} = \{ e^{-i (\gamma_1)_{j,-\mathbf{k},\mathbf{G}}} | j = 1 \ldots N_{\mathrm{occ}} \}, \\
	&\{ (\gamma_1)_{j,\mathbf{k},\mathbf{G}} | j = 1 \ldots N_{\mathrm{occ}} \} \text{ mod } 2\pi = \{ -(\gamma_1)_{j,-\mathbf{k},\mathbf{G}} | j = 1 \ldots N_{\mathrm{occ}} \} \text{ mod } 2\pi, \label{eq:I_constraints_on_gamma_1_j} \\
	&(\gamma_1)_{\mathbf{k},\mathbf{G}} \text{ mod } 2\pi = -(\gamma_1)_{-\mathbf{k},\mathbf{G}} \text{ mod } 2\pi, \\
	& \left( (\gamma_1)_{\mathbf{k}_{TRIM},\mathbf{G}} \mod \pi \right) = 0, \label{eq:general_notation_gamma_1_I_constraint}
\end{align}
where 
\begin{equation}
    (\gamma_1)_{\mathbf{k},\mathbf{G}} \equiv \sum_{j=1}^{N_{\mathrm{occ}}}(\gamma_1)_{j,\mathbf{k},\mathbf{G}} \text{ mod } 2\pi, \label{eq:summed-gamma-1kG-general-notation}
\end{equation}
and $(\gamma_1)_{j,\mathbf{k},\mathbf{G}}$ are the eigenphases of the $\widehat{\mathbf{G}}$-directed $P$-Wilson loop operator $\mathcal{W}_{1,\mathbf{k},\mathbf{G}}$ satisfying 
\begin{equation}
    [\mathcal{I}]\mathcal{W}_{1,\mathbf{k},\mathbf{G}} [\mathcal{I}]^{\dagger} = \mathcal{W}_{1,-\mathbf{k},\mathbf{G}}^{\dagger}. \label{eq:general-notation-inversion-constraint-on-P-Wilson-loop-operator}
\end{equation}
SEq.~\eqref{eq:general-notation-inversion-constraint-on-P-Wilson-loop-operator} is true provided that the matrix projector onto the occupied energy bands is inversion-symmetric as specified in SEq.~\eqref{eq:inversion-act-on-P}.
Finally, since $(\gamma_1)_{j,\mathbf{k},\mathbf{G}}$ is independent of the momentum component $\mathbf{k} \cdot \mathbf{a}$, where $\mathbf{a}$ is the real-space primitive lattice vector dual to the primitive reciprocal lattice vector $\mathbf{G}$, $\mathbf{k}_{TRIM}$  in SEq.~\eqref{eq:general_notation_gamma_1_I_constraint} should be interpreted as a $\mathbf{k}$-vector with $(\mathbf{k}_{TRIM} \cdot \mathbf{a'}) \mod \pi = 0$ for each primitive lattice vector $\mathbf{a'}\neq\mathbf{a}$.

\subsection{Unitary $\mathcal{I}$ Constraint on the Nested $P$-Wilson Loop} \label{appendix:I-constraint-on-nested-P-wilson}

Recall that to construct the nested $P$-Wilson loop operator from SN~\ref{sec:nested_P_Wilson_loop}, we must first specify two primitive reciprocal lattice vectors $\mathbf{G}_1$ and $\mathbf{G}_2$, respectively.
We then solve the eigenvalue equation for the $\widehat{\mathbf{G}}_1$-directed $P$-Wilson loop operator [SEq.~\eqref{eq:def-W1}],
\begin{equation}
    \mathcal{W}_{1,\mathbf{k},\mathbf{G}_1} | w_{j}(\mathbf{k}) \rangle = e^{i\gamma_{1,j}(k_2,k_3)}| w_{j}(\mathbf{k}) \rangle, \label{eq:nu-j-as-eigvector-of-W1}
\end{equation}
where the eigenvectors $\{ | w_{j}(\mathbf{k}) \rangle | j = 1 \ldots N_{\mathrm{occ}} \}$ with unimodular eigenvalues form an orthonormal set such that
\begin{equation}
	\langle w_{j}(\mathbf{k}) | w_{j'}(\mathbf{k}) \rangle = \delta_{jj'}.
\end{equation}
In SEq.~\eqref{eq:nu-j-as-eigvector-of-W1}, the Wannier band basis vector $| w_{j}(\mathbf{k}) \rangle$ lies in the image of $[P(\mathbf{k})]$ and is an eigenvector of $\mathcal{W}_{1,\mathbf{k},\mathbf{G}_1}$ with unimodular eigenvalue $e^{i\gamma_{1,j}(k_2,k_3)}$~\cite{benalcazar2017electric}.
Although the eigenvalue $e^{i\gamma_{1,j}(k_2,k_3)}$ of $\mathcal{W}_{1,\mathbf{k},\mathbf{G}_1}$ [SEq.~\eqref{eq:nu-j-as-eigvector-of-W1}] is independent of $k_1$ (the component of the momentum along $\mathbf{G}_1$), the eigenvector $| w_{j}(\mathbf{k}) \rangle$ depends on all momentum components $k_1$, $k_2$, and $k_3$ of $\mathbf{k}$.
In particular, if we expand $| w_{j}(\mathbf{k}) \rangle$ in terms of the occupied energy eigenvectors $\{| u_{m,\mathbf{k}} \rangle | m=1 \ldots N_{\mathrm{occ}} \}$ as
\begin{equation}
	| w_{j}(\mathbf{k}) \rangle = \sum_{m=1}^{N_{\mathrm{occ}}} [\nu_{j}(\mathbf{k})]_m | u_{m,\mathbf{k}} \rangle,
\end{equation}
then as discussed in SN~\ref{sec:nested_P_Wilson_loop} the $N_{\mathrm{occ}}$-component vector $| \nu_{j}(\mathbf{k}) \rangle$ satisfies the parallel transport condition in SEq.~\eqref{eq:parallel-transport-nu-j-k-G}. 

Let us consider an inversion-symmetric system such that the $\widehat{\mathbf{G}}_1$-directed $P$-Wilson loop operator satisfies SEq.~\eqref{eq:I_transform_W1}.
Acting with $[\mathcal{I}]$, the unitary matrix representative of $\mathcal{I}$ in the tight-binding basis states, on both sides of the eigenvalue equation in SEq.~\eqref{eq:nu-j-as-eigvector-of-W1}, we have
\begin{equation}
	[\mathcal{I}] \mathcal{W}_{1,\mathbf{k},\mathbf{G}_1}  [\mathcal{I}]^{\dagger} [\mathcal{I}] | w_{j}(\mathbf{k}) \rangle = e^{i\gamma_{1,j}(k_2,k_3)} [\mathcal{I}]|  w_{j}(\mathbf{k}) \rangle, \label{eq:act-I-on-nu-j-as-eigvector-of-W1}
\end{equation}
where we have inserted the identity matrix $[\mathcal{I}]^{\dagger}[\mathcal{I}]$ since $[\mathcal{I}]$ is unitary.
Combining SEqs.~\eqref{eq:I_transform_W1} and \eqref{eq:act-I-on-nu-j-as-eigvector-of-W1}, we obtain
\begin{equation}
	\mathcal{W}_{1,-\mathbf{k},\mathbf{G}_1}^{\dagger}  [\mathcal{I}] | w_{j}(\mathbf{k}) \rangle = e^{i\gamma_{1,j}(k_2,k_3)} [\mathcal{I}]|  w_{j}(\mathbf{k}) \rangle. \label{eq:act-I-on-nu-j-as-eigvector-of-W1-2ndstep}
\end{equation}
Recall that the eigenvector of $\mathcal{W}_{1,\mathbf{k},\mathbf{G}_1}$ with eigenvalue $e^{i \gamma_{1,j}(k_2,k_3)}$ is also the eigenvector of $\mathcal{W}_{1,\mathbf{k},\mathbf{G}_1}^{\dagger}$ with eigenvalue $e^{-i \gamma_{1,j}(k_2,k_3)}$ (see SN~\ref{appendix:eigval_eigvec_W} for more details). 
From SEq.~\eqref{eq:act-I-on-nu-j-as-eigvector-of-W1-2ndstep}, we then deduce that for a given eigenvector $| w_{j}(\mathbf{k}) \rangle$ of $\mathcal{W}_{1,\mathbf{k},{\mathbf{G}}_1}$ with eigenvalue $e^{i\gamma_{1,j}(k_2,k_3)}$, $[\mathcal{I}] | w_{j}(\mathbf{k}) \rangle$ is an eigenvector of $\mathcal{W}_{1,-\mathbf{k},{\mathbf{G}}_1}$ with eigenvalue $e^{-i\gamma_{1,j}(k_2,k_3)}$.
In other words, we have
\begin{equation}
	\mathcal{W}_{1,-\mathbf{k},\mathbf{G}_1}  [\mathcal{I}] | w_{j}(\mathbf{k}) \rangle = e^{-i\gamma_{1,j}(k_2,k_3)} [\mathcal{I}]|  w_{j}(\mathbf{k}) \rangle. \label{eq:act-I-on-nu-j-as-eigvector-of-W1-3rdstep}
\end{equation}
In addition to the effective particle-hole symmetry of the $\widehat{\mathbf{G}}_1$-directed $P$-Wannier bands that we have already established in SEq.~\eqref{eq:I-constraint-on-theta-1j}, SEq.~\eqref{eq:act-I-on-nu-j-as-eigvector-of-W1-3rdstep} implies that the eigenvectors $\{  \ket{w_{j}(\mathbf{k})} | j=1\ldots N_{\mathrm{occ}} \}$ and $\{  \ket{w_{j}(-\mathbf{k})} | j=1\ldots N_{\mathrm{occ}} \}$ of $\mathcal{W}_{1,\mathbf{k},{\mathbf{G}}_1}$ and $\mathcal{W}_{1,-\mathbf{k},{\mathbf{G}}_1}$, respectively, are related to each other by $[\mathcal{I}]$.

Suppose that the $\widehat{\mathbf{G}}_1$-directed $P$-Wannier bands can be separated into disjoint groupings and we choose a grouping of $N_{W}$ bands described by
\begin{equation}
	\{ \gamma_{1,j}(k_2,k_3) | j=1\ldots N_{W} \} \label{eq:grouping_of_band_eqn}
\end{equation}
to form the projector 
\begin{align}
    [\widetilde{P}(\mathbf{k})] = \sum_{j=1}^{N_W} |  w_{j}(\mathbf{k}) \rangle \langle  w_{j}(\mathbf{k}) |, \label{eq:proj-Wannier-basis}
\end{align}
which projects onto the vector space spanned by the $N_W$ Wannier band eigenstates $| w_{j} (\mathbf{k})\rangle$ from SEq.~\eqref{eq:nu-j-as-eigvector-of-W1}.
By definition, the matrix projector $[\widetilde{P}(\mathbf{k})] $ in SEq.~\eqref{eq:proj-Wannier-basis} is Hermitian, and therefore satisfies
\begin{equation}
    [\widetilde{P}(\mathbf{k})]^{\dagger}=[\widetilde{P}(\mathbf{k})]. \label{eq:WannierbandbasisprojectorisHermitian}
\end{equation}
For an inversion-symmetric system, an isolated grouping of Wannier bands [SEq.~\eqref{eq:grouping_of_band_eqn}] centered around an inversion-invariant eigenphase $\gamma_1 \mod \pi =0$ can be chosen to satisfy
\begin{equation}
	[\mathcal{I}] [\widetilde{P}(\mathbf{k})] [\mathcal{I}]^{\dagger} = [\widetilde{P}(-\mathbf{k})]. \label{eq:inversion-act-on-tilde-P}
\end{equation}
Combining SEqs.~\eqref{eq:nu-j-as-eigvector-of-W1}, \eqref{eq:act-I-on-nu-j-as-eigvector-of-W1-3rdstep}, and \eqref{eq:inversion-act-on-tilde-P}, such a grouping of $N_W$ Wannier bands [SEq.~\eqref{eq:grouping_of_band_eqn}] must satisfy
\begin{equation}
	\{ \gamma_{1,j}(k_2,k_3) | j=1\ldots N_{W} \} \text{ mod } 2\pi = \{ -\gamma_{1,j}(-k_2,-k_3) | j=1\ldots N_{W} \} \text{ mod } 2\pi. \label{eq:NWbandsparticleholesymmetric}
\end{equation}
In this work we will always choose $[\widetilde{P}(\mathbf{k})]$ for inversion-symmetric systems such that SEqs.~\eqref{eq:inversion-act-on-tilde-P} and \eqref{eq:NWbandsparticleholesymmetric} hold.
Similar to the effective particle-hole symmetry of the entire $\widehat{\mathbf{G}}_1$-directed $P$-Wannier band structure described in SEq.~\eqref{eq:I-constraint-on-theta-1j}, the grouping of $N_W$ $\widehat{\mathbf{G}}_1$-directed $P$-Wannier bands chosen in a manner that satisfies SEq.~\eqref{eq:inversion-act-on-tilde-P} also has an effective particle-hole symmetry described by SEq.~\eqref{eq:NWbandsparticleholesymmetric}~\cite{wieder2018axion}.

Before we move on, we emphasize that the eigenvectors $|w_{j}(\mathbf{k}) \rangle$  satisfy the boundary condition
\begin{equation}
	|w_{j}(\mathbf{k}+\mathbf{G}) \rangle = [V(\mathbf{G})]^{-1} |w_{j}(\mathbf{k}) \rangle = [V(\mathbf{G})]^{\dagger} |w_{j}(\mathbf{k}) \rangle, \label{eq:bc-nu-j}
\end{equation}
where $\mathbf{G}$ is a reciprocal lattice vector and $[V(\mathbf{G})]$ is given in SEq.~\eqref{eq:def-VG}.
To verify SEq.~\eqref{eq:bc-nu-j}, we note that under a shift of the base point from $\mathbf{k}$ to $\mathbf{k} + \mathbf{G}$, the $\widehat{\mathbf{G}}_1$-directed $P$-Wilson loop operator $\mathcal{W}_{1,\mathbf{k},\mathbf{G}_1}$ [SEq.~\eqref{eq:def-W1}] transforms according to
\begin{align}
    \mathcal{W}_{1,\mathbf{k}+\mathbf{G},\mathbf{G}_1} & = [V(\mathbf{G}_1)] [P(\mathbf{k}+\mathbf{G}+\mathbf{G}_1)] \cdots [P(\mathbf{k}+\mathbf{G})] \nonumber \\
    & = [V(\mathbf{G}_1)] [V(\mathbf{G})]^{-1} [P(\mathbf{k}+\mathbf{G}_1)] [V(\mathbf{G})] \cdots [V(\mathbf{G})]^{-1} [P(\mathbf{k})] [V(\mathbf{G})] \nonumber \\
    & = [V(\mathbf{G})]^{-1} [V(\mathbf{G}_1)] [P(\mathbf{k}+\mathbf{G}_1)]  \cdots  [P(\mathbf{k})] [V(\mathbf{G})] \nonumber \\
    & = [V(\mathbf{G})]^{-1} \mathcal{W}_{1,\mathbf{k},\mathbf{G}_1}[V(\mathbf{G})], \label{eq:W-upon-G-translation-in-ky}
\end{align}
where $[P(\mathbf{k}_f)] \cdots [P(\mathbf{k}_i)]$ is a shorthand notation for the product of projection matrices along a straight path in $\mathbf{k}$-space from $\mathbf{k}_i$ to $\mathbf{k}_f$. 
In simplifying SEq.~\eqref{eq:W-upon-G-translation-in-ky} we have used the boundary condition SEq.~\eqref{eq:bcs} on the energy eigenvectors $| u_{n,\mathbf{k}} \rangle$, as well as the commutation relation SEq.~\eqref{eq:Vgs_commute} between the matrices $[V(\mathbf{G})]$ and $[V(\mathbf{G}_1)]$. 
SEq.~\eqref{eq:W-upon-G-translation-in-ky} indicates that $\mathcal{W}_{1,\mathbf{k}+\mathbf{G},\mathbf{G}_1}$ and $\mathcal{W}_{1,\mathbf{k},\mathbf{G}_1}$ are related to each other by a similarity transformation, hence implies that their eigenvectors $| w_{j}(\mathbf{k}+\mathbf{G}) \rangle$ and $| w_{j}(\mathbf{k}) \rangle$ can be chosen to satisfy SEq.~\eqref{eq:bc-nu-j}.

Using the definition in SEq.~\eqref{eq:proj-Wannier-basis} of the matrix projector $[\widetilde{P}(\mathbf{k})]$  onto the $N_W$ Wannier band eigenfunctions with the eigenphases in SEq.~\eqref{eq:grouping_of_band_eqn}, we can construct the following $N_{\mathrm{sta}} \times N_{\mathrm{sta}}$ $\widetilde{P}$-Wilson loop operator along a closed loop parallel to $\mathbf{G}_2$,
\begin{equation}
	\mathcal{W}_{2,\mathbf{k},\mathbf{G}_1,\mathbf{G}_2} = [V(\mathbf{G}_2)] \lim_{N \to \infty} \left( [\widetilde{P}(k_1,k_2+2\pi,k_3)] [\widetilde{P}(k_1,k_2+\frac{2\pi(N-1)}{N},k_3)] \cdots [\widetilde{P}(k_1,k_2+\frac{2\pi}{N},k_3)] [\widetilde{P}(k_1,k_2,k_3)] \right). \label{eq:def-nested-P-wilson-loop}
\end{equation}
$\mathcal{W}_{2,\mathbf{k},\mathbf{G}_1,\mathbf{G}_2}$ in SEq.~\eqref{eq:def-nested-P-wilson-loop} is then the nested $P$-Wilson loop operator constructed by first computing the holonomy along a closed loop parallel to $\mathbf{G}_1$, followed by computing the holonomy along a closed loop parallel to $\mathbf{G}_2$. $\mathcal{W}_{2,\mathbf{k},\mathbf{G}_1,\mathbf{G}_2}$ has $N_{\mathrm{sta}}-N_{W}$ zero and $N_W$ unimodular eigenvalues that are independent of $k_2$~\cite{alexandradinata2014wilsonloop,vanderbilt2018berry,lecture_notes_on_berry_phases_and_topology_bb}. 
For simplicity, we denote the set of unimodular eigenvalues of $\mathcal{W}_{2,\mathbf{k},\mathbf{G}_1,\mathbf{G}_2}$ as $\{ e^{i\gamma_{2,j}(k_1,k_3)} | j=1\ldots N_W \}$ where $j$ is the nested Wannier band index. 
Under an $\mathcal{I}$ transformation, $\mathcal{W}_{2,\mathbf{k},\mathbf{G}_1,\mathbf{G}_2}$ in SEq.~\eqref{eq:def-nested-P-wilson-loop} transforms according to
\begin{align}
    [\mathcal{I}] \mathcal{W}_{2,\mathbf{k},\mathbf{G}_1,\mathbf{G}_2} [\mathcal{I}]^{\dagger} & = [\mathcal{I}] [V(\mathbf{G}_2)] [\widetilde{P}(k_1,k_2 + 2\pi,k_3)] \cdots [\widetilde{P}(k_1,k_2,k_3)] [\mathcal{I}]^{\dagger} \nonumber \\
    & = [\mathcal{I}] [V(\mathbf{G}_2)] [\mathcal{I}]^{\dagger} [\mathcal{I}] [\widetilde{P}(k_1,k_2 + 2\pi,k_3)] [\mathcal{I}]^{\dagger} \cdots [\mathcal{I}] [\widetilde{P}(k_1,k_2,k_3)] [\mathcal{I}]^{\dagger} \nonumber \\
    & = [V(-\mathbf{G}_2)] [\widetilde{P}(-k_1,-k_2-2\pi,-k_3)] \cdots [\widetilde{P}(-k_1,-k_2,-k_3)] \nonumber \\
    & = [V(-\mathbf{G}_2)] [V(-\mathbf{G}_2)]^{-1} [\widetilde{P}(-k_1,-k_2,-k_3)] [V(-\mathbf{G}_2)] \cdots [V(-\mathbf{G}_2)]^{-1} [\widetilde{P}(-k_1,-k_2+2\pi,-k_3)] [V(-\mathbf{G}_2)] \nonumber \\
    & = [\widetilde{P}(-k_1,-k_2,-k_3)]  \cdots  [\widetilde{P}(-k_1,-k_2+2\pi,-k_3)] [V(-\mathbf{G}_2)] \nonumber \\
    & = \left( [V(-\mathbf{G}_2)]^{\dagger} [\widetilde{P}(-k_1,-k_2+2\pi,-k_3)] \cdots [\widetilde{P}(-k_1,-k_2,-k_3) ] \right)^{\dagger} \nonumber \\
    & = \left( [V(\mathbf{G}_2)] [\widetilde{P}(-k_1,-k_2+2\pi,-k_3)] \cdots [\widetilde{P}(-k_1,-k_2,-k_3)] \right)^{\dagger} \nonumber \\
    & = \mathcal{W}_{2,-\mathbf{k},\mathbf{G}_1,\mathbf{G}_2}^{\dagger}, \label{eq:W-2-k-G1-G2-I-transform}
\end{align}
where $[\widetilde{P}(\mathbf{k}_{f})] \cdots [\widetilde{P}(\mathbf{k}_{i})] $ represents a product of Wannier band projectors [SEq.~\eqref{eq:proj-Wannier-basis}] along a straight path in $\mathbf{k}$-space from $\mathbf{k}_{i}$ to $\mathbf{k}_{f}$. 
In deriving SEq.~\eqref{eq:W-2-k-G1-G2-I-transform} we have also used SEqs.~\eqref{eq:def-VG}, \eqref{eq:inversion-act-on-VG}, \eqref{eq:proj-Wannier-basis}, \eqref{eq:inversion-act-on-tilde-P}, and \eqref{eq:bc-nu-j}, and the unitarity of the matrix $[\mathcal{I}]$. 
SEq.~\eqref{eq:W-2-k-G1-G2-I-transform} implies that $\mathcal{W}_{2,\mathbf{k},\mathbf{G}_1,\mathbf{G}_2}$ and $\mathcal{W}_{2,-\mathbf{k},\mathbf{G}_1,\mathbf{G}_2}^{\dagger}$ are isospectral since they are related to each other by a unitary transformation.
According to SEq.~\eqref{eq:W-2-k-G1-G2-I-transform} and using the fact that the unimodular eigenvalues of $\mathcal{W}_{2,-\mathbf{k},\mathbf{G}_1,\mathbf{G}_2}^{\dagger}$ are the complex conjugates of the unimodular eigenvalues of $\mathcal{W}_{2,-\mathbf{k},\mathbf{G}_1,\mathbf{G}_2}$ (see SN~\ref{appendix:eigval_eigvec_W} for more details), we have 
\begin{equation}
    \{ e^{i\gamma_{2,j}(k_1,k_{3})} | j=1\ldots N_W \} = \{ e^{-i\gamma_{2,j}(-k_1,-k_3)} | j = 1 \ldots N_W \}. \label{eq:I-constraint-on-exp-theta-2j}
\end{equation}
In terms of the set of eigenphases $\gamma_{2,j}(k_1,k_3)$, SEq.~\eqref{eq:I-constraint-on-exp-theta-2j} implies that
\begin{equation}
    \{ \gamma_{2,j}(k_1,k_3) |j=1\ldots N_W\} \text{ mod } 2\pi = \{ -\gamma_{2,j}(-k_1,-k_3) | j = 1 \ldots N_W \} \text{ mod } 2\pi. \label{eq:I-constraint-on-theta-2j}
\end{equation}
From the $\mathcal{I}$ constraint in SEq.~\eqref{eq:I-constraint-on-theta-2j}, we identify an effective particle-hole symmetry in the set of nested $P$-Wilson loop eigenphases. 
A more operational way to interpret SEq.~\eqref{eq:I-constraint-on-theta-2j} is that the nested $P$-Wannier spectrum $\{\gamma_{2,j}(k_1,k_3) | j=1\ldots N_W \}$ is invariant under a simultaneous sign-change of the  momenta $(k_1,k_3)\rightarrow(-k_1,-k_3)$ and the phase  $\gamma_2\rightarrow-\gamma_2$. 
Recall from SN~\ref{sec:nested_P_Wilson_loop} that the eigenphases [SEq.~\eqref{eq:I-constraint-on-theta-2j}] of the nested $P$-Wilson loop operator [SEq.~\eqref{eq:def-nested-P-wilson-loop}] correspond to the localized positions of the hybrid Wannier functions formed from a group of $P$-Wannier bands [SEq.~\eqref{eq:grouping_of_band_eqn}]~\cite{benalcazar2017electric,benalcazar2017quantized}. 
With this in mind, we can understand SEq.~\eqref{eq:I-constraint-on-theta-2j} as a consequence of the fact that inversion flips both position and momentum of a hybrid Wannier function.

We can also compute the sum over $j$ of the $\gamma_{2,j}(k_1,k_3)$, which defines the nested Berry phase~\cite{wang2019higherorder,wieder2018axion,varnava2020axion}
\begin{equation}
    \gamma_{2}(k_1,k_3) \equiv \sum_{j=1}^{N_W} \gamma_{2,j}(k_1,k_3) \text{ mod } 2\pi, \label{eq:def-total-theta-2}
\end{equation}
where SEq.~(\ref{eq:I-constraint-on-theta-2j}) indicates that
\begin{equation}
    \gamma_{2}(k_1,k_3) \text{ mod } 2\pi = -\gamma_{2}(-k_1,-k_3) \text{ mod } 2\pi. \label{eq:I-constraint-on-total-theta-2}
\end{equation}
Notice that $\mathcal{W}_{2,\mathbf{k},\mathbf{G}_1,\mathbf{G}_2}$ and $\mathcal{W}_{2,\mathbf{k}+\mathbf{G},\mathbf{G}_1,\mathbf{G}_2}$ are isospectral for any reciprocal lattice vector $\mathbf{G}$ since $\mathcal{W}_{2,\mathbf{k},\mathbf{G}_1,\mathbf{G}_2}$ and $\mathcal{W}_{2,\mathbf{k}+\mathbf{G},\mathbf{G}_1,\mathbf{G}_2}$ are related to each other by a similarity transformation:
\begin{align}
	\mathcal{W}_{2,\mathbf{k}+\mathbf{G},\mathbf{G}_1,\mathbf{G}_2} & = [V(\mathbf{G}_2)] [\widetilde{P}(\mathbf{k}+\mathbf{G}+\mathbf{G}_2)] \cdots [\widetilde{P}(\mathbf{k}+\mathbf{G})] \nonumber \\
	& = [V(\mathbf{G}_2)] [V(\mathbf{G})]^{-1} [\widetilde{P}(\mathbf{k}+\mathbf{G}_2)]  \cdots [\widetilde{P}(\mathbf{k}+\mathbf{G})] [V(\mathbf{G})] \nonumber \\
	& = [V(\mathbf{G})]^{-1} [V(\mathbf{G}_2)] [\widetilde{P}(\mathbf{k}+\mathbf{G}_2)]  \cdots [\widetilde{P}(\mathbf{k}+\mathbf{G})] [V(\mathbf{G})] \nonumber \\
	& = [V(\mathbf{G})]^{-1} \mathcal{W}_{2,\mathbf{k},\mathbf{G}_1,\mathbf{G}_2} [V(\mathbf{G})],\label{eq:W2-upon-G-translation-in-ky} 
\end{align}
where again $[\widetilde{P}(\mathbf{k}_{f})] \cdots [\widetilde{P}(\mathbf{k}_{i})] $ is a shorthand notation for a product of Wannier band projectors [SEq.~\eqref{eq:proj-Wannier-basis}] along a straight path in $\mathbf{k}$-space from $\mathbf{k}_{i}$ to $\mathbf{k}_{f}$, and we have used SEqs.~\eqref{eq:def-VG}, \eqref{eq:proj-Wannier-basis}, and \eqref{eq:bc-nu-j}. 
Therefore, SEq.~\eqref{eq:I-constraint-on-total-theta-2} implies that at the four TRIMs $(k_1^{TRIM},k_3^{TRIM})=(0,0)$, $(\pi,0)$, $(0,\pi)$, and $(\pi,\pi)$, we have
\begin{equation}
    \left(\gamma_{2}(k_1^{TRIM},k_3^{TRIM})\mod \pi\right) =0. \label{eq:I-quantization-of-theta-2}
\end{equation}

Since inversion symmetry treats all momentum components $k_i$ in the same way---it flips the sign of all the components $k_i$---the results of this section can be generalized to the nested $P$-Wilson loop for any primitive reciprocal lattice vectors $\mathbf{G}$ for the direction of the first loop and $\mathbf{G}'$ for the direction of the second loop. 
The  results of SEqs.~\eqref{eq:I-constraint-on-exp-theta-2j}, \eqref{eq:I-constraint-on-theta-2j}, \eqref{eq:I-constraint-on-total-theta-2}, and \eqref{eq:I-quantization-of-theta-2}, written in the general notation in SN~\ref{sec:nested_P_Wilson_loop}, generalize to
\begin{align}
	& \{ e^{i (\gamma_2)_{j,\mathbf{k},\mathbf{G},\mathbf{G}'}} | j = 1 \ldots N_{W}\} = \{ e^{-i (\gamma_2)_{j,-\mathbf{k},\mathbf{G},\mathbf{G}'}} | j = 1 \ldots N_{W} \}, \\
	&\{ (\gamma_2)_{j,\mathbf{k},\mathbf{G},\mathbf{G}'} | j = 1 \ldots N_{W} \} \text{ mod } 2\pi = \{ -(\gamma_2)_{j,-\mathbf{k},\mathbf{G},\mathbf{G}'} | j = 1 \ldots N_{W} \} \text{ mod } 2\pi, \label{eq:general_notation_gamma_2j_I_constraint} \\
	&(\gamma_2)_{\mathbf{k},\mathbf{G},\mathbf{G}'} \text{ mod } 2\pi = -(\gamma_2)_{-\mathbf{k},\mathbf{G},\mathbf{G}'} \text{ mod } 2\pi, \\
	& \left((\gamma_2)_{\mathbf{k}_{TRIM},\mathbf{G},\mathbf{G}'} \mod \pi \right)  = 0,\label{eq:general_notation_gamma_2_I_constraint}
\end{align}
where 
\begin{equation}
	(\gamma_2)_{\mathbf{k},\mathbf{G},\mathbf{G}'} \equiv \sum_{j=1}^{N_{W}}(\gamma_2)_{j,\mathbf{k},\mathbf{G},\mathbf{G}'} \text{ mod } 2\pi, \label{eq:summed-gamma-2-def-general}
\end{equation}
and $(\gamma_2)_{j,\mathbf{k},\mathbf{G},\mathbf{G}'}$ are the eigenphases of the nested $P$-Wilson loop operator $\mathcal{W}_{2,\mathbf{k},\mathbf{G},\mathbf{G}'}$ satisfying 
\begin{equation}
	[\mathcal{I}]\mathcal{W}_{2,\mathbf{k},\mathbf{G},\mathbf{G}'} [\mathcal{I}]^{\dagger} = \mathcal{W}_{2,-\mathbf{k},\mathbf{G},\mathbf{G}'}^{\dagger}. \label{eq:I-constraint-on-nested-P-Wilson-loop-operator-1}
\end{equation}
There are two requirements in order for SEq.~\eqref{eq:I-constraint-on-nested-P-Wilson-loop-operator-1} to hold.
First, the Hamiltonian must be inversion-symmetric such that SEq.~\eqref{eq:inversion-act-on-P} holds, which implies that the $\widehat{\mathbf{G}}$-directed $P$-Wilson loop operator satisfies SEq.~\eqref{eq:general-notation-inversion-constraint-on-P-Wilson-loop-operator}.
Second, we must choose an inversion-symmetric grouping of isolated $N_W$ $\widehat{\mathbf{G}}$-directed $P$-Wannier bands in SEq.~\eqref{eq:P_nested_Wilson_loop_2nd_projector},  namely 
\begin{equation}
	[\mathcal{I}][\widetilde{P}_{\mathbf{G}}(\mathbf{k})][\mathcal{I}]^{\dagger} = [\widetilde{P}_{\mathbf{G}}(-\mathbf{k})], \label{eq:I-act-on-tilde-P-G-k}
\end{equation}
which implies that the chosen $N_W$ $\widehat{\mathbf{G}}$-directed $P$-Wannier bands satisfy
\begin{equation}
	\{ (\gamma_1)_{j,\mathbf{k},\mathbf{G}} | j=1\ldots N_{W}\} \text{ mod } 2\pi = \{ -(\gamma_1)_{j,-\mathbf{k},\mathbf{G}} | j=1\ldots N_{W}\} \text{ mod } 2\pi.
\end{equation}
Finally, since $(\gamma_2)_{j,\mathbf{k},\mathbf{G},\mathbf{G}'}$ is independent of the momentum component $\mathbf{k} \cdot \mathbf{a}'$, where $\mathbf{a}'$ is the primitive lattice vector dual to the primitive reciprocal lattice vector $\mathbf{G}'$, $\mathbf{k}_{TRIM}$ in SEq.~\eqref{eq:general_notation_gamma_2_I_constraint} should be interpreted as a $\mathbf{k}$-vector with $(\mathbf{k}_{TRIM}\cdot\mathbf{a})\mod \pi=0$ for each primitive lattice vector $\mathbf{a}\neq\mathbf{a'}$ .

\subsection{Antiunitary $\mathcal{T}$ Constraint on the $P$-Wilson Loop} \label{appendix:T-constraint-on-P-wilson}

We will next examine the constraints that antiunitary $\mathcal{T}$ symmetry places on the $P$-Wilson loop. 
For concreteness, we will specifically analyze the $\widehat{\mathbf{G}}_1$-directed $P$-Wilson loop operator $\mathcal{W}_{1,\mathbf{k},\mathbf{G}_1}$ with base point $(k_1,k_2,k_3)$ given in SEq.~\eqref{eq:def-W1}, although our results generalize straightforwardly to $P$-Wilson loops taken along any primitive reciprocal lattice direction. 
If the Hamiltonian has $\mathcal{T}$ symmetry, then the projection matrix $[P(\mathbf{k})]$ onto the occupied states defined in SEq.~(\ref{eq:def-Pk}) satisfies
\begin{align}
    [\mathcal{T}] [P(\mathbf{k})] [\mathcal{T}]^{-1} = [P(-\mathbf{k})], \label{eq:T-act-on-Pk}
\end{align}
where $[\mathcal{T}]$ is the antiunitary representative of $\mathcal{T}$ explicitly constructed in SN~\ref{sec:proof-TVGT-inverse}.
Note that $[P(\mathbf{k})]$ must project onto a set of energy bands that is mapped into itself under $\mathcal{T}$ in order for SEq.~(\ref{eq:T-act-on-Pk}) to be satisfied. 
This is guaranteed if $[P(\mathbf{k})]$ projects onto a set of gapped energy bands for a $\mathcal{T}$-symmetric Hamiltonian. 
Acting with $\mathcal{T}$ on the definition of $\mathcal{W}_{1,\mathbf{k},\mathbf{G}_1}$ [SEq.~\eqref{eq:def-W1}], we find
\begin{align}
\hspace{-0.5cm}
    [\mathcal{T}] \mathcal{W}_{1,\mathbf{k},\mathbf{G}_1} [\mathcal{T}]^{-1} & = [\mathcal{T}] [V(\mathbf{G}_1)] [P(k_{1}+2\pi,k_2,k_3)] \cdots [P(k_{1},k_2,k_3)] [\mathcal{T}]^{-1}\nonumber \\ 
    & = [\mathcal{T}] [V(\mathbf{G}_1)] [\mathcal{T}]^{-1} [\mathcal{T}] [P(k_{1}+2\pi,k_2,k_3)] [\mathcal{T}]^{-1} \cdots [\mathcal{T}] [P(k_{1},k_2,k_3)] [\mathcal{T}]^{-1} \nonumber\\
    & = [V(-\mathbf{G}_1)] [P(-k_{1}-2\pi,-k_2,-k_3)] \cdots  [P(-k_{1},-k_2,-k_3)]\nonumber  \\
    & = [V(-\mathbf{G}_1)] [V(-\mathbf{G}_1)]^{-1} [P(-k_{1},-k_2,-k_3)] [V(-\mathbf{G}_1)] \cdots  [V(-\mathbf{G}_1)]^{-1} [P(-k_{1}+2\pi,-k_2,-k_3)] [V(-\mathbf{G}_1)]\nonumber \\
    & =  [P(-k_{1},-k_2,-k_3)] \cdots  [P(-k_{1}+2\pi,-k_2,-k_3)] [V(-\mathbf{G}_1)] \nonumber\\
    & = \left( [V(-\mathbf{G}_1)]^{\dagger} [P(-k_{1}+2\pi,-k_2,-k_3)] \cdots [P(-k_{1},-k_2,-k_3)] \right)^{\dagger}\nonumber \\
    & = \left( [V(\mathbf{G}_1)] [P(-k_{1}+2\pi,-k_2,-k_3)] \cdots [P(-k_{1},-k_2,-k_3)] \right)^{\dagger}\nonumber \\
    & = \mathcal{W}_{1,-\mathbf{k},\mathbf{G}_1}^{\dagger}, \label{eq:T_transform_W1}
\end{align}
where we have made use of SEqs.~\eqref{eq:bc-of-u-nk}, \eqref{eq:T-act-on-Pk}, and \eqref{eq:after-arriving-conclusion-of-TVT-inv} (proved in SN~\ref{sec:proof-TVGT-inverse}). 
While this is formally equivalent to SEq.~\eqref{eq:I_transform_W1}, the fact that $\mathcal{T}$ is antiunitary implies that SEq.~\eqref{eq:T_transform_W1} places distinct constraints on the eigenvalues of $\mathcal{W}_{1,\mathbf{k},\mathbf{G}_1}$. 
In particular, suppose that $\ket{w_j(\mathbf{k})}$ is an eigenvector of $\mathcal{W}_{1,\mathbf{k},\mathbf{G}_1}$ with unimodular eigenvalue $e^{i\gamma_{1,j}(k_2,k_3)}$, such that
\begin{equation}
\mathcal{W}_{1,\mathbf{k},\mathbf{G}_1}\ket{w_j(\mathbf{k})} = e^{i\gamma_{1,j}(k_2,k_3)}\ket{w_j(\mathbf{k})}.\label{eq:wj-is-an-evec-of-w1}
\end{equation}
Acting with $\mathcal{T}$ on both sides of SEq.~\eqref{eq:wj-is-an-evec-of-w1} and using SEq.~\eqref{eq:T_transform_W1} we find that
\begin{align}
[\mathcal{T}]\mathcal{W}_{1,\mathbf{k},\mathbf{G}_1}\ket{w_j(\mathbf{k})} &= [\mathcal{T}]e^{i\gamma_{1,j}(k_2,k_3)}\ket{w_j(\mathbf{k})}, \label{eq:w1_t_evec_1}\\
[\mathcal{T}]\mathcal{W}_{1,\mathbf{k},\mathbf{G}_1}[\mathcal{T}]^{-1}[\mathcal{T}]\ket{w_j(\mathbf{k})} & = e^{-i\gamma_{1,j}(k_2,k_3)}[\mathcal{T}]\ket{w_j(\mathbf{k})}, \label{eq:w1_t_evec_2}\\
\mathcal{W}_{1,-\mathbf{k},\mathbf{G}_1}^\dag\left([\mathcal{T}]\ket{w_j(\mathbf{k})}\right) &= e^{-i\gamma_{1,j}(k_2,k_3)}\left([\mathcal{T}]\ket{w_j(\mathbf{k})}\right),\label{eq:w1_t_evec_3}
\end{align}
where in going from SEq.~\eqref{eq:w1_t_evec_1} to SEq.~\eqref{eq:w1_t_evec_2} we made use of the antiunitarity of $[\mathcal{T}]$. 
SEq.~\eqref{eq:w1_t_evec_3} shows that if $\ket{w_j(\mathbf{k})}$ is an eigenvector of $\mathcal{W}_{1,\mathbf{k},\mathbf{G}_1}$ with unimodular eigenvalue $e^{i\gamma_{1,j}(k_2,k_3)}$, then $[\mathcal{T}]\ket{w_j(\mathbf{k})}$ is an eigenvector of $\mathcal{W}_{1,-\mathbf{k},\mathbf{G}_1}^\dag$ with unimodular eigenvalue $e^{-i\gamma_{1,j}(k_2,k_3)}$.
Using the fact that the unimodular eigenvalues of $\mathcal{W}_{1,-\mathbf{k},\mathbf{G}_1}^{\dagger}$ are the complex conjugates of the unimodular eigenvalues of $\mathcal{W}_{1,-\mathbf{k},\mathbf{G}_1}$ (see SN~\ref{appendix:eigval_eigvec_W} for more details), we have that the set of unimodular eigenvalues of  $\mathcal{W}_{1,\mathbf{k},\mathbf{G}_1}$ satisfies 
\begin{equation}
    \{ e^{i\gamma_{1,j}(k_2,k_{3})} | j=1\ldots N_{\mathrm{occ}} \} = \{ e^{i\gamma_{1,j}(-k_2,-k_3)} | j = 1 \ldots N_{\mathrm{occ}} \}. \label{eq:T-constraint-on-exp-theta-1j}
\end{equation}
In terms of the set of eigenphases $\gamma_{1,j}(k_2,k_3)$, SEq.~\eqref{eq:T-constraint-on-exp-theta-1j} implies
\begin{equation}
    \{ \gamma_{1,j}(k_2,k_3) |j=1\ldots N_{\mathrm{occ}}\} \text{ mod } 2\pi = \{ \gamma_{1,j}(-k_2,-k_3) | j = 1 \ldots N_{\mathrm{occ}} \} \text{ mod } 2\pi. \label{eq:T-constraint-on-theta-1j}
\end{equation}
SEq.~\eqref{eq:T-constraint-on-theta-1j} indicates specifically that the $P$-Wilson loop eigenphase spectrum (\emph{i.e.} the  $P$-Wannier band structure) is invariant under a reversal of the sign of the crystal momentum $(k_2,k_3)\rightarrow(-k_2,-k_3)$ .
Recall from SN~\ref{sec:P_Wilson_loop} that the eigenphases [SEq.~\eqref{eq:T-constraint-on-theta-1j}] of the $P$-Wilson loop operator [SEq.~\eqref{eq:def-W1}] correspond to the localized positions of hybrid Wannier functions formed from the set of occupied states~\cite{alexandradinata2014wilsonloop,benalcazar2017electric,benalcazar2017quantized,lecture_notes_on_berry_phases_and_topology_bb,vanderbilt2018berry}. 
With this in mind, SEq.~\eqref{eq:T-constraint-on-theta-1j} then follows from the fact that time-reversal flips the momentum, but not the position, of a hybrid Wannier function.

We can also examine the $\mathcal{T}$ constraint on the sum over $j$ of the eigenphases $\gamma_{1,j}(k_2,k_3)$, as defined in SEq.~\eqref{eq:def-total-theta-1}.  
SEq.~\eqref{eq:T-constraint-on-theta-1j} indicates that 
\begin{equation}
    \gamma_{1}(k_2,k_3) \text{ mod } 2\pi  = \gamma_{1}(-k_2,-k_3) \text{ mod } 2\pi. \label{eq:T-constraint-on-total-theta-1}
\end{equation}
Notice that SEq.~(\ref{eq:T-constraint-on-total-theta-1}) {\it does not} lead to the quantization of $\gamma_{1}(k_2,k_3)$ at the four TRIMs $(k_2^{TRIM},k_3^{TRIM})=(0,0)$, $(\pi,0)$, $(0,\pi)$, and $(\pi,\pi)$.

As we mentioned at the start of this section (SN~\ref{appendix:symmetry-constraints-on-Wilson-loop}), the SEqs.~\eqref{eq:T_transform_W1}, \eqref{eq:T-constraint-on-exp-theta-1j}, \eqref{eq:T-constraint-on-theta-1j}, and \eqref{eq:T-constraint-on-total-theta-1} hold for both spinless and spinful systems. 
Specializing now to spinful systems with $\mathcal{T}^2=-1$, we have that SEqs.~\eqref{eq:wj-is-an-evec-of-w1}--\eqref{eq:w1_t_evec_3}, together with the related SEq.~\eqref{eq:w1_t_evec_4} explained at the beginning of the next section (SN~\ref{sec:T-constraint-on-nested-P}), imply that the eigenphases $\gamma_{1,j}(k_2,k_3)$ are twofold degenerate at the four TRIMs $(k_2^{TRIM},k_3^{TRIM})=(0,0)$, $(\pi,0)$, $(0,\pi)$, and $(\pi,\pi)$. 
This follows from the periodicity [SEq.~\eqref{eq:W-upon-G-translation-in-ky}] of the Wilson loop operator $\mathcal{W}_{1,\mathbf{k},\mathbf{G}_1}$ and the fact that since at TRIMs, namely at $\mathbf{k} = \mathbf{G}/2$ where $\mathbf{G}$ is a reciprocal lattice vector, $\ket{w_j(\mathbf{k})}$ and $[V(\mathbf{G})]^\dag[\mathcal{T}]\ket{w_j(\mathbf{k})}$ are linear combinations of the Bloch states $|u_{n,\mathbf{k}}\rangle$ at the same momentum (where the base point $k_1$, of which $\gamma_{1,j}(k_2,k_3)$ is independent, is taken to be $0$ or $\pi$). 
Additionally, by Kramers' theorem we find that $\ket{w_j(\mathbf{k})}$ and $[V(\mathbf{G})]^\dag[\mathcal{T}]\ket{w_j(\mathbf{k})}$ are orthogonal. 
Note that we have also used the property that if $[\mathcal{T}]$ is antiunitary and satisfies $[\mathcal{T}]^2 = -1$, then $[V(\mathbf{G})]^\dag [\mathcal{T}]$ is also antiunitary and satisfies $\left( [V(\mathbf{G})]^\dag [\mathcal{T}]\right)^2 = -1$, which can be deduced by using SEq.~\eqref{eq:after-arriving-conclusion-of-TVT-inv} proved in SN~\ref{sec:proof-TVGT-inverse}. 
Hence Kramers' theorem can be applied to deduce the orthogonality between $\ket{w_j(\mathbf{k})}$ and $[V(\mathbf{G})]^\dag[\mathcal{T}]\ket{w_j(\mathbf{k})}$ at the BZ boundary TRIMs $\mathbf{k} = \mathbf{G}/2$.

Since time-reversal symmetry treats all momentum components $k_i$ in the same way---it flips the sign of all the components $k_i$---the results of this section can be generalized to $\widehat{\mathbf{G}}_2$- and $\widehat{\mathbf{G}}_3$-directed $P$-Wilson loops. 
The results of SEqs.~\eqref{eq:T-constraint-on-exp-theta-1j}, \eqref{eq:T-constraint-on-theta-1j}, and \eqref{eq:T-constraint-on-total-theta-1}, written in the general notation in SN~\ref{sec:P_Wilson_loop}, therefore generalize to
\begin{align}
    & \{ e^{i (\gamma_1)_{j,\mathbf{k},\mathbf{G}}} | j = 1 \ldots N_{\mathrm{occ}}\} = \{ e^{i (\gamma_1)_{j,-\mathbf{k},\mathbf{G}}} | j = 1 \ldots N_{\mathrm{occ}} \}, \\
    &\{ (\gamma_1)_{j,\mathbf{k},\mathbf{G}} | j = 1 \ldots N_{\mathrm{occ}} \} \text{ mod } 2\pi = \{ (\gamma_1)_{j,-\mathbf{k},\mathbf{G}} | j = 1 \ldots N_{\mathrm{occ}} \} \text{ mod } 2\pi, \label{eq:general_notation_T_constraint_on_gamma_1jkG} \\
    & (\gamma_1)_{\mathbf{k},\mathbf{G}} \text{ mod } 2\pi= (\gamma_1)_{-\mathbf{k},\mathbf{G}} \text{ mod } 2\pi
\end{align}
where $(\gamma_1)_{\mathbf{k},\mathbf{G}}$ is defined in SEq.~\eqref{eq:summed-gamma-1kG-general-notation},
and $(\gamma_1)_{j,\mathbf{k},\mathbf{G}}$ are the eigenphases of the $\widehat{\mathbf{G}}$-directed $P$-Wilson loop operator $\mathcal{W}_{1,\mathbf{k},\mathbf{G}}$ satisfying 
\begin{equation}
    [\mathcal{T}]\mathcal{W}_{1,\mathbf{k},\mathbf{G}} [\mathcal{T}]^{-1} = \mathcal{W}_{1,-\mathbf{k},\mathbf{G}}^{\dagger}. \label{eq:general-notation-T-constraint-on-P-Wilson-loop-operator}
\end{equation}
SEq.~\eqref{eq:general-notation-T-constraint-on-P-Wilson-loop-operator} is true provided that the matrix projector onto the occupied energy bands is time-reversal-symmetric as specified in SEq.~\eqref{eq:T-act-on-Pk}.
Finally, since $(\gamma_1)_{j,\mathbf{k},\mathbf{G}}$ is independent of the momentum component $\mathbf{k}\cdot\mathbf{a}$ (where $\mathbf{a}$ is the primitive lattice vector dual to $\mathbf{G}$), Kramers' theorem implies that if $\mathcal{T}^2=-1$ then the eigenphases $(\gamma_1)_{j,\mathbf{k},\mathbf{G}}$ are twofold degenerate at TRIMs $\mathbf{k}_{TRIM}$ satisfying $(\mathbf{k}_{TRIM}\cdot\mathbf{a'})\mod \pi = 0$ for each primitive lattice vector $\mathbf{a'}\neq\mathbf{a}$. 

\subsection{Antiunitary $\mathcal{T}$ Constraint on the Nested $P$-Wilson Loop} \label{sec:T-constraint-on-nested-P}

We next investigate the constraints that antiunitary $\mathcal{T}$ symmetry places on the nested $P$-Wilson loop. 
To begin, we consider the eigenvector $| w_{j}(\mathbf{k}) \rangle$ of the $\widehat{\mathbf{G}}_1$-directed $P$-Wilson loop operator $\mathcal{W}_{1,\mathbf{k},\mathbf{G}_1}$ with eigenvalue $e^{i\gamma_{1,j}(k_2,k_3)}$ as shown in SEq.~\eqref{eq:nu-j-as-eigvector-of-W1}. 
From SEqs.~\eqref{eq:w1_t_evec_1}--\eqref{eq:w1_t_evec_3} and the surrounding discussion, we know that $[\mathcal{T}] | w_{j}(\mathbf{k}) \rangle$ is an eigenvector of $\mathcal{W}_{1,-\mathbf{k},\mathbf{G}_1}$ with eigenvalue $e^{i\gamma_{1,j}(k_2,k_3)}$, namely
\begin{equation}
    \mathcal{W}_{1,-\mathbf{k},\mathbf{G}_1} [\mathcal{T}] | w_{j}(\mathbf{k}) \rangle = e^{i\gamma_{1,j}(k_2,k_3)} [\mathcal{T}] | w_{j}(\mathbf{k}) \rangle. \label{eq:w1_t_evec_4}
\end{equation}
According to SEqs.~\eqref{eq:wj-is-an-evec-of-w1} and \eqref{eq:w1_t_evec_4}, there is then a one-to-one correspondence between the eigenvectors of $\mathcal{W}_{1,\mathbf{k},\mathbf{G}_1}$ and $\mathcal{W}_{1,-\mathbf{k},\mathbf{G}_1}$ with unimodular eigenvalues. 
In particular, if we choose a grouping of $N_{W}$ $P$-Wannier bands [denoted as in SEq.~\eqref{eq:grouping_of_band_eqn}] that respects the time-reversal symmetry constraint 
\begin{equation}
    [\mathcal{T}] [\widetilde{P}(\mathbf{k})] [\mathcal{T}]^{-1} = [\widetilde{P}(-\mathbf{k})], \label{eq:T-act-on-tilde-P}
\end{equation}
where $[\widetilde{P}(\mathbf{k})]$ [defined as in SEq.~\eqref{eq:proj-Wannier-basis}] is the $N_{\mathrm{sta}}\times N_{\mathrm{sta}}$ matrix projector onto the $N_{W}$ Wannier band eigenstates $| w_{j}(\mathbf{k}) \rangle$, then such a grouping of $N_W$ Wannier bands must satisfy
\begin{equation}
    \{ \gamma_{1,j}(k_2,k_3) | j=1\ldots N_W \} \text{ mod } 2\pi = \{ \gamma_{1,j}(-k_2,-k_3) | j=1\ldots N_W \} \text{ mod }2\pi. \label{eq:NWbandsTsymmetric}
\end{equation}
In this work we will always choose $[\widetilde{P}(\mathbf{k})]$ for time-reversal-symmetric systems in a manner that satisfies SEqs.~\eqref{eq:T-act-on-tilde-P} and \eqref{eq:NWbandsTsymmetric}.

We now consider the $N_{\mathrm{sta}}\times N_{\mathrm{sta}}$ nested $P$-Wilson loop operator $\mathcal{W}_{2,\mathbf{k},\mathbf{G}_1,\mathbf{G}_2}$ given in SEq.~\eqref{eq:def-nested-P-wilson-loop}. 
Acting with $\mathcal{T}$, we find that
\begin{align}
\hspace{-0.5cm}
    [\mathcal{T}] \mathcal{W}_{2,\mathbf{k},\mathbf{G}_1,\mathbf{G}_2} [\mathcal{T}]^{-1} & = [\mathcal{T}] [V(\mathbf{G}_2)] [\widetilde{P}(k_{1},k_{2} + 2\pi,k_3)] \cdots [\widetilde{P}(k_{1},k_{2},k_3)] [\mathcal{T}]^{-1} \nonumber\\
    & = [\mathcal{T}] [V(\mathbf{G}_2)] [\mathcal{T}]^{-1} [\mathcal{T}] [\widetilde{P}(k_{1},k_{2} + 2\pi,k_3)] [\mathcal{T}]^{-1} \cdots [\mathcal{T}][\widetilde{P}(k_{1},k_{2},k_3)] [\mathcal{T}]^{-1}\nonumber \\
    & = [V(-\mathbf{G}_2)]  [\widetilde{P}(-k_{1},-k_{2} - 2\pi,-k_3)]  \cdots [\widetilde{P}(-k_{1},-k_{2},-k_3)]\nonumber \\
    & = [V(-\mathbf{G}_2)] [V(-\mathbf{G}_2)]^{-1} [\widetilde{P}(-k_{1},-k_{2},-k_3)] [V(-\mathbf{G}_2)]  \cdots [V(-\mathbf{G}_2)]^{-1} [\widetilde{P}(-k_{1},-k_{2}+2\pi,-k_3)] [V(-\mathbf{G}_2)]\nonumber \\
    & = [\widetilde{P}(-k_{1},-k_{2},-k_3)]  \cdots  [\widetilde{P}(-k_{1},-k_{2}+2\pi,-k_3)] [V(-\mathbf{G}_2)]\nonumber \\
    & = \left( [V(-\mathbf{G}_2 )]^{\dagger}[\widetilde{P}(-k_{1},-k_{2}+2\pi,-k_3)] \cdots [\widetilde{P}(-k_{1},-k_{2},-k_3)]  \right)^{\dagger} \nonumber\\
    & = \left( [V(\mathbf{G}_2)] [\widetilde{P}(-k_{1},-k_{2}+2\pi,-k_3)] \cdots [\widetilde{P}(-k_{1},-k_{2},-k_3)]  \right)^{\dagger}\nonumber \\
    & = {\mathcal{W}}_{2,-\mathbf{k},\mathbf{G}_1,\mathbf{G}_2}^{\dagger}, \label{eq:tilde-W-T-transform}
\end{align}
where we have made use of SEqs.~\eqref{eq:def-VG}, \eqref{eq:proj-Wannier-basis}, \eqref{eq:bc-nu-j}, \eqref{eq:T-act-on-tilde-P}, and \eqref{eq:after-arriving-conclusion-of-TVT-inv} (proved in SN~\ref{sec:proof-TVGT-inverse}). 
Following the same logic we used in SEqs.~\eqref{eq:T_transform_W1}--\eqref{eq:T-constraint-on-exp-theta-1j}, SEq.~\eqref{eq:tilde-W-T-transform} implies that the set of unimodular eigenvalues $e^{i\gamma_{2,j}(k_{1},k_3)} $ of $\mathcal{W}_{2,\mathbf{k},\mathbf{G}_1,\mathbf{G}_2}$ satisfies
\begin{equation}
    \{ e^{i\gamma_{2,j}(k_{1},k_3)}| j=1 \cdots N_W  \} = \{ e^{i\gamma_{2,j}(-k_{1},-k_3)}| j=1 \cdots N_W  \}\label{eq:T-constraint-on-exp-theta-2j}
\end{equation}
due to $\mathcal{T}$ symmetry. 
In terms of the set of eigenphases $\gamma_{2,j}(k_{1},k_3)$, SEq.~\eqref{eq:T-constraint-on-exp-theta-2j} implies that
\begin{equation}
    \{ \gamma_{2,j}(k_{1},k_3)| j=1 \cdots N_W  \}\text{ mod } 2\pi = \{ \gamma_{2,j}(-k_{1},-k_3)| j=1 \cdots N_W  \} \text{ mod } 2\pi. \label{eq:T-constraint-on-theta-2j}
\end{equation}
SEq.~\eqref{eq:T-constraint-on-theta-2j} specifically implies that the nested $P$-Wilson loop eigenphase spectrum, \emph{i.e.} the nested $P$-Wannier band structure, is invariant under a reversal of the sign of the crystal momentum $(k_1,k_3)\rightarrow(-k_1,-k_3)$.
Recall from SN~\ref{sec:nested_P_Wilson_loop} that the eigenphases in SEq.~\eqref{eq:T-constraint-on-theta-2j} of the nested $P$-Wilson loop operator [SEq.~\eqref{eq:def-nested-P-wilson-loop}] correspond to the localized positions of the hybrid Wannier functions formed from a group of $P$-Wannier bands [denoted as in SEq.~\eqref{eq:grouping_of_band_eqn}]~\cite{benalcazar2017electric,benalcazar2017quantized}. 
With this in mind, we can understand SEq.~\eqref{eq:T-constraint-on-theta-2j} as a consequence of the fact that time-reversal flips the momentum, but not the position, of a hybrid Wannier function.

We can also examine the $\mathcal{T}$ constraint on the sum over $j$ of the eigenphases $\gamma_{2,j}(k_{1},k_3)$, as defined in SEq.~(\ref{eq:def-total-theta-2}). 
SEq.~(\ref{eq:T-constraint-on-theta-2j}) indicates that
\begin{equation}
    \gamma_{2}(k_{1},k_3)\text{ mod } 2\pi =  \gamma_{2}(-k_{1},-k_3) \text{ mod } 2\pi. \label{eq:T-constraint-on-total-theta-2}
\end{equation}
Notice that SEq.~(\ref{eq:T-constraint-on-total-theta-2}) {\it does not} lead to the quantization of $\gamma_{2}(k_{1},k_3)$ at the four TRIMs $(k_{1}^{TRIM},k_3^{TRIM})=(0,0)$, $(\pi,0)$, $(0,\pi)$, and $(\pi,\pi)$.

As we mentioned at the start of this section (SN~\ref{appendix:symmetry-constraints-on-Wilson-loop}), the SEqs.~\eqref{eq:tilde-W-T-transform}, \eqref{eq:T-constraint-on-exp-theta-2j}, \eqref{eq:T-constraint-on-theta-2j}, and \eqref{eq:T-constraint-on-total-theta-2} hold for both spinless and spinful systems. 
Specializing now to spinful systems with $\mathcal{T}^2=-1$, using SEq.~\eqref{eq:tilde-W-T-transform} and following the same logic as SEqs.~\eqref{eq:wj-is-an-evec-of-w1}--\eqref{eq:w1_t_evec_3} and \eqref{eq:w1_t_evec_4}, 
we have that 
\begin{align}
    \mathcal{W}_{2,\mathbf{k},\mathbf{G}_1,\mathbf{G}_2} |w_{2,j}(\mathbf{k})\rangle &= e^{i\gamma_{2,j}(k_1,k_3)} |w_{2,j}(\mathbf{k})\rangle, \label{eq:w2_t_evec_1} \\
    \mathcal{W}_{2,-\mathbf{k},\mathbf{G}_1,\mathbf{G}_2} [\mathcal{T}] |w_{2,j}(\mathbf{k})\rangle &= e^{i\gamma_{2,j}(k_1,k_3)} [\mathcal{T}] |w_{2,j}(\mathbf{k})\rangle, \label{eq:w2_t_evec_2}
\end{align}
where $|w_{2,j}(\mathbf{k})\rangle$ is the eigenstate of $\mathcal{W}_{2,\mathbf{k},\mathbf{G}_1,\mathbf{G}_2}$ with unimodular eigenvalue $e^{i\gamma_{2,j}(k_1,k_3)}$.
As in SN~\ref{appendix:T-constraint-on-P-wilson}, SEqs.~\eqref{eq:w2_t_evec_1} and \eqref{eq:w2_t_evec_2} imply that the eigenphases $\gamma_{2,j}(k_1,k_3)$ are twofold degenerate at the four TRIMs $(k_1^{TRIM},k_3^{TRIM})=(0,0)$, $(\pi,0)$, $(0,\pi)$, and $(\pi,\pi)$. 
This follows from the periodicity [SEq.~\eqref{eq:W2-upon-G-translation-in-ky}] of the nested Wilson loop and the fact that since at TRIMs, namely at $\mathbf{k} = \mathbf{G}/2$ where $\mathbf{G}$ is a reciprocal lattice vector, the eigenvectors $\ket{w_{2,j}(\mathbf{k})}$ and $[V(\mathbf{G})]^\dag[\mathcal{T}]\ket{w_{2,j}(\mathbf{k})}$ of $\mathcal{W}_{2,\mathbf{k},\mathbf{G}_1,\mathbf{G}_2}$ are linear combinations of Bloch states $|u_{n,\mathbf{k}}\rangle$ at the same momentum (where the base point $k_2$, of which $\gamma_{2,j}(k_1,k_3)$ is independent, is taken to be $0$ or $\pi$). 
Additionally, by Kramers' theorem we find that $\ket{w_{2,j}(\mathbf{k})}$ and $[V(\mathbf{G})]^\dag[\mathcal{T}]\ket{w_{2,j}(\mathbf{k})}$ are orthogonal.

Since time-reversal symmetry treats all momentum components $k_i$ in the same way---it flips the sign of all the components $k_i$---the results of this section can be generalized to the nested $P$-Wilson loop for any primitive reciprocal lattice vectors $\mathbf{G}$ for the direction of the first loop and $\mathbf{G}'$ for the direction of the second loop.
The results of SEqs.~\eqref{eq:T-constraint-on-exp-theta-2j}, \eqref{eq:T-constraint-on-theta-2j}, and \eqref{eq:T-constraint-on-total-theta-2}, written in the general notation in SN~\ref{sec:nested_P_Wilson_loop}, therefore generalize to
\begin{align}
    & \{ e^{i (\gamma_2)_{j,\mathbf{k},\mathbf{G},\mathbf{G'}}} | j = 1 \ldots N_{W}\} = \{ e^{i (\gamma_2)_{j,-\mathbf{k},\mathbf{G},\mathbf{G}'}} | j = 1 \ldots N_{W} \}, \\
    &\{ (\gamma_2)_{j,\mathbf{k},\mathbf{G},\mathbf{G'}} | j = 1 \ldots N_{W} \} \text{ mod } 2\pi = \{ (\gamma_2)_{j,-\mathbf{k},\mathbf{G},\mathbf{G'}} | j = 1 \ldots N_{W} \} \text{ mod } 2\pi, \label{eq:general_notation_T_constraint_on_gamma_2jkGGp} \\
    & (\gamma_2)_{\mathbf{k},\mathbf{G},\mathbf{G}'} \text{ mod } 2\pi = (\gamma_2)_{-\mathbf{k},\mathbf{G},\mathbf{G}'} \text{ mod } 2\pi,
\end{align}
where $(\gamma_2)_{\mathbf{k},\mathbf{G},\mathbf{G}'}$ is defined in SEq.~\eqref{eq:summed-gamma-2-def-general},
and $(\gamma_2)_{j,\mathbf{k},\mathbf{G},\mathbf{G'}}$ are the eigenphases of the nested $P$-Wilson loop operator $\mathcal{W}_{2,\mathbf{k},\mathbf{G},\mathbf{G}'}$ satisfying 
\begin{equation}
    [\mathcal{T}]\mathcal{W}_{2,\mathbf{k},\mathbf{G},\mathbf{G}'} [\mathcal{T}]^{-1} = \mathcal{W}_{2,-\mathbf{k},\mathbf{G},\mathbf{G}'}^{\dagger}. \label{eq:general-notation-T-constraint-on-nested-P-Wilson-loop-operator}
\end{equation}
SEq.~\eqref{eq:general-notation-T-constraint-on-nested-P-Wilson-loop-operator} is true provided that the projector $[\widetilde{P}_{\mathbf{G}}(\mathbf{k})]$ [SEq.~\eqref{eq:P_nested_Wilson_loop_2nd_projector}] onto the isolated grouping of $N_W$ $\widehat{\mathbf{G}}$-directed $P$-Wannier bands satisfies the time-reversal-symmetric constraint
\begin{equation}
    [\mathcal{T}][\widetilde{P}_{\mathbf{G}}(\mathbf{k})][\mathcal{T}]^{-1} = [\widetilde{P}_{\mathbf{G}}(-\mathbf{k})]. \label{eq:T-act-on-tilde-P-G-k}
\end{equation}
Finally, since $(\gamma_2)_{j,\mathbf{k},\mathbf{G},\mathbf{G'}}$ is independent of the momentum component $\mathbf{k}\cdot\mathbf{a'}$ (where $\mathbf{a'}$ is the primitive lattice vector dual to $\mathbf{G'}$), Kramers' theorem implies that if $\mathcal{T}^2=-1$ then the eigenphases $(\gamma_2)_{j,\mathbf{k},\mathbf{G},\mathbf{G'}}$ are twofold degenerate at TRIMs $\mathbf{k}_{TRIM}$ satisfying $(\mathbf{k}_{TRIM}\cdot\mathbf{a})\mod \pi = 0$ for each primitive lattice vector $\mathbf{a}\neq\mathbf{a'}$.

\subsection{Unitary $\mathcal{I}$ Constraint on the $P_{\pm}$-Wilson Loop} 
\label{appendix:I-constraint-on-P-pm}

In this section, we will extend the analysis of SN~\ref{appendix:I-constraint-on-P-wilson} to investigate the constraints that unitary $\mathcal{I}$ places on the $P_{\pm}$-Wilson loop (SN~\ref{sec:P_pm_Wilson_loop}) for a system with spin-$1/2$ degrees of freedom.
We will make the physically relevant assumption that since spin is a $\mathcal{T}$-odd pseudovector, inversion does not act on spin, implying that
\begin{equation}
	[\mathcal{I}] s = s [\mathcal{I}], \label{eq:s-invariant-upon-I}
\end{equation}
where $s = \mathbf{s} \cdot \hat{\mathbf{n}}$ is the spin operator along the direction $\hat{\mathbf{n}}$ and $\mathbf{s} = (s_x,s_y,s_z)$ is given in SEq.~\eqref{eq:appendix-def-s-i}.
Acting with inversion on the projected spin operator $[s(\mathbf{k})] \equiv [P(\mathbf{k})]s[P(\mathbf{k})]$ [SEq.~\eqref{eq:appendix-def-PsP}], we have that for an inversion-symmetric system 
\begin{align}
	[\mathcal{I}] [s(\mathbf{k})] [\mathcal{I}]^{\dagger} &= [\mathcal{I}] [P(\mathbf{k})] [\mathcal{I}]^{\dagger} [\mathcal{I}] s [\mathcal{I}]^{\dagger} [\mathcal{I}] [P(\mathbf{k})] [\mathcal{I}]^{\dagger}\nonumber \\
    & =  [P(-\mathbf{k})] [\mathcal{I}] s [\mathcal{I}]^{\dagger} [P(-\mathbf{k})]\nonumber \\ 
    & = [P(-\mathbf{k})] s [P(-\mathbf{k})]\nonumber \\
    & = [s(-\mathbf{k})], \label{eq:IsIdag}
\end{align}
where we have used SEqs.~\eqref{eq:inversion-act-on-P}, \eqref{eq:s-invariant-upon-I}, and the unitarity of $[\mathcal{I}]$.
Next, we assume that the spin bands, defined as the eigenvalues of $[s(\mathbf{k})]$ with eigenvectors in the image of $[P(\mathbf{k})]$, can be separated into disjoint upper ($+$) and lower ($-$) groupings.
The corresponding eigenvalue equation for $[s(\mathbf{k})]$ is then
\begin{equation}
	[s(\mathbf{k})] | u^{\pm}_{n,\mathbf{k}} \rangle = \lambda^{\pm}_{n,\mathbf{k}} | u^{\pm}_{n,\mathbf{k}} \rangle, \label{eq:eigvaleqnsk}
\end{equation}
where the eigenstates $| u^{\pm}_{n,\mathbf{k}} \rangle \in \mathrm{Image}([P(\mathbf{k})])$ of $[s(\mathbf{k})]$ satisfy the boundary condition in SEq.~\eqref{eq:P_pm_Wilson_loop_BC_for_u_pm}.
We notice that, with only inversion symmetry, the numbers $N_{\mathrm{occ}}^\pm$ of upper and lower spin bands are not necessarily the same.
For instance, if we consider an inversion-symmetric magnetic system where the electronic ground state contains states with spins nearly aligned along $\hat{\mathbf{n}}$, then the upper and lower spin bands can both have $\lambda^\pm_{n\mathbf{k}}>0$. 
However, we always have that $N_{\mathrm{occ}}^+ + N_{\mathrm{occ}}^- = N_{\mathrm{occ}}$, the number of occupied electronic energy bands.
Using SEqs.~\eqref{eq:IsIdag}, \eqref{eq:eigvaleqnsk}, and the assumption that the upper and lower spin bands are disjoint from each other, we can deduce that 
\begin{equation}
	[\mathcal{I}] [P_{\pm}(\mathbf{k})] [\mathcal{I}]^\dag = [P_{\pm}(-\mathbf{k})], \label{eq:I-act-on-P-pm}
\end{equation}
where $[P_{\pm}(\mathbf{k})]$ [defined in SEq.~\eqref{eq:P_pm_k_projector}]
is the projector onto the upper/lower spin bands.
From SEq.~\eqref{eq:I-act-on-P-pm}, we see that the upper and lower spin bands transform under inversion independently of each other, which is a consequence of SEq.~\eqref{eq:s-invariant-upon-I}.
Using the projector $[P_{\pm}(\mathbf{k})]$ onto the upper/lower spin bands, the $\widehat{\mathbf{G}}$-directed $P_{\pm}$-Wilson loop operator $\mathcal{W}^\pm_{1,\mathbf{k},\mathbf{G}}$ can be constructed as 
\begin{equation}
    \mathcal{W}^\pm_{1,\mathbf{k},\mathbf{G}} = [V(\mathbf{G})] \lim_{N\to \infty} \left( [P_{\pm}(\mathbf{k}+\mathbf{G})][P_{\pm}(\mathbf{k}+\frac{N-1}{N}\mathbf{G})] \cdots [P_{\pm}(\mathbf{k}+\frac{1}{N}\mathbf{G})] [P_{\pm}(\mathbf{k})]\right) \label{eq:def-W1-pm}
\end{equation}
along the closed loop parallel to $\mathbf{G}$ analogous to SEq.~\eqref{eq:def-W1}.
$\mathcal{W}^\pm_{1,\mathbf{k},\mathbf{G}}$ has $N_{\mathrm{sta}}-N_{\mathrm{occ}}^\pm$ zero eigenvalues corresponding to the number of unoccupied eigenstates. 
More importantly, $\mathcal{W}^\pm_{1,\mathbf{k},\mathbf{G}}$ has $N_{\mathrm{occ}}^\pm$ unimodular eigenvalues that are independent of the momentum component $\mathbf{k} \cdot \mathbf{a}$ where $\mathbf{a}$ is the real-space primitive lattice vector dual to the primitive reciprocal lattice vector $\mathbf{G}$~\cite{alexandradinata2014wilsonloop,vanderbilt2018berry,lecture_notes_on_berry_phases_and_topology_bb}. 
Using SEqs.~\eqref{eq:I-act-on-P-pm} and \eqref{eq:def-W1-pm}, and following the same logic as SEq.~\eqref{eq:I_transform_W1}, we have that $\mathcal{W}^\pm_{1,\mathbf{k},\mathbf{G}}$ transforms under inversion according to
\begin{equation}
	[\mathcal{I}] \mathcal{W}^\pm_{1,\mathbf{k},\mathbf{G}} [\mathcal{I}]^\dag = (\mathcal{W}^\pm_{1,-\mathbf{k},\mathbf{G}})^\dag. \label{eq:I-act-on-W-1pm}
\end{equation}
As in SN~\ref{appendix:I-constraint-on-P-wilson}, SEq.~\eqref{eq:I-act-on-W-1pm} implies the following $\mathcal{I}$-symmetry constraints on the $P_\pm$-Wilson loop eigenphases:
\begin{align}
	& \{ e^{i (\gamma_1^\pm)_{j,\mathbf{k},\mathbf{G}}} | j = 1 \ldots N_{\mathrm{occ}}^\pm\} = \{ e^{-i (\gamma_1^\pm)_{j,-\mathbf{k},\mathbf{G}}} | j = 1 \ldots N_{\mathrm{occ}}^\pm \}, \\
	&\{ (\gamma_1^\pm)_{j,\mathbf{k},\mathbf{G}} | j = 1 \ldots N_{\mathrm{occ}}^\pm \} \text{ mod } 2\pi = \{ -(\gamma_1^\pm)_{j,-\mathbf{k},\mathbf{G}} | j = 1 \ldots N_{\mathrm{occ}}^\pm \} \text{ mod } 2\pi, \label{eq:I_constraints_on_gamma_1pm_j}  \\
	&(\gamma_1^\pm)_{\mathbf{k},\mathbf{G}} \text{ mod } 2\pi = -(\gamma_1^\pm)_{-\mathbf{k},\mathbf{G}} \text{ mod } 2\pi, \\
	& \left( (\gamma_1^\pm)_{\mathbf{k}_{TRIM},\mathbf{G}} \mod \pi \right) = 0, \label{eq:general_notation_gamma_1_pm_I_constraint}
\end{align}
where 
\begin{equation}
    (\gamma_1^\pm)_{\mathbf{k},\mathbf{G}} \equiv \sum_{j=1}^{N_{\mathrm{occ}}^\pm}(\gamma_1^\pm)_{j,\mathbf{k},\mathbf{G}} \text{ mod } 2\pi, \label{eq:summed-gamma-1pmkG-general-notation}
\end{equation}
and $(\gamma_1^\pm)_{j,\mathbf{k},\mathbf{G}}$ are the eigenphases ($P_\pm$-Wannier bands) of the $\widehat{\mathbf{G}}$-directed $P_\pm$-Wilson loop operator $\mathcal{W}_{1,\mathbf{k},\mathbf{G}}^\pm$ satisfying SEq.~\eqref{eq:I-act-on-W-1pm}.
Recall from SN~\ref{sec:P_pm_Wilson_loop} that the eigenphases $(\gamma_1^\pm)_{j,\mathbf{k},\mathbf{G}}$ of the $P_\pm$-Wilson loop operator [SEq.~\eqref{eq:def-W1-pm}] correspond to the localized positions of spin-resolved hybrid Wannier functions formed from the set of upper/lower spin bands~\cite{alexandradinata2014wilsonloop,benalcazar2017electric,benalcazar2017quantized,lecture_notes_on_berry_phases_and_topology_bb,vanderbilt2018berry}. 
SEq.~\eqref{eq:I_constraints_on_gamma_1pm_j} is then a consequence of the fact that inversion flips both the position and momentum, but does not flip the spin, of a spin-resolved hybrid Wannier function.
Since $(\gamma_1^\pm)_{j,\mathbf{k},\mathbf{G}}$ is independent of the momentum component $\mathbf{k} \cdot \mathbf{a}$, where $\mathbf{a}$ is the real-space primitive lattice vector dual to the primitive reciprocal lattice vector $\mathbf{G}$, $\mathbf{k}_{TRIM}$  in SEq.~\eqref{eq:general_notation_gamma_1_pm_I_constraint} should be interpreted as a $\mathbf{k}$-vector with $(\mathbf{k}_{TRIM} \cdot \mathbf{a'}) \mod \pi = 0$ for each primitive lattice vector $\mathbf{a'}\neq\mathbf{a}$.

\subsection{Unitary $\mathcal{I}$ Constraint on the Nested $P_{\pm}$-Wilson Loop} 
\label{appendix:I-constraint-on-nested-P-pm}

Following SN~\ref{appendix:I-constraint-on-P-pm},  we will in this section extend the analysis in SN~\ref{appendix:I-constraint-on-nested-P-wilson} to derive the constraints that unitary $\mathcal{I}$ places on the nested $P_{\pm}$-Wilson loop (SN~\ref{sec:nested_P_pm_Wilson_loop}) for a system with spin-$1/2$ degrees of freedom.
Since inversion does not act on spin [see SEq.~\eqref{eq:s-invariant-upon-I} and the surrounding text], then the analysis for the nested $P_{+}$- and nested $P_{-}$-Wilson loops are independent of each other.
Suppose that the $\widehat{\mathbf{G}}$-directed $P_\pm$-Wannier bands ($\{ (\gamma_1^\pm)_{j,\mathbf{k},\mathbf{G}} | j=1\ldots N_{\mathrm{occ}}^\pm \}$) can be separated into disjoint groupings and that we choose an isolated grouping of $N_{W}^\pm$ bands described by $\{ (\gamma_1^\pm)_{j,\mathbf{k},\mathbf{G}} | j=1\ldots N_{W}^\pm \}$ (where $N_{W}^\pm \leq N_{\mathrm{occ}}^\pm$) onto which the operator $[\widetilde{P}^\pm_{\mathbf{G}}(\mathbf{k})]$ constructed in SEq.~\eqref{eq:P_pm_nested_Wilson_loop_2nd_projector} projects.
We now follow the same logic as SN~\ref{appendix:I-constraint-on-nested-P-wilson} and focus on either the $P_+$- or $P_{-}$-Wannier bands. 
For an inversion-symmetric system, the projector $[\widetilde{P}^\pm_{\mathbf{G}}(\mathbf{k})]$ onto an isolated grouping of $N_W^\pm$ $\widehat{\mathbf{G}}$-directed $P_\pm$-Wannier bands centered around an inversion-invariant eigenphase $\gamma_1^\pm \mod \pi =0$ can be chosen to satisfy
\begin{equation}
	[\mathcal{I}] [\widetilde{P}^\pm_{\mathbf{G}}(\mathbf{k})] [\mathcal{I}]^{\dagger} = [\widetilde{P}^\pm_{\mathbf{G}}(-\mathbf{k})]. \label{eq:inversion-act-on-tilde-P-pm}
\end{equation}
SEqs.~\eqref{eq:I-act-on-W-1pm} and \eqref{eq:inversion-act-on-tilde-P-pm} imply that the isolated grouping of $N_W^\pm$ $\widehat{\mathbf{G}}$-directed $P_\pm$-Wilson loop eigenphases satisfies
\begin{equation}
	\{ (\gamma_1^\pm)_{j,\mathbf{k},\mathbf{G}} | j=1\ldots N_{W}^\pm \} \text{ mod } 2\pi = \{ -(\gamma_1^\pm)_{j,-\mathbf{k},\mathbf{G}} | j=1\ldots N_{W}^\pm \} \text{ mod } 2\pi. \label{eq:NWpmbandsparticleholesymmetric}
\end{equation}
Similar to the effective particle-hole symmetry of the entire $P_\pm$-Wannier band structure described in SEq.~\eqref{eq:I_constraints_on_gamma_1pm_j}, the isolated grouping of $N_W^\pm$ $P_\pm$-Wannier satisfying SEq.~\eqref{eq:inversion-act-on-tilde-P-pm} also has an effective particle-hole symmetry described by SEq.~\eqref{eq:NWpmbandsparticleholesymmetric}.
From the matrix projector $[\widetilde{P}^\pm_{\mathbf{G}}(\mathbf{k})]$, we can construct the $\widetilde{P}^\pm_{\mathbf{G}}$-Wilson loop operator along a closed loop parallel to $\mathbf{G}'$ as
\begin{equation}
    \mathcal{W}^\pm_{2,\mathbf{k},\mathbf{G},\mathbf{G}'} = [V(\mathbf{G}')] \lim_{N\to \infty} \left( [\widetilde{P}^{\pm}_{\mathbf{G}}(\mathbf{k}+\mathbf{G}')][\widetilde{P}^{\pm}_{\mathbf{G}}(\mathbf{k}+\frac{N-1}{N}\mathbf{G}')] \cdots [\widetilde{P}^{\pm}_{\mathbf{G}}(\mathbf{k}+\frac{1}{N}\mathbf{G}')] [\widetilde{P}^{\pm}_{\mathbf{G}}(\mathbf{k})]\right). \label{eq:def-W2-pm}
\end{equation}
$\mathcal{W}^\pm_{2,\mathbf{k},\mathbf{G},\mathbf{G}'}$ in SEq.~\eqref{eq:def-W2-pm} is the nested $P_{\pm}$-Wilson loop operator with the first and second closed loop parallel to $\mathbf{G}$ and $\mathbf{G}'$ respectively.
$\mathcal{W}^\pm_{2,\mathbf{k},\mathbf{G},\mathbf{G}'}$ has $N_{\mathrm{sta}}-N_{W}^\pm$ zero eigenvalues corresponding to the number of states annihilated by $\widetilde{P}^\pm_{\mathbf{G}}(\mathbf{k})$.  
More importantly, $\mathcal{W}^\pm_{2,\mathbf{k},\mathbf{G},\mathbf{G}'}$ has $N_{W}^\pm$ unimodular eigenvalues that are independent of the momentum component $\mathbf{k} \cdot \mathbf{a}'$ where $\mathbf{a}'$ is the real-space primitive lattice vector dual to the primitive reciprocal lattice vector $\mathbf{G}'$~\cite{alexandradinata2014wilsonloop,vanderbilt2018berry,lecture_notes_on_berry_phases_and_topology_bb}. 

Using SEqs.~\eqref{eq:inversion-act-on-tilde-P-pm} and \eqref{eq:def-W2-pm}, and following the same logic as SEq.~\eqref{eq:W-2-k-G1-G2-I-transform}, we have that $\mathcal{W}^\pm_{2,\mathbf{k},\mathbf{G},\mathbf{G}'}$ transforms under inversion according to
\begin{equation}
	[\mathcal{I}] \mathcal{W}^\pm_{2,\mathbf{k},\mathbf{G},\mathbf{G}'} [\mathcal{I}]^\dag = (\mathcal{W}^\pm_{2,-\mathbf{k},\mathbf{G},\mathbf{G}'})^\dag. \label{eq:W-2-pm-k-G1-G2-I-transform}
\end{equation}
As in SN~\ref{appendix:I-constraint-on-nested-P-wilson}, SEq.~\eqref{eq:W-2-pm-k-G1-G2-I-transform} allows us to deduce the following constraints on the nested $P_{\pm}$-Wilson loop eigenphases:
\begin{align}
	& \{ e^{i (\gamma_2^\pm)_{j,\mathbf{k},\mathbf{G},\mathbf{G}'}} | j = 1 \ldots N_{W}^\pm\} = \{ e^{-i (\gamma_2^\pm)_{j,-\mathbf{k},\mathbf{G},\mathbf{G}'}} | j = 1 \ldots N_{W}^\pm \}, \\
	&\{ (\gamma_2^\pm)_{j,\mathbf{k},\mathbf{G},\mathbf{G}'} | j = 1 \ldots N_{W}^\pm \} \text{ mod } 2\pi = \{ -(\gamma_2^\pm)_{j,-\mathbf{k},\mathbf{G},\mathbf{G}'} | j = 1 \ldots N_{W}^\pm \} \text{ mod } 2\pi, \label{eq:I_constraints_on_gamma_2pm_j} \\
	&(\gamma_2^\pm)_{\mathbf{k},\mathbf{G},\mathbf{G}'} \text{ mod } 2\pi = -(\gamma_2^\pm)_{-\mathbf{k},\mathbf{G},\mathbf{G}'} \text{ mod } 2\pi, \\
	& \left((\gamma_2^\pm)_{\mathbf{k}_{TRIM},\mathbf{G},\mathbf{G}'} \mod \pi \right)  = 0,\label{eq:general_notation_gamma_2_pm_I_constraint}
\end{align}
where 
\begin{equation}
	(\gamma_2^\pm)_{\mathbf{k},\mathbf{G},\mathbf{G}'} \equiv \sum_{j=1}^{N_{W}^\pm}(\gamma_2^\pm)_{j,\mathbf{k},\mathbf{G},\mathbf{G}'} \text{ mod } 2\pi, \label{eq:summed-gamma-2-pm-def-general}
\end{equation}
and $(\gamma_2^\pm)_{j,\mathbf{k},\mathbf{G},\mathbf{G}'}$ are the eigenphases (nested $P_\pm$-Wannier bands) of the nested $P_\pm$-Wilson loop operator $\mathcal{W}^\pm_{2,\mathbf{k},\mathbf{G},\mathbf{G}'}$ satisfying SEq.~\eqref{eq:W-2-pm-k-G1-G2-I-transform}.
Recall from SN~\ref{sec:nested_P_pm_Wilson_loop} that the eigenphases $(\gamma_2^\pm)_{j,\mathbf{k},\mathbf{G},\mathbf{G}'}$ of the nested $P_\pm$-Wilson loop operator [SEq.~\eqref{eq:def-W2-pm}] correspond to the localized positions of the spin-resolved hybrid Wannier functions formed from a group of $P_\pm$-Wannier bands
~\cite{benalcazar2017electric,benalcazar2017quantized}. 
SEq.~\eqref{eq:I_constraints_on_gamma_2pm_j} is then a consequence of the fact that inversion flips both position and momentum, but does not flip the spin of a spin-resolved hybrid Wannier function.
Since $(\gamma_2^\pm)_{j,\mathbf{k},\mathbf{G},\mathbf{G}'}$ is independent of the momentum component $\mathbf{k} \cdot \mathbf{a}'$, where $\mathbf{a}'$ is the primitive lattice vector dual to the primitive reciprocal lattice vector $\mathbf{G}'$, $\mathbf{k}_{TRIM}$ in SEq.~\eqref{eq:general_notation_gamma_2_pm_I_constraint} should be interpreted as a $\mathbf{k}$-vector with $(\mathbf{k}_{TRIM}\cdot\mathbf{a})\mod \pi=0$ for each primitive lattice vector $\mathbf{a}\neq\mathbf{a'}$ .

\subsection{Antiunitary $\mathcal{T}$ Constraint on the $P_{\pm}$-Wilson Loop} \label{appendix:T-constraint-on-P-pm}

In this section, we will extend the analysis in SN~\ref{appendix:T-constraint-on-P-wilson} to investigate the constraints that spinful antiunitary $\mathcal{T}$ places on the $P_{\pm}$-Wilson loop (SN~\ref{sec:P_pm_Wilson_loop}) for a system with spin-$1/2$ degrees of freedom and with $\mathcal{T}^2 = -1$.
In the presence of spinful $\mathcal{T}$ symmetry, we denote the $N_{\mathrm{occ}}/2$ spin bands with largest $PsP$ eigenvalue as the upper spin bands; we denote the $N_{\mathrm{occ}}/2$ spin bands with smallest $PsP$ eigenvalue as the lower spin bands. 
We assume that the upper and lower spin bands are disjoint from each other [see also SFig.~\ref{fig:schematic_spin_bands_with_I_T_and_IT}(b)]. 
The projectors $[P_{\pm}(\mathbf{k})]$  onto the upper~($+$)/lower~($-$) spin bands defined in SEq.~\eqref{eq:P_pm_k_projector} then transforms under spinful $\mathcal{T}$ according to
\begin{equation}
    [\mathcal{T}]  [P_{\pm}(\mathbf{k})] [\mathcal{T}]^{-1}  = [P_{\mp}(-\mathbf{k})]. \label{eq:TR-act-on-P-plus-minus}
\end{equation}
Consequently, the upper and lower $PsP$ eigenvalues [$\lambda_{n,\mathbf{k}}^\pm$ in SEq.~\eqref{eq:eigvaleqnsk}] are related by
\begin{equation}
	\{ \lambda_{n,\mathbf{k}}^+ | j=1\ldots N_{\mathrm{occ}}^+ \} = \{ -\lambda_{n,-\mathbf{k}}^- | j=1\ldots N_{\mathrm{occ}}^- \}.
\end{equation}
SEq.~\eqref{eq:TR-act-on-P-plus-minus} follows from the eigenvalue equation for $[s(\mathbf{k})]$ in SEq.~\eqref{eq:eigvaleqnsk}, and the fact that time-reversal flips spins, namely
\begin{equation}
	[\mathcal{T}] s = -s [\mathcal{T}], \label{eq:s-flip-upon-T}
\end{equation}
such that the projected spin operator $[s(\mathbf{k})]$ transforms under spinful $\mathcal{T}$ according to
\begin{align}
	[\mathcal{T}] [s(\mathbf{k})] [\mathcal{T}]^{-1} & = [\mathcal{T}] [P(\mathbf{k})] [\mathcal{T}]^{-1} [\mathcal{T}] s [\mathcal{T}]^{-1} [\mathcal{T}] [P(\mathbf{k})] [\mathcal{T}]^{-1} \nonumber \\
	& =  [P(-\mathbf{k})] [\mathcal{T}] s [\mathcal{T}]^{-1} [P(-\mathbf{k})] \nonumber \\ &= -[P(-\mathbf{k})] s [P(-\mathbf{k})]\nonumber \\ & = -[s(-\mathbf{k})], \label{eq:TsTinverse}
\end{align}
where we have also used SEq.~\eqref{eq:T-act-on-Pk}. 
Different from the analysis in SN~\ref{appendix:I-constraint-on-P-pm}, we here specialize to the case in which $N_{\mathrm{occ}}^\pm=N_{\mathrm{occ}}/2$, which is required by $\mathcal{T}$ symmetry. 
Following the same logic as in SEq.~\eqref{eq:T_transform_W1} and using SEq.~\eqref{eq:TR-act-on-P-plus-minus}, the $P_{\pm}$-Wilson loop operator [SEq.~\eqref{eq:def-W1-pm}] transforms under spinful $\mathcal{T}$ according to
\begin{equation}
	[\mathcal{T}] \mathcal{W}^\pm_{1,\mathbf{k},\mathbf{G}} [\mathcal{T}]^{-1} = (\mathcal{W}^\mp_{1,-\mathbf{k},\mathbf{G}})^\dag. \label{eq:T-act-on-W-1pm}
\end{equation}
As in SN~\ref{appendix:T-constraint-on-P-wilson}, SEq.~\eqref{eq:T-act-on-W-1pm} allows us to deduce the following spinful time-reversal constraints on the $P_{\pm}$-Wilson loop eigenphases:
\begin{align}
    & \{ e^{i (\gamma_1^\pm)_{j,\mathbf{k},\mathbf{G}}} | j = 1 \ldots N_{\mathrm{occ}}^\pm\} = \{ e^{i (\gamma_1^\mp)_{j,-\mathbf{k},\mathbf{G}}} | j = 1 \ldots N_{\mathrm{occ}}^\mp \}, \\
    &\{ (\gamma_1^\pm)_{j,\mathbf{k},\mathbf{G}} | j = 1 \ldots N_{\mathrm{occ}}^\pm \} \text{ mod } 2\pi = \{ (\gamma_1^\mp)_{j,-\mathbf{k},\mathbf{G}} | j = 1 \ldots N_{\mathrm{occ}}^\mp \} \text{ mod } 2\pi, \label{eq:T_constraints_on_gamma_1pm_j} \\
    & (\gamma_1^\pm)_{\mathbf{k},\mathbf{G}} \text{ mod } 2\pi= (\gamma_1^\mp)_{-\mathbf{k},\mathbf{G}} \text{ mod } 2\pi
\end{align}
where $N_{\mathrm{occ}}^+=N_{\mathrm{occ}}^-=N_{\mathrm{occ}}/2$,  $(\gamma_1^\pm)_{\mathbf{k},\mathbf{G}}$ is defined in SEq.~\eqref{eq:summed-gamma-1pmkG-general-notation}, and $(\gamma_1^\pm)_{j,\mathbf{k},\mathbf{G}}$ are the eigenphases ($P_\pm$-Wannier bands) of the $\widehat{\mathbf{G}}$-directed $P_\pm$-Wilson loop operator $\mathcal{W}^\pm_{1,\mathbf{k},\mathbf{G}}$ satisfying SEq.~\eqref{eq:T-act-on-W-1pm}.
Recall from SN~\ref{sec:P_pm_Wilson_loop} that the eigenphases $(\gamma_1^\pm)_{j,\mathbf{k},\mathbf{G}}$ of the $P_\pm$-Wilson loop operator [SEq.~\eqref{eq:def-W1-pm}] correspond to the localized positions of spin-resolved hybrid Wannier functions formed from the set of upper/lower spin bands~\cite{alexandradinata2014wilsonloop,benalcazar2017electric,benalcazar2017quantized,lecture_notes_on_berry_phases_and_topology_bb,vanderbilt2018berry}. 
SEq.~\eqref{eq:T_constraints_on_gamma_1pm_j} is then a consequence of the fact that spinful time-reversal flips both momentum and spin, but not the position, of a spin-resolved hybrid Wannier function.

\subsection{Antiunitary $\mathcal{T}$ Constraint on the Nested $P_{\pm}$-Wilson Loop} 
\label{appendix:T-constraint-on-nested-P-pm}

Following SN~\ref{appendix:T-constraint-on-P-pm}, in this section, we will extend the analysis in SN~\ref{sec:T-constraint-on-nested-P} to deduce the constraints that antiunitary spinful $\mathcal{T}$ places on the nested $P_{\pm}$-Wilson loop (SN~\ref{sec:nested_P_pm_Wilson_loop}) for a system with spin-$1/2$ degrees of freedom and for which $\mathcal{T}^2 = -1$.
For a spinful $\mathcal{T}$-symmetric system, the projector $[\widetilde{P}^\pm_{\mathbf{G}}(\mathbf{k})]$ [SEq.~\eqref{eq:P_pm_nested_Wilson_loop_2nd_projector}] onto an isolated grouping of $N_W^\pm$ $\widehat{\mathbf{G}}$-directed $P_\pm$-Wannier bands can be chosen to satisfy
\begin{equation}
	[\mathcal{T}] [\widetilde{P}^\pm_{\mathbf{G}}(\mathbf{k})] [\mathcal{T}]^{-1} = [\widetilde{P}^\mp_{\mathbf{G}}(-\mathbf{k})], \label{eq:T-act-on-tilde-P-pm}
\end{equation}
which combining with SEq.~\eqref{eq:T-act-on-W-1pm} implies that the isolated grouping of $N_W^\pm$ $\widehat{\mathbf{G}}$-directed $P_\pm$-Wilson loop eigenphases satisfies
\begin{equation}
	\{ (\gamma_1^\pm)_{j,\mathbf{k},\mathbf{G}} | j=1\ldots N_{W}^\pm \} \text{ mod } 2\pi = \{ (\gamma_1^\mp)_{j,-\mathbf{k},\mathbf{G}} | j=1\ldots N_{W}^\mp \} \text{ mod } 2\pi, \label{eq:NWpmbandsTRsymmetric}
\end{equation}
which is similar to the constraints imposed by spinful $\mathcal{T}$ on the entire $P_{\pm}$-Wannier band structure [SEq.~\eqref{eq:T_constraints_on_gamma_1pm_j}].
Notice that, as a consequence of enforcing SEqs.~\eqref{eq:T-act-on-tilde-P-pm} and \eqref{eq:NWpmbandsTRsymmetric}, we have that $N_W^+ = N_W^- \leq N_{\mathrm{occ}}/2$.
Following the same logic as in SEq.~\eqref{eq:tilde-W-T-transform} and using SEq.~\eqref{eq:T-act-on-tilde-P-pm}, the nested $P_{\pm}$-Wilson loop operator [SEq.~\eqref{eq:def-W2-pm}] transforms under spinful $\mathcal{T}$ according to
\begin{equation}
	[\mathcal{T}] \mathcal{W}^\pm_{2,\mathbf{k},\mathbf{G},\mathbf{G}'} [\mathcal{T}]^{-1} = ({\mathcal{W}}^\mp_{2,-\mathbf{k},\mathbf{G},\mathbf{G}'})^{\dagger}. \label{eq:W2pm-T-transform}
\end{equation}
As in SN~\ref{sec:T-constraint-on-nested-P}, SEq.~\eqref{eq:W2pm-T-transform} allows us to deduce the following spinful time-reversal constraints on the nested $P_\pm$-Wilson loop eigenphases:
\begin{align}
    & \{ e^{i (\gamma_2^\pm)_{j,\mathbf{k},\mathbf{G},\mathbf{G'}}} | j = 1 \ldots N_{W}^\pm\} = \{ e^{i (\gamma_2^\mp)_{j,-\mathbf{k},\mathbf{G},\mathbf{G}'}} | j = 1 \ldots N_{W}^\mp \}, \\
    &\{ (\gamma_2^\pm)_{j,\mathbf{k},\mathbf{G},\mathbf{G'}} | j = 1 \ldots N_{W}^\pm \} \text{ mod } 2\pi = \{ (\gamma_2^\mp)_{j,-\mathbf{k},\mathbf{G},\mathbf{G'}} | j = 1 \ldots N_{W}^\mp \} \text{ mod } 2\pi, \label{eq:T_constraints_on_gamma_2pm_j}\\
    & (\gamma_2^\pm)_{\mathbf{k},\mathbf{G},\mathbf{G}'} \text{ mod } 2\pi = (\gamma_2^\mp)_{-\mathbf{k},\mathbf{G},\mathbf{G}'} \text{ mod } 2\pi,
\end{align}
where $(\gamma_2^\pm)_{\mathbf{k},\mathbf{G},\mathbf{G}'}$ is defined in SEq.~\eqref{eq:summed-gamma-2-pm-def-general} and $(\gamma_2^\pm)_{j,\mathbf{k},\mathbf{G},\mathbf{G'}}$ are the eigenphases (nested $P_\pm$-Wannier bands) of the nested $P$-Wilson loop operator $\mathcal{W}^\pm_{2,\mathbf{k},\mathbf{G},\mathbf{G}'}$ satisfying SEq.~\eqref{eq:W2pm-T-transform}.
Recall from SN~\ref{sec:nested_P_pm_Wilson_loop} that the eigenphases $(\gamma_2^\pm)_{j,\mathbf{k},\mathbf{G},\mathbf{G}'}$ of the nested $P_\pm$-Wilson loop operator [SEq.~\eqref{eq:def-W2-pm}] correspond to the localized positions of the spin-resolved hybrid Wannier functions formed from a group of $P_\pm$-Wannier bands
~\cite{benalcazar2017electric,benalcazar2017quantized}. 
SEq.~\eqref{eq:T_constraints_on_gamma_2pm_j} is then a consequence of the fact that spinful time-reversal flips both momentum and spin, but not the position, of a spin-resolved hybrid Wannier function.

\subsection{A Summary of Symmetry Constraints on $P$-, $P_{\pm}$-, Nested $P$-, and Nested $P_{\pm}$-Wilson Loop} \label{app:short_summary_for_symmetry_constraints}

In this section, we present two tables [Supplementary Tables \ref{tab:inversion-symmetry-constraint-on-wilson-loops} and \ref{tab:time-reversal-symmetry-constraint-on-wilson-loops}] summarizing the symmetry constraints derived in SN~\ref{appendix:I-constraint-on-P-wilson}--\ref{appendix:T-constraint-on-nested-P-pm} on the projectors and the eigenphases of the Wilson loops constructed from the product of these projectors along a closed loop.
Supplementary Tables \ref{tab:inversion-symmetry-constraint-on-wilson-loops} and \ref{tab:time-reversal-symmetry-constraint-on-wilson-loops} summarize the symmetry constraints from unitary inversion ($\mathcal{I}$) and antiunitary time-reversal ($\mathcal{T}$) symmetries, respectively.
The expressions in Supplementary Tables~\ref{tab:inversion-symmetry-constraint-on-wilson-loops} and~\ref{tab:time-reversal-symmetry-constraint-on-wilson-loops} use the general notation employed in SN~\ref{sec:P_Wilson_loop}, \ref{sec:P_pm_Wilson_loop}, \ref{sec:nested_P_Wilson_loop}, and \ref{sec:nested_P_pm_Wilson_loop}.

\begin{table}[H]
\centering
\begin{tabular}{|c|c|c|}
\hline
  \multicolumn{3}{|c|}{Unitary Inversion ($\mathcal{I}$) Symmetry}   \\
\hline
\hline
& Projector & Wilson Loop Eigenphases  \\
\hline
$P$-Wilson Loop & $[\mathcal{I}][P(\mathbf{k})][\mathcal{I}]^{\dagger} = [P(-\mathbf{k})]$   & $\{ (\gamma_1)_{j,\mathbf{k},\mathbf{G}} | j=1\ldots N_{\mathrm{occ}} \}$ mod $2\pi$  \\
(SN~\ref{appendix:I-constraint-on-P-wilson}) & [SEq.~\eqref{eq:inversion-act-on-P}] & $=  \{ -(\gamma_1)_{j,-\mathbf{k},\mathbf{G}} | j=1\ldots N_{\mathrm{occ}} \}$ mod $2\pi$ [SEq.~\eqref{eq:I_constraints_on_gamma_1_j}] \\
\hline
$P_{\pm}$-Wilson Loop  & $[\mathcal{I}][P_{\pm}(\mathbf{k})][\mathcal{I}]^{\dagger} = [P_{\pm}(-\mathbf{k})]$ & $\{ (\gamma_1^\pm)_{j,\mathbf{k},\mathbf{G}} | j=1\ldots N_{\mathrm{occ}}^\pm \} $ mod $2\pi$  \\
(SN~\ref{appendix:I-constraint-on-P-pm}) & [SEq.~\eqref{eq:I-act-on-P-pm}] & $= \{ -(\gamma_1^\pm)_{j,-\mathbf{k},\mathbf{G}} | j=1\ldots N_{\mathrm{occ}}^\pm \}$ mod $2\pi$ [SEq.~\eqref{eq:I_constraints_on_gamma_1pm_j}] \\
\hline
Nested $P$-Wilson Loop & $[\mathcal{I}][\widetilde{P}_{\mathbf{G}}(\mathbf{k})][\mathcal{I}]^{\dagger} = [\widetilde{P}_{\mathbf{G}}(-\mathbf{k})]$  & $\{ (\gamma_2)_{j,\mathbf{k},\mathbf{G},\mathbf{G}'} | j=1\ldots N_{W} \}$ mod $2\pi$ \\
(SN~\ref{appendix:I-constraint-on-nested-P-wilson}) &   [SEq.~\eqref{eq:I-act-on-tilde-P-G-k}]  & $ = \{ -(\gamma_2)_{j,-\mathbf{k},\mathbf{G},\mathbf{G}'} | j=1\ldots N_{W} \}$ mod $2\pi$ [SEq.~\eqref{eq:general_notation_gamma_2j_I_constraint}]  \\
\hline
Nested $P_{\pm}$-Wilson Loop & $[\mathcal{I}][\widetilde{P}^\pm_{\mathbf{G}}(\mathbf{k})][\mathcal{I}]^{\dagger} = [\widetilde{P}^\pm_{\mathbf{G}}(-\mathbf{k})]$ & $\{ (\gamma_2^\pm)_{j,\mathbf{k},\mathbf{G},\mathbf{G}'} | j=1\ldots N_{W}^\pm \}$ mod $2\pi$  \\
(SN~\ref{appendix:I-constraint-on-nested-P-pm}) & [SEq.~\eqref{eq:inversion-act-on-tilde-P-pm}]  & $= \{ -(\gamma_2^\pm)_{j,-\mathbf{k},\mathbf{G},\mathbf{G}'} | j=1\ldots N_{W}^\pm \}$ mod $2\pi$ [SEq.~\eqref{eq:I_constraints_on_gamma_2pm_j}] \\
\hline
\end{tabular}
\caption{Constraints from unitary inversion ($\mathcal{I}$) symmetry on the projectors and on the eigenphases of Wilson loops.
$[P(\mathbf{k})]$ [SEq.~\eqref{eq:P_projector}] is a projector onto $N_{\mathrm{occ}}$ occupied energy bands. 
$[P_{\pm}(\mathbf{k})]$ [SEq.~\eqref{eq:P_pm_k_projector}] is a projector onto $N_{\mathrm{occ}}^\pm$ upper/lower spin bands.
$[\widetilde{P}_{\mathbf{G}}(\mathbf{k})]$ [SEq.~\eqref{eq:P_nested_Wilson_loop_2nd_projector}] is a projector onto an isolated grouping of $N_W$ $\widehat{\mathbf{G}}$-directed $P$-Wannier bands.
$[\widetilde{P}^{\pm}_{\mathbf{G}}(\mathbf{k})]$ [SEq.~\eqref{eq:P_pm_nested_Wilson_loop_2nd_projector}] is a projector onto an isolated grouping of $N_W^\pm$ $\widehat{\mathbf{G}}$-directed $P_{\pm}$-Wannier bands.
$[\mathcal{I}]$ is the unitary matrix representative of $\mathcal{I}$.}
\label{tab:inversion-symmetry-constraint-on-wilson-loops}
\end{table}

\begin{table}[H]
\centering
\begin{tabular}{|c|c|c|}
\hline
 \multicolumn{3}{|c|}{Antiunitary Time-Reversal ($\mathcal{T}$) Symmetry}   \\
\hline
\hline
& Projector & Wilson Loop Eigenphases  \\
\hline
$P$-Wilson Loop  & $[\mathcal{T}][P(\mathbf{k})][\mathcal{T}]^{-1} = [P(-\mathbf{k})]$   & $\{ (\gamma_1)_{j,\mathbf{k},\mathbf{G}} | j=1\ldots N_{\mathrm{occ}} \}  $ mod $2\pi$  \\
(SN~\ref{appendix:T-constraint-on-P-wilson}) & [SEq.~\eqref{eq:T-act-on-Pk}]  & $= \{ (\gamma_1)_{j,-\mathbf{k},\mathbf{G}} | j=1\ldots N_{\mathrm{occ}} \}$ mod $2\pi$ [SEq.~\eqref{eq:general_notation_T_constraint_on_gamma_1jkG}] \\
& & (Kramers' degeneracy at TRIMs when $\mathcal{T}^2=-1$)  \\
\hline
$P_{\pm}$-Wilson Loop  & $[\mathcal{T}][P_{\pm}(\mathbf{k})][\mathcal{T}]^{-1} = [P_{\mp}(-\mathbf{k})]$  & $\{ (\gamma_1^\pm)_{j,\mathbf{k},\mathbf{G}} | j=1\ldots N_{\mathrm{occ}}^\pm \}$  mod $2\pi$ \\
(SN~\ref{appendix:T-constraint-on-P-pm}) & [SEq.~\eqref{eq:TR-act-on-P-plus-minus}] & $= \{ (\gamma_1^\mp)_{j,-\mathbf{k},\mathbf{G}} | j=1\ldots N_{\mathrm{occ}}^\mp \}$ mod $2\pi$ [SEq.~\eqref{eq:T_constraints_on_gamma_1pm_j}] \\
& & (assuming  $\mathcal{T}^2 = -1$) \\
\hline
Nested $P$-Wilson Loop  & $[\mathcal{T}][\widetilde{P}_{\mathbf{G}}(\mathbf{k})][\mathcal{T}]^{-1} = [\widetilde{P}_{\mathbf{G}}(-\mathbf{k})]$  & $\{ (\gamma_2)_{j,\mathbf{k},\mathbf{G},\mathbf{G}'} | j=1\ldots N_{W} \}$  mod $2\pi$  \\
(SN~\ref{sec:T-constraint-on-nested-P})    & [SEq.~\eqref{eq:T-act-on-tilde-P-G-k}] & $ = \{ (\gamma_2)_{j,-\mathbf{k},\mathbf{G},\mathbf{G}'} | j=1\ldots N_{W} \}$ mod $2\pi$ [SEq.~\eqref{eq:general_notation_T_constraint_on_gamma_2jkGGp}] \\
& & (Kramers' degeneracy at TRIMs when $\mathcal{T}^2=-1$) \\
\hline
Nested $P_{\pm}$-Wilson Loop & $[\mathcal{T}][\widetilde{P}^\pm_{\mathbf{G}}(\mathbf{k})][\mathcal{T}]^{-1} = [\widetilde{P}^\mp_{\mathbf{G}}(-\mathbf{k})]$  & $\{ (\gamma_2^\pm)_{j,\mathbf{k},\mathbf{G},\mathbf{G}'} | j=1\ldots N_{W}^\pm \}$ mod $2\pi$    \\
 (SN~\ref{appendix:T-constraint-on-nested-P-pm}) & [SEq.~\eqref{eq:T-act-on-tilde-P-pm}] & $ = \{ (\gamma_2^\mp)_{j,-\mathbf{k},\mathbf{G},\mathbf{G}'} | j=1\ldots N_{W}^\mp \}$ mod $2\pi$  [SEq.~\eqref{eq:T_constraints_on_gamma_2pm_j}]  \\
& & (assuming $\mathcal{T}^2 = -1$) \\
\hline
\end{tabular}
\caption{Constraints from antiunitary time-reversal ($\mathcal{T}$) symmetry on the projectors and on the eigenphases of the Wilson loops.
$[P(\mathbf{k})]$ [SEq.~\eqref{eq:P_projector}] is a projector onto $N_{\mathrm{occ}}$ occupied energy bands. 
$[P_{\pm}(\mathbf{k})]$ [SEq.~\eqref{eq:P_pm_k_projector}] is a projector onto $N_{\mathrm{occ}}^\pm$ upper/lower spin bands.
$[\widetilde{P}_{\mathbf{G}}(\mathbf{k})]$ [SEq.~\eqref{eq:P_nested_Wilson_loop_2nd_projector}] is a projector onto an isolated grouping of $N_W$ $\widehat{\mathbf{G}}$-directed $P$-Wannier bands.
$[\widetilde{P}^{\pm}_{\mathbf{G}}(\mathbf{k})]$ [SEq.~\eqref{eq:P_pm_nested_Wilson_loop_2nd_projector}] is a projector onto an isolated grouping of $N_W^\pm$ $\widehat{\mathbf{G}}$-directed $P_{\pm}$-Wannier bands.
$[\mathcal{T}]$ is the antiunitary representative of $\mathcal{T}$.}
\label{tab:time-reversal-symmetry-constraint-on-wilson-loops}
\end{table}

\subsection{Eigenvalues and Eigenvectors of Wilson Loop Operators and Their Hermitian Conjugates} \label{appendix:eigval_eigvec_W}

Suppose an operator $\mathcal{W}$ has spectral decomposition
\begin{equation}
	\mathcal{W} = \sum_{j} e^{i\gamma_j} |w_j\rangle \langle w_j|, \label{eq:decompose-W}
\end{equation}
where $\langle w_j| = (|w_j\rangle)^\dag$, $\langle w_i | w_j \rangle = \delta_{ij}$, and $\gamma_j \in \mathbb{R}$.
For example, all of the Wilson loop operators that we considered in this section (SN~\ref{appendix:symmetry-constraints-on-Wilson-loop}), including the $P$-Wilson loop operator [SEq.~\eqref{eq:def-W1}], the nested $P$-Wilson loop operator [SEq.~\eqref{eq:def-nested-P-wilson-loop}], the $P_{\pm}$-Wilson loop operator [SEq.~\eqref{eq:def-W1-pm}], and the nested $P_\pm$-Wilson loop operator [SEq.~\eqref{eq:def-W2-pm}], satisfy SEq.~\eqref{eq:decompose-W}. 
Taking the Hermitian conjugate of SEq.~\eqref{eq:decompose-W}, we obtain
\begin{equation}
    \mathcal{W}^{\dagger} = \sum_{j} e^{-i\gamma_{j}} |w_{j}\rangle \langle w_{j} |. \label{eq:decompose-W-dag}
\end{equation}
SEqs.~\eqref{eq:decompose-W} and \eqref{eq:decompose-W-dag} imply that the eigenvector $|w_j \rangle$ of $\mathcal{W}$ with unimodular eigenvalue $e^{i\gamma_j}$ is also an eigenvector of $\mathcal{W}^\dag$ with eigenvalue $e^{-i\gamma_j}$.
In other words, the sets of unimodular eigenvalues of $\mathcal{W}$ and $\mathcal{W}^\dag$ are complex conjugates of each other; in addition, $\mathcal{W}$ and $\mathcal{W}^\dag$  have the same set of orthonormal eigenvectors.

\subsection{Transformation of the $[V(\mathbf{G})]$ Matrix Under Unitary $\mathcal{I}$} \label{sec:proof-IVGI-dag}

In this section, we will derive the representative $[\mathcal{I}]$ of $\mathcal{I}$ in the tight-binding basis states.
We will denote by $\mathcal{I}$ the unitary inversion operator that acts on the second-quantized electron operators in SEqs.~\eqref{eq:TB_convention_c_dagger} and~\eqref{eq:TB_convention_c}.
We will use $[\mathcal{I}]$ to denote the $N_\mathrm{sta}\times N_\mathrm{sta}$ unitary representative of inversion in the tight-binding basis states.

In this section, we will specifically prove that
\begin{equation}
[\mathcal{I}][V(\mathbf{G})][\mathcal{I}]^{\dagger}=[V(-\mathbf{G})],
\end{equation}
where $[V(\mathbf{G})]$ is the $N_{\text{sta}} \times N_{\text{sta}}$ matrix defined in SEq.~\eqref{eq:def-VG} that encodes the positions of the tight-binding basis states.

Let us consider the creation operators [SEq.~\eqref{eq:TB_convention_c_dagger}] for the Bloch basis states
\begin{equation}
    c^{\dagger}_{\mathbf{k},\alpha} = \frac{1}{\sqrt{N}}\sum_{\mathbf{R}}e^{i\mathbf{k}\cdot (\mathbf{R}+\mathbf{r}_{\alpha})} c^{\dagger}_{\mathbf{R},\alpha}.
\end{equation}
Upon a displacement $\mathbf{k} \to \mathbf{k}+\mathbf{G}$ where $\mathbf{G}$ is a reciprocal lattice vector, we have
\begin{equation}
    c^{\dagger}_{\mathbf{k}+\mathbf{G},\alpha} = \frac{1}{\sqrt{N}}\sum_{\mathbf{R}} e^{i(\mathbf{k}+\mathbf{G})\cdot(\mathbf{R}+\mathbf{r}_{\alpha})}c^{\dagger}_{\mathbf{R},\alpha}=e^{i\mathbf{G}\cdot \mathbf{r}_{\alpha}}  \frac{1}{\sqrt{N}}\sum_{\mathbf{R}}e^{i\mathbf{k}\cdot (\mathbf{R}+\mathbf{r}_{\alpha})} c^{\dagger}_{\mathbf{R},\alpha} = e^{i\mathbf{G}\cdot \mathbf{r}_{\alpha}} c^{\dagger}_{\mathbf{k},\alpha}
\end{equation}
where we have made use of $e^{i\mathbf{G}\cdot \mathbf{R}}=1$.
Grouping the Bloch basis state creation operators into a creation-operator-valued row vector
\begin{equation}
    \psi^{\dagger}_{\mathbf{k}} = [c^{\dagger}_{\mathbf{k},1},\cdots,c^{\dagger}_{\mathbf{k},N_{\text{sta}}}], \label{eq:row_psi_k}
\end{equation}
we have that
\begin{equation}
    \psi^{\dagger}_{\mathbf{k}+\mathbf{G}} = \psi^{\dagger}_{\mathbf{k}} [V(\mathbf{G})]. \label{eq:row_psi_k_plus_G}
\end{equation}
Next we consider the inversion operation acting on the creation operators of the tight-binding basis states~\cite{wieder2018wallpaper}
\begin{equation}
    \mathcal{I} c^{\dagger}_{\mathbf{R},\alpha} \mathcal{I}^{\dagger} = c^{\dagger}_{\mathbf{R}',\beta} [\mathcal{I}]_{\beta \alpha} \label{eq:def-I-matrix-rep}
\end{equation}
where $\mathbf{R}'=-(\mathbf{R}+\mathbf{r}_{\alpha})-\mathbf{r}_{\beta}$ and the $\beta$ index is implicitly summed over.
$[\mathcal{I}]$ in SEq.~\eqref{eq:def-I-matrix-rep} is then the unitary matrix representative of inversion, which can be obtained from the position-space symmetry data of the tight-binding basis states of the system.
Notice that if $-(\mathbf{R}+\mathbf{r}_{\alpha})-\mathbf{r}_{\beta}$ is not a linear combination of the Bravais lattice vectors with integral coefficients, we automatically have that the corresponding matrix element $[\mathcal{I}]_{\beta \alpha}=0$.
Using SEqs.~\eqref{eq:TB_convention_c_dagger} and~\eqref{eq:def-I-matrix-rep}, we can obtain 
\begin{align}
    \mathcal{I} c^{\dagger}_{\mathbf{k},\alpha} \mathcal{I}^{\dagger} & =  \frac{1}{\sqrt{N}}\sum_{\mathbf{R}}e^{i\mathbf{k}\cdot (\mathbf{R}+\mathbf{r}_{\alpha})} \mathcal{I}c^{\dagger}_{\mathbf{R},\alpha} \mathcal{I}^{\dagger} = \frac{1}{\sqrt{N}}\sum_{\mathbf{R}}e^{i\mathbf{k}\cdot (\mathbf{R}+\mathbf{r}_{\alpha})} c^{\dagger}_{\mathbf{R}',\beta} [\mathcal{I}]_{\beta \alpha} \nonumber \\
    & = \frac{1}{\sqrt{N}}\sum_{\mathbf{R}}e^{i\mathbf{k}\cdot (\mathbf{R}+\mathbf{r}_{\alpha})} c^{\dagger}_{-(\mathbf{R}+\mathbf{r}_{\alpha})-\mathbf{r}_{\beta},\beta} [\mathcal{I}]_{\beta \alpha} \nonumber \\
    & = \frac{1}{\sqrt{N}}\sum_{\mathbf{R}}e^{i(-\mathbf{k})\cdot (-(\mathbf{R}+\mathbf{r}_{\alpha})-\mathbf{r}_{\beta}+\mathbf{r}_{\beta})} c^{\dagger}_{-(\mathbf{R}+\mathbf{r}_{\alpha})-\mathbf{r}_{\beta},\beta} [\mathcal{I}]_{\beta \alpha} = \frac{1}{\sqrt{N}}\sum_{\mathbf{R}'}e^{i(-\mathbf{k})\cdot (\mathbf{R}'+\mathbf{r}_{\beta})} c^{\dagger}_{\mathbf{R}',\beta} [\mathcal{I}]_{\beta \alpha} \nonumber \\
    & = c^{\dagger}_{-\mathbf{k},\beta} [\mathcal{I}]_{\beta \alpha}.
\end{align}
In other words, under inversion symmetry, $\psi^{\dagger}_{\mathbf{k}}$ in SEq.~\eqref{eq:row_psi_k} transforms according to
\begin{equation}
    \mathcal{I} \psi^{\dagger}_{\mathbf{k}} \mathcal{I}^{\dagger} = \psi^{\dagger}_{-\mathbf{k}} [\mathcal{I}]. \label{eq:row_psi_k_I_transform}
\end{equation}
Now, let us consider the expression
\begin{equation}
    \mathcal{I} \psi^{\dagger}_{\mathbf{k}+\mathbf{G}} \mathcal{I}^{\dagger}. \label{eq:IpsikIdag}
\end{equation}
There are two ways we can rewrite SEq.~\eqref{eq:IpsikIdag}.
First, using SEq.~\eqref{eq:row_psi_k_plus_G} and then SEq.~\eqref{eq:row_psi_k_I_transform} we can write
\begin{align}
    \mathcal{I} \psi^{\dagger}_{\mathbf{k}+\mathbf{G}} \mathcal{I}^{\dagger} &= \mathcal{I} \psi^{\dagger}_{\mathbf{k}} [V(\mathbf{G})] \mathcal{I}^{\dagger} \nonumber \\
    &= \mathcal{I} \psi^{\dagger}_{\mathbf{k}}  \mathcal{I}^{\dagger} [V(\mathbf{G})] \nonumber \\
    &=  \psi^{\dagger}_{-\mathbf{k}} [\mathcal{I}] [V(\mathbf{G})], \label{eq:rewrite-IpsikIdag-1}
\end{align}
where, importantly, notice that $\mathcal{I}$ {\it does not} act on the numerical matrix $[V(\mathbf{G})]$. 
Second, using SEq.~\eqref{eq:row_psi_k_I_transform} and then SEq.~\eqref{eq:row_psi_k_plus_G}, we can also write SEq.~\eqref{eq:IpsikIdag} as
\begin{align}
    \mathcal{I} \psi^{\dagger}_{\mathbf{k}+\mathbf{G}} \mathcal{I}^{\dagger} &= \psi^{\dagger}_{-\mathbf{k}-\mathbf{G}} [\mathcal{I}]\nonumber\\
    & = \psi^{\dagger}_{-\mathbf{k}}[V(-\mathbf{G})] [\mathcal{I}]. \label{eq:rewrite-IpsikIdag-2}
\end{align}
By comparing SEqs.~(\ref{eq:rewrite-IpsikIdag-1}) and (\ref{eq:rewrite-IpsikIdag-2}), we have
\begin{equation}
    \psi^{\dagger}_{-\mathbf{k}} [\mathcal{I}] [V(\mathbf{G})] = \psi^{\dagger}_{-\mathbf{k}}[V(-\mathbf{G})] [\mathcal{I}]. \label{eq:before-arriving-conclusion-of-IVI-dag}
\end{equation}
We next assume that all of the tight-binding basis states are orthogonal to each other. 
Therefore, SEq.~(\ref{eq:before-arriving-conclusion-of-IVI-dag}) can only be satisfied if
\begin{equation}
    [\mathcal{I}] [V(\mathbf{G})] = [V(-\mathbf{G})] [\mathcal{I}]. \label{eq:IVVI-1}
\end{equation}
Since $[\mathcal{I}]$ is unitary, we hence conclude that
\begin{equation}
    [\mathcal{I}] [V(\mathbf{G})] [\mathcal{I}]^{\dagger} = [V(-\mathbf{G})]. \label{eq:appendix-final-result-IVI-inverse}
\end{equation}

\subsection{Transformation of the $[V(\mathbf{G})]$ Matrix Under Antiunitary $\mathcal{T}$} \label{sec:proof-TVGT-inverse}

In this section, we will derive the transformation of $[V(\mathbf{G})]$ under antiunitary $\mathcal{T}$.
As in SN~\ref{sec:proof-IVGI-dag}, we will first define the antiunitary representative $[\mathcal{T}]$ of $\mathcal{T}$.
Let us consider the action of $\mathcal{T}$ on the creation operators of the tight-binding basis states~\cite{wieder2018wallpaper}
\begin{equation}
    \mathcal{T} c^{\dagger}_{\mathbf{R},\alpha} \mathcal{T}^{-1} = c^{\dagger}_{\mathbf{R}',\beta} [U_{\mathcal{T}}]_{\beta\alpha} \label{eq:T-act-on-c-dagger}
\end{equation}
where $\mathbf{R}'=(\mathbf{R}+\mathbf{r}_{\alpha})-\mathbf{r}_{\beta}$, and $\beta$ is summed over.
$[U_{\mathcal{T}}]$ in SEq.~\eqref{eq:T-act-on-c-dagger} is the unitary part of the representative of the antiunitary $\mathcal{T}$.
Since $\mathcal{T}$ acts trivially in position space, $[U_{\mathcal{T}}]_{\beta\alpha}$ has nonzero matrix elements only when $\mathbf{r}_{\beta}=\mathbf{r}_{\alpha}$.
With SEq.~(\ref{eq:T-act-on-c-dagger}), we can deduce that the Bloch basis state creation operator [SEq.~\eqref{eq:TB_convention_c_dagger}] transforms under $\mathcal{T}$ according~to
\begin{align}
    \mathcal{T} c^{\dagger}_{\mathbf{k},\alpha} \mathcal{T}^{-1} & = \mathcal{T} \frac{1}{\sqrt{N}}\sum_{\mathbf{R}}e^{i\mathbf{k}\cdot (\mathbf{R}+\mathbf{r}_{\alpha})} c^{\dagger}_{\mathbf{R},\alpha} \mathcal{T}^{-1} =  \frac{1}{\sqrt{N}}\sum_{\mathbf{R}}e^{-i\mathbf{k}\cdot (\mathbf{R}+\mathbf{r}_{\alpha})} \mathcal{T}c^{\dagger}_{\mathbf{R},\alpha} \mathcal{T}^{-1} = \frac{1}{\sqrt{N}}\sum_{\mathbf{R}}e^{-i\mathbf{k}\cdot (\mathbf{R}+\mathbf{r}_{\alpha})} c^{\dagger}_{\mathbf{R}',\beta} [U_{\mathcal{T}}]_{\beta\alpha} \nonumber \\
    &  =\frac{1}{\sqrt{N}} \sum_{\mathbf{R}} e^{-i \mathbf{k} \cdot (\mathbf{R}+\mathbf{r}_{\alpha})} c^{\dagger}_{\mathbf{R}+\mathbf{r}_{\alpha}-\mathbf{r}_{\beta},\beta}[U_{\mathcal{T}}]_{\beta \alpha} =\frac{1}{\sqrt{N}} \sum_{\mathbf{R}} e^{-i \mathbf{k} \cdot (\mathbf{R}+\mathbf{r}_{\alpha}+\mathbf{r}_{\beta}-\mathbf{r}_{\beta})} c^{\dagger}_{\mathbf{R}+\mathbf{r}_{\alpha}-\mathbf{r}_{\beta},\beta}[U_{\mathcal{T}}]_{\beta \alpha} \nonumber  \\
    & = \frac{1}{\sqrt{N}}\sum_{\mathbf{R}'}e^{-i\mathbf{k}\cdot (\mathbf{R}'+\mathbf{r}_{\beta})} c^{\dagger}_{\mathbf{R}',\beta} [U_{\mathcal{T}}]_{\beta\alpha} = c^{\dagger}_{-\mathbf{k},\beta} [U_{\mathcal{T}}]_{\beta\alpha},
\end{align}
where we have also used the fact that
 $\mathcal{T}$ is antiunitary. 
In terms of the creation-operator-valued row vector $\psi^{\dagger}_{\mathbf{k}}$ in SEq.~\eqref{eq:row_psi_k}, we then have
\begin{equation}
    \mathcal{T} \psi^{\dagger}_{\mathbf{k}} \mathcal{T}^{-1} = \psi^{\dagger}_{-\mathbf{k}} [U_{\mathcal{T}}]. \label{eq:row_psi_k_T_transform}
\end{equation}
As in SN~\ref{sec:proof-IVGI-dag}, we then consider the expression
\begin{equation}
    \mathcal{T} \psi^{\dagger}_{\mathbf{k}+\mathbf{G}} \mathcal{T}^{-1}. \label{eq:TpsikTinverse}
\end{equation}
First, using SEq.~\eqref{eq:row_psi_k_plus_G} and then SEq.~\eqref{eq:row_psi_k_T_transform}, we can write SEq.~\eqref{eq:TpsikTinverse} as
\begin{align}
    \mathcal{T} \psi^{\dagger}_{\mathbf{k}+\mathbf{G}} \mathcal{T}^{-1} &= \mathcal{T} \psi^{\dagger}_{\mathbf{k}} [V(\mathbf{G})] \mathcal{T}^{-1} \nonumber \\
    &= \mathcal{T} \psi^{\dagger}_{\mathbf{k}} \mathcal{T}^{-1} \mathcal{T} [V(\mathbf{G})] \mathcal{T}^{-1}\nonumber \\
    & = \mathcal{T} \psi^{\dagger}_{\mathbf{k}} \mathcal{T}^{-1}  [V(\mathbf{G})]^{*} \nonumber \\
    &= \psi^{\dagger}_{-\mathbf{k}} [U_{\mathcal{T}}][V(\mathbf{G})]^{*}. \label{eq:T-act-on-psi-k-G-1}
\end{align}
Notice that in SEq.~(\ref{eq:T-act-on-psi-k-G-1}) we have also used 
\begin{equation}
    \mathcal{T} [V(\mathbf{G})] \mathcal{T}^{-1}=[V(\mathbf{G})]^{*} \label{eq:T-complex-conj-V}
\end{equation}
which is a consequence of the antiunitarity of $\mathcal{T}$. 

Second, using SEq.~\eqref{eq:row_psi_k_T_transform} and then SEq.~\eqref{eq:row_psi_k_plus_G}, we can also write SEq.~(\ref{eq:TpsikTinverse}) as
\begin{align}
    \mathcal{T} \psi^{\dagger}_{\mathbf{k}+\mathbf{G}} \mathcal{T}^{-1} &= \psi^{\dagger}_{-\mathbf{k}-\mathbf{G}} [U_{\mathcal{T}}] \nonumber \\
    &= \psi^{\dagger}_{-\mathbf{k}} [V(-\mathbf{G})] [U_{\mathcal{T}}]. \label{eq:T-act-on-psi-k-G-2}
\end{align}
Comparing SEqs.~(\ref{eq:T-act-on-psi-k-G-1}) and~(\ref{eq:T-act-on-psi-k-G-2}), we find that
\begin{equation}
    \psi^{\dagger}_{-\mathbf{k}} [U_{\mathcal{T}}][V(\mathbf{G})]^{*} = \psi^{\dagger}_{-\mathbf{k}} [V(-\mathbf{G})] [U_{\mathcal{T}}]. \label{eq:before-arriving-conclusion-of-TVT-inv}
\end{equation}
As in SN~\ref{sec:proof-IVGI-dag}, we use the orthogonality of the tight-binding basis states to deduce that SEq.~(\ref{eq:before-arriving-conclusion-of-TVT-inv}) can only be satisfied if
\begin{equation}
    [U_{\mathcal{T}}][V(\mathbf{G})]^{*} = [V(-\mathbf{G})] [U_{\mathcal{T}}]. \label{eq:before-arriving-conclusion-of-TVT-inv-2}
\end{equation}
In the first-quantized matrix formalism, we can introduce the antiunitary representative
\begin{equation}
    [\mathcal{T}] = [U_{\mathcal{T}}] \mathcal{K} \label{eq:antiunitary-matrix-representative-of-T}
\end{equation}
of $\mathcal{T}$, where $[U_{\mathcal{T}}]$ is the unitary matrix in SEq.~\eqref{eq:T-act-on-c-dagger} that can be obtained from the position-space symmetry data of the tight-binding basis orbitals, and $\mathcal{K}$ is the complex conjugation operator. 
SEq.~\eqref{eq:antiunitary-matrix-representative-of-T} also implies that $[\mathcal{T}]^{-1} = \mathcal{K} [U_{\mathcal{T}}]^{\dagger}$. 
Using the antiunitary representative $[\mathcal{T}]$ of time-reversal, SEq.~\eqref{eq:before-arriving-conclusion-of-TVT-inv-2} can then be written as
\begin{equation}
    [\mathcal{T}] [V(\mathbf{G})] [\mathcal{T}]^{-1} = [V(-\mathbf{G})]. \label{eq:after-arriving-conclusion-of-TVT-inv}
\end{equation}

\section{Bulk Spin Hall Conductivity}\label{sec:bulk_spin_hall_conductivity}

In this section, we will use the Kubo formula of SRef.~\cite{Monaco2022shc_nonconserved_spin} and SEqs.~\eqref{eq:shckubo1}--\eqref{eq:intrinsicspinhall} to compute the bulk contribution to the spin Hall conductivity in the clean limit for models and materials considered in this work. 
We start in SN~\ref{sec:shc_comp} by reviewing the computational techniques used to evaluate the Kubo formula. 
Next, in SN~\ref{sec:2dshc} we compute the spin-$s_z$ Hall conductivity for the two-dimensional spin-stable $C_s=4$ quantum spin Hall insulator introduced in SN~\ref{sec:spin_stable_topology_2d_fragile_TI}, where in here and SN~\ref{sec:2dshc} we denote the spin Chern number as $C_s$ instead of $C^s_{\gamma_1}$ in SN~\ref{sec:spin_stable_topology_2d_fragile_TI}, for simplicity.
We will pay special attention to how the strength of $s_z$-conservation-breaking SOC affects the spin Hall conductivity. 
Next, in SN~\ref{sec:3dshc} we will introduce the \emph{layer}-resolved spin Hall conductivity for three-dimensional systems. 
We will compute the layer-resolved spin Hall conductivity for the T-DAXI model introduced in SEq.~\eqref{eq:helical_HOTI_TB_model}, and compare it with the layer-resolved spin Chern number computed in SN~\ref{sec:layer-resolved-Cs-of-a-helical-HOTI}.

\subsection{Computational Details}\label{sec:shc_comp}

The spin conductivity tensor $\sigma^{s,i}_{\mu\nu}$ parametrizes the linear response of the spin current $\mathbf{J}^{s,i}$ to an applied DC electric field $\mathbf{E}$, via
\begin{equation}
\langle J^{s,i}_\mu\rangle = \sum_\nu \sigma^{s,i}_{\mu\nu}E_\nu.
\end{equation}
Here $\mu$ and $\nu$ index spatial coordinates (with respect to, \emph{e.g.} a Cartesian basis, as we will elaborate on in Sec.~\ref{sec:3dshc}), and $i=x,y,z$ indexes the spin direction. 
The spin conductivity can be evaluated using the standard Kubo formula
\begin{equation}\label{eq:spinhallkubo}
\sigma^{s,i}_{\mu\nu} = \lim_{\epsilon\rightarrow 0} \int_0^\infty dt \langle \left[J^{s,i}_\mu(t), X_\nu(0)\right]\rangle e^{-\epsilon t},
\end{equation}
where $X_\nu$ is the $\nu$ component of  the position operator (which couples to the external electric field in the Hamiltonian), the time-dependence of operators is evaluated in the Heisenberg picture using the unperturbed ($\mathbf{E}=0$) Hamiltonian $H_0$, and the average is with respect to the  unperturbed ground state. 
There are two main obstacles to the direct evaluation of SEq.~\eqref{eq:spinhallkubo} for a tight-binding model of a system. 
First, we must identify the spin current operator $\mathbf{J}^{s,i}$. 
As proposed in SRefs.~\cite{shi2006proper,Monaco2022shc_nonconserved_spin,tokatly2006magnetoelasticity,gorini2012onsager}, we adopt the definition
\begin{equation}
J^{s,i}_{\mu} = \frac{\partial}{\partial t}\left(X_\mu s^i\right) = i\left[H_0, X_\mu s^i\right].\label{eq:spincurrentdef}
\end{equation}
Defining the spin current via SEq.~\eqref{eq:spincurrentdef} ensures that the spin conductivity satisfies the Onsager reciprocity relations~\cite{gorini2012onsager}. 
This is crucial for relating the spin conductivity to experimental observables, since Onsager reciprocity relates the spin conductivity [SEq.~\eqref{eq:spinhallkubo}] to the inverse spin conductivity that is measured by injecting a spin current into a sample and measuring the induced voltage.
The definition [SEq.~\eqref{eq:spincurrentdef}] corresponds to adding to the ``conventional spin current'' $\mathbf{J}^{s,i}_\mathrm{conv}=1/2\{\mathbf{v},s^i\}$ the curl-free contributions to the spin torque~\cite{shi2006proper}. 

Having defined the spin current operator, we can attempt to evaluate SEq.~\eqref{eq:spinhallkubo}. 
One difficulty that arises is that the Kubo formula involves the position operator, both explicitly in SEq.~\eqref{eq:spinhallkubo} and implicitly via the definition [SEq.~\eqref{eq:spincurrentdef}] of the spin current. 
We must take care then in applying SEq.~\eqref{eq:spinhallkubo} to infinite or periodic systems, where the position operator may not be well-defined. 
SRef.~\cite{Monaco2022shc_nonconserved_spin} gives a careful exposition of these issues. 
In particular, since $X_\nu(0)$ appears in the commutator in SEq.~\eqref{eq:spinhallkubo}, only the off-diagonal matrix elements of $X_\nu$, which are well-defined even in infinite or periodic system: the off-diagonal matrix elements of $X_\nu$ are expressible in terms of the Berry connection.  
Thus, when evaluating $\sigma_{\mu\nu}^{s,i}$ using SEq.~\eqref{eq:spinhallkubo} we are free to take our system to be infinite in the $\nu$ direction. 

More subtle is the position operator appearing in SEq.~\eqref{eq:spincurrentdef} for the spin current. 
When the $i$-component of spin is conserved, then SEq.~\eqref{eq:spincurrentdef} only involves off-diagonal components of the position operator $X_\mu$, and we have no issues. 
However, when $s^i$ is not conserved, $J_\mu^{s,i}$ involves both diagonal and off-diagonal matrix elements of $X_\mu$, and hence is not manifestly well-defined for periodic or infinite systems. 
Nevertheless, SRef.~\cite{Monaco2022shc_nonconserved_spin} showed that for insulators in the thermodynamic limit, the average in SEq.~\eqref{eq:spinhallkubo} is given by a constant times the number of unit cells of the system, and so can be evaluated by taking a trace over the degrees of freedom in a single unit cell. 
Thus, we can evaluate the Kubo formula for $\sigma_{\mu\nu}^{s,i}$ when $\mu\neq \nu$ by considering a system finite in the $\mu$-direction (consisting of $N_\mu$ unit cells with periodic boundary conditions) and infinite in the $\nu$ direction. 
Provided we choose the origin for $X_\mu$ to lie at the center of the finite system, then we recover the bulk spin conductivity for $N_\mu$ sufficiently large compared to the inverse bulk energy gap. 

Concretely, we will consider $d$-dimensional tight-binding Hamiltonians defined on a cylinder finite and periodic in the $\mu$ direction (with $d=2$ or $3$). 
The Bloch Hamiltonian for such a system can be written as $H(\mathbf{k})$, where $\mathbf{k}$ is a ($d-1$)-dimensional vector. 
The basis states for the Hilbert space can be written as $\ket{R_\mu\mathbf{k}\alpha s}$ where $R_\mu$ indexes the unit cells in the $X_\mu$ direction, $\alpha$ indexes the orbital degrees of freedom, and $s$ indexes the spin. 
In this basis the energy eigenstates are $\ket{u_{n\mathbf{k}}}=\sum_{R_\mu\alpha s}u_{R_\mu\alpha sn\mathbf{k}}\ket{R_\mu\mathbf{k}\alpha s}$ with energies $\epsilon_{n\mathbf{k}}$. 
We can then rewrite SEq.~\eqref{eq:spinhallkubo} for the off-diagonal components of the spin conductivity tensor as
\begin{equation}\label{eq:shc_implementable}
\sigma^{s,i}_{\mu\neq\nu} = \frac{i}{(2\pi)^{d-1}}\int d^{d-1}k\sum_{\alpha s}\bra{0\mathbf{k}\alpha s} \left[H(\mathbf{k}),X_\mu s^i\right] \Pi_\nu(\mathbf{k})\ket{0\mathbf{k}\alpha s},
\end{equation}
where $H(\mathbf{k})$ is the Bloch Hamiltonian and $\Pi_\nu$ is the first-order correction to the projection operator given by
\begin{equation}\label{eq:p1def}
\Pi_\nu(\mathbf{k}) = \sum_{\substack{n \in \mathrm{occ} \\ m \in  \mathrm{unocc}}} \frac{i}{\epsilon_{m\mathbf{k}}-\epsilon_{n\mathbf{k}}}\left(
\ket{u_{n\mathbf{k}}}\bra{u_{n\mathbf{k}}}\ket{\partial_\nu u_{m\mathbf{k}}}\bra{u_{m\mathbf{k}}} + \ket{u_{m\mathbf{k}}}\bra{u_{m\mathbf{k}}}\ket{\partial_\nu u_{n\mathbf{k}}}\bra{u_{n\mathbf{k}}}
\right),
\end{equation}
where $\partial_\nu \equiv \partial / \partial k_\nu$.
To compute $\Pi_\nu$ numerically, we note that the numerator can be computed in a gauge invariant form,
\begin{equation}\label{eq:projderivs}
\sum_{\substack{n \in \mathrm{occ} \\ m \in \mathrm{unocc}}}\left(
\ket{u_{n\mathbf{k}}}\bra{u_{n\mathbf{k}}}\ket{\partial_\nu u_{m\mathbf{k}}}\bra{u_{m\mathbf{k}}} + \ket{u_{m\mathbf{k}}}\bra{u_{m\mathbf{k}}}\ket{\partial_\nu u_{n\mathbf{k}}}\bra{u_{n\mathbf{k}}}
\right) = Q\partial_\nu PP - P\partial_\nu P Q.
\end{equation}
where $P(\mathbf{k})$ is the projector onto the occupied states at $\mathbf{k}$, and $Q(\mathbf{k})=1-P(\mathbf{k})$.
SEq.~\eqref{eq:projderivs} can be evaluated using a symmetric finite difference approximation to the derivative $\partial_\nu$. 
We can then isolate the off-diagonal matrix elements and divide by the energy denominators in SEq.~\eqref{eq:p1def} to compute $\Pi_\nu$.

In the subsequent sections, we will apply SEq.~\eqref{eq:shc_implementable} to compute the spin Hall conductivity for models of spin-stable topological phases in two and three dimensions. 
We begin by computing the spin Hall conductivity $\sigma^{s,z}_{xy}$ for the fragile two-dimensional topological insulator analyzed in SN~\ref{sec:spin_stable_topology_2d_fragile_TI}.

\subsection{Spin Hall Conductivity of the 2D Spin-Stable Quantum Spin Hall Insulator}\label{sec:2dshc}

To begin, we consider the square lattice fragile TI with Hamiltonian in SEq.~\eqref{eq:H_2d_fragile_original} and with additional trivial bands coupled via SEq.~\eqref{eq:V_C_coupled_to_HF}. 
As shown in SN~\ref{sec:spin_stable_topology_2d_fragile_TI}, this model realizes a spin-stable topological phase with spin-$s_z$ Chern number $C_s=4$. 
When the spin-orbit coupling parameter $v_{M_z}$ in SEq.~\eqref{eq:H_2d_fragile_original} is zero, $s_z$ is conserved and the model has two pairs of counterpropagating edge states and a bulk Wilson loop with nonzero helical winding~\cite{wieder2020strong}. 
When $v_{M_z}\neq 0$ however, we showed how the spin Chern number could be computed from the net winding of the spin-resolved Wilson loop (SFig.~\ref{fig:P6_wilson_loop_2d_fragile}). 

The fact that $C_s=4$ even when $v_{M_z}\neq 0$ suggests that the fragile TI should have a large spin-Hall conductivity even in the absence of spin-$s_z$ conservation. 
To justify this, we use SEq.~\eqref{eq:shc_implementable} to compute the spin-$s_z$ Hall conductivity $\sigma^{s,z}_{xy}$ as a function of $v_{M_z}$. 
Note that due to the $4mm$ symmetry of the Hamiltonian [SEqs.~\eqref{eq:H_2d_fragile_original} and \eqref{eq:V_C_coupled_to_HF}], the spin conductivity tensor satisfies
\begin{equation}
\sigma^{s,z}_{xy}=-\sigma^{s,z}_{yx}. \label{eq:spinhallantisymmetry}
\end{equation}
This follow from the fact that $m_{110}$ symmetry (which is an element of wallpaper group $p4mm$) maps $x\rightarrow -y, y\rightarrow -x,$ and $s_z\rightarrow -s_z$. 
Applying $m_{110}$ to the Kubo formula [SEq.~\eqref{eq:spinhallkubo}] yields the antisymmetry condition [SEq.~\eqref{eq:spinhallantisymmetry}].

\begin{figure}[ht]
\includegraphics[width=0.5\textwidth]{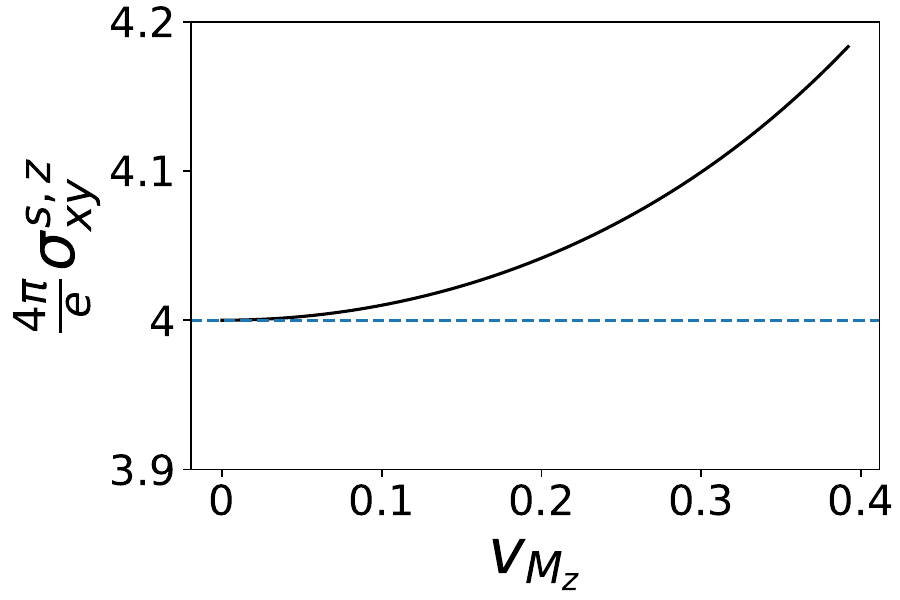}
\caption{Spin-$s_z$ Hall conductivity $\sigma^{s,z}_{xy}$ for the fragile TI model given by SEqs.~\eqref{eq:H_2d_fragile_original} and \eqref{eq:V_C_coupled_to_HF}, plotted as a function of $s_z$-nonconserving SOC strength $v_{M_z}$.}\label{fig:fragile_shc}
\end{figure}

 We show the computed spin Hall conductivity for the fragile TI model in SFig.~\ref{fig:fragile_shc}. 
 When $v_{M_z}=0$, we see that the spin Hall conductivity is given by
$\sigma_{xy}^{s,z}=e/(4\pi) C_s = 4e/(4\pi)$ as expected from SEq.~\eqref{eq:intrinsicspinhall}. 
Away from $v_{M_z}=0$, we see that the spin Hall conductivity increases quadratically as a function of $v_{M_z}$, consistent with the analysis of SRef.~\cite{Monaco2022shc_nonconserved_spin} on other two-dimensional topological systems. 
For weak spin-orbit coupling, we thus expect the spin Hall conductivity to be perturbatively close to the quantized value of $e/(4\pi)C_s$ for spin-stable topological phases in two dimensions.

\subsection{Layer-Resolved Spin Hall Conductivity of T-DAXIs in 3D}\label{sec:3dshc}

Next, we examine the spin Hall conductivity of helical HOTIs in the T-DAXI regime. 
To do so, let us return to SEq.~\eqref{eq:shc_implementable} for the off-diagonal spin conductivity tensor. 
In analogy with Secs.~\ref{sec:layer-resolved-P-Chern-number} and \ref{sec:layer-resolved-Cs-of-a-helical-HOTI}, for a quasi-2D slab we can organize the sum over orbitals $\alpha,s$ in the unit cell in SEq.~\eqref{eq:shc_implementable} to identify the \emph{layer-resolved} spin conductivity. 

Concretely, we consider a tight-binding model for a three-dimensional system with primitive Bravais lattice vectors $\mathbf{a}_1,\mathbf{a}_2,$ and $\mathbf{a}_3$. 
We consider a two-dimensional slab finite along $\mathbf{a}_3$ with $N_3$ unit cells. 
We take there to be $N_\mathrm{sta}=2N_\mathrm{orb}$ tight-binding basis states per unit cell. 
We will compute the spin conductivity in reduced coordinates, defined as
\begin{equation}
\sigma_{a\neq b}^{s,i} = \frac{i}{(2\pi)}\int dk_b \sum_{\alpha s}\bra{0k_b\alpha s} [H(k_b),\mathbf{G}_a\cdot\mathbf{X}s^i]\mathbf{G}_b\cdot\Pi(k_b)\ket{0k_b\alpha s},\label{eq:spinhallprimitive}
\end{equation}
where $\{\mathbf{G}_a\}$ are the reciprocal lattice vectors dual to $\{\mathbf{a}_a\}$. 
The sum over $\alpha$ runs over the $N_\mathrm{orb}$ orbitals in each of the $N_3$ unit cells, allowing us to write
\begin{equation}
\ket{0k_b\alpha s} = \ket{0k_bn_3\beta s},
\end{equation}
where $n_3=1,\dots N_3$ indexes the unit cell in the $\mathbf{a}_3$ direction of the slab, and $\beta=1,\dots N_\mathrm{orb}$ indexes the orbitals in each unit cell. 
Inserting this parametrization into SEq.~\eqref{eq:spinhallprimitive}, we have
\begin{align}
\sigma_{a\neq b}^{s,i} &= \sum_{n_3} \int dk_b \frac{i}{(2\pi)}\sum_{\beta}\bra{0k_bn_3 \beta s} [H(k_b),\mathbf{G}_a\cdot\mathbf{X}s^i]\mathbf{G}_b\cdot\Pi(k_b)\ket{0k_bn_3\beta s} \\
&= \sum_{n_3} \sigma_{a\neq b}^{s,i}(n_3),
\end{align}
where we have defined the layer-resolved spin conductivity
\begin{equation}\label{eq:lrshcdef}
\sigma_{a\neq b}^{s,i}(n_3) = \int dk_b \frac{i}{(2\pi)}\sum_{\beta}\bra{0k_bn_3 \beta s} [H(k_b),\mathbf{G}_a\cdot\mathbf{X}s^i]\mathbf{G}_b\cdot\Pi(k_b)\ket{0k_bn_3\beta s}.
\end{equation}
We will focus in particular on the layer-resolved spin Hall conductivity
\begin{equation}
\sigma_{H}^{s,i}(n_3)=\frac{1}{2}\left(\sigma_{12}^{s,i}(n_3)-\sigma_{21}^{s,i}(n_3)\right),
\end{equation}
which gives the contribution of layer $n_3$ to the spin Hall conductivity
\begin{equation}
\sigma_{H}^{s,i} = \sum_{n_3}\sigma_{H}^{s,i}(n_3)
\end{equation}
of a 2D slab. 

Let us now employ SEq.~\eqref{eq:lrshcdef} to compute the layer-resolved spin Hall conductivity for the model of a symmetry-indicated helical HOTI with inversion and time-reversal symmetry introduced in SRef.~\cite{wieder2018axion} and analyzed in SN~\ref{sec:numerical-section-of-nested-P-pm} and \ref{sec:layer-resolved-Cs-of-a-helical-HOTI}. 
In SN~\ref{sec:numerical-section-of-nested-P-pm} we demonstrated that this model has a spin gap and realizes the T-DAXI spin stable topological phase. 
In SN~\ref{sec:layer-resolved-Cs-of-a-helical-HOTI} and Supplementary Table~\ref{tab:summed_layer_resolved_partial_Chern_numbers} we further showed that the two-dimensional surfaces of the model realized an anomalous odd integer spin Chern number, and we argued that this implied an anomalous topological contribution to the surface spin Hall conductivity. 
We can now substantiate this argument by computing the layer-resolved spin Hall conductivity for the T-DAXI slab.

We begin by constructing a semi-infinite slab of a T-DAXI. 
Recall that our model has an orthorhombic lattice.
Normalizing the lattice constants to $1$, we take the three position-space primitive lattice vectors to be $\mathbf{a}_1 = \widehat{x}$, $\mathbf{a}_2 = \widehat{y}$, and $\mathbf{a}_3 = \widehat{z}$. 
The dual primitive reciprocal lattice vectors are $\mathbf{G}_1 = 2\pi \widehat{x}$, $\mathbf{G}_2 = 2\pi \widehat{y}$, and $\mathbf{G}_3 = 2\pi \widehat{z}$, satisfying $\mathbf{a}_i \cdot \mathbf{G}_j = 2\pi \delta_{ij}$ ($i,j=1\ldots 3$).
Our model has eight bands, for which the matrix Bloch Hamiltonian is given explicitly in SEq.~\eqref{eq:helical_HOTI_TB_model}.
The tight-binding parameters, specified in SEq.~\eqref{eq:parameter_choice_of_helical_HOTI}, together with spin-non-conserving SOC term [$A_{\mathrm{spin-mixing}}$ in SEq.~\eqref{eq:helical_HOTI_TB_model}] are chosen such that the energy spectrum of the surfaces with normal vectors $\pm \widehat{x}$, $\pm \widehat{y}$, and $\pm \widehat{z}$ are all gapped.
We construct a 2D inversion- and time-reversal-symmetric helical HOTI slab that is finite along $\mathbf{a}_3$ with $15$ unit cells and infinite along $\mathbf{a}_1$ and $\mathbf{a}_2$. 
We then compute the layer-resolved spin-$s_z$ Hall conductivity $\sigma_H^{s,z}(n_3)$ for the helical HOTI slab with both $s_z$ conservation ($A_{\mathrm{spin-mixing}}=0$) and with large spin non-conserving SOC ($A_{\mathrm{spin-mixing}}=0.5$). 

\begin{figure}[ht]
\centering\captionsetup[subfloat]{labelformat=simple,labelfont={large,bf},position=t,singlelinecheck=off,justification=raggedright}
\subfloat[]{
	\includegraphics[width=0.4\textwidth]{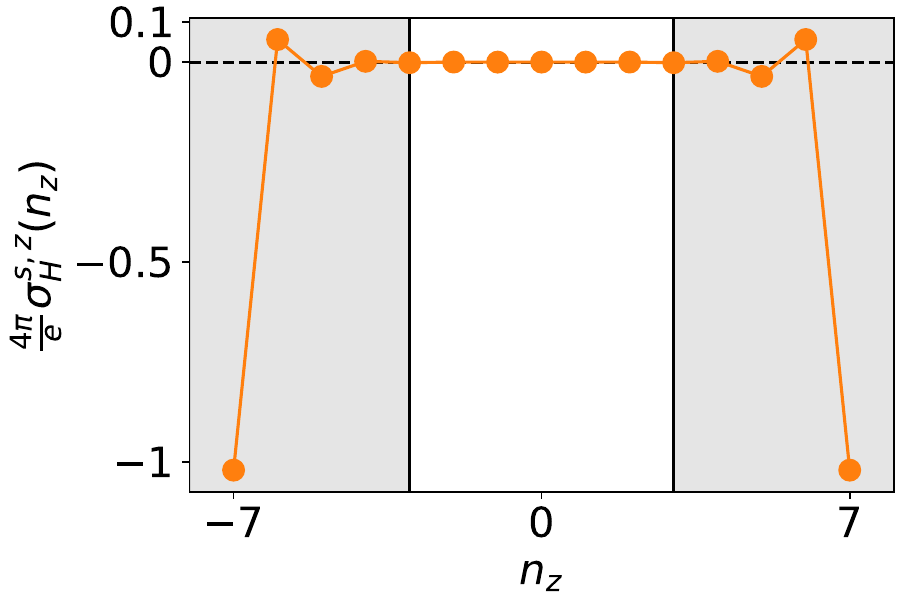}
}\quad
\subfloat[]{
	\includegraphics[width=0.4\textwidth]{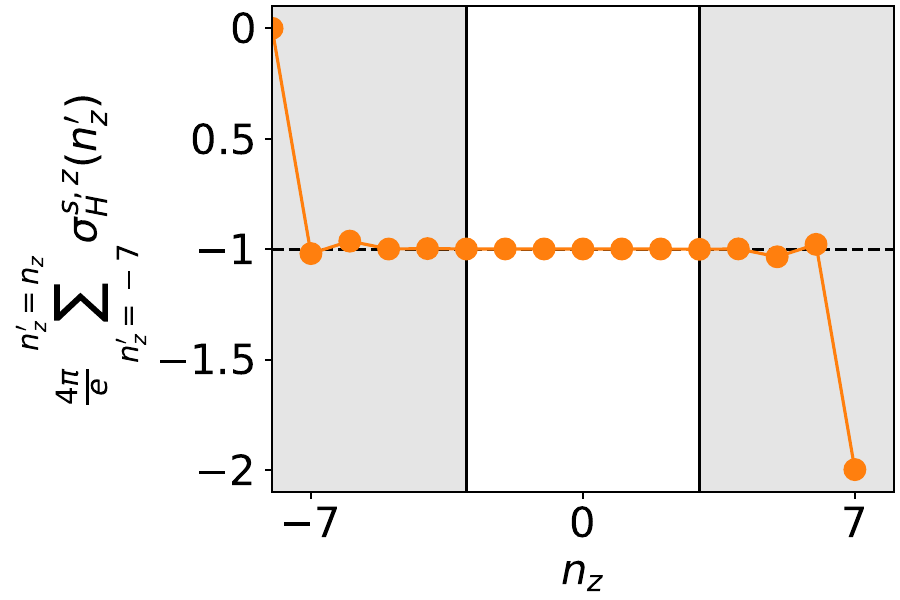}
}
\caption{Layer-resolved spin Hall conductivity $\sigma_{H}^{s,z}(n_z)$ of a 2D slab formed from our model of an $\mathcal{I}$- and $\mathcal{T}$-symmetric helical HOTI [SEq.~\eqref{eq:helical_HOTI_TB_model}] in the T-DAXI regime (see SN~\ref{sec:numerical-section-of-nested-P-pm}) with spin $s_z$ conservation ($A_{\mathrm{spin-mixing}}=0.0$). 
(a) shows the layer-resolved spin Hall conductivity $\sigma_{H}^{s,z}(n_z)$ for a 2D slab finite along $z$ with $15$ unit cells and infinite along $x$ and $y$.
(b) shows the cumulative spin Hall conductivity $\sum_{n_{z}'=-7}^{n_{z}'=n_{z}} \sigma_{H}^{s,z}(n'_z)$ as a function of $n_z$ beginning from the bottom layer in (a).
As we can see in (a), the nonzero values of $\sigma_{H}^{s,z}(n_z)$ are concentrated around the top and bottom layers. 
In addition, (b) demonstrates the appearance of anomalous odd-integer spin Hall conductivity (in units of $e/(4\pi)$) around the gapped surfaces of the T-DAXI, as the cumulative values of $\sigma_{H}^{s,z}(n_z)$ beginning from the bottom layer quickly converge to $-e/4\pi$ and remain constant in the bulk.}\label{fig:tdaxi_shc_a0}
\end{figure}

SFig.~\ref{fig:tdaxi_shc_a0} shows the layer-resolved spin Hall conductivity in the $s_z$-conserving limit $A_{\mathrm{spin-mixing}}=0$, in units of $e/(4\pi)$. 
In this limit, the spin Hall conductivity is given entirely by the topological contribution defined in SEq.~\eqref{eq:intrinsicspinhall}. 
We see from SFig.~\ref{fig:tdaxi_shc_a0}(a) that the layer-resolved spin Hall conductivity is zero deep in the bulk of the slab and nonzero near the surfaces. 
This is consistent with the spin-resolved layer construction of the T-DAXI presented in SN~\ref{sec:defs_partial_sis}. 
In SFig.~\ref{fig:tdaxi_shc_a0}(b) we show the cumulative spin Hall conductivity $\sum_{n_z'=-7}^{n_z'=n_z}\sigma_{H}^{s,z}(n_z)$. 
Summing over the five layers closest to the surface, we find that each surface contributes 
\begin{equation}
\sigma_{H,\mathrm{surface}}^{s,z}=\sum_{n_z'=-7}^{n_z'=-3}\sigma_{H}^{s,z}(n_z)=-0.999e/(4\pi)\approx (-1)e/(4\pi)
\end{equation}
to the spin Hall conductivity. 
This is consistent with our computation in SN~\ref{sec:layer-resolved-Cs-of-a-helical-HOTI}, where we showed that the surface spin-$s_z$ Chern number for this helical HOTI slab is given by $C_{xy}^{+}-C_{xy}^{-} = -1$. 
We thus see that in the $s_z$-conserving limit, the odd-integer surface spin Chern number of the T-DAXI implies a quantized odd-integer (in units of $e/(4\pi)$) surface spin Hall conductivity. 
The total spin Hall conductivity $\sigma_{H}^{s,z}=\sum_{n_z'=-7}^{n_z'=7}\sigma_{H}^{s,z}(n_z)=(-2)e/(4\pi)$, consistent with the partial Chern number of the T-DAXI slab given in Supplementary Table~\ref{tab:summed_layer_resolved_partial_Chern_numbers}. 
In comparing SFig.~\ref{fig:tdaxi_shc_a0} for the layer-resolved spin Hall conductivity with SFig.~\ref{fig:layer-resolved-partial-Chern-number-of-helical-HOTI} for the layer-resolved partial Chern number, we see that although the cumulative spin Hall conductivity for the top surface is proportional to the surface spin Chern number, the layer-resolved spin Hall conductivity is not proportional to the layer-resolved spin Chern number even in this spin-conserving limit. 
In particular, we see from SFig.~\ref{fig:tdaxi_shc_a0} that the layer-resolved spin Hall conductivity oscillates between positive and negative values near each surface, while the layer-resolved spin Chern number in SFig.~\ref{fig:layer-resolved-partial-Chern-number-of-helical-HOTI} decreases monotonically to zero near each surface. 
To understand this discrepancy, we recall from SEqs.~\eqref{eq:shckubo1}--\eqref{eq:intrinsicspinhall} that the spin Hall conductivity can be written as the sum of the spin Chern number and a correction term $\sigma_{II}^s$. 
Although $\sigma_{II}^s$ vanishes in the $s_z$-conserving limit when summed over all layers, it still gives a non-vanishing contribution to the layer-resolved spin Hall conductivity. 
Physically, this is because each layer of the slab is not an isolated 2D system: $s_z$-per layer is not a good quantum number since electrons can hop into and out of each layer. 
Nevertheless, we see that the surface contribution to the spin Hall conductivity (summed over the layers near the surface) are given by $e/(4\pi)$ times the surface contribution to the spin Chern number [SEq.~\eqref{eq:surface_partial_Chern_numbers_def}]. 

\begin{figure}[ht]
\centering\captionsetup[subfloat]{labelformat=simple,labelfont={large,bf},position=t,singlelinecheck=off,justification=raggedright}
\subfloat[]{
	\includegraphics[width=0.4\textwidth]{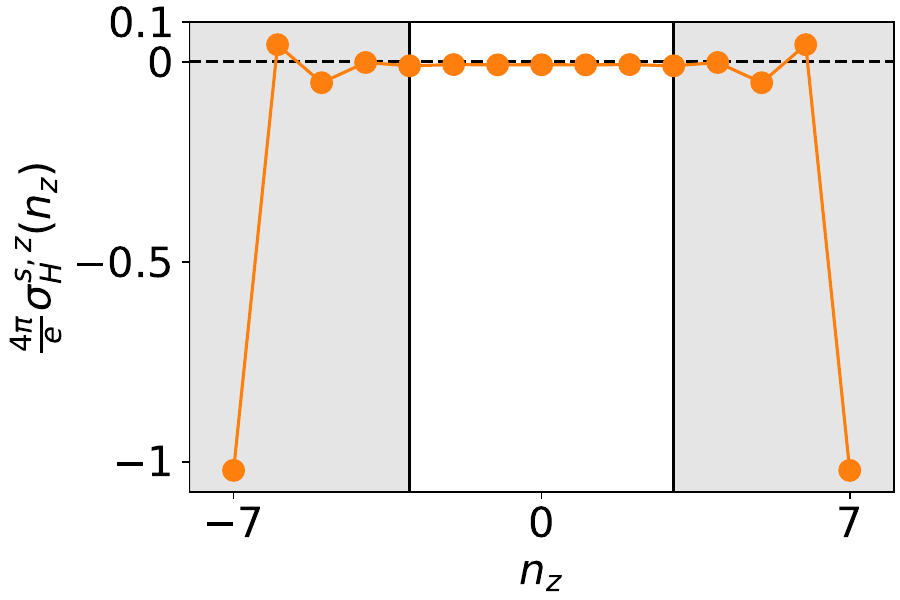}
}\quad
\subfloat[]{
	\includegraphics[width=0.4\textwidth]{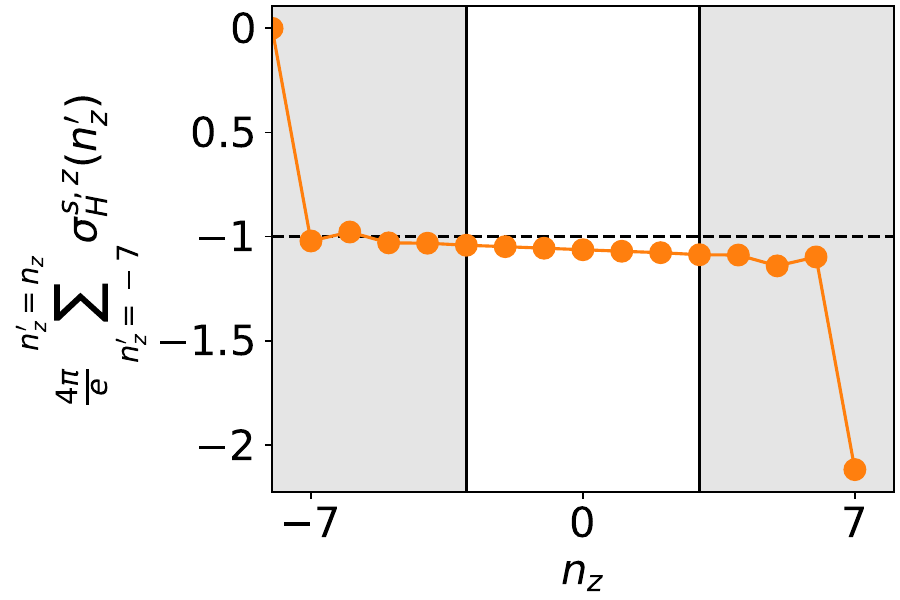}
}
\caption{Layer-resolved spin Hall conductivity $\sigma_{H}^{s,z}(n_z)$ of a 2D slab formed from our model of an $\mathcal{I}$- and $\mathcal{T}$-symmetric helical HOTI [SEq.~\eqref{eq:helical_HOTI_TB_model}] in the T-DAXI regime (see SN~\ref{sec:numerical-section-of-nested-P-pm}) with strong $s_z$ non-conserving SOC ($A_{\mathrm{spin-mixing}}=0.5$). 
(a) shows the layer-resolved spin Hall conductivity $\sigma_{H}^{s,z}(n_z)$ for a 2D slab finite along $z$ with $15$ unit cells and infinite along $x$ and $y$.
(b) shows the cumulative spin Hall conductivity $\sum_{n_{z}'=-7}^{n_{z}'=n_{z}} \sigma_{H}^{s,z}(n'_z)$ as a function of $n_z$ beginning from the bottom layer in (a).
As we can see in (a), the nonzero values of $\sigma_{H}^{s,z}(n_z)$ are largely concentrated around the top and bottom layers, although the spin Hall conductivity remains small but nonzero in the bulk of the system. 
This becomes clearer in (b), where we see both large anomalous topological contributions to the spin Hall conductivity around the gapped surfaces of the T-DAXI, as well as a residual small nontopological contribution from the bulk of the system.}\label{fig:tdaxi_shc_a05}
\end{figure}

When $A_{\mathrm{spin-mixing}}=0.5$, $s_z$ is no longer conserved, and so we expect the spin Hall conductivity to receive both topological and non-topological contributions per SEqs.~\eqref{eq:shckubo1}--\eqref{eq:intrinsicspinhall}. 
In SFig.~\ref{fig:tdaxi_shc_a05} we show the layer-resolved spin Hall conductivity $\sigma_{H}^{s,z}(n_z)$ [panel (a)] and the layer-summed cumulative spin Hall conductivity $\sum_{n_z'=-7}^{n_z'=n_z}\sigma_{H}^{s,z}(n_z)$ [panel (b)] For the T-DAXI model with $A_{\mathrm{spin-mixing}}=0.5$. 
Since $s_z$ is no longer conserved, the bulk spin Hall conductivity is no longer zero, but receives small corrections; we find that $\sigma_{H}^{s,z}(n_z=0)\approx-0.007e/(4\pi)$ in the center of the slab. 
Similarly, the total spin Hall conductivity of the slab deviates from the spin Chern number, and is given by $\sigma_{H}^{s,z}=\sum_{n_z'=-7}^{n_z'=7}\sigma_{H}^{s,z}(n_z)\approx -2.11 e/(4\pi)$. 
Nevertheless, we see from SFig.~\ref{fig:tdaxi_shc_a05}(a) that the largest contributions to the spin Hall conductivity come from the anomalous gapped surfaces. 
Summing over the five layers closest to the surface, we find that the surface contribution to the spin Hall conductivity is
\begin{equation}
\sigma_{H,\mathrm{surface}}^{s,z}=\sum_{n_z'=-7}^{n_z'=-3}\sigma_{H}^{s,z}(n_z)=-1.042e/(4\pi).
\end{equation}

\section{Spin-Resolved Topology in Real Materials}
\label{app:materials}

Importantly, the spin-resolved Wilson loop and nested spin-resolved Wilson loop formalisms developed in SN~\ref{app:Wilson} and~\ref{app:w2} are applicable beyond toy models, and can be applied without modification to deduce the spin-resolved topology of real materials.  Additionally, when there is a spin gap, the spin-resolved topology can then be compared to the intrinsic bulk contribution to the spin Hall conductivity, which can be computed through the Kubo formula as shown in SN~\ref{sec:bulk_spin_hall_conductivity}. 
Below, in SN~\ref{app:mote2} and \ref{sec:bibr}, we will present detailed analyses of the spin-resolved topology and physical observables for two experimentally accessible candidate helical HOTI materials: $\beta$-MoTe$_2$~[\href{https://www.topologicalquantumchemistry.com/#/detail/14349}{ICSD 14349}, space group (SG) $P2_{1}/m1'$ (\#11.51) in magnetic (Shubnikov) notation, SG 11 ($P 2_1 / m$) in nonmagnetic notation]~\cite{wang2019higherorder,tang2019efficient,elcoro2021magnetic} and $\alpha$-BiBr~[\href{https://www.topologicalquantumchemistry.com/#/detail/1560}{ICSD 1560}, SG $C2/m1'$ (\#12.59) in magnetic notation, SG $12$ ($C2/m$) in nonmagnetic notation]~\cite{liu2016weak,tang2019efficient,SYBiBr,BiBrFanHOTI,noguchi2021evidence,FanZahidRoomTempBiBrExp}. 
In the remainder of this section, we will briefly summarize our application of spin-resolved topological analysis to real materials and numerical demonstration of the associated physical signatures of nontrivial spin-resolved topology; complete details are provided below in SN~\ref{app:mote2} and \ref{sec:bibr}.

First, in SN~\ref{sec:mote2_spin_resolved_topology}, we will show that $\beta$-MoTe$_2$ realizes a spin-Weyl state (SN~\ref{app:response_QSHI_TDAXI}) for all choices of the spin resolution direction. 
We will then demonstrate in SN~\ref{app:mote2_fermi_arc_strong_zeeman} that $\beta$-MoTe$_2$, when subjected to a large external Zeeman field, exhibits $(001)$-surface topological Fermi arcs originating from bulk spin-Weyl points (see SN~\ref{sec:zeeman} for the underlying theoretical details).
Next, in SN~\ref{sec:spin_resolved_topology_of_alpha_bibr}, we will show that $\alpha$-BiBr remarkably exhibits a large spin gap across a significant range of spin resolution directions, and specifically hosts both spin-stable 3D QSHI and T-DAXI states (SN~\ref{sec:defs_partial_sis} and~\ref{app:response_QSHI_TDAXI}).
Lastly, in SN~\ref{sec:bibrshc}, we will numerically demonstrate that the spin-gapped 3D QSHI and T-DAXI states in $\alpha$-BiBr respectively carry nearly quantized and nearly vanishing bulk spin Hall conductivities per unit cell, respectively (see SN~\ref{sec:bulk_spin_hall_conductivity}).    

In our topological analysis of real materials, we will work primarily with Wannier-based approximate tight-binding models. 
We will compute (spin-resolved) Wilson loops and transport coefficients in terms of the tight-binding eigenstates. 
We take care to note that in doing so, we will in our analysis ignore contributions to Wilson loops and transport coefficients arising from the finite extent and spatial profile of the Wannier functions themselves. 
In particular, Wilson loops computed using tight-binding eigenstates can have a different eigenspectrum than Wilson loops computed using exact Bloch eigenstates due to off-diagonal matrix elements of the position operator in the Wannier-function basis~\cite{alexandradinata2014wilsonloop,bradlyn2021lecture}. 
Similarly, off-diagonal matrix elements of the position operator in the Wannier-function basis lead to modifications of the current and spin-current operators that can alter the numerical value of computed transport coefficients~\cite{parker2019diagrammatic}. 
Since the discrepancy between tight-binding and ab-initio derived quantities originates from the spatial extent of the Wannier functions, we expect that the discrepancies will be exponentially small provided that the Wannier functions are well-localized. 
In particular, for a model with exponentially localized Wannier functions that are spin eigenstates, the (partial) Chern numbers computed from the tight-binding eigenstates will coincide with the (partial) Chern numbers computed from the full ab-initio wave functions. 
In our analysis of real materials detailed below, we have taken care to include sufficient bands in the tight-binding Hilbert space to ensure that the Wannier functions of our tight-binding model are symmetric, exponentially localized, and spin eigenstates.
As such, we expect that the (partial) Chern numbers that we obtain from Wannier-based tight-binding calculations accurately reflect the band topology of the ab-initio wave functions, and we similarly expect our tight-binding-derived numerical calculation of the spin Hall conductivity to be a good approximation of the intrinsic spin Hall conductivity of real materials (here $\alpha$-BiBr).

\section{First-Principles Analysis of $\beta$-$\mathrm{MoTe}_2$}
\label{app:mote2}

In this section, we will compute the spin-resolved topology of 3D $\beta$-phase $\mathrm{MoTe}_2$, which was identified in SRefs.~\cite{wang2019higherorder,tang2019efficient} as a helical HOTI with both inversion ($\mathcal{I}$) and time-reversal ($\mathcal{T}$) symmetries (specifically characterized by a nontrivial $\mathbb{Z}_4$-invariant $z_4=2$ and vanishing weak indices $z_{2i}=0$). 

Below, we will begin our spin-resolved topological analysis of 3D $\beta$-MoTe$_2$ by detailing in SN~\ref{sec:mote2_dft_details} the density-functional-theory (DFT) calculations that we employed to obtain a symmetric, Wannier-based tight-binding model of $\beta$-MoTe$_2$.
In SN~\ref{sec:mote2_spin_resolved_topology}, we will then show that 3D $\beta$-MoTe$_2$ generically lies in the spin-Weyl regime of a helical HOTI [SN~\ref{app:response_QSHI_TDAXI}] for all choices of spin direction by computing the spectrum of the projected spin operator [SEq.~\eqref{eq:appendix-def-PsP}] and spin-resolved Wilson loops [SN~\ref{sec:P_pm_Wilson_loop}].
Finally in SN~\ref{app:mote2_fermi_arc_strong_zeeman}, we will demonstrate the appearance of $(001)$-surface topological Fermi arcs in $\beta$-MoTe$_2$ subjected to a strong (spin-) Zeeman field, which represent a physical signature of its bulk spin-Weyl points (see SN~\ref{sec:zeeman}).

\begin{figure}[ht]
\includegraphics[width=\textwidth]{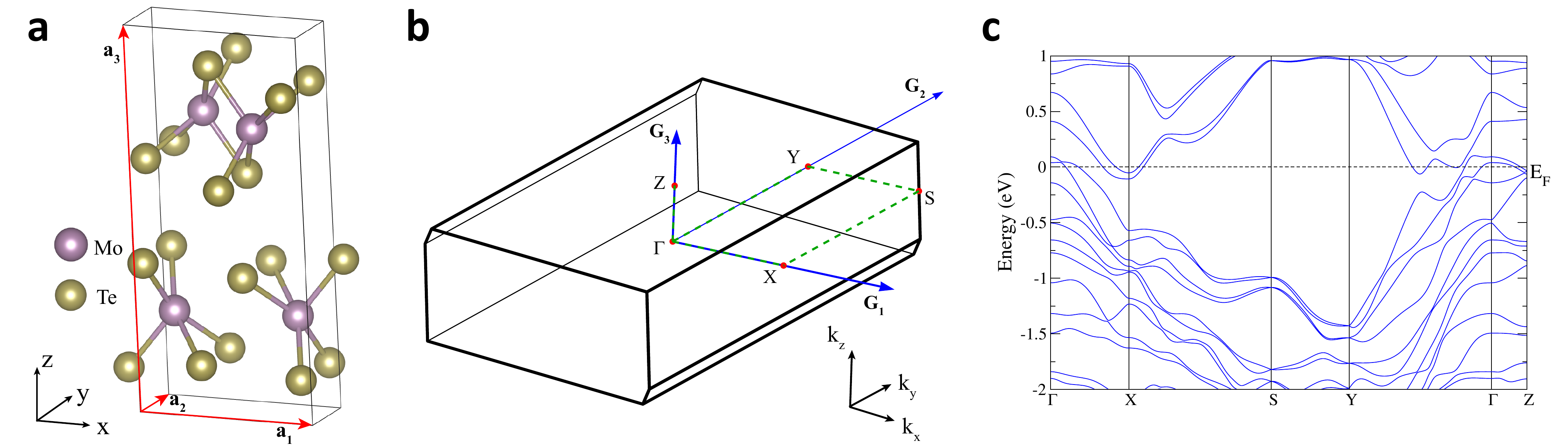}
\caption{First-principles electronic structure of $\beta$-MoTe$_2$.
(a) The crystal structure of $\beta$-MoTe$_2$, which respects the symmetries of space group (SG) $P2_1/m1'$ ($\# 11.51$)~\cite{MoTe2Structure} .
There are four Mo atoms and eight Te atoms in each primitive (unit) cell.
In (a), the red (black) arrows denote the primitive-cell lattice (Cartesian-unit) vectors [SEq.~(\ref{eq:mote2lattice})].
(b) The bulk Brillouin zone (BZ) for $\beta$-MoTe$_2$; the primitive reciprocal lattice vectors $\mathbf{G}_{i}$ (blue arrows) are given in SEq.~(\ref{eq:mote2ReciprocalLattice}), and notably differ from the Cartesian reciprocal unit vectors (black arrows).
In (b), the time-reversal-invariant ${\bf k}$ (TRIM) points are labeled for consistency using the convention previously employed in SRef.~\cite{wang2019higherorder}, which we note differs from the labeling convention for SG $P2_1/m1'$ ($\# 11.51$) on the Bilbao Crystallographic Server~\cite{aroyo2006bilbaoa,aroyo2006bilbao,aroyo2011crystallography}.
(c) The first-principles- (DFT-) obtained electronic structure of $\beta$-MoTe$_2$ plotted along the green dashed ${\bf k}$-path in (b). The Fermi energy is denoted as $E_{F}$.}
\label{fig:abinit}
\end{figure}

\subsection{Details of Density Functional Theory Calculations on $\beta$-MoTe$_2$}
\label{sec:mote2_dft_details}

The 3D $\beta$ phase of MoTe$_2$ crystallizes a centrosymmetric structure that respects the symmetries of nonmagnetic SG $P2_1/m1'$ (\#11.51). 
Each primitive (unit) cell of $\beta$-MoTe$_2$ contains four Mo atoms and eight Te atoms [SFig.~\ref{fig:abinit}(a)]. 
The primitive Bravais lattice vectors of $\beta$-MoTe$_2$ are given by
\begin{eqnarray}
\mathbf{a}_1 &=& \mathbf{a} = (6.3299999 \text{ \AA}) \hat{\mathbf{x}} = a \hat{\mathbf{x}}, \nonumber \\
\mathbf{a}_2 &=& \mathbf{b} = (3.4690001 \text{ \AA}) \hat{\mathbf{y}} = b \hat{\mathbf{y}}, \nonumber \\
\mathbf{a}_3 &=& \mathbf{c} = -(0.9467946 \text{ \AA}) \hat{\mathbf{x}} + (13.8276235 \text{ \AA}) \hat{\mathbf{z}} = c \cos{\beta} \hat{\mathbf{x}} + c \sin{\beta} \hat{\mathbf{z}},
\label{eq:mote2lattice}
\end{eqnarray}
where $a$, $b$, and $c$ denote the conventional-cell lattice parameters~\cite{MoTe2Structure} 
\begin{equation}
a = |\mathbf{a}| = 6.3299999 \text{ \AA},\ b= |\mathbf{b}| = 3.4690001 \text{ \AA},\ c= |\mathbf{c}| = 13.8599997 \text{ \AA},
\end{equation}
such that for $\beta$-MoTe$_2$, $a$, $b$, and $c$ respectively coincide  with $|\mathbf{a}_1|$, $|\mathbf{a}_2|$, and $|\mathbf{a}_3|$, because nonmagnetic SG $P2_1/m1'$ (\#11.51) is primitive ($P$) monoclinic.  In SEq.~(\ref{eq:mote2lattice}), $\beta \approx 93.91699963419269^{\circ}$ is the angle between $\mathbf{a}_1$ and $\mathbf{a}_3$.
The primitive reciprocal lattice vectors [SFig.~\ref{fig:abinit}(b)] for $\beta$-MoTe$_2$ are correspondingly given by
\begin{align}
& \mathbf{G}_1 = \frac{2\pi}{a}\left( \hat{\mathbf{x}}-\cot{\beta}\hat{\mathbf{z}}\right), \nonumber \\
& \mathbf{G}_2 = \frac{2\pi}{b} \hat{\mathbf{y}}, \nonumber \\
& \mathbf{G}_3 = \frac{2\pi}{c} \csc{\beta} \hat{\mathbf{z}}.
\label{eq:mote2ReciprocalLattice}
\end{align}

We use ab-initio (DFT) calculations incorporating the effects of SOC to compute the electronic band structure of $\beta$-MoTe$_2$. 
Our first-principles calculations were specifically performed within the DFT framework using the projector-augmented wave (PAW) method~\cite{blochl1994improved,kresse1994norm} as implemented in the Vienna ab-initio simulation package (VASP)~\cite{kresse1996efficiency,kresse1996efficient}. 
In our first-principles calculations, we adopted the Perdew-Burke-Ernzerhof (PBE) generalized gradient approximation exchange-correlation functional~\cite{perdew1996generalized}, and 
SOC was incorporated self-consistently. 
The cutoff energy for the plane-wave expansion was 400 eV, and $0.03 \times 2\pi \text{ \AA}^{-1}$ $\mathbf{k}$-point sampling grids were used in the self-consistent process. 
In SFig.~\ref{fig:abinit}(c) we show the ab-initio band structure for $\beta$-MoTe$_2$ computed along high-symmetry lines.  We note that throughout this work, high-symmetry BZ (TRIM) points in the bulk electronic and spin spectra of $\beta$-MoTe$_2$ are labeled for consistency using the convention previously employed in SRef.~\cite{wang2019higherorder}. The TRIM point labels for $\beta$-MoTe$_2$ in this work hence differ from the standard labels for SG $P2_1/m1'$ (\#11.51) listed on the Bilbao Crystallographic Server~\cite{aroyo2006bilbao,aroyo2006bilbaoa,aroyo2011crystallography,bilbaocrystallogrserver2017bandrep,bilbaocrystallogrserver2018check,elcoro2017double,elcoro2021magnetic}.

\begin{figure}[t]
\includegraphics[width=0.9\textwidth]{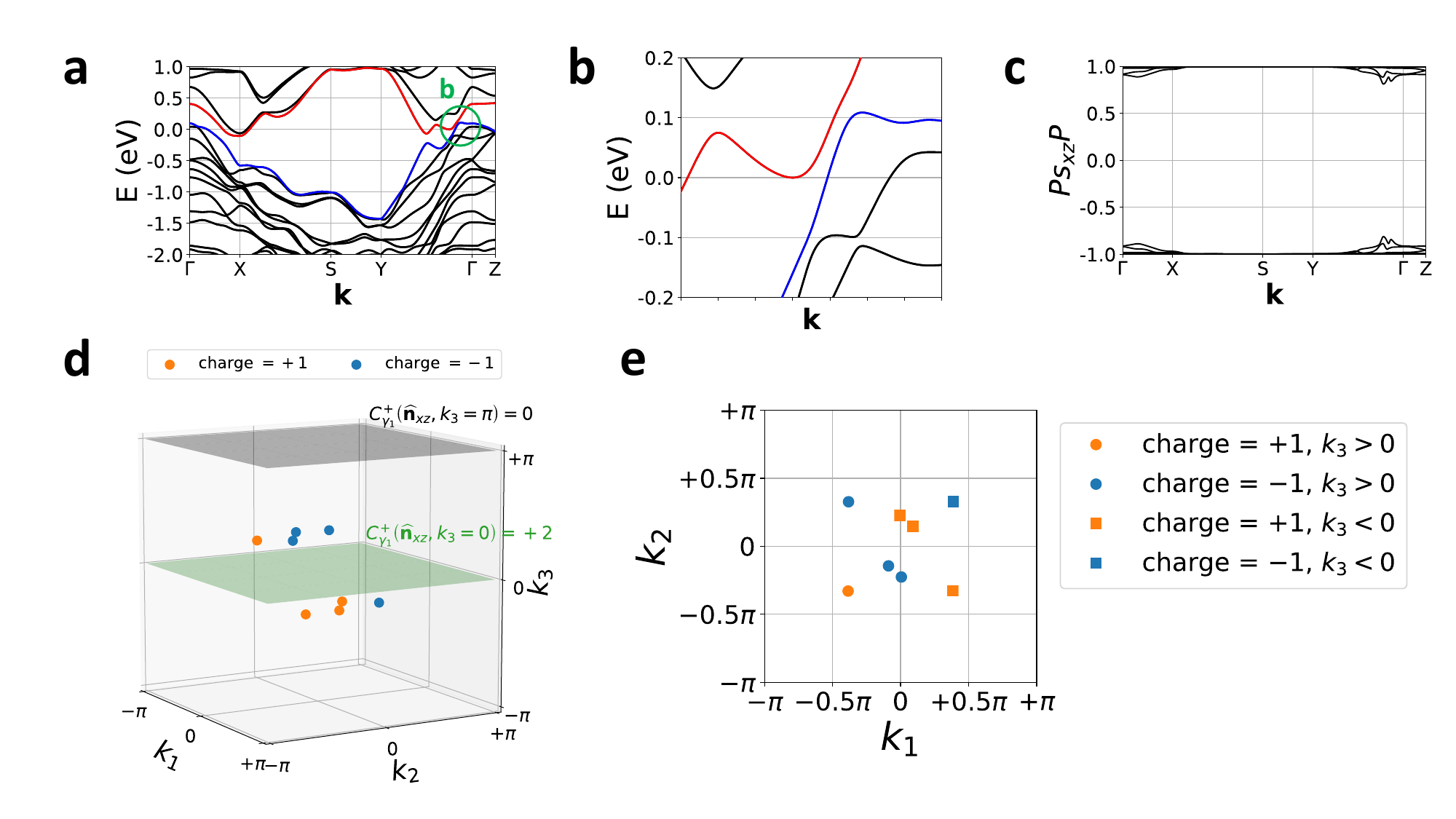}
\caption{Energy and spin spectrum for the ab-initio-derived symmetric Wannier-based tight-binding model of $\beta$-MoTe$_2$. 
(a) shows the band structure for our Wannier-based tight-binding model along high-symmetry lines in the BZ.  
The high-symmetry BZ (TRIM) points are labeled employing the convention in SFig.~\ref{fig:abinit}(c), which follows the notation employed in SRef.~\cite{wang2019higherorder}.
Panel (b) shows an enlarged view of the band structure along the $\Gamma-Y$ line highlighted with a green circle in (a). 
In (b), there is an avoided crossing between the highest valence (blue) and lowest conduction (red) bands, which is a remnant of a monopole-charged Dirac nodal line that is gapped by SOC~\cite{wang2019higherorder}.
The energy bands in (a,b) are doubly degenerate due to the presence of both inversion and spinful time-reversal symmetries~\cite{wieder2016spinorbit}. 
(c) The spin spectrum of $P(s_x+s_z)P/\sqrt{2}\equiv Ps_{xz}P$ plotted along the same high-symmetry BZ path as in (a) [see SFig.~\ref{fig:abinit}(b)]. 
Despite being gapped along the high-symmetry ${\bf k}$-path in (c), the $Ps_{xz}P$ spectrum is in fact gapless, specifically exhibiting eight spin-Weyl nodes at $Ps_{xz}P$ eigenvalue $0$. 
(d) A 3D plot of the locations of the spin-Weyl nodes in the $Ps_{xz}P$ spectrum of $\beta$-MoTe$_2$.  
Each node in (d) is a monopole source of partial Chern number $C^+_{\gamma_1}$ for the positive spin bands [see SN~\ref{appendix:3D-TI-with-and-without-inversion} for the definition of the partial monopole (chiral) charge of a spin-Weyl node].
In (d), spin-Weyl nodes with a positive (negative) partial chiral charge $C^+_{\gamma_1}=1$ ($C^+_{\gamma_1}=-1$) are shown in orange (blue). 
(e) shows the projection of the spin-Weyl nodes from (d) onto the $k_1-k_2$ BZ plane. 
In (e), spin-Weyl nodes in the $k_3>0$ half of the BZ are labeled with circles, and spin-Weyl nodes in the  $k_3<0$ half of the BZ are labeled with squares. 
In the $k_3>0$ half of the BZ in (d,e), there are three spin-Weyl nodes with partial chiral charge $-1$ and one spin-Weyl node with partial chiral charge $+1$. 
Due to time-reversal symmetry, which flips the sign of the chiral charge of spin-Weyl points (see SN~\ref{appendix:explicit-calculation-of-BHZ-model}), there are therefore in the $k_3<0$ half of the BZ three spin-Weyl nodes with partial chiral charge $+1$ and one spin-Weyl node with partial chiral charge $-1$.
The calculations detailed in this figure were performed using the freely available Python package~\href{https://github.com/kuansenlin/nested_and_spin_resolved_Wilson_loop}{\textsc{nested\_and\_spin\_resolved\_Wilson\_loop}}~\cite{lin2023nestedWilsonLib}, which represents an extension of the~\href{https://www.physics.rutgers.edu/pythtb/}{PythTB} open-source Python tight-binding package~\cite{coh2013python} that was implemented and utilized for the preparation of SRefs.~\cite{wieder2018axion,wieder2020strong} and the present work.
}\label{fig:mote2_main_fig}
\end{figure}

Next, to analyze the spin-resolved band topology, we constructed a symmetric, Wannier-based tight-binding model fit to the electronic structure of $\beta$-MoTe$_2$ obtained from our DFT calculations.
We constructed symmetric Wannier functions for the bands near the Fermi energy $E_{F}$ in $\beta$-MoTe$_2$ by using the Wannier90 package~\cite{pizzi2020wannier90} for the Mo $4d$ and the Te $5p$ orbitals,  and then performing a subsequent SG symmetrization using WannierTools~\cite{WU2018405}. 
For the following discussion, we denote the Hamiltonian of the Wannier-based tight-binding model as $[H_{\mathrm{MoTe}_2}]$. 
The single-particle Hilbert space of $[H_{\mathrm{MoTe}_2}]$ consists of 44 spinful Wannier functions per unit cell; 
the Bloch Hamiltonian $[H_{\mathrm{MoTe}_2}(\mathbf{k})]$ is therefore an $88\times 88$ matrix. 
To reduce the computational resources required for our spin-resolved and Wilson loop calculations, we next truncated $[H_\mathrm{MoTe_2}]$ to only contain hopping terms with an absolute magnitude greater than or equal to $0.001$eV. 
We have confirmed that this truncation affects neither the band ordering nor the qualitative features of the band structure near the Fermi energy. Specifically in SFig.~\ref{fig:mote2_main_fig}(a,b), we show the band structure of $\beta$-MoTe$_2$ computed from the Wannier-based tight-binding model $[H_\mathrm{MoTe_2}]$, which shows good agreement with the DFT-obtained band structure in SFig.~\ref{fig:abinit}(c), demonstrating that our truncated Wannier-based tight-binding model well-approximates the electronic structure of $\beta$-MoTe$_2$.   
In both the DFT-obtained and Wannier tight-binding band structures, there is an avoided crossing between the valence and conduction bands along the $\Gamma-Y$ line [enlarged view in SFig.~\ref{fig:mote2_main_fig}(b)], which is a remnant of a monopole-charged Dirac nodal line that is gapped by SOC~\cite{wang2019higherorder}.

In 3D $\beta$-MoTe$_2$, there are $56$ valence electrons per unit cell.  
To analyze the (spin-resolved) topology of the valence electrons in $\beta$-MoTe$_2$, we will hence in the discussion below take the lowest $56$ electronic states in energy of $[H_\mathrm{MoTe_2}]$ to be separately occupied at each ${\bf k}$ point.
Though $\beta$-MoTe$_2$ is a (semi)metal [\emph{i.e.} valence and conduction bands cross $E_{F}$ (set to zero) in SFig.~\ref{fig:mote2_main_fig}(a,b)], the topology of the 56 valence bands of $\beta$-MoTe$_2$ is nevertheless a gauge-invariant quantity.
Specifically, the $56$ occupied (valence) bands in $\beta$-MoTe$_2$ are separated from the unoccupied (conduction) bands by an energy gap at each $\mathbf{k}$ point.
Because there is an energy gap between the 56$^\text{th}$ and 57$\text{th}$ Bloch states at each ${\bf k}$ point [SFig.~\ref{fig:mote2_main_fig}(a)], we can therefore uniquely and consistently define a projector onto the valence bands at each ${\bf k}$ point to characterize their topology across the 3D BZ~\cite{wieder2021topological,wieder2018wallpaper}.
Hence throughout this work, we will analyze $\beta$-MoTe$_2$---despite the existence of electron and hole pockets at $E_{F}$---as if it is an insulator with 56 occupied valence bands.
Lastly, by computing the SIs of our Wannier-based tight-binding model, we find that $\beta$-MoTe$_2$ is specifically a helical HOTI with $(z_4,z_{21},z_{22},z_{23})=(2,0,0,0)$, consistent with the findings of previous works~\cite{wang2019higherorder,tang2019efficient}. Below in SN~\ref{sec:mote2_spin_resolved_topology}, we will next analyze the spin-resolved topology of the $56$ occupied valence bands in $\beta$-MoTe$_2$.

\subsection{Spin-Resolved Topology of $\beta$-MoTe$_2$}
\label{sec:mote2_spin_resolved_topology}

In this section, we will show that the helical HOTI phase of $\beta$-MoTe$_2$ carries the spin-resolved topology of a spin-Weyl semimetal state (SN~\ref{app:response_QSHI_TDAXI}) with an even number of spin-Weyl nodes in each half of the BZ for every choice of spin resolution direction.

\begin{figure}[t]
\includegraphics[width=1.0\textwidth]{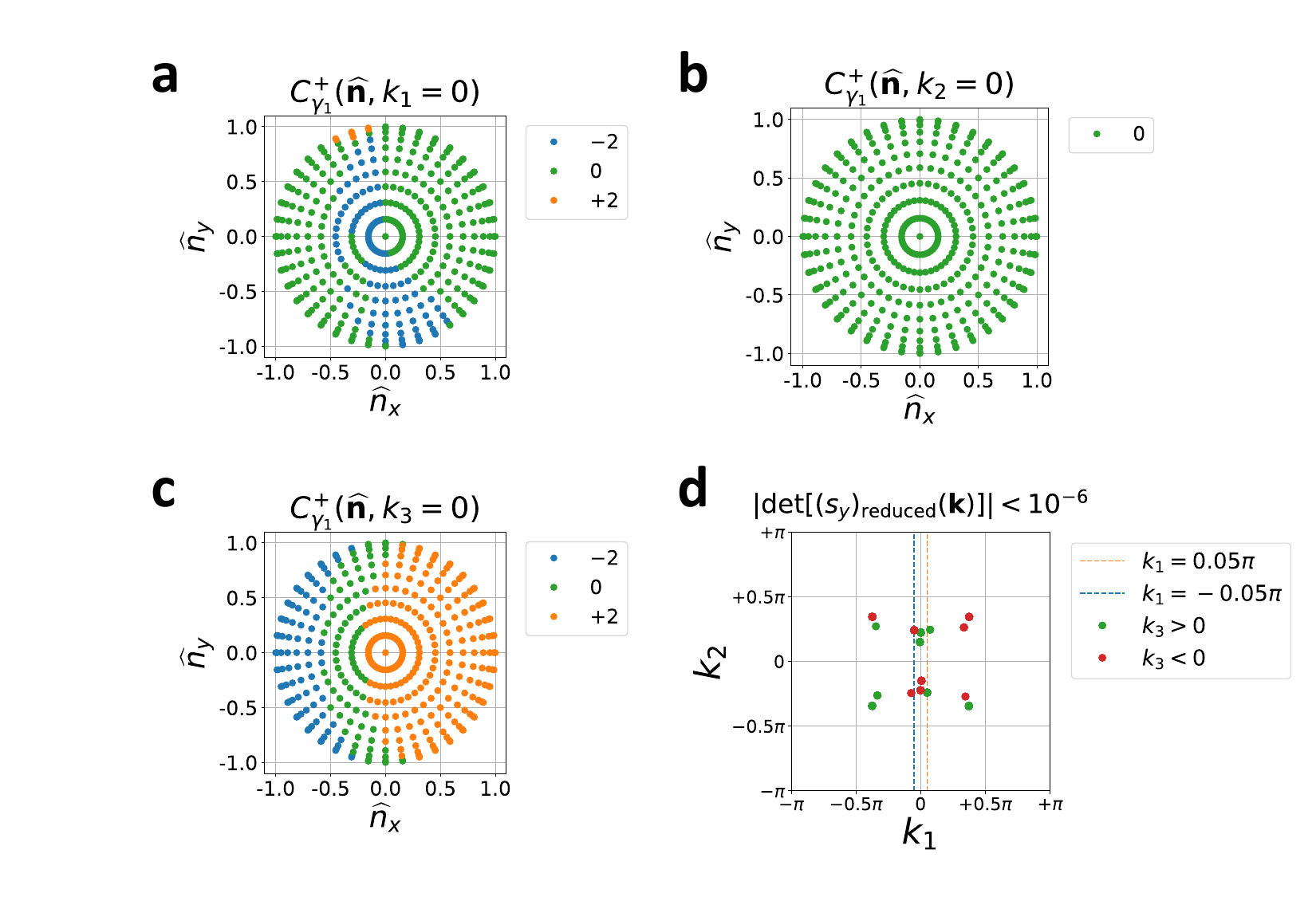}
\caption{Partial Chern numbers for the occupied bands of $\beta$-MoTe$_2$ as a function of the spin resolution direction $\hat{\mathbf{n}}$ in $P\hat{\mathbf{n}}\cdot\mathbf{s}P$.  
The direction $\hat{\mathbf{n}}$ of the spin operator is taken to lie in the upper hemisphere $\hat{n}_z\geq 0$ without loss of generality, which is sufficient to determine the spin-resolved topology (here partial Chern numbers) for all $\hat{\mathbf{n}}$ in time-reversal- ($\mathcal{T}$-) invariant insulators, because $\mathcal{T}$ symmetry relates the $\pm \hat{\mathbf{n}}$ spin resolution directions.
The $x$- and $y$-axes in panels (a-c) respectively correspond to the $\hat{n}_x$ and $\hat{n}_y$ components of the spin-resolution unit vector $\hat{\mathbf{n}}$. 
First, we find that $C^+_{\gamma_1}(\hat{\mathbf{n}}, k_i=\pi)=0$ for all $\hat{\mathbf{n}}$; we hence do not show in this figure the partial Chern numbers in the $k_i=\pi$ planes. 
(a) shows the partial Chern number $C^+_{\gamma_1}(\hat{\mathbf{n}}, k_1=0)$ in the $k_1=0$ plane as a function of $\hat{\mathbf{n}}$.  In (a), $C^+_{\gamma_1}(\hat{\mathbf{n}}, k_1=0)$ oscillates between $-2,0$ and $2$ as $\hat{\mathbf{n}}$ is varied. 
In particular, when $\hat{n}_x<0$ we find that the regions of constant $C^+_{\gamma_1}(\hat{\mathbf{n}}, k_1=0)$ are discontinuous. 
This is a signature of spin-Weyl nodes lying very close to the $k_1=0$ plane, for which the representative example of $\hat{\mathbf{n}}=\hat{\mathbf{y}}$ is shown in (d). 
(b) shows the partial Chern number $C^+_{\gamma_1}(\hat{\mathbf{n}}, k_2=0)$ in the $k_2=0$ plane as a function of $\hat{\mathbf{n}}$. 
We see that $C^+_{\gamma_1}(\hat{\mathbf{n}}, k_2=0)=0$ for all spin directions. 
(c) shows the partial Chern number $C^+_{\gamma_1}(\hat{\mathbf{n}}, k_3=0)$ in the $k_3=0$ plane as a function of $\hat{\mathbf{n}}$. 
We see that there are well-defined (smooth and continuous) regions in spin-resolution parameter ($\hat{\mathbf{n}}$-) space for which $C^+_{\gamma_1}(\hat{\mathbf{n}}, k_3=0)=0,\pm 2$. 
In (d), we plot the 2D projection onto the $k_1-k_2$ plane of the $\mathbf{k}$ points at which the absolute values of the determinant of the reduced spin $s_y$ matrix [SEq.~\eqref{eq:P_pm_Wilson_loop_s_reduced_def} for $\hat{\mathbf{n}}=\hat{\mathbf{y}}$] is smaller than $10^{-6}$, which numerically defines the locations of the spin-Weyl nodes for the spin resolution direction $\hat{\mathbf{n}}=\hat{\mathbf{y}}$.  
We find numerous (8) additional spin-Weyl points lying close to $k_{1}=0$ for $\hat{\mathbf{n}}=\hat{\mathbf{y}}$ in (d), consistent with the numerical oscillations of $C^+_{\gamma_1}(\hat{\mathbf{n}}, k_1=0)$ in (a).
The calculations detailed in this figure were performed using the freely available Python package~\href{https://github.com/kuansenlin/nested_and_spin_resolved_Wilson_loop}{\textsc{nested\_and\_spin\_resolved\_Wilson\_loop}}~\cite{lin2023nestedWilsonLib}, which represents an extension of the~\href{https://www.physics.rutgers.edu/pythtb/}{PythTB} open-source Python tight-binding package~\cite{coh2013python} that was implemented and utilized for the preparation of SRefs.~\cite{wieder2018axion,wieder2020strong} and the present work.}
\label{fig:mote2_partial_chern_numbers}
\end{figure}

To begin, we recall that the $P_\pm$-Wilson loops defined in SN~\ref{sec:P_pm_Wilson_loop} depend on the choice of spin operator $s=\hat{\mathbf{n}}\cdot\mathbf{s}$ used to separate occupied states into positive and negative $PsP$ eigenspaces.
To analyze the spin-resolved topology of $\beta$-MoTe$_2$, we first compute the partial Chern numbers $C^\pm_{\gamma_1}(\hat{\mathbf{n}},k_i)$ [SEq.~\eqref{eq:partial_chern_def}] through the winding numbers of the $P_\pm$-Wilson loops of the occupied states in constant-$k_i$ BZ planes ($i=1,2,3$).
For $\hat{\mathbf{n}}$ sampled on a uniform (in spherical coordinates) mesh of $451$ points in the upper hemisphere of the unit sphere (to exclude spin directions related by $\mathcal{T}$),
we show in SFig.~\ref{fig:mote2_partial_chern_numbers}(a,b) the winding numbers of the $P_\pm$-Wilson loop spectra for the positive spin bands in the $\mathcal{T}$-invariant $k_{i} = 0$ and $k_{i} = \pi$ planes.
We first find that for all choices of spin direction $\hat{\mathbf{n}}$ and for all $i=1,2,3$, the partial Chern numbers of the occupied bands in the $k_{i}=\pi$ planes vanish [$C^\pm_{\gamma_1}(\hat{\mathbf{n}},k_i=\pi)=0$]. 
This implies that the occupied bands of the Hamiltonian $[H_{\mathrm{MoTe}_2}]$ restricted to each $k_i=\pi$ plane are equivalent to a 2D $\mathcal{I}$- and $\mathcal{T}$-symmetric Hamiltonian with trivial spin-resolved stable topology. 
Next, in SFig.~\ref{fig:mote2_partial_chern_numbers}(a--c) we show the partial Chern number $C^+_{\gamma_1}(\hat{\mathbf{n}},k_i=0)$ for $i=1,2,3$ computed as a function of $\hat{\mathbf{n}}$. 
We see that $C^\pm_{\gamma_1}(\hat{\mathbf{n}},k_i=0)$ takes the values $0$ or $\pm 2$ for different choices of spin direction when $i\neq 2$. 
Recalling from SN~\ref{appendix:explicit-calculation-of-BHZ-model} and \ref{sec:main-text-3D-TI-P-pm} that the partial Chern number can only change as a function of $k_i$ due to the presence of spin-Weyl nodes (chirally-charged 3D spin-gap closing points), we deduce that for the spin resolution directions $\hat{\mathbf{n}}$ for which $C^\pm_{\gamma_1}(\hat{\mathbf{n}},k_i=0)=\pm 2$, $\beta$-MoTe$_2$ must host an even number of spin-Weyl nodes per half BZ.

If we focus in particular on constant-$k_3$ planes, we see from SFig.~\ref{fig:mote2_partial_chern_numbers}(c) that there is a set of $\hat{\mathbf{n}}$ for which the partial Chern numbers $C^\pm_{\gamma_1}(\hat{\mathbf{n}},k_3=0) =C^\pm_{\gamma_1}(\hat{\mathbf{n}},k_3=\pi)=0$. 
From this, one might assume that there are no spin-Weyl nodes for these $\hat{\mathbf{n}}$, for example $\hat{\mathbf{n}} = \hat{\mathbf{y}}$. 
However comparing with SFig.~\ref{fig:mote2_partial_chern_numbers}(a), we see that when $C^\pm_{\gamma_1}(\hat{\mathbf{n}},k_3=0)=0$, the partial Chern numbers $C^\pm_{\gamma_1}(\hat{\mathbf{n}},k_1=0)$ are poorly conditioned: if the spin gap were open near the $k_1=0$ plane, we would expect the partial Chern number $C^\pm_{\gamma_1}(\hat{\mathbf{n}},k_1=0)$ to be constant in $\hat{\mathbf{n}}$ taken over continuous patches of the upper spin hemisphere. 
However in SFig.~\ref{fig:mote2_partial_chern_numbers}(a), we see that the partial Chern number oscillates rapidly between $-2$, $0$, and $2$ as $\hat{\mathbf{n}}$ is varied. 
Through an explicit numerical computation of the $PsP$ spectrum, we verify that this rapid oscillation of $C^+_{\gamma_1}$ is due to the presence of spin-Weyl nodes lying close to the $k_{1}=0$ plane. 
Specifically, as done previously in SN~\ref{sec:main-text-3D-TI-P-pm}, we determine the ${\bf k}$-space locations of spin-Weyl nodes by minimizing the absolute value of the determinant of the reduced spin matrix $[s_\mathrm{reduced}]$ [SEq.~\eqref{eq:P_pm_Wilson_loop_s_reduced_def}]. 
This provides a reasonable indicator of the locations of the spin-Weyl nodes in 3D systems like $\beta$-MoTe$_2$ with $\mathcal{I}$ and $\mathcal{T}$ symmetries, which together restrict spin-Weyl nodes to only occur when two eigenvalues of $[s_\mathrm{reduced}]$ go to zero (see SN~\ref{sec:main-text-3D-TI-P-pm}).   
We show a representative configuration of spin-Weyl points in $\beta$-MoTe$_2$ for $\hat{\mathbf{n}}=\hat{\mathbf{y}}$ in SFig.~\ref{fig:mote2_partial_chern_numbers}(d), in which we find that indeed numerous spin-Weyl nodes lie within the small BZ region $|k_1|<0.05\pi$. 
Below in SN~\ref{sec:mote_2spin_gap_minimization}, we will further perform an extensive numerical sampling and minimization of the spin gap for all spin directions $\hat{\mathbf{n}}$, the results of which demonstrate the absence of a discernible numerical spin gap in any spin resolution direction in $\beta$-MoTe$_2$.
This calculation allows us to conclude that even for the values of $\hat{\mathbf{n}}$ in SFig.~\ref{fig:mote2_partial_chern_numbers}(c) for which $C^\pm_{\gamma_1}(\hat{\mathbf{n}},k_3=0) = C^\pm_{\gamma_1}(\hat{\mathbf{n}},k_3=\pi)=0$, $\beta$-MoTe$_2$ realizes a spin-Weyl semimetal state with an even number of spin-Weyl nodes per half BZ. 
To summarize, through spin-resolved Wilson loop calculations (supported by direct spin gap calculations that will be detailed in SN~\ref{sec:mote_2spin_gap_minimization}), we have demonstrated that for all choices of $\hat{\mathbf{n}}$, $\beta$-MoTe$_2$ carries an even number of spin-Weyl nodes per half BZ whose positions continuously evolve as a function of the choice of spin direction $\hat{\mathbf{n}}$, but not in a manner in which all spin-Weyl points are annihilated for any particular $\hat{\mathbf{n}}$.  
As shown in SRef.~\cite{po2017symmetry} and discussed in the main text, simply doubling---or ``stacking''---a model of a 3D strong TI is one way of constructing a model of a helical HOTI, a construction that the authors of SRef.~\cite{po2017symmetry} specifically termed a ``doubled strong TI'' (DSTI).  Because a 3D TI hosts an odd number of spin-Weyl nodes per half BZ (SN~\ref{sec:main-text-3D-TI-P-pm}), then a helical HOTI in the DSTI regime necessarily hosts an even number of spin-Weyl points per half BZ.
As shown in SN~\ref{app:response_QSHI_TDAXI}, the DSTI regime of a helical HOTI with $\mathcal{I}$ and $\mathcal{T}$ symmetries can be understood as an intermediate critical spin-stable (spin-gapless) phase separating spin-gapped $\mathcal{T}$-doubled axion insulator [SFig.~\ref{fig:layer-construction-QSHI-DAXI}(b)] and 3D quantum spin Hall insulator [SFig.~\ref{fig:layer-construction-QSHI-DAXI}(a)] states (see the main text and SN~\ref{app:comparison-spin-stable-and-symmetry-indicated-topology}).
From the above analysis, we determine $\beta$-MoTe$_2$ to be a helical HOTI that lies in the DSTI regime for all choices of spin resolution direction $\hat{\mathbf{n}}$.

We will next fix a choice of $\hat{\mathbf{n}}$ and investigate the spin spectrum in more detail.  
As we will show below, the spin spectrum (spin-Weyl point distribution) in $\beta$-MoTe$_2$ is particularly simple for $\hat{\mathbf{n}} =\hat{\mathbf{n}}_{xz} = (\hat{\mathbf{x}}+\hat{\mathbf{z}})/\sqrt{2}$, to which we will hence specialize in the analysis below.
This specialization to $\hat{\mathbf{n}}_{xz}$ is further justified because $\hat{\mathbf{n}}_{xz}$ is invariant under the $m_y\times \mathcal{T}$ (magnetic~\cite{elcoro2021magnetic}) reflection symmetry of $\beta$-MoTe$_2$ [nonmagnetic SG $P2_1/m1'$ (\#11.51)], and
recent studies have observed preferential spin-electromagnetic responses for reflection-invariant spin directions in devices based on transition metal dichalcogenides like MoTe$_2$~\cite{zhao2021determination,garcia2020canted,kurebayashi2022magnetism,safeer2019large,zhao2021determination,weber2018spin}.

First, from our analysis in SFig.~\ref{fig:mote2_partial_chern_numbers}(c), we see that the partial Chern numbers $C^\pm_{\gamma_1}(\hat{\mathbf{n}}_{xz},k_3=0)=\pm 2$, while $C^\pm_{\gamma_1}(\hat{\mathbf{n}}_{xz},k_3=\pi)=0$. 
This is precisely what one would expect from a naive doubling of the partial Chern numbers of the adapted BHZ model of a 3D strong TI (SN~\ref{sec:main-text-3D-TI-P-pm}), consistent with our above determination of $\beta$-MoTe$_2$ as a DSTI.

We next explicitly compute the configuration of spin-Weyl nodes in $\beta$-MoTe$_2$ for $\hat{\mathbf{n}}=\hat{\mathbf{n}}_{xz}$. 
We find that there are a total of four spin-Weyl nodes in each half of the BZ. 
We plot the locations of the spin-Weyl nodes in the 3D BZ for $\hat{\mathbf{n}}=\hat{\mathbf{n}}_{xz}$ in SFig.~\ref{fig:mote2_main_fig}(d,e). 
Focusing on the upper half  of the BZ ($k_3\geq 0$), we see that there are three spin-Weyl nodes with partial Chern number (or partial monopole chiral charge) $-1$, and one spin-Weyl node with partial Chern number $+1$. 
From this, we deduce that the change in partial Chern number $C^+_{\gamma_1}(\hat{\mathbf{n}}_{xz},k_3)$ between $k_3=0$ and $k_3=\pi$ is given by 
\begin{equation}
C^+_{\gamma_1}(\hat{\mathbf{n}}_{xz},k_3=\pi) - C^+_{\gamma_1}(\hat{\mathbf{n}}_{xz},k_3=0) = -1-1-1+1 = -2.
\end{equation}
This is consistent with our calculation in SFig.~\ref{fig:mote2_partial_chern_numbers}(c) showing that $C^+_{\gamma_1}(\hat{\mathbf{n}}_{xz},k_3=\pi) = 0 $ and $C^+_{\gamma_1}(\hat{\mathbf{n}}_{xz},k_3=0)  = +2$. 
We therefore conclude that while $\beta$-MoTe$_2$ realizes a DSTI for all choices of $\hat{\mathbf{n}}$, $\beta$-MoTe$_2$ is not a \emph{minimal} DSTI for $\hat{\mathbf{n}}=\hat{\mathbf{n}}_{xz}$.
Specifically, in a helical HOTI lying in the minimal DSTI regime, there are only two spin-Weyl nodes of the same partial chiral charge in each half of the 3D BZ. 
Instead, for $\hat{\mathbf{n}}=\hat{\mathbf{n}}_{xz}$ in $\beta$-MoTe$_2$, an additional dipole of spin-Weyl nodes of opposite partial Chern numbers is present in each half of the BZ. 

\subsubsection{Searching for a Spin Gap in $\beta$-MoTe$_2$ at Generic Spin Resolution Directions}
\label{sec:mote_2spin_gap_minimization}

\begin{figure}[t]
\hspace*{-2cm}\includegraphics[width=0.6\columnwidth]{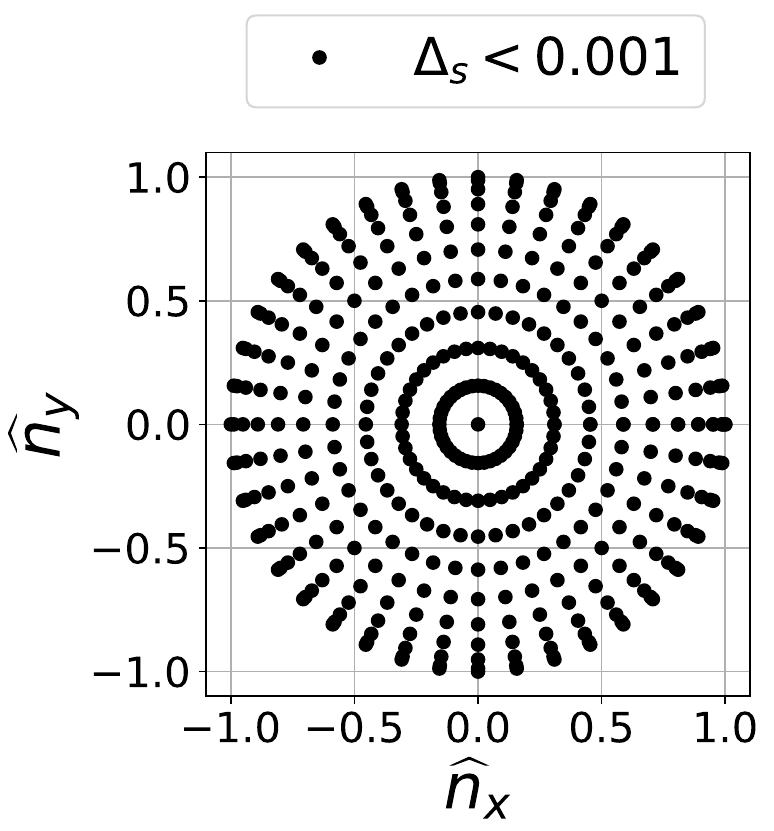}
\caption{Numerical spin gap $\Delta_{s = \hat{\mathbf{n}} \cdot \mathbf{s}}$ as a function of the spin resolution direction $\hat{\mathbf{n}}$ for 3D $\beta$-MoTe$_2$. 
The angular parameterization of $\hat{\mathbf{n}}$ is defined in SEq.~\eqref{eq:hat_n_around_z_axis}, and our calculations were performed over the spin-direction hemisphere of $\hat{\mathbf{n}}$ defined by $\vartheta \in [0,0.5\pi]$ and $\phi \in [0,2\pi]$.
For our calculations, the angular variables $(\vartheta,\phi)$ were respectively sampled using the numerical resolution of $\Delta \vartheta = 0.05\pi$ and $\Delta \phi = 0.05\pi$.
By performing a Nelder-Mead minimization~\cite{pandey2022py} on the spin gap function $\Delta_s (\mathbf{k})$ individually for $100$ $\mathbf{k}$ points randomly sampled from the 3D Brillouin zone as the initial points, we define for each $\hat{\mathbf{n}}$ the \textit{numerical spin gap} $\Delta_{s=\hat{\mathbf{n}}\cdot \mathbf{s}}$ as the minimal value (in the units of $\hbar/2$) of the $100$ minimization results for the fixed value of $\hat{\mathbf{n}}$.
In this figure, we use black dots to indicate the values of $\hat{\mathbf{n}}$ for which the numerical spin gap $\Delta_{s = \hat{\mathbf{n}} \cdot \mathbf{s}}$ in $\beta$-MoTe$_2$ is less than $10^{-3}=0.001$.
From the uniform (projected) spacing of the black dots in this figure, we conclude that $\beta$-MoTe$_2$ is in fact spin-gapless (within numerical precision) for \emph{all} values of $\hat{\mathbf{n}}$.
For completeness, we note that within the resolution of the angular variables $(\vartheta,\phi)$ sampled over the spin hemisphere considered in our calculations, the maximum numerical spin gap $\Delta_s \approx 1.401393465896699 \times 10^{-4}$ lies at $(\vartheta,\phi) = (0.25\pi,0.05\pi)$, and the minimal numerical spin gap $\Delta_s \approx 2.4070658354675936 \times 10^{-6}$ lies at $(\vartheta,\phi) = (0.05\pi,1.75\pi)$.
The calculations detailed in this figure were performed using the freely available Python package~\href{https://github.com/kuansenlin/nested_and_spin_resolved_Wilson_loop}{\textsc{nested\_and\_spin\_resolved\_Wilson\_loop}}~\cite{lin2023nestedWilsonLib}, which represents an extension of the~\href{https://www.physics.rutgers.edu/pythtb/}{PythTB} open-source Python tight-binding package~\cite{coh2013python} that was implemented and utilized for the preparation of SRefs.~\cite{wieder2018axion,wieder2020strong} and the present work.}
\label{fig:mote2_numerical_spin_gap_for_different_spin_directions}
\end{figure}

\begin{figure}[t]
\includegraphics[width=0.9\textwidth]{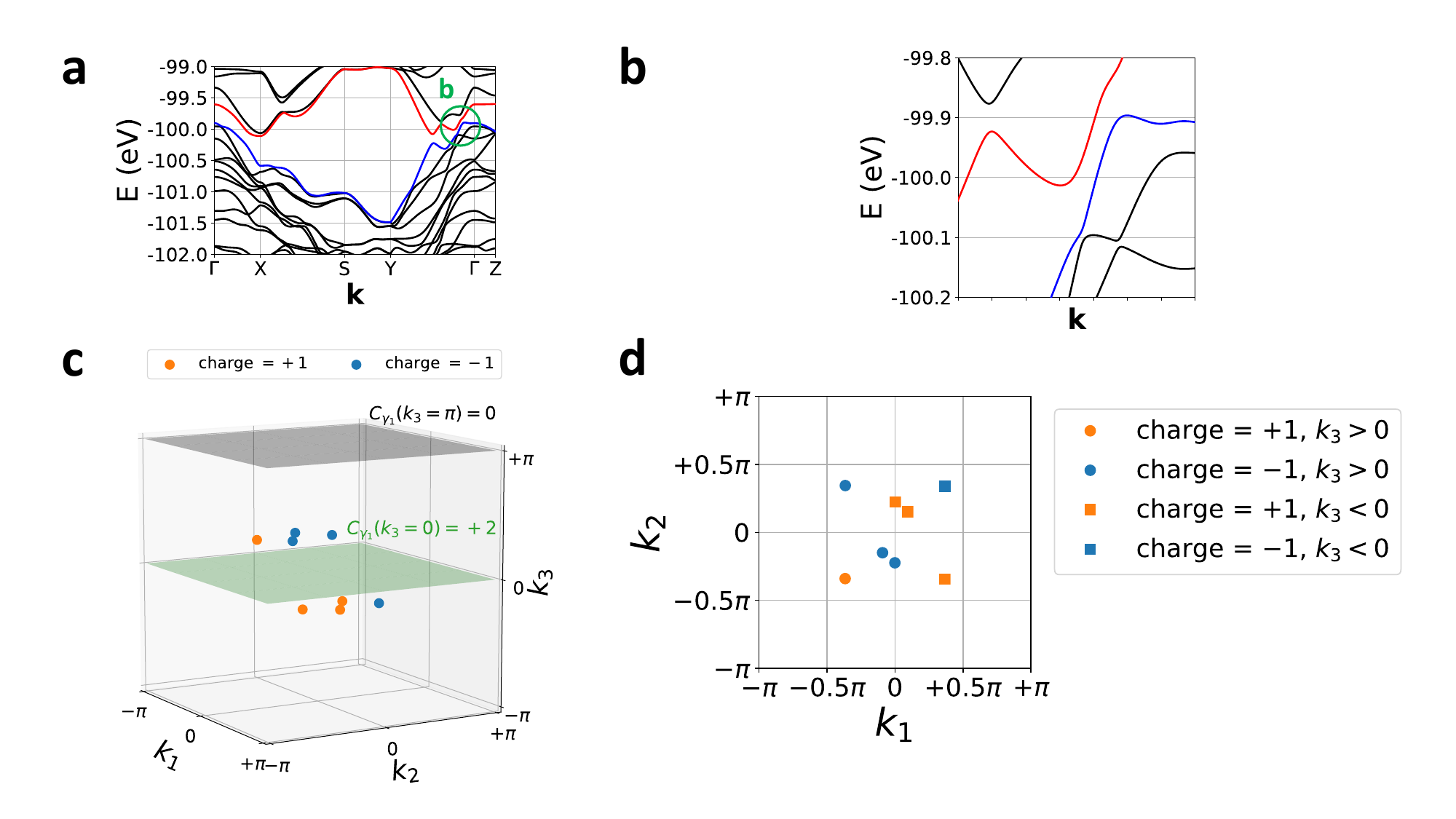}
\caption{Energy spectrum and Weyl nodes for $\beta$-MoTe$_2$ in a strong ($\hat{\mathbf{x}} + \hat{\mathbf{z}}$)-directed Zeeman field.
(a) The band structure of $[H_Z]$ in SEq.~\eqref{eq:mote2_HZ}, which was obtained by applying a strong ($\hat{\mathbf{x}}+\hat{\mathbf{z}}$)-directed (spin-) Zeeman field to a Wannier-based tight-binding model of $\beta$-MoTe$_2$ obtained from DFT calculations (SN~\ref{sec:mote2_dft_details}).
Near $E\approx -B=-100$eV in (a), the electronic band structure appears qualitatively similar to the zero-field bulk band structure in SFig.~\ref{fig:mote2_main_fig}(a), albeit with half the number of bands [the remaining bands have energies $E\approx +B$, and so do not appear (a)]. 
Because the product of inversion and time-reversal symmetries ($\mathcal{I}\times\mathcal{T}$) is broken by the large Zeeman field, the energy bands in (a) are singly degenerate at generic ${\bf k}$ points, as opposed to the doubly-degenerate bulk bands in $\beta$-MoTe$_2$ in the absence of an external field [SFig.~\ref{fig:mote2_main_fig}(a)].
(b) shows an enlarged view of the energy band structure along the $\Gamma-Y$ line near the green circle in (a). 
In (b), there is an avoided crossing between the red and blue bands.
Nevertheless, upon closer examination, we find that the red and blue bands in (a,b) do indeed cross and form Weyl points at lower-symmetry ${\bf k}$ points throughout the 3D BZ.
(c) shows the positions and monopole chiral charges of the eight Weyl nodes in the energy spectrum of $[H_Z]$.
The locations and chiral charges of the Weyl points in (c) are nearly identical to the locations and chiral charges of the $\hat{\mathbf{n}}_{xz}$ spin-Weyl nodes in $\beta$-MoTe$_2$ [SFig.~\ref{fig:mote2_main_fig}(d,e)]. 
(d) shows a projection of the Weyl nodes in (c) onto the $k_1-k_2$ plane; Weyl nodes with positive (negative) chiral charges are shown in orange (blue).
In the $k_3>0$ half of the BZ in (c,d), there are three Weyl nodes with charge $-1$ and one Weyl node with charge $+1$. 
As required by inversion symmetry, which is preserved by an external Zeeman field, there are three Weyl nodes with charge $+1$ and one Weyl node with charge $-1$ in the $k_3<0$ half of the BZ. 
The calculations detailed in this figure were performed using the freely available Python package~\href{https://github.com/kuansenlin/nested_and_spin_resolved_Wilson_loop}{\textsc{nested\_and\_spin\_resolved\_Wilson\_loop}}~\cite{lin2023nestedWilsonLib}, which represents an extension of the~\href{https://www.physics.rutgers.edu/pythtb/}{PythTB} open-source Python tight-binding package~\cite{coh2013python} that was implemented and utilized for the preparation of SRefs.~\cite{wieder2018axion,wieder2020strong} and the present work.
}\label{fig:zeeman_bulk}
\end{figure}

\begin{figure}[t]
\includegraphics[width=0.95\textwidth]{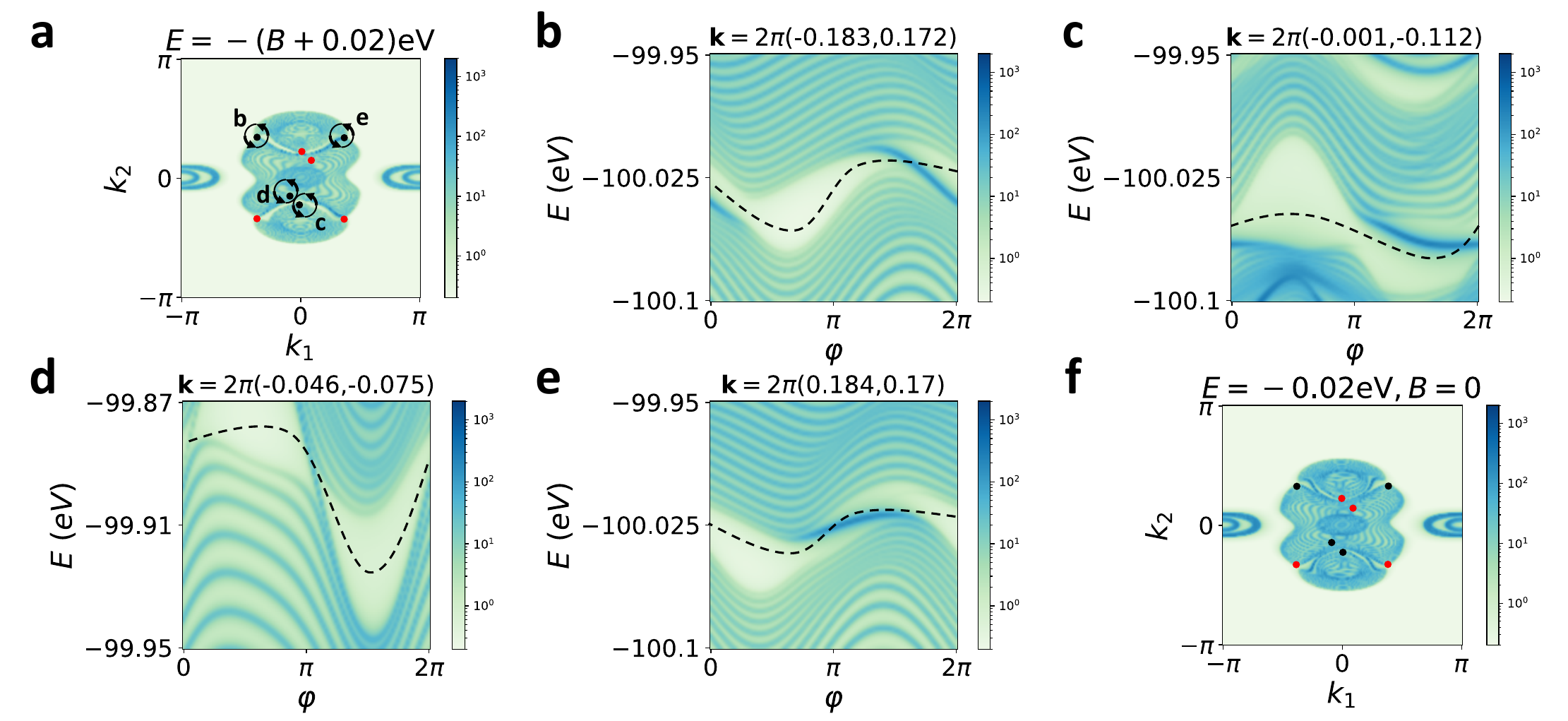}
\caption{Spectral function for the $(001)$-surface of $\beta$-MoTe$_2$ in a strong $(\hat{\mathbf{x}}+\hat{\mathbf{z}})$-directed Zeeman field. 
(a) shows the surface spectral function at $E=(-B-0.02)$eV as a function of $k_1$ and $k_2$ for an $(001)$-oriented slab of the tight-binding model $[H_Z]$ in SEq.~\eqref{eq:mote2_HZ}, which consists of a DFT-obtained, Wannier-based tight-binding model of $\beta$-MoTe$_2$ (SN~\ref{sec:mote2_dft_details}) subjected to a $B=100$eV (spin-) Zeeman field. 
In (a), red and black dots mark the positions of the Weyl nodes in the energy spectrum with positive and negative chiral charges, respectively. 
Arc-like surface features can be seen emanating from the surface projections of the bulk Weyl points. 
(b--e) show the surface spectral function computed on circles surrounding the negatively-charged Weyl nodes [indicated in panel (a)] plotted as functions of $E$ and the polar angle $\varphi$ of the circles in (a). 
$\varphi=0$ corresponds to the positive $k_1$ axis, and the circles are traversed counterclockwise in (b--e) as indicated by the black arrows in (a). 
In (b--e), gap-traversing chiral modes indicative of topological surface Fermi arcs can be seen emanating towards the center of the surface BZ. 
In panel (b), we see that the Fermi-arc chiral mode crosses the dashed black line near $\varphi\approx 3\pi/2$, indicating that the Fermi arc is directed towards the origin in (a). 
In (b), there is specifically one chiral surface state with a negative slope crossing the dashed line, in correspondence with the $-1$ charge of the encircled Weyl node in (a). 
(c) and (d) show surface spectral functions surrounding the surface projections of the two Weyl nodes closest to the origin in the $k_2<0$ half of (a). 
The Weyl points in (c,d) are embedded into the continuum of bulk states at lower energies. 
In both (c,d), a single chiral surface state with negative slope crosses the dashed line, corresponding again to the $-1$ charge of the encircled Weyl nodes in (a).
(e) shows the surface spectral function surrounding the Weyl node in the upper right quadrant of (a). 
In (e), there is a Fermi-arc surface state near $\varphi=\pi$, which corresponds to an arc emanating towards the center of the surface BZ in (a). 
A single chiral surface state with negative slope crosses the dashed line in (e), corresponding to the $-1$ charge of the encircled Weyl-node projection in (a). 
To draw comparison with the Zeeman-field energy spectrum in (a), in (f), we plot the zero-field $(001)$-surface spectral function of $\beta$-MoTe$_2$ at $E=-0.02$eV and indicate the surface projections and partial chiral charges of the bulk $\hat{\mathbf{n}}_{xz}$ spin-Weyl nodes.  The zero-field spin-Weyl points in (f) lie at almost identical ${\bf k}$ positions and carry the same chiral charges as the Zeeman-induced Weyl points in (a).}
\label{fig:zeeman_surface}
\end{figure}

As shown in SFigs.~\ref{fig:mote2_main_fig} and \ref{fig:mote2_partial_chern_numbers}, by calculating the partial Chern numbers of occupied states in high-symmetry ($\mathcal{T}$-invariant) BZ planes, 3D $\beta$-MoTe$_2$ can be determined to lie in the DSTI regime of a helical HOTI~\cite{po2017symmetry,khalaf2018symmetry,wang2019higherorder,tang2019efficient} for a large range of spin resolution directions $\hat{\mathbf{n}}$.  
However, as shown in SFig.~\ref{fig:mote2_partial_chern_numbers}(a,d), there exist several regions of $\hat{\mathbf{n}}$-parameter space (such as in the vicinity of $\hat{\mathbf{n}}=\hat{\mathbf{y}}$) in which the partial Chern number is poorly conditioned (\emph{i.e.} unstably oscillating).
To further characterize the spin-resolved topology of $\beta$-MoTe$_2$ in the spin directions for which high-symmetry-plane ($k_{1}=0$) partial Chern numbers cannot be computed, we in this section directly numerically investigate the presence (or absence) of a spin gap throughout the full 3D BZ for each spin direction $\hat{\mathbf{n}}$ in the $P (\hat{\mathbf{n}} \cdot \mathbf{s}) P$ spectrum of $\beta$-MoTe$_2$.
As shown below, our calculations show that $\beta$-MoTe$_2$ realizes a spin-Weyl semimetal (spin-gapless) state for \emph{all} choices of spin resolution direction $\hat{\mathbf{n}}$.

To begin, we parameterize the spin direction $\hat{\mathbf{n}}$ as a 3D unit vector described by
\begin{equation}
	\hat{\mathbf{n}} = (\sin{\vartheta}\cos{\phi},\sin{\vartheta}\sin{\phi},\cos{\vartheta}). \label{eq:hat_n_around_z_axis}
\end{equation}
The parameterization in SEq.~\eqref{eq:hat_n_around_z_axis} is chosen such that the angular variables $(\vartheta,\phi)=(0,0)$, $(0.5\pi,0)$, and $(0.5\pi,0.5\pi)$ correspond to the Cartesian unit vectors $\hat{\mathbf{z}}$, $\hat{\mathbf{x}}$, and $\hat{\mathbf{y}}$, respectively.
Because 3D $\beta$-MoTe$_2$ has both inversion ($\mathcal{I}$) and spinful time-reversal ($\mathcal{T}$) symmetries, its $P s P$ spectrum exhibits an antiunitary chiral symmetry due to the relation
\begin{equation}
	[\mathcal{I}] [\mathcal{T}][P(\mathbf{k})]s[P(\mathbf{k})] ([\mathcal{I}] [\mathcal{T}])^{-1} = - [P(\mathbf{k})]s[P(\mathbf{k})], \label{eq:mote2_section_IT_on_PsP}
\end{equation}
where $[\mathcal{I}]$ and $[\mathcal{T}]$ are the unitary and antiunitary matrix representatives of $\mathcal{I}$ and spinful $\mathcal{T}$ acting on the $88 \times 88$ Wannier-based tight-binding Bloch Hamiltonian matrix $[H_{\mathrm{MoTe}_2}(\mathbf{k})]$ that we obtained from DFT calculations of the electronic structure of 3D $\beta$-MoTe$_2$ (SN~\ref{sec:mote2_dft_details}).
In SEq.~\eqref{eq:mote2_section_IT_on_PsP}, $[P(\mathbf{k})]$ denotes the matrix projector $[P(\mathbf{k})]=\sum_{n=1}^{56}|u_{n,\mathbf{k}}\rangle \langle u_{n,\mathbf{k}} |$ onto the $56$ occupied (valence) bands of $\beta$-MoTe$_2$ (see SN~\ref{sec:mote2_dft_details}), where $|u_{n,\mathbf{k}}\rangle$ is the $n^{\mathrm{th}}$ eigenvector of $[H_{\mathrm{MoTe}_2}(\mathbf{k})]$.
As discussed in SN~\ref{appendix:properties-of-the-projected-spin-operator}, the presence of $\mathcal{I}\times\mathcal{T}$ symmetry [SEq.~\eqref{eq:mote2_section_IT_on_PsP}] further implies that 
\begin{equation}
	\mathrm{Spec}\left( [P(\mathbf{k})]s[P(\mathbf{k})] \right) = -\mathrm{Spec}\left( [P(\mathbf{k})]s[P(\mathbf{k})] \right), 
\label{eq:mote2_chiral_PsP_spec}
\end{equation}
in which $\mathrm{Spec} (\mathcal{O})$ denotes the spectrum of the operator $\mathcal{O}$.
From SEq.~\eqref{eq:mote2_chiral_PsP_spec}, we can then define the spin gap function $\Delta_s (\mathbf{k})$ as the difference between the smallest positive and largest negative $PsP$ eigenvalues at each crystal momentum $\mathbf{k}$.
From the values of $\Delta_s (\mathbf{k})$ taken across the BZ for a given spin operator $s = \hat{\mathbf{n}} \cdot \mathbf{s}$, we then define the spin gap $\Delta_s$ as $\Delta_s \equiv \min_{\mathbf{k} \in \mathrm{BZ}} [ \Delta_s (\mathbf{k})]$, namely the minimal value of $\Delta_s (\mathbf{k})$ over the entire BZ.

Having established the spin-direction parameterization in SEq.~\eqref{eq:hat_n_around_z_axis} as a function of the angular variables $(\vartheta,\phi)$ and defined the spin gap $\Delta_s \equiv \min_{\mathbf{k} \in \mathrm{BZ}} [ \Delta_s (\mathbf{k})]$ in the text following SEq.~(\ref{eq:mote2_chiral_PsP_spec}), we next compute $\Delta_{\hat{\mathbf{n}} \cdot \mathbf{s}}$ as a function of the spin direction $\hat{\mathbf{n}}$ through numerical minimization.
Practically, for each spin direction $\hat{\mathbf{n}}$, we perform a Nelder-Mead minimization~\cite{pandey2022py} on the spin gap function $\Delta_s (\mathbf{k})$ for $100$ $\mathbf{k}$ points randomly sampled across the 3D BZ as the initial points.
We then define the \textit{numerical spin gap} $\Delta_s$ as the minimal value of the $100$ minimization results for each spin direction $\hat{\mathbf{n}}$.

In SFig.~\ref{fig:mote2_numerical_spin_gap_for_different_spin_directions}, we show the numerical spin gap $\Delta_s$ (in the units of $\hbar/2$) for $\hat{\mathbf{n}}$ (projected onto the $xy$ plane) sampled over the spin hemisphere defined by $\vartheta \in [0,0.5\pi]$ and $\phi \in [0,2\pi]$ using the angular resolutions $\Delta \vartheta = 0.05\pi$ and $\Delta \phi = 0.05\pi$.
Because $\Delta_s = \Delta_{-s}$ due to $\mathcal{T}$ symmetry, this sampling of the angular variables $(\vartheta,\phi)$ contains all independent values of the numerical spin gap $\Delta_{s}$ over the full sphere of spin resolution directions $\hat{\mathbf{n}}$. 
In SFig.~\ref{fig:mote2_numerical_spin_gap_for_different_spin_directions}, we use black dots to indicate the spin directions $\hat{\mathbf{n}}$ for which $\Delta_{s}$ is smaller than $10^{-3}$=0.001.
From the regular (projected) spacing of black dots in SFig.~\ref{fig:mote2_numerical_spin_gap_for_different_spin_directions}, we conclude that the numerical spin gap $\Delta_s$ of 3D $\beta$-MoTe$_2$ is smaller than $10^{-3}$ for all $\hat{\mathbf{n}}$, which implies that within numerical precision, $\beta$-MoTe$_2$ is spin-gapless for \emph{all} choices of $\hat{\mathbf{n}}$.  
Within the spin spectrum, the only nodal degeneracies that can appear in 3D insulators with only bulk $\mathcal{I}$ and $\mathcal{T}$ symmetry are 3D spin-Weyl points (SN~\ref{app:response_QSHI_TDAXI}); other degeneracies, such as spin-nodal-lines, would require higher crystal symmetries.
In addition to $\mathcal{I}$ and $\mathcal{T}$ symmetries, the crystal structure of $\beta$-MoTe$_2$ [SG $P2_1/m1'$ (\#11.51)], respects $m_{y}$ mirror and $s_{2y}$ screw symmetries, which are both broken at generic spin resolution directions. 
Specifically, the spin spectrum of $\beta$-MoTe$_2$ is only constrained by symmetries other than $\mathcal{I}$ and $\mathcal{T}$ for $\hat{\mathbf{n}}=\hat{\mathbf{y}}$ ($m_{y}$-preserving) and $\hat{\mathbf{n}}$ lying in the $xz$-plane ($m_{y}\times\mathcal{T}$-preserving).
Together with our previous observation of numerically stable partial Chern numbers for $\hat{\mathbf{n}}$ lying in the $xz$-plane [SFig.~\ref{fig:mote2_partial_chern_numbers}(a,b,c)], and our direct computation of the spin spectrum for $\hat{\mathbf{n}}=\hat{\mathbf{y}}$ [the spin-Weyl state shown in SFig.~\ref{fig:mote2_partial_chern_numbers}(d)], the spin-gaplessness of $\beta$-MoTe$_2$ in SFig.~\ref{fig:mote2_numerical_spin_gap_for_different_spin_directions} hence specifically indicates that 3D $\beta$-MoTe$_2$ realizes a spin-Weyl state for all choices of spin resolution direction $\hat{\mathbf{n}}$.

\subsection{Physical Observables of the Spin-Weyl State in $\beta$-MoTe$_2$: Surface Fermi Arcs in a Strong Zeeman Field}\label{app:mote2_fermi_arc_strong_zeeman}

To demonstrate physical signatures of the bulk spin-Weyl points $\beta$-MoTe$_2$, we begin with the relationship, previously established in SN~\ref{sec:zeeman}, between the $PsP$ spectrum and the spectrum of the Hamiltonian in a strong Zeeman field.
Specifically, as shown in SN~\ref{sec:zeeman}, if the bulk of an insulator hosts a spin-Weyl semimetal state for a particular spin resolution direction $\hat{\mathbf{n}}$, then under the application of a large Zeeman field parallel to $\hat{\mathbf{n}}$, the \emph{energy} spectrum of the system will develop Weyl points whose positions in ${\bf k}$ and chiral charges lie close to those of the spin-Weyl points in the original zero-field insulator.
For $\hat{\mathbf{n}} =\hat{\mathbf{n}}_{xz} = (\hat{\mathbf{x}}+\hat{\mathbf{z}})/\sqrt{2}$ in $\beta$-MoTe$_2$ [SFig.~\ref{fig:mote2_main_fig}(c,d,e)], we specifically expect that the low-energy ($E\sim-B$) bands of 
\begin{align}
[H_Z] &= [H_{\mathrm{MoTe}_2}] - \frac{B}{\sqrt{2}}([s_x] + [s_z]) \nonumber \\
&=[H_{\mathrm{MoTe}_2}]-Bs_{xz} 
\label{eq:mote2_HZ}
\end{align}
will host Weyl nodes that are adiabatically connected to the spin-Weyl nodes in the $Ps_{xz}P$ spectrum. 
To verify this, in SFig.~\ref{fig:zeeman_bulk}(a,b) we compute the energy band structure for $[H_Z]$ in SEq.~\eqref{eq:mote2_HZ} with the external Zeeman field strength set to $B=100$eV. 
Along high-symmetry BZ lines, we see that the electronic band structure at $E\approx -B \approx -100$eV in the presence of a strong ($\hat{\mathbf{x}}+\hat{\mathbf{z}}$)-directed Zeeman field structure strongly resembles that of the original field-free Hamiltonian [shown in SFig.~\ref{fig:mote2_main_fig}(a,b)], albeit with half the number of bands [the remaining bands lie at energies $E\approx +B$, and hence do not appear in SFig.~\ref{fig:zeeman_bulk}(a)].
As was done previously in SN~\ref{sec:mote_2spin_gap_minimization}, we next perform a numerical minimization using the Nelder-Mead method~\cite{pandey2022py} to search for degeneracies between the red and blue bands near $E=-100$eV in SFig.~\ref{fig:zeeman_bulk}(a,b).
We find that, like the $\hat{\mathbf{n}}_{xz}$ spin spectrum of $\beta$-MoTe$_2$ [SFig.~\ref{fig:mote2_main_fig}(d,e)], the \emph{energy} spectrum of $[H_Z]$ contains four Weyl nodes per half-BZ. 
In SFig.~\ref{fig:zeeman_bulk}(c,d) we plot the locations and chiral charges of the Weyl nodes in the energy spectrum of $[H_Z]$ [SEq.~(\ref{eq:mote2_HZ})] in the vicinity of $E=-100$eV.
We find that as expected from the analytic analysis in SN~\ref{sec:zeeman}, the energy Weyl nodes in  SFig.~\ref{fig:zeeman_bulk}(c,d) lie at very similar positions to the $\hat{\mathbf{n}}_{xz}$ spin-Weyl nodes in $\beta$-MoTe$_2$ [SFig.~\ref{fig:mote2_main_fig}(d,e)].

Finally, as established in SN~\ref{sec:zeeman} and \ref{sec:spin_entanglement_spectrum}, the presence of Weyl nodes in the spectrum of $[H_Z]$ allows us to infer the existence of topological surface Fermi arcs on surfaces of $\beta$-MoTe$_2$ subjected to a strong (spin-) Zeeman field. 
To verify this intuition, we next construct an $L=21$ unit-cell-thick slab of the tight-binding model $[H_Z]$ with a surface normal vector pointing in the experimentally-accessible $(001)$-direction of $\beta$-MoTe$_2$~\cite{WTe2Arpes1,WTe2Arpes2,WTe2Arpes3,MoTe2Arpes1,MoTe2Arpes2,WTe2STM,WTe2QPI2018,MoTe2STM}. 
To remove dangling-bond surface states, we have added a constant chemical potential offset for atoms in the quarter of the unit cell closest to each surface to passivate the outermost (fractional) van der Waals layer of the tight-binding model $[H_Z]$. 
In SFig.~\ref{fig:zeeman_surface}, we plot top- [$(001)$-] surface spectrum of $[H_Z]$ obtained from surface Green's functions.
SFig.~\ref{fig:zeeman_surface}(b--e) specifically show the $(001)$-surface spectral function as a function of energy and position plotted along circles surrounding the surface projections of the four energy Weyl points with $-1$ chiral charges. 
In all four panels [SFig.~\ref{fig:zeeman_surface}(b--e)], we observe a chiral surface mode traversing the (direct) gap between the bands surrounding $E=-100$eV, confirming the presence of topological surface Fermi arcs. 
In SFig.~\ref{fig:zeeman_surface}(a), we show the $(001)$-surface spectral function of $[H_Z]$ for a fixed energy $(-B-0.02)$eV, plotted as a function of $k_1$ and $k_2$. 
The Fermi arcs in SFig.~\ref{fig:zeeman_surface}(b--e) can be seen extending towards the center of the BZ. 
To draw comparison with the energy spectrum of $[H_Z]$, in SFig.~\ref{fig:zeeman_surface}(f), we show the surface spectral function of $[H_{\mathrm{MoTe}_2}]$ in the absence of a Zeeman field and label the $(001)$-surface projections and partial chiral charges of the $\hat{\mathbf{n}}_{xz}$ spin-Weyl points.  The zero-field spin-Weyl points in SFig.~\ref{fig:zeeman_surface}(f) lie at almost identical ${\bf k}$ positions and carry the same chiral charges as the Zeeman-induced Weyl points in SFig.~\ref{fig:zeeman_surface}(a) [see also SFigs.~\ref{fig:mote2_main_fig}(d,e) and~\ref{fig:zeeman_bulk}(c,d)]).

To conclude, we have hence crucially demonstrated that the spin-resolved topological analysis techniques developed in this work can be applied to ab-initio calculations of the electronic structure of real materials, here specifically $\beta$-MoTe$_2$.  Below, in SN~\ref{sec:bibr}, we will next apply the machinery of spin-resolved topology to the candidate helical HOTI $\alpha$-BiBr~\cite{tang2019efficient,SYBiBr,BiBrFanHOTI}, which unlike $\beta$-MoTe$_2$ exhibits a bulk topological spin gap for large regions of $\hat{\mathbf{n}}$ spin-resolution parameter space.

\section{First-Principles Analysis of $\alpha$-$\mathrm{Bi}\mathrm{Br}$}
\label{sec:bibr}

In this section, we will compute the spin-resolved topology of 3D $\alpha$-BiBr, which was theoretically identified in SRefs.~\cite{tang2019efficient,SYBiBr,BiBrFanHOTI} as a candidate helical higher-order topological insulator (HOTI) with both inversion ($\mathcal{I}$) and time-reversal ($\mathcal{T}$) symmetries.
As previously with $\beta$-MoTe$_2$ in SN~\ref{app:mote2}, the helical HOTI phase predicted in $\alpha$-BiBr is characterized by a nontrivial $\mathbb{Z}_4$-invariant $z_4=2$ and vanishing weak indices $z_{2i}=0$.
Recent experimental studies have also revealed evidence of helical higher-order topology in $\alpha$-BiBr, including signatures of 1D helical hinge states in laser angle-resolved photoemission experiments~\cite{noguchi2021evidence}, and scanning tunneling microscopy signatures of hinge-localized, $\mathcal{T}$-protected 1D gapless (helical) conducting channels that persist up to room temperature~\cite{FanZahidRoomTempBiBrExp}.  
Further experiments have also reported spectroscopic and transport signatures of helical hinge modes in $\alpha$-BiBr~\cite{BiBrHingeExp1,BiBrHingeExp2,BiBrFacetDependent,BiBrTempLifshitz,BiBrNanowireSubstrate,BiBrNanobelt,BiBrQuantumTransport,BiBrOpticalDichotomy}.

We will begin our spin-resolved topological analysis of $\alpha$-BiBr below by first in SN~\ref{sec:bibr_dft_details} detailing the DFT calculations that we performed to obtain a symmetric, Wannier-based tight-binding model of $\alpha$-BiBr.
We then in SN~\ref{sec:spin_resolved_topology_of_alpha_bibr} compute the $PsP$ spin spectrum for $\alpha$-BiBr over the full range of spin-resolution directions $\hat{\mathbf{n}}$.
In SN~\ref{sec:spin_resolved_topology_of_alpha_bibr}, we will specifically use (nested) spin-resolved Wilson loops to show that unlike previously for $\beta$-MoTe$_2$ (SN~\ref{app:mote2}), $\alpha$-BiBr exhibits a bulk topological spin gap over a large range of spin resolution directions $\hat{\mathbf{n}}$, and hosts both spin-stable 3D quantum spin Hall insulator (QSHI) states as well as, remarkably, the $\mathcal{T}$-doubled axion insulating (T-DAXI) state introduced in this work.  
Lastly, in SN~\ref{sec:bibrshc}, we will numerically compute the bulk intrinsic contribution to the spin Hall conductivity (per layered unit cell) in $\alpha$-BiBr, which we find to be nearly quantized in the 3D QSHI regime, and nearly vanishing for the spin-stable T-DAXI state.

\subsection{Details of Density Functional Theory Calculations on $\alpha$-BiBr}
\label{sec:bibr_dft_details}

\begin{figure}[t]
\includegraphics[width=1.0\columnwidth]{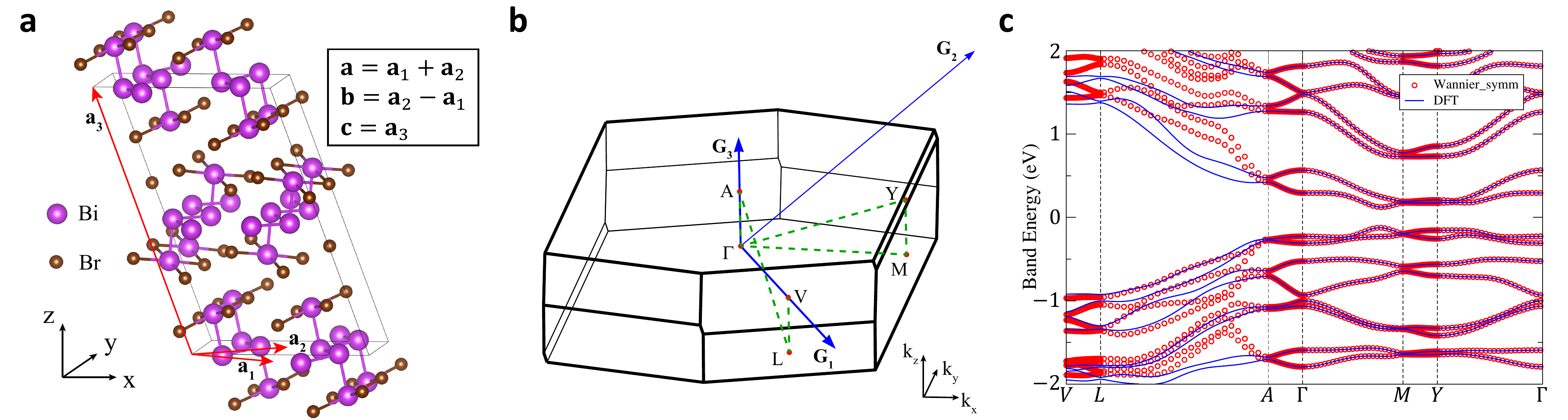}
\caption{First-principles and Wannier-based tight-binding electronic structure of $\alpha$-BiBr.  
(a) The crystal structure of $\alpha$-BiBr, which respects the symmetries of SG $C 2 / m 1'$ ($\#12.59$). 
There are eight Bi atoms and eight Br atoms per primitive (unit) cell.  
However, because $\alpha$-BiBr is $C$-face-centered, it is also frequently characterized by its larger conventional primitive monoclinic supercell.  
In (a), the red (black) arrows denote the primitive-cell lattice (Cartesian-unit) vectors [SEq.~(\ref{eq:bibrlatticevectorall})] and the inset box provides the relationship between the primitive lattice vectors ${\bf a}_{1,2,3}$ and the conventional supercell lattice vectors ${\bf a}$, ${\bf b}$, and ${\bf c}$ [SEq.~(\ref{eq:bibr_conventional_lattice_all})].
(b) The bulk BZ for $\alpha$-BiBr; the primitive reciprocal lattice vectors ${\bf G}_{i}$ (blue) arrows are given in SEq.~(\ref{eq:bibrreciprocallatticevectorall}), and importantly differ from the Cartesian reciprocal unit vectors (black arrows).
In (b), the TRIM points are labeled using the convention of the Bilbao Crystallographic Server~\cite{aroyo2006bilbao,aroyo2006bilbaoa,aroyo2011crystallography,bilbaocrystallogrserver2017bandrep,bilbaocrystallogrserver2018check,elcoro2017double,elcoro2021magnetic}, and their specific coordinates in reduced ${\bf k}$ units are provided in SEq.~(\ref{eq:BiBrTRIMpoints}).
(c) The DFT-obtained (``DFT,'' blue lines) and symmetric Wannier-based (``Wannier\_symm,'' red circles) electronic band structure of $\alpha$-BiBr plotted along the green dashed ${\bf k}$-path in (b).  
Although there are some qualitative discrepancies between the DFT and Wannier-based tight-binding band structures along $\overline{LA}$ in (c), the two band structures are otherwise in quite close agreement, and we have confirmed that $\alpha$-BiBr exhibits the same band ordering and bulk stable topology at $E_{F}$ (set to zero) in both calculations.
We will hence employ the Wannier-based tight-binding model shown with red circles in (c) to below compute the spin-resolved topology (SN~\ref{sec:spin_resolved_topology_of_alpha_bibr}) and spin-electromagnetic response (SN~\ref{sec:bibrshc}) of $\alpha$-BiBr.}
\label{fig:bibr_dft_band}
\end{figure}

$\alpha$-BiBr crystallizes in a centrosymmetric structure that respects the symmetries of nonmagnetic SG $C 2 / m 1'$ ($\#12.59$).
Each primitive (unit) cell of $\alpha$-BiBr contains eight Bi atoms and eight Br atoms [SFig.~\ref{fig:bibr_dft_band}(a)].
The primitive lattice vectors of $\alpha$-BiBr are given by
\begin{align}
	& \mathbf{a}_1 = (6.5320001 \text{ \AA}) \hat{\mathbf{x}}  - (2.1689999 \text{ \AA})  \hat{\mathbf{y}}, \nonumber  \\
	& \mathbf{a}_2 = (6.5320001 \text{ \AA}) \hat{\mathbf{x}} +     (2.1689999 \text{ \AA})  \hat{\mathbf{y}}, \nonumber \\
	& \mathbf{a}_3 = -(6.0057388 \text{ \AA}) \hat{\mathbf{x}}  +    (19.1409210 \text{ \AA}) \hat{\mathbf{z}}, \label{eq:bibrlatticevectorall}
\end{align}
SG $C2/m 1'$ ($\#12.59$) is generated by $C$-face-centered 3D lattice translations [SEq.~\eqref{eq:bibrlatticevectorall}], as well as $C_{2y}$ ($180^{\circ}$ rotation about the Cartesian $y$ axis), $\mathcal{I}$, and $\mathcal{T}$ symmetries.
In particular, $C_{2y}$ and $\mathcal{I}$ act on the lattice vectors ${\bf a}_{i}$ as
\begin{align}
	& C_{2y}: (\mathbf{a}_1,\mathbf{a}_2,\mathbf{a}_3) \to (-\mathbf{a}_2,-\mathbf{a}_1,-\mathbf{a}_3) ,\\
	& \mathcal{I}: (\mathbf{a}_1,\mathbf{a}_2,\mathbf{a}_3) \to (-\mathbf{a}_1,-\mathbf{a}_2,-\mathbf{a}_3).
\end{align}
The primitive reciprocal lattice vectors ${\bf G}_{i}$ [SFig.~\ref{fig:bibr_dft_band}(b)] dual to the lattice vectors ${\bf a}_{i}$ in SEq.~\eqref{eq:bibrlatticevectorall} satisfy $\mathbf{a}_i \cdot \mathbf{G}_j = 2\pi \delta_{ij}$, and are specifically given by 
\begin{align}
    & \mathbf{G}_1 = \frac{2\pi \left( \mathbf{a}_2 \times \mathbf{a}_3 \right)}{\mathbf{a}_1 \cdot \left( \mathbf{a}_2 \times \mathbf{a}_3 \right)} = 2\pi \left[ \left( 0.07654623 \text{ \AA}^{-1} \right) \hat{\mathbf{x}} - \left( 0.23052099 \text{ \AA}^{-1} \right) \hat{\mathbf{y}} + \left( 0.02401748 \text{ \AA}^{-1} \right) \hat{\mathbf{z}} \right] , \nonumber \\
    & \mathbf{G}_2 = \frac{2\pi \left( \mathbf{a}_3 \times \mathbf{a}_1 \right)}{\mathbf{a}_2 \cdot \left( \mathbf{a}_3 \times \mathbf{a}_1 \right)} = 2\pi \left[ \left( 0.07654623 \text{ \AA}^{-1} \right) \hat{\mathbf{x}} + \left( 0.23052099 \text{ \AA}^{-1} \right) \hat{\mathbf{y}} + \left( 0.02401748 \text{ \AA}^{-1} \right) \hat{\mathbf{z}} \right] , \nonumber \\
    & \mathbf{G}_3 = \frac{2\pi \left( \mathbf{a}_1 \times \mathbf{a}_2 \right)}{\mathbf{a}_3 \cdot \left( \mathbf{a}_1 \times \mathbf{a}_2 \right)} = 2\pi \left[ 0.05224409 \text{ \AA}^{-1} \right] \hat{\mathbf{z}}. \label{eq:bibrreciprocallatticevectorall}
\end{align}
To provide symmetry intuition for the ${\bf k}$-space electronic structure and spin-orbital texture of $\alpha$-BiBr, we note that $C_{2y}$ and $\mathcal{I}$ act on the primitive reciprocal lattice vectors $\{ \mathbf{G}_1 , \mathbf{G}_2 , \mathbf{G}_3 \}$ in SEq.~\eqref{eq:bibrreciprocallatticevectorall} as
\begin{align}
	& C_{2y}: (\mathbf{G}_1 , \mathbf{G}_2 , \mathbf{G}_3) \to (-\mathbf{G}_2 , -\mathbf{G}_1 , -\mathbf{G}_3), \nonumber \\
	& \mathcal{I}: (\mathbf{G}_1 , \mathbf{G}_2 , \mathbf{G}_3) \to (-\mathbf{G}_1 , -\mathbf{G}_2 , -\mathbf{G}_3).
\label{eq:BiBrC2yI}
\end{align}
Because SG $C 2 / m 1'$ ($\#12.59$) is $C$-face centered, and not primitive monoclinic, $\alpha$-BiBr is sometimes instead characterized by its conventional (primitive supercell) lattice vectors [SFig.~\ref{fig:bibr_dft_band}(a)]
\begin{align}
& \mathbf{a} = \mathbf{a}_1 + \mathbf{a}_2 = a \hat{\mathbf{x}}, \nonumber \\
& \mathbf{b} = -\mathbf{a}_1 + \mathbf{a}_2 = b \hat{\mathbf{y}}, \nonumber \\
& \mathbf{c} = \mathbf{a}_3 = c \cos{\beta} \hat{\mathbf{x}} + c \sin{\beta} \hat{\mathbf{z}}, 
\label{eq:bibr_conventional_lattice_all}
\end{align}
where $a$, $b$, and $c$ denote the conventional-cell lattice parameters~\cite{BiBrStructure}
\begin{equation}
    a = |\mathbf{a}| = 13.0640002 \text{ \AA},\ b = |\mathbf{b}| = 4.3379998 \text{ \AA},\ c = |\mathbf{c}| = 20.0610009 \text{ \AA}.
\end{equation}
In SEq.~\eqref{eq:bibr_conventional_lattice_all}, $\beta \approx 107.41999821484573^{\circ}$ is the angle between $\mathbf{a}$ and $\mathbf{c}$~\cite{BiBrStructure}.
The conventional reciprocal lattice vectors of $\alpha$-BiBr are correspondingly given by~\cite{SYBiBr}
\begin{align}
& \mathbf{a}^{*} = \frac{2\pi}{a}\left( \hat{\mathbf{x}} -\cot{\beta}\hat{\mathbf{z}}\right), \nonumber \\
& \mathbf{b}^{*} = \frac{2\pi}{b} \hat{\mathbf{y}}, \nonumber \\
& \mathbf{c}^{*} = \frac{2\pi}{c} \csc{\beta} \hat{\mathbf{z}}.
\end{align}

In order to analyze the spin-resolved bulk topology, we first use DFT incorporating the effects of SOC to compute the electronic band structure of $\alpha$-BiBr.  
We specifically performed first-principles calculations within the DFT framework using the PAW method~\cite{blochl1994improved,kresse1994norm} as implemented in VASP~\cite{kresse1996efficiency,kresse1996efficient}.  
In our DFT calculations, we adopted the PBE generalized gradient approximation exchange-correlations functional~\cite{perdew1996generalized}, and SOC was incorporated self-consistently. 
The cutoff energy for the plane-wave expansion was 400 eV, and $0.03 \times 2\pi \text{ \AA}^{-1}$ $\mathbf{k}$-point sampling grids were used in the self-consistent process. 
In SFig.~\ref{fig:bibr_dft_band}(c), we show the ab-initio band structure (labeled as ``DFT'') for $\alpha$-BiBr computed along high-symmetry BZ lines connecting TRIM points labeled using the Bilbao Crystallographic Server convention for SG $C 2 / m 1'$ ($\#12.59$) (see SFig.~\ref{fig:bibr_dft_band}(b) and SRefs.~\cite{aroyo2006bilbao,aroyo2006bilbaoa,aroyo2011crystallography,bilbaocrystallogrserver2017bandrep,bilbaocrystallogrserver2018check,elcoro2017double,elcoro2021magnetic}).
In our topological (Wilson-loop, see SN~\ref{sec:general_properties_nested_ppm}) analysis of $\alpha$-BiBr below, we will used reduced ${\bf k}$ coordinates $(k_{1},k_{2},k_{3})$ defined by $k_i = \mathbf{k} \cdot \mathbf{a}_i$, such that:
\begin{equation}
{\bf k} = \frac{1}{2\pi} \left( k_{1}{\bf G}_{1} + k_{2}{\bf G}_{2} + k_{3}{\bf G}_{3} \right),
\label{eq:kReduced}
\end{equation}
where each ${\bf G}_{i}$ is defined in SEq.~(\ref{eq:bibrreciprocallatticevectorall}).
In the reduced ${\bf k}=(k_{1},k_{2},k_{3})$ coordinates of SEq.~(\ref{eq:kReduced}), the TRIM points in $\alpha$-BiBr lie at the positions [SFig.~\ref{fig:bibr_dft_band}(b)]:
\begin{equation}
V=(\pi ,0, 0),\ L=(\pi ,0, -\pi),\ A=(0 ,0 ,\pi),\ \Gamma=(0, 0 ,0),\ M=(\pi ,\pi, -\pi),\ Y=(\pi ,\pi ,0).
\label{eq:BiBrTRIMpoints}
\end{equation}

To analyze the spin-resolved band topology of $\alpha$-BiBr, we next constructed a symmetric, Wannier-based tight-binding model fit to the electronic structure of $\alpha$-BiBr obtained from our DFT calculations.
We specifically constructed symmetric Wannier functions for the bands near $E_{F}$ in $\alpha$-BiBr by using the Wannier90 package~\cite{pizzi2020wannier90} for the Bi $6p$ and the Br $4p$ orbitals, and then performing a subsequent SG symmetrization using WannierTools~\cite{WU2018405}. 
Here and below, we denote the tight-binding Hamiltonian of the Wannier-based tight-binding model as $H_{\mathrm{Bi}\mathrm{Br}}$.
The single-particle Hilbert space of $H_{\mathrm{Bi}\mathrm{Br}}$ consists of 48 spinful Wannier functions per primitive (unit) cell; the Bloch Hamiltonian $[H_{\mathrm{Bi}\mathrm{Br}}(\mathbf{k})]$ is therefore a $96\times 96$ matrix, 
To reduce the computational resources required for our spin-resolved and Wilson loop tight-binding calculations, we have truncated $[H_{\mathrm{Bi}\mathrm{Br}}(\mathbf{k})]$ to only contain hopping terms with an absolute magnitude greater than or equal to $0.001$ eV. 
In SFig.~\ref{fig:bibr_dft_band}(c) we show the band structure of $[H_{\mathrm{Bi}\mathrm{Br}}(\mathbf{k})]$ in the vicinity of $E_{F}$ using red circles (denoted as ``Wannier\_symm'') overlaying the DFT-obtained electronic band structure (blue lines, denoted as ``DFT'').
We have confirmed that the truncated Wannier-based tight-binding model exhibits the same band ordering and qualitative features as the first-principles electronic structure.

\begin{figure}[t]
\includegraphics[width=\columnwidth]{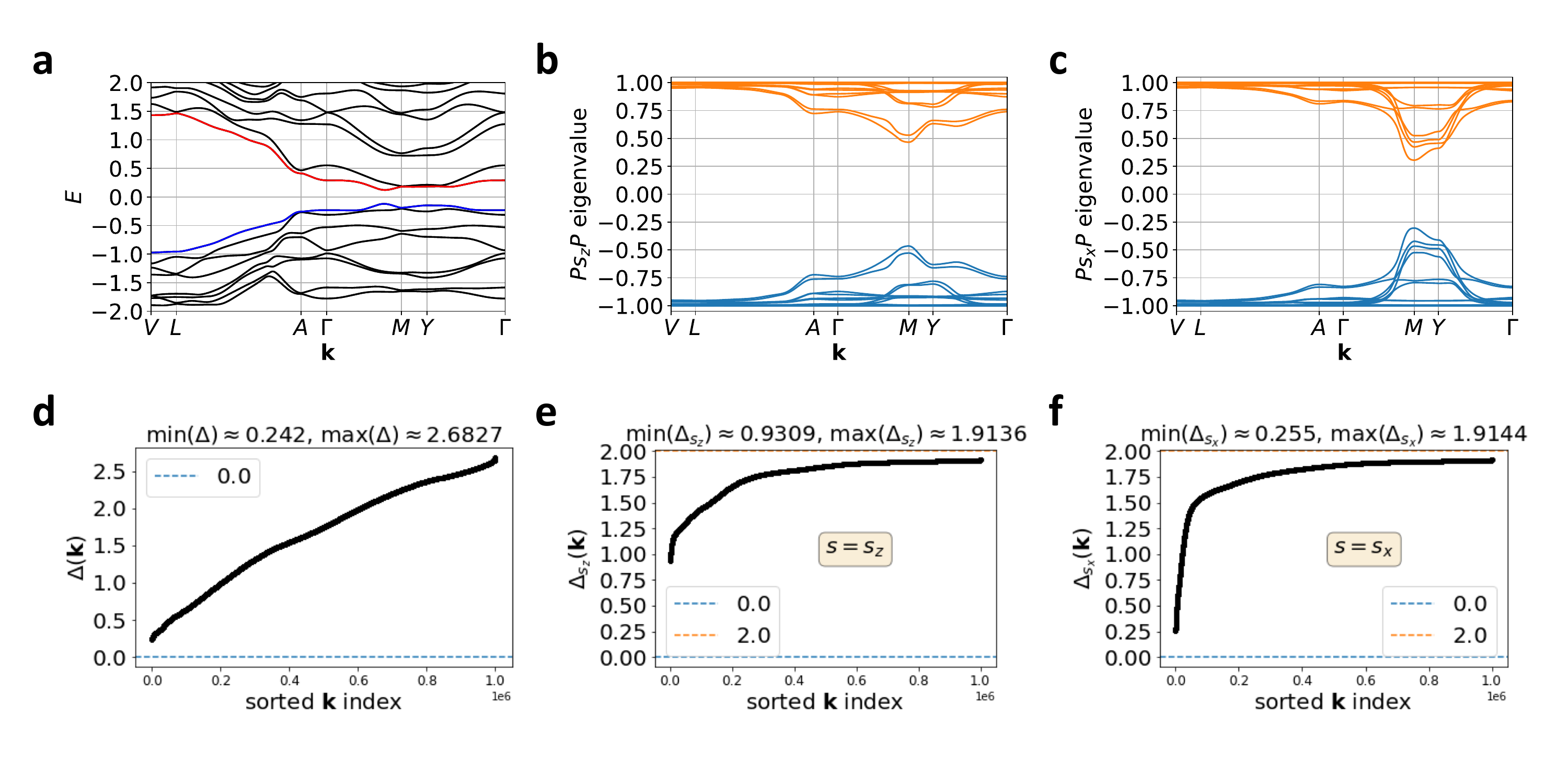}
\caption{Energy bands, energy gaps, $PsP$ spin bands, and $PsP$ spin gaps for the Wannier-based tight-binding model of $\alpha$-BiBr. 
(a) Tight-binding energy band structure along high-symmetry BZ lines in $\alpha$-BiBr [dashed green lines in SFig.~\ref{fig:bibr_dft_band}(b)].
In (a), the bands are doubly degenerate due to bulk $\mathcal{I}\times\mathcal{T}$ symmetry~\cite{wieder2016spinorbit}, and the highest valence (lowest conduction) bands are colored in blue (red).
(b,c) Respectively the $P s_z P$ and $P s_x P$ band structures computed along the same high-symmetry BZ lines as the electronic band structure in (a), with the positive (negative) spin bands colored in orange (blue).
(d) Direct energy gaps $\Delta({\bf k})$ of the Wannier-based tight-binding model in (a) at $\mathbf{k}$ points sampled from a uniformly spaced [in the reduced ${\bf k}$ coordinates defined in SEq.~(\ref{eq:kReduced}) and the surrounding text] $100 \times 100 \times 100$ grid over the 3D BZ of $\alpha$-BiBr.
We find that the direct energy gap takes the minimal value $\Delta({\bf k})\approx 0.24196355481095755$ eV. 
(e,f) Respectively the direct $P s_z P$ and $P s_x P$ spin gaps $\Delta_{s}({\bf k})$ (in the units of $\hbar/2$) computed over the uniform (in reduced coordinates) ${\bf k}$-grid employed in (d).
For the spin direction $\hat{\mathbf{n}}=\hat{\mathbf{z}}$ in panel (e) [$\hat{\mathbf{n}}=\hat{\mathbf{x}}$ in panel (f)], we find that the direct spin gap takes the minimal value $\Delta_{s_{z}}({\bf k})\approx 0.9309028798325673$ [$\Delta_{s_{x}}({\bf k})\approx 0.2550432063802285$].
The calculations detailed in this figure were performed using the freely available Python package~\href{https://github.com/kuansenlin/nested_and_spin_resolved_Wilson_loop}{\textsc{nested\_and\_spin\_resolved\_Wilson\_loop}}~\cite{lin2023nestedWilsonLib}, which represents an extension of the~\href{https://www.physics.rutgers.edu/pythtb/}{PythTB} open-source Python tight-binding package~\cite{coh2013python} that was implemented and utilized for the preparation of SRefs.~\cite{wieder2018axion,wieder2020strong} and the present work.}
\label{fig:bibr_Wannier_energy_bands_spin_sz_sx_bands}
\end{figure}

In our DFT calculations, $\alpha$-BiBr is an insulator with 64 occupied spinful valence bands, which appear in doubly-degenerate pairs due to bulk $\mathcal{I}\times\mathcal{T}$ symmetry~\cite{wieder2016spinorbit}.  
In SFig.~\ref{fig:bibr_Wannier_energy_bands_spin_sz_sx_bands}(a), we again plot the band structure of the Wannier-based tight-binding model $[H_{\mathrm{Bi}\mathrm{Br}}(\mathbf{k})]$, coloring the highest occupied (lowest unoccupied) pair of bands in blue (red).  
To confirm that the truncated form of $[H_{\mathrm{Bi}\mathrm{Br}}(\mathbf{k})]$ exhibits a band gap at $E_{F}$ at all ${\bf k}$ points, we have computed the energy gap $\Delta (\mathbf{k})$ between the 64$^\text{th}$ (highest occupied valence) and 65$^\text{th}$ (lowest unoccupied conduction) bands of $[H_{\mathrm{Bi}\mathrm{Br}}(\mathbf{k})]$ using a $100 \times 100 \times 100$ grid uniformly spaced over the 3D BZ in the reduced ${\bf k}$ coordinates $(k_{1},k_{2},k_{3})$ defined in SEq.~(\ref{eq:kReduced}) and the surrounding text.
As shown in SFig.~\ref{fig:bibr_Wannier_energy_bands_spin_sz_sx_bands}(d), we find that the band gap at $E_{F}$ at all ${\bf k}$ points is nonzero, and takes values greater than $\Delta({\bf k})\approx 0.24196355481095755$ eV. 
As a final test that $[H_{\mathrm{Bi}\mathrm{Br}}(\mathbf{k})]$ is gapped at $E_{F}$, we further performed a numerical Nelder-Mead minimization~\cite{pandey2022py} on the direct energy band gap $\Delta (\mathbf{k})$ using uniformly sampled $20 \times 20 \times 20$ ${\bf k}$ grids (again in reduced ${\bf k}$ coordinates) as initial points of minimization.
For the truncated form of $[H_{\mathrm{Bi}\mathrm{Br}}(\mathbf{k})]$, we found after 8000 Nelder-Mead minimizations that $\Delta (\mathbf{k})\geq 0.24181320342543444$eV, essentially identical to the value obtained from uniform BZ sampling [SFig.~\ref{fig:bibr_Wannier_energy_bands_spin_sz_sx_bands}(d)].
Hence for our calculations below of the spin-resolved topology (SN~\ref{sec:spin_resolved_topology_of_alpha_bibr}) and spin-electromagnetic response (SN~\ref{sec:bibrshc}) of $\alpha$-BiBr, we have established that the projector onto the occupied (valence) bands, $[P(\mathbf{k})] = \sum_{n=1}^{64} | u_{n,\mathbf{k}}\rangle \langle u_{n,\mathbf{k}} |$ where $| u_{n,\mathbf{k}}\rangle$ is the $n^{th}$ eigenvector of $[H_{\mathrm{Bi}\mathrm{Br}}(\mathbf{k})]$, is numerically well-defined over the full 3D BZ.

\subsection{Spin-Resolved Topology of $\alpha$-BiBr}
\label{sec:spin_resolved_topology_of_alpha_bibr}

In this section we will analyze the spin-resolved topology of the candidate helical HOTI $\alpha$-BiBr.  
As we will show below, we find that $\alpha$-BiBr hosts a bulk spin gap over a large range of spin resolution directions $\hat{\mathbf{n}}$, and specifically hosts both 3D QSHI (\emph{e.g.} for $\hat{\mathbf{n}}=\hat{\mathbf{z}}$) and T-DAXI (\emph{e.g.} for $\hat{\mathbf{n}}=\hat{\mathbf{x}}$) spin-stable states.

\textit{Spin Spectrum of $\alpha$-BiBr.}  We will begin by computing the spin gap of $\alpha$-BiBr over the full range of spin resolution directions $\hat{\mathbf{n}}$.  
The spin band structure is defined as the eigenspectrum of $PsP \equiv [P(\mathbf{k})]s[P(\mathbf{k})]$ computed as a function of $\mathbf{k}$.
As previously with $\beta$-MoTe$_2$ in SN~\ref{sec:mote2_spin_resolved_topology}, $\alpha$-BiBr has both $\mathcal{I}$ and spinful $\mathcal{T}$ symmetries. Hence, the $PsP$ spectrum of $\alpha$-BiBr exhibits an antiunitary chiral symmetry due to the relation
\begin{equation}
	[\mathcal{I}] [\mathcal{T}][P(\mathbf{k})]s[P(\mathbf{k})] ([\mathcal{I}] [\mathcal{T}])^{-1} = - [P(\mathbf{k})]s[P(\mathbf{k})],
 \label{eq:BiBrChiral}
\end{equation}
where $[\mathcal{I}]$ and $[\mathcal{T}]$ are the matrix representatives of $\mathcal{I}$ and spinful $\mathcal{T}$ symmetries acting on the $96\times 96$ Wannier-based tight-binding Bloch Hamiltonian matrix $[H_{\mathrm{Bi}\mathrm{Br}} (\mathbf{k})]$ that we obtained from DFT calculations of the electronic structure of $\alpha$-BiBr (SN~\ref{sec:bibr_dft_details}), and where
$[P(\mathbf{k})]$ denotes the matrix projector $[P(\mathbf{k})] = \sum_{n=1}^{64} | u_{n,\mathbf{k}}\rangle \langle u_{n,\mathbf{k}} |$ onto the 64 occupied (valence) bands of $\alpha$-BiBr.
As discussed in SN~\ref{appendix:properties-of-the-projected-spin-operator}, the presence of bulk $\mathcal{I}\times\mathcal{T}$ symmetry [SEq.~\eqref{eq:BiBrChiral}] further implies that 
\begin{equation}
	\mathrm{Spec}\left( [P(\mathbf{k})]s[P(\mathbf{k})] \right) = -\mathrm{Spec}\left( [P(\mathbf{k})]s[P(\mathbf{k})] \right),
\label{eq:BiBrSpecO}
\end{equation}
where $\mathrm{Spec} (\mathcal{O})$ denotes the spectrum of the operator $\mathcal{O}$.
From SEq.~(\ref{eq:BiBrSpecO}), we then again define the spin gap function $\Delta_s (\mathbf{k})$ as the difference between the smallest positive and the largest negative $PsP$ eigenvalues at each ${\bf k}$ point.

\begin{figure}[t]
\includegraphics[width=0.65\columnwidth]{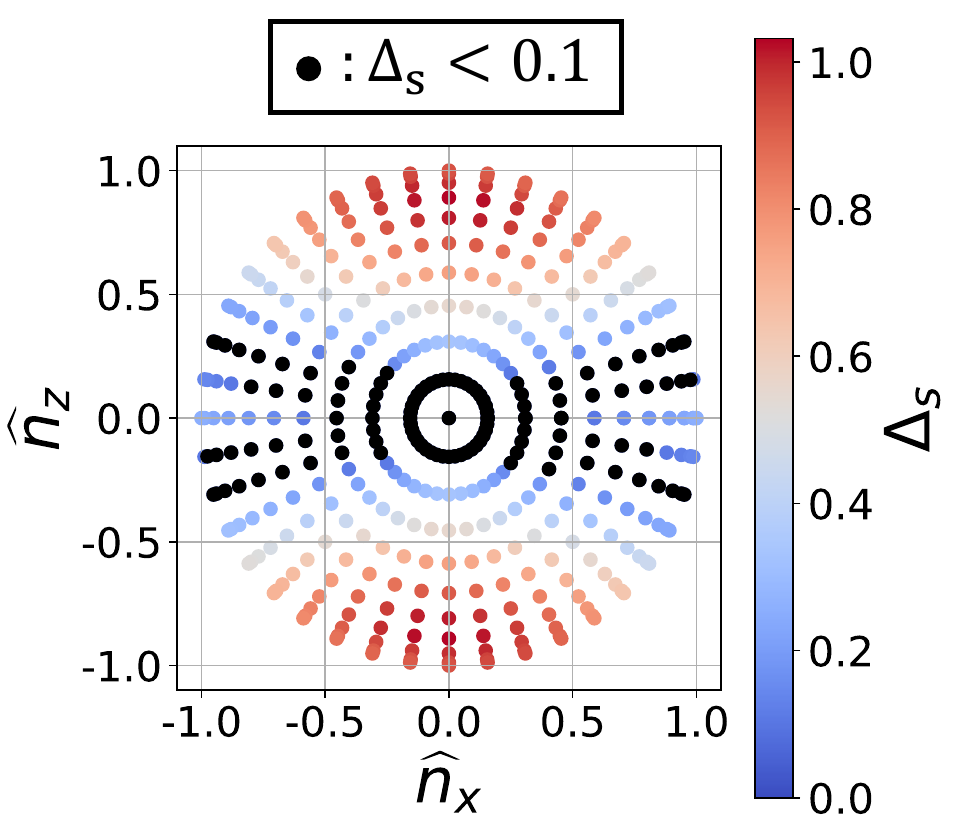}
\caption{Numerical spin gap $\Delta_{s = \hat{\mathbf{n}} \cdot \mathbf{s}}$ as a function of the spin resolution direction $\hat{\mathbf{n}}$ for $\alpha$-BiBr. 
The angular parameterization of $\hat{\mathbf{n}}$ is defined in SEq.~\eqref{eq:hat_n_around_y_axis}, and our calculations were performed over the spin-direction hemisphere of $\hat{\mathbf{n}}$ defined by $\vartheta \in [0,0.5\pi]$ and $\phi \in [0,2\pi]$.
For our calculations, the angular variables $(\vartheta,\phi)$ were respectively sampled using the numerical resolution of $\Delta \vartheta = 0.05\pi$ and $\Delta \phi = 0.05\pi$.
By performing a Nelder-Mead minimization~\cite{pandey2022py} on the spin gap function $\Delta_s (\mathbf{k})$ taking as the initial points $100$ $\mathbf{k}$ points randomly sampled from the 3D BZ in the reduced ${\bf k}$ coordinates $(k_{1},k_{2},k_{3})$ defined in SEq.~(\ref{eq:kReduced}), we define for each $\hat{\mathbf{n}}$ the numerical spin gap $\Delta_{s = \hat{\mathbf{n}} \cdot \mathbf{s}}$ as the minimal value (in the units of $\hbar/2$) of the 100 minimization results for the fixed value of $\hat{\mathbf{n}}$.
Unlike previously for $\beta$-MoTe$_2$ [SFig.~\ref{fig:mote2_numerical_spin_gap_for_different_spin_directions}], we find that $\alpha$-BiBr exhibits a significant spin gap ($\Delta_s > 0.1$) for most values of $\hat{\mathbf{n}}$, and becomes larger than $0.8$ (40\% of the maximum possible value $\Delta_{s}=2$) over a significant range of spin resolution directions roughly centered around $\hat{\mathbf{n}}=\pm\hat{\mathbf{z}}$ (light and dark red circles).
We find specifically that the global spin gap in $\alpha$-BiBr peaks at a large value [$\Delta_{s}\approx 0.9479813926905263$] and lies at $\hat{\mathbf{n}} = (\hat{n}_x,\hat{n}_y,\hat{n}_z) = \pm ( -0.2486898871648553, 0, 0.968583161128631 )$, which is surprisingly oriented within $\approx 3.019998214845685 $ degrees of the $\mathbf{a}_3\parallel {\bf c}$ lattice vector [see SEqs.~\eqref{eq:bibrlatticevectorall} and~\eqref{eq:bibr_conventional_lattice_all} and SFig.~\ref{fig:bibr_dft_band}(a)]. 
This indicates that the bulk spin-orbital texture in $\alpha$-BiBr is dominated by contributions that are almost entirely polarized along the $\mathbf{a}_{3} \parallel {\bf c}$ axis.
We note that there also exists a narrower pair of spin-gapped region with smaller values of $\Delta_{s}$ centered around $\hat{\mathbf{n}}=\pm\hat{\mathbf{x}}$ (light blue circles), and that in particular $\Delta_s (\mathbf{k})\geq 0.2550432063802285$ for $\hat{\mathbf{n}}=\pm\hat{\mathbf{x}}$ [see SFig.~\ref{fig:bibr_Wannier_energy_bands_spin_sz_sx_bands}(f)].  As we will show later in this section, all four spin-gapped regions of $\alpha$-BiBr exhibit nontrivial spin-resolved topology, with the $\pm\hat{\mathbf{z}}$- ($\pm\hat{\mathbf{x}}$-) centered regions specifically hosting 3D QSHI (T-DAXI) spin-stable states.
The calculations detailed in this figure were performed using the freely available Python package~\href{https://github.com/kuansenlin/nested_and_spin_resolved_Wilson_loop}{\textsc{nested\_and\_spin\_resolved\_Wilson\_loop}}~\cite{lin2023nestedWilsonLib}, which represents an extension of the~\href{https://www.physics.rutgers.edu/pythtb/}{PythTB} open-source Python tight-binding package~\cite{coh2013python} that was implemented and utilized for the preparation of SRefs.~\cite{wieder2018axion,wieder2020strong} and the present work.}
\label{fig:bibr_numerical_spin_gap_for_different_spin_directions}
\end{figure}

Before performing the more intensive calculation of the global (direct) spin gap for each spin resolution direction $\hat{\mathbf{n}}$, we have first computed the full spin spectrum (band structure) of $\alpha$-BiBr for the high-symmetry $\hat{\mathbf{n}}=\hat{\mathbf{z}}$ and $\hat{\mathbf{n}}=\hat{\mathbf{x}}$ spin directions. 
As shown in SFig.~\ref{fig:bibr_Wannier_energy_bands_spin_sz_sx_bands}(b,c), we find that the $P s_z P$ and $P s_x P$ spin spectra of $\alpha$-BiBr are gapped along high-symmetry BZ lines.
Next, to search for the existence of spin gap closures at lower-symmetry ${\bf k}$ points in the 3D BZ interior (\emph{e.g.} spin-Weyl points), we computed the local spin gap $\Delta_s (\mathbf{k})$ over a $100 \times 100 \times 100$ grid uniformly spaced over the 3D BZ in the reduced ${\bf k}$ coordinates $(k_{1},k_{2},k_{3})$ defined in SEq.~(\ref{eq:kReduced}).
For both $\hat{\mathbf{n}}=\hat{\mathbf{z}}$ and $\hat{\mathbf{n}}=\hat{\mathbf{x}}$, we observe a nonzero $\Delta_s ({\bf k})$ at all sampled ${\bf k}$ points, with the specific local spin gap values of $\Delta_{s_z} (\mathbf{k})\geq 0.9309028798325673$ for $\hat{\mathbf{n}}=\hat{\mathbf{z}}$ [SFig.~\ref{fig:bibr_Wannier_energy_bands_spin_sz_sx_bands}(e)] and $\Delta_{s_x} (\mathbf{k})\geq 0.2550432063802285$ for $\hat{\mathbf{n}}=\hat{\mathbf{x}}$ [SFig.~\ref{fig:bibr_Wannier_energy_bands_spin_sz_sx_bands}(f)].

To gain a more comprehensive understanding of the spin gap structure in $\alpha$-BiBr, we next computed the global (minimum) spin gap $\Delta_s \equiv \min_{\mathbf{k} \in \mathrm{BZ}} [ \Delta_s (\mathbf{k})]$ for each spin direction $\hat{\mathbf{n}}$ in the $P(\hat{\mathbf{n}}\cdot\mathbf{s})P$ spectrum of $\alpha$-BiBr.
As we will show below, we find that unlike $\beta$-MoTe$_2$ (SN~\ref{sec:mote2_spin_resolved_topology}), $\alpha$-BiBr in fact hosts a topological spin gap over a large range of $\hat{\mathbf{n}}$.
We begin by re-expressing the spin direction $\hat{\mathbf{n}}$ as a 3D unit vector parameterized in rotated spherical coordinates as
\begin{equation}
	\hat{\mathbf{n}} = (\sin{\vartheta}\cos{\phi},\cos{\vartheta},-\sin{\vartheta}\sin{\phi}). 
\label{eq:hat_n_around_y_axis}
\end{equation}
The parameterization in SEq.~\eqref{eq:hat_n_around_y_axis} is chosen such that $(\vartheta,\phi)=(0,0)$, $(0.5\pi,0)$, and $(0.5\pi,-0.5\pi)$ respectively correspond to the Cartesian unit vectors $\hat{\mathbf{y}}$, $\hat{\mathbf{x}}$, and $\hat{\mathbf{z}}$, a choice that is motivated by the $C_{2y}$ rotation symmetry of $\alpha$-BiBr [SEq.~(\ref{eq:BiBrC2yI})].
For each $(\vartheta,\phi)$, we then performed a Nelder-Mead minimization~\cite{pandey2022py} on the spin gap function $\Delta_s (\mathbf{k})$ taking for the initial points $100$ $\mathbf{k}$ points randomly sampled from the 3D BZ in the reduced ${\bf k}$ coordinates $(k_{1},k_{2},k_{3})$ defined in SEq.~(\ref{eq:kReduced}).
We next define the \textit{numerical spin gap} $\Delta_s$ as the minimal value of the $100$ minimization results for each spin direction $\hat{\mathbf{n}}$.
In SFig.~\ref{fig:bibr_numerical_spin_gap_for_different_spin_directions}, we show the numerical spin gap $\Delta_s$ (in the units of $\hbar/2$) for $\hat{\mathbf{n}}$ (projected into the $xz$ plane) sampled over the spin hemisphere defined by $\vartheta \in [0,0.5\pi]$ and $\phi \in [0,2\pi]$ [SEq.~\eqref{eq:hat_n_around_y_axis}] using the angular resolutions $\Delta \vartheta = 0.05\pi$ and $\Delta \phi = 0.05\pi$.
Such a sampling of $(\vartheta,\phi)$ specifically corresponds to the upper hemisphere of a unit sphere with the north pole is identified as $\hat{\mathbf{y}}$.
Because $\Delta_s = \Delta_{-s}$ due to $\mathcal{T}$ symmetry, this sampling of the angular variables $(\vartheta,\phi)$ contains all independent values of the numerical spin gap $\Delta_{s}$ over the full sphere of spin resolution directions $\hat{\mathbf{n}}$.

As shown in SFig.~\ref{fig:bibr_numerical_spin_gap_for_different_spin_directions}, the numerical spin gap $\Delta_s > 0.1$ for most values of $\hat{\mathbf{n}}$, and is larger than $0.8$ (40\% of the maximum possible value $\Delta_{s}=2$) over a significant range of spin resolution directions roughly centered around $\hat{\mathbf{n}}=\pm\hat{\mathbf{z}}$ (light and dark red circles in SFig.~\ref{fig:bibr_numerical_spin_gap_for_different_spin_directions}).
For our subsequent calculations, it is important to note that there also exists a narrower pair of spin-gapped regions in $\alpha$-BiBr in the vicinity of $\hat{\mathbf{n}}=\pm\hat{\mathbf{x}}$ with smaller values of $\Delta_{s}$ (light blue circles in SFig.~\ref{fig:bibr_numerical_spin_gap_for_different_spin_directions}) relative to those in the vicinity of $\hat{\mathbf{n}}=\pm\hat{\mathbf{z}}$. 
The $\pm s_{z}$ spin gap in $\alpha$-BiBr is hence large ($\Delta_{s_{z}}\approx 0.9309028798325673$), and is much larger than the $\pm s_{x}$ spin gap ($\Delta_{s_{x}}\approx 0.2550432063802285$).
This is consistent with earlier first-principles investigations of $\alpha$-BiBr, which found the spin-electromagnetic (Rashba-Edelstein) response of its $(010)$-surface states to be strongly polarized in the $z$-direction relative to the $x$-direction~\cite{SYBiBr}.
We further crucially observe in SFig.~\ref{fig:bibr_numerical_spin_gap_for_different_spin_directions} a pair of effectively spin-gapless ($\Delta_{s} < 0.1$) lines separating the four spin-gapped regions, suggesting the possibility that the $\hat{\mathbf{n}}=\pm\hat{\mathbf{z}}$ and $\hat{\mathbf{n}}=\pm\hat{\mathbf{x}}$ spin-gapped regions of $\alpha$-BiBr are separated by spin-resolved topological phase transitions.
Below, we will shortly confirm this intuition, finding specifically through (nested) spin-resolved Wilson-loop calculations that $\alpha$-BiBr is a 3D QSHI [T-DAXI] for $\hat{\mathbf{n}}=\pm\hat{\mathbf{z}}$ [$\hat{\mathbf{n}}=\pm\hat{\mathbf{x}}$].

Overall, we find that the global spin gap in $\alpha$-BiBr peaks at a large value ($\Delta_{s}\approx 0.9479813926905263$) and lies at $\hat{\mathbf{n}} = (\hat{n}_x,\hat{n}_y,\hat{n}_z) = \pm ( -0.2486898871648553, 0, 0.968583161128631 )$, which is surprisingly oriented within $\approx 3.019998214845685^\circ $ of the lattice vector $\mathbf{a}_3$ [SEq.~\eqref{eq:bibrlatticevectorall}]. 
This indicates that the bulk spin-orbital texture in $\alpha$-BiBr is dominated by contributions that are almost entirely polarized along the $\mathbf{a}_{3} \parallel {\bf c}$ lattice vector [SFig.~\ref{fig:bibr_dft_band}(a)].
In 2D materials, such as superconducting transition-metal dichalcogenides, analogous spin-orbital textures polarized along high-symmetry crystallographic axes have also been observed, and have been termed ``Ising SOC''~\cite{wang2021isingSC,IsingSCexp1,IsingSCexp2,IsingSCexp3,BiaoIsingSCsearch}.
The appearance of a large spin gap nearly locked to a crystallographic axis in SFig.~\ref{fig:bibr_numerical_spin_gap_for_different_spin_directions} suggests that it would be intriguing to investigate the microscopic mechanism of the SOC in $\alpha$-BiBr in future theoretical studies, and to study the spin-resolved response of $\alpha$-BiBr in future photoemission and transport experiments, which may exhibit an unusually high degree of spin polarization relative to other strongly spin-orbit-coupled 3D materials.

To conclude our analysis of the spin spectrum of $\alpha$-BiBr, we last performed a Nelder-Mead minimization~\cite{pandey2022py} for $\hat{\mathbf{n}}=\hat{\mathbf{y}}$ on the spin gap function $\Delta_s (\mathbf{k})$, taking for the initial points uniformly sampled ${\bf k}$ points drawn from $10 \times 10 \times 10$ grids in the 3D BZ in the reduced ${\bf k}$ coordinates $(k_{1},k_{2},k_{3})$ defined in SEq.~(\ref{eq:kReduced}).
We find that the numerical spin gap $\Delta_{s_y}\approx 7.981038733049717 \times 10^{-6}$, such that the $P s_y P$ spectrum of $\alpha$-BiBr can be identified as spin-gapless within numerical precision.
By direct computation, we have confirmed that like $\beta$-MoTe$_2$ for all spin-resolution directions (SN~\ref{sec:mote2_spin_resolved_topology}), $\alpha$-BiBr for $\hat{\mathbf{n}}=\hat{\mathbf{y}}$ is specifically a spin-Weyl semimetal with two spin-Weyl points within each half of the 3D BZ, as expected for a helical HOTI lying in the critical DSTI (spin-Weyl) regime that separates 3D QSHI and T-DAXI spin-stable states (see SN~\ref{app:response_QSHI_TDAXI}).

\begin{figure}[t]
\includegraphics[width=\columnwidth]{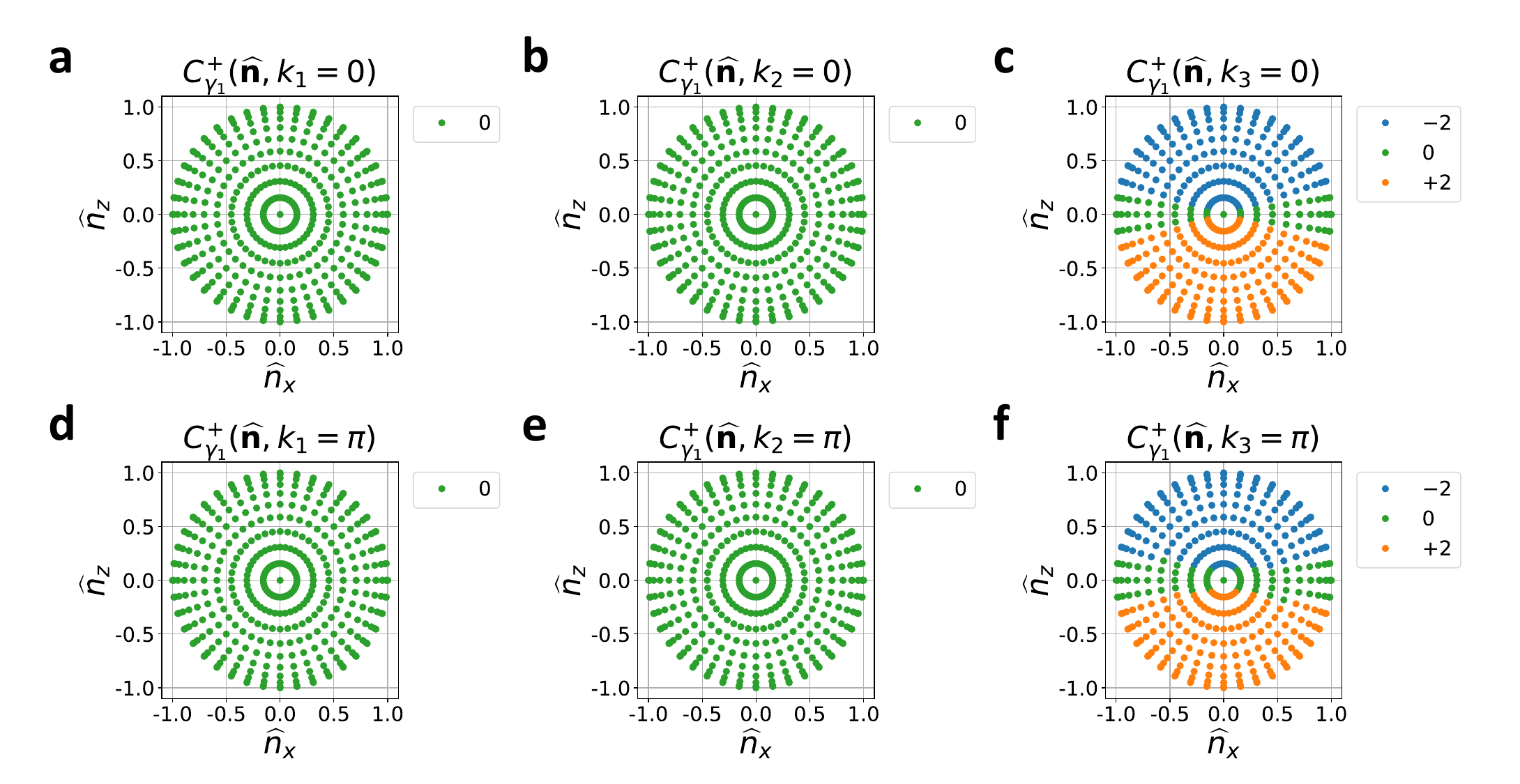}
\caption{Partial Chern numbers of the occupied bands in $\alpha$-BiBr.
In each panel, we show the partial Chern numbers $C^+_{\gamma_1}(\hat{\mathbf{n}}, k_i )$ in the high-symmetry ($\mathcal{T}$-invariant) $k_{i}=0,\pi$ ($i=1,2,3)$ BZ planes.
Because $\mathcal{T}$ symmetry enforces that $C^+_{\gamma_1}(-\hat{\mathbf{n}}, k_i )=-C^+_{\gamma_1}(\hat{\mathbf{n}}, k_i )$, and because $C^-_{\gamma_1}(\hat{\mathbf{n}}, k_i )=-C^+_{\gamma_1}(\hat{\mathbf{n}}, k_i )$ at $k_i = 0,\pi$, then we in this figure only show $C^+_{\gamma_1}(\hat{\mathbf{n}}, k_i )$ sampled over the positive spin hemisphere defined in SEq.~(\ref{eq:hat_n_around_y_axis}) for which $\hat{\mathbf{n}}=\hat{\mathbf{y}}$ is taken to be the north pole.
At each point in each panel in this figure, we show the value of the partial Chern number $C^+_{\gamma_1}(\hat{\mathbf{n}}, k_i )$ plotted at the position of $\hat{\mathbf{n}}$ projected onto the $xz$-plane.
As shown in (a,b,d,e), $C^+_{\gamma_1}(\hat{\mathbf{n}}, k_i )=0$ at all $\hat{\mathbf{n}}$ for $i=1,2$ and $k_{i}=0,\pi$.
Conversely for $k_{3}=0,\pi$, there are large regions of $\hat{\mathbf{n}}$ parameter space (centered around $\hat{\mathbf{n}}=\pm\hat{\mathbf{z}}$) in which the partial Chern numbers are nontrivial $C^+_{\gamma_1}(\hat{\mathbf{n}}, k_3 )=\pm 2$.
Though the regions with $C^+_{\gamma_1}(\hat{\mathbf{n}}, k_3) = \pm 2$ in (c,f) [blue and orange circles] appear identical, we note that there exist small numerical differences in the shape and range of the intermediate $C^+_{\gamma_1}(\hat{\mathbf{n}}, k_3) = 0$ regions in (c,f) [green circles].
Along with the nonvanishing global spin gap for most $\hat{\mathbf{n}}$ in $\alpha$-BiBr established in SFig.~\ref{fig:bibr_numerical_spin_gap_for_different_spin_directions}, and the nontrivial (partial) SIs of $\alpha$-BiBr (see SN~\ref{sec:defs_partial_sis} and SRefs.~\cite{tang2019efficient,SYBiBr,BiBrFanHOTI}), the data in this figure indicate that $\alpha$-BiBr is a $\bm{\nu}^{\pm}=(\nu^{\pm}_1,\nu^{\pm}_2,\nu^{\pm}_3) = (0,0,\mp 2)$ or $(0,0,\pm 2)$ 3D QSHI over the large spin-gapped regions in (c,f) with $C^+_{\gamma_1}(\hat{\mathbf{n}}, k_3) = -2$ or $+2$, and is remarkably a T-DAXI for the smaller spin-gapped region in (c,f) with $C^+_{\gamma_1}(\hat{\mathbf{n}}, k_3) = 0$ [see SFig.~\ref{fig:bibr_numerical_spin_gap_for_different_spin_directions} for comparison].
We will shortly confirm this conclusion by direct computation of the nested partial Wilson loop spectrum of $\alpha$-BiBr.
The calculations detailed in this figure were performed using the freely available Python package~\href{https://github.com/kuansenlin/nested_and_spin_resolved_Wilson_loop}{\textsc{nested\_and\_spin\_resolved\_Wilson\_loop}}~\cite{lin2023nestedWilsonLib}, which represents an extension of the~\href{https://www.physics.rutgers.edu/pythtb/}{PythTB} open-source Python tight-binding package~\cite{coh2013python} that was implemented and utilized for the preparation of SRefs.~\cite{wieder2018axion,wieder2020strong} and the present work.}
\label{fig:bibr_partial_Chern_numbers_different_spin_directions}
\end{figure}

\textit{Spin-Resolved Topology of $\alpha$-BiBr: Spin-Resolved Wilson Loop.} Having established that $\alpha$-BiBr hosts a bulk spin gap over a large range of spin resolution directions $\hat{\mathbf{n}}$, we will next focus on analyzing the bulk (spin-resolved) topology of $\alpha$-BiBr from several perspectives.
We begin by first investigating the partial Chern numbers $C^{\pm}_{\gamma_1}(\hat{\mathbf{n}},k_i)$ of the occupied $64$ occupied valence bands in constant-$k_i$ BZ planes ($i=1,2,3$).
We first recall that if a 3D insulator hosts a global spin gap for a spin direction $\hat{\mathbf{n}}$, then $C^{\pm}_{\gamma_1}(\hat{\mathbf{n}},k_i)$ evaluated in \emph{any} BZ plane of constant $k_{i}$ will take the same values in all other BZ planes of constant $k_{i}$ (keeping $i$ fixed).
This can be seen by recognizing that $C^{\pm}_{\gamma_1}(\hat{\mathbf{n}},k_i)$ can only change via the closing and reopening of a spin gap that manifests as a nodal degeneracy in the 3D spin spectrum with a nonvanishing partial chiral charge, such as a spin-Weyl point [see SN~\ref{sec:main-text-3D-TI-P-pm} and \ref{app-relation-nested-P-pm-and-partial-weak-Chern}].
In this section and below in SN~\ref{sec:bibrshc} we will, for simplicity, express the partial Chern vector $\bm{\nu}^{\pm}$ in the reduced coordinate of the reciprocal lattice vectors $\{\mathbf{G}_1 , \mathbf{G}_2 , \mathbf{G}_3 \}$.
As discussed in SN~\ref{sec:general_properties_of_winding_num_of_P_pm_Wilson} and \ref{app-relation-nested-P-pm-and-partial-weak-Chern}, in spin-gapped 3D insulators, the partial Chern vectors $\bm{\nu}^{\pm}$ indicate the topological contributions $\sigma^{H}_{s=\hat{\mathbf{n}}\cdot \mathbf{s},\mathrm{top},ij}$ to the bulk \emph{non-quantized} spin Hall conductivity $\sigma^{H}_{s=\hat{\mathbf{n}}\cdot \mathbf{s},ij}$ ($i \neq j $ and $i,j \in \{1,2,3\}$).
As shown in SEq.~\eqref{eq:3dshc}, $\sigma^{H}_{s=\hat{\mathbf{n}}\cdot \mathbf{s},ij}$ specifically satisfies the relation
\begin{equation}
	\sigma^{H}_{s=\hat{\mathbf{n}}\cdot \mathbf{s},\mathrm{top},ij} = \frac{e}{4\pi} \sum_{k=1}^{3} \epsilon_{ijk} \left[ \sum_{l=1}^{3} (\nu_l^+ - \nu_l^-)\mathbf{G}_l \right]_k.
\end{equation}
In $\mathcal{T}$-invariant (nonmagnetic) insulators like $\alpha$-BiBr, $\nu^{-}_i = -\nu^{+}_i$, such that
\begin{equation}
	\sigma^{H}_{s=\hat{\mathbf{n}}\cdot \mathbf{s},\mathrm{top},ij} = \frac{2e}{4\pi} \sum_{k=1}^{3} \epsilon_{ijk} \left[ \sum_{l=1}^{3} \nu_l^+ \mathbf{G}_l \right]_k. \label{eq:bibr_sec_topo_shc}
\end{equation}

As introduced in this work, the partial Chern numbers $C^{\pm}_{\gamma_1}(\hat{\mathbf{n}},k_i)$ can be numerically obtained via the winding numbers of the spin-resolved Wilson loop spectra of the occupied bands in constant-$k_i$ BZ planes, here specifically $k_{i}=0,\pi$ (see SN~\ref{sec:general_properties_of_winding_num_of_P_pm_Wilson}).
As discussed above, if $C^{\pm}_{\gamma_1}(\hat{\mathbf{n}},k_i)$ in the $k_i = 0,\pi$ planes take different values, then there must exist spin gap closing points with nonvanishing partial chiral charges (such as spin-Weyl points) between $k_{i}=0,\pi$.
Lastly as seen in SN~\ref{sec:mote2_spin_resolved_topology}, diagnosing the bulk spin-resolved topology via high-symmetry ($k_{i}=0,\pi$) BZ-plane partial Chern numbers is only a numerically stable calculation when spin-gap closing points do not lie close to the $k_{i}=0,\pi$ BZ planes.
However, as shown earlier in this section [SFig.~\ref{fig:bibr_numerical_spin_gap_for_different_spin_directions}], $\alpha$-BiBr is in fact spin-gapped for most values of the spin resolution direction $\hat{\mathbf{n}}$.

To compute the high-symmetry $k_{i}=0,\pi$ partial Chern numbers of $\alpha$-BiBr, we begin by again parameterizing $\hat{\mathbf{n}}$ using SEq.~\eqref{eq:hat_n_around_y_axis} with $\vartheta$ and $\phi$ sampled over the angular range $\vartheta \in [0,0.5\pi]$ and $\phi \in [0,2\pi]$ using the angular resolution $\Delta \vartheta = 0.05\pi$ and $\Delta \phi = 0.05\pi$.
In SFig.~\ref{fig:bibr_partial_Chern_numbers_different_spin_directions}, we show the partial Chern numbers for the occupied bands in $\alpha$-BiBr $C^{+}_{\gamma_1}(\hat{\mathbf{n}},k_i)$ computed over all $\hat{\mathbf{n}}$ for $k_{i}=0,\pi$.
We specifically in SFig.~\ref{fig:bibr_partial_Chern_numbers_different_spin_directions} only show the upper spin hemisphere parameterized by $(\vartheta,\phi)$ in SEq.~(\ref{eq:hat_n_around_y_axis}), relying on the result that in the $\mathcal{T}$-invariant BZ planes $k_{i}=0,\pi$,
\begin{equation}
	C^{\pm}_{\gamma_1}(\hat{\mathbf{n}},k_i) = -C^{\mp}_{\gamma_1}(\hat{\mathbf{n}},k_i),
\label{eq:bibr_partial_Chern_symmetry}
\end{equation}
which implies that $C^{\pm}_{\gamma_1}(\hat{\mathbf{n}},k_i) = -C^{\pm}_{\gamma_1}(-\hat{\mathbf{n}},k_i)$ at $k_{i}=0,\pi$, due the general property that  $C^{\pm}_{\gamma_1}(\hat{\mathbf{n}},k_i) = C^{\mp}_{\gamma_1}(-\hat{\mathbf{n}},k_i)$.

As shown in SFig.~\ref{fig:bibr_partial_Chern_numbers_different_spin_directions}(a,b,d,e), we find that
\begin{equation}
	C^{\pm}_{\gamma_1}(\hat{\mathbf{n}},k_1 = 0) = C^{\pm}_{\gamma_1}(\hat{\mathbf{n}},k_1 = \pi) = C^{\pm}_{\gamma_1}(\hat{\mathbf{n}},k_2 = 0) = C^{\pm}_{\gamma_1}(\hat{\mathbf{n}},k_2 = \pi) = 0,
\label{eq:bibr_partial_Chern_k1_k2_planes_generic_n}
\end{equation}
for all $\hat{\mathbf{n}}$ in $\alpha$-BiBr.
Conversely in the $k_{3}=0,\pi$ BZ planes, $C^{\pm}_{\gamma_1}(\hat{\mathbf{n}},k_3 = 0)$ and $C^{\pm}_{\gamma_1}(\hat{\mathbf{n}},k_3 = \pi)$ take nonvanishing values ($\pm 2$) over a large range of $\hat{\mathbf{n}}$ centered around $\hat{\mathbf{n}}=\hat{\mathbf{z}}$, and vanish (while remaining numerically stable) over a smaller range of $\hat{\mathbf{n}}$ centered around $\hat{\mathbf{n}}=\hat{\mathbf{x}}$, such that
\begin{align}
	& C^{\pm}_{\gamma_1}(\hat{\mathbf{z}},k_3 = 0)=C^{\pm}_{\gamma_1}(\hat{\mathbf{z}},k_3 = \pi)=\mp 2, \label{eq:bibr_partial_Chern_k3_plane_sz}  \\
	& C^{\pm}_{\gamma_1}(\hat{\mathbf{x}},k_3 = 0)=C^{\pm}_{\gamma_1}(\hat{\mathbf{x}},k_3 = \pi)= 0.
\label{eq:bibr_partial_Chern_k3_plane_sx}
\end{align}
Along with the finite numerical spin gap $\Delta_s$ established in SFig.~\ref{fig:bibr_numerical_spin_gap_for_different_spin_directions} for the spin resolution directions $\hat{\mathbf{n}}$ centered around $\hat{\mathbf{n}}=\hat{\mathbf{z}}$ and $\hat{\mathbf{n}}=\hat{\mathbf{x}}$, we conclude from SEqs.~\eqref{eq:bibr_partial_Chern_k1_k2_planes_generic_n}, \eqref{eq:bibr_partial_Chern_k3_plane_sz}, and \eqref{eq:bibr_partial_Chern_k3_plane_sx} that $\alpha$-BiBr hosts the partial weak Chern vectors
\begin{align}
	& \bm{\nu}^{\pm} = (0,0,\mp 2) \mathrm{\ for\ } s_z \text{ }(\text{and surrounding }\hat{\mathbf{n}}) , \nonumber \\
	& \bm{\nu}^{\pm} = (0,0,0) \mathrm{\ for\ } s_x \text{ }(\text{and surrounding }\hat{\mathbf{n}}). 
\label{eq:bibr_partial_chern_vector_sx}
\end{align}
Because $\alpha$-BiBr is a symmetry-indicated helical HOTI~\cite{tang2019efficient,SYBiBr,BiBrFanHOTI}, SEq.~(\ref{eq:bibr_partial_chern_vector_sx}), along with the $s_{z}$ and $s_{x}$ numerical spin gaps, imply that $\alpha$-BiBr realizes a 3D QSHI (T-DAXI) state for $\hat{\mathbf{n}}=\hat{\mathbf{z}}$ ($\hat{\mathbf{n}}=\hat{\mathbf{x}}$), due to its bulk spin-resolved (partial) SIs (SN~\ref{sec:defs_partial_sis}).  
We will below in this section shortly confirm this conclusion via the direct computation of the nested spin-resolved Wilson loop spectrum of $\alpha$-BiBr.
In SN~\ref{sec:bibrshc}, we will also compare the partial weak indices in SEq.~(\ref{eq:bibr_partial_chern_vector_sx}) to the bulk intrinsic spin Hall conductivity in $\alpha$-BiBr for $s_{z}$ and $s_{x}$ spins, which we will show provides a physically measurable signature of its spin-gapped bulk topology.

\textit{Higher-Order Spectral Flow in $\alpha$-BiBr: Nested Wilson Loop Spectrum.} We will next pause from analyzing spin-resolved quantities to compute the ordinary (nested) Wilson loop of $\alpha$-BiBr.
This calculation will provide us with a reference hybrid Wannier (nested Wilson) spectrum for the nested spin-resolved Wilson loop of $\alpha$-BiBr, which will be computed later in this section.
Furthermore as we will discuss below, confirming helical nested Wilson loop flow in the ab-initio-derived electronic structure of a candidate helical HOTI itself represents a significant result; such calculations remain exceedingly rare, with a noteworthy previous example being the nested Wilson loop identification of a non-symmetry-indicated helical HOTI state in $\gamma$-MoTe$_2$~\cite{wang2019higherorder}.
Here and throughout this work, our (spin-resolved) nested Wilson loop calculations have been performed using the freely available Python package~\href{https://github.com/kuansenlin/nested_and_spin_resolved_Wilson_loop}{\textsc{nested\_and\_spin\_resolved\_Wilson\_loop}}~\cite{lin2023nestedWilsonLib}, which represents an extension of the~\href{https://www.physics.rutgers.edu/pythtb/}{PythTB} open-source Python tight-binding package~\cite{coh2013python} that was implemented and utilized for the preparation of SRefs.~\cite{wieder2018axion,wieder2020strong}, and was then greatly expanded for the present work.

\begin{figure}[t]
\includegraphics[width=\columnwidth]{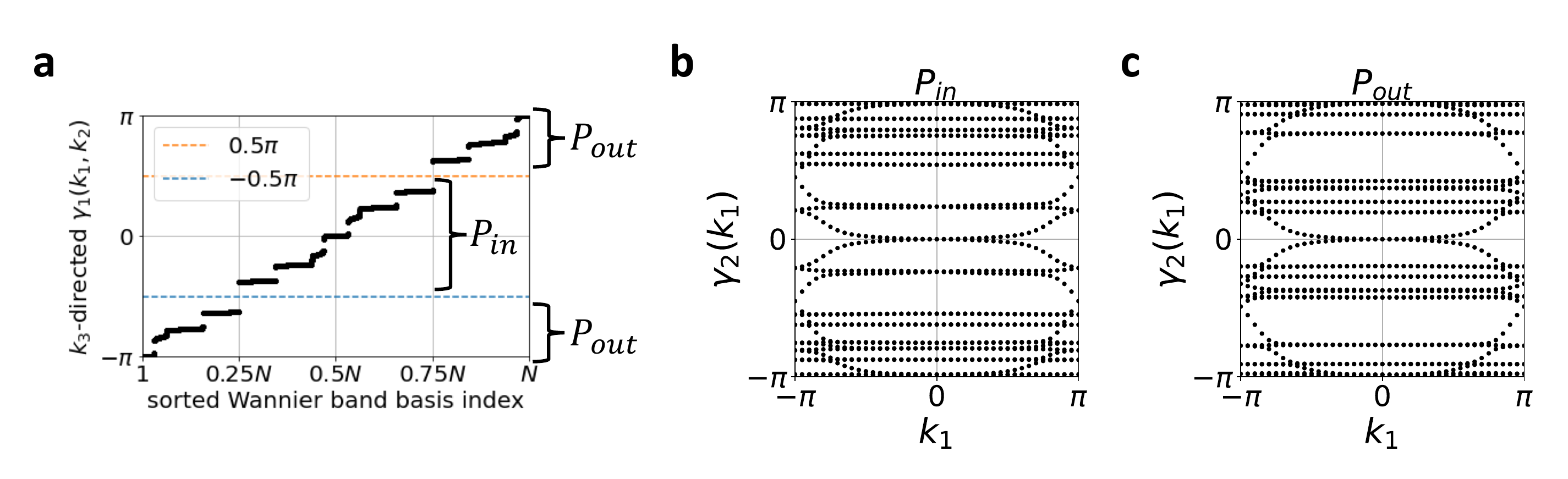}
\caption{Nested Wilson loop spectrum of $\alpha$-BiBr.
In this figure, we compute the Wilson loop (a) and nested Wilson loop (b,c) spectra of $\alpha$-BiBr.
The calculations shown in this figure were performed using a $41 \times 41 \times 41$ ${\bf k}$-grid uniformly spaced over the 3D BZ in the reduced ${\bf k}$ coordinates defined in SEq.~(\ref{eq:kReduced}).
(a) The eigenphases of the $k_{3}$-directed [see SEq.~(\ref{eq:kReduced})] $P$-Wilson loop spectrum of $\alpha$-BiBr.  
Like that of the candidate helical HOTI $\beta$-MoTe$_2$~\cite{wang2019higherorder}, the Wilson loop spectrum in (a) exhibits clear gaps at $\gamma_{1}=\pm \pi/2=0.5\pi$.
We have also confirmed that the gaps in the $k_{3}$-directed $P$-Wilson loop spectrum remain present for all values of $k_{2}$.
We may hence define in (a) the $\mathcal{I}$- and $\mathcal{T}$-invariant (see SRefs.~\cite{wieder2018axion,wang2019higherorder,WiederDefect} and SN~\ref{appendix:symmetry-constraints-on-Wilson-loop}) nested Wilson projectors $P_{in}$ and $P_{out}$ [SEq.~\eqref{eq:P_nested_Wilson_loop_2nd_projector}].
(b,c) The $k_{2}$-directed nested $P$-Wilson loop spectrum (SN~\ref{sec:nested_P_Wilson_loop}) of $\alpha$-BiBr respectively computed over the inner ($P_{in}$) and outer ($P_{out}$) Wilson band projectors in (a).
In both (b,c), the nested Wilson spectrum exhibits the characteristic odd helical winding of a $\mathcal{I}$- and $\mathcal{T}$-protected helical HOTI (see SRef.~\cite{wang2019higherorder} and SN~\ref{sec:numerical-section-of-nested-P-pm}).
The calculations detailed in this figure were performed using the freely available Python package~\href{https://github.com/kuansenlin/nested_and_spin_resolved_Wilson_loop}{\textsc{nested\_and\_spin\_resolved\_Wilson\_loop}}~\cite{lin2023nestedWilsonLib}, which represents an extension of the~\href{https://www.physics.rutgers.edu/pythtb/}{PythTB} open-source Python tight-binding package~\cite{coh2013python} that was implemented and utilized for the preparation of SRefs.~\cite{wieder2018axion,wieder2020strong} and the present work.}
\label{fig:bibr_nested_Wilson_loop}
\end{figure}

To investigate the hybrid Wannier spectrum of $\alpha$-BiBr, we first compute the $k_3$-directed Wilson loop eigenphases [Wannier band energies] $\gamma_1 (k_1,k_2)$ for the $64$ occupied (valence) bands [SFig.~\ref{fig:bibr_nested_Wilson_loop}(a)].
The Wilson loop spectrum in [SFig.~\ref{fig:bibr_nested_Wilson_loop}(a)] exhibits clear gaps at $\gamma_1 (k_1,k_2)=\pm \pi/2 = \pm 0.5 \pi$, reminiscent of the Wilson loop spectrum of the candidate helical HOTI $\beta$-MoTe$_2$~\cite{wang2019higherorder}, whose spin-resolved topology was previously analyzed in SN~\ref{sec:mote2_spin_resolved_topology}. 
Following the procedure introduced in SRefs.~\cite{wieder2018axion,wang2019higherorder} and discussed in SN~\ref{sec:numerical-section-of-nested-P-pm}, we then divide the Wannier (Wilson) bands in SFig.~\ref{fig:bibr_nested_Wilson_loop}(a) into two $\mathcal{I}$- and $\mathcal{T}$-symmetric groupings: an inner set centered around $\gamma_1 = 0$ and an outer set centered around $\gamma_1 = \pi$.  We crucially further find that for all perpendicular momenta $k_{2}$, the $k_{3}$-directed Wannier spectrum of $\alpha$-BiBr remains gapped in the vicinity of $\gamma_1 = \pm \pi/2$.

We may hence compute the $\mathcal{I}$- and $\mathcal{T}$-invariant \emph{nested} Wilson spectrum of $\alpha$-BiBr (see SRefs.~\cite{wieder2018axion,wang2019higherorder,WiederDefect} and SN~\ref{appendix:symmetry-constraints-on-Wilson-loop}). 
To perform the nested Wilson loop calculation, we first define the inner and outer Wannier- (Wilson-) band projectors as $P_{in}$ and $P_{out}$ [SFig.~\ref{fig:bibr_nested_Wilson_loop}(a)].
We then separately compute the $k_2$-directed nested Wilson loop eigenvalues $\gamma_2 (k_1)$ for the inner [SFig.~\ref{fig:bibr_nested_Wilson_loop}(b)] and outer [SFig.~\ref{fig:bibr_nested_Wilson_loop}(c)] Wannier bands.
In both the inner and outer nested Wilson loop spectra in SFig.~\ref{fig:bibr_nested_Wilson_loop}(b,c), we observe the odd helical winding characteristic of an $\mathcal{I}$- and $\mathcal{T}$-protected helical HOTI (see SRef.~\cite{wang2019higherorder} and SN~\ref{sec:numerical-section-of-nested-P-pm}).
Specifically, due to the correspondence between the Wilson spectrum and the hybrid Wannier spectrum (see SN~\ref{appendix:TB-notation} and~\ref{sec:P_Wilson_loop}, and SRefs.~\cite{benalcazar2017electric,alexandradinata2014wilsonloop}), the nested Wilson spectra in SFig.~\ref{fig:bibr_nested_Wilson_loop}(b,c) indicate that the hybrid Wannier spectrum of $\alpha$-BiBr can be deformed without breaking $\mathcal{I}$ or $\mathcal{T}$ symmetry (or closing a bulk energy gap) to the hybrid Wannier spectrum of the minimal layer construction of a helical HOTI~\cite{song2018mapping}.
More specifically, the $k_{3}$-directed hybrid Wannier spectrum of $\alpha$-BiBr can be symmetrically deformed without closing an energy gap to a hybrid Wannier spectrum consisting of 2D TI layers with normal vectors parallel to ${\bf a}_{3}$ pinned by $\mathcal{I}$ symmetry to the origin and ${\bf a}_{3}/2$, equivalent to that of the minimal layer construction of a $\mathcal{I}$- and $\mathcal{T}$-protected helical HOTI (see SN~\ref{app:comparison-spin-stable-and-symmetry-indicated-topology}).
We have hence shown, for the first time, that the hybrid Wannier spectrum of the candidate helical HOTI $\alpha$-BiBr~\cite{tang2019efficient,SYBiBr,BiBrFanHOTI,noguchi2021evidence,FanZahidRoomTempBiBrExp} exhibits higher-order helical spectral flow.

\textit{Spin-Resolved Topology of $\alpha$-BiBr: Nested Spin-Resolved Wilson Loop.} Having confirmed that $\alpha$-BiBr hosts large spin-gapped regions in spin-direction ($\hat{\mathbf{n}}$) parameter space (SFig.~\ref{fig:bibr_numerical_spin_gap_for_different_spin_directions}), and that its occupied bands exhibit the characteristic nested Wilson spectrum of an $\mathcal{I}$- and $\mathcal{T}$-protected helical HOTI (SFig.~\ref{fig:bibr_nested_Wilson_loop}), we will next directly apply the (nested) spin-resolved Wilson loop method developed in this work (SN~\ref{sec:P_pm_Wilson_loop} and~\ref{sec:nested_P_pm_Wilson_loop}) to extract the spin-resolved topology of $\alpha$-BiBr.
First, earlier in this section, we showed that the spin spectrum of $\alpha$-BiBr can be divided into four spin-gapped regions respectively centered around $\hat{\mathbf{n}}=\pm \hat{\mathbf{z}}$ and $\hat{\mathbf{n}}=\pm \hat{\mathbf{x}}$ (SFig.~\ref{fig:bibr_numerical_spin_gap_for_different_spin_directions}).
For this reason, we will here focus on computing the spin-resolved topology of $\alpha$-BiBr in just the $\hat{\mathbf{n}}=\hat{\mathbf{z}}$ and $\hat{\mathbf{n}}=\hat{\mathbf{x}}$ spin resolution directions, noting that the $\mathcal{I}$-protected spin-resolved topology of $\alpha$-BiBr will be the same for other values of $\hat{\mathbf{n}}$ that are related to $\hat{\mathbf{n}}=\hat{\mathbf{z}},\hat{\mathbf{x}}$ by (adiabatic) paths in $\hat{\mathbf{n}}$ parameter space along which the spin gap does not close (see SN~\ref{sec:pspperturbation} and SFig.~\ref{fig:bibr_numerical_spin_gap_for_different_spin_directions}).  
As noted earlier in this section, the (nested) spin-resolved Wilson loop calculations detailed below were performed using the freely available Python package~\href{https://github.com/kuansenlin/nested_and_spin_resolved_Wilson_loop}{\textsc{nested\_and\_spin\_resolved\_Wilson\_loop}}~\cite{lin2023nestedWilsonLib}, which was developed for SRefs.~\cite{wieder2018axion,wieder2020strong} and the present work.

\begin{figure}[t]
\includegraphics[width=\columnwidth]{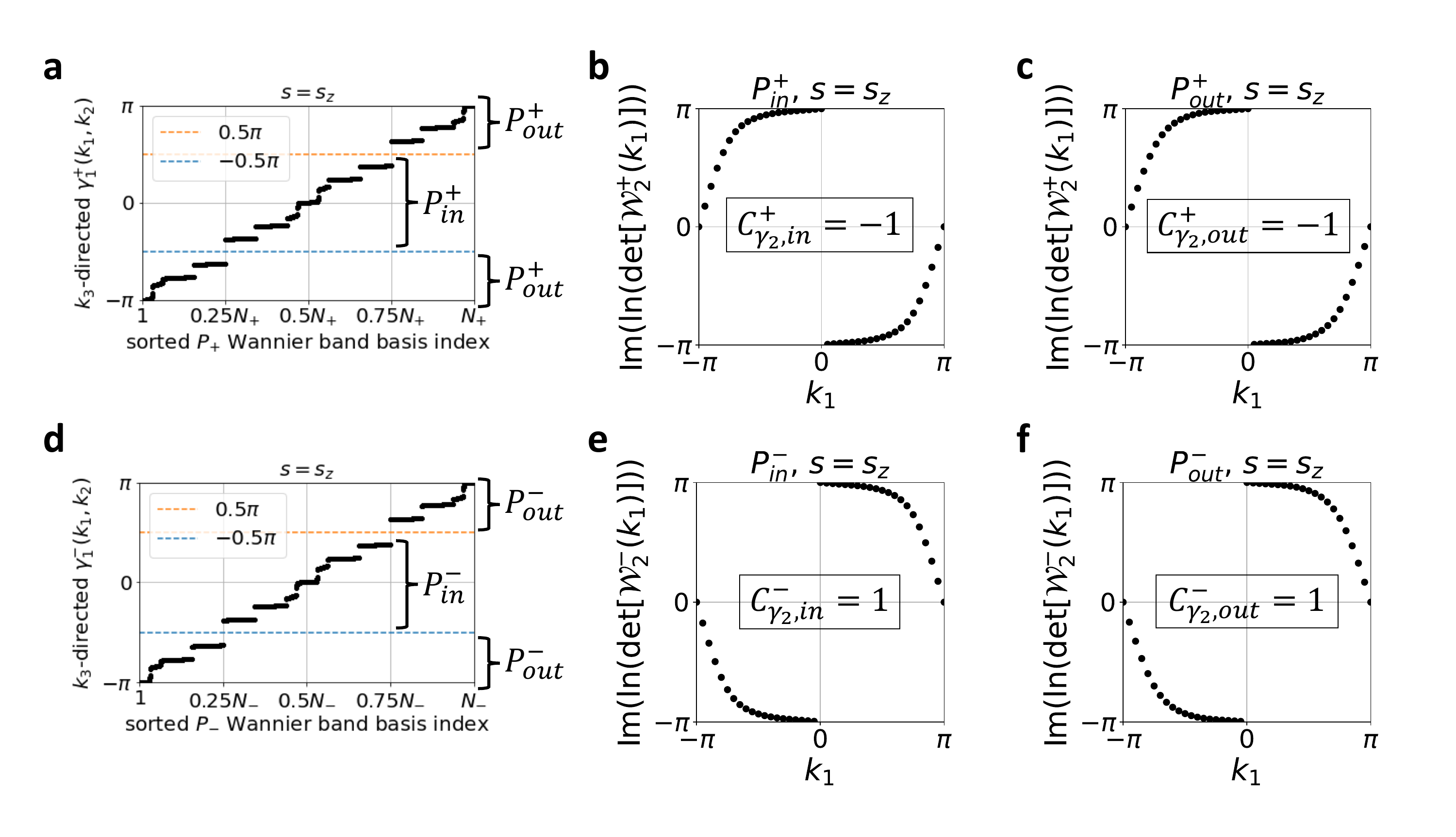}
\caption{$\hat{\mathbf{n}}=\hat{\mathbf{z}}$ nested spin-resolved Wilson loop spectrum of $\alpha$-BiBr.
The calculations shown in this figure were performed using a $41 \times 41 \times 41$ ${\bf k}$-grid uniformly spaced over the 3D BZ in the reduced ${\bf k}$ coordinates defined in SEq.~(\ref{eq:kReduced}).
(a,d) Respectively the eigenphases of the $k_3$-directed [see SEq.~(\ref{eq:kReduced})] $s_{z}$ spin-resolved $P_{+}$- and $P_{-}$-Wilson loop spectra of $\alpha$-BiBr.
The $s_{z}$ spin-resolved Wilson loop spectra in (a,d) exhibit clear gaps at $\gamma_{1}^{\pm}=\pi/2=0.5\pi$.  
We have also confirmed that the gaps in the $k_{3}$-directed $s_{z}$ spin-resolved Wilson loop spectra remain present for all values of $k_{2}$.
We may hence define in (a) the $\mathcal{I}$-invariant (see SRef.~\cite{wieder2018axion} and SN~\ref{appendix:I-constraint-on-P-pm}) nested Wilson projectors $P_{in}^{+}$ and $P_{out}^{+}$ [SEq.~\eqref{eq:P_pm_nested_Wilson_loop_2nd_projector}], and similarly define in (d) the $\mathcal{I}$-invariant nested Wilson projectors $P_{in}^{-}$ and $P_{out}^{-}$ [SEq.~\eqref{eq:P_pm_nested_Wilson_loop_2nd_projector}].
(b,c) Respectively the determinants of the $k_2$-directed nested $P_{+}$-Wilson loop matrix $[\mathcal{W}_2^{+}(k_1)]$ for the inner and outer $s_{z}$ spin-resolved $P_{+}$-Wilson bands in (a), plotted as functions of the remaining crystal momentum $k_{1}$.  
Following the sign convention established in SEq.~\eqref{eq:partial_chern_def}, the nested $s_{z}$ spin-resolved $P_{+}$-Wilson loop winding numbers in (b,c) respectively indicate the nested partial Chern numbers [SEq.~\eqref{eq:C_gamma_2_n_pm_k_i}] $C_{\gamma_2 , in}^{+} = -1$ and $C_{\gamma_2 , out}^{+} = -1$.
(e,f) Respectively the determinants of the $k_2$-directed nested $P_{-}$-Wilson loop matrix $[\mathcal{W}_2^{-}(k_1)]$ for the inner and outer $s_{z}$ spin-resolved $P_{-}$-Wilson bands in (d), plotted as functions of the remaining crystal momentum $k_{1}$.  
Following the sign convention established in SEq.~\eqref{eq:partial_chern_def}, the nested $s_{z}$ spin-resolved $P_{-}$-Wilson loop winding numbers in (e,f) respectively indicate the nested partial Chern numbers [SEq.~\eqref{eq:C_gamma_2_n_pm_k_i}] $C_{\gamma_2 , in}^{-} = 1$ and $C_{\gamma_2 , out}^{-} = 1$, which carry opposite signs from those in (b,c) due to the action of $\mathcal{T}$ symmetry on the nested spin-resolved Wilson spectrum (SN~\ref{appendix:T-constraint-on-nested-P-pm}).
Because the nested $s_{z}$ spin-resolved partial Chern numbers in (b,c) are the same [and opposite to those in (e,f)], this implies that for $\hat{\mathbf{n}}=\hat{\mathbf{z}}$, $\alpha$-BiBr has nonvanishing ${\bf G}_{3}$ partial weak Chern numbers [SEq.~(\ref{eq:partial_Chern_vector_def})] $\nu^{\pm}_{3}$.
This is consistent with our earlier $P_{\pm}$-Wilson loop calculations demonstrating that for $\hat{\mathbf{n}}=\hat{\mathbf{z}}$, $\alpha$-BiBr realizes a 3D QSHI state with $\nu^{\pm}_{3}=\mp 2$ [see SN~\ref{sec:defs_partial_sis} and SEq.~(\ref{eq:bibr_partial_chern_vector_sx}) and the surrounding text].
The calculations detailed in this figure were performed using the freely available Python package~\href{https://github.com/kuansenlin/nested_and_spin_resolved_Wilson_loop}{\textsc{nested\_and\_spin\_resolved\_Wilson\_loop}}~\cite{lin2023nestedWilsonLib}, which represents an extension of the~\href{https://www.physics.rutgers.edu/pythtb/}{PythTB} open-source Python tight-binding package~\cite{coh2013python} that was implemented and utilized for the preparation of SRefs.~\cite{wieder2018axion,wieder2020strong} and the present work.}
\label{fig:bibr_nested_spin_resolved_wilson_loop_sz}
\end{figure}

We begin by reconsidering the $\hat{\mathbf{n}}=\hat{\mathbf{z}}$ $P(\hat{\mathbf{n}}\cdot\mathbf{s})P$ [$Ps_{z}P$] spectrum of $\alpha$-BiBr.
Because there is a spin gap at all ${\bf k}$ points for $\hat{\mathbf{n}}=\hat{\mathbf{z}}$ [SFig.~\ref{fig:bibr_Wannier_energy_bands_spin_sz_sx_bands}(b,e)], then we may compute the spin-resolved $P_\pm$-Wilson loop on the positive and negative spin bands (SN~\ref{sec:P_pm_Wilson_loop}).
In SFig.~\ref{fig:bibr_nested_spin_resolved_wilson_loop_sz}(a) [SFig.~\ref{fig:bibr_nested_spin_resolved_wilson_loop_sz}(d)], we show the $k_{3}$-directed $P_{+}$- [$P_{-}$-] Wilson loop spectrum for the positive [negative] $s_{z}$ spin bands of $\alpha$-BiBr.  
As with the ordinary $k_{3}$-directed $P$-Wilson loop in $\alpha$-BiBr [SFig.~\ref{fig:bibr_nested_Wilson_loop}(a)], the spin-resolved $k_{3}$-directed $P_{\pm}$-Wilson bands are gapped, and form $\mathcal{I}$-symmetric groupings (see SN~\ref{appendix:I-constraint-on-P-pm}) centered around $\gamma_1^{\pm}(k_1 , k_2)=0,\pi$ that are separated by pronounced spin-resolved Wilson gaps at $\gamma_1^{\pm}(k_1 , k_2)=\pm \pi/2 = \pm 0.5\pi$.
We have also numerically confirmed that the $k_{3}$-directed $P_{\pm}$-Wilson bands in $\alpha$-BiBr remain gapped at $\gamma_1^{\pm}(k_1 , k_2)=\pm \pi/2$ for all values of $k_{2}$, allowing us to consistently define $\mathcal{I}$-symmetric inner [$P^{\pm}_{in}$] and outer [$P^{\pm}_{out}$] $s_{z}$ nested spin-resolved Wilson projectors  [SFig.~\ref{fig:bibr_nested_spin_resolved_wilson_loop_sz}(a,d)].
Using $P^{\pm}_{in}$ and $P^{\pm}_{out}$, we then respectively compute the inner and outer $k_{2}$-directed nested spin-resolved Wilson loop (SN~\ref{sec:nested_P_pm_Wilson_loop}) of $\alpha$-BiBr for both the positive and negative $s_{z}$ spin bands, whose determinants $\mathrm{Im}\left( \mathrm{log} \left( \mathrm{det} [ \mathcal{W}_2^{\pm} (k_1)] \right) \right)$ are plotted in  SFig.~\ref{fig:bibr_nested_spin_resolved_wilson_loop_sz}(b,c,e,f) as functions of the remaining crystal momentum $k_{1}$.
For both the positive and negative spin bands, the inner and outer nested $s_{z}$ spin-resolved Wilson bands exhibit the same odd winding numbers [SEq.~\eqref{eq:C_gamma_2_n_pm_k_i}].
Specifically, we respectively find that $C^{\pm}_{\gamma_2 , in} = \mp 1$ [$C^{\pm}_{\gamma_2 , out} = \mp 1$] for the inner [outer] $s_{z}$ $P_{\pm}$-Wilson bands in $\alpha$-BiBr, employing the sign convention established in SEq.~\eqref{eq:partial_chern_def} in which the \textit{negative} winding number of the $k_2$-directed nested spin-resolved Wilson loop as a function of $k_1$ gives the nested partial Chern number.
Our observation that the $s_{z}$ spin-resolved $k_{3}$-directed hybrid Wannier sheets (Wilson bands) in $\alpha$-BiBr have the same Chern numbers $C^{\pm}_{\gamma_2 , in}=C^{\pm}_{\gamma_2 , out} = \mp 1$ implies that $\alpha$-BiBr overall has nonvanishing ${\bf G}_{3}$ partial weak Chern numbers $\nu^{\pm}_{3}$ [SEq.~(\ref{eq:partial_Chern_vector_def})], consistent with our earlier $P_{\pm}$-Wilson loop calculations demonstrating that for $\hat{\mathbf{n}}=\hat{\mathbf{z}}$, $\alpha$-BiBr realizes a 3D QSHI state with $\nu^{\pm}_{3}=\mp 2$ [see SN~\ref{sec:defs_partial_sis} and SEq.~(\ref{eq:bibr_partial_chern_vector_sx}) and the surrounding text].

\begin{figure}[t]
\includegraphics[width=\columnwidth]{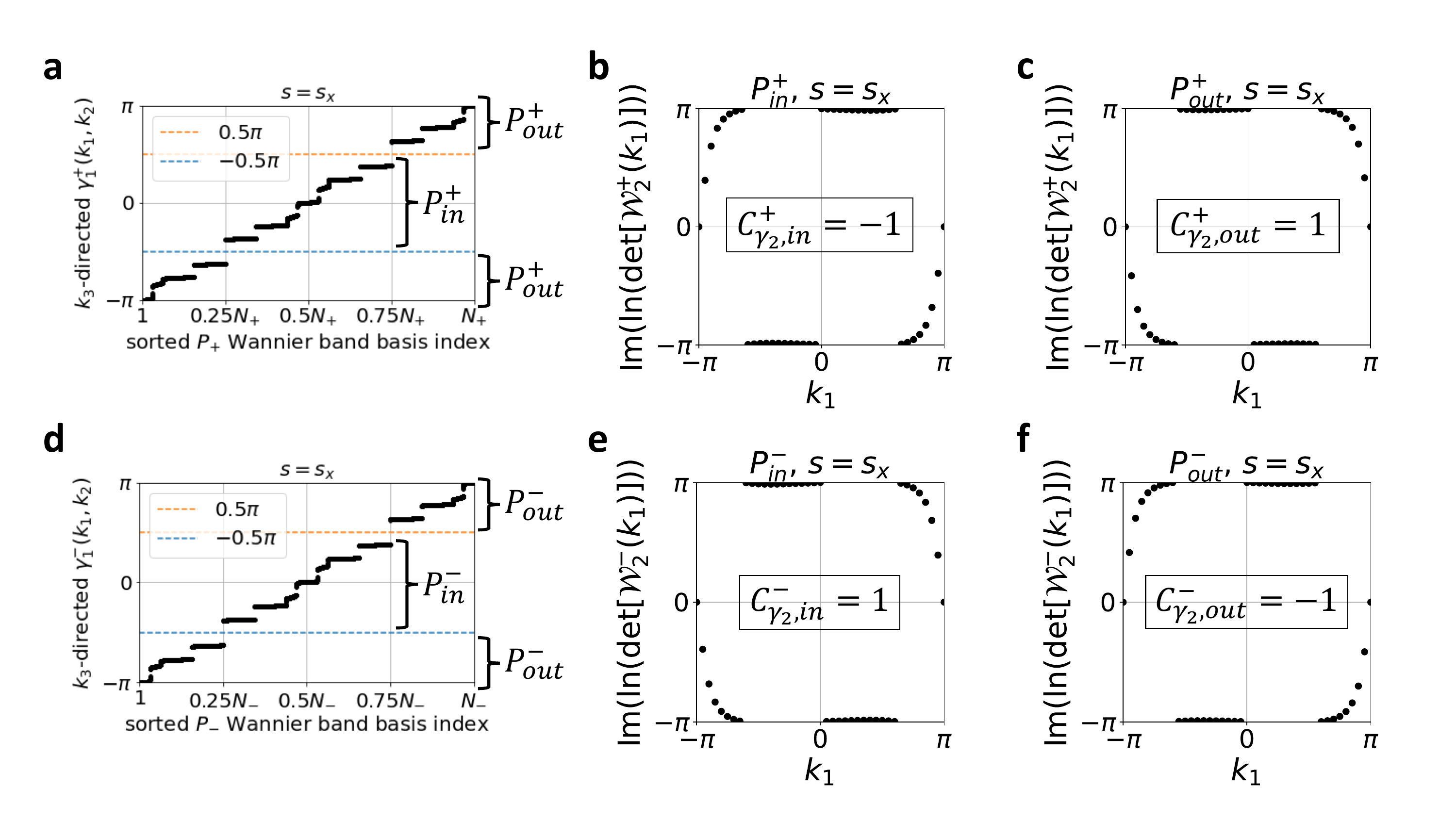}
\caption{$\hat{\mathbf{n}}=\hat{\mathbf{x}}$ nested spin-resolved Wilson loop spectrum of $\alpha$-BiBr.
The calculations shown in this figure were performed using a $41 \times 41 \times 41$ ${\bf k}$-grid uniformly spaced over the 3D BZ in the reduced ${\bf k}$ coordinates defined in SEq.~(\ref{eq:kReduced}).
(a,d) Respectively the eigenphases of the $k_3$-directed [see SEq.~(\ref{eq:kReduced})] $s_{x}$ spin-resolved $P_{+}$- and $P_{-}$-Wilson loop spectra of $\alpha$-BiBr.
The $s_{x}$ spin-resolved Wilson loop spectra in (a,d) exhibit clear gaps at $\gamma_{1}^{\pm}=\pi/2=0.5\pi$.  
We have also confirmed that the gaps in the $k_{3}$-directed $s_{x}$ spin-resolved Wilson loop spectra remain present for all values of $k_{2}$.
We may hence define in (a) the $\mathcal{I}$-invariant (see SRef.~\cite{wieder2018axion} and SN~\ref{appendix:I-constraint-on-P-pm}) nested Wilson projectors $P_{in}^{+}$ and $P_{out}^{+}$ [SEq.~\eqref{eq:P_pm_nested_Wilson_loop_2nd_projector}], and similarly define in (d) the $\mathcal{I}$-invariant nested Wilson projectors $P_{in}^{-}$ and $P_{out}^{-}$ [SEq.~\eqref{eq:P_pm_nested_Wilson_loop_2nd_projector}].
(b,c) Respectively the determinants of the $k_2$-directed nested $P_{+}$-Wilson loop matrix $[\mathcal{W}_2^{+}(k_1)]$ for the inner and outer $s_{x}$ spin-resolved $P_{+}$-Wilson bands in (a), plotted as functions of the remaining crystal momentum $k_{1}$.  
Following the sign convention established in SEq.~\eqref{eq:partial_chern_def}, the nested $s_{x}$ spin-resolved $P_{+}$-Wilson loop winding numbers in (b,c) respectively indicate the nested partial Chern numbers [SEq.~\eqref{eq:C_gamma_2_n_pm_k_i}] $C_{\gamma_2 , in}^{+} = -1$ and $C_{\gamma_2 , out}^{+} = 1$.
(e,f) Respectively the determinants of the $k_2$-directed nested $P_{-}$-Wilson loop matrix $[\mathcal{W}_2^{-}(k_1)]$ for the inner and outer $s_{x}$ spin-resolved $P_{-}$-Wilson bands in (d), plotted as functions of the remaining crystal momentum $k_{1}$.  
Following the sign convention established in SEq.~\eqref{eq:partial_chern_def}, the nested $s_{x}$ spin-resolved $P_{-}$-Wilson loop winding numbers in (e,f) respectively indicate the nested partial Chern numbers [SEq.~\eqref{eq:C_gamma_2_n_pm_k_i}] $C_{\gamma_2 , in}^{-} = 1$ and $C_{\gamma_2 , out}^{-} = -1$, which carry opposite signs from those in (b,c) due to the action of $\mathcal{T}$ symmetry on the nested spin-resolved Wilson spectrum (SN~\ref{appendix:T-constraint-on-nested-P-pm}).
Because the nested $s_{x}$ spin-resolved partial Chern numbers in (b,c) are the opposite [and opposite to those in (e,f)], this implies that for $\hat{\mathbf{n}}=\hat{\mathbf{x}}$, $\alpha$-BiBr has vanishing ${\bf G}_{3}$ partial weak Chern numbers [SEq.~(\ref{eq:partial_Chern_vector_def})] $\nu^{\pm}_{3}=0$.
Instead, the opposite, odd nested partial Chern numbers in (b,c) [and in (e,f)] are identical to those of an $\mathcal{I}$-protected AXI~\cite{wieder2018axion}, indicating that for $\hat{\mathbf{n}}=\hat{\mathbf{x}}$, $\alpha$-BiBr carries nontrivial, origin-independent, $\mathcal{I}$-quantized partial axion angles $\theta^{\pm}=\pi$ (SN~\ref{sec:defs_partial_sis}).
$\alpha$-BiBr for $\hat{\mathbf{n}}=\hat{\mathbf{x}}$ hence represents a real-material realization of the spin-stable T-DAXI state introduced in this work.
The calculations detailed in this figure were performed using the freely available Python package~\href{https://github.com/kuansenlin/nested_and_spin_resolved_Wilson_loop}{\textsc{nested\_and\_spin\_resolved\_Wilson\_loop}}~\cite{lin2023nestedWilsonLib}, which represents an extension of the~\href{https://www.physics.rutgers.edu/pythtb/}{PythTB} open-source Python tight-binding package~\cite{coh2013python} that was implemented and utilized for the preparation of SRefs.~\cite{wieder2018axion,wieder2020strong} and the present work.}
\label{fig:bibr_nested_spin_resolved_wilson_loop_sx}
\end{figure}

We next revisit the $Ps_{x}P$ spectrum of $\alpha$-BiBr.
Because there is a spin gap at all ${\bf k}$ points for $\hat{\mathbf{n}}=\hat{\mathbf{x}}$ [SFig.~\ref{fig:bibr_Wannier_energy_bands_spin_sz_sx_bands}(c,f)], then we may compute the spin-resolved $P_\pm$-Wilson loop on the positive and negative spin bands (SN~\ref{sec:P_pm_Wilson_loop}).
In SFig.~\ref{fig:bibr_nested_spin_resolved_wilson_loop_sx}(a) [SFig.~\ref{fig:bibr_nested_spin_resolved_wilson_loop_sx}(d)], we show the $k_{3}$-directed $P_{+}$- [$P_{-}$-] Wilson loop spectrum for the positive [negative] $s_{x}$ spin bands of $\alpha$-BiBr.  
As with the ordinary $k_{3}$-directed $P$-Wilson loop [SFig.~\ref{fig:bibr_nested_Wilson_loop}(a)] and the $s_{z}$ spin-resolved $k_{3}$-directed $P_{\pm}$-Wilson bands [SFig.~\ref{fig:bibr_nested_spin_resolved_wilson_loop_sz}(a,d)], the $s_{x}$ spin-resolved $k_{3}$-directed $P_{\pm}$-Wilson bands are gapped, and form $\mathcal{I}$-symmetric groupings (see SN~\ref{appendix:I-constraint-on-P-pm}) centered around $\gamma_1^{\pm}(k_1 , k_2)=0,\pi$ that are separated by spin-resolved Wilson gaps at $\gamma_1^{\pm}(k_1 , k_2)=\pm \pi/2 = \pm 0.5\pi$.
We have also numerically confirmed that the $k_{3}$-directed $P_{\pm}$-Wilson bands in $\alpha$-BiBr remain gapped at $\gamma_1^{\pm}(k_1 , k_2)=\pm \pi/2$ for all values of $k_{2}$, allowing us to consistently define $\mathcal{I}$-symmetric inner [$P^{\pm}_{in}$] and outer [$P^{\pm}_{out}$] nested $s_{x}$ spin-resolved Wilson projectors  [SFig.~\ref{fig:bibr_nested_spin_resolved_wilson_loop_sx}(a,d)].
Using $P^{\pm}_{in}$ and $P^{\pm}_{out}$, we then respectively compute the inner and outer $k_{2}$-directed nested spin-resolved Wilson loop (SN~\ref{sec:nested_P_pm_Wilson_loop}) of $\alpha$-BiBr for both the positive and negative $s_{x}$ spin bands, whose determinants $\mathrm{Im}\left( \mathrm{log} \left( \mathrm{det} [ \mathcal{W}_2^{\pm} (k_1)] \right) \right)$ are plotted in  SFig.~\ref{fig:bibr_nested_spin_resolved_wilson_loop_sx}(b,c,e,f) as functions of the remaining crystal momentum $k_{1}$.
For both the positive and negative spin bands, the inner and outer nested $s_{x}$ spin-resolved Wilson bands exhibit \emph{opposite} winding numbers [SEq.~\eqref{eq:C_gamma_2_n_pm_k_i}].
Specifically, we respectively find that $C^{\pm}_{\gamma_2 , in} = \mp 1$ [$C^{\pm}_{\gamma_2 , out} = \pm 1$] for the inner [outer] $s_{x}$ $P_{\pm}$-Wilson bands in $\alpha$-BiBr, employing the sign convention established in SEq.~\eqref{eq:partial_chern_def} in which the \textit{negative} winding number of the $k_2$-directed nested spin-resolved Wilson loop as a function of $k_1$ gives the nested partial Chern number.
Our observation that the $s_{x}$ spin-resolved $k_{3}$-directed hybrid Wannier sheets (Wilson bands) in $\alpha$-BiBr have opposite Chern numbers $C^{\pm}_{\gamma_2 , in}= - C^{\pm}_{\gamma_2 , out} = \mp 1$ implies that $\alpha$-BiBr overall has vanishing ${\bf G}_{3}$ partial weak Chern numbers $\nu^{\pm}_{3}$ [SEq.~(\ref{eq:partial_Chern_vector_def})], consistent with our earlier $P_{\pm}$-Wilson loop calculations demonstrating that $\nu^{\pm}_{3}=0$ for $\hat{\mathbf{n}}=\hat{\mathbf{x}}$ in $\alpha$-BiBr [see SEq.~(\ref{eq:bibr_partial_chern_vector_sx}) and the surrounding text].
Unlike the 3D QSHI state realized for $\hat{\mathbf{n}}=\hat{\mathbf{z}}$ in $\alpha$-BiBr [SFig.~\ref{fig:bibr_nested_spin_resolved_wilson_loop_sz}(b,c,e,f)], the oppositely signed nested partial Chern numbers of the inner and outer Wilson bands in SFig.~\ref{fig:bibr_nested_spin_resolved_wilson_loop_sx}(b,c,e,f) indicate that for $\hat{\mathbf{n}}=\hat{\mathbf{x}}$, $\alpha$-BiBr instead has nontrivial, origin-independent, $\mathcal{I}$-quantized partial axion angles $\theta^{\pm}=\pi$ (see SRef.~\cite{wieder2018axion} and~SN~\ref{sec:defs_partial_sis}), and hence realizes the spin-stable T-DAXI state introduced in this work.

\begin{figure}[t]
\includegraphics[width=\columnwidth]{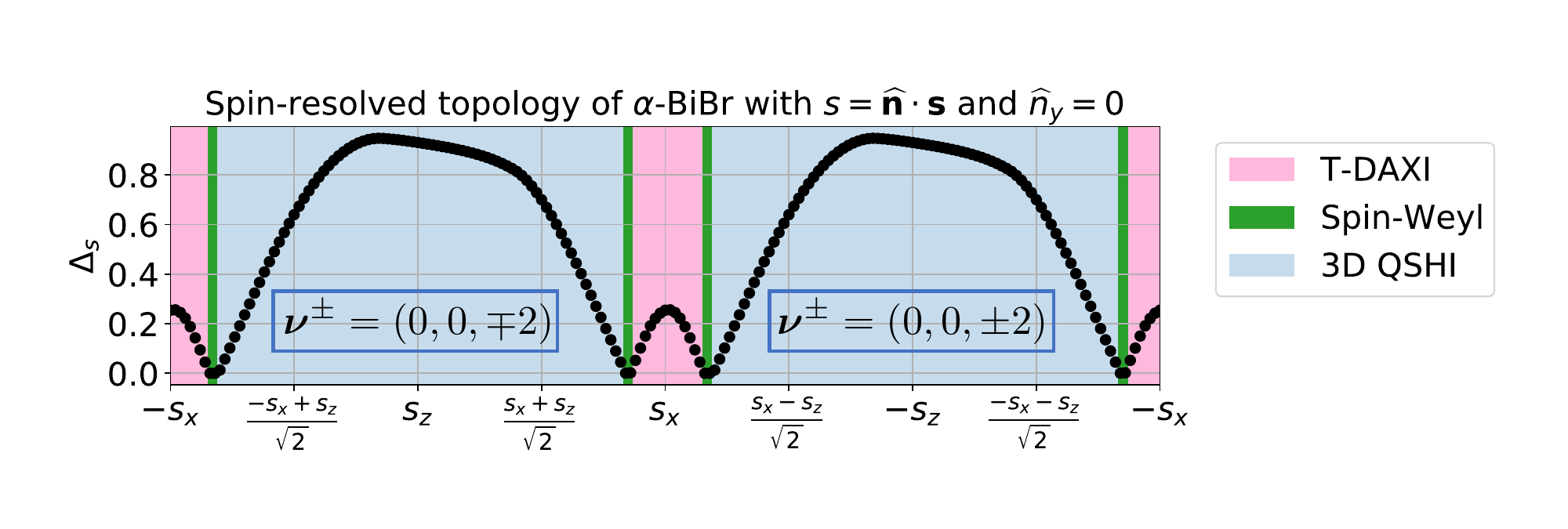}
\caption{High-resolution numerical spin gap $\Delta_{s = \hat{\mathbf{n}} \cdot \mathbf{s}}$ in $\alpha$-BiBr computed as a function of the spin resolution direction $\hat{\mathbf{n}}$.
We take $\hat{\mathbf{n}}$ to lie in the $xz$-plane and then angular parameterize $\hat{\mathbf{n}}$ using SEq.~\eqref{eq:hat_n_around_y_axis} with fixed $\vartheta = 0.5\pi$ and varying $\phi \in [-\pi , \pi]$. 
Unlike the previous numerical (global) spin gap calculation in SFig.~\ref{fig:bibr_numerical_spin_gap_for_different_spin_directions}, we here take a finer angular resolution $\Delta \phi = 0.01\pi$ in order to determine the spin-resolved topological phase boundaries in $\alpha$-BiBr.
By performing a Nelder-Mead minimization~\cite{pandey2022py} on the direct spin gap function $\Delta_s (\mathbf{k})$ taking as the initial points 100 $\mathbf{k}$ points randomly sampled from the 3D BZ in the reduced ${\bf k}$ coordinates defined in SEq.~(\ref{eq:kReduced}), we define for each $\hat{\mathbf{n}}$ the numerical spin gap $\Delta_{s = \hat{\mathbf{n}} \cdot \mathbf{s}}$ as the minimal value (in the units of $\hbar/2$) of the 100 minimization results for the fixed value of $\hat{\mathbf{n}}$.
We note that the right half of this figure was generated directly from the left half for numerical expediency by relying on the fact that $\Delta_{s}=\Delta_{-s}$, which follows from the definition of the projected spin operator $PsP$ (see SN~\ref{appendix:properties-of-the-projected-spin-operator}). 
We identify four narrow regions (green background) that are spin-gapless to within numerical precision. 
By directly computing the spin spectrum for selected $\hat{\mathbf{n}}$ within the green regions, we confirm that they represent spin-Weyl states that separate four spin-gapped regions in $\hat{\mathbf{n}}$ parameter space (blue and pink regions).
By comparing the regions in this plot to the (nested) spin-resolved Wilson loop calculations performed earlier in this section [SFigs.~\ref{fig:bibr_partial_Chern_numbers_different_spin_directions},~\ref{fig:bibr_nested_spin_resolved_wilson_loop_sz}, and~\ref{fig:bibr_nested_spin_resolved_wilson_loop_sx}], as well as by performing additional nested spin-resolved Wilson loop calculations for selected additional $\phi$, we confirm that the blue (pink) regions represent spin-stable 3D QSHI (T-DAXI) regimes of $\alpha$-BiBr.
Specifically, the blue 3D QSHI region centered around $\hat{\mathbf{n}}=\hat{\mathbf{z}}$ [$\hat{\mathbf{n}}=-\hat{\mathbf{z}}$] is characterized by the partial weak Chern numbers $\bm{\nu}^{\pm} = (0,0,\mp 2)$ [$\bm{\nu}^{\pm} = (0,0,\pm 2)$], and the two pink T-DAXI regions centered around $\hat{\mathbf{n}}=\pm\hat{\mathbf{x}}$ are characterized by nontrivial, origin-independent, $\mathcal{I}$-quantized partial axion angles $\theta^{\pm}=\pi$ (see SN~\ref{app:comparison-spin-stable-and-symmetry-indicated-topology}).
The calculations detailed in this figure were performed using the freely available Python package~\href{https://github.com/kuansenlin/nested_and_spin_resolved_Wilson_loop}{\textsc{nested\_and\_spin\_resolved\_Wilson\_loop}}~\cite{lin2023nestedWilsonLib}, which represents an extension of the~\href{https://www.physics.rutgers.edu/pythtb/}{PythTB} open-source Python tight-binding package~\cite{coh2013python} that was implemented and utilized for the preparation of SRefs.~\cite{wieder2018axion,wieder2020strong} and the present work.}
\label{fig:bibr_spin_resolved_phase_diagram}
\end{figure}

\textit{Spin-Resolved Topology of $\alpha$-BiBr: Phase Diagram.} We have in this section performed extensive spin gap and (nested, partial) Wilson loop calculations to demonstrate that $\alpha$-BiBr exhibits a bulk spin gap over a large range of $\hat{\mathbf{n}}$, realizes a 3D QSHI state for $\hat{\mathbf{n}}=\hat{\mathbf{z}}$, and realizes a T-DAXI state for $\hat{\mathbf{n}}=\hat{\mathbf{x}}$.
As established in SN~\ref{sec:pspperturbation}, the spin spectrum varies adiabatically under small changes in the system Hamiltonian, which can be treated as equivalent to small changes in the spin resolution direction $\hat{\mathbf{n}}$. 
This indicates that for small deviations in $\hat{\mathbf{n}}$ from $\hat{\mathbf{n}}=\hat{\mathbf{z}}$ [$\hat{\mathbf{n}}=\hat{\mathbf{x}}$], $\alpha$-BiBr will still also realize a spin-stable 3D QSHI [T-DAXI] state.
However we have not yet determined the exact phase boundaries in $\hat{\mathbf{n}}$ parameter space of the 3D QSHI and T-DAXI spin-stable states in $\alpha$-BiBr.
To create a \emph{spin-resolved phase diagram} for $\alpha$-BiBr, we begin by taking $\hat{\mathbf{n}}$ to lie in the $xz$-plane, such that $\hat{\mathbf{n}} = (\hat{n}_x,0,\hat{n}_z)$.
We next again parameterize $\hat{\mathbf{n}}$ using the $(\vartheta,\phi)$ angular parameterization in SEq.~\eqref{eq:hat_n_around_y_axis}, but here fixing $\vartheta=\pi/2$ while varying $\phi$ with an angular resolution of $\Delta \phi = 0.01\pi$ [a finer resolution than the full spin (hemi-) sphere $\Delta_{s}$ calculations for $\alpha$-BiBr in SFig.~\ref{fig:bibr_numerical_spin_gap_for_different_spin_directions}].
For each $\phi$, we then perform a Nelder-Mead minimization~\cite{pandey2022py} on the direct spin gap function $\Delta_s (\mathbf{k})$ taking $100$ $\mathbf{k}$ points randomly sampled from the 3D BZ as the initial points.
Lastly, we define the \textit{numerical spin gap} $\Delta_{s=\hat{\mathbf{n}}\cdot \mathbf{s}}$ as the minimal value of the $100$ minimization results for each spin direction $\hat{\mathbf{n}}$.

In SFig.~\ref{fig:bibr_spin_resolved_phase_diagram} we plot the numerical (global) spin gap in $\alpha$-BiBr for $\hat{n}_{y}=0$ [$\vartheta=\pi/2$ in SEq.~\eqref{eq:hat_n_around_y_axis}].
Over the spin resolution circle parameterized by $\vartheta=\pi/2$ in $\alpha$-BiBr [SEq.~\eqref{eq:hat_n_around_y_axis}], there are only four very narrow regions in which the global spin gap closes within numerical precision [green regions in SFig.~\ref{fig:bibr_spin_resolved_phase_diagram}].
By computing the full spin spectrum of $\alpha$-BiBr for representative and continuous values of $\hat{\mathbf{n}}$ within the green (spin-gapless) regions in SFig.~\ref{fig:bibr_spin_resolved_phase_diagram}, we have confirmed that the spin-gapless regions correspond to spin-Weyl semimetal states.
Combined with the extensive (nested) spin-resolved Wilson loops performed in this section, we hence conclude that for $\vartheta=\pi/2$, $\alpha$-BiBr hosts four spin-gapped regions: a $\bm{\nu}^{\pm} = (0,0,\mp 2)$ 3D QSHI region centered around $\hat{\mathbf{n}}=\hat{\mathbf{z}}$, a $\bm{\nu}^{\pm} = (0,0,\pm 2)$ 3D QSHI region centered around $\hat{\mathbf{n}}=-\hat{\mathbf{z}}$, and two T-DAXI regions centered around $\hat{\mathbf{n}}=\pm\hat{\mathbf{x}}$.
$\alpha$-BiBr hence realizes for varying $\hat{\mathbf{n}}$ \emph{all three} of the spin resolutions of a helical HOTI uncovered in this work: 3D QSHI, spin-Weyl semimetal, and T-DAXI states.

\subsection{Physical Signatures of Spin-Gapped States in $\alpha$-BiBr: Bulk Spin Hall Conductivity}
\label{sec:bibrshc}

To demonstrate physical signatures of the spin-gapped states in $\alpha$-BiBr, we will in this section compute the intrinsic bulk contribution to the (non-quantized) spin Hall conductivity.  
We will specifically compute the non-quantized bulk spin Hall conductivity (per unit cell) for $s_{z}$ and $s_{x}$, and will then compare the results to the \emph{quantized} bulk spin-resolved topology.
Previously in SN~\ref{sec:bulk_spin_hall_conductivity}, we used the Kubo formula to derive the spin Hall conductivity tensor $\sigma^{s,i}_{\mu \nu}$, which we emphasize is generically nonquantized due to $s$-nonconserving SOC.
As discussed in SEq.~(\ref{eq:3dshc}) and the surrounding text, in spin-gapped states, the quantized partial weak Chern vector $\bm{\nu}^{\pm} = (\nu_1^\pm , \nu_2^\pm , \nu_3^\pm)$ indicates the bulk topological contribution to the intrinsic spin Hall conductivity per layered unit cell.
However in real materials like $\alpha$-BiBr---even those with bulk spin gap---one might expect that SOC destroys any discernible relationship between $\bm{\nu}^{\pm}$ and $\sigma^{s,i}_{\mu \nu}$.  
However as we will show below, we find remarkable agreement between $\bm{\nu}^{\pm}$ and $\sigma^{s,i}_{\mu \nu}$ in $\alpha$-BiBr for $s_{z}$ and $s_{x}$ spins.

To compute the spin Hall conductivity per unit cell of $\alpha$-BiBr, we use our implementation [SEq.~\eqref{eq:spinhallprimitive}] of the Kubo formula and focus on the $s_{z}$ and $s_{x}$ spin Hall conductivities.
Previously in SN~\ref{sec:spin_resolved_topology_of_alpha_bibr}, we found that $\alpha$-BiBr is a $\bm{\nu}^{\pm} = (0,0,\mp 2)$ 3D QSHI [$\bm{\nu}^{\pm} = \mathbf{0}$ T-DAXI] for $\hat{\mathbf{n}}=\hat{\mathbf{z}}$ [$\hat{\mathbf{n}}=\hat{\mathbf{x}}$] spin resolution direction.  By direct numerical computation, we first find that the $s_{z}$ spin Hall conductivity (per unit cell) in $\alpha$-BiBr---crucially allowing for spin-nonconserving SOC---is given by
\begin{align}
\sigma^{s,z}_{12} &= \frac{e}{4\pi}(-3.62), \nonumber \\
\sigma^{s,z}_{31} &= \frac{e}{4\pi}(-0.0001), \nonumber \\
\sigma^{s,z}_{23} &= \frac{e}{4\pi}(-0.009).
\label{eq:shc_bibr_sz_23}
\end{align}
SEq.~(\ref{eq:shc_bibr_sz_23}) shows remarkable agreement with the partial weak Chern numbers of $\alpha$-BiBr.
Specifically as discussed in SEq.~(\ref{eq:3dshc}) and the surrounding text, an $s$-conserving, $\mathcal{T}$-invariant quantum spin Hall state will carry a bulk spin Hall conductivity of $[e/(4\pi)]\times 2\nu^{+}$ (noting that $\nu^{+}=-\nu^{-}$ due to $\mathcal{T}$ symmetry).
This implies that if $\alpha$-BiBr had perfect $s_{z}$ spin symmetry, it would carry the quantized spin Hall conductivities (per cell) of $\sigma^{s,z}_{12}=[e/(4\pi)]\times -4$, $\sigma^{s,z}_{31}=\sigma^{s,z}_{23}=0$ (\emph{i.e.} the spin Hall conductivity would be entirely given by the topological contribution). 
Hence even though the $s_{z}$ spin gap in $\alpha$-BiBr $\Delta_{s_{z}}\approx 0.9309028798325673$ is only $\approx 46\%$ of the maximum possible value $\Delta_{s}=2$ [see SFig.~\ref{fig:bibr_Wannier_energy_bands_spin_sz_sx_bands}(e)], indicating the presence of non-negligible $s_{z}$-nonconserving SOC, the bulk intrinsic spin Hall conductivity (per unit cell) in SEq.~(\ref{eq:shc_bibr_sz_23}) only deviates slightly from the quantized topological contribution that originates from its nontrivial spin-resolved topology.

We next perform the analogous computation of the $s_{x}$ spin Hall conductivity per unit cell.
We find that for $\alpha$-BiBr
\begin{align}
\sigma^{s,x}_{12} &= \frac{e}{4\pi}(-0.026),\nonumber \\
\sigma^{s,x}_{31} &= \frac{e}{4\pi}(-0.00002),\nonumber \\
\sigma^{s,x}_{23} &= \frac{e}{4\pi}(-0.023).
\label{eq:shc_bibr_sx_23}
\end{align}
Previously in SN~\ref{sec:spin_resolved_topology_of_alpha_bibr}, we found that $\alpha$-BiBr realizes a $\bm{\nu}^{\pm} = \mathbf{0}$ T-DAXI for $\hat{\mathbf{n}}=\hat{\mathbf{x}}$.
In a T-DAXI state, the bulk topological contribution to the spin Hall conductivity vanishes (though the bulk nontrivial 3D partial axion angles $\theta^{\pm}=\pi$ give rise to an anomalous 2D surface spin Hall conductivity, see SN~\ref{app:response_QSHI_TDAXI}, \ref{app:layer-resolved}, and \ref{sec:3dshc}). 
Even though the $s_{x}$ spin gap in $\alpha$-BiBr is even smaller than its $s_{z}$ spin gap [$\Delta_{s_{x}}\approx 0.2550432063802285$, $\approx 12\%$ of the maximum possible value $\Delta_{s}=2$, see SFig.~\ref{fig:bibr_Wannier_energy_bands_spin_sz_sx_bands}(f)], we find that the $s_{x}$ spin Hall conductivity nevertheless nearly matches the vanishing value expected for an idealized $s_{x}$-conserving T-DAXI state.

We have hence demonstrated that for $s_{z}$ and $s_{x}$ spins, the nonquantized bulk intrinsic spin Hall conductivity in $\alpha$-BiBr lies close to the quantized topological contribution from its nontrivial spin-resolved bulk topology.  
Practically, the calculations performed in this section suggest a highly anisotropic spin Hall response in $\alpha$-BiBr that interpolates between a large, extensive bulk contribution for $s_{z}$ spin transport to a small, surface-dominated contribution for $s_{x}$ spin transport.  
Given that $\alpha$-BiBr is readily synthesizable~\cite{BiBrStructure,noguchi2021evidence, FanZahidRoomTempBiBrExp,BiBrHingeExp1,BiBrHingeExp2,BiBrFacetDependent,BiBrTempLifshitz,BiBrNanowireSubstrate,BiBrNanobelt,BiBrQuantumTransport,BiBrOpticalDichotomy}, the anisotropic spin-electromagnetic response predicted in this work should be accessible through straightforward (inverse) spin Hall measurements that are achievable within a short timeframe.

\bibliographystyle{naturemag}
\bibliography{refs}